\documentclass[useAMS,usenatbib]{mn2e}
\usepackage{subfiles}
\usepackage[english]{babel}
\usepackage{natbib}
\usepackage{graphics}
\usepackage{graphicx}
\usepackage{hyperref}
\usepackage{amsmath}
\usepackage{array}

\renewcommand{\vec}[1]{\boldsymbol{#1}}
\newcommand{\tens}[1]{\boldsymbol{\mathsf{#1}}}
\DeclareMathOperator\arctanh{arctanh}

\title[AutoLens]
{\textit{AutoLens}: Automated Modeling of a Strong Lens's Light, Mass and Source}
\author[Nightingale, Dye \& Massey]
{\parbox{\textwidth}{J. W. Nightingale,$^{1,2}$\thanks{E-mail:james.w.nightingale@durham.ac.uk}
S. Dye$,^{2}$
Richard J. Massey$^{1}$
}
\vspace{4mm}\\
$^{1}$Centre for Extragalactic Astronomy, Department of Physics, Durham University, South Road, Durham, DH1 3LE, UK\\
$^{2}$School of Physics and Astronomy, Nottingham University,
University Park, Nottingham, NG7 2RD, UK\\
}

\begin{document}

\bibliographystyle{mn2e}
\bibpunct{(}{)}{;}{a}{}{;}
\date{}
\pagerange{\pageref{firstpage}--\pageref{lastpage}} 
\pubyear{2018}
\maketitle
\label{firstpage}

\begin{abstract}
This work presents {\tt AutoLens}, the first entirely automated modeling suite for the analysis of galaxy-scale strong gravitational lenses. {\tt AutoLens} simultaneously models the lens galaxy's light and mass whilst reconstructing the extended source galaxy on an adaptive pixel-grid. The method's approach to source-plane discretization is amorphous, adapting its clustering and regularization to the intrinsic properties of the lensed source. The lens's light is fitted using a superposition of Sersic functions, allowing {\tt AutoLens} to cleanly deblend its light from the source. Single component mass models representing the lens's total mass density profile are demonstrated, which in conjunction with light modeling can detect central images using a centrally cored profile. Decomposed mass modeling is also shown, which can fully decouple a lens's light and dark matter and determine whether the two component are geometrically aligned. The complexity of the light and mass models are automatically chosen via Bayesian model comparison. These steps form {\tt AutoLens}'s automated analysis pipeline, such that all results in this work are generated without any user-intervention. This is rigorously tested on a large suite of simulated images, assessing its performance on a broad range of lens profiles, source morphologies and lensing geometries. The method's performance is excellent, with accurate light, mass and source profiles inferred for data sets representative of both existing Hubble imaging and future Euclid wide-field observations.  
\end{abstract}

\begin{keywords}
gravitational lensing - galaxies: structure
\end{keywords}

\section{INTRODUCTION}

Strong gravitational lensing offers a unique means of measuring the mass distribution and composition of galaxies within our Universe. Through the intricate analysis of a lensed source's extended light profile one can robustly infer the lens galaxy's density profile, a technique that has been exploited to provide observations in the fields of dark matter substructure \citep{Vegetti2009, Vegetti2009b, Vegetti2012, Vegetti2014}, stellar dynamics \citep{Barnabe2007, Barnabe2009, Barnabe2011} and cosmology \citep{Suyu2013, Collett2014, Suyu2016, Wong2016}. Equally, this analysis provides a full reconstruction of the highly-magnified source galaxy and therefore offers an unprecedented view of the high redshift Universe \citep{Shirazi2013, Dye2014, Dye2015, Rybak2015, Swinbank2015}. 

However, unlike the works above, the majority of strong lensing studies exploit just one lensing observable, the Einstein Mass, $M_{\rm  \rm Ein}$, which is widely accepted as a robust mass estimator that is essentially independent of the density profile assumed for the lens. $M_{\rm  Ein}$ is constrained by the first derivative of the lens's potential, therefore it is the \textit{position} of the lensed source in the image-plane that is key. Measuring $M_{\rm  Ein}$ therefore requires relatively simple lens modeling methodology (e.g. \citealt{Bolton2008, Sonnenfeld2013b}) and has already been performed on the majority of known strong lenses over the past decade \citep{Bolton2008, Koopmans2009, Brewer2012a, Dutton2012, Sonnenfeld2013c, Bolton2012, Sonnenfeld2015}.

When an extended source is lensed, light rays emanating from different regions of the source trace through different regions of the lens galaxy. Therefore, the lensed source's \textit{extended surface-brightness profile} contains a wealth of additional information, that if exploited can be used to measure the lens potential's second derivative, its density profile. Extracting this signal requires more sophisticated lens modeling capable of both reconstructing the source's intrinsic light profile and modeling the lens galaxy's mass distribution \citep{Warren2003,Dye2005,Suyu2006,Vegetti2009,Tagore2014, Birrer2015b, Tessore2016}. Unfortunately, the involved nature of extended source modeling has seen it struggle to scale up to large samples, with most analyses focusing on samples of just one to ten objects (e.g. \citealt{Dye2014, Vegetti2014, Birrer2015a, Suyu2016, Dye2017}).

The aim of this paper is to rectify this, by demonstrating a fully automated approach to extended source modeling which performs all analysis and generates all results without any user-intervention after a brief initial set up.  This is well motivated, given archival lens data-sets have several hundred HST-quality images warranting such an analysis \citep{Bolton2006, Auger2009, Sonnenfeld2013b}. Furthermore, with ongoing and future surveys such as the Large Syntopic Survey Telesccope and Euclid set to find of order one hundred thousand strong lenses \citep{Oguri2012, Collett2015}, an automated pipeline is paramount to fully exploit the expansive incoming datasets. 

This paper builds upon the adaptive semi-linear inversion method developed by \citep[][N15 hereafter]{Nightingale2015} with a new comprehensive and automated modeling process which we have named {\tt Autolens}. We test this on an extensive suite of simulated imaging, which is paramount given we are in a regime where the detailed inspection of results on an individual case-by-case basis is not feasible. This also includes data representative of Euclid imaging, thus giving first insights into the type of modeling and observations that may (and may not) be possible with direct analysis of wide-field imaging.

This work is performed using a {\tt Fortran} version of {\tt AutoLens}. A project is now underway to redevelop {\tt AutoLens} in {\tt Python} and make it publicly available and open-source software for the community. The latest status of this project can be found at {\url https://github.com/Jammy2211/PyAutoLens}.

Table \ref{table:Parameters} at the end of the script lists parameters and symbols used in this work.

This paper is structured as follows; section \ref{Overview} gives an overview of {\tt AutoLens}'s key features. Section \ref{Simulation} describes the light profiles, mass profiles and simulated images used to test {\tt AutoLens}. Section \ref{Method} presents in detail {\tt AutoLens}'s lens and source analysis, including the method's adaptive source analysis and variance scaling, which section \ref{AdaptDemo} demonstrates. Section \ref{SLPipeline} describes the method's development into an automated analysis pipeline. Section \ref{Results} demonstrates this on the simulated image suite and section \ref{Diss3} discusses the results and summarizes the paper.

\section{Overview of AutoLens}\label{Overview}

{\tt AutoLens} is described fully in section \ref{Method}. Here, an overview of the method's key features is given.

{\tt AutoLens} brings about a number of improvements over the source analysis of N15, who demonstrated the use of an adaptive pixel-grid to reconstruct the source galaxy. This computed a unique source pixelization in a completely stochastic manner for every lens model, a feature which was key to removing previously unknown systematics associated with the discrete nature of source reconstruction. {\tt AutoLens} now adapts its pixel-grid and regularization scheme to the morphology of the lensed source galaxy, in a manner that significantly improves lens modeling within the Bayesian framework of \citet{Suyu2006}. This ensures the method can handle the diverse range of strongly lensed sources that are in existing lens samples (e.g. \citep{Newton2011, Dye2015, Shu2016, Oldham2016, Enia2018}.

{\tt AutoLens} now fits the lens galaxy's light, thereby unifying the modeling of the lens's mass and light and the reconstruction of the source galaxy into one coherent framework. This is in contrast to other methods in the literature, which typically subtract the lens galaxy's light before performing lens modeling (e.g. \citealt{Bolton2006}) therefore discarding the information that it contains. Light profile fitting with {\tt AutoLens} supports both single and multi-component models and the complexity of the light model is chosen within the framework of Bayesian model comparison, ensuring an appropriate light profile is fitted for lenses of different morphological classes. This achieves a clean separation of the lens and source light (which is not possible when modeled independently \citealt{Marshall2007,Biernaux2016}) and measures the lens galaxy's light profile, a quantity routinely measured to study the structure of large samples of galaxies \citep{Hoyos2011, Bruce2012, Vulcani2014, Vika2013, Bluck2014, Bruce2014, Bruce2014b, Vika2014}.

Three approaches to mass modeling are demonstrated, the first invoking the same mass model as N15, a power-law density profile representing the lens's total mass distribution. This is the model assumed in most strong lensing works (e.g. \citet{Dye2014, Vegetti2014}) and has been fitted to over one-hundred Early-Type Galaxy (ETG) lenses from surveys such as the Sloan Lens ACS Survey (SLACS) \citep{Bolton2006, Auger2009}, the Strong Lensing in the Legacy Survey (SL2S) \citep{Sonnenfeld2013b} and the BOSS Emission-Line Lens Survey (BELLS) \citep{Brownstein2012}. This has revealed that the inner mass distribution of ETG's are accurately approximated by a nearly isothermal density profile \citep{Gavazzi2007, Koopmans2009, Barnabe2011, Sonnenfeld2013c, Bolton2012, Sonnenfeld2015} and has been termed the `bulge-halo conspiracy', given that neither the light matter component, the bulge, nor the dark component, the halo, have this profile and yet their combination conspires to produce one. However, these large samples measure only $M_{\rm  Ein}$ from the lensing data, using it as an additional constraint on stellar dynamical modeling. The extended source modeling used in this work can infer the lens's density profile without any kinematic data, offering a complementary measurement to these previous studies. {\tt AutoLens} also models the contribution of large-scale structure via an external shear term, the inclusion of which is subject to Bayesian model comparison given that incorrectly assuming a shear can potentially bias the lens model inferred \citep{Balmes2013}.

By simultaneously fitting and subtracting the lens's light, {\tt AutoLens} can potentially reveal faint features in a lensed source that previous approaches to lens analysis may have missed, due to, for example, falsely over-subtracting these features before lens modeling or masking them in the subsequent lens analysis. Therefore, {\tt AutoLens}'s second approach to mass modeling attempts to detect a source's central features, like a third or fifth image \citep{Rusin2000, Mao2000, Keeton2003} or radial arcs, by invoking the cored power-law density profile. {\tt AutoLens} is able to detect these features provided they are present in the image and sufficiently extended and again uses Bayesian model comparison to ascertain whether such features are residuals resulting from the subtraction of an overly-simplistic lens light profile or whether they are genuine lensed image components. {\tt AutoLens} thus brings a new capability of searching for central images at optical and UV wavelengths to compete with existing efforts in the submm and radio where the lens light is typically not detected \citep{Winn2003, Hezaveh2015, Quinn2016}. A promising aspect of searching at shorter wavelengths is the possibility that sources posses a flatter and more extended light profile, lessening the central image's demagnification.

Fitting of the lens's light plays another crucial role, allowing {\tt AutoLens} to advocate decomposed mass profiles which separately treat the lens's light and dark matter. The final approach to mass modeling thus incorporates the light profile into the mass model, exploiting the fact that by tracing the lens's underlying stellar mass distribution it offers additional information about approximately half of the lens's overall density profile, information which other methods omit. In doing so, {\tt AutoLens} is able to make unique measurements about both components, for example the light component's mass-to-light ratio (independent of stellar population synthesis) and the dark matter's ellipticity, as well as comparing how the two are distributed relative to one another. {\tt AutoLens} makes no assumptions about the geometric alignment of the light and dark matter, determining whether there is a positional and / or rotational offset between the components via Bayesian Model comparison, allowing {\tt AutoLens} to offer a first observational insight into the geometry of light and dark matter \citep{Navarro1991, Piontek2009, Bett2010, Sales2012, Velliscig2015, Schaller2015b, Liao2016}.

{\tt AutoLens} is tested on a large suite of simulated images chosen to rigorously test {\tt AutoLens} on a range of lens light and mass profiles, source morphologies and strong lens geometries representative of forthcoming large lens samples. Each simulated lens has two images generated, one at the resolution and signal-to-noise level of currently available HST imaging data and one where these properties are in-line with what can be anticipated from Euclid imaging. Finally, the reader should note that, although not shown in this paper, {\tt AutoLens}'s development was performed in conjunction with testing on images of real strong lenses and some of the method's design choices reflect circumstances not tested by the simulated data-set. These design choices are discussed in this work whenever relevant.

\section{SIMULATED DATA}\label{Simulation}

\subsection{Light Profiles}\label{FGModeling}

Light profiles for the lens galaxy are computed using Sersic functions, which have elliptical coordinates $\xi_{\rm l} = \sqrt{{x_{\rm l}}^2 + y_{\rm l}^2/q_{\rm l}^2}$, such that the intensity at a given coordinate is given by
\begin{equation}
\label{eqn:Sersickap}
I_{\rm  Ser} (\xi_{\rm l}) = I_{\rm  l} \exp \bigg\{ -k_{\rm l} \bigg[ \bigg( \frac{\xi_{\rm l}}{R_{\rm  l}} \bigg)^{\frac{1}{n_{\rm l}}} - 1 \bigg] \bigg\} ,
\end{equation}
which has seven parameters: $(x_{\rm l},y_{\rm l})$, the light centre, $q_{\rm l}$, the axis ratio, $\theta_{\rm l}$, the orientation angle (defined counter-clockwise from the positive x-axis), $I_{\rm l}$, the intensity at the effective radius $R_{\rm l}$ and $n_{\rm l}$, the Sersic index. $k_{\rm l}$ is a function of $n_{\rm l}$. In general, a subscript `l' signifies that a parameter belongs to the light model. The de Vaucouleurs light profile $I_{\rm  Dev} (\xi_{\rm l})$ corresponds to $n_{\rm l} = 4$ and the exponential light profile $I_{\rm  Exp} (\xi_{\rm l})$ corresponds to $n_{\rm l} = 1$, respectively. The resulting two-dimensional light profile is then convolved with the instrumental Point spread function (PSF).

Composite light models are calculated by summing individual component intensity maps. When multiple light-components are used, each component's parameters are labeled with an additional numeric subscript (e.g. $n_{\rm  l1}$, $n_{\rm  l2}$, etc.). All multi-component light models are assumed to share the same center and rotation angle. 

An adaptive over-sampling routine is used to ensure light profiles are computed in both an accurate and efficient manner. This is described in appendix \ref{AppCalc}. 
\subsection{Mass Profiles}\label{LensProfs}

To generate a lensed source each image pixel must be traced from the image-plane to the source-plane via the lens equation. This is performed using the deflection angles computed from the lens convergence profile (see below), kappa($\vec{\xi}$). Like the light profile, this is a function of the elliptical radius $\xi = \sqrt{{x}^2 + y^2/q^2}$. This has center $(x, y)$, projected axis ratio $q$ and is rotated by an angle $\theta$ defined counter-clockwise from the positive x-axis. In general, all parameters associated with the lens's total mass profile have no subscript, whereas those associated with a dark matter component have subscript `d' and a light matter component a subscript `l'. 

N15 used a Singular Power-Law Ellipsoid ($SPLE$) lens model with volume mass density profile of the form $ \rho(r) = \rho_o (r / r_o)^{-\alpha}$. $\vec{{\alpha}}_{\rm x,y}$ was computed following K01, where the lens mass normalization was given by the equivalent velocity dispersion $\sigma$. However, this parameterization is uncommon within the literature and to ease future comparison the formalism of \citet{Suyu2012} is used hereafter, where the elliptical power-law surface density is given by
\begin{equation}
\label{eqn:SPLEkap}
\kappa_{\rm  pl} (\xi) = \frac{(3 - \alpha)}{1 + q} \bigg( \frac{\theta_{\rm  E}}{\xi + S^2} \bigg)^{\alpha - 1} .
\end{equation}
Here $\theta_E$ is the model Einstein radius in arc seconds. The core radius is given by $S$, which is set to zero for singular power-law models. Profiles that include a core are referred to as `$PL\textsubscript{Core}$'. The factors $(3 - \alpha)$ and $1 + q$ rescale $\theta_E$ to give the same mass normalization for a changing density slope $\alpha$ or axis ratio $q$. The potential and deflection angles are computed from equation (\ref{eqn:SPLEkap}) using the method of \citep{Barkana1998}. The case $\alpha = 2$ again corresponds to the Singular Isothermal Ellipsoid (SIE) lens profile. Although parameterized differently, the $\kappa_{\rm  pl} (\vec{\xi})$ profile given in (\ref{eqn:SPLEkap}) and used in N15 are identical, therefore the $\sigma$-$q$-$\alpha$ degeneracy described in N15 is again present, however now between $\theta_{\rm  Ein}$, q and $\alpha$ (although the rescalings by $(3 - \alpha)$ and $1 + q$ make these degeneracies appear more orthogonal). 

The inclusion of an external shear field is supported, which introduces two additional parameters with subscript `sh', the shear strength $\gamma_{\rm  sh}$ and orientation of the semi-major axis measured counter-clockwise from east, $\theta_{\rm  sh}$. To numerically compute its deflection angles this shear requires a center, which assumes either the mass profile's center ($x$ and $y$) or dark matter's center ($x_{\rm d}$ and $y_{\rm d}$) for a decomposed model.

Decomposed mass profiles assume separate density profiles for the light and dark matter components. The light component uses the elliptical Sersic profile given by equation (\ref{eqn:Sersickap}), converting it to a mass profile as
\begin{equation}
\label{eqn:Sersickap}
\kappa_{\rm  Ser} (\xi_{\rm l}) = \Psi_{\rm l}  I_{\rm  Ser} (\xi_{\rm l}) ,
\end{equation}
therefore sharing the same parameters as the light profile, but with an additional parameter $\Psi_{\rm l}$, the mass-to-light ratio. Simulated lenses with multiple light components are generated using the same $\Psi_{\rm l}$ for each component.

The dark matter component is given by an elliptical Navarro-Frenk-White ($NFW$) profile, which represents the universal density profile predicted for dark matter halos by cosmological N-body simulations \citep{Navarro1996, Zhao1996, Navarro1997}. This has volume mass density given by 
\begin{equation}
\label{eqn:NFWrho}
\rho = \frac{\rho_{\rm s}}{(r/r_{\rm s}) (1 + r/r_{\rm s})^{2}},
\end{equation}
where $\rho_{\rm s}$ gives the halo normalization and $r_{\rm s}$ the scale radius, which is fixed throughout this work to the value $r_{\rm s} = 30$ kpc \citep{Bullock2001}. Coordinates for the $NFW$ profile are scaled by $r_{\rm s}$, giving the scaled elliptical coordinate $\eta_{\rm d} = \xi_{\rm d} / r_{\rm s}$.

Analytic solutions for the $NFW$ model are given in \citep{Golse2002} and are given by

\begin{equation}
\label{eqn:DMkap}
\kappa_{\rm  NFW} (\eta_{\rm d}) = 2 \kappa_{\rm  d} \frac{1 - \mathcal{F}(\eta_{\rm d})}{\eta_{\rm d}^2 – 1} ,
\end{equation}
where
\begin{equation}
\label{eqn:DMFunc}
\mathcal{F}(\eta_{\rm d}) = \left\{
  \begin{array}{lr}
    \frac{1}{\sqrt{\eta_{\rm d}^2 - 1}} \arctan \sqrt{\eta_{\rm d}^2 - 1} & : \eta_{\rm d} > 0 \\
    \frac{1}{\sqrt{1 - \eta_{\rm d}^2}} \arctanh \sqrt{1 - \eta_{\rm d}^2} & : \eta_{\rm d} < 0 \\
     1 & : \eta_{\rm d} = 1 
  \end{array}
\right.
\end{equation}
where $\kappa_{\rm  d}$ is related to the lens halo normalization by $\kappa_{\rm  d} = \rho_{\rm s} r_{\rm s} / \Sigma_{\rm  cr}$ and $\Sigma_{\rm cr}$ is the critical surface density. A spherical NFW profile may also be used, which removes the axis ratio $q_d$ and rotation angle $\theta_d$ as free parameters and has the title $NFWSph$.

Unlike the $PL\textsubscript{Core}$ profile, the light and dark matter profiles above do not a have a prescription to include a centrally cored profile. Therefore, the decomposed mass profiles used in this work are not equipped to produce the type of source features indicative of a cored central density, like a central image or radial arc. Cored models for both the light (e.g. cored-Sersic \citealt{Dullo2013, Dullo2014} or Nuker \citealt{Faber1997} profiles) and dark matter (e.g. a generalized NFW profile \citealt{Zhao1996}) will be considered if and when objects with central source features are detected first using the $PL\textsubscript{Core}$ profile.

The deflection of light at the center of each image pixel is computed by integrating the lens's convergence profile $\kappa$ using the equation
\begin{equation}
\label{eqn:KapDefl}
\vec{{\alpha}}_{\rm x,y} (\vec{x}) = \frac{1}{\pi} \int \frac{\vec{x} - \vec{x'}}{\left | \vec{x} - \vec{x'} \right |^2} \kappa(\vec{x'}) d\vec{x'} \, ,
\end{equation}
where $\vec{x}$ is the image-plane coordinate. The method of \citep[][K01 hereafter]{Keeton2001} is followed to compute the two-dimensional deflection angle map $\vec{{\alpha}}_{\rm x,y}$ from equation \ref{eqn:KapDefl}. An adaptive numerical integrator following the method of K01 is used to compute deflection angles and is described in appendix \ref{AppCalc}.

\subsection{Source Profiles}

The intrinsic source surface brightness profiles used in this work follow one or a summation of several elliptical Sersic functions using the elliptical radius $\xi_{\rm s} = \sqrt{{x_{\rm s}}^2 + y_{\rm s}^2/q_{\rm s}^2}$. All parameters associated with the source have the subscript `s'. The lensed image of the source is convolved with the instrumental PSF.
\subsection{Simulation Suite}
{\tt AutoLens} is tested using a suite of fifty four simulated images which are generated using the light, mass and source profiles described above. Such an extensive library of images is necessary to explore the diverse range of lensing geometries, mass profiles and lens and source morphologies that are possible in any strong lens sample, as well as for ensuring that {\tt AutoLens}'s Bayesian model comparison features (see section \ref{SLPipeline}) correctly choose the lens model complexity. The images have been chosen to span a broad range of image resolutions, signal-to-noise ratios and source morphologies, the three key attributes in determining how accurately a lens model can be constrained for a given image \citep{Lagattuta2012, Vegetti2014}. The highest quality images simulated in this work are of a comparatively lower S/N and resolution than currently available (e.g. SLACS, SL2S) strong lens observations, thus the precision of the results may be viewed as conservative.

Images are representative of either Hubble Space Telescope (HST) strong lens imaging or that which is anticipated from Euclid optical imaging. HST simulated images are generated with a pixel scale of 0.06", are convolved with a circularly symmetric Gaussian PSF of size 0.085", include read noise of $4$ e-, a flat background sky of 1500 e- and Poisson noise (including the background sky). Euclid images have a pixel scale 0.1", a PSF of size 0.125", read noise of $4$ e-, a background sky of 300 e- and Poisson noise. Details of how sky subtraction is performed are given in the next section. The signal-to-noise (S/N) ratio of the lensed source component in each image is summarized using the S/N value of its brightest pixel, which is located by scanning the model lensed source during the image simulation (after the PSF convolution step but before noise is added). The S/N of the lens's light profile is computed in an analogous way using its model light profile. In line with current lens datasets, these S/N values range between 10-50 for the source and 40-80 for the lens. By using just the brightest pixel in each component, more concentrated source and lens profiles have fewer high S/N pixels compared to flatter profiles.

Lens profiles are chosen using parameters consistent with fiducial redshifts of $z_{\rm lens} = 0.5$ for the lens and $z_{\rm src} = 1.0$ for the source. The lens light profile is generated using the adaptive oversampling routine to a fractional accuracy of $10^{-7}$. Each lensed source pixel is computed with oversampling of degree $20 \times 20$. Images are representative of a single exposure, therefore omitting dithered observing strategies and multidrizzling of images to a common frame. Thus, effects such as correlated noise are not present in the simulated data-set and are not considered in this work.

To generate the 54 simulated images, 19 different combinations of light, mass and source profiles are used, which are shown in table \ref{table:SimModels}. These lens and source models have been chosen to test specific aspects of {\tt AutoLens}. Their model names reflect these chosen aspect (e.g. the $\textbf{LensMassShear}$ model tests mass modeling with an external shear). Each model is then used to generate multiple images at different image resolutions and S/N ratios, with table \ref{table:SimModels} also listing the images generated from each model. An image is then referred to by its model name and a subscript describing the image properties as follows; an `H' or `E' for HST or Euclid resolution, `S\#\#' for the source S/N, `L\#\#' for the lens S/N and either `Disk', `Bulge', `BD' (Bulge-Disk), `Cusp' or `Multi' (multiple sources) to describe the source morphology. The suite of images can be characterized in more detail as follows:
\begin{itemize}
\item Twenty four images generated without a lens light component (`NL' in tag replacing `L\#\#'), aimed primarily at testing {\tt AutoLens}'s source analysis. Of these twenty-four images, four unique $SPLE$ lens models are used. Therefore, each lens model comprises six images with a S/N of $50$, $30$, $10$ and representing either HST of Euclid imaging. Example image names are $\textbf{SrcBD}_{\rm  HS30NLBD}$ (tests modeling a source with a bulge-disk morphology) or $\textbf{SrcDisk}_{\rm  ES10NLDisk}$ (tests a source with only a disk). 

\item Fourteen images generated with a lens light component and a $SPLE$ mass profile, aimed at testing lens light modeling for different lens and source morphologies. Seven unique lens models are used, using sources with S/N ranging from 20-30, lenses with S/N ranging from 40-80 and with every model generating a Hubble and Euclid resolution image. Included in this set are images which test a shear component in the mass model, multi-component light profiles and how light modeling fares with either cuspy or flat source morphologies. Example image names are $\textbf{LensSrcBulge}_{\rm  HS25L50BD}$ (tests light modeling with a bulge-like source) and $\textbf{LensMassShear}_{\rm  ES50L80Flat}$ (tests shear modeling).

\item Six images generated with a lens light component and a $PL\textsubscript{Core}$ mass profile, aimed at testing cored-mass modeling and central image detection. Three unique lens models are used, using sources with S/N ranging from 20-30 and lenses ranging from 40-80, with every model again generating a Hubble and Euclid image. Example image names are $\textbf{CoreSrcDisk}_{\rm  HS35L70Disk}$ (tests cored modeling for a flat source profile) and $\textbf{CoreSrcDouble}_{\rm  ES25L50BD}$ (tests cored modeling for a doubly imaged source).

\item Ten images generated using a decomposed mass model, aimed at testing light and dark matter modeling as well as the detection of light / dark matter alignments. Five unique lens models are used, generating sources and lenses with the same S/N ranges as before and with every model again being used to generate a Hubble and Euclid image. Included in this set are images which test a small rotational offset and aligned light and dark components. Example image names are $\textbf{LMDMRot}_{\rm  HS30L40BD}$ (tests the detection of a rotational offset) and $\textbf{LMDMShear}_{\rm  ES25L80Cusp}$ (tests decomposed modeling with an external shear).
 \end{itemize}
\begin{figure*}
\centering
\includegraphics[width=0.24\textwidth]{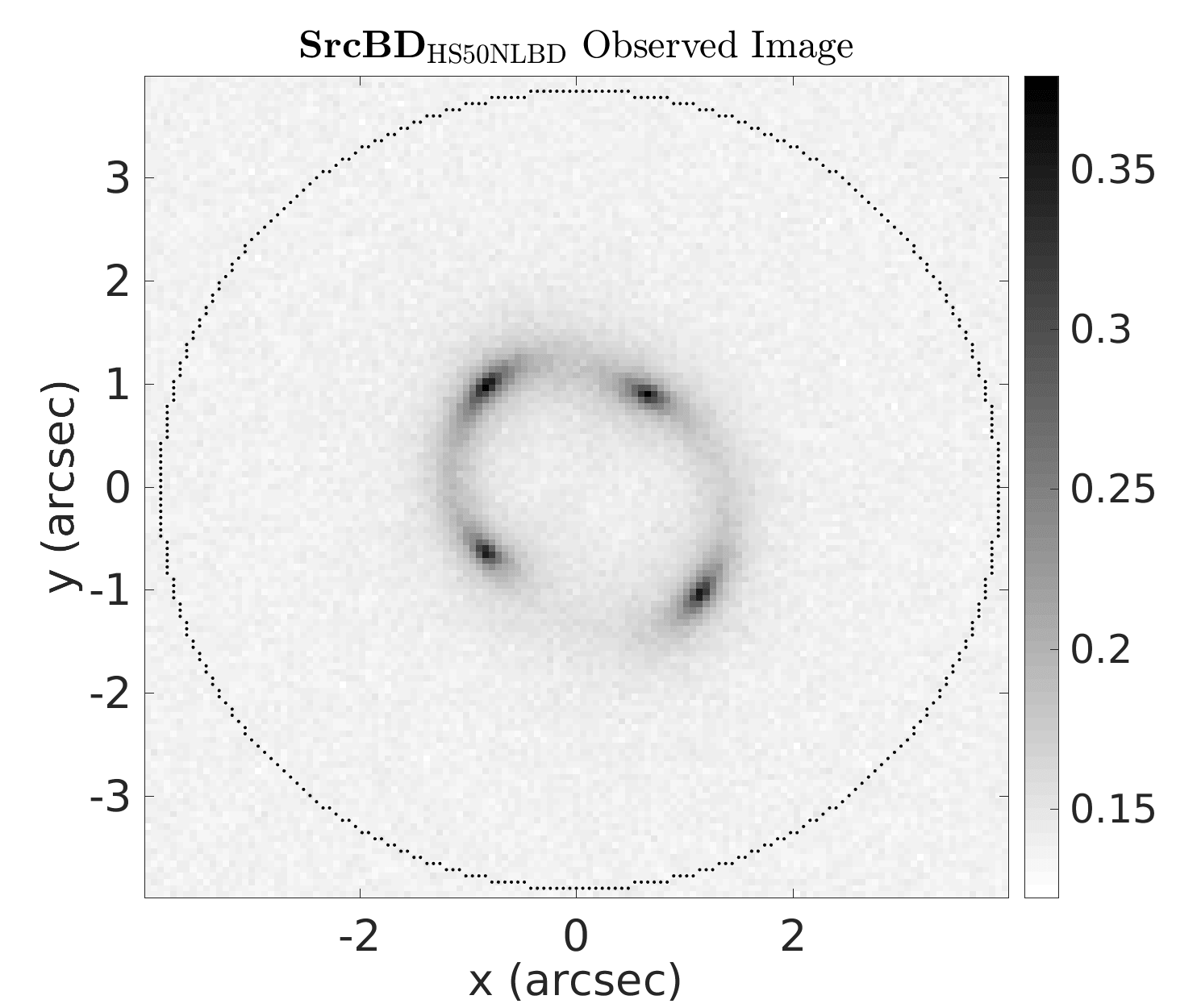}
\includegraphics[width=0.24\textwidth]{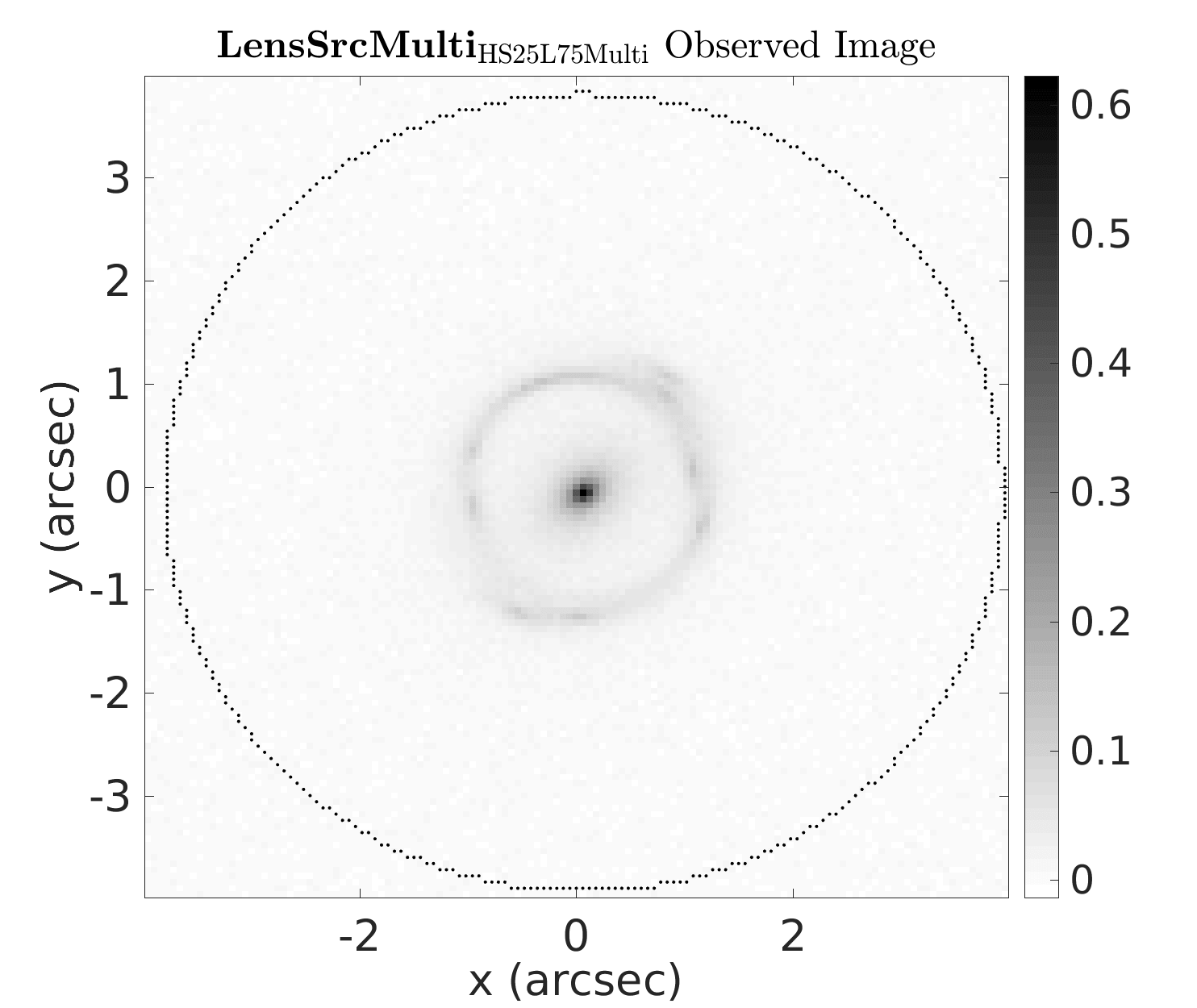}
\includegraphics[width=0.24\textwidth]{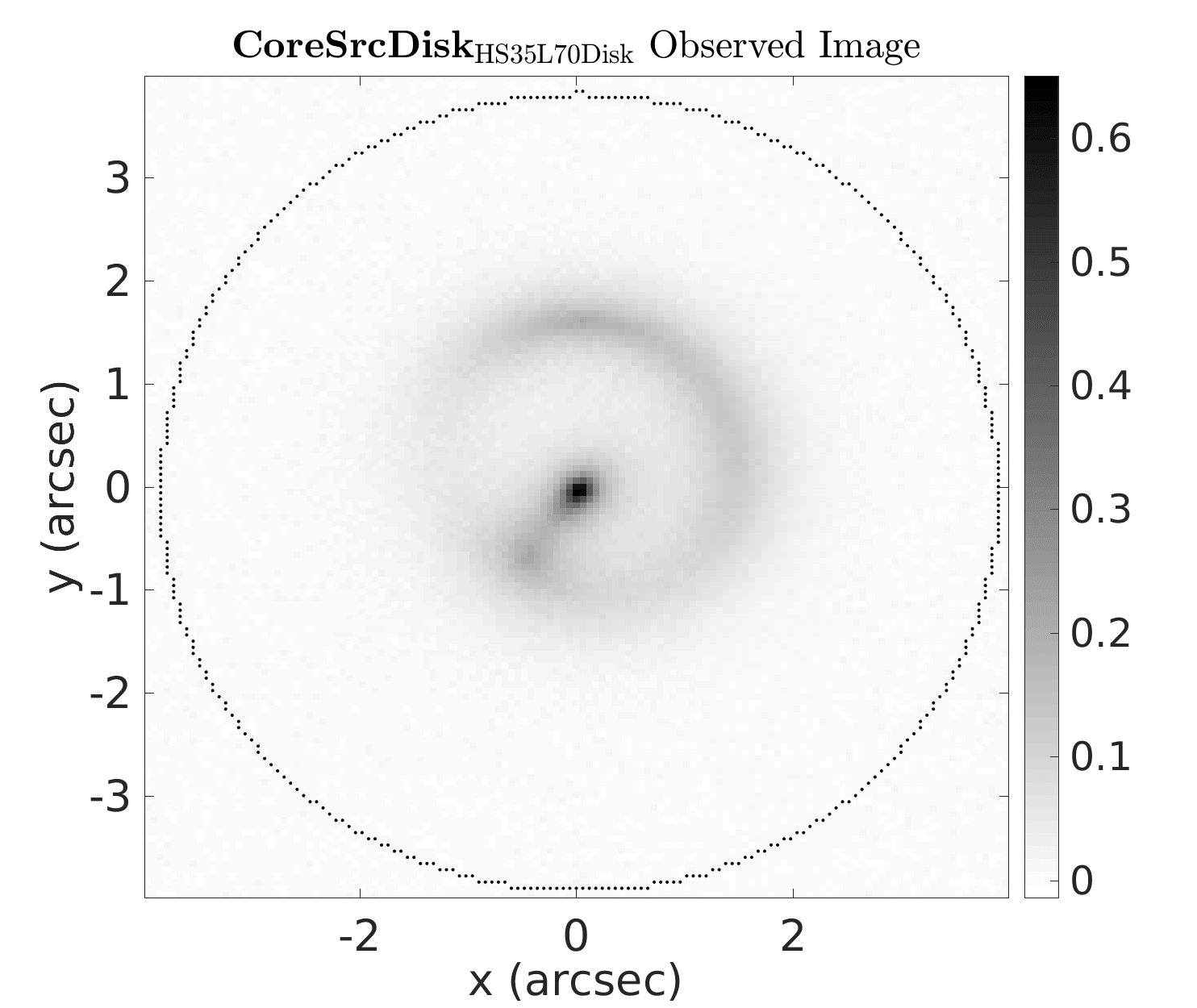}
\includegraphics[width=0.24\textwidth]{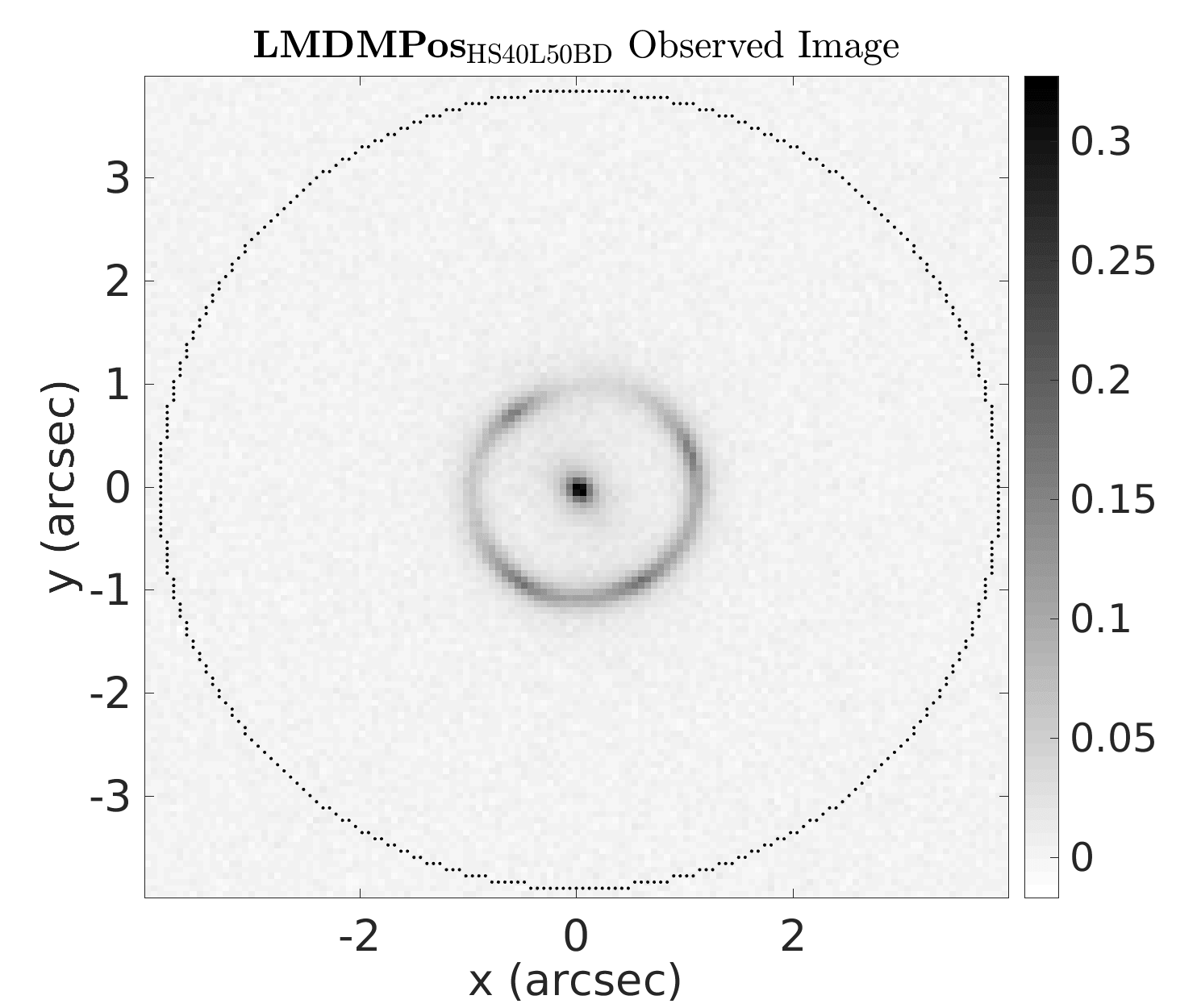}
\includegraphics[width=0.24\textwidth]{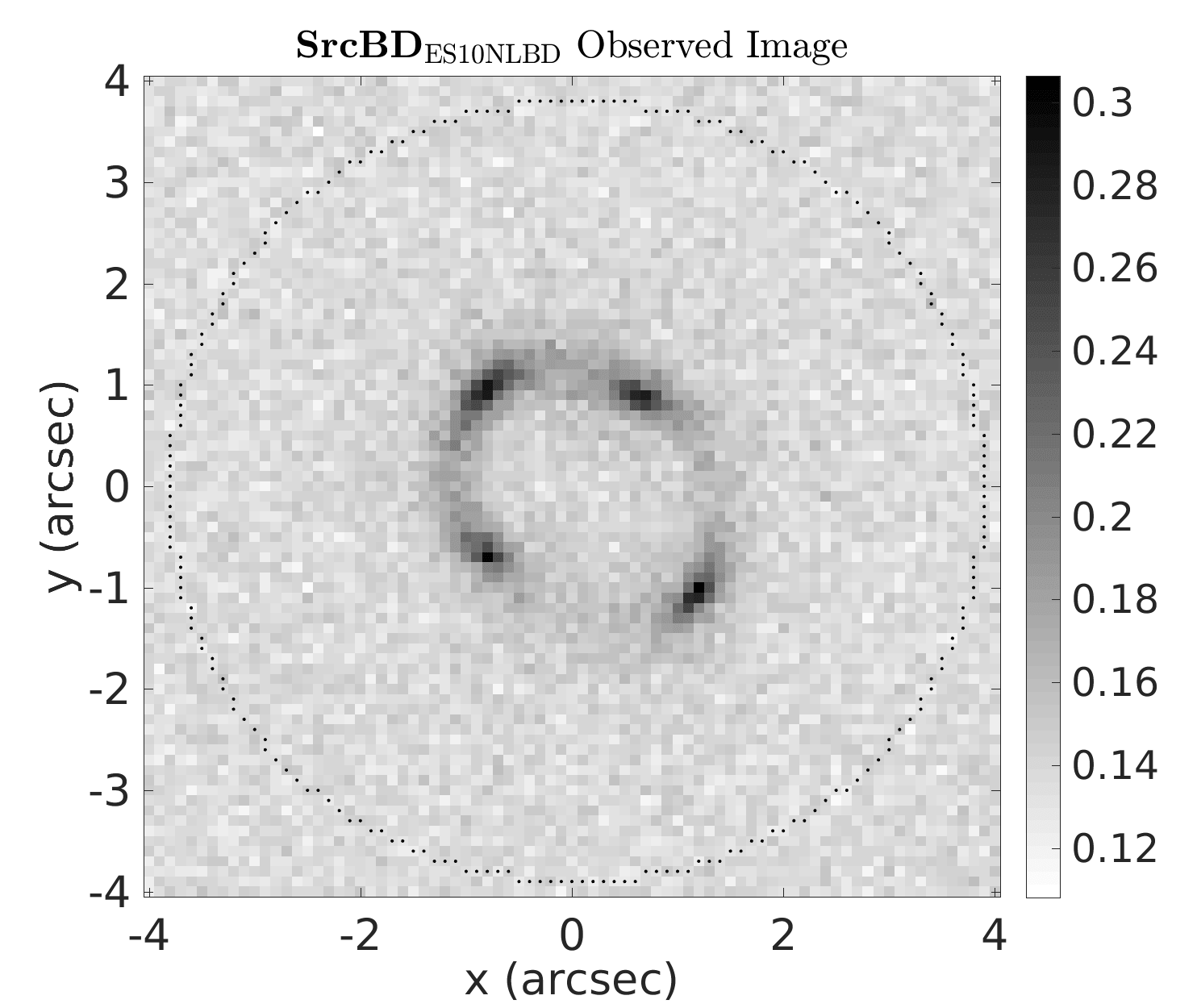}
\includegraphics[width=0.24\textwidth]{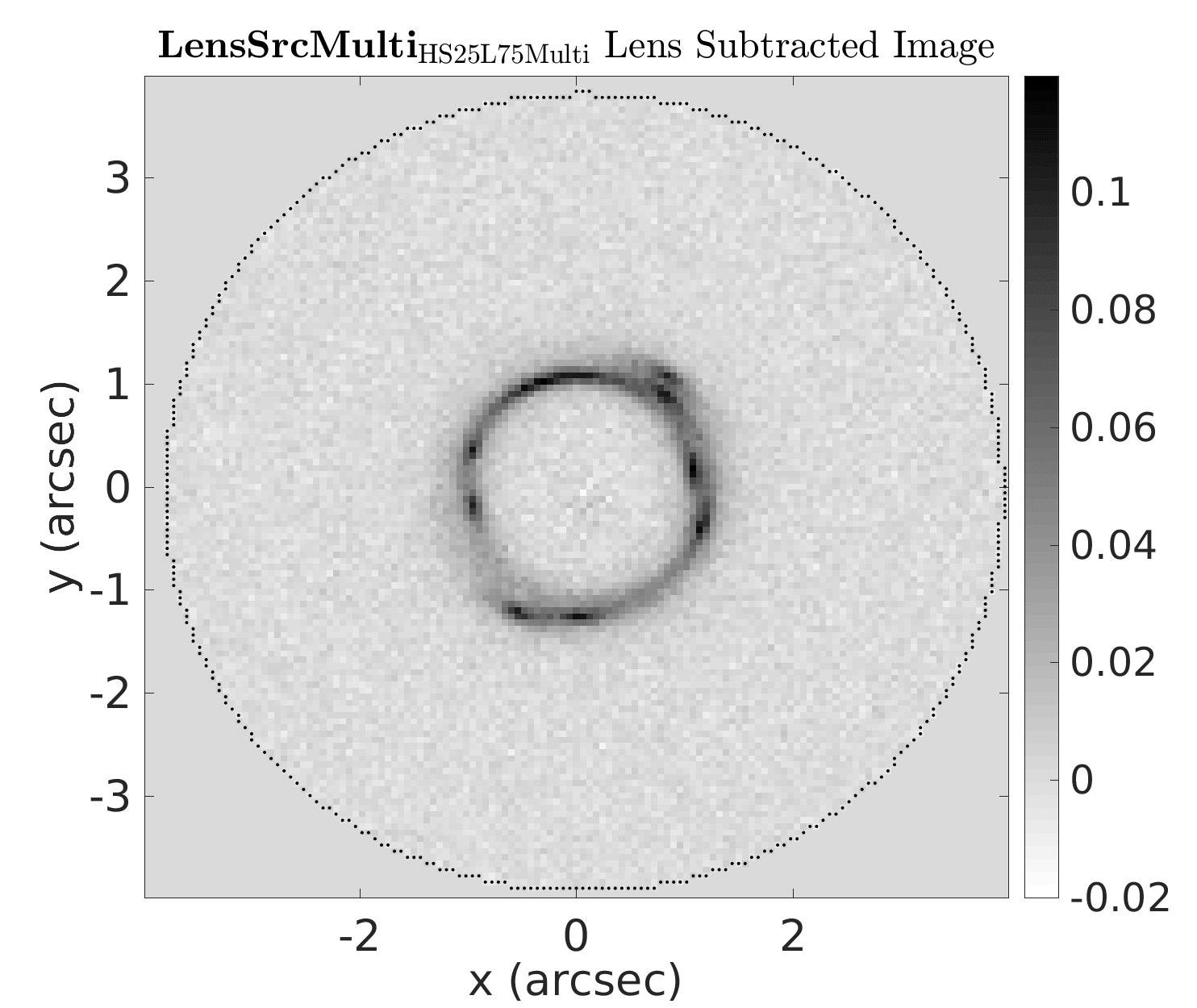}
\includegraphics[width=0.24\textwidth]{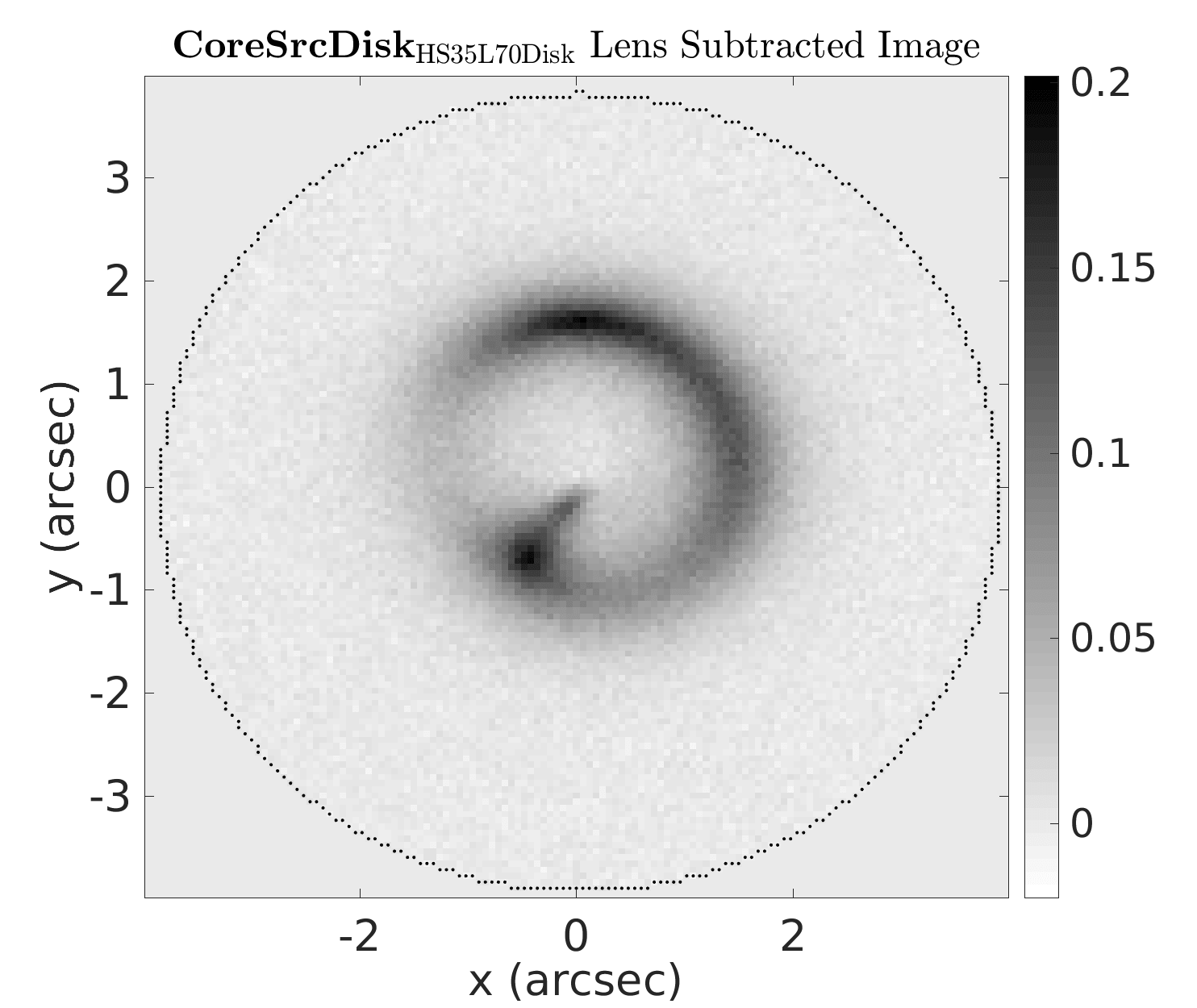}
\includegraphics[width=0.24\textwidth]{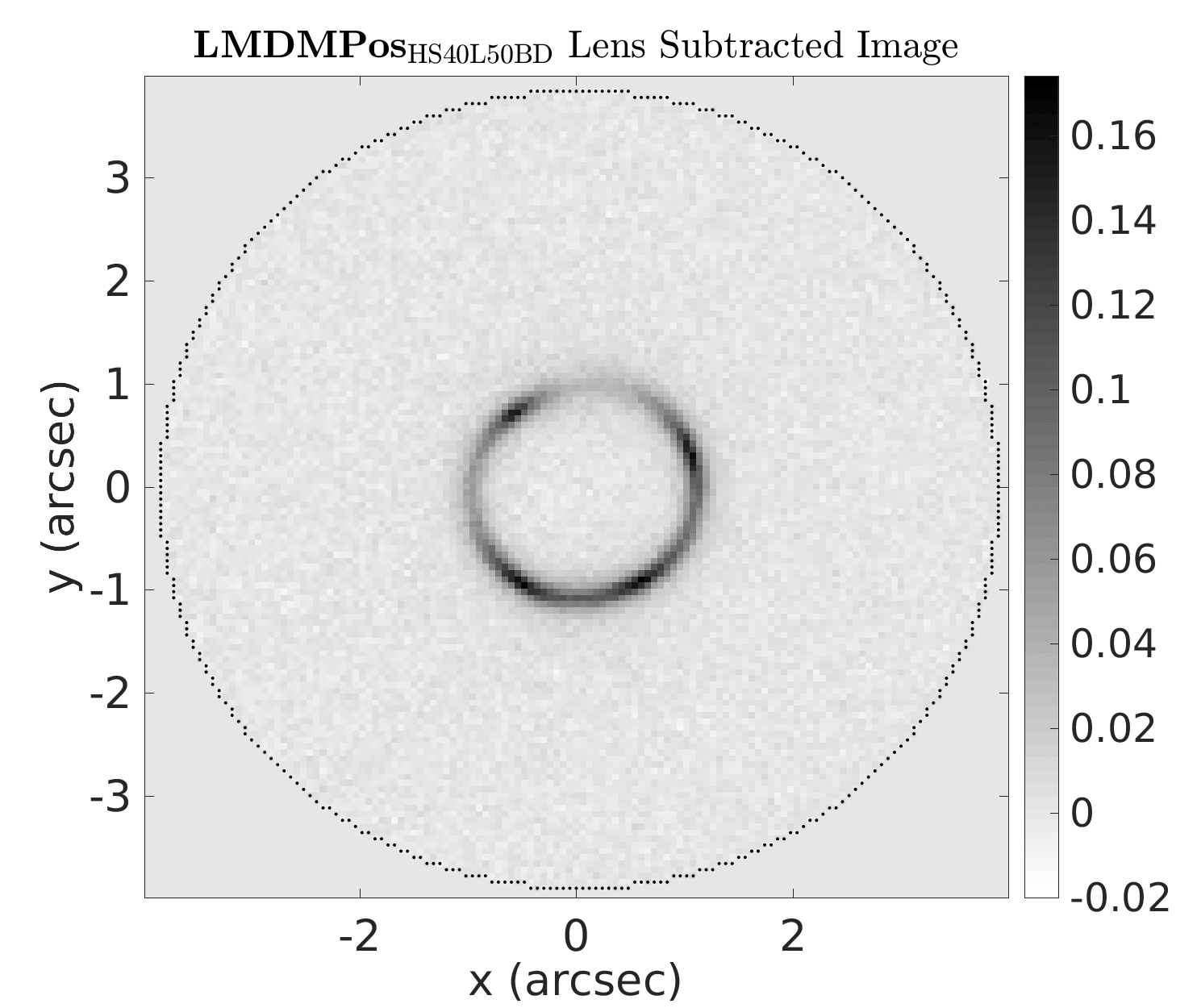}
\includegraphics[width=0.24\textwidth]{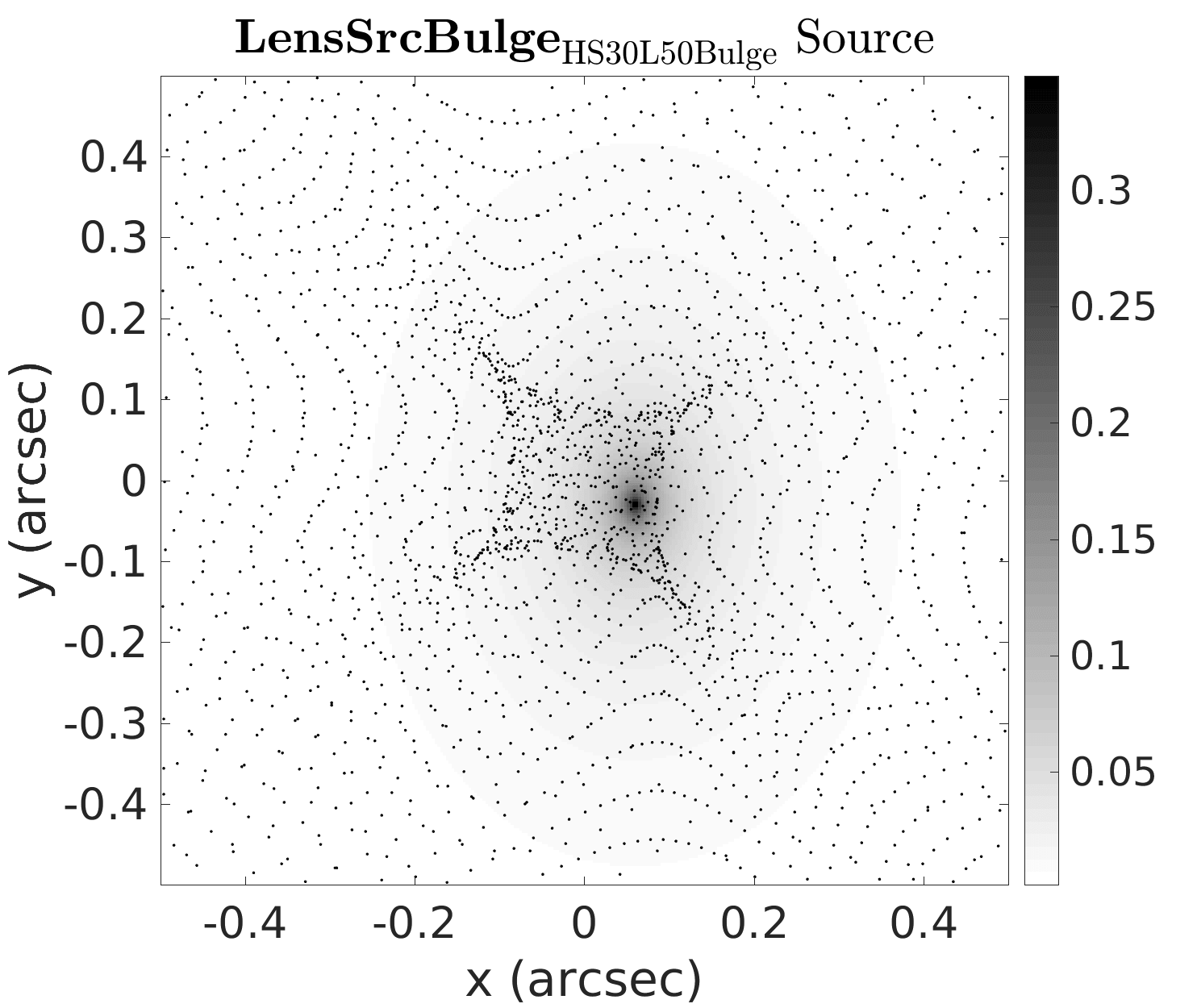}
\includegraphics[width=0.24\textwidth]{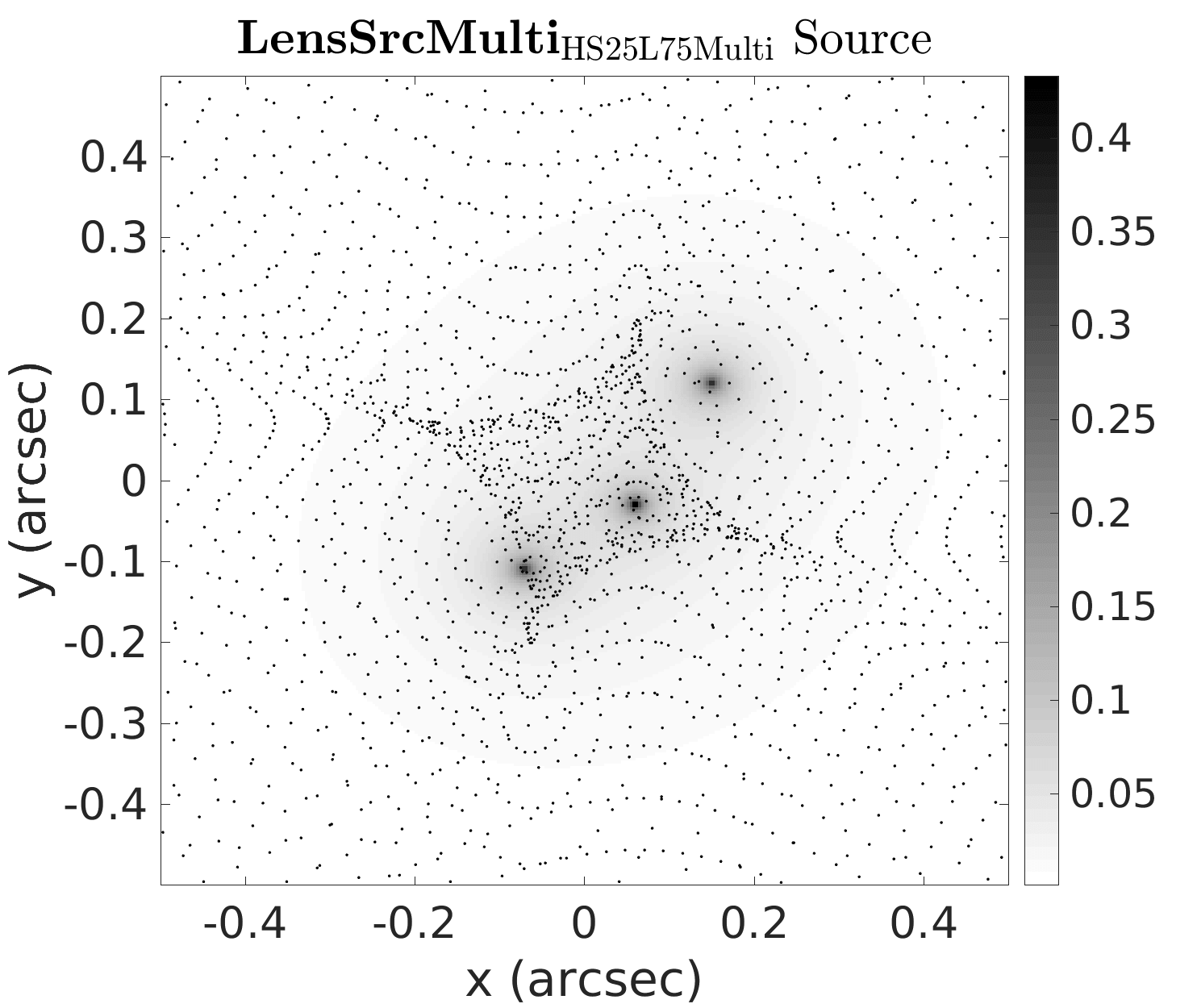}
\includegraphics[width=0.24\textwidth]{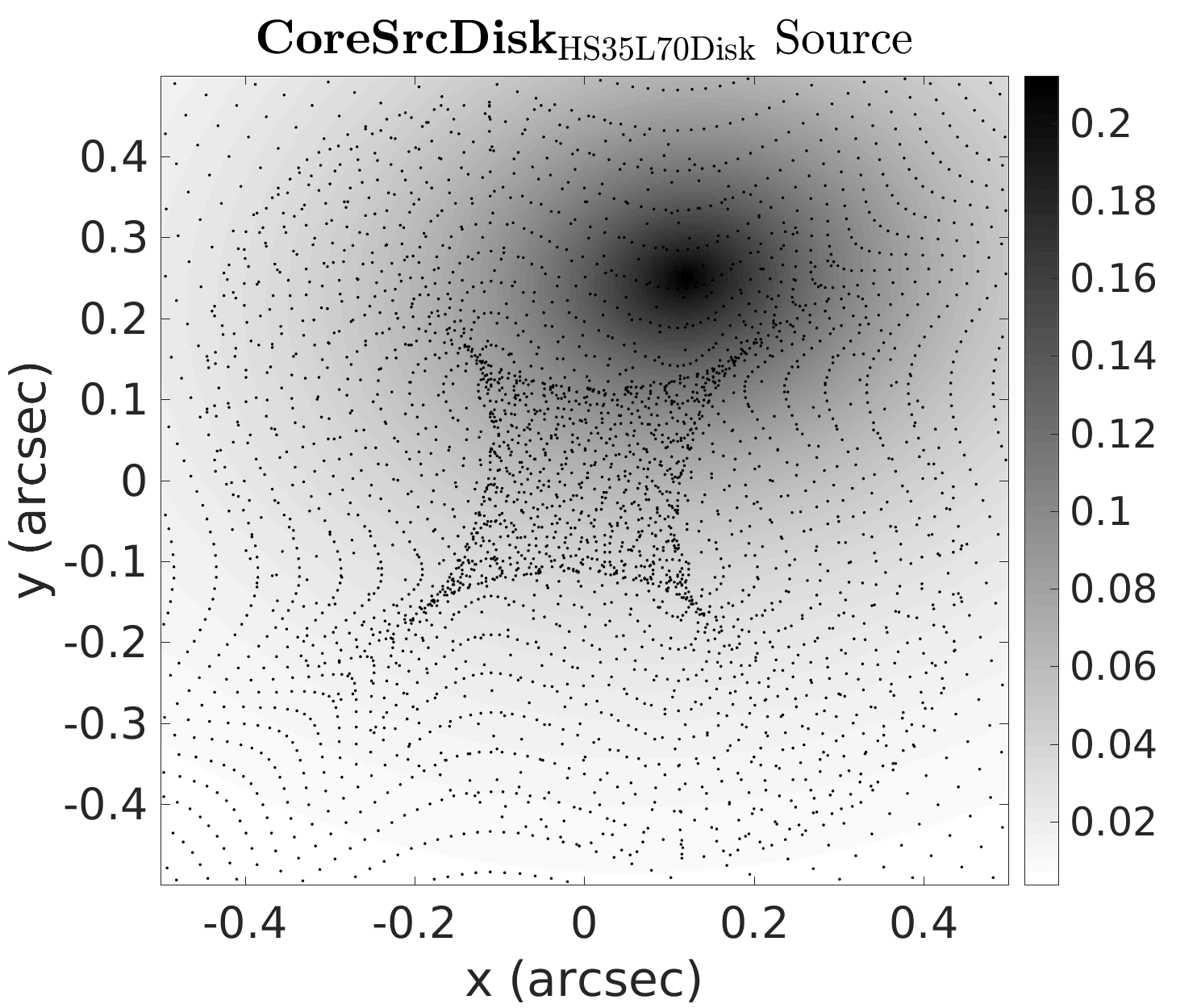}
\includegraphics[width=0.24\textwidth]{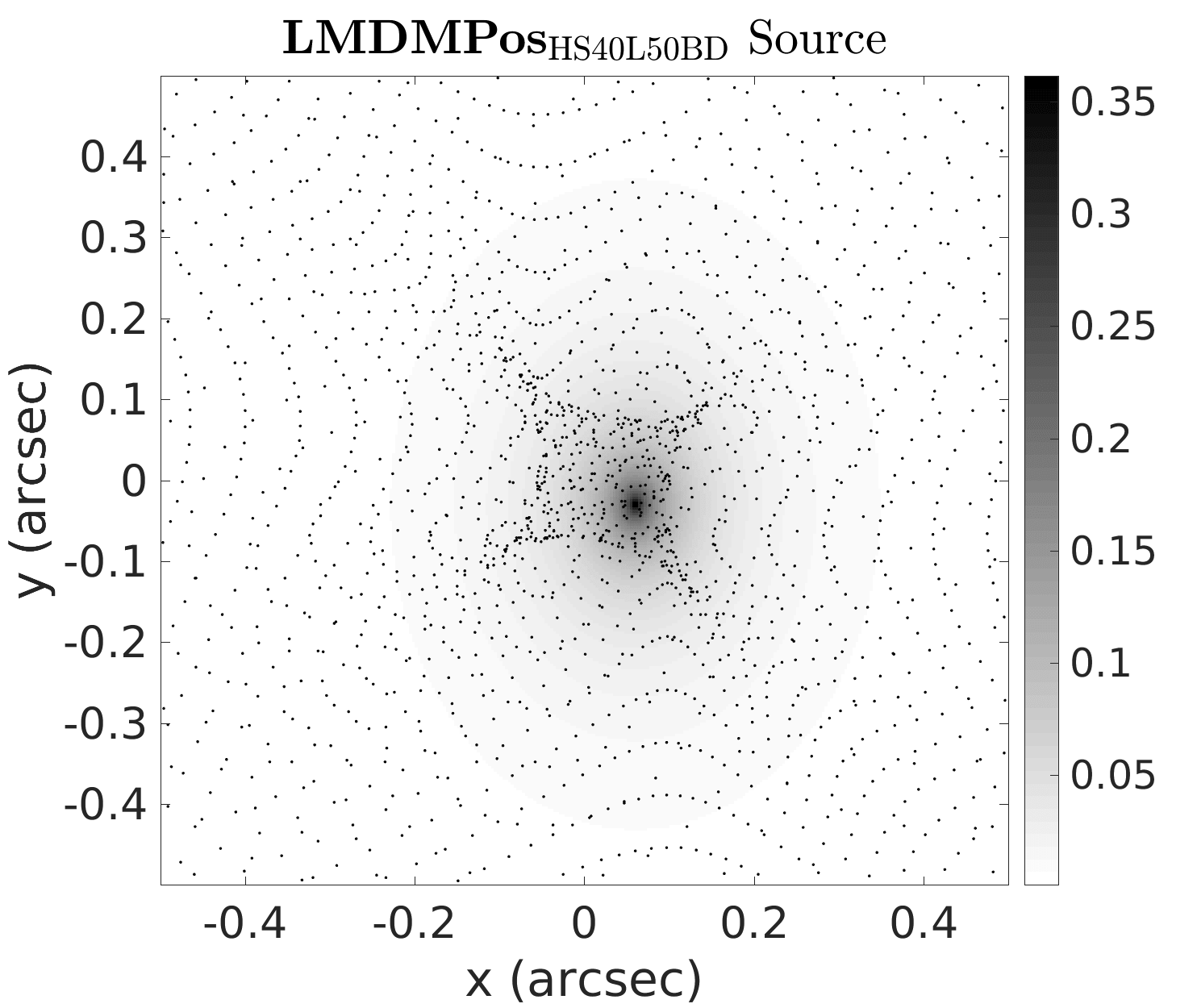}
\caption{A sub-set of simulated images corresponding to the images $\textbf{SrcBD}_{\rm  HS50NLBulge}$ (top left panel), $\textbf{SrcBD}_{\rm  ES10NLBulge}$ (middle left panel), $\textbf{LensSrcMulti}_{\rm  HS25L75Multi}$ (second column), $\textbf{CoreSrcDisk}_{\rm  HS35L70Disk}$ (third column) and $\textbf{LMDMPos}_{\rm  HS50L40BD}$ (fourth column), which are given by the third, ninth, fourteenth and seventeenth models listed in table \ref{table:SimModels} respectively. The top row shows the simulated images, the middle row lens subtracted images (or a second simulated image at lower resolution and S/N in the left column) and third rows the simulated parametric source of each image, where black points are the traced image pixel coordinates.} 
\label{figure:SimNL}
\end{figure*}

\begin{table*}
\tiny
\resizebox{\linewidth}{!}{
\begin{tabular}{ l | l | l | l l l l l l l } 
\multicolumn{1}{p{1.3cm}|}{\centering \textbf{Model} \\ \textbf{Title}} 
& \multicolumn{1}{p{1.1cm}|}{\centering \textbf{Comp- \\ onent}} 
& \multicolumn{1}{p{1.2cm}|}{\centering \textbf{Model}} 
& \multicolumn{1}{p{1.2cm}}{\textbf{Parameters}} 
& \multicolumn{1}{p{1.1cm}}{} 
& \multicolumn{1}{p{1.1cm}}{} 
& \multicolumn{1}{p{1.1cm}}{} 
& \multicolumn{1}{p{1.1cm}}{} 
& \multicolumn{1}{p{1.1cm}}{} 
& \multicolumn{1}{p{1.1cm}}{} 
\\ \hline
& & & &  & & & & & \\[-4pt]
$\textbf{SrcBulge}$                             & Light  & None     & & & & & & &  \\[2pt]
\tiny HS50NLBulge, HS30NLBulge                  & Mass   & $SPLE$     & $x = 0.00"$     & $y =  0.00"$     & $\theta = 127\,^{\circ}$     & $\theta_{\rm  E} = 1.2$    &  $q = 0.8$       & $\alpha = 2.0$    & \\[2pt]
\tiny HS10NLBulge, ES50NLBulge                  & Source & $Sersic$   & $x_{\rm  s} = 0.06"$ & $y_{\rm  s} =  -0.03"$& $\theta_{\rm  s} = 90\,^{\circ}$  & $I_{\rm  s} = 0.015$        &  $R_{\rm  s} = 0.4"$  & $n_{\rm s} = 2.5$       & $q_{\rm s} = 0.7$   \\[2pt]
\tiny ES30NLBulge, ES10NLBulge                  & &  & & & & & & &  \\[2pt]
\hline
& & & & & & & & & \\[-4pt]
$\textbf{SrcDisk}$                                  & Light  & None      & & & & & & &  \\[2pt]
\tiny HS50NLDisk, HS30NLDisk                        & Mass   & $SPLE$      & $x = 0.00"$      & $y =  0.00"$     & $\theta = 75\,^{\circ}$     & $\theta_{\rm  E} = 1.2$   &  $q = 0.75$      & $\alpha = 2.3$ & \\[2pt]
\tiny HS10NLDisk, ES50NLDisk                        & Source & $Sersic$    & $x_{\rm  s} = 0.02"$ & $y_{\rm  s} =  0.6"$  & $\theta_{\rm  s} = 10\,^{\circ}$ & $I_{\rm  s} = 0.033$        &  $R_{\rm  s} = 0.9"$ & $n_{\rm s} = 1.0$       & $q_{\rm s} = 0.80$   \\[2pt]
\tiny ES30NLDisk, ES10NLDisk                        & &  & & & & & & &  \\[2pt]
\hline
& & & & & & & & & \\[-4pt]
$\textbf{SrcBD}$                                & Light    & None      & & & & & & &  \\[2pt]
\tiny HS50NLBD, HS30NLBD                        & Mass     & $SPLE$      & $x = 0.00"$     & $y =  0.00"$       & $\theta = 45\,^{\circ}$     & $\theta_{\rm  E} = 1.0$    &  $q = 0.7$      & $\alpha = 1.7$  & \\[2pt]
\tiny HS10NLBD, ES50NLBD                        & Source 1 & $Sersic$    & $x_{\rm  s1} = 0.06"$ & $y_{\rm  s1} =  -0.03"$  & $\theta_{\rm  s1} = 0\,^{\circ}$  & $I_{\rm  s1} = 0.006$    &  $R_{\rm  s1} = 0.4"$ & $n_{\rm  s1} = 2.5$ & $q_{\rm  s1} = 0.7$   \\[2pt]
\tiny ES30NLBD, ES10NLBD                        & Source 2 & $Sersic$    & $x_{\rm  s2} = 0.06"$ & $y_{\rm  s2} =  -0.03"$  & $\theta_{\rm  s2} = 0\,^{\circ}$  & $I_{\rm  s2} = 0.0003$    &  $R_{\rm  s2} = 0.9"$ & $n_{\rm  s2} = 1.0$ & $q_{\rm  s2} = 0.80$   \\[2pt]
\hline
& & & & & & & & & \\[-4pt]
$\textbf{SrcMulti}$                             & Light    & None      & & & & & & &  \\[2pt]
\tiny HS50NLMulti, HS30NLMulti                  & Mass     & $SPLE$     & $x = 0.0"$     & $y =  0.0"$       & $\theta = 160\,^{\circ}$     & $\theta_{\rm  E} = 1.0$  &  $q = 0.75$      & $\alpha = 2.1$      & \\[2pt]
\tiny HS10NLMulti, ES50NLMulti                  & Source 1 & $Sersic$   & $x_{\rm  s1} = 0.06"$ & $y_{\rm  s1} =  -0.03"$ & $\theta_{\rm  s1} = 0\,^{\circ}$ & $I_{\rm  s1} = 0.0013$      &  $R_{\rm  s1} = 0.3"$ & $n_{\rm  s1} = 4.0$      & $q_{\rm  s1} = 0.9$   \\[2pt]
\tiny ES30NLMulti, ES10NLMulti                  & Source 2 & $Sersic$   & $x_{\rm  s2} = 0.06"$ & $y_{\rm  s2} =  -0.03"$ & $\theta_{\rm  s2} = 0\,^{\circ}$ & $I_{\rm  s2} = 0.0003$     &  $R_{\rm  s2} = 0.9"$ & $n_{\rm  s2} = 1.0$      & $q_{\rm  s2} = 0.9$   \\[2pt]
                                                & Source 3 & $Sersic$   & $x_{\rm  s3} = 0.15"$ & $y_{\rm  s3} =  0.12"$  & $\theta_{\rm  s3} = 0\,^{\circ}$ & $I_{\rm  s3} = 0.0036$      &  $R_{\rm  s3} = 0.3"$ & $n_{\rm  s3} = 3.0$      & $q_{\rm  s3} = 0.9$   \\[2pt]
                                                & Source 4 & $Sersic$   & $x_{\rm  s4} = -0.07"$& $y_{\rm  s4} =  -0.11"$ & $\theta_{\rm  s4} = 0\,^{\circ}$ & $I_{\rm  s4} = 0.0036$      &  $R_{\rm  s4} = 0.3"$ & $n_{\rm  s4} = 3.0$      & $q_{\rm  s4} = 0.9$   \\[2pt]
\hline
& & & & & & & & & \\[-4pt]
$\textbf{LensSrcBulge}$        & Light  & $Sersic$    & $x_{\rm  l} = 0.00"$ & $y_{\rm  l} =  0.00"$  & $\theta_{\rm  l} = 127\,^{\circ}$  & $I_{\rm  l} = 0.0085$         &  $R_{\rm  l} = 0.6"$ & $n_{\rm  l} = 4.0$  & $q_{\rm  l} = 0.72$ \\[2pt]
\tiny HS30L50Bulge             & Mass   & $SPLE$      & $x = 0.00"$     & $y =  0.00"$      & $\theta = 127\,^{\circ}$      & $\theta_{\rm  E} = 1.2$    &  $q = 0.8$      & $\alpha = 2.0$ &   \\[2pt]
\tiny ES30L50Bulge             & Source 1 \& 2 & $Sersic$ & \multicolumn{7}{p{11cm}}{Identical to $\textbf{SrcBulge}$} \\[2pt]
\hline
& & & & & & & & & \\[-4pt]
$\textbf{LensSrcDisk}$         & Light  & $Sersic$    & $x_{\rm  l} = 0.00"$ & $y_{\rm  l} =  0.00"$  & $\theta_{\rm  l} = 127\,^{\circ}$  & $I_{\rm  l} = 0.0085$      &  $R_{\rm  l} = 0.6"$ & $n_{\rm  l} = 4.0$  & $q_{\rm  l} = 0.72$ \\[2pt]
\tiny HS30L50Disk             & Mass   & $SPLE$      & $x = 0.00"$     & $y =  0.00"$      & $\theta = 127\,^{\circ}$      & $\theta_{\rm  E} = 1.2$    &  $q = 0.75$      & $\alpha = 2.3$ &   \\[2pt]
\tiny ES30L50Disk             & Source 1 \& 2 & $Sersic$ & \multicolumn{7}{p{11cm}}{Identical to $\textbf{SrcDisk}$} \\[2pt]
\hline
& & & & & & & & & \\[-4pt]
$\textbf{LensSrcCusp}$ & Light   & $Sersic$  & $x_{\rm  l} = -0.03"$ & $y_{\rm  l} =  0.04"$   & $\theta_{\rm  l} = 30\,^{\circ}$  & $I_{\rm  l} = 0.1377$     &  $R_{\rm  l} = 1.2"$  & $n_{\rm  l} = 1.25$   & $q_{\rm  l} = 0.6$ \\[2pt]
\tiny HS20L60Cusp                & Mass     & $SPLE$    & $x = -0.03"$     & $y =  0.04"$       & $\theta = 30\,^{\circ}$      & $\theta_{\rm  E} = 1.4$   &  $q = 0.7$       & $\alpha = 2.35$ &  \\[2pt]
\tiny ES20L60Cusp                & Source 1 & $Sersic$ & $x_{\rm  s1} = 0.07"$ & $y_{\rm  s1} =  0.04"$  & $\theta_{\rm  s1} = 125\,^{\circ}$  & $I_{\rm  s1} = 0.001$    &  $R_{\rm  s1} = 0.4"$ & $n_{\rm  s1} = 4.0$ & $q_{\rm  s1} = 0.8$   \\[2pt]
                                 & Source 2 & $Sersic$ & $x_{\rm  s2} = 0.07"$ & $y_{\rm  s2} =  0.04"$  & $\theta_{\rm  s2} = 125\,^{\circ}$  & $I_{\rm  s2} = 0.00004$    &  $R_{\rm  s2} = 0.8"$ & $n_{\rm  s2} = 1.0$ & $q_{\rm  s2} = 0.7$   \\[2pt]
\hline
& & & & & & & & & \\[-4pt]
$\textbf{LensSrcDouble}$   & Light   & $Sersic$ & $x_{\rm  l} = -0.03"$ & $y_{\rm  l} =  -0.08"$   & $\theta_{\rm  l} = 127\,^{\circ}$  & $I_{\rm  l} = 0.024$      &  $R_{\rm  l} = 0.9"$  & $n_{\rm  l} = 2.0$   & $q_{\rm  l} = 0.8$ \\[2pt]
\tiny HS25L60BD              & Mass    & $SPLE$   & $x = 0.03"$      & $y =  -0.08"$       & $\theta = 127\,^{\circ}$      & $\theta_{\rm  E} = 1.2$    &  $q = 0.75$      & $\alpha = 2.1$ &  \\[2pt]
\tiny ES25L60BD              & Source 1 \& 2 & $Sersic$ & \multicolumn{7}{p{11cm}}{Identical to $\textbf{SrcBD}$ except $x_{\rm  s1} = x_{\rm  s2} = 0.25"$ and $y_{\rm  s1} = y_{\rm  s2} = 0.15"$} \\[2pt]
\hline
& & & & & & & & & \\[-4pt]
$\textbf{LensSrcMulti}$ & Light    & $Sersic$   & $x_{\rm  l} = 0.04"$ & $y_{\rm  l} =  -0.03"$   & $\theta_{\rm  l} = 45\,^{\circ}$ & $I_{\rm  l} = 0.017$    &  $R_{\rm  l} = 0.8"$  & $n_{\rm l} = 3.0$         & $q_{\rm l} = 0.7$   \\[2pt]
\tiny HS25L75BD         & Mass     & $SPLE$     & $x = 0.0"$     & $y =  0.0"$       & $\theta = 160\,^{\circ}$     & $\theta_{\rm  E} = 1.0$  &  $q = 0.75$      & $\alpha = 2.1$      & \\[2pt]
\tiny ES25L75BD         & Source 1, 2, 3 \& 4 & $Sersic$ & \multicolumn{7}{p{11cm}}{Identical to $\textbf{SrcMulti}$} \\[2pt]\hline
& & & & & & & & & \\[-4pt]
$\textbf{LensMassShear}$   & Light   & $Sersic$    & $x_{\rm  l} = 0.03"$ & $y_{\rm  l} =  0.05"$   & $\theta_{\rm  l} = 60\,^{\circ}$ & $I_{\rm  l} = 0.02$                   &  $R_{\rm  l} = 1.5"$ & $n_{\rm  l} = 2.5$ & $q_{\rm  l} = 0.6$ \\[2pt]
\tiny HS40L80Disk              & Mass    & $SPLE$      & $x = 0.03"$      & $y =  0.05"$      & $\theta = 60\,^{\circ}$     & $\theta_{\rm  E} = 1.15$                 &  $q = 0.95$      & $\alpha = 1.92$ &  \\[2pt]
\tiny ES40L80Disk              & Mass    & $Shear$     & $x_{\rm  sh} = 0.03"$   & $y_{\rm  sh} =  0.05"$                             & $\theta_{\rm  sh} = 40\,^{\circ}$      & $\gamma_{\rm  sh} = 0.03$   &  & &  \\[2pt]
                       & Source  & $Sersic$    & $x_{\rm  s} = 0.06"$ & $y_{\rm  s} =  -0.07"$& $\theta_{\rm  s} = 30\,^{\circ}$ & $I_{\rm  s} = 0.016$        &  $R_{\rm  s} = 0.5"$ & $n_{\rm s} = 1.0$       & $q_{\rm s} = 0.6$   \\[2pt]
\hline
& & & & & & & & & \\[-4pt]
$\textbf{LensLightBD}$       & Light 1  & $Dev$    & $x_{\rm  l1} = 0.00"$ & $y_{\rm  l1} =  0.00"$ & $\theta_{\rm  l1} = 90\,^{\circ}$    & $I_{\rm  l1} = 0.012$    &  $R_{\rm  l1} = 0.4"$ & $n_{\rm  l1} = 3.0$ & $q_{\rm  l1} = 0.74$ \\[2pt]
\tiny HS25L50BD              & Light 2  & $Exp$    & $x_{\rm  l2} = 0.00"$ & $y_{\rm  l2} =  0.00"$ & $\theta_{\rm  l2} = 90\,^{\circ}$    & $I_{\rm  l2} = 0.026$    &  $R_{\rm  l2} = 1.15"$ & $n_{\rm  l2} = 1.0$ & $q_{\rm  l2} = 0.8$  \\[2pt]
\tiny ES25L50BD              & Mass     & $SPLE$   & $x = 0.00"$      & $y =  0.00"$      & $\theta = 90\,^{\circ}$         & $\theta_{\rm  E} = 1.2$  &  $q = 0.8$      & $\alpha = 2.05$ &  \\[2pt]
                        & Source 1 & $Sersic$ & $x_{\rm  s1} = -0.04"$ & $y_{\rm  s1} =  -0.07"$  & $\theta_{\rm  s1} = 90\,^{\circ}$ & $I_{\rm  s1} = 0.0036$     &  $R_{\rm  s1} = 0.4"$ & $n_{\rm  s1} = 2.5$ & $q_{\rm  s1} = 0.7$   \\[2pt]
                        & Source 2 & $Sersic$ & $x_{\rm  s2} = 0.1"$ & $y_{\rm  s2} =  -0.1"$     & $\theta_{\rm  s2} = 0\,^{\circ}$  & $I_{\rm  s2} = 0.00036$     &  $R_{\rm  s2} = 0.9"$ & $n_{\rm  s2} = 1.0$ & $q_{\rm  s2} = 0.8$   \\[2pt]
\hline
& & & & & & & & & \\[-4pt]
$\textbf{CoreSrcDisk}$         & Light  & $Sersic$                       & $x_{\rm  l} = 0.00"$ & $y_{\rm  l} =  0.00"$ & $\theta_{\rm  l} = 40\,^{\circ}$ & $I_{\rm  l} = 0.027$       &  $R_{\rm  l} = 0.48"$ & $n_{\rm  l} = 2.5$ & $q_{\rm  l} = 0.6$ \\[2pt]
\tiny HS35L70Disk             & Mass   & $PL\textsubscript{Core}$     & $x = 0.00"$      & $y =  0.00"$      & $\theta = 40\,^{\circ}$      & $\theta_{\rm  E} = 1.4$   &  $q = 0.8$      & $\alpha = 1.85$ & $s = 0.2"$   \\[2pt]
\tiny ES35L70Disk             & Source & $Sersic$                       & $x_{\rm  s} = 0.12"$ & $y_{\rm  s} =  0.25"$& $\theta_{\rm  s} = 10\,^{\circ}$ & $I_{\rm  s} = 0.04$        &  $R_{\rm  s} = 0.4"$ & $n_{\rm s} = 1.0$       & $q_{\rm s} = 0.8$   \\[2pt]
\hline
& & & & & & & & & \\[-4pt]
$\textbf{CoreSrcQuad}$ & Light 1  & $Sersic$                     & $x_{\rm  l1} = 0.00"$ & $y_{\rm  l1} =  0.00"$ & $\theta_{\rm  l1} = 110\,^{\circ}$    & $I_{\rm  l1} = 0.045$    &  $R_{\rm  l1} = 0.25"$ & $n_{\rm  l1} = 2.5$ & $q_{\rm  l1} = 0.77$ \\[2pt]
\tiny HS40L60BD                & Light 2  & $Exp$                        & $x_{\rm  l2} = 0.00"$ & $y_{\rm  l2} =  0.00"$ & $\theta_{\rm  l2} = 110\,^{\circ}$    & $I_{\rm  l2} = 0.03$    &  $R_{\rm  l2} = 1.35"$ & $n_{\rm  l2} = 1.0$ & $q_{\rm  l2} = 0.6$  \\[2pt]
\tiny ES40L60BD                & Mass     & $PL\textsubscript{Core}$   & $x = 0.00"$      & $y =  0.00"$      & $\theta = 110\,^{\circ}$      & $\theta_{\rm  E} = 1.0$   &  $q = 0.7$      & $\alpha = 1.75$ & $s = 0.3"$   \\[2pt]
                                & Source 1 \& 2 & $Sersic$ & \multicolumn{7}{p{11cm}}{Identical to $\textbf{SrcBD}$ except $x_{\rm  s1} = x_{\rm  s2} = 0.01"$ and $y_{\rm  s1} = y_{\rm  s2} = -0.01"$} \\[2pt]
\hline
& & & & & & & & & \\[-4pt]
$\textbf{CoreSrcDouble}$  & Light  & $Sersic$                       & $x_{\rm  l} = 0.05"$ & $y_{\rm  l} =  -0.06"$ & $\theta_{\rm  l} = 170\,^{\circ}$ & $I_{\rm  l} = 0.0067$       &  $R_{\rm  l} = 1.5"$ & $n_{\rm  l} = 3.5$ & $q_{\rm  l} = 0.7$ \\[2pt]
\tiny HS25L50BD             & Mass  & $PL\textsubscript{Core}$     & $x = 0.05"$      & $y =  -0.06"$      & $\theta = 170\,^{\circ}$      & $\theta_{\rm  E} = 1.3$   &  $q = 0.8$      & $\alpha = 1.65$ & $s = 0.25"$   \\[2pt]
\tiny ES25L50BD             & Source 1 & $Sersic$ & $x_{\rm  s1} = 0.2"$ & $y_{\rm  s1} =  -0.1"$     & $\theta_{\rm  s1} = 90\,^{\circ}$  & $I_{\rm  s1} = 0.0075$     &  $R_{\rm  s1} = 0.4"$ & $n_{\rm  s1} = 2.0$ & $q_{\rm  s1} = 0.9$   \\[2pt]
                             & Source 2 & $Sersic$ & $x_{\rm  s2} = 0.2"$ & $y_{\rm  s2} =  -0.1"$     & $\theta_{\rm  s2} = 90\,^{\circ}$  & $I_{\rm  s2} = 0.0025$     &  $R_{\rm  s2} = 0.8"$ & $n_{\rm  s2} = 1.0$ & $q_{\rm  s2} = 0.8$   \\[2pt]
\hline
& & & & & & & & & \\[-4pt]
$\textbf{LMDMAlign}$       & Light  & $Sersic$    & $x_{\rm  l} = 0.00"$ & $y_{\rm  l} =  0.00"$  & $\theta_{\rm  l} = 127\,^{\circ}$  & $I_{\rm  l} = 0.0085$      &  $R_{\rm  l} = 0.6"$ & $n_{\rm  l} = 4.0$  & $q_{\rm  l} = 0.72$ \\[2pt]
\tiny HS50L40BD             & Mass   & $NFW$ + $\Psi_{\rm l}$      & $x_{\rm d} = 0.00"$     & $y_{\rm d} =  0.00"$      & $\theta_{\rm d} = 127\,^{\circ}$      & $\kappa_{\rm  d} = 0.13"$    &  $q = 0.82$      & $\Psi_{\rm l} = 25.0$ &  $\Psi_{\rm l} = 6.73$   \\[2pt]
\tiny ES50L40BD             & Source 1 \& 2 & $Sersic$ & \multicolumn{7}{p{11cm}}{Identical to $\textbf{SrcBD}$} \\[2pt]
\hline
& & & & & & & & & \\[-4pt]
$\textbf{LMDMRot}$       & Light  & $Sersic$    & \multicolumn{7}{p{11cm}}{Identical to $\textbf{LMDMAlign}$} \\[2pt]
\tiny HS50L40BD             & Mass   & $NFW$ + $\Psi_{\rm l}$ & \multicolumn{7}{p{11cm}}{Identical to $\textbf{LMDMAlign}$ except $\theta_{\rm d} = 132\,^{\circ}$} \\[2pt]
\tiny ES50L40BD             & Source 1 \& 2 & $Sersic$ & \multicolumn{7}{p{11cm}}{Identical to $\textbf{SrcBD}$} \\[2pt]
\hline
& & & & & & & & & \\[-4pt]
$\textbf{LMDMPos}$       & Light  & $Sersic$   & \multicolumn{7}{p{11cm}}{Identical to $\textbf{LMDMAlign}$} \\[2pt]
\tiny HS50L40BD             & Mass   & $NFW$ + $\Psi_{\rm l}$     & \multicolumn{7}{p{11cm}}{Identical to $\textbf{LMDMAlign}$ except $x_{\rm d} = 0.05"$} \\[2pt]
\tiny ES50L40BD             & Source 1 \& 2 & $Sersic$ & \multicolumn{7}{p{11cm}}{Identical to $\textbf{SrcBD}$} \\[2pt]
\hline
& & & & & & & & & \\[-4pt]
$\textbf{LMDMRot90}$       & Light  & $Sersic$   & \multicolumn{7}{p{11cm}}{Identical to $\textbf{LMDMAlign}$} \\[2pt]
\tiny ES30L65Multi       & Mass   & $NFW$ + $\Psi_{\rm l}$      & $x_{\rm d} = 0.00"$     & $y_{\rm d} =  0.00"$      & $\theta_{\rm d} = 17\,^{\circ}$      & $\kappa_{\rm  d} = 0.13"$    &  $q = 0.82$      & $\Psi_{\rm l} = 42.0$ & $\Psi_{\rm l} = 6.73$  \\[2pt]
\tiny ES50L40BD             & Source 1 \& 2 & $Sersic$ & \multicolumn{7}{p{11cm}}{Identical to $\textbf{SrcBD}$} \\[2pt]
\hline
& & & & & & & & & \\[-4pt]
$\textbf{LMDMShear}$     & Light   & $Sersic$    & $x_{\rm  l} = 0.00"$ & $y_{\rm  l} =  0.00"$   & $\theta_{\rm  l} = 100\,^{\circ}$ & $I_{\rm  l} = 0.033$                   &  $R_{\rm  l} = 0.75"$ & $n_{\rm  l} = 2.5$ & $q_{\rm  l} = 0.75$ \\[2pt]
\tiny HS35L80Cusp        & Mass   & $NFW$ + $\Psi_{\rm l}$      & $x_{\rm d} = 0.03"$     & $y_{\rm d} =  0.03"$      & $\theta_{\rm d} = 90\,^{\circ}$      & $\kappa_{\rm  d} = 0.13"$    &  $q = 0.8$      & $\Psi_{\rm l} = 11.5$ & $\Psi_{\rm l} = 2.0$  \\[2pt]
\tiny ES35L80Cusp        & Mass    & $Shear$     & $x_{\rm  sh} = 0.03"$   & $y_{\rm  sh} =  0.03"$                             & $\theta_{\rm  sh} = 150\,^{\circ}$      & $\gamma_{\rm  sh} = 0.03$   &  & &  \\[2pt]
                         & Source 1 \& 2 & $Sersic$ & \multicolumn{7}{p{11cm}}{Identical to $\textbf{LensSrcCusp}$ except $R_{\rm s2} = 0.3"$} \\[2pt]
\end{tabular}
}
\caption[Simulated image's light, mass and source parameters]{The lens light, mass and source profiles used to create each image of the simulation suite. The left column of the table gives the title of each model, the name of which signifies the aspect of ${\tt AutoLens}$ that model has been made to test (e.g. the $\textbf{SrcBulge}$ model tests modeling sources with a bulge morphology). Each model is used to generate multiple images and in the first column underneath each model name are the tags describing those images, which can be read as follows: ‘H' or ‘E' for Hubble or Euclid resolution, ‘S\#\#' the source S/N, ‘L\#\#' the lens S/N (NL for no lens) and ‘Bulge', ‘Disk', ‘BD', ‘Cusp' or ‘Multi' to describe the source morphology. The second and third columns list each component and its corresponding profile. The remaining columns show the input lens parameters of each model. The values of $I_{\rm l}$  and $I_{\rm s}$ correspond to the values used for the highest S/N Hubble resolution image of each lens model and their values are reduced for generating each model's lower S/N and Euclid resolution images. $\Psi_{\rm l}$ is therefore also reduced for these images, with its scaled values given in the right-most column of each light-matter mass profiles.}
\label{table:SimModels}
\end{table*}

Figure \ref{figure:SimNL} shows postage-stamp cut-outs of a small sub-set of images and their source-plane configurations. Comparison between the different cut-outs shows the broad range in image resolution, S/N and source and lens morphologies the simulation suite covers. 

\section{Image Analysis, Lens Modeling and Source Reconstruction}\label{Method}

\subsection{Extended Source Modeling}\label{SLTheory}

This section gives a brief overview of the theory relevant for modeling strongly lensed extended sources. A more detailed description of this overview can be found in \citet{Schneider1992} and \citet{Keeton2003}.

As discussed in the introduction, extended source modeling offers information about the second derivative of the lens's potential. However, this signal is encoded into the lensed source's extended surface-brightness distribution and is therefore only available wherever the lensed source is actually observed, around $R_{\rm  Ein}$, the Einstein radius. The extension of this measurement to smaller radii (where there is typically no source light) is therefore something of an extrapolation \citep{Sonnenfeld2012, Schneider2014, Xu2015}, albeit one aided by how the mass model's overall normalization must still give an accurate $M_{\rm  Ein}$. The constraints that a lens offers therefore varies from system to system, depending on the source size, lensing geometry and source and lens redshifts, with the most exceptional examples spanning over $15$ kpc in extent \citep{Gavazzi2006, Sonnenfeld2012, Eichner2012}. Thus, for many lenses, this measurement does not require large extrapolations, except in the very central regions. By fitting the lens's light profile, {\tt AutoLens} partly constrains these central regions, through both the detection or absence of the lensed source's faint central features and by incorporating the lens's light profile into the mass model. 

N15 illustrated the nature of extended source analysis. First, a clear degeneracy emerges between the parameters governing the lens's mass distribution, which for the power-law density profile used in N15 was its mass normalization, ellipticity and density slope (see also \citealt{Suyu2012, Suyu2013}). This degenerate sub-set of mass models all integrate to give approximately the same $M_{\rm  Ein}$, with the different models varying only how they distribute this mass. The favoured model from this sub-set is then whichever best reconstructs the extended source. N15 also demonstrated how these degenerate lens models are fully degenerate with the source-plane magnification (see also \citealt{Birrer2015a}), such that more centrally concentrated mass profiles result in a more spatially expanded source reconstruction (i.e. lower total magnification, see figure 4 of N15). This requires specific care to ensure that the inferred lens model is not biased (section \ref{CenIms}) and as such, {\tt AutoLens} adapts to and scales with this phenomenon (section \ref{AdaptFeats}).

There is an important caveat to lens modeling of this nature, associated with the form of $\kappa$ assumed for the lens. If the allowed (parametric) form of $\kappa$ is unable to accurately follow the actual mass distribution, the sub-set of lens models which integrate to give the correct $M_{\rm  Ein}$ will offer only an approximate match to the lens's actual mass profile. They may still provide a good fit to the lensing data, but can misestimate a number of the lens's properties, like the lens's true slope at $R_{\rm  Ein}$. This is a manifestation of the much studied mass-sheet transformation (MST) and source position transformation \citep{Falco1985, Schneider2013, Schneider2014, Schneider2014b, Schneider2014c, Xu2015, Tagore2017}. This work circumvents this issue by using the same density profile for both the modelling and creation of each simulated image, as was performed in N15. Use of the lens's light profile to trace its underlying stellar matter profile may reduce the freedom of the MST, however a more detailed investigation of this is beyond this paper's scope. 

\subsection{Semi-linear Inversion}\label{SLIRecap}

The semi-linear inversion (SLI) method simultaneously reconstructs the surface brightness distribution of a strongly lensed source and models the lens galaxy mass distribution. It was first presented in \citep[][WD03 hereafter]{Warren2003}, placed within a Bayesian framework by \citep[][S06 hereafter]{Suyu2006} and developed into adaptive SLI in \citep[][N15 hereafter]{Nightingale2015}. An outline of the SLI method is given here but readers are referred to these publications for comprehensive details.

The SLI method assumes a pixelized source-plane, computing the linear superposition of PSF-smeared source pixel images which best fits the observed image, for a given lens model. This is done via the matrix $f_{\rm  ij}$, which maps the $j$th pixel of each lensed image to each source pixel $i$ and produces the source pixel surface brightness vector $\vec{s}$. Finally, the $j$th pixel of the model image is computed as $\sum_{\rm i} s{\rm _i} f_{\rm  ij}$, which is subtracted from the observed image with flux values $d{\rm _j}$ and statistical uncertainties $\sigma{\rm _j}$. In the original implementation of the SLI method, the values $d{\rm _j}$ have had a pre-computed foreground light model subtracted. The sum of the squared significances of the residuals between the observed and model images then gives a $\chi^2$ statistic. 

The pixelization used by the SLI method may be discretized into pixels of arbitrary shape or tessellation. In N15, the source-plane pixelization was derived using an h-means clustering algorithm, which defined source pixels as clusters of traced image pixels. The same clustering methodology is used here to compute source pixels, however switching instead to a weighted k-means clustering algorithm (see \citep{Hartigan1979}). This allows clustering to be weighted, thus enabling the source pixelization to adapt to the source's surface brightness (see section \ref{AdaptFeats}), unlike N15 which adapted to the mass model magnification. K-means clustering also produces more uniform and regular source-plane pixelizations (albeit still stochastic enough to sample and overcome discretization biases). The randomisation of the clustering which N15 showed to remove discreteness biases has also been slightly modified to ensure that even the exact same lens model parameterization gives a different source-plane pixelization (the reason for this is described in section \ref{HyperParams}).

Due to the ill-posed nature of the matrix inversion used by the SLI method the solution must be regularized using a linear regularization matrix, which is described in WD03 and appendix \ref{AppReg}. Regularization acts as a prior on the source reconstruction, imposing a smooth source solution. {\tt AutoLens} follows a Voronoi regularization scheme which is scale-independent, such that regularization is the same for a larger or smaller source, a property key to handling the source rescaling that emerges during lens modeling. This Voronoi grid is also used to visualize source reconstructions. In N15, regularization was controlled by the hyper-parameter $\lambda$, which set the degree to which smoothness is imposed on the solution following the Bayesian framework of S06. Section \ref{HyperParams} presents {\tt AutoLens}'s new approach to source-plane regularization.

\subsection{Lens Light and Mass Modeling}\label{FGModeling}

Fitting and subtraction of the lens's light is fully integrated into {\tt AutoLens}, with all parameters associated with the lens's light model sampled within the same non-linear parameter space as those governing the mass model. Therefore, for each iteration of the method, before reconstruction of the lensed source, {\tt AutoLens} first computes a model two-dimensional light distribution using one or more elliptial Sersic functions. The resulting two-dimensional light model is then convolved with the instrumental PSF and subtracted from the observed image. 

The mass model is then used to compute the deflection angle map $\vec{{\alpha}}_{\rm x,y}$ and trace image-pixel to the source plane. The source reconstruction outlined above is then performed. N15 showed that, due to aliasing effects, the source reconstruction benefits from oversampling (termed subgridding in N15), which splits each image-pixel into a set of square sub-pixels, which are each individually traced to the source plane and used by the inversion. Appendix \ref{AppCalc} describes a bilinear interpolation scheme used to speed this calculation up, allowing higher levels of oversampling ($8$ $\times$ $8$) to be used in this work. Appendix \ref{AppCalc} also describes how the positions of the image's brightest pixels are used to speed up mass modeling, by discarding models where they do not trace close to one another. 

The incorporation of lens light fitting into {\tt AutoLens} only slightly changes the modeling formalism given in N15 and the previous section. All pixels within the masked region retain the subscript $j$, with the definition of terms $f_{\rm  ij}$ and $\sigma{\rm _j}$ unchanged. However, $d{\rm _j}$ is now defined to be the observed flux in pixel $j$ including the lens flux contribution which is denoted as $b{\rm _j}$. The quantity $\vec{D}{\rm _i}$ used in WD03 and N15 must therefore also change to 
\begin{equation}
\label{eqn:FD2}
\vec{D}_{\rm  i} = \sum_{\rm  j=1}^{J}f_{\rm  ij}(d_{\rm  j} - b{\rm _j})/\sigma{\rm _j}^2 \, \, .
\end{equation}
$\chi^2$ is therefore given by
\begin{equation}
\label{eqn:ChiSqSrc}
\chi^2 = \sum_{\rm  j=1}^{J} \bigg[
\frac{(\sum_{\rm  i=1}^{I} s{\rm _i} f_{\rm  ij}) + b{\rm _j} - d{\rm _j}}{\sigma{\rm _j}}
\bigg]^2   \, \, .
\end{equation}
This is identical to before, except for the change in the definition of $d{\rm _j}$ and inclusion of the $b{\rm _j}$ term.  The overall likelihood function follows the same Bayesian framework used in N15 and is given in section \ref{HyperParams}. 

The determination of the lens model parameters is a standard non-linear search problem, performed using the {\tt MultiNest} algorithm \citep{Feroz2007,Feroz2009}, based on the nested sampling Monte Carlo technique of \citet{Skilling2006}. As described in N15, the random nature of {\tt AutoLens}'s source-plane discretization results in a noisy likelihood function which can rapidly fluctuate over small scales in parameter space, ill suited to Markov Chain Monte Carlo analysis. {\tt MultiNest}'s approach of first mapping out parameter space over large scales, followed by convergence toward the more noisy, higher evidence small scales, is therefore well suited. The implementation of {\tt MultiNest} in {\tt AutoLens} uses constant efficiency sampling mode. This tunes the reduction of {\tt MultiNest}'s elliptical contours such that the acceptance ratio is kept at a target level, which is set to 10 per cent for the final analysis of each image. Importance sampling is also employed, which as discussed in \citet{Feroz2013} improves the accuracy of sampling in constant efficiency mode, especially the estimation of the Bayesian evidence.

\subsection{Masking}\label{Masking}

Before performing the lens analysis the image is masked, removing the regions in the image which only contain background sky (or unwanted contaminants like foreground stars). In the early stages of development, {\tt AutoLens} used a dual-masking scheme. The first mask encapsulated both the lens and source and had only the lens's light profile subtracted from it. The second was then tailored to contain only the lensed source galaxy, with the source reconstruction performed only within this smaller second mask. The motivation behind this was that the source reconstruction is the most computationally demanding aspect of the analysis, thus a much faster run-time is possible by performing it exclusively on a smaller masked region. Unfortunately, testing of this masking scheme found it biased the lens's light model, as the omission of the source reconstruction in the first mask meant it dominated the overall $\chi^2$ value. Attempts to circumvent this by, for example, weighting the likelihood of each mask never led to satisfactory results. Therefore, it was concluded that the lens and source must be analysed within the same masked region and that the approach used in N15 of tailoring a hand-drawn mask around the lensed source was no longer viable. Thus, comparatively wide and extensive masking possibly extending well beyond the lensed source as well as encompassing the entire region within the $R_{\rm  Ein}$ (where typically no source light is present) is now necessary. 

This masking scheme offers a number of benefits to lens modeling. For example, it ensures that if a lens model incorrectly places extraneous images within the image reconstruction they are not masked out and ignored. Equally, faint source features which may have been masked previously will now be detected and modeled. The drawback (and reason why source-only masks are generally used in other studies e.g. \citealt{Dye2014, Vegetti2014}) is that the overall run-time of a lens analysis scales directly with the number of image pixels. For this more extensive masking scheme the number of image pixels increases by a factor of $2-4$, leading to an increase of {\tt AutoLens}'s overall run-time by the same factor or more. This provides a significant computational challenge and motivates the new source-plane analysis features described in section \ref{AdaptFeats}.

In this work, a circular mask of radius $3.9"$ is used to model all lenses, which is sufficiently large to fully capture the source and lens of every simulated image. The use of a circular mask provides a regular and symmetric source-plane pixelization. This gives the adaptive source-plane features described in section \ref{AdaptFeats} better control of the source reconstruction and reduces the discretization effects discussed in N15. In contrast, masks tailored to the lensed source produce irregular edges in the source pixelization which, as discussed next, have the potential to bias the analysis. During an early initialization phase of the pipeline, an annulus mask is used instead of a circular mask.

\subsection{Central Image Pixels}\label{CenIms}

\begin{figure*}
\centering
\includegraphics[width=0.32\textwidth]{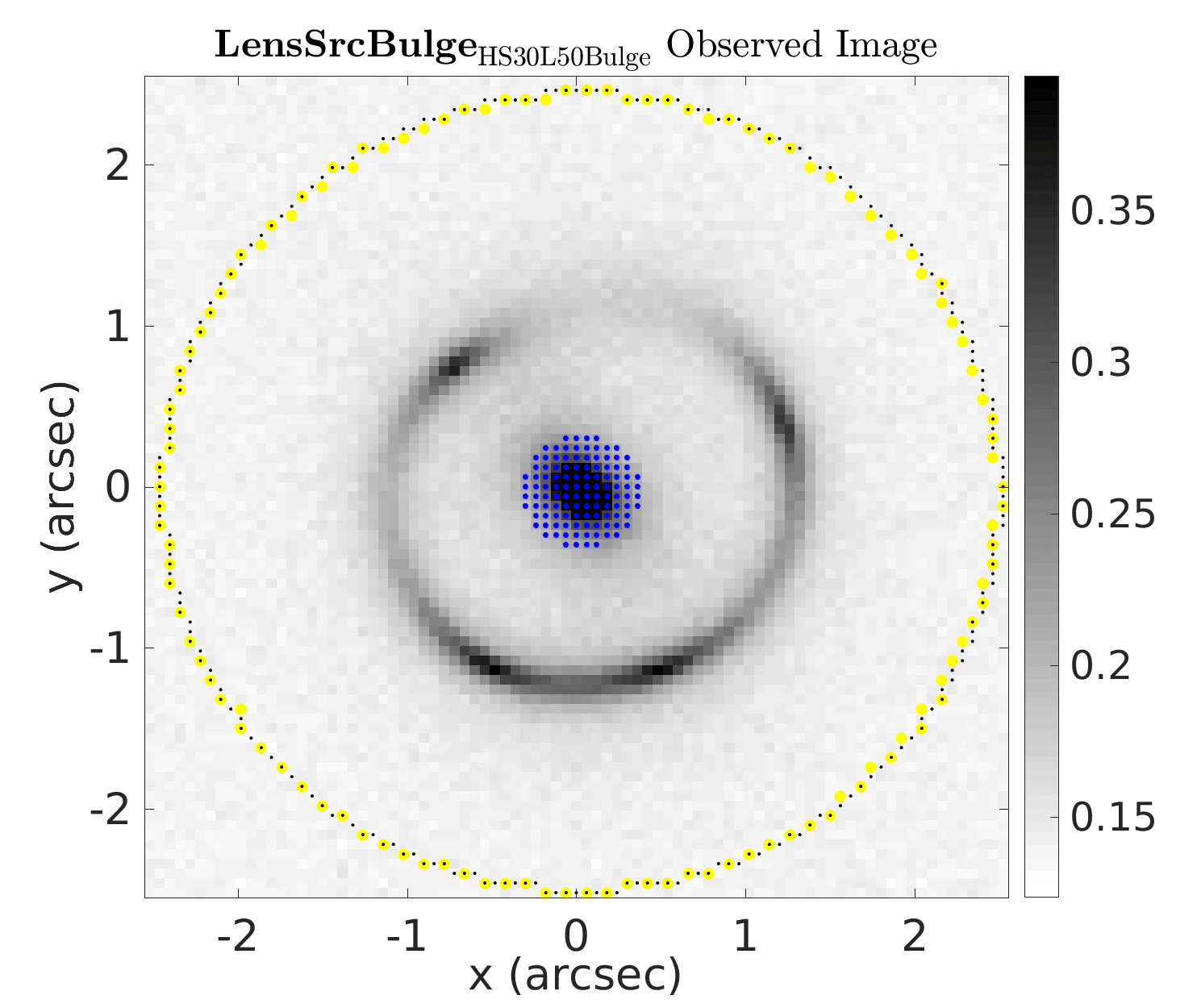}
\includegraphics[width=0.32\textwidth]{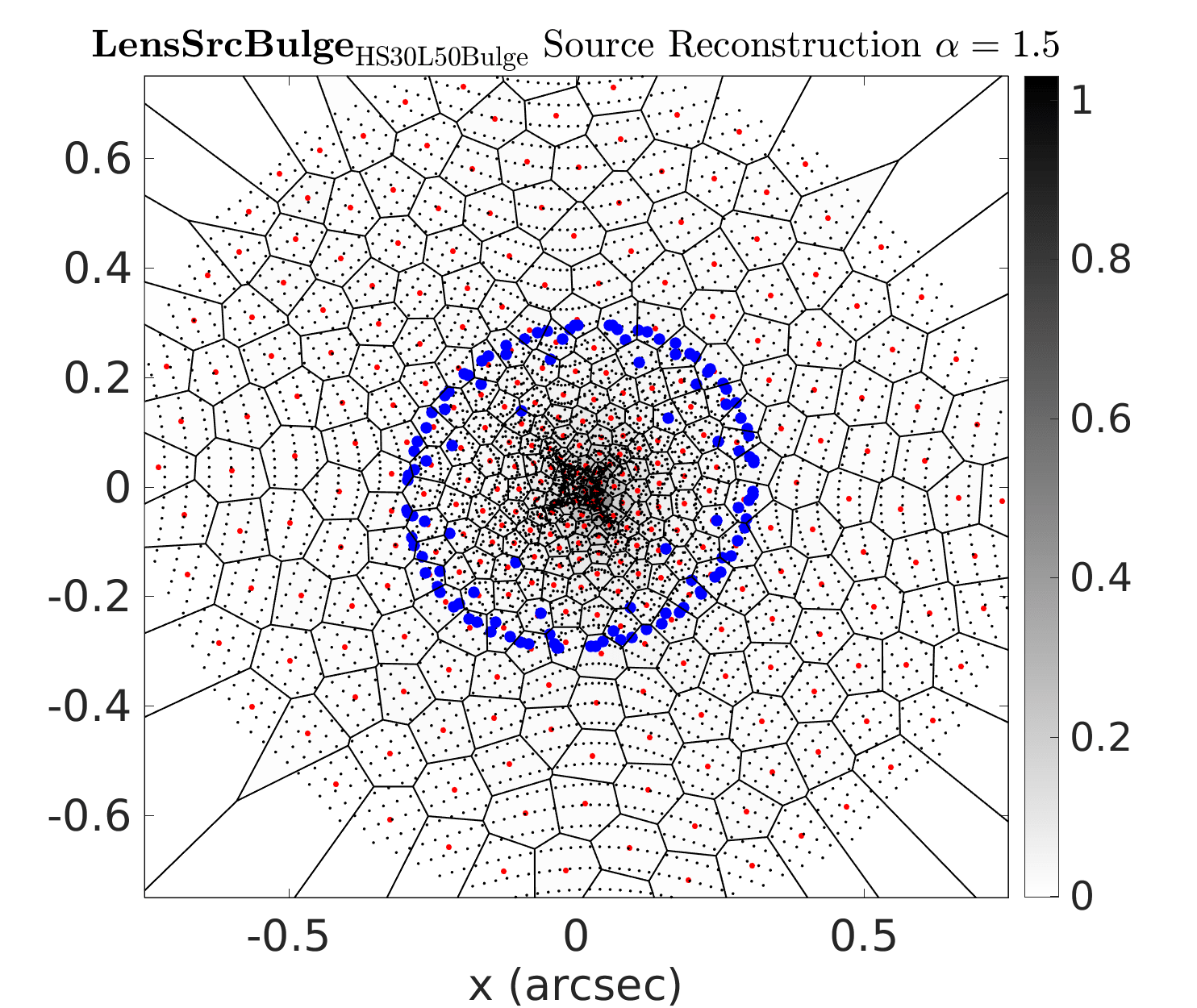}
\includegraphics[width=0.32\textwidth]{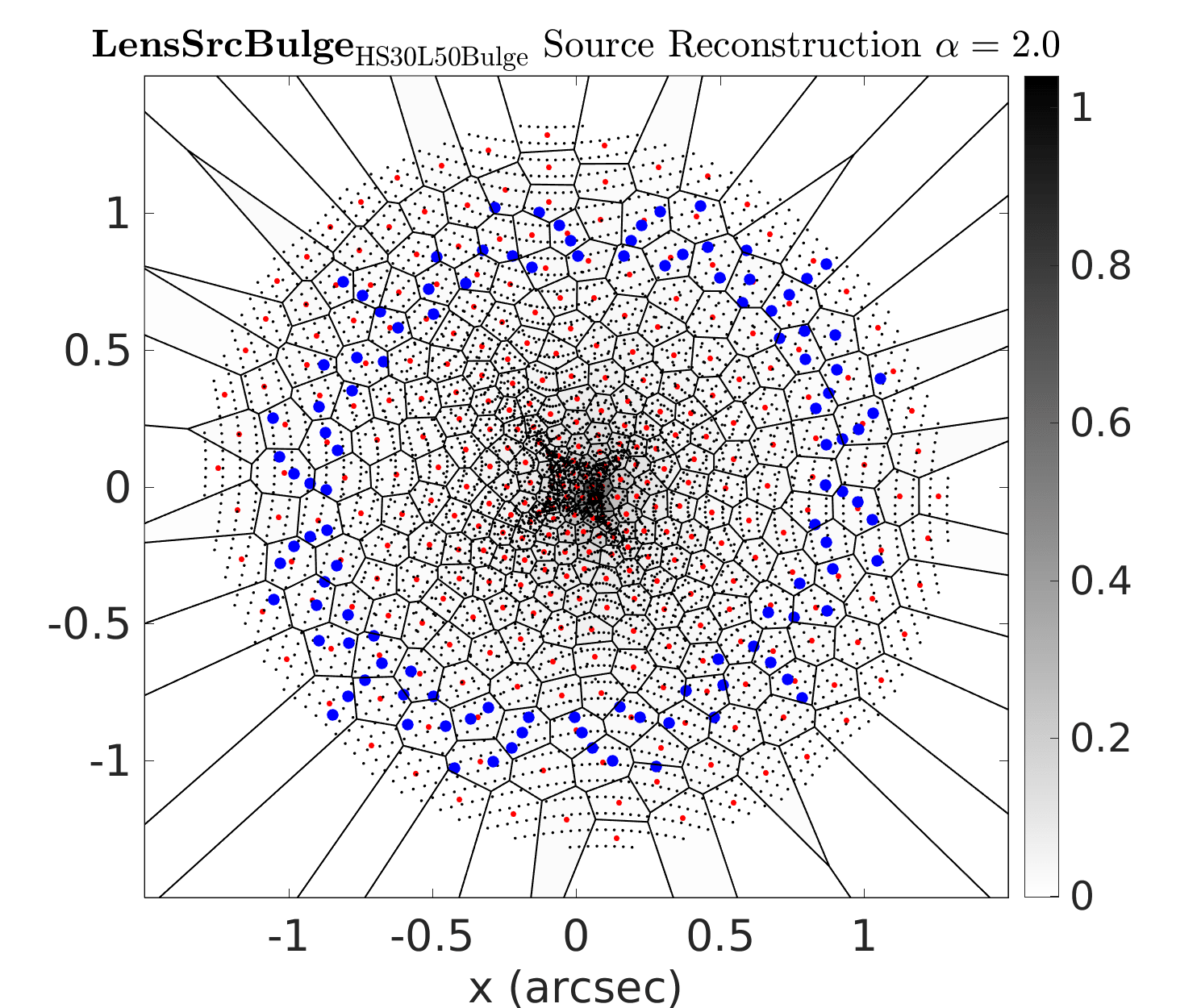}
\includegraphics[width=0.32\textwidth]{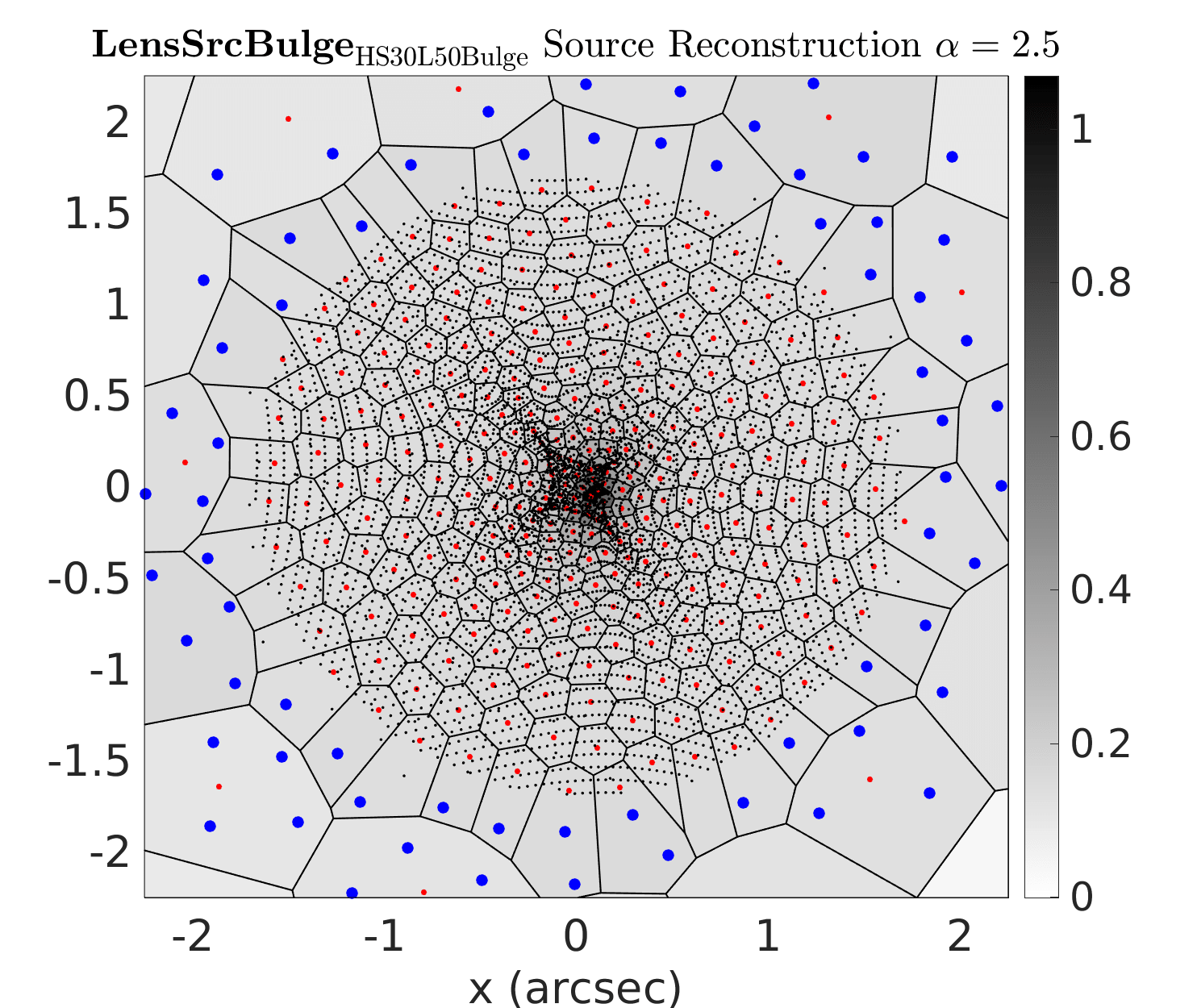}
\includegraphics[width=0.32\textwidth]{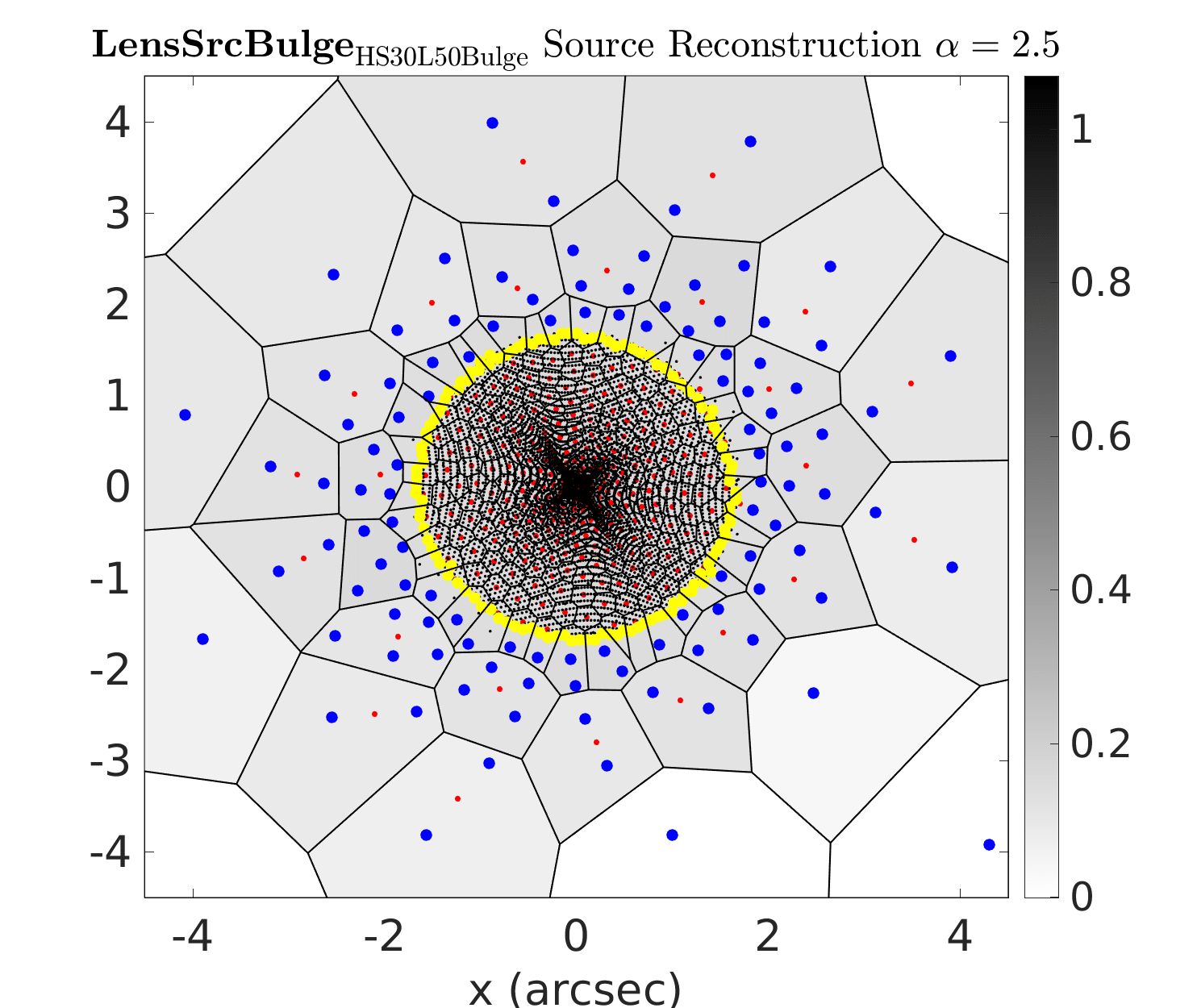}
\includegraphics[width=0.32\textwidth]{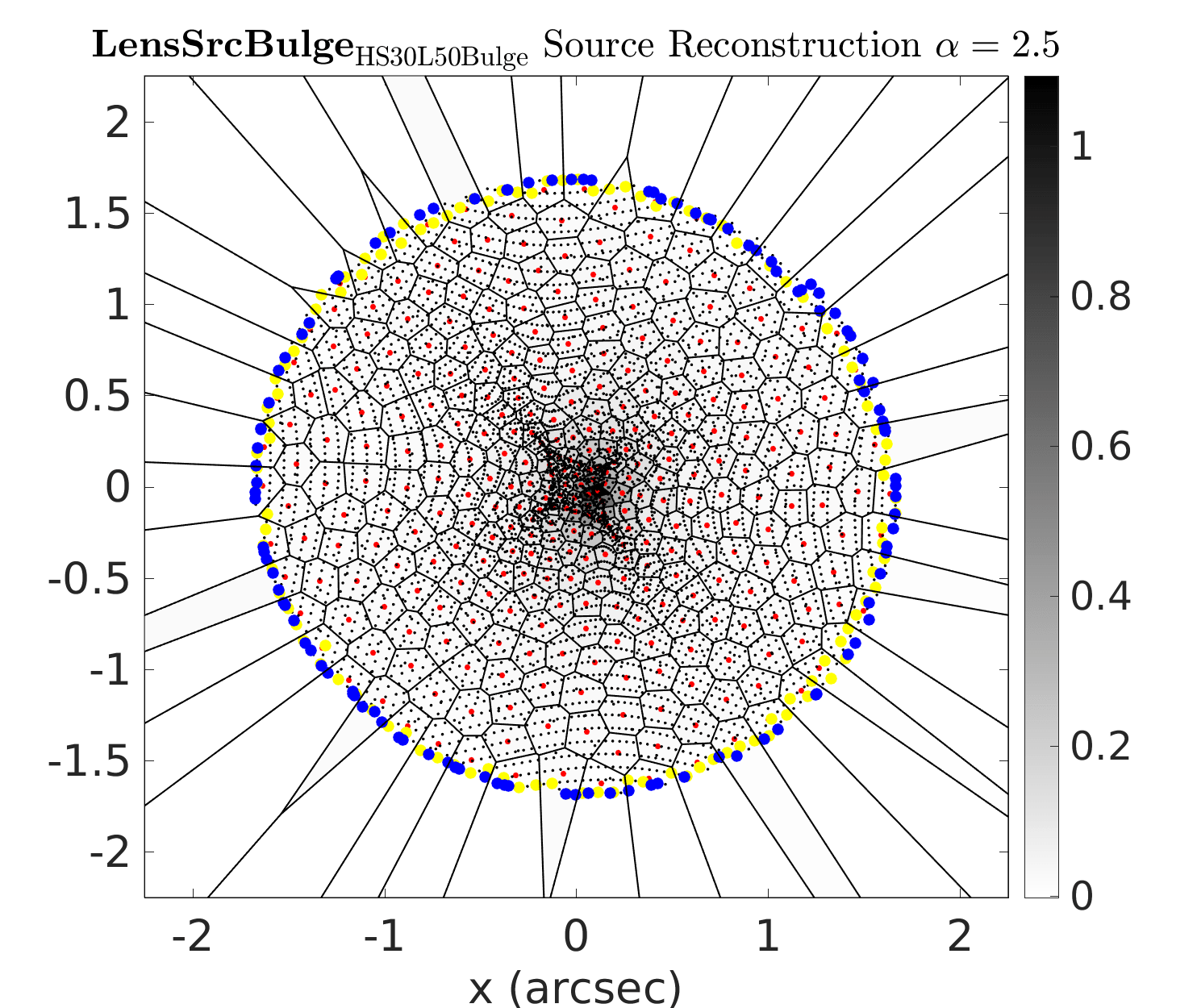}
\caption{A demonstration of where central image pixels trace in the source-plane for mass models with varying density-slope, using the image $\textbf{LensSrcBulge}_{\rm  HS30L50Bulge}$, which is pictured in the top left panel. The remaining panels show source reconstructions for this image, using three different lens models, which all use the image's input light model (see table \ref{table:SimModels}) and input mass model geometric parameters ($x=0.0"$, $y=0.0"$ and $\theta = 127\,^{\circ}$). Three different sets of mass profile parameters are used: $\textit{top-centre}$ ($\alpha = 1.5$, $\theta_{\rm  E} = 1.16$, $q = 0.93$);  $\textit{top-right}$ ($\alpha = 2.0$, $\theta_{\rm  E} = 1.2$, $q = 0.73$); $\textit{bottom row}$ ($\alpha = 2.5$, $\theta_{\rm  E} = 1.36$, $q = 0.56$). These parameters were determined by fixing $\alpha$ to $1.5$, $2.0$ and $2.5$ and computing $\theta_{\rm  Ein}$ and $q$ via a full {\tt AutoLens} analysis. Black dots depict the locations of traced image-pixels and red dots are the centres of each Voronoi source pixel. Central image pixels are defined as pixels whose image-plane coordinates are within 6 pixels ($0.36"$) of the mass model centre and are marked using blue dots, depicted both in the image-plane (top left) and source-plane (remaining panels). Their source-plane locations can be seen to depend critically on $\alpha$, tracing further away from the source as $\alpha$ increases. The overall size of the source-plane increases from $1.5" \times 1.5"$ for $\alpha = 1.5$ to $4.5" \times 4.5"$ for $\alpha = 2.5$, demonstrating the source-plane scaling effect discussed in N15. The image-plane mask and border is marked with yellow dots in the top-left panel and the location this border traces to in the source plane is shown in the bottom-centre and bottom-right panels (for the $\alpha = 2.5$ mass model), forming a ring of pixels outside of which only demagnified central image pixels trace. Outside this border, central image pixels can be seen to form their own source pixels, an effect which can prove problematic for the source reconstruction. Therefore, the scheme shown in the bottom-right panel is used, which relocates central image pixels which trace beyond this border to its edge.}
\label{figure:CenIms1}
\end{figure*}

A consequence of the masking scheme above is that the lens's central image pixels are now traced to the source-plane and included as part of the image and source reconstruction. This is acceptable for modeling a a cored mass profile (like the $PL\textsubscript{Core}$ profile), as these pixels are expected to trace near the source. However, for singular mass profiles these central image pixels may be significantly demagnified and trace to exterior regions of the source-plane that negatively impact the resulting source pixelization. This is illustrated in figure \ref{figure:CenIms1} for the image $\textbf{LensSrcSersic}_{\rm  HS50L40BD}$, where central image pixels are marked as blue dots and in the top-left panel can be seen to correspond to the regions where the lens's light is brightest. The remaining panels show source-reconstructions using three different $SPLE$ mass models with density slopes ($\alpha$) of $1.5$, $2.0$ (this image's input value) and $2.5$ (see the figure's caption for how the overall lens model is computed). This figure reaffirms the source-plane scaling discussed in N15, noting that the source-plane axis increases from $1.5" \times 1.5"$ for the $\alpha = 1.5$ mass model to $4.5" \times 4.5"$ for $\alpha = 2.5$. For $\alpha = 2.0$ (the top-right panel) the lens model matches the image's input model. Thus, the lens's light is subtracted perfectly (not shown) and the central pixels trace to regions where the source is very faint, therefore having no impact on the source and image reconstruction. 

During testing, it emerged that this ideal scenario was not always reached and central pixels could bias lens modelling in two different ways. The first is due to the interplay between the location to which central image pixels trace in the source-plane and the mass model's density slope $\alpha$, illustrated in figure \ref{figure:CenIms1}. For mass profiles with a lower value of $\alpha$, central image pixels are less demagnified and thus trace closer to the source, giving them the potential to impact the source reconstruction. Indeed, the central image pixels shown in the top-centre panel (for $\alpha = 1.5$) trace within the source's faint extended envelope and are therefore allocated a low-level of extraneous flux by the source reconstruction. Thus, it is possible that the source reconstruction wrongly places extraneous flux in the image reconstruction's central regions, which can potentially bias lens modeling in two different ways:
\begin{itemize}
\item When the lens light subtraction leaves significant residuals. In this instance, the mass model may be biased towards lower $\alpha$ solutions that allow the source reconstruction to fit these residuals.

\item When the value of $\alpha$ assumed for the mass model is lower than the true value. In this case, the source-reconstruction may fit some of the flux in the central image pixels, leading to an inaccurate lens light model.
\end{itemize}
At the beginning of a lens analysis care must therefore be taken to ensure these biases are circumvented. {\tt AutoLens} achieves this by assuming an $SPLE$ mass model with a fixed value of $\alpha = 2.2$ early in the analysis. Later in the analysis, once the lens subtraction is accurate, $\alpha$ can safely be treated as free. The second problem is counteracted because a slope $\alpha = 2.2$ is steeper than most strong lenses \citep{Koopmans2009}, thus ensuring that central image pixels trace well away from the source. Whilst this may not be sufficient for all lenses (e.g. those with a very steep density profile or very extended source) it has proven adequate for all test-cases thus far. Sanity checks flag up when central image pixels receive extraneous flux, ensuring this bias will be spotted on large lens samples. 

The second problem is also illustrated in figure \ref{figure:CenIms1}, particularly the bottom-centre and bottom-right panels, which depict where central image pixels trace relative to the `image border' (yellow dots), the ring of image pixels located at the edge of the image-plane mask. These panels show that for the $\alpha = 2.5$ mass model (chosen to exaggerate this effect) central image pixels trace well beyond this image-plane border (the yellow ring of dots) in the source-plane, forming their own source-pixels and offering the reconstruction an unphysical means by which to fit the lens subtraction's residuals or noise. Thus, the mass model may be biased to high $\alpha$ solutions which allow these exterior source pixels to form. To counteract this, {\tt AutoLens} relocates all central image pixels which trace beyond the image border in the source-plane to its edge, as shown in the bottom-right panel of figure \ref{figure:CenIms1}. This prevents central image pixels forming their own source pixels and therefore removes their potential to bias the lens model in a computationally efficient manner.

\subsection{Bayesian Framework}\label{HyperParams}

{\tt AutoLens}'s source and image analysis is based on the Bayesian framework for interpolation, model comparison and regularization presented in \citet{MacKay1992} (in particular chapters 2 and 6), which S06 generalized to lens modeling. Many other methods in the literature are also based on this (e.g. \citealt{Dye2008, Vegetti2009, Collett2014, Tagore2014}). This framework objectively ranks every image and source reconstruction that is produced by {\tt AutoLens}'s linear inversion step. For every light model, mass model and source reconstruction, the overall probability is given by the Bayesian evidence, $\epsilon$,
\begin{eqnarray}
\label{eqn:evidence2}
-2 \,{  \mathrm{ln}} \, \epsilon &=& \chi^2 + \vec{s}^{T}\tens{H}_{\rm  \Lambda}\vec{s}
+{ \mathrm{ln}} \, \left[ { \mathrm{det}} (\tens{F}+\tens{H}_{\rm  \Lambda})\right]
-{ \mathrm{ln}} \, \left[ { \mathrm{det}} (\tens{H}_{\rm  \Lambda})\right]
\nonumber \\
& &
+ \sum_{\rm  j=1}^{J}
{ \mathrm{ln}} \left[2\pi (\sigma{\rm _j})^2 \right]  \, .
\end{eqnarray}
This expression was derived in \citet{Dye2008} from S06, is used for all modeling presented in this work and is equivalent to the expression used in N15. However, the regularization matrix $\tens{H}_{\rm  \Lambda}$ has been redefined such that {\tt AutoLens} can now also apply a non-constant regularization scheme to the source reconstruction, as described next. The mathematical formalism for this non-constant regularization is given in appendix \ref{AppReg}.

Equation \ref{eqn:evidence2} quantifies three aspects of the image and source reconstruction, the first being the quality of the image reconstruction. Because the source reconstruction is a linear inversion which takes as an input the image-data when reconstructing it, it is in principle able to perfectly reconstruct the image regardless of the image's noise or the accuracy of the lens model (e.g. at infinite source resolution without regularization). This is why the problem is `ill-posed' and why regularization is necessary. However, this still raises the question of what constitutes a `good' solution? Equation \ref{eqn:evidence2} defines this by assuming that the image data consist of independent Gaussian noise in each image pixel, defining a `good' solution as one whose $\chi^2$ residuals are consistent with Gaussian noise, therefore producing a reduced $\chi^2 \sim 1$. Solutions which give a reduced $\chi^2 < 1$ are penalized for being overly complex and fitting the image's noise, whereas those with a reduced $\chi^2 > 1$ are penalized for not invoking a more complex source model when the data supports it. In both circumstances, these penalties lead to a reduction in ln$\epsilon$.

The second aspect of the analysis which equation \ref{eqn:evidence2} quantifies is the complexity of the source reconstruction. This uses terms two, three and four of this expression (those containing the regularization matrix $\tens{H}_{\rm  \Lambda}$), which from here on are collectively referred to as the `regularization terms'. These terms estimate the number of source pixels that are used to reconstruct the image, after accounting for their correlation with one another due to regularization. Solutions that require fewer correlated source pixels collectively decrease the total value of these regularization terms, increasing the value of ln$\epsilon$. Thus, simpler and less complex source reconstructions are favoured by this expression. 

Finally, equation \ref{eqn:evidence2} favours models which fit higher S/N realizations of the observed imaging data (where the S/N is determined using the image-pixel variances, the $\sigma_{\rm j}$'s found in the $\chi^2$, $\tens{F}$ and $\sum_{\rm  j=1}^{J} {\mathrm{ln}} \left[2\pi (\sigma{ _j})^2 \right]$ terms of equation \ref{eqn:evidence2}). If fixed variances are assumed throughout the analysis this aspect of equation \ref{eqn:evidence2} has no impact on modeling. However, a number of methods have invoked scaling the image pixel variances wherever the image reconstruction fits the data poorly (e.g. \citealt{Suyu2012}), an approach {\tt AutoLens} follows. 

The premise is that whilst increasing the variances of image pixels lowers their S/N values and therefore also decreases ln$\epsilon$ by increasing  $\sum_{\rm  j=1}^{J} { \mathrm{ln}} \left[2\pi (\sigma{ _j})^2 \right]$, doing so may produce a net increase in ln$\epsilon$ by decreasing $\chi^2$ and $\tens{F}$. This occurs when the $\chi^2$ values of the image pixels whose variances are increased were initially very high and therefore fit poorly by the lens model. Conversely, variances cannot be reduced to arbitrarily low values, as doing so will inflate their $\chi^2$ contribution (again decreasing ln$\epsilon$). In fact, {\tt AutoLens} does not allow a pixel's variance to be scaled below its `baseline' value, the value that is expected from a consideration of instrumental noise sources like Poisson counts and read noise. 

In summary, ln$\epsilon$ is maximized for solutions which most accurately reconstruct the highest S/N realization of the observed image, without over-fitting its noise and using the fewest correlated source pixels. By employing this framework throughout, {\tt AutoLens} objectively determines the final lens model following the principles of Bayesian analysis and Occam's Razor.

The simplest application of the Bayesian evidence was shown in N15, where it was used to set the regularization coefficient $\lambda$, a hyper-parameter which controls the degree of smoothing applied to the source reconstruction ($\lambda$ is included in equation \ref{eqn:evidence2} via the matrix $\tens{H}_{\rm  \Lambda}$). This amounted to fixing the lens model (and the source pixelization, regularization scheme, etc.) and iterating over the value of $\lambda$ until the peak value of ln$\epsilon$ is reached. This peak value strikes a balance: too high values of $\lambda$ over-smooth the source reconstruction and thus give a poor overall fit to the data (decreasing ln$\epsilon$ by increasing $\chi^2$), whereas too low values give a source reconstruction that accurately reconstructs the image but also fits large portions of its noise (decreasing ln$\epsilon$ by increasing the regularization terms). The optimum value of $\lambda$ therefore again corresponded to the solution which gives an overall reduced $\chi^2$ of approximately one. Section \ref{AdaptDemo} demonstrates there are many scenarios where this simple scheme does not produce a satisfactory fit to the image data, motivating the features introduced below.

\subsection{Adaptive Image and Source Reconstruction}\label{AdaptFeats}

In addition to the lens model, the source and image analysis therefore determine the value of ln$\epsilon$. For instance, ln$\epsilon$ depends on the source-plane pixelization (see N15 and also \citet{Tagore2014}), the degree of regularization applied to it and the regularization scheme that is applied (e.g. zeroth order, gradient, curvature, see WD03). The observed image's variances (which can now be scaled) also determine ln$\epsilon$. Thus, the setup of the source and image analysis will determine the lens model that is inferred. To determine what is objectively the most probable lens model, one must therefore find the model which maximizes ln$\epsilon$ including all these aspects of the analysis. {\tt AutoLens} achieves this by changing its source pixelization, regularization and image variances, in conjunction with the lens model, throughout the analysis. Other methods follow a similar approach for choosing the source-plane regularization scheme or resolution (e.g. \citealt{Suyu2012, Vegetti2014}), but do not do so in a fully automated or self-consistent way.

\begin{figure*}
\centering
\includegraphics[width=0.4\textwidth]{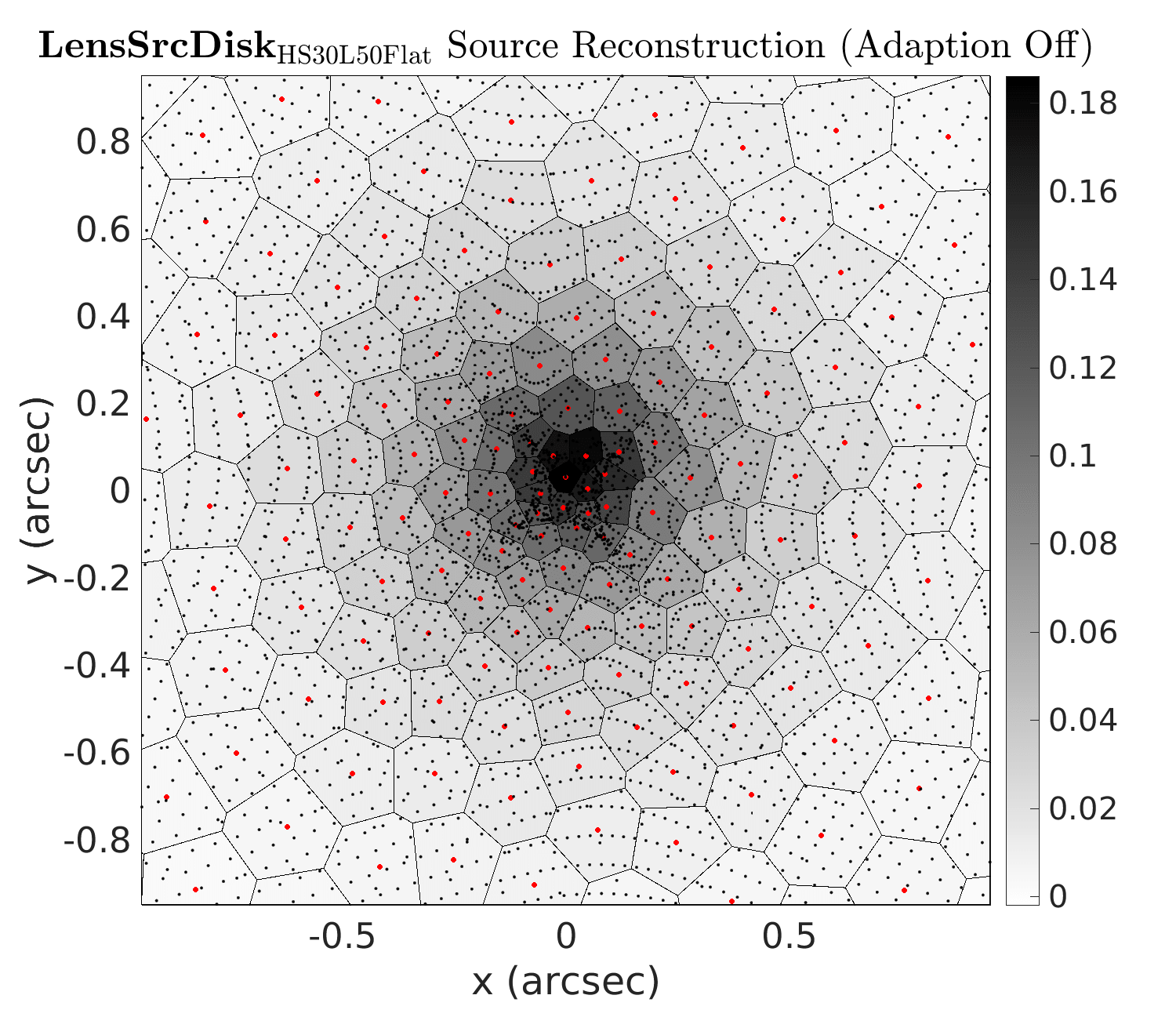}
\includegraphics[width=0.4\textwidth]{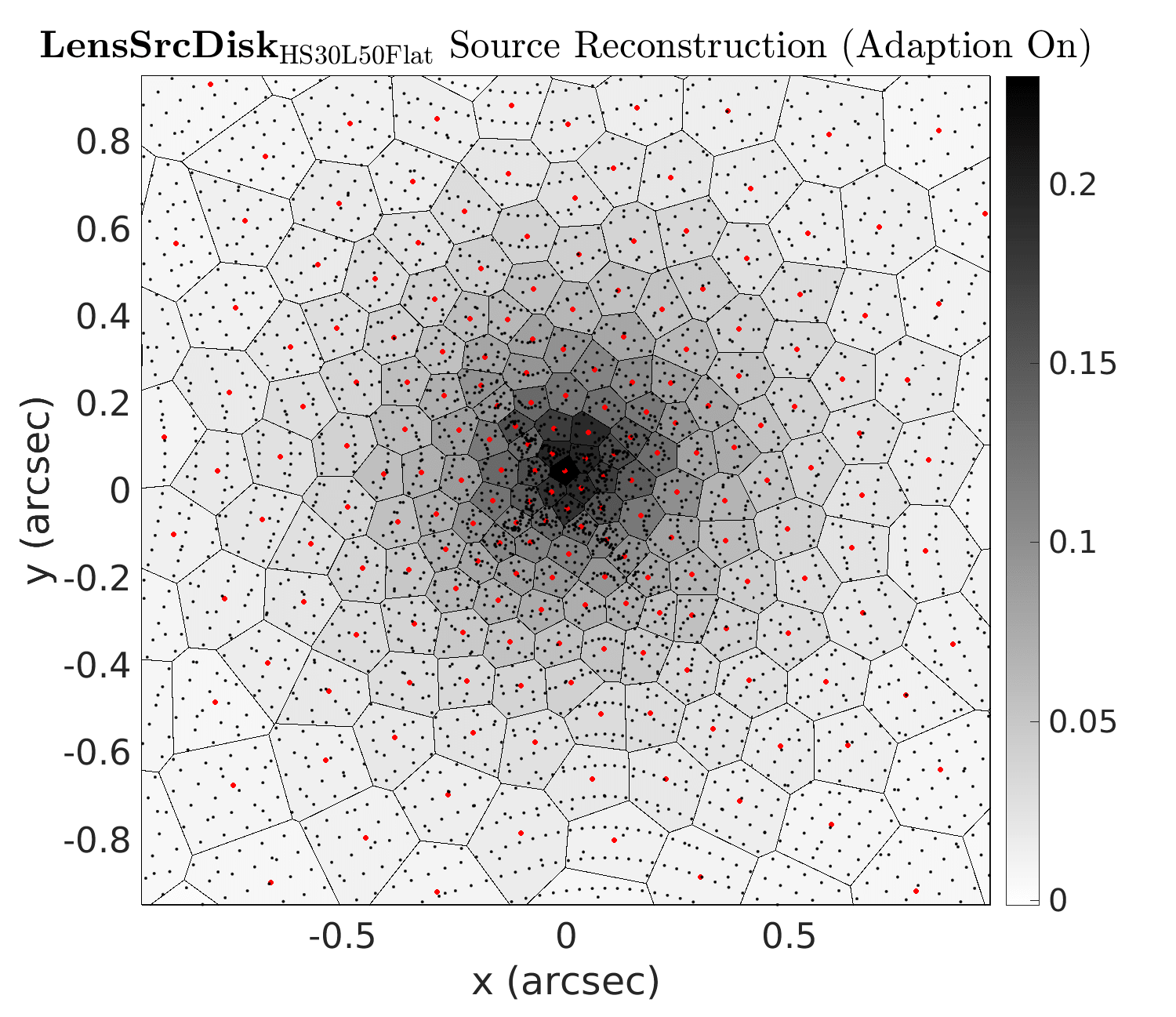}
\includegraphics[width=0.4\textwidth]{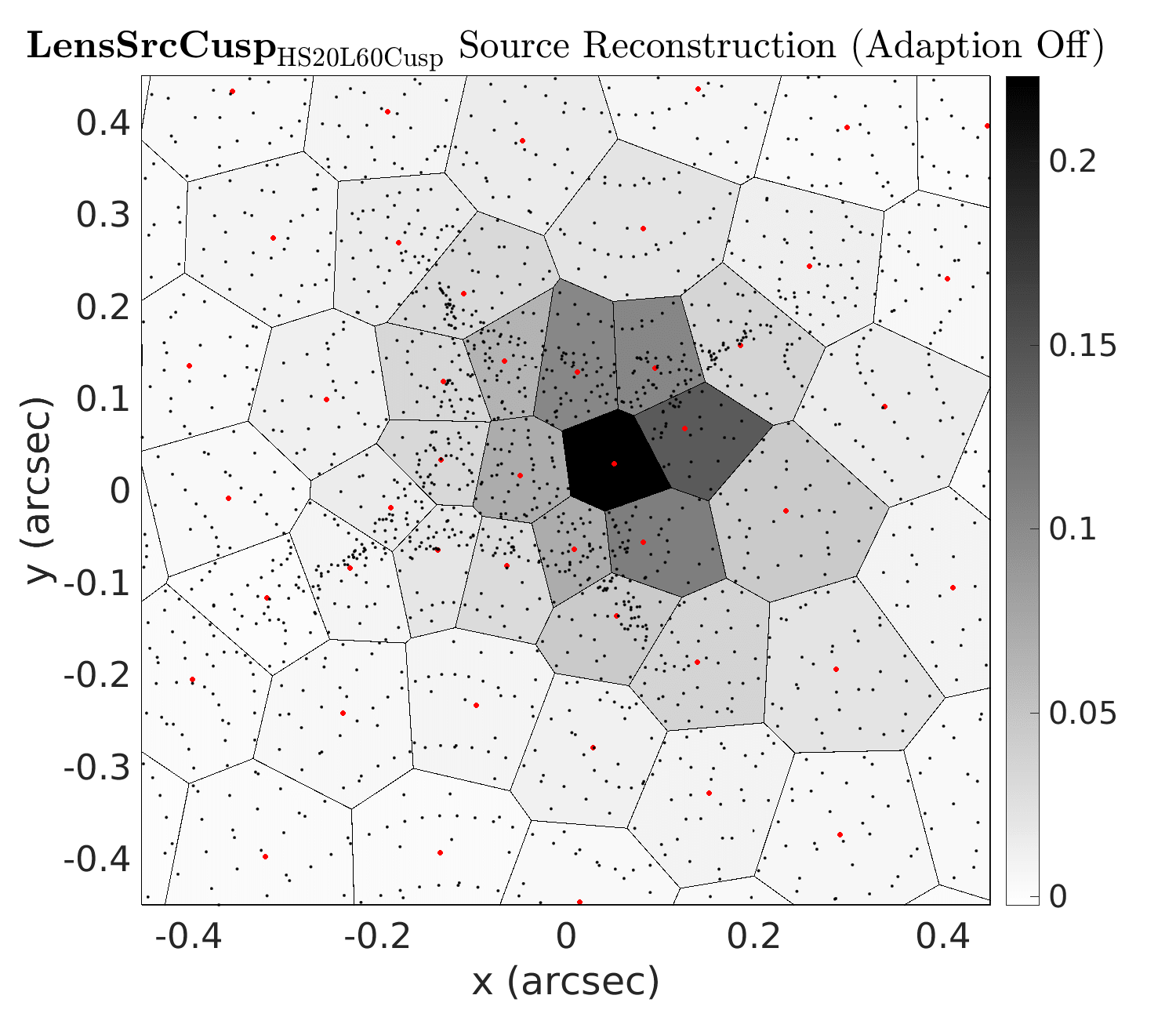}
\includegraphics[width=0.4\textwidth]{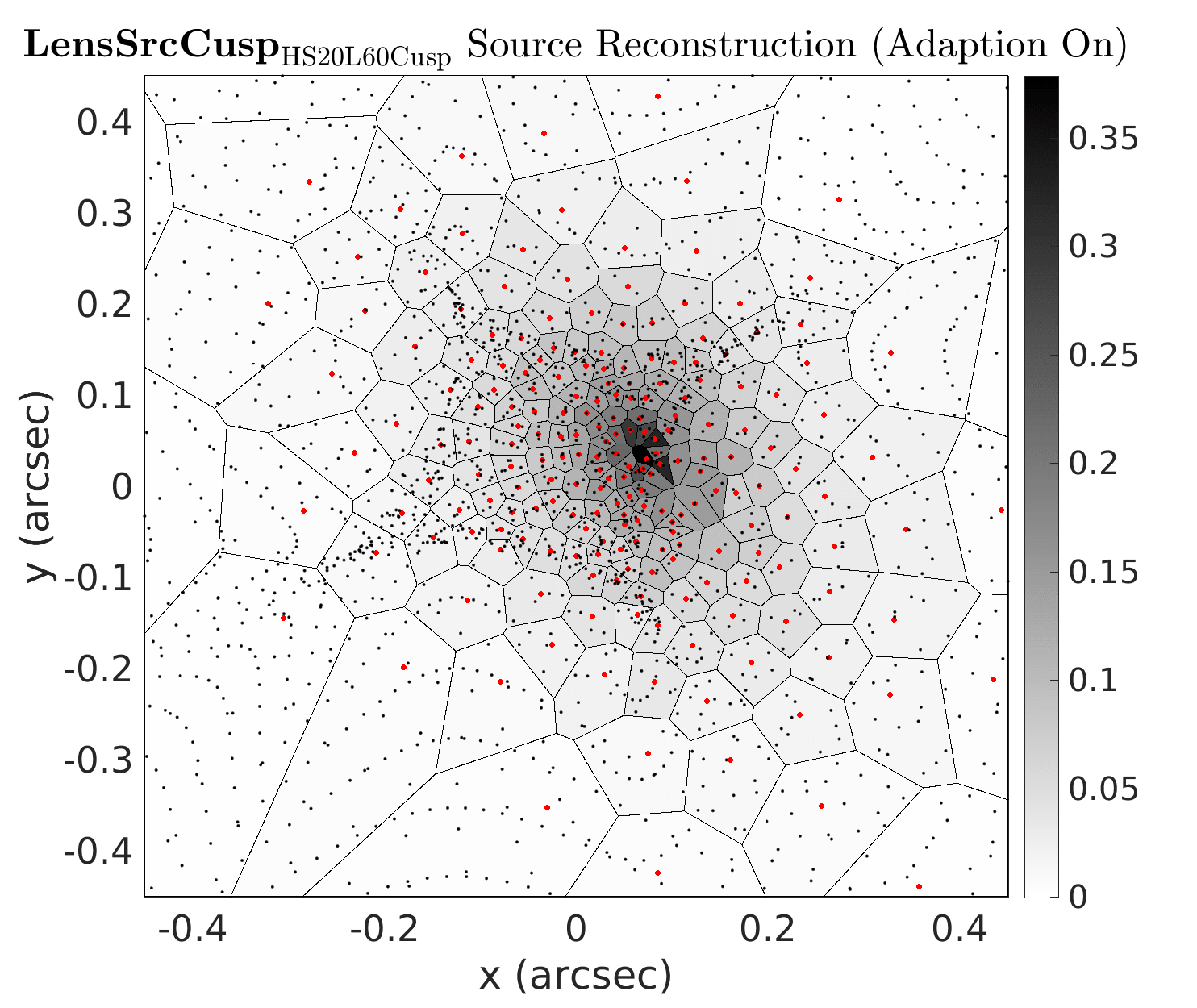}
\caption{An illustration of source surface-brightness adaptation (see section \ref{LightAdapt}) using the images $\textbf{LensSersicDisk}_{\rm  HS30L50Disk}$ (top row) and $\textbf{LensCuspySrc}_{\rm  HS20L60Cusp}$ (bottom row). All panels depict the result of using each image's input lens model to optimize the hyper-parameters, with the left panels adapting to the mass model's magnification and optimizing just $\lambda$ ($N_{\rm s} = 500$) and right panels adapting to the source's surface-brightness by optimizing all hyper-parameters described in this section, including $N_{\rm s}$, $L_{\rm  Clust1}$ and $L_{\rm  Clust2}$. Black dots depict the locations of traced image-pixels and red dots are the centres of each Voronoi source pixel. For both images surface-brightness adaptation congregates more source pixels in the source's brighter central regions, leading to an increase in ln$\epsilon$.}
\label{figure:LightAdapts}
\end{figure*} 

To adapt the source reconstruction and scale the image's variances, pre-computed model images of the lens galaxy's light profile and reconstructed lensed source (in the image-plane) are used (using lens light and mass models that have already been estimated in earlier phases of the automated analysis pipeline described in section \ref{SLPipeline}). These model images are stored in vectors of $J$ image pixels, where the lensed source model in each image pixel is given by $\Xi{\rm _j} = \sum_{\rm i} s_{\rm i} f_{\rm ij}$ and the lens light model in each image pixel by $\L{ \rm _j} = b_{\rm j}$. These vectors are updated throughout the automated analysis pipeline and correspond to the highest likelihood model that has previously been estimated.

In the following sub-sections, we describe each adaptive feature of the source and image analysis, alongside their associated hyper-parameters.

\subsubsection{Source Pixelization}\label{LightAdapt}

Three new hyper-parameters are associated with the source pixelization. The first is simply the number of source-plane pixels, $N_{\rm s}$. The second and third control the source plane clustering. The weighted k-means clustering algorithm used for source plane pixels minimises the sum of cluster `energies', $E$, given by
\begin{equation}
E = \sum^{I}_{\rm  i=1} e{\rm _i} = \sum^{I}_{\rm  i=1} \sum^{K}_{\rm  k=1} \bigg( {\frac{r{\rm _k}}{W{\rm _k}}} \bigg)^2 \, ,
\label{eqn:clusterE}
\end{equation}
where a cluster energy $e{ _i}$ is the quadrature sum of the distances $r{ _k}$ of its $K$ associated traced image-plane pixels to its center divided by each pixel's weight $W{ _k}$. In N15, all traced image pixels were given unit weighting, leading the method to adapt to the mass model's magnification pattern. Instead, the method now adapts to the surface brightness of the lensed source, using the weight vector $\vec{W}$, which calculates the weight $W{ _j}$ of each image pixel using the pre-computed source model vector as
\begin{equation}
W_{\rm  j} = \bigg[  \bigg({\frac{\Xi{\rm _j} - \Xi_{\rm  min}}{ \Xi_{\rm  max} - \Xi_{\rm  min}}} \bigg) + L_{\rm  Clust1} \bigg] ^{L_{\rm  Clust2}}  ,
\label{eqn:SrcWeight}
\end{equation}
where $\Xi_{\rm  min}$ and $\Xi_{\rm  max}$ are the maximum and minimum values of $\vec{\Xi}$ such that the first term of the right-hand side of this equation ranges between zero and one. As the hyper-parameter $L_{\rm  Clust2}$ increases, the separation between the lowest and highest $W_{\rm  j}$ values increases, such that minimization of the statistic $E$ prefers a source-plane clustering which places a greater number of smaller source pixels within the source's brightest regions. Conversely, as larger values of the hyper-parameter $L_{\rm Clust1}$ are added, the resulting distribution of $W_{\rm  j}$ values is flattened, such that minimization of the statistic $E$ places more source pixels away from the source. Together, $L_{\rm  Clust1}$ and $L_{\rm  Clust2}$ give {\tt AutoLens} complete control of its source pixeliation. For $L_{\rm  Clust2} = 0$, all $W_{\rm  j} = 1$ and source-plane adaptation reverts to pure magnification scaling as in N15. Negative values of $L_{\rm  Clust1}$ and $L_{\rm  Clust2}$ are not permitted, which would lead the source-plane to adapt to regions of background sky.

Figure \ref{figure:LightAdapts} illustrates surface brightness adaptation for the simulated images $\textbf{LensSrcDisk}_{\rm  HS30L50Disk}$ and $\textbf{LensSrcCusp}_{\rm  HS20L60Cusp}$. The left panels show source reconstructions not using this feature, equivalent to using the analysis of N15 or a value $L_{\rm  Clust2} = 0$. The right panels show the result of including $N_{\rm s}$, $L_{\rm  Clust1}$ and $L_{\rm  Clust2}$ as free-parameters in the hyper-parameter optimization. For the image $\textbf{LensSrcCusp}_{\rm  HS20L60Cusp}$ surface-brightness adaptation can be seen to have a significant effect, congregating a large number of source pixels around the cuspy source's bright central regions. For the image $\textbf{LensSersicDisk}_{\rm  HS30N50Flat}$, it plays a lesser role, owing to the source's flatter light profile.

\subsubsection{Source and Lens Contribution Maps}

The remaining adaptive image and source features require an estimate of how much of the flux in each image pixel can be attributed to the source and lens. To achieve this, two `flux contribution maps' are generated, $\vec{\Omega_{\rm  Src}}$ and  $\vec{\Omega_{\rm  Lens}}$. To compute these vectors, the total flux in each image pixel that can be accounted for by the pre-computed source and lens light models is first computed as
\begin{equation}
 T{\rm _j} = \Xi{\rm _j} + \L{\rm _j} .
\label{eqn:Rem}
\end{equation}
The contribution of flux that can be attributed to the source light in each image pixel is then estimated as
\begin{equation}
\Omega_{\rm  Src,j} = \frac { \Xi{\rm _j} } {T{\rm _j} + \omega_{\rm  SrcFrac}} ,
\label{eqn:FracSrc}
\end{equation}
where values of $\vec{\Omega_{\rm  Src}}$ below $0.02$ are set to $0$ to remove residual features in the source reconstruction. The contribution of flux from the lens is then given as
\begin{equation}
\Omega_{\rm  Lens,j} = \frac {\L{\rm _j}}{T{\rm _j} + \omega_{\rm  LensFrac}} .
\label{eqn:FracLens}
\end{equation}
Both vectors are then divided by their maximum values, such that they range between values just above $0$ and $1$. $\vec{\Omega_{\rm  Src}}$ will therefore contain values close to $1$ where only the source is present and close to $0$ where it is not, whereas $\vec{\Omega_{\rm  Lens}}$ will behave analogously for the lens. The above expressions also include the hyper-parameters $\omega_{\rm  SrcFrac}$ and $\omega_{\rm  LensFrac}$, the practical role of which is to allow the source and lens contribution maps to attribute more pixels to values closer to $1$. Without these hyper-parameters only the brightest pixels are able to obtain a value near $1$, limiting the applicability of the contribution maps for the features discussed next.

\begin{figure*}
\centering
\includegraphics[width=0.32\textwidth]{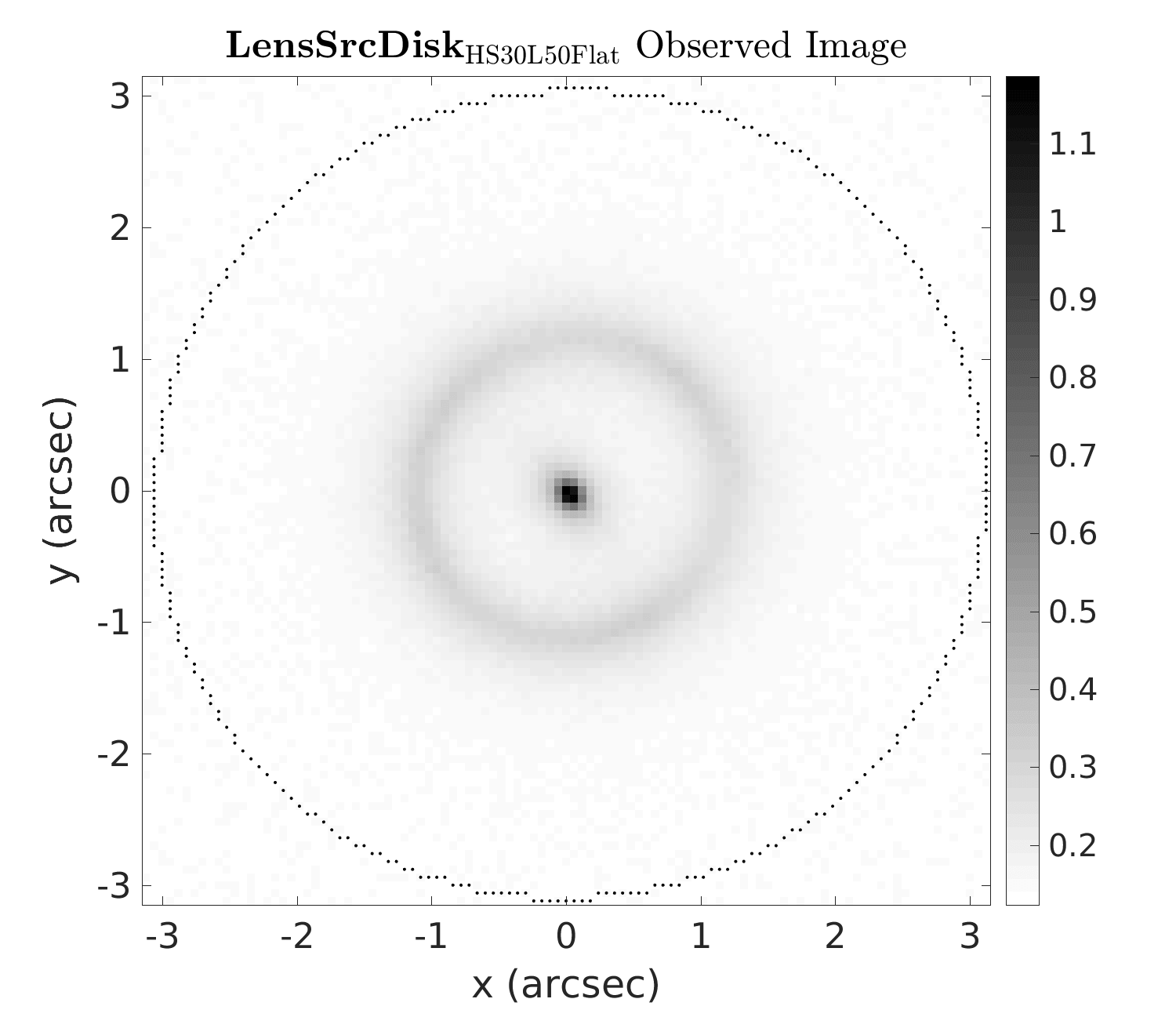}
\includegraphics[width=0.32\textwidth]{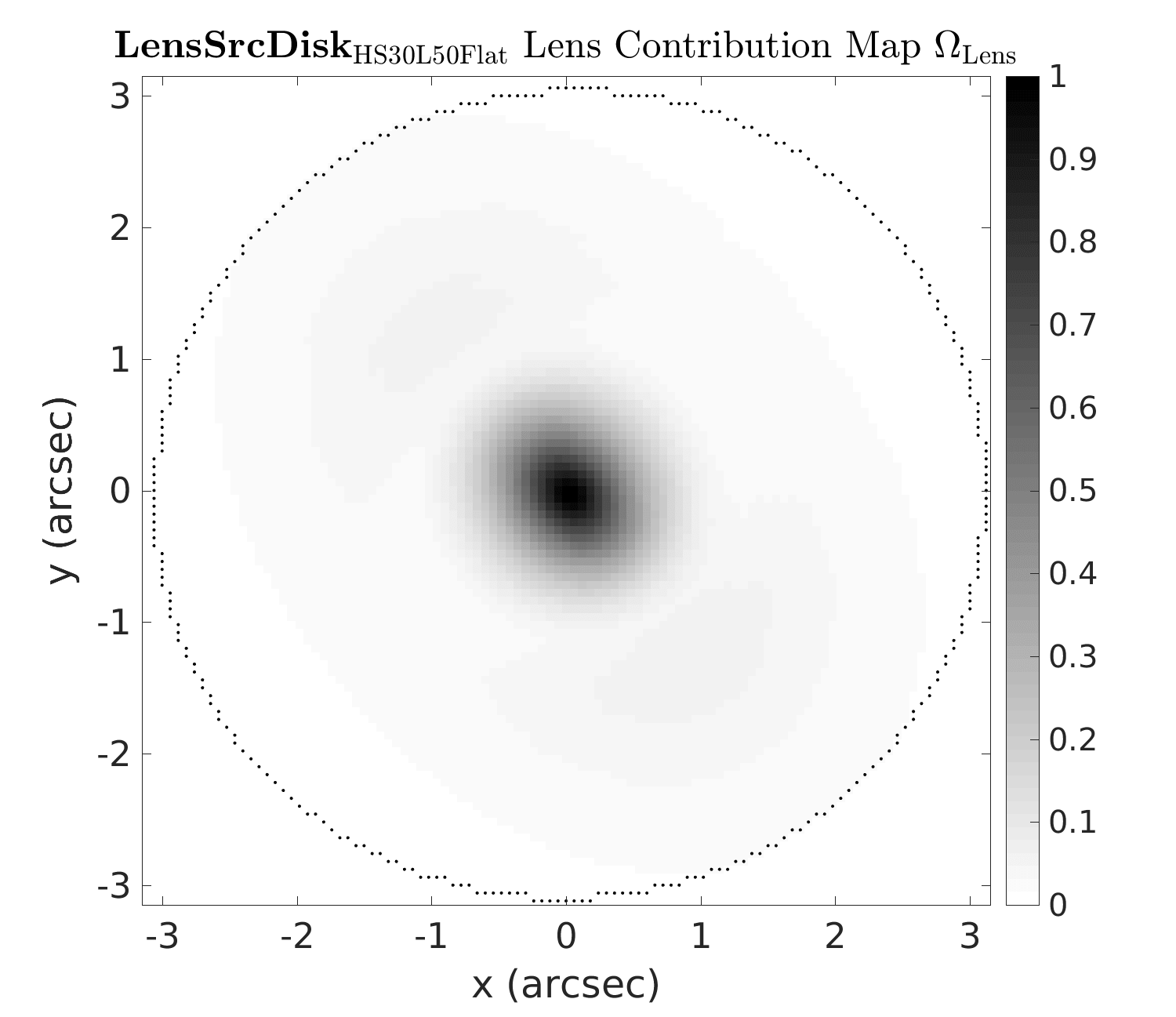}
\includegraphics[width=0.32\textwidth]{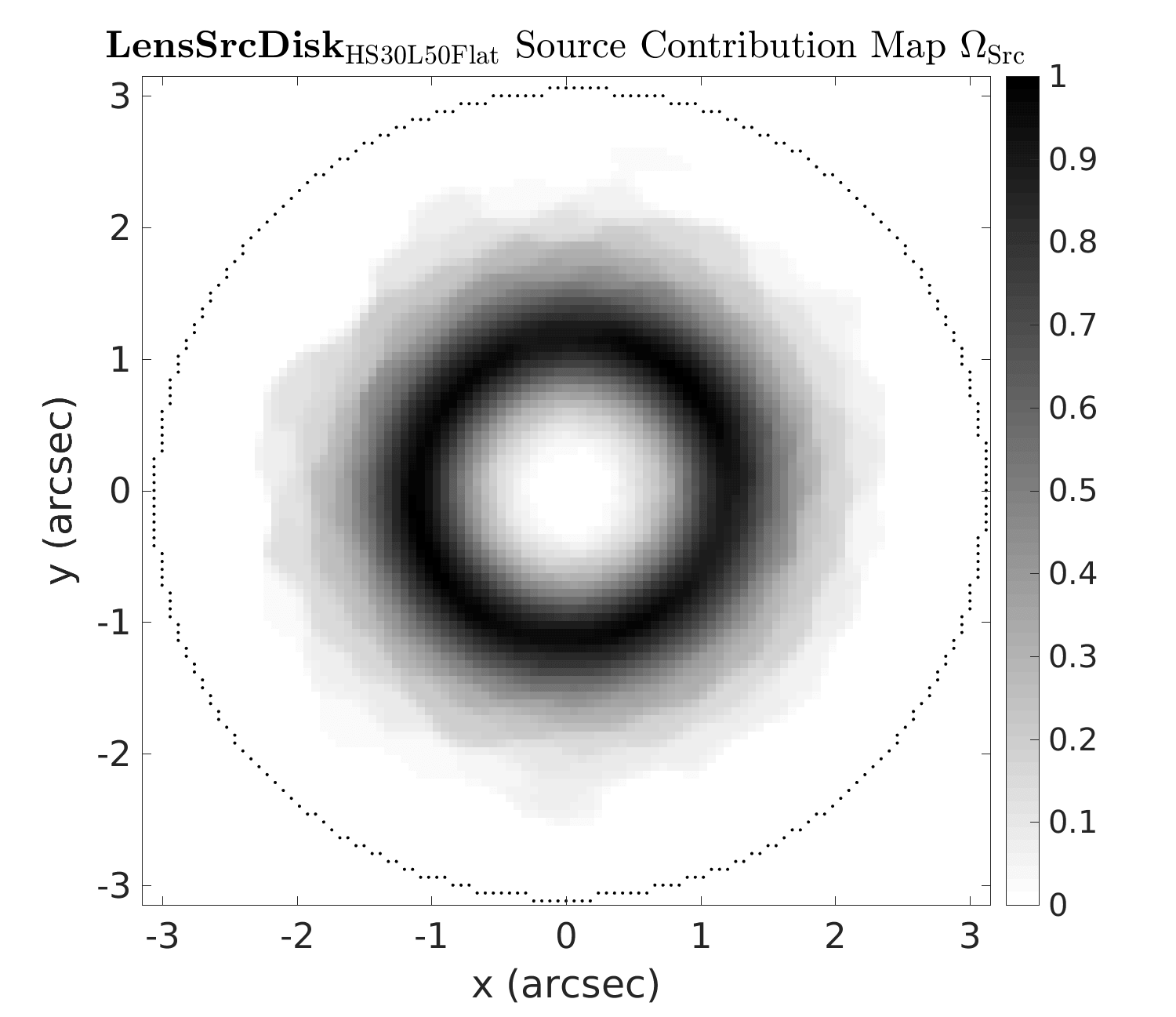}
\includegraphics[width=0.32\textwidth]{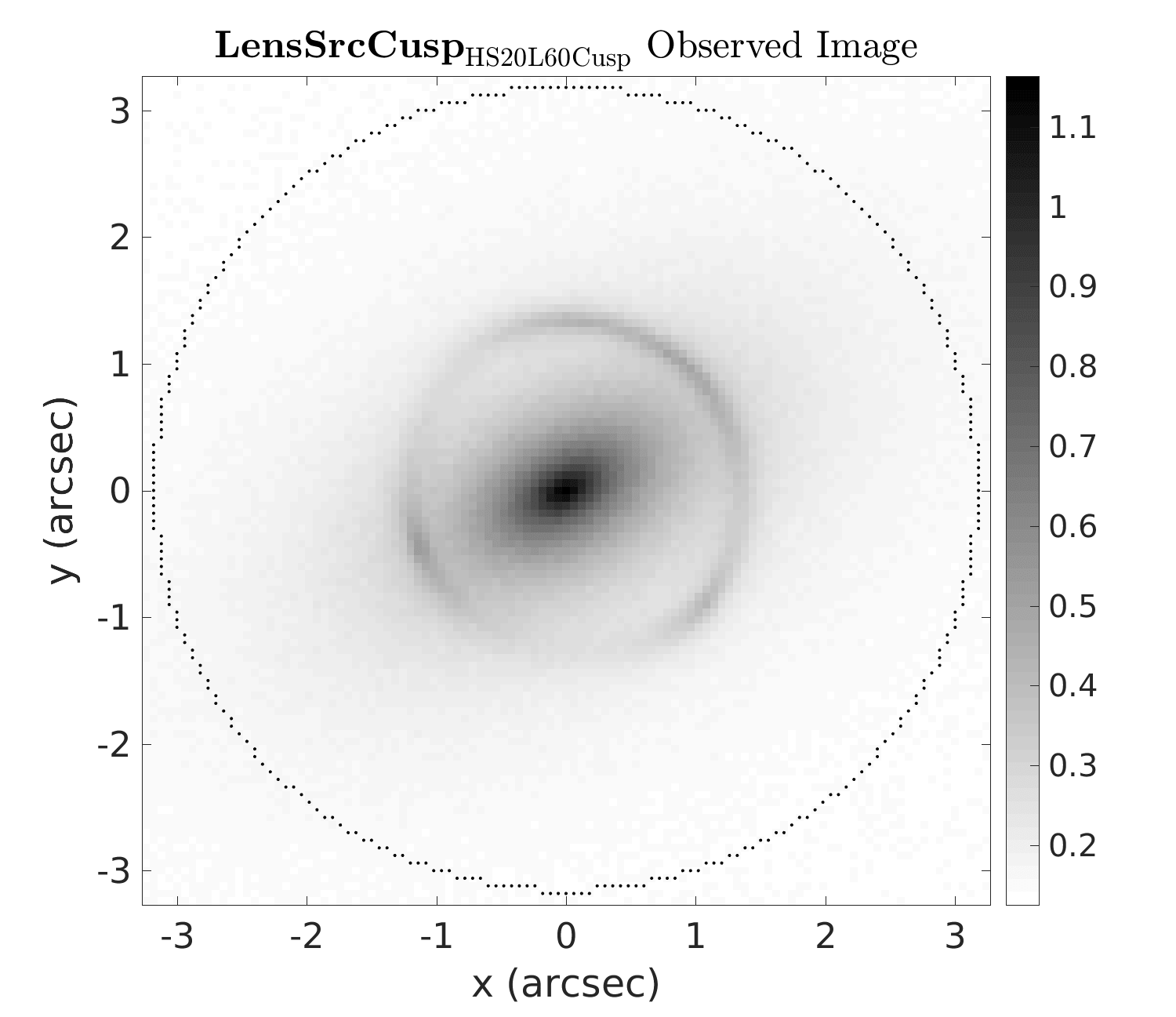}
\includegraphics[width=0.32\textwidth]{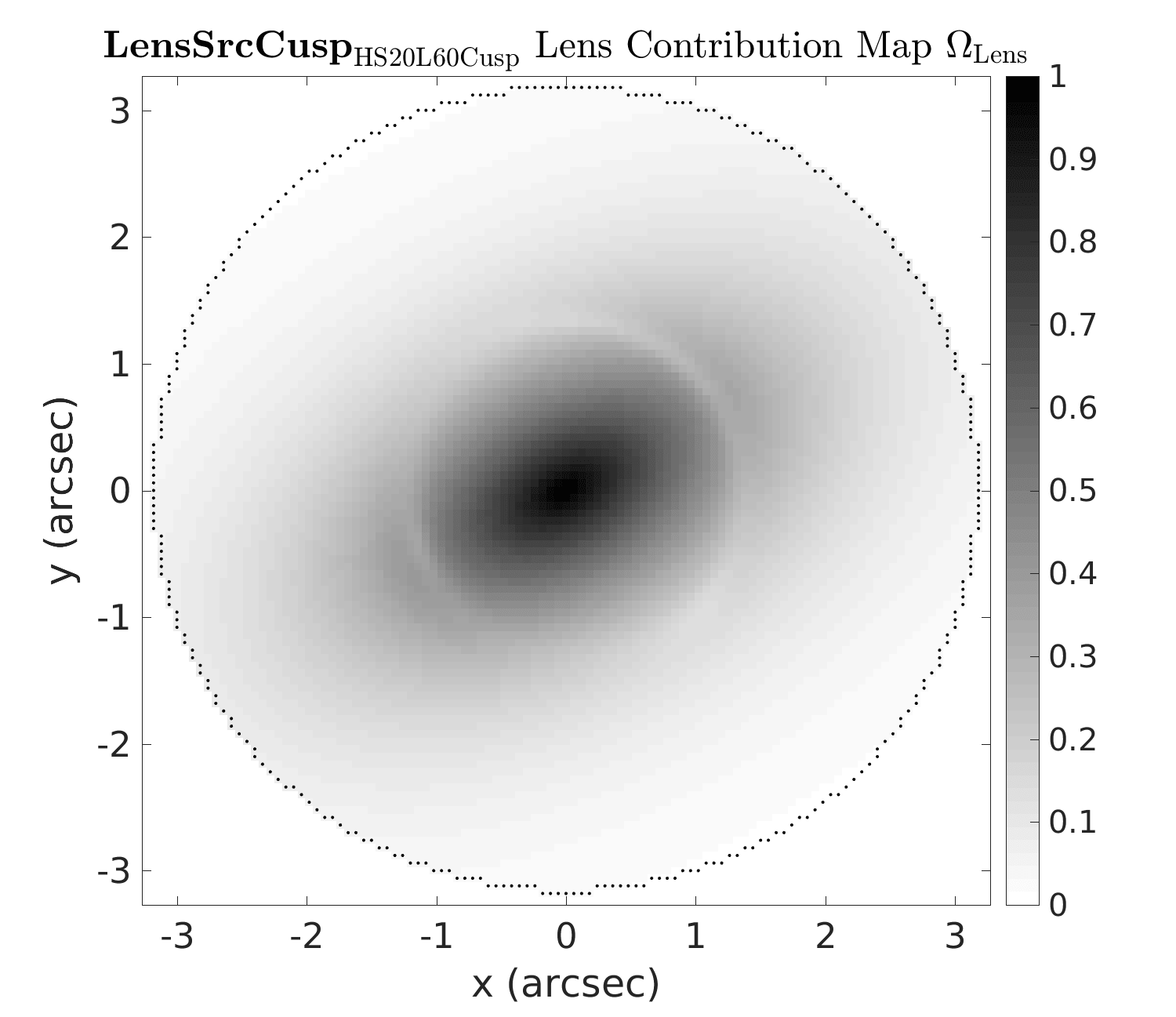}
\includegraphics[width=0.32\textwidth]{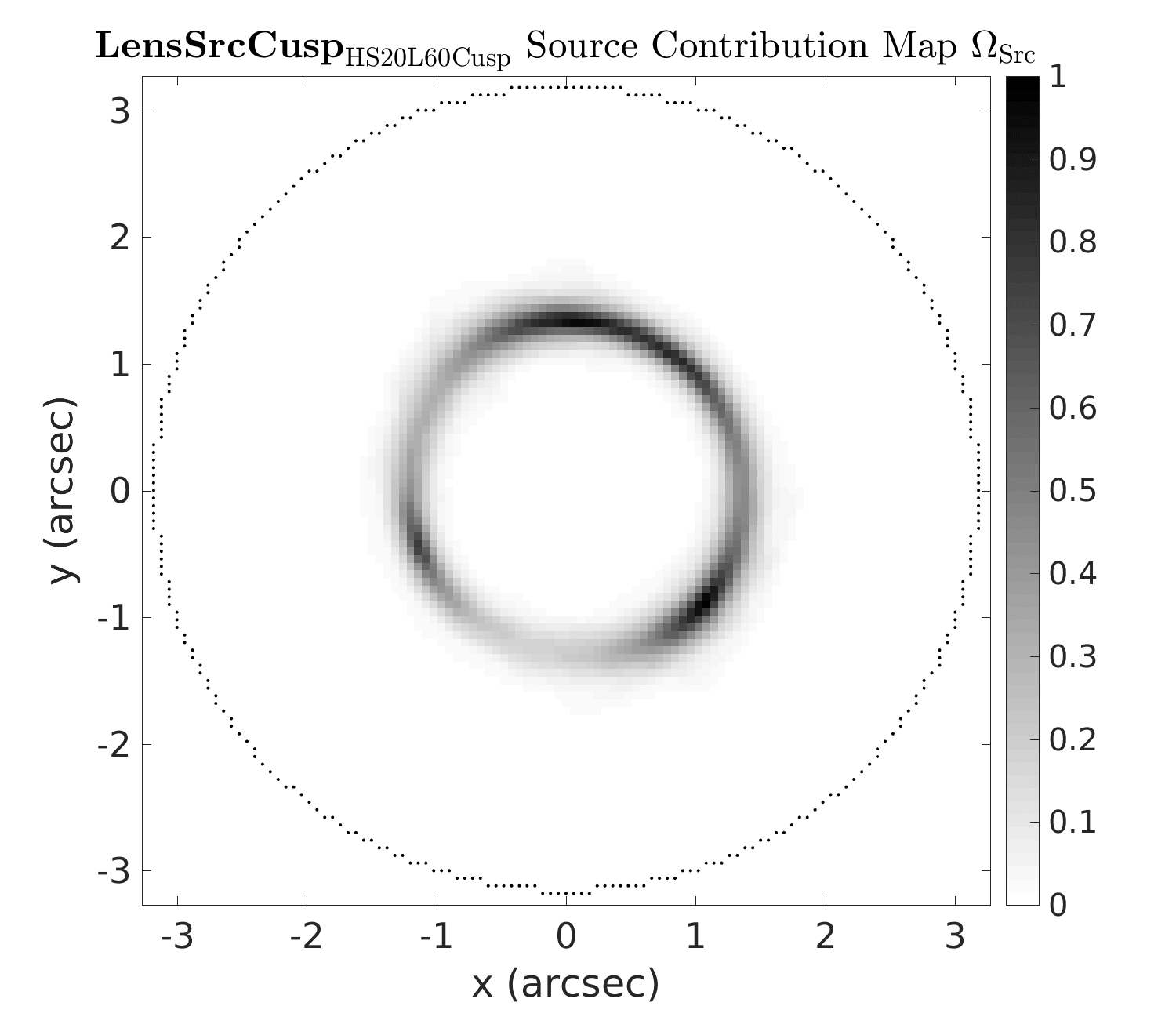}
\caption{The observed images (left panels), lens flux contribution maps $\vec{\Omega_{\rm  Lens}}$ (central panels) and source flux contribution maps $\vec{\Omega_{\rm  Src}}$ (right panels) described in section \ref{ErrorScale} for the images $\textbf{LensSrcDisk}_{\rm  HS30L50Disk}$ (top row) and $\textbf{LensSrcCusp}_{\rm  HS20L60Cusp}$ (bottom row). Figures show the result of using each image's input lens model to optimize all of the hyper-parameters described in this section, including $\omega_{\rm SrcFrac}$ and $\omega_{\rm  LensFrac}$. The contribution maps successfully split each pixel's flux contribution between the source and lens.}
\label{figure:FracAdapts}
\end{figure*}

Figure \ref{figure:FracAdapts} shows the flux contribution maps of the images $\textbf{LensSrcCusp}_{\rm  HS20L60Cusp}$ and $\textbf{LensSrcDisk}_{\rm  HS50L100Flat}$, where both, as expected, correctly trace either the source or lens.

\subsubsection{Luminosity-weighted Regularization}\label{RegAdapt}
The next three hyper-parameters introduce a luminosity-weighted regularization scheme, using the redefined regularization matrix $\vec{{H}_{\rm  \Lambda}}$ given in equation \ref{eqn:evidence2} and described in appendix \ref{AppReg}. A similar scheme is employed by \citet{Suyu2013} and \citet{Vegetti2014}.

To weight regularization by the lensed source's flux, each source pixel requires some measure of how much of the source's flux it contains before the actual source reconstruction is performed. To do this, $\vec{\Omega_{\rm  Src}}$ is used, summing over the $K$ image pixels allocated to each source pixel to compute the vector $\vec{v}$ as
\begin{equation}
v_{\rm  i} = \frac{\sum^{K}_{\rm  k=1} \Omega_{\rm  Src,k}}{K}, 
\label{eqn:LumReg1}
\end{equation}
where $i$ is again the source pixel number. Each element in $\vec{v}$ is divided by $K$ to normalize for the number of allocated image pixels, which can vary due to the k-means algorithm. The vector $\vec{V}$ is then computed, where each element is given by
\begin{equation}
V_{\rm  i} = \bigg[ \frac{v_{\rm  i}}{v_{\rm  max}} \bigg]^{L_{\rm  Lum}} .
\label{eqn:LumReg2}
\end{equation}
Once again, each element is divided by the maximum value of $\vec{v}$ to scale all values between zero and one and raised to the power of the hyper-parameter $L_{\rm  Lum}$. $\vec{V}$ is then used to compute the luminosity-weighted regularization value of each source pixel (see appendix \ref{AppReg}) as
\begin{equation}
\Lambda_{\rm  i} = \lambda_{\rm  Src} V_{\rm  i} + \lambda_{\rm  BG} (1-V_{\rm  i}) ,
\label{eqn:LumReg3}
\end{equation}
therefore leading to two regularization coefficients $\lambda_{\rm  Src}$ and $\lambda_{\rm  BG}$. 

The importance of luminosity-weighted regularization is that it divides source-plane regularization into two regions: (i) pixels that map to the lensed source; (ii) pixels that map to the background sky or central regions of the lens galaxy. The hyper-parameter $L_{\rm  Lum}$ controls the smoothness of the transition between these two regions, whereby higher values give a sharper transition and $L_{\rm  Lum} = 0$ reverts to the constant regularization scheme of N15. By using two regularization coefficients ($\lambda_{\rm  Src}$ and $\lambda_{\rm  BG}$) each region therefore receives its own level of regularization. 

\begin{figure*}
\centering
\includegraphics[width=0.47\textwidth]{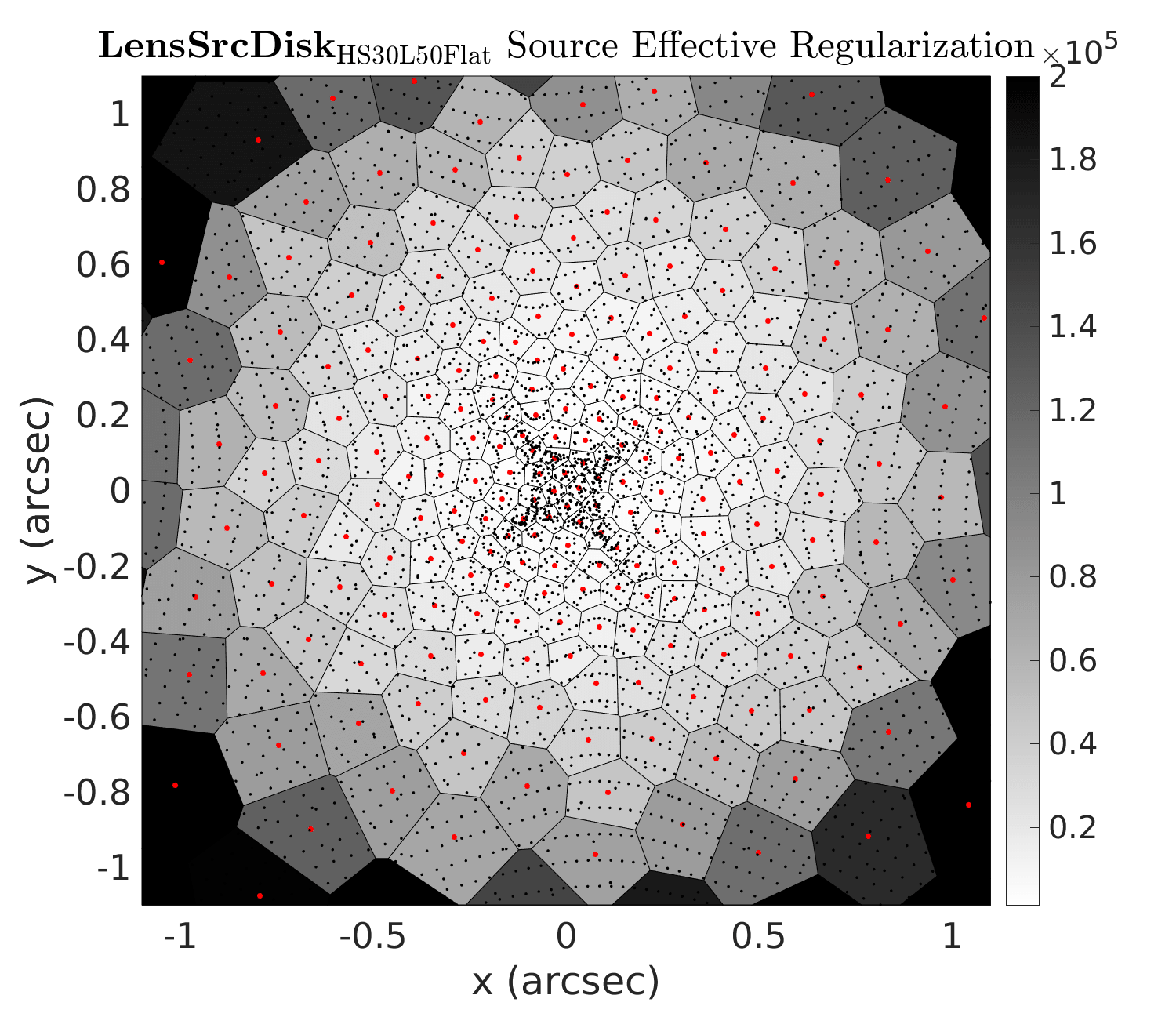}
\includegraphics[width=0.47\textwidth]{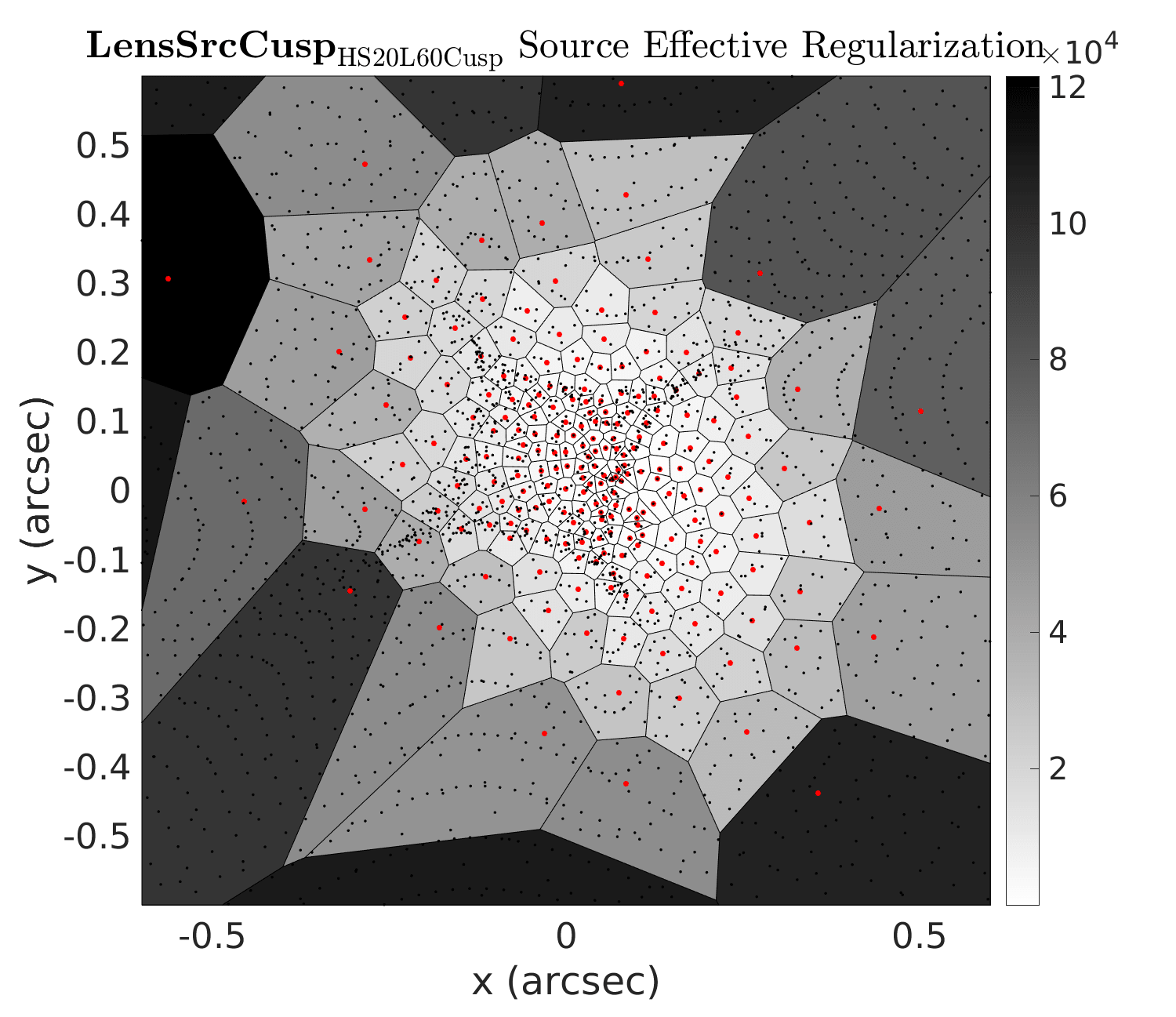}
\caption{An illustration of the luminosity-weighted regularization scheme described in section \ref{RegAdapt} using the images $\textbf{LensSersicDisk}_{\rm  HS30L50Disk}$ (left panel) and $\textbf{LensCuspySrc}_{\rm  HS20L60Cusp}$ (right panel). Black dots depict the locations of traced image-pixels and red dots are the centres of each Voronoi source pixel. Both panels correspond to the result of optimizing the hyper-parameters, including $\lambda_{\rm  Src}$, $\lambda_{\rm  Bg}$ and $L_{\rm  Lum}$, with the input lens model, thereby producing a non-constant regularization scheme. Grey scaling shows the effective regularization coefficient $\lambda_{\rm  eff}$ of each source pixel, which as described in appendix \ref{AppReg} represents the effective degree of regularization applied to that pixel. The constant regularization scheme that is produced by using only the hyper-parameter $\lambda$ is not shown, but gives all pixels the same value of $\lambda_{\rm  eff} = 610$ for the image $\textbf{LensSersicDisk}_{\rm  HS30L50Disk}$ and $\lambda_{\rm eff} = 230$ for $\textbf{LensCuspySrc}_{\rm  HS20L60Cusp}$. Luminosity-weighted regularization can be seen to reduce the degree of regularization applied to the source's brightest regions and increase it in its exterior regions, both contributing to an increase in ln$\epsilon$.}
\label{figure:RegAdapts}
\end{figure*}

Figure \ref{figure:RegAdapts} illustrates this, by showing the effective regularization coefficient $\lambda_{\rm  eff}$ applied to each source pixel (see appendix \ref{AppReg} for the exact definition of this quantity). The constant regularization scheme, which gives each source pixel the same value of $\lambda_{\rm eff}$ is not shown, but computes values of $\lambda_{\rm eff} = 610$ and $\lambda_{\rm eff} = 230$ for $\textbf{LensSersicDisk}_{\rm  HS30L50Disk}$ and $\textbf{LensCuspySrc}_{\rm  HS20L60Cusp}$ respectively. The luminosity-weighted scheme reduces regularization in the central regions of the source, facilitating a more detailed reconstruction of its light. Simultaneously, it increases the regularization of the source-plane's exterior pixels (where the source isn't located), allowing the method to fully correlate the exterior source pixels which only reconstruct the background sky. The constant regularization scheme is used in the early stages of the automated pipeline, before non-constant regularization has been appropriately set up.

\subsubsection{Sky Background}\label{SkyBG}
The data-vector $\vec{d}$ does not include the background sky, which is subtracted using each image's background-sky flux $f_{\rm  BG}$. However, for real data, $f_{\rm  BG}$ is estimated using the image itself (e.g. by taking a median value of a region of sky), which introduces uncertainty in $f_{\rm  BG}$. Therefore, to include this uncertainty, sky subtraction is controlled by the hyper-parameter $\omega_{\rm  Sky}$, where
\begin{equation}
\vec{d} = \vec{d'} + \omega_{\rm  Sky} f_{\rm  BG} ,
\label{eqn:BGSky}
\end{equation}
noting that $\vec{d}$ and $\vec{d'}$ must therefore be in units of electrons per second. Early in the analysis $\omega_{\rm  Sky}$ is fixed to zero. The prior assigned to $\omega_{\rm  Sky}$ can be chosen to match the uncertainties found by the sky estimation. The motivation behind incorporating the sky subtraction into the analysis is well documented, whereby a poor or uncertain sky subtraction can make it difficult to quantify the faint regions of a galaxy's extended light profile \citep{Haussler2013}. 

\subsubsection{Variance Scaling}\label{ErrorScale}

A baseline variance is assigned to each pixel as the quadrature sum of Gaussian background noise and Poisson photon error
\begin{equation}
 \sigma_{\rm  j,base} = \sqrt{ \sigma^2_{\rm  BG}  + d{\rm _j}} ,
\label{eqn:ErrorStd}
\end{equation}
where $\sigma_{\rm  BG}$ is the value of the overall background noise in counts (in this work the background sky and read noise) and where $d{\rm _j}$ has been converted from electrons per second to counts. This gives the image's `baseline variance-map', with its corresponding $\chi^2$ values termed $\chi_{\rm base}^2$. For real imaging data, equation \ref{eqn:ErrorStd} could contain additional terms due to other aspects of the data reduction (e.g. hot pixels, cosmic rays, dithering etc). The initial stages of an {\tt AutoLens} analysis use the baseline variance map. However, after these initial stages, the variances may be increased in regions of the image where a poor fit is obtained.

The next set of hyper-parameters thus offer {\tt AutoLens} the ability to perform this variance scaling, therefore producing a `scaled variance-map' with corresponding scaled $\chi_{\rm scale}^2$ values. The variances are scaled separately for the background, source and lens, by using the flux contributions maps and the expression
\begin{eqnarray}
\label{eqn:NoiseMap}
 \sigma_{\rm  j, scale} = \sigma_{\rm  j,base} + \omega_{\rm  BG} \sigma_{\rm  BG} 
+  \omega_{\rm  Src} \sqrt{d{ _j}} (\Omega_{\rm  Src,j})^{\omega_{\rm  Src2}} + \nonumber \\
\omega_{\rm  Lens} \sqrt{d{ _j}} (\Omega_{\rm  Lens,j})^{\omega_{\rm  Lens2}} ,
\end{eqnarray}
where $\sigma_{\rm  BG}$ and $d{\rm _j}$ are again in counts. The $\omega$ terms are all hyper-parameters which scale the variances of image pixels in different regions of the observed image. If the lens's light is not modeled the corresponding $\omega_{\rm  Lens}$ terms are not included in equation \ref{eqn:NoiseMap} or the hyper-parameter optimization. The scaled variances cannot go below their baseline values because the method requires that $\omega_{\rm  BG} \geq 0$, $\omega_{\rm  Src} \geq 0$ and $\omega_{\rm  Lens} \geq 0$. The background sky variances are scaled because there is uncertainty in the background sky subtraction.

\begin{table*}
\resizebox{\linewidth}{!}{
\begin{tabular}{ l | l | l l l | l | l } 
\multicolumn{1}{p{1.6cm}|}{\centering \textbf{Image}} 
& \multicolumn{1}{p{2.3cm}|}{\centering \textbf{Implementation}} 
& \multicolumn{1}{p{2.2cm}}{\textbf{Parameters ($3 \sigma$)}}  
& \multicolumn{1}{p{2.2cm}}{} 
& \multicolumn{1}{p{2.2cm}|}{} 
& \multicolumn{1}{p{2.2cm}}{\textbf{Parameters ($1 \sigma$)}} 
& \multicolumn{1}{|p{1.2cm}}{ln$\epsilon$} 
\\ \hline
& & & & & \\[-4pt]
$\textbf{SrcBD}_{\rm  HS50NLBD}$  & \textbf{Basic} & $\theta_E = 1.2184^{+0.0672} _{\rm  -0.1011}$ & $q = 0.6802^{+0.1296} _{\rm  -0.1527}$ & $\alpha = 1.7342^{+0.02004} _{\rm  -0.2202}$ & $\alpha = 1.7342^{+0.0468} _{\rm  -0.0725}$ & 53297 \\[2pt]
$\textbf{SrcBD}_{\rm  HS50NLBD}$  & \textbf{Adaptive}  & $\theta_E = 1.2047^{+0.01520} _{\rm  -0.01700}$ & $q = 0.7000^{+0.0279} _{\rm  -0.0261}$ & $\alpha = 1.7082^{+0.0386} _{\rm  -0.0442}$ & $\alpha = 1.7082^{+0.01390} _{\rm  -0.0078}$ & 55518 \\[2pt]
\hline
& & & & & \\[-4pt]
$\textbf{SrcBD}_{\rm  HS50NLBD}$  & \textbf{Basic} & $\theta_E = 1.3785^{+0.0408} _{\rm  -0.0456}$ & $q = 0.4925^{+0.0535} _{\rm  -0.0547}$ & $\alpha = 2.0$ & $\alpha = 2.0$ & 53292 \\[2pt]
$\textbf{SrcBD}_{\rm  HS50NLBD}$  & \textbf{Adaptive} & $\theta_E = 1.3899^{+0.0121} _{\rm  -0.0201}$ & $q = 0.4818^{+0.0203} _{\rm  -0.0181}$ & $\alpha = 2.0$ & $\alpha = 2.0$ & 53455 \\[2pt]
$\textbf{SrcBD}_{\rm  HS50NLBD}$  & \textbf{Adaptive} & $\theta_E = 1.3899^{+0.0121} _{\rm  -0.0201}$ & $q = 0.4818^{+0.0203} _{\rm  -0.0181}$ & $\alpha = 2.0$ & $\alpha = 2.0$ & 54300 \\[2pt]
\hline
& & & & & \\[-4pt]
$\textbf{LensLightBD}_{\rm  HS50NLBD}$  & \textbf{Basic} & $n_{\rm l} = 3.7567^{+0.0125} _{\rm  -0.0222}$ & $R_{\rm l} = 4.5132^{+0.0021} _{\rm  -0.0083}$    & & $R_{\rm l} = 4.5132^{+0.0001} _{\rm  -0.0003}$  & 54375 \\[2pt]
$\textbf{LensLightBD}_{\rm  HS50NLBD}$  & \textbf{Basic} & $\theta_E = 1.2275^{+0.0053} _{\rm  -0.0031}$ & $q = 0.7333^{+0.0081} _{\rm  -0.0074}$ & $\alpha = 2.0$ & & 54375 \\[2pt]
$\textbf{LensLightBD}_{\rm  HS50NLBD}$  & \textbf{Adaptive}  & $n_{\rm l} = 2.3453^{+0.0175} _{\rm  -0.0199}$ & $R_{\rm l} = 3.3554^{+0.1052} _{\rm  -0.1532}$    & & $R_{\rm l} = 3.3554^{+0.0784} _{\rm  -0.1193}$  & 55075 \\[2pt]
$\textbf{LensLightBD}_{\rm  HS50NLBD}$  & \textbf{Adaptive} & $\theta_E = 1.2298^{+0.0021} _{\rm  -0.0013}$ & $q = 0.7411^{+0.0051}_{\rm  -0.0064}$ & $\alpha = 2.0$ & & 55075 \\[2pt]
\end{tabular}
}
\caption{A sub-set of lens model parameters calculated for the images $\textbf{SrcBD}_{\rm  HS50NLBD}$ (top five rows) and $\textbf{LensLightBD}_{\rm  HS40L60BD}$ (bottom four rows), using either the `basic implementation' (which omits the adaptive image and source analysis features described above) or the `adaptive implementation' (which uses them). The input mass model parameters for the image $\textbf{SrcBD}_{\rm  HS50NLBD}$ are $\theta_{\rm Ein} = 1.2"$, $q = 0.7$ and $\alpha = 1.7$ and for $\textbf{LensLightBD}_{\rm  HS40L60BD}$ are $\theta_{\rm Ein} = 1.0"$, $q = 0.8$ and $\alpha = 2.05$. In the top two rows, $\textbf{SrcBD}_{\rm  HS50NLBD}$ is fitted with a $SPLE$ profile. The next three rows show the results of fitting it with an $SIE$ profile, to demonstrate the effect of assuming an approximate mass model, with the fourth row showing results where variance scaling is manually switched off. The final four rows show fitting the image $\textbf{LensLightBD}_{\rm  HS40L60BD}$, generated using a $Sersic$ + $Exp$ + $SPLE$ model (see table \ref{table:SimModels}), with a $Sersic$ + $SPLE$ model, to demonstrate the effects of fitting a mismatched light profile. The second column displays whether the basic or adaptive implementation was used. Columns three to six give parameter estimates. The final column gives the value of Bayesian evidence that results from the analysis, which can be seen to increase for the adaptive implementation in all cases. These models are visualized in figures \ref{figure:AdaptDemoSrc}, \ref{figure:AdaptDemoSrcWrong} and \ref{figure:AdaptDemoLens}. }
\label{table:AdaptDemoModels}
\end{table*}

During testing, it emerged that the method must not be allowed to scale variances to arbitrarily high values. This prevents the S/N of image pixels that actually contain significant flux from being unrealistically small. To implement this, first, the maximum values of $\chi_{\rm base}^2$ are determined using the cuts $\mathbf{\Omega_{\rm Src}} > 0.75$ and $\mathbf{\Omega_{\rm  Lens}} > 0.75$. These provide an indication of how well the lens model currently fits the source and lens galaxies. The highest values allowed for the hyper-parameters $\omega_{\rm  Src}$ and $\omega_{\rm  Lens}$ are then set such that their corresponding $\chi_{\rm scale}^2$ values cannot be scaled below a target value $\chi^2$ value, which we set to $10$. If either $\chi_{\rm base}^2$ value is already below $10$ (because the lens model is already fitting the data accurately), variance scaling for that component is switched off and its corresponding hyper-parameters are omitted. For this reason, variance scaling is not required in cases where the lens model that created the simulated image matches the model used to fit it. We therefore defer its demonstration until the next subsection when this is not the case.

\subsubsection{Implementation}

Whereas N15 used just one hyper-parameter, {\tt AutoLens} now uses up to 14 simultaneously. Setting these parameters by maximizing ln$\epsilon$ can therefore no longer rely on simple iteration as it did in N15. Instead, a fully non-linear {\tt MultiNest} search is performed, which treats every hyper-parameter as a free parameter. This again uses constant efficiency mode and importance sampling and the priors used for each hyper-parameter are discussed in appendix \ref{AppPrior}. The figures shown in the previous section were generated by using this non-linear optimization.

In general, the lens model and hyper-parameters are sampled separately from one another. That is, a lens model is estimated which is used to optimize the hyper-parameters, which are next fixed to improve the lens model, and so on. Bayesian inference therefore retains the three-level structure described in \citet{Dye2008}, where the linear source inversion forms the inner-most level. However, {\tt AutoLens} can include the hyper-parameters in the non-linear search of the lens model, thus sampling any of the parameters described above alongside the mass and light models. This feature is used at various points throughout the analysis pipeline and the importance of this will be discussed in section \ref{SLPipeline}.

It is here that the fully randomized source-plane discretization discussed in section \ref{SLIRecap} is important, where the random seed of source-plane clustering was updated to always produce a different source-plane discretization. When the lens's mass model is fixed (as it is for a hyper-parameter optimization), its corresponding deflection angles are also fixed, therefore also fixing the source-plane coordinates of the traced image pixels. Therefore, for the implementation of clustering used in N15 the exact same source-plane discretization would have been used throughout the entire hyper-parameter optimization. This is problematic, as a particular source-plane discretization favors a particular combination of hyper-parameters, which in turn favor a particular lens model, leading the overall analysis to end up biased towards a specific parameter set. Thus, by fully randomizing the source-plane discretization, this biasing is removed.

\section{Demonstration}\label{AdaptDemo}

The importance of these features is now demonstrated by analysing two simulated images with two different hyper-parameter strategies: (i) the `basic implementation', which uses only $N_{\rm s}$ and $\lambda$ as hyper-parameters and therefore omits the adaptive image and source features above. This closely resembles N15, with the only difference being that N15 used a fixed value of $N_{\rm s}$. (ii) The `adaptive implementation', which uses all of the features above, giving up to fourteen free hyper-parameters if variance scaling is activated for both components. Both implementations use a {\tt MultiNest} search to set the hyper-parameters and the analysis is performed using the automated analysis pipeline introduced in the next section. Whilst readers are not yet aware of how this operates, the specific details are not important for providing a simple demonstration of the adaptive image and source features. However, it is worth noting that: (i) this analysis does not use a fixed lens model, but determines the lens model via multiple {\tt MutliNest} runs and (ii) initial runs assume a simplified lens model (a $Sersic$ light profile and $SIE$ mass profile). The second point is of particular importance for demonstrating variance scaling.

Figures \ref{figure:AdaptDemoSrc}, \ref{figure:AdaptDemoSrcWrong} and \ref{figure:AdaptDemoLens} show the results of this analysis. These figures follow the same format, showing the reconstructed model image, residuals, $\chi^2$ image (residuals divided by the baseline or scaled variances squared, equation \ref{eqn:ChiSqSrc}) and reconstructed source (figure \ref{figure:AdaptDemoSrc}) or baseline / scaled variance maps (figures \ref{figure:AdaptDemoSrcWrong} and \ref{figure:AdaptDemoLens}) for the basic implementation (top rows) and adaptive implementation (middle and bottom rows). The input lens models for both images are given in table \ref{table:SimModels}.

\subsection{Source Modeling - Correct Mass Model}

\begin{figure*}
\centering
\includegraphics[width=0.24\textwidth]{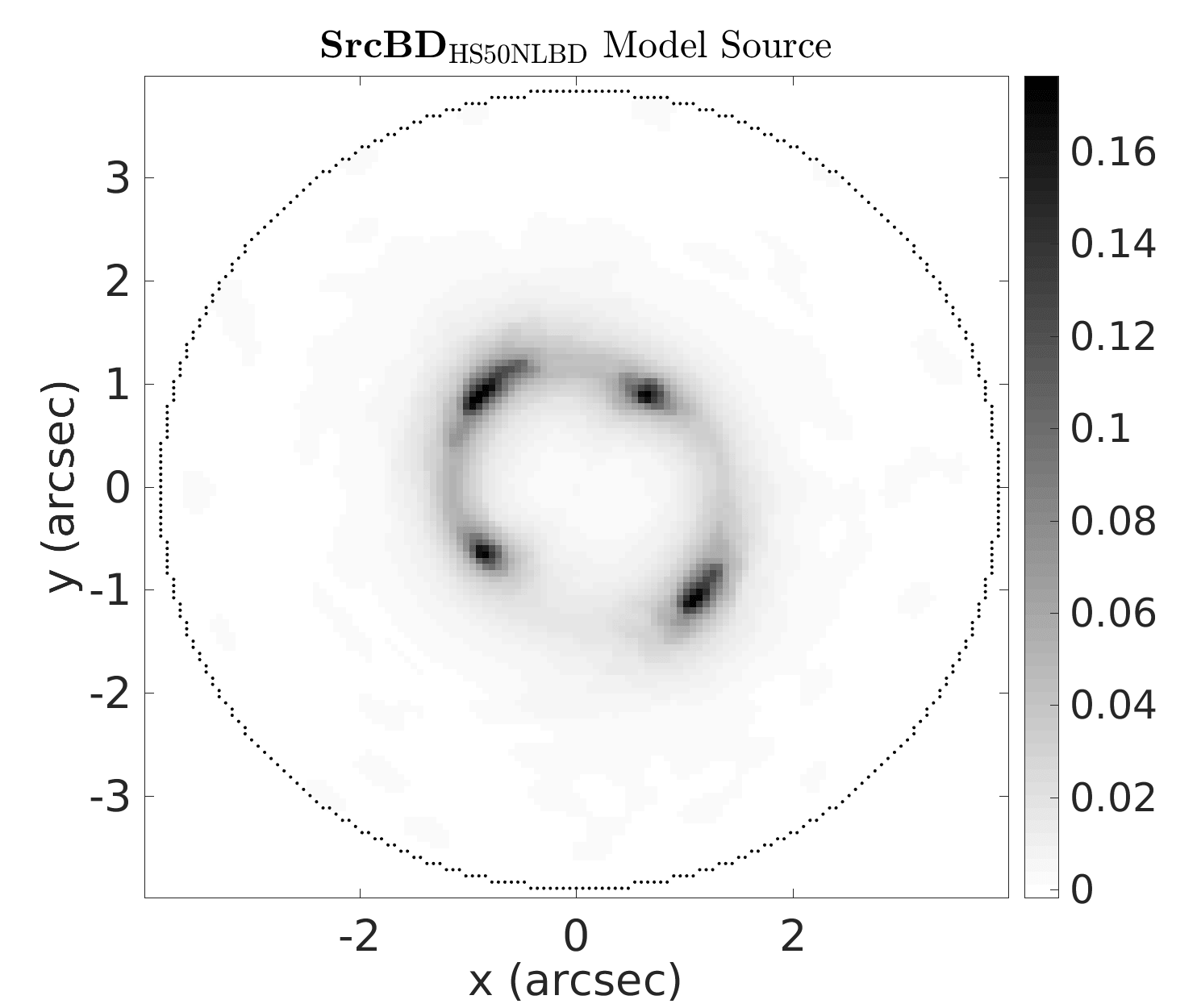}
\includegraphics[width=0.24\textwidth]{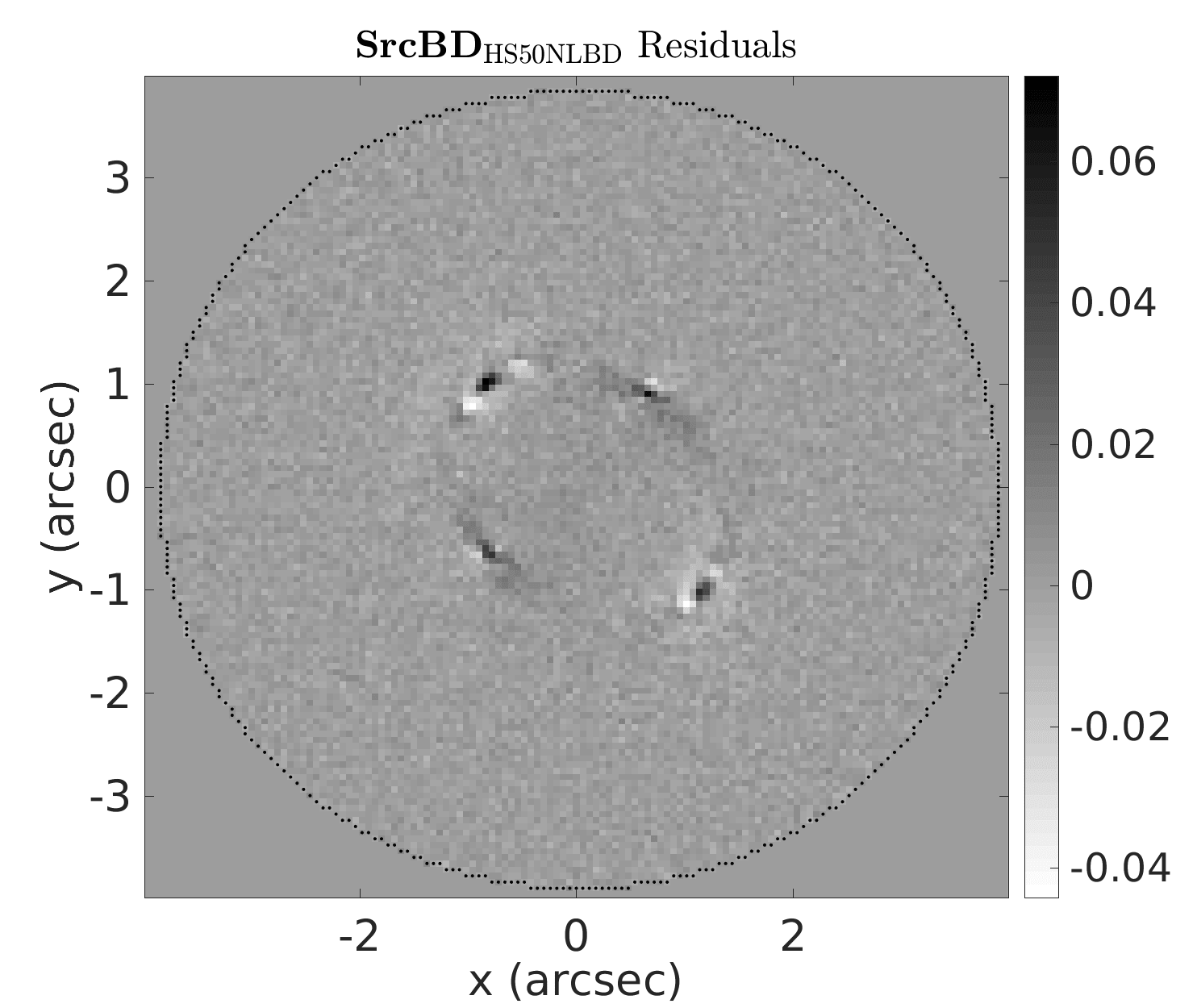}
\includegraphics[width=0.24\textwidth]{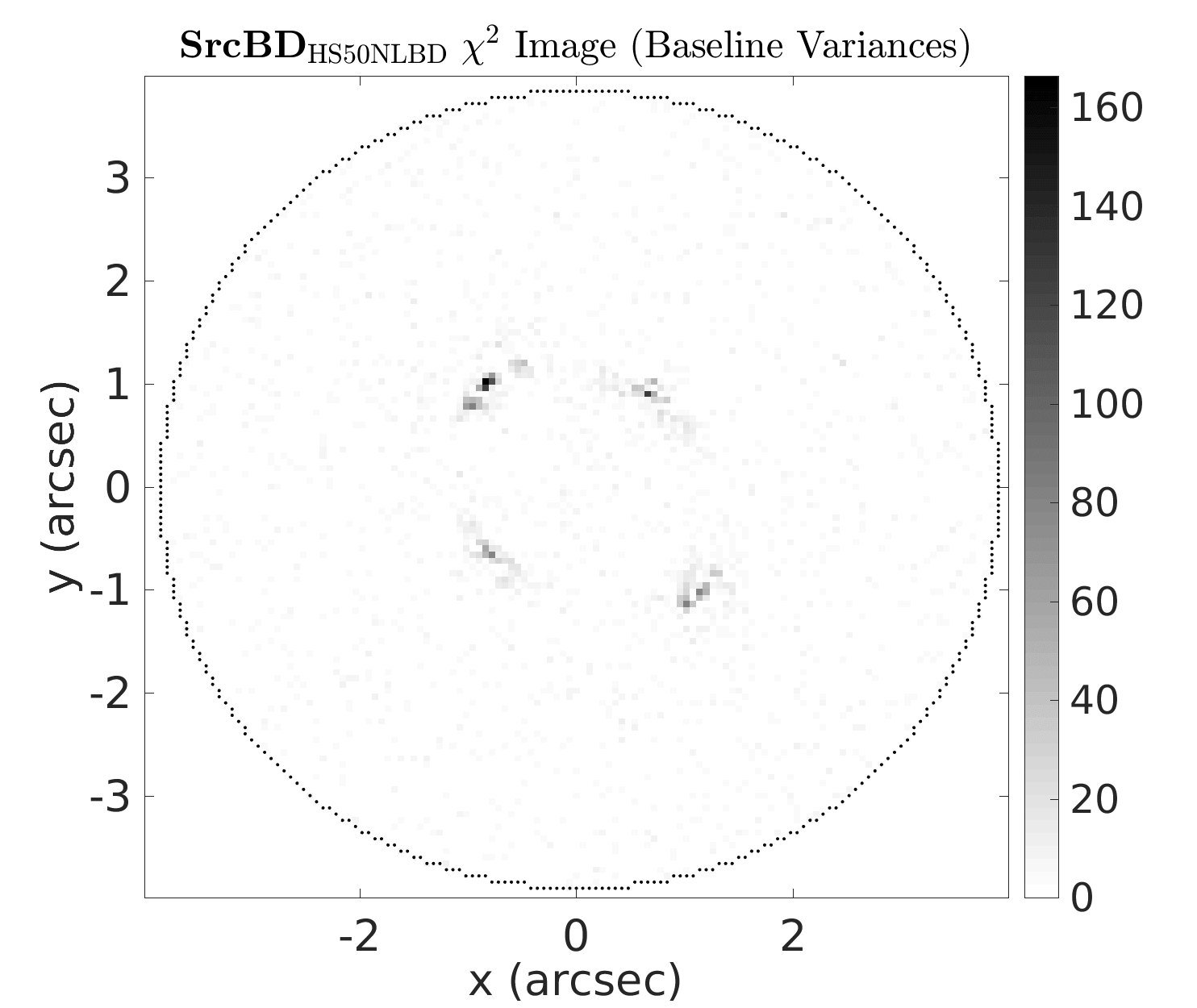}
\includegraphics[width=0.232\textwidth]{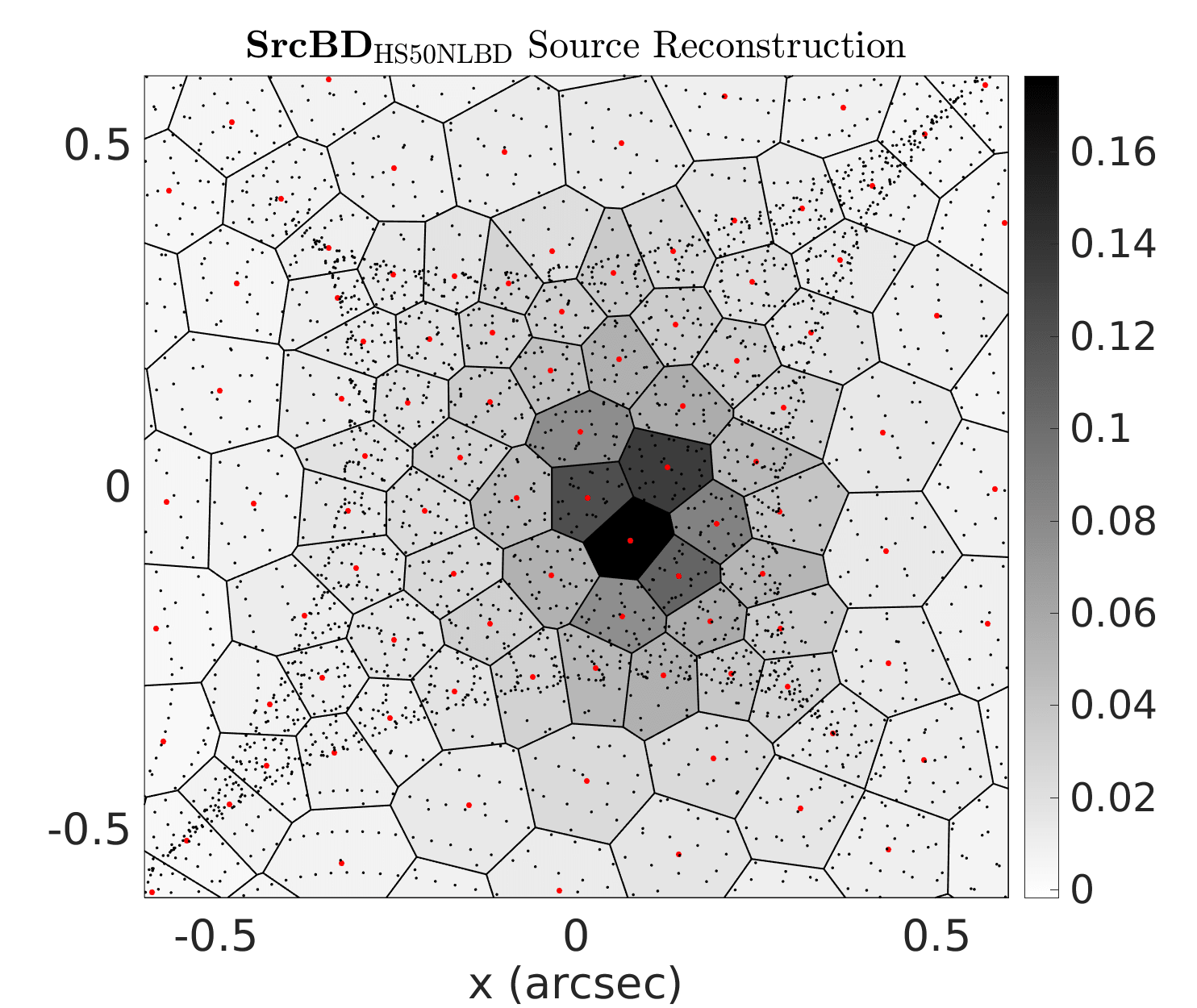}
\includegraphics[width=0.24\textwidth]{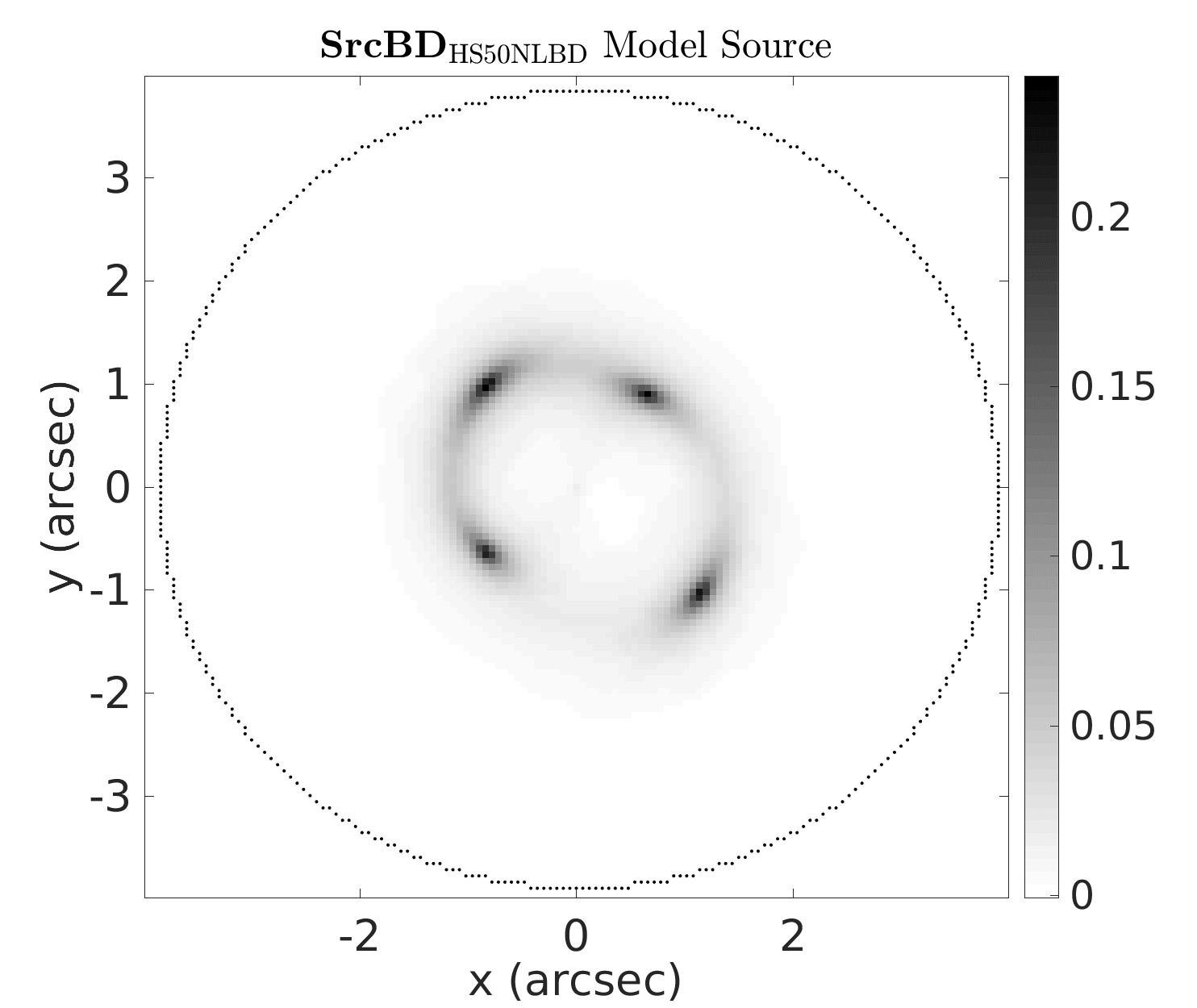}
\includegraphics[width=0.24\textwidth]{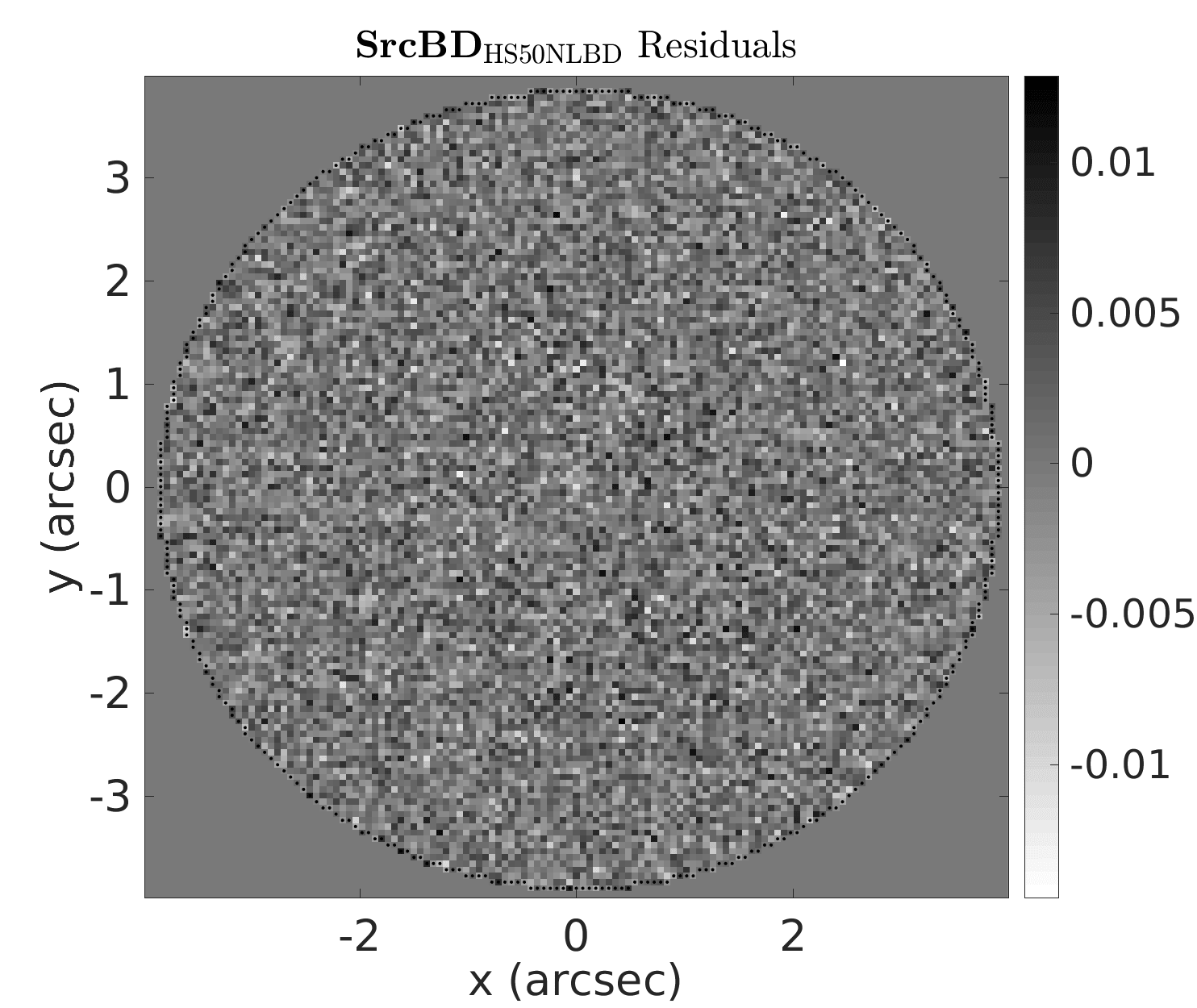}
\includegraphics[width=0.24\textwidth]{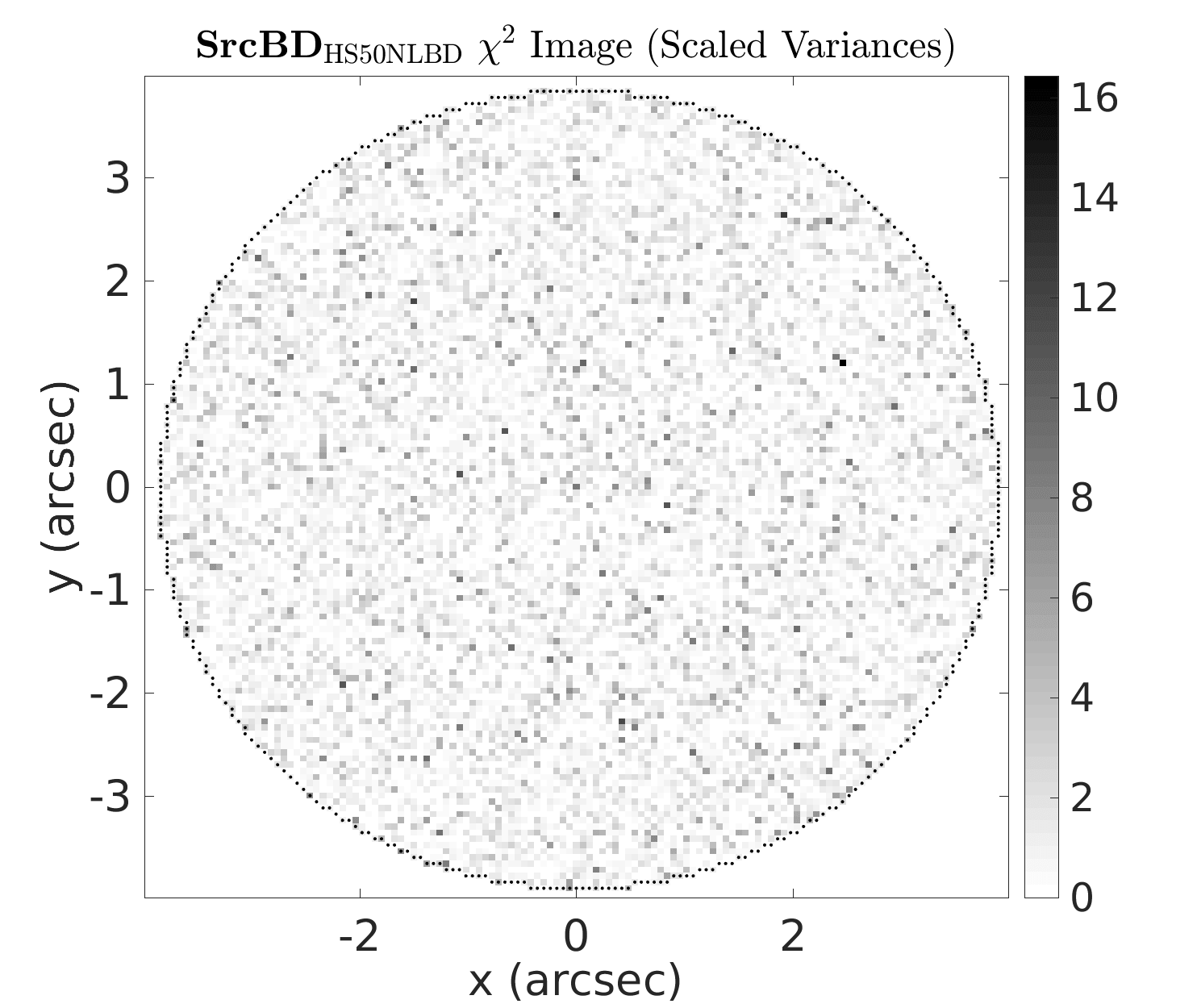}
\includegraphics[width=0.232\textwidth]{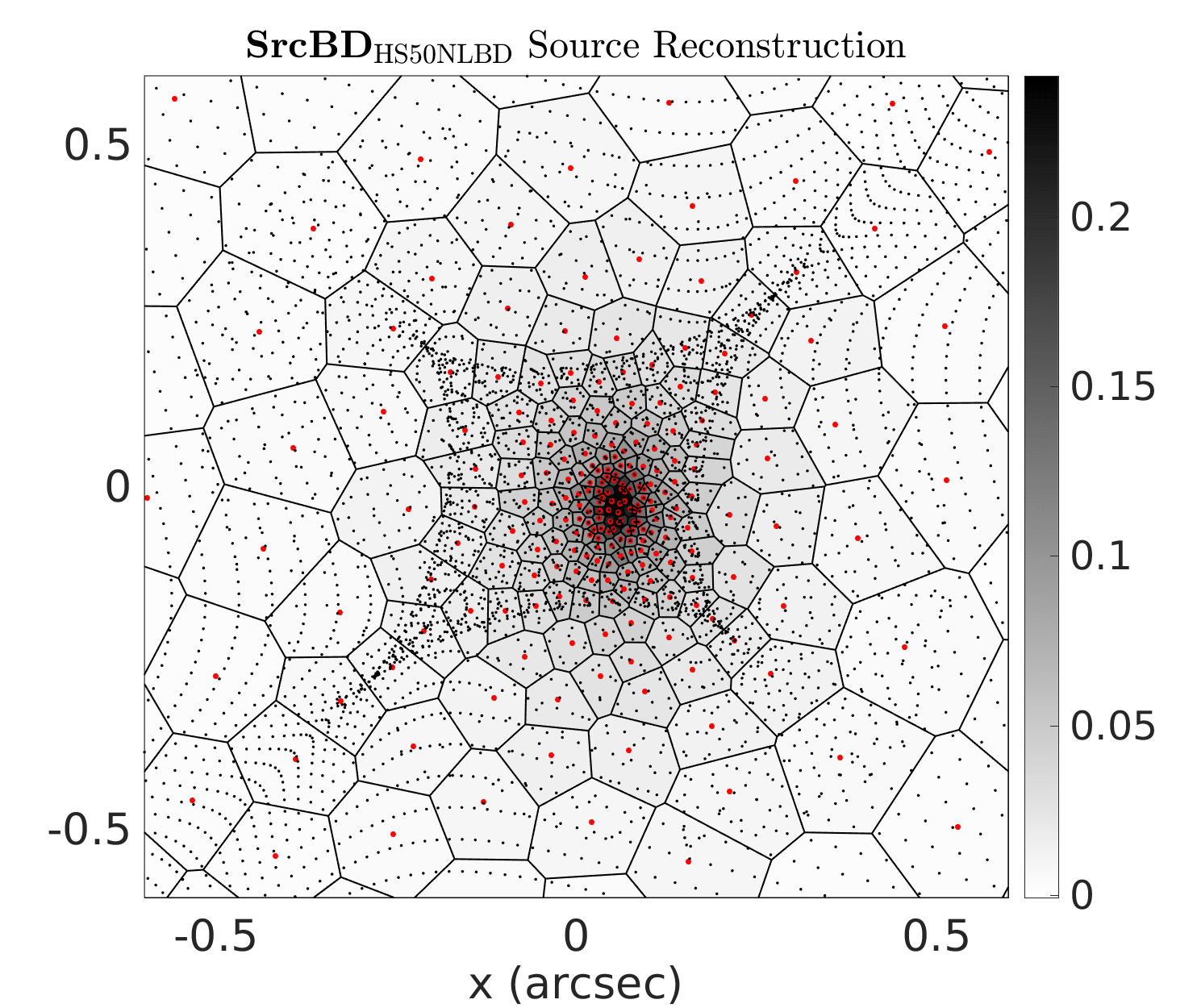}
\caption{The model source, residuals, $\chi^2$ images and source reconstructions for the analysis of the image $\textbf{SrcBD}_{\rm  HS50NLBulge}$, using either the basic implementation (top row, adaptive image and source analysis switched off) or adaptive implementation (bottom row, adaptive image and source analysis turned on). The resulting lens models are given by the first two rows of table \ref{table:AdaptDemoModels}. The basic implementation can be seen to fit the image poorly, leaving noticeable residuals in each multiple image, which dominate the model's overall $\chi^2$ value. The adaptive implementation gives featureless residuals and a Gaussian $\chi^2$ image.} 
\label{figure:AdaptDemoSrc}
\end{figure*}

The first issue arises when modeling sources with a cuspy and rapidly changing light profile. This is illustrated in figure \ref{figure:AdaptDemoSrc} using the image $\textbf{SrcBD}_{\rm  HS50NLBD}$. Significant residuals can be seen in the basic implementation's reconstruction of the image's bright, high S/N pixels, which causes a small sub-set of image pixels to obtain large $\chi_{\rm base}^2$ values. This means that the overall $\chi^2$ is constrained by only a small portion of data; approximately $5$ per cent of image pixels contribute to over $90$ per cent of the overall $\chi^2$ value. Whilst this is clearly not ideal for any form of data analysis, table \ref{table:AdaptDemoModels} shows that the basic implementation still computes the correct lens model, suggesting that the poor residuals and skewed $\chi^2$ distribution does not bias lens modeling. However, as discussed next, this holds only for simulated images where the mass model matches exactly the lens's true mass profile. For real imaging this is not the case, thus these issues must be corrected. 

The issue arises because of the basic implementation's source-plane pixelization and regularization scheme. By adapting to the magnification, pixels in the source's central regions (where its intrinsic light profile is most rapidly changing) cover roughly the same area as those further out (where its light profile is flatter). However, although the pixels in these central regions are reconstructing a more rapidly declining light profile, they are regularized with the same $\lambda$ as those further out. Therefore, when setting $\lambda$, the method has to compromise between a $\lambda$ low enough to accurately reconstruct the source's central regions but also high enough to correlate the source pixels further out. A compromised and intermediate value of $\lambda$ is ultimately calculated. This `over smooths' the reconstructed source's central light, producing the residuals seen in figure \ref{figure:AdaptDemoSrc} (where the fact these are the highest S/N image pixels inflates their $\chi^2$ contribution). The exterior regions of the source-plane (which map to background sky in the image) are simultaneously `under-regularized', in the sense that an unnecessarily high number of correlated source pixels are used to fit the regions of the image where the source is not present. Altogether, a reduced value of ln$\epsilon$ is inferred. The fact $N_s$ is a free parameter for the basic implementation demonstrates that a higher source-plane resolution by itself cannot alleviate these problems.

The bottom row of figure  \ref{figure:AdaptDemoSrc} shows that the adaptive implementation removes this problem, producing nearly featureless residuals and a $\chi^2$ image fully consistent with Gaussian noise. Table \ref{table:AdaptDemoModels} confirms that this comes with an increase in ln$\epsilon$ and shows a reduction in the lens model parameter errors, suggesting the basic implementation's skewed $\chi^2$ distribution produced over-estimated errors because it constrained the model with a small subset of the available image data. For this analysis, variance scaling was switched off, as the baseline $\chi^2$ values in the lensed source model were all below $10$.

In this example, the improved pixelization and regularization scheme both contribute to this. The congregation of smaller source pixels around the source's central cusp of light provides a better spatial sampling of its rapidly changing surface brightness, meaning it reconstructs the source more accurately. Non-constant regularization ensures that each source pixel is subject to an appropriate level of regularization, regardless of its location in the source-plane. The interplay between both of these features gives {\tt AutoLens} complete freedom in how it reconstructs the source and ensures it finds the simplest solution possible (in a Bayesian sense). These solutions, by definition, use the fewest number of correlated source pixels and the adaptive implementation is indeed found to assume lower $N_{\rm s}$ values, offering significant gains in run-time efficiency. Model comparison with {\tt AutoLens} therefore has no bias or preference to sources of a specific morphology or smoothness profile, which is not necessarily the case for approaches using a fixed pixelization (see S06).

\subsection{Source Modeling - Incorrect Mass Model}
\begin{figure*}
\centering
\includegraphics[width=0.24\textwidth]{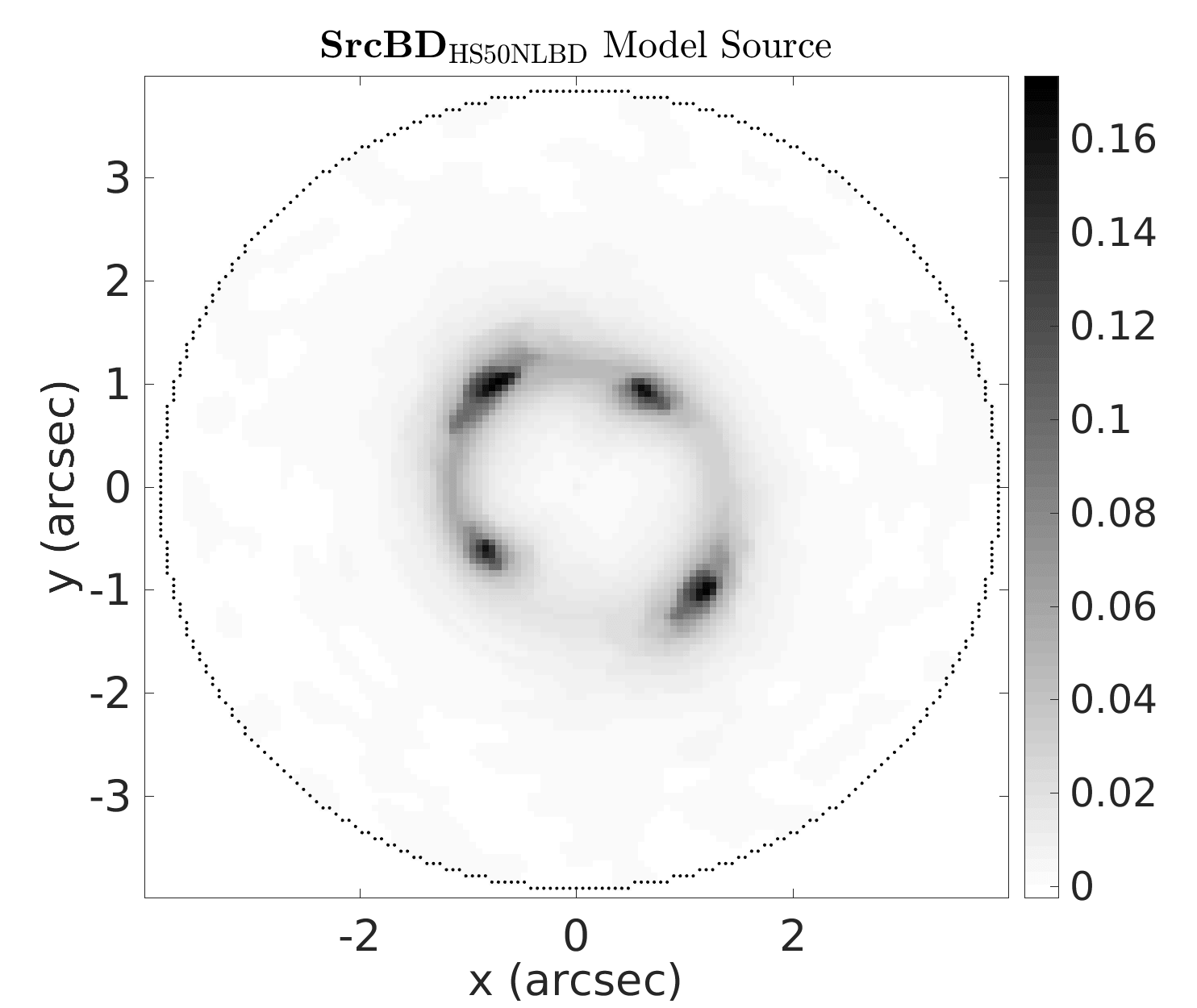}
\includegraphics[width=0.24\textwidth]{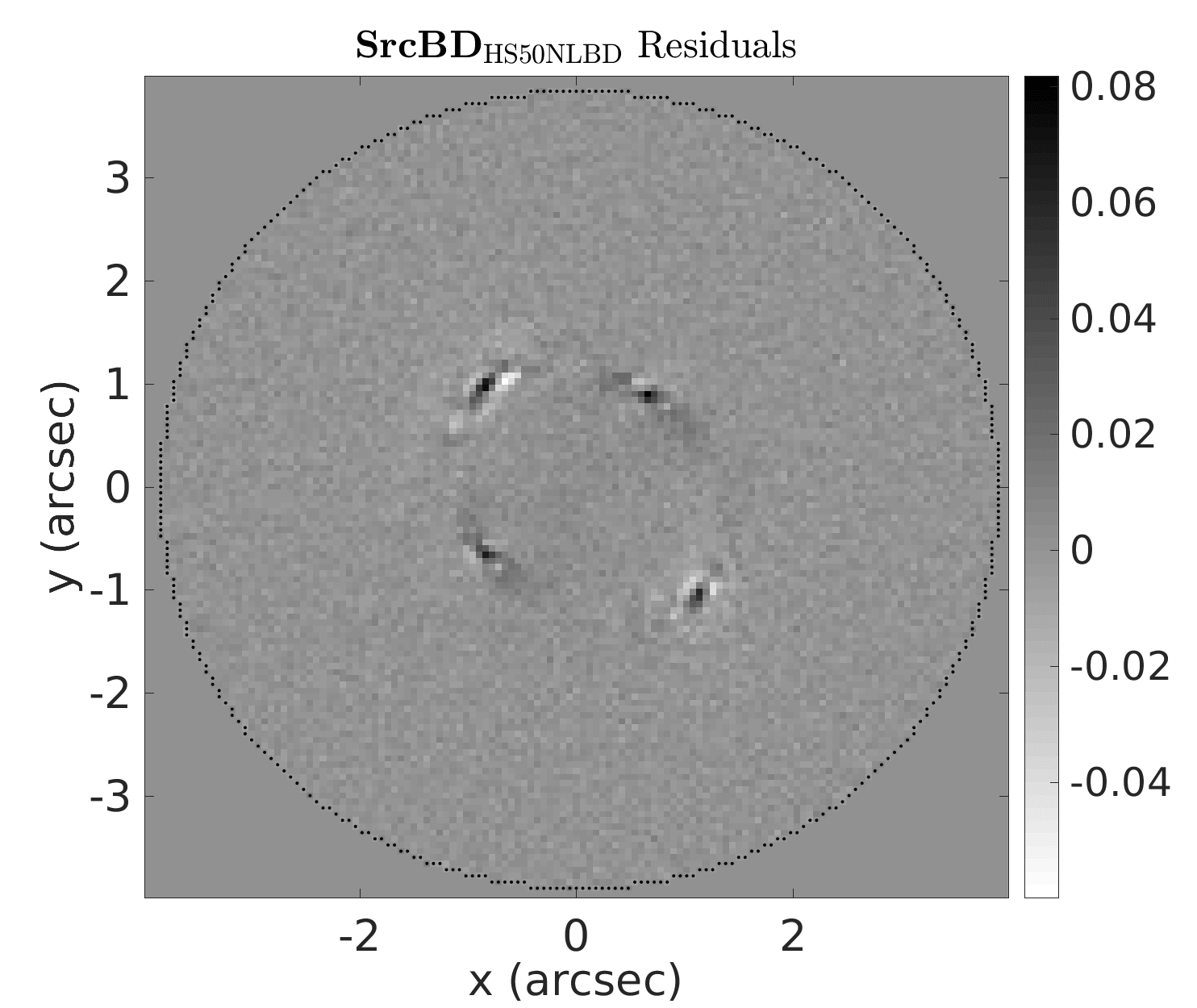}
\includegraphics[width=0.24\textwidth]{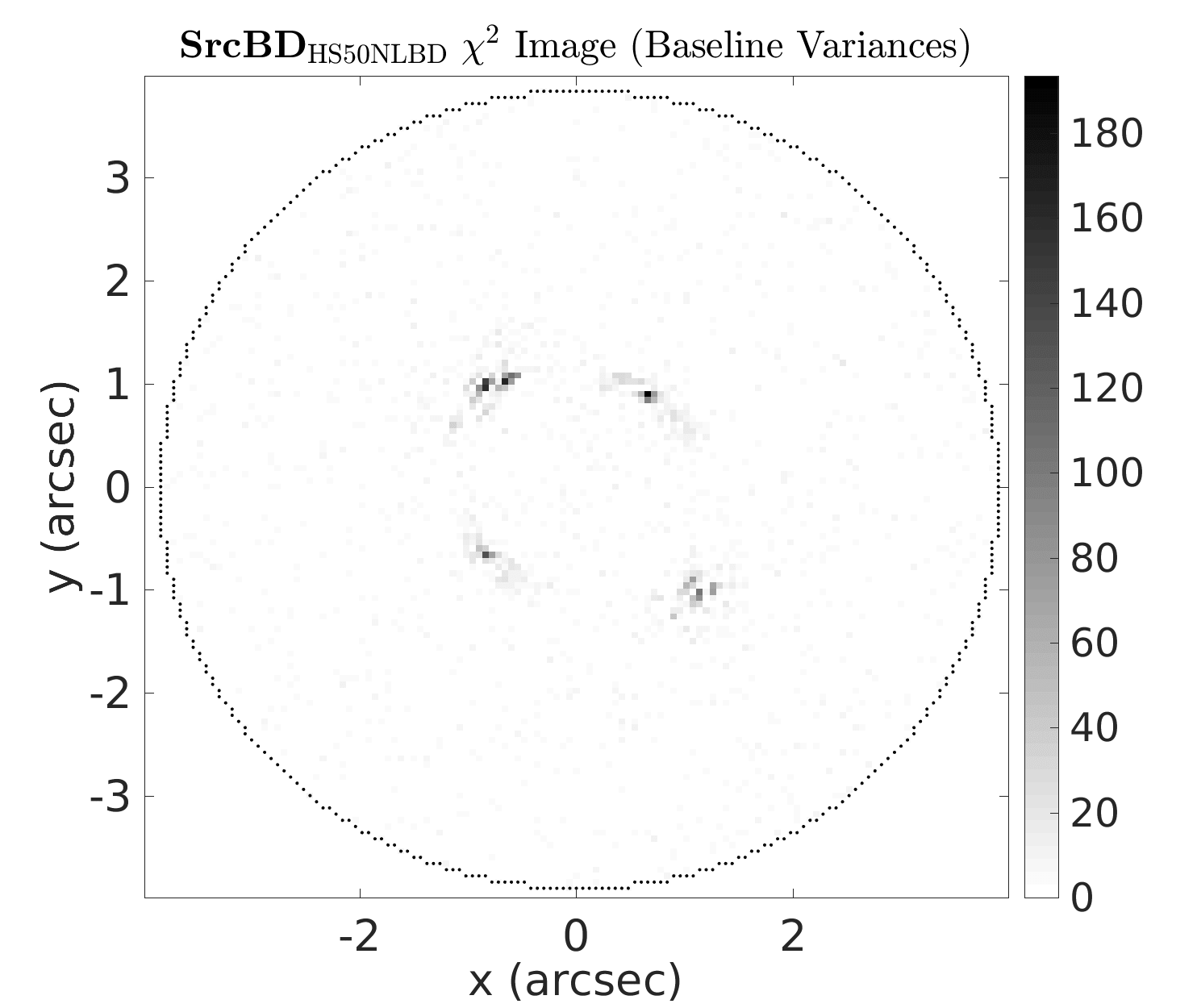}
\includegraphics[width=0.232\textwidth]{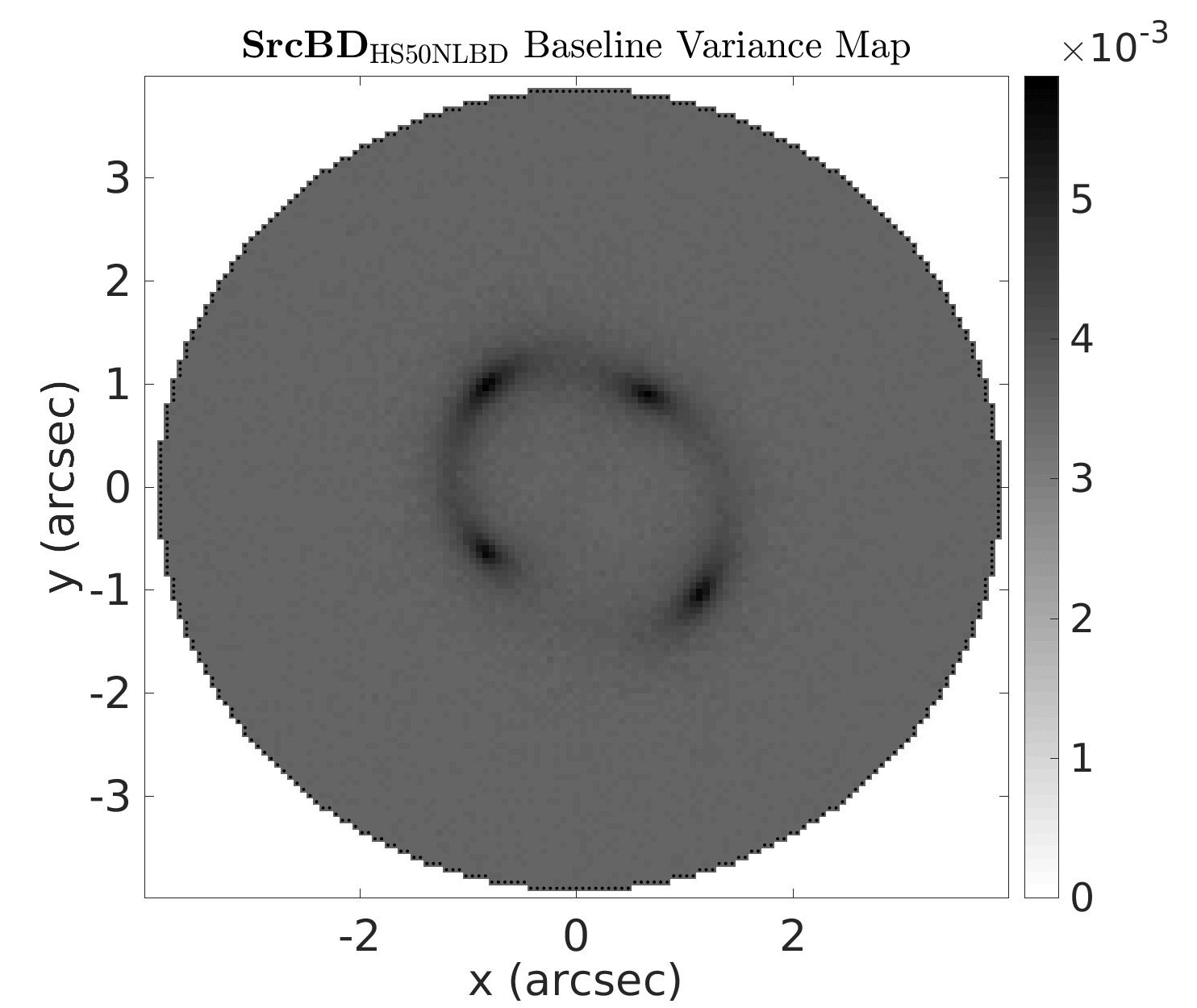}
\includegraphics[width=0.24\textwidth]{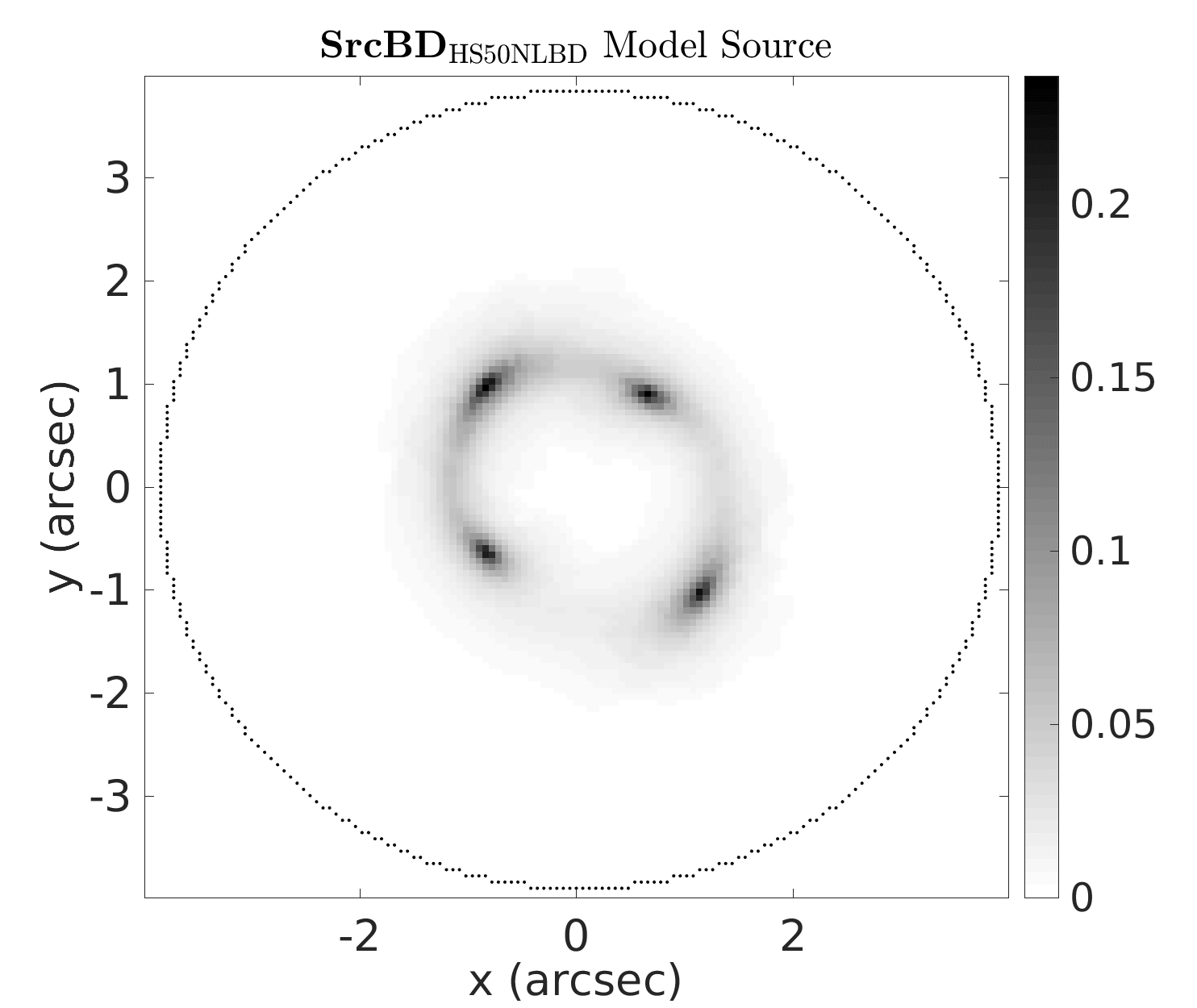}
\includegraphics[width=0.24\textwidth]{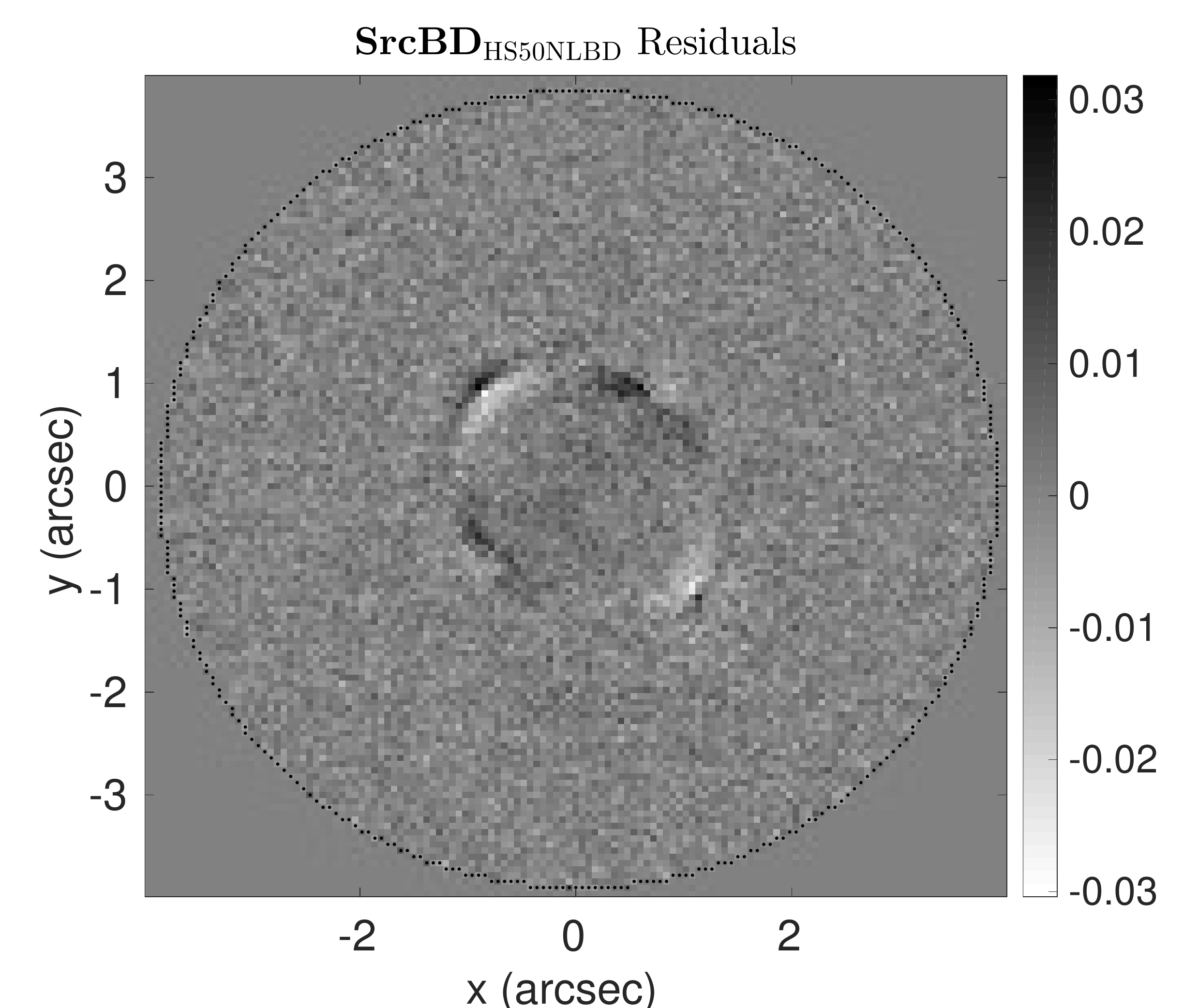}
\includegraphics[width=0.24\textwidth]{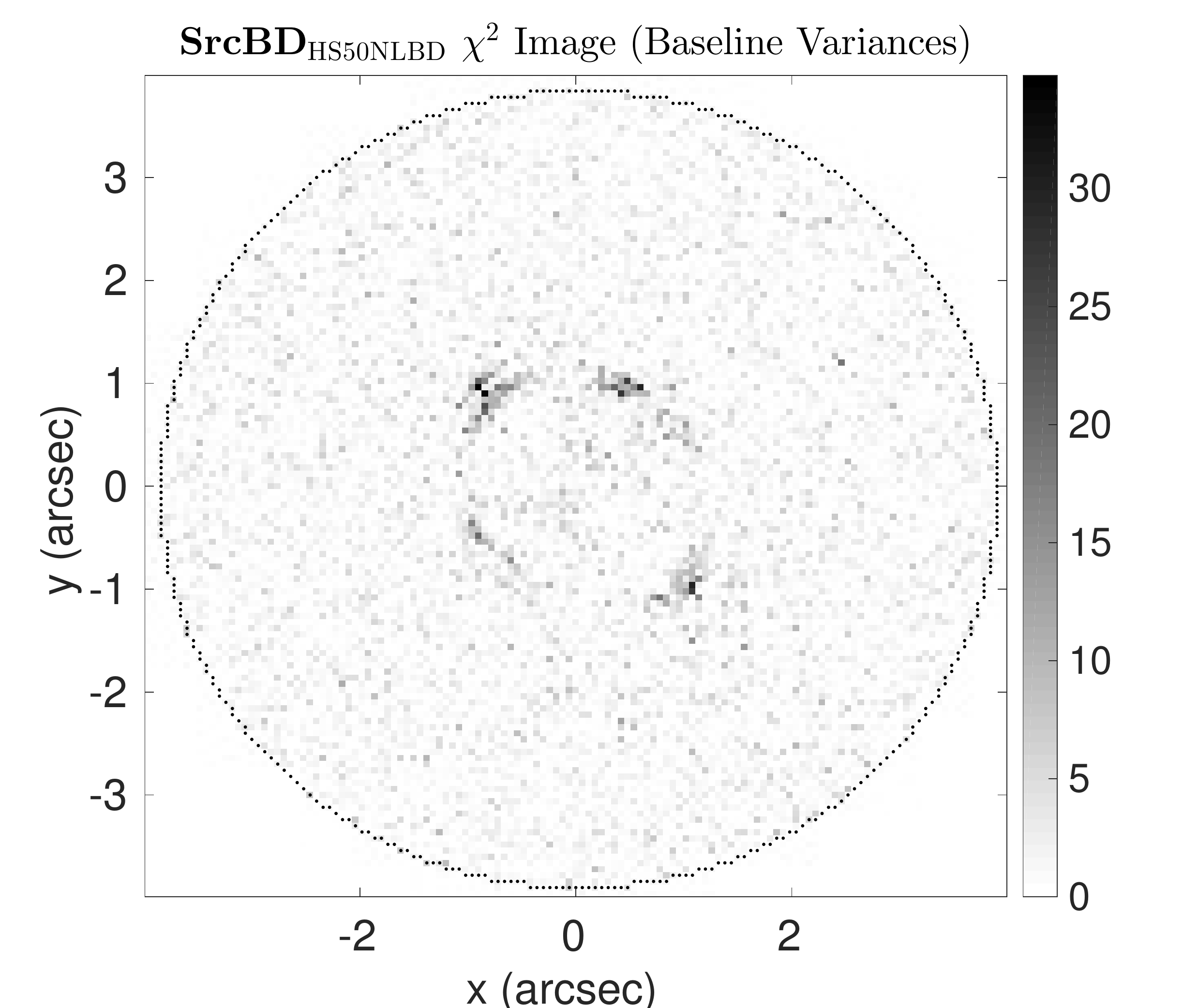}
\includegraphics[width=0.232\textwidth]{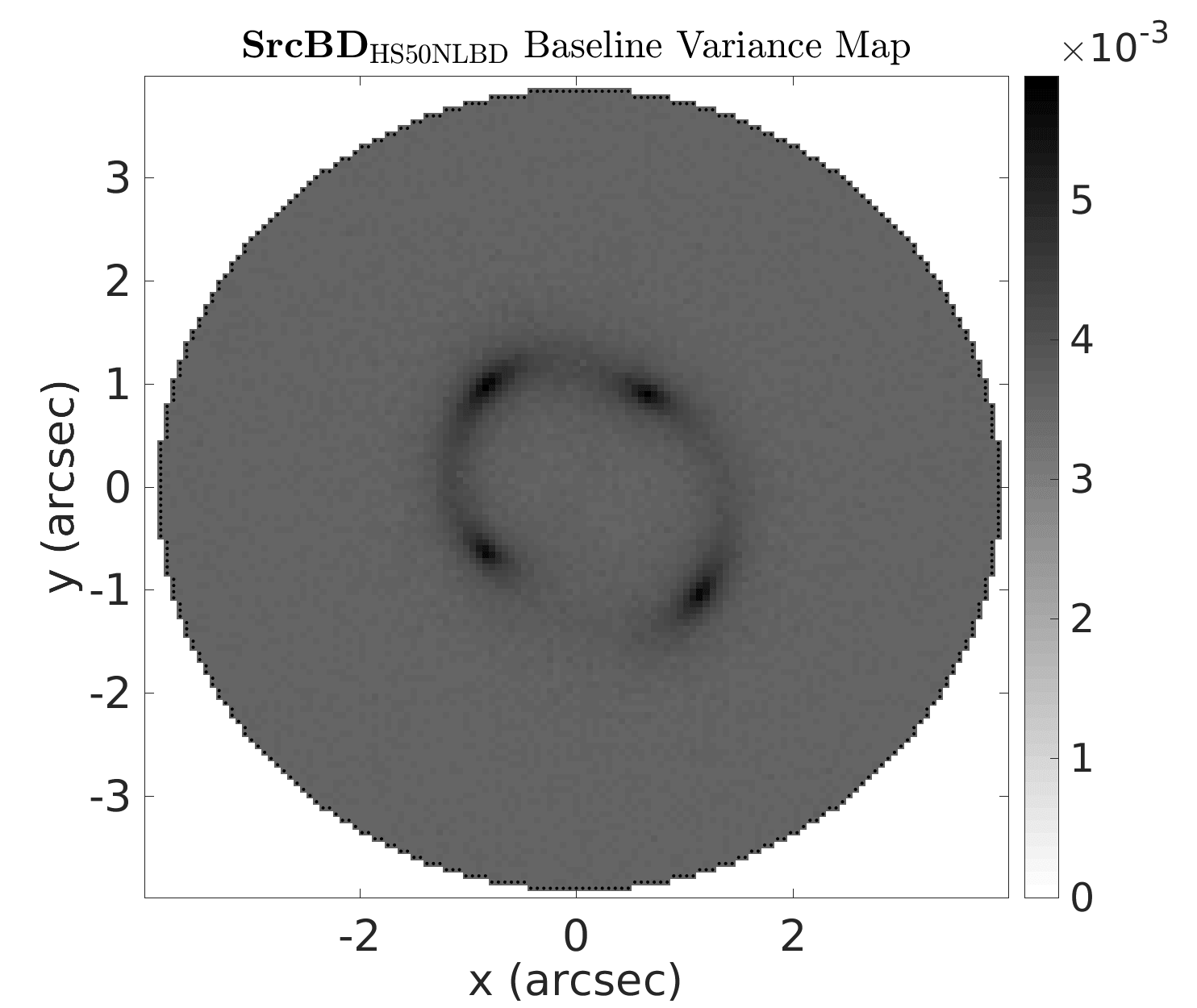}
\includegraphics[width=0.24\textwidth]{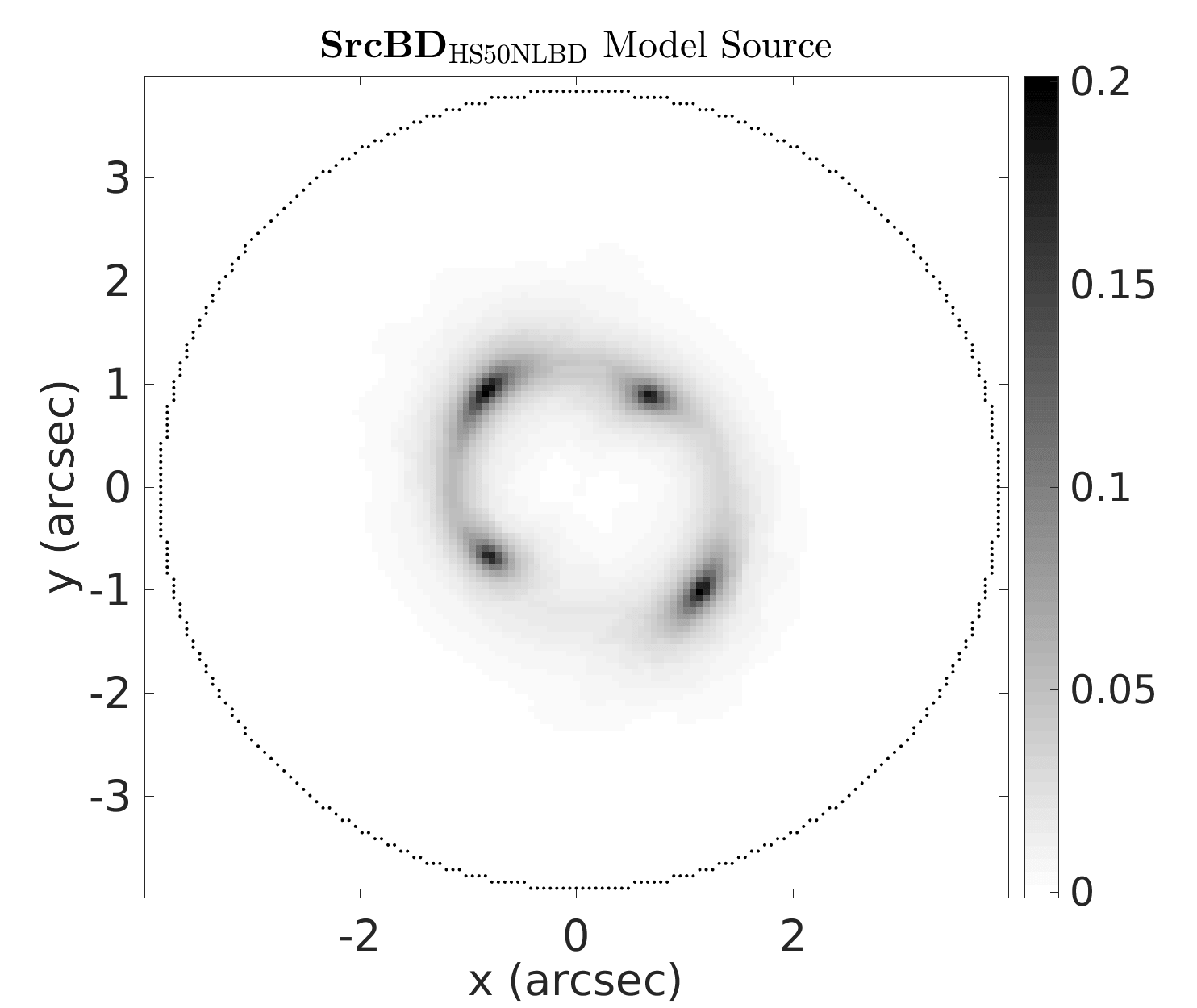}
\includegraphics[width=0.24\textwidth]{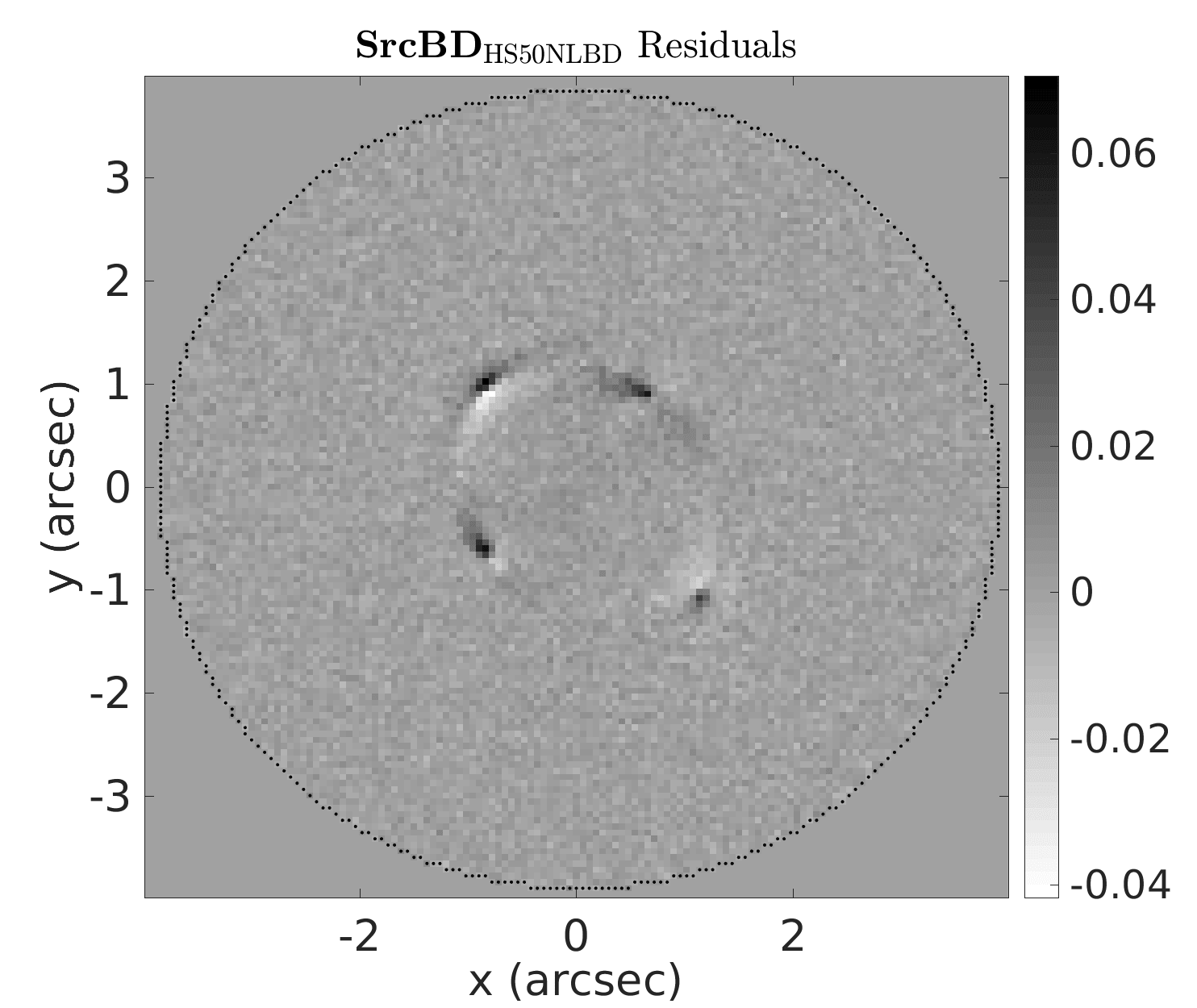}
\includegraphics[width=0.24\textwidth]{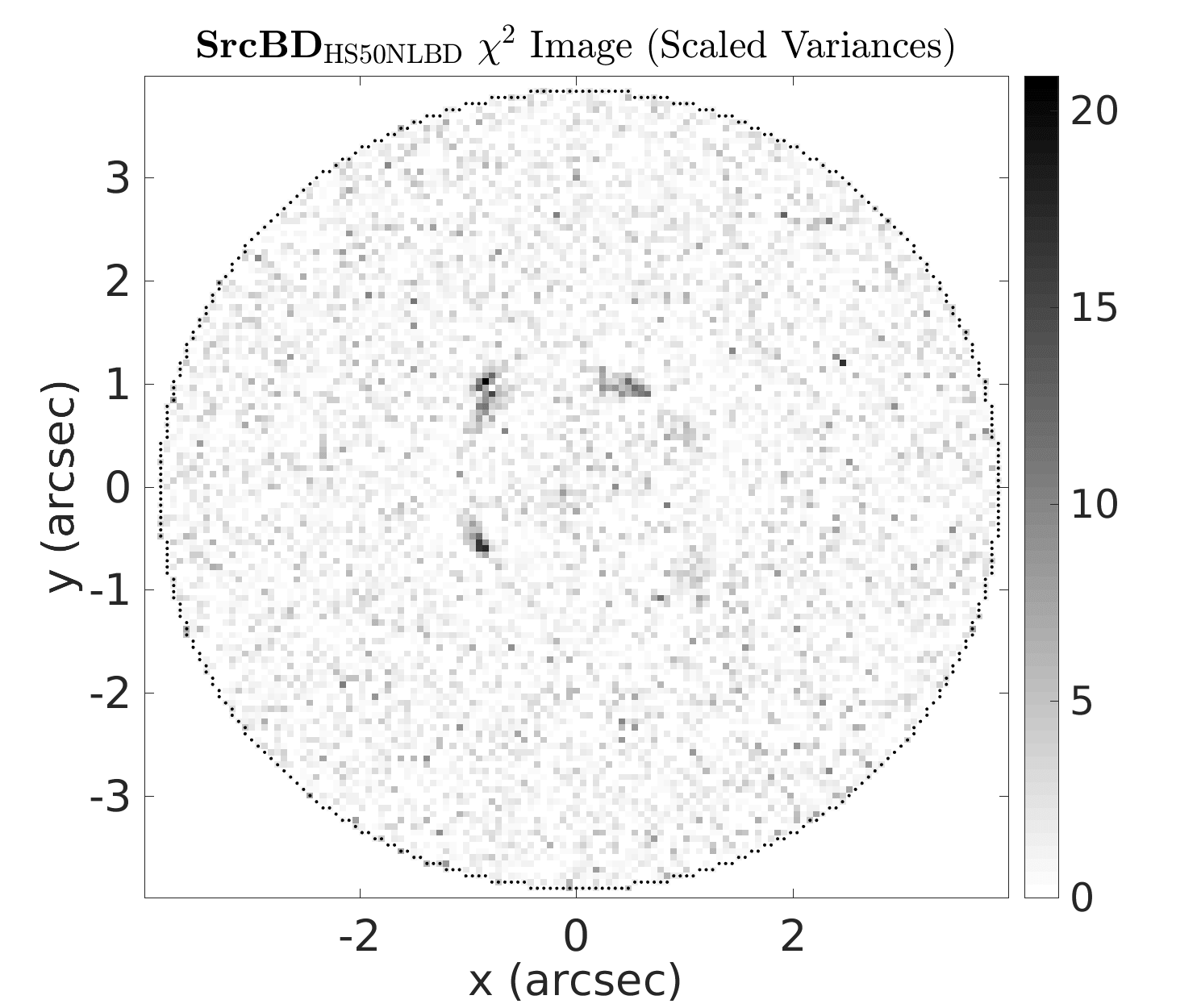}
\includegraphics[width=0.232\textwidth]{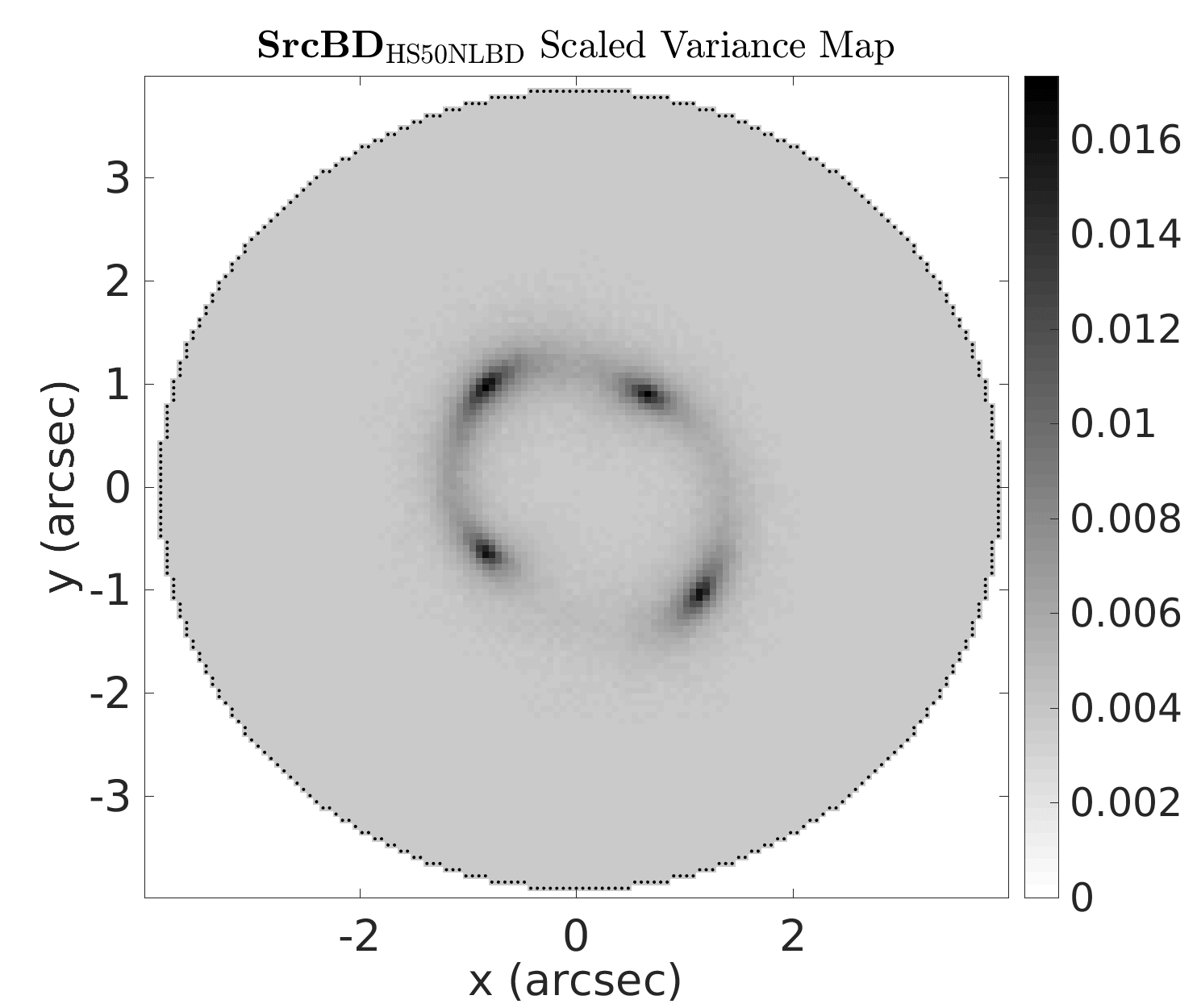}
\caption{The model source, residuals, $\chi^2$ images and variance maps for the analysis of the image $\textbf{SrcBD}_{\rm  HS50NLBulge}$, using either the basic implementation (top row, adaptive image and source analysis switched off) or adaptive implementation with variance scaling turned off (middle row) or turned on (bottom row) and a lens model which incorrectly assumes that $\alpha=2.0$ (the true value is $\alpha=1.7$). The resulting lens models are given by the third, fourth and fifth rows of table \ref{table:AdaptDemoModels}. The basic implementation can be seen to fit the image poorly, leaving noticeable residuals in each multiple image, which dominate the model's overall $\chi^2$ value. The adaptive implementation also gives poor residuals, given that the lens model is incorrect, however the $\chi^2$ image shows a far less skewed distribution, which the right panels show is the result of variance map scaling.} 
\label{figure:AdaptDemoSrcWrong}
\end{figure*}
Figure \ref{figure:AdaptDemoSrcWrong} shows three independent analyses of the same image above, $\textbf{SrcBD}_{\rm  HS50NLBD}$, but where its $SPLE$ mass model has been fixed to an incorrect power-law slope $\alpha = 2.0$ (its input value is $\alpha = 1.7$). Three analyses are performed using: the basic implementation (top row), the adaptive implementation with variance-scaling manually switched off (middle row) and with variance scaling on (bottom row). The basic implementation suffers the same issues as before; noticeable residuals and a skewed $\chi^2$ distribution. However, the same is also now true for the adaptive implementation. This is because, even with the improved source pixelization and regularization, the mismatch between the assumed mass model and true lens profile means only a poor fit is obtainable. When variance scaling is turned on, the residuals are equally poor. After all, scaling the variances can't change the fact that this mass model simply does not provide a good fit to the observed image. However, the $\chi^2$ image shows fewer pixels with high $\chi^2$ values and $\chi^2$ values that are lower. As discussed for the previous issue, this is the more desirable solution, as it uses all of the data that is available to constrain the lens model. Table \ref{table:AdaptDemoModels} shows this in turn provides the smallest errors and highest Bayesian evidence. However, the scaled $\chi^2$ values shown in the bottom row of figure \ref{figure:AdaptDemoSrcWrong} are not fully consistent with Gaussian noise, because the limits of variance scaling prevent $\chi^2_{\rm sca}$ values to go below $10$. The reasoning behind this set up is discussed next in section \ref{VarOverSca}.

In this example, the issue could easily be fixed by allowing $\alpha$ to be a free parameter. For real lenses, however, the mass model will always (to some degree) be `incorrect', because the mass models assumed during a lens analysis are a simplified representation of any galaxy's true underlying mass distribution (see discussions by \citet{Brewer2012a, Suyu2012}). Noticeable residuals and non-uniformly distributed $\chi^2$ images (without variance scaling) are therefore commonplace when analysing real strong lenses, which will negatively impact lens modeling by over-fitting a small fraction of the available imaging data. This can lead to over-estimated parameter errors or biased parameter estimates. Instead, it is more desirable that the lens model is constrained using all of the data that is available by fitting the image in an equally weighted manner, especially once other uncertainties like a poor PSF-sampling are considered. Variance scaling ensures that this is the case and in conjunction with the adaptive source features offers a natural means to test more complex mass models using Bayesian model comparison.

\subsection{Simultaneous Source and Lens Modeling - Incorrect Light Model}
\begin{figure*}
\centering
\includegraphics[width=0.24\textwidth]{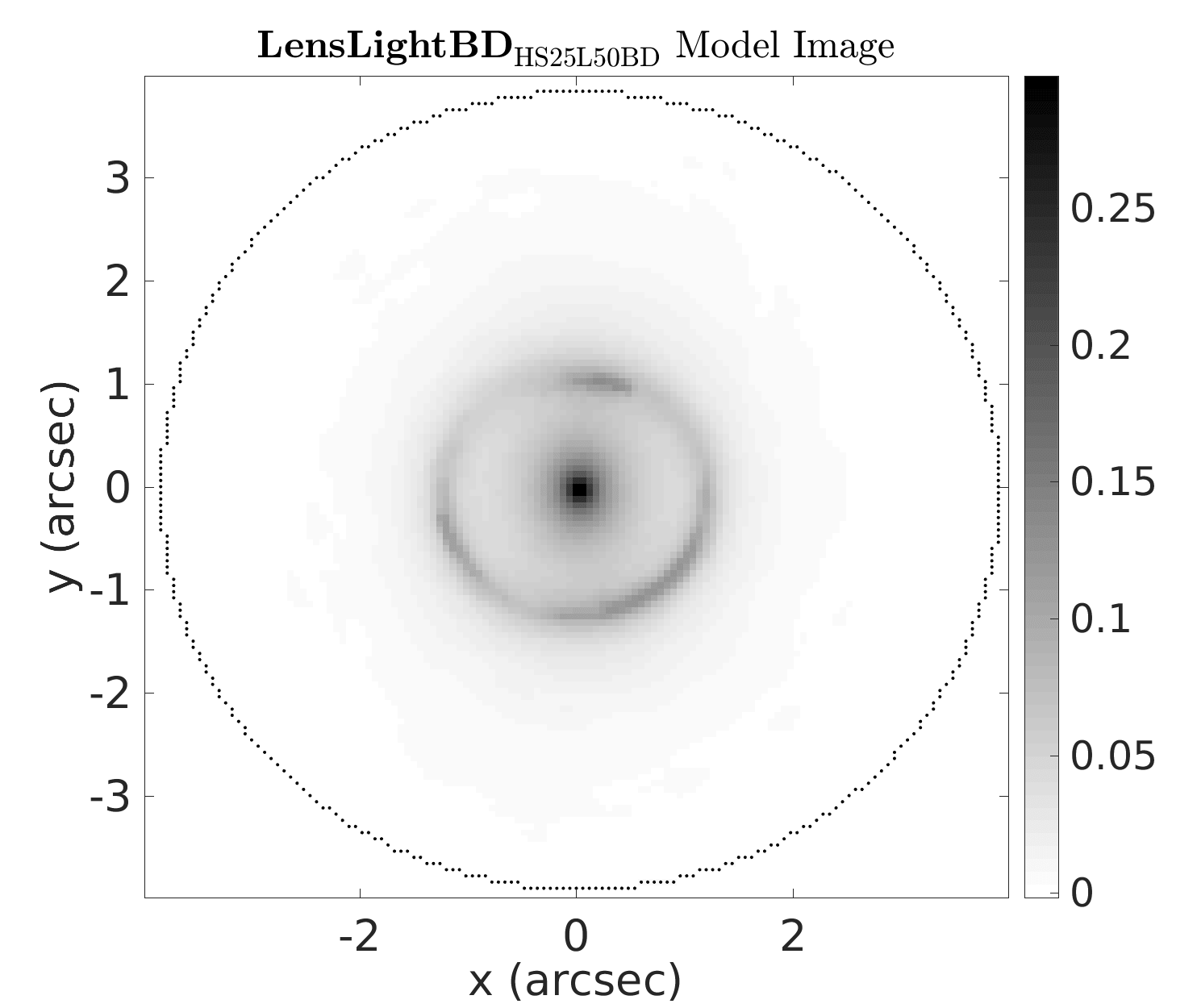}
\includegraphics[width=0.24\textwidth]{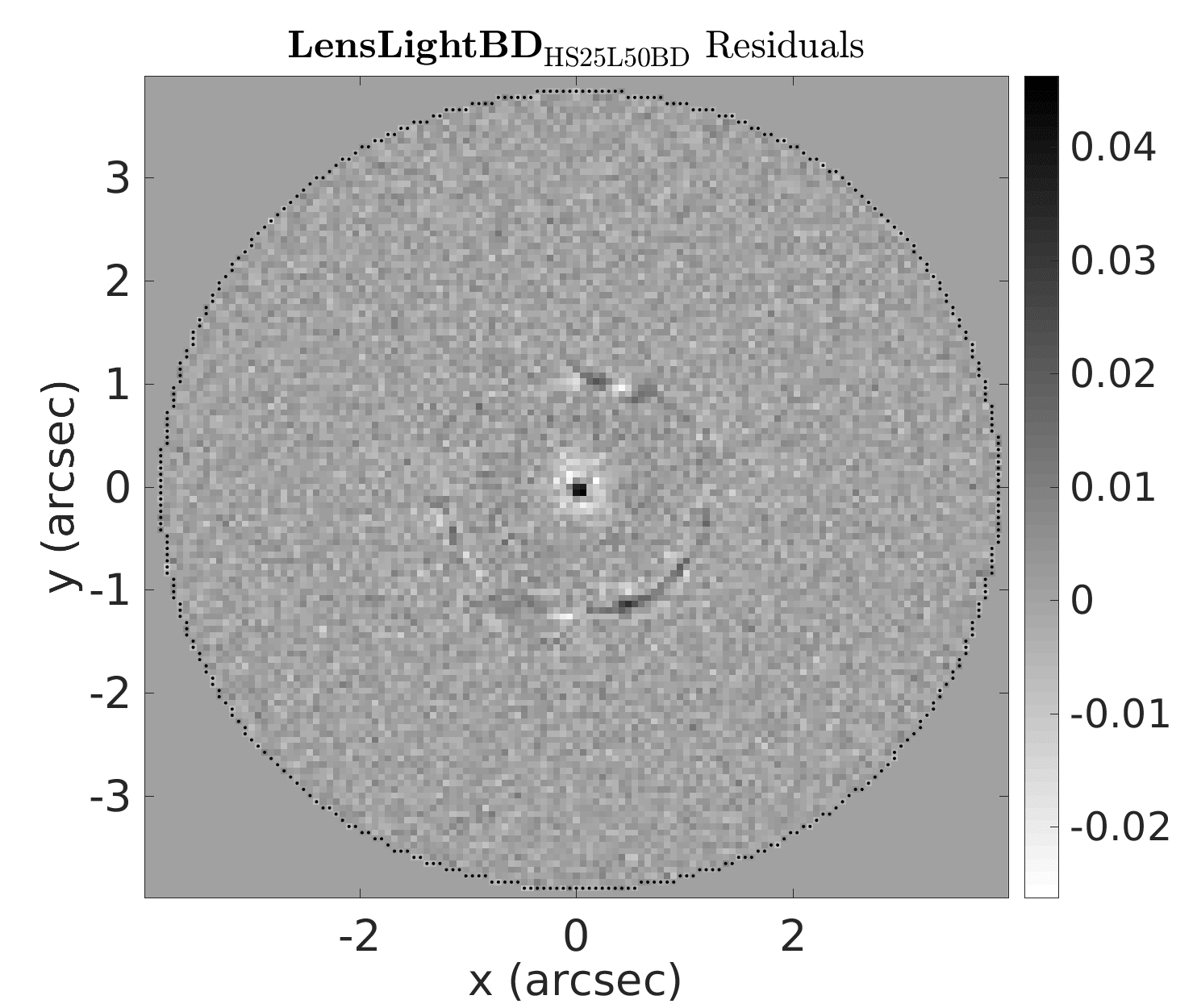}
\includegraphics[width=0.24\textwidth]{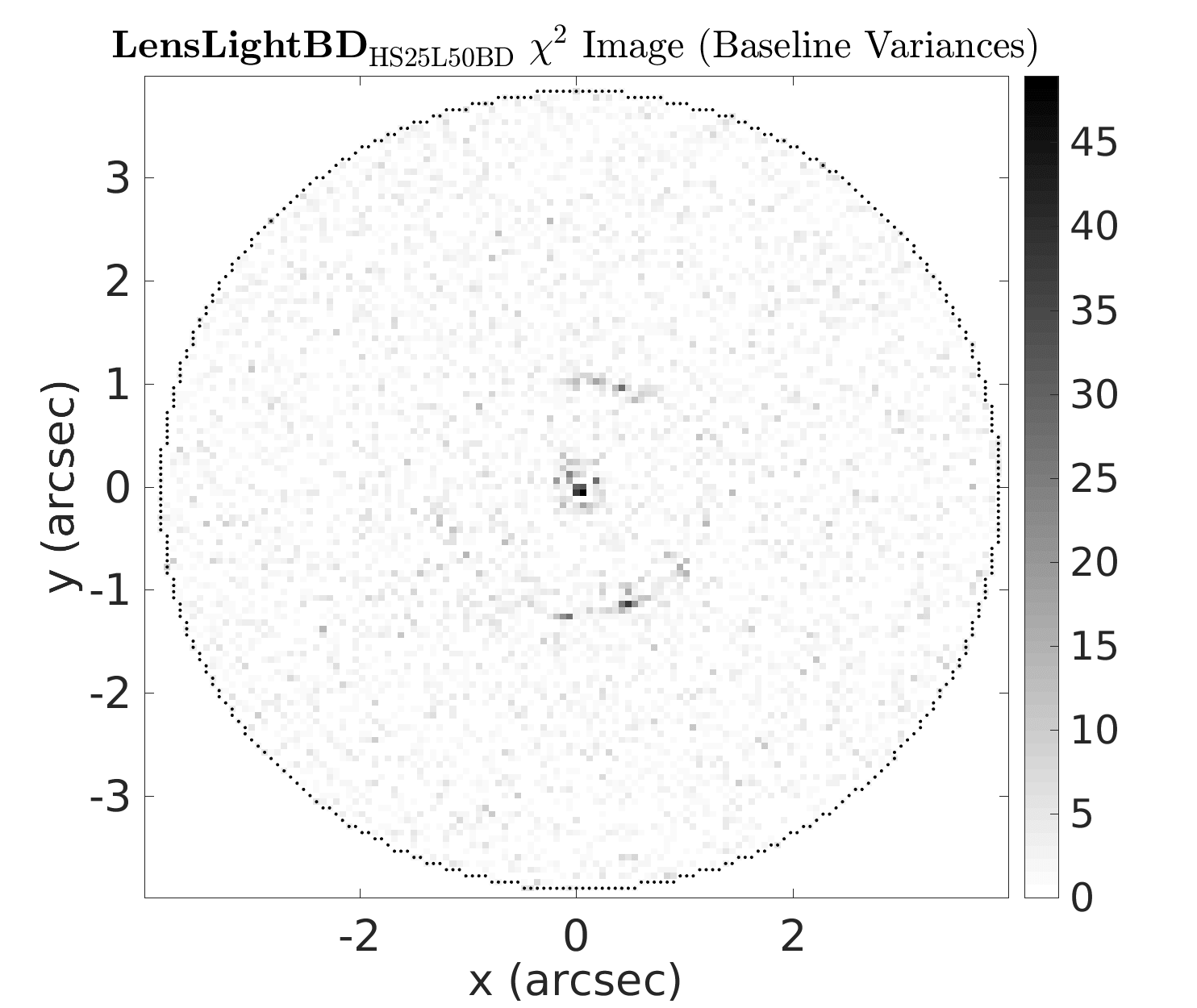}
\includegraphics[width=0.232\textwidth]{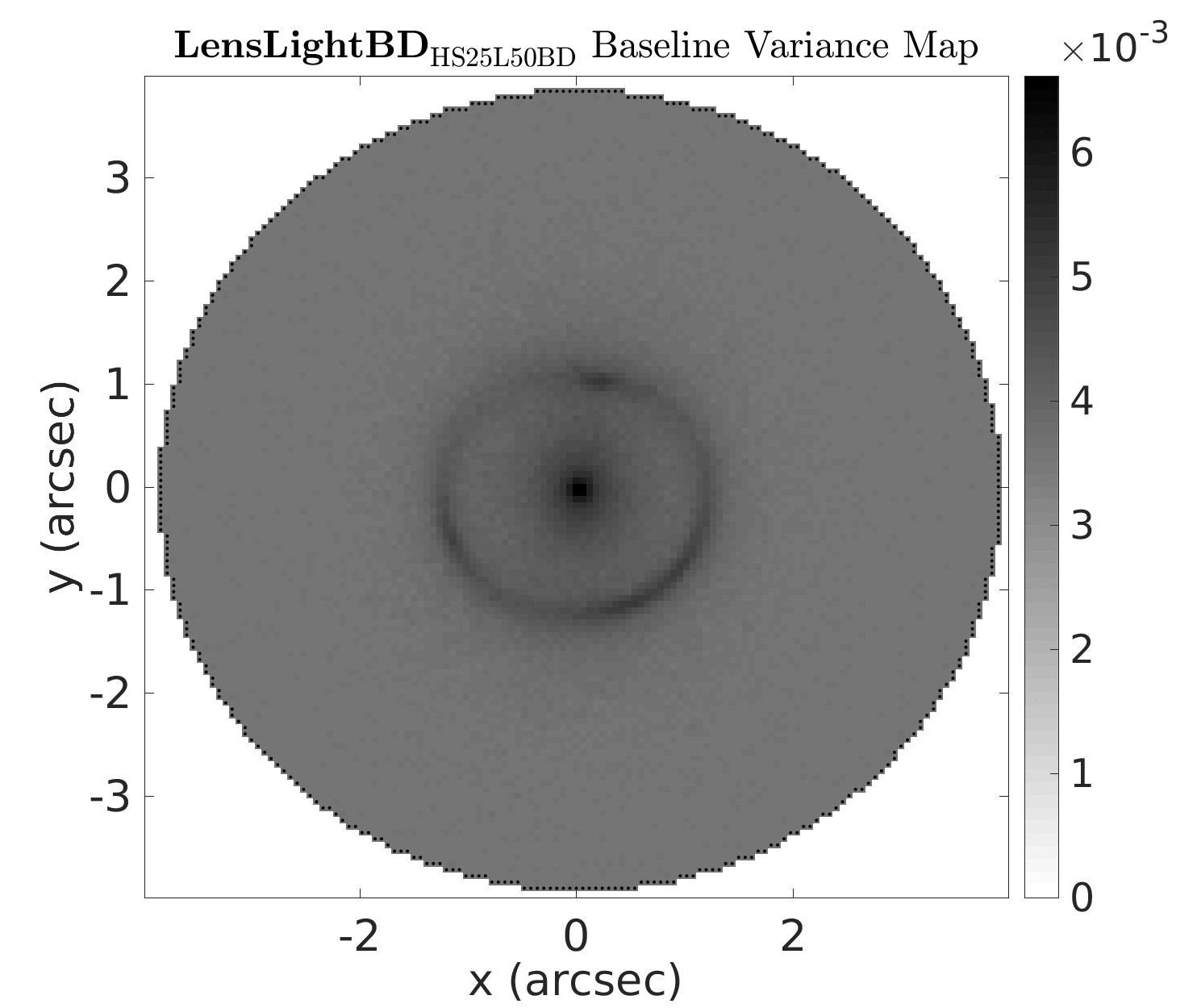}
\includegraphics[width=0.24\textwidth]{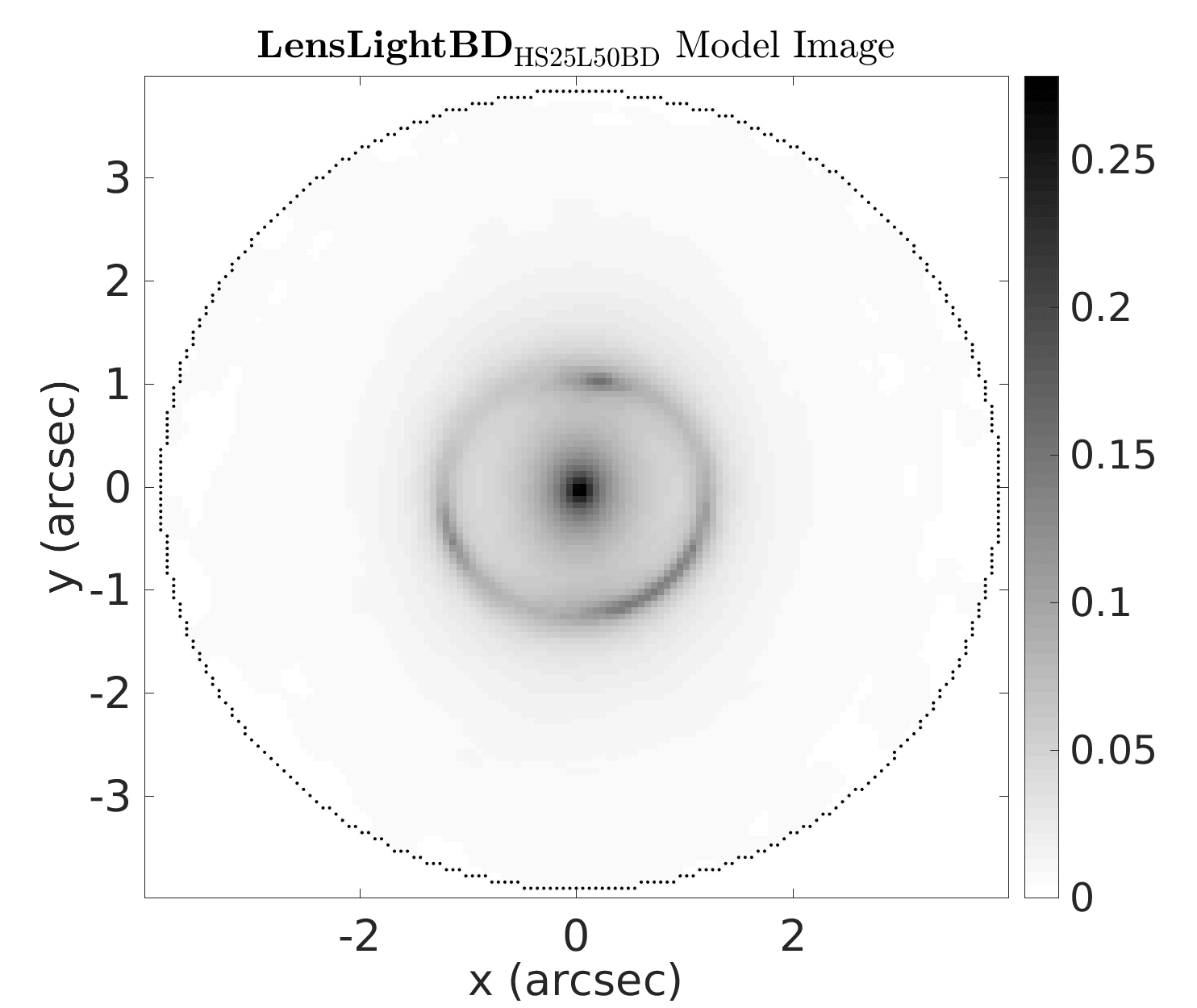}
\includegraphics[width=0.24\textwidth]{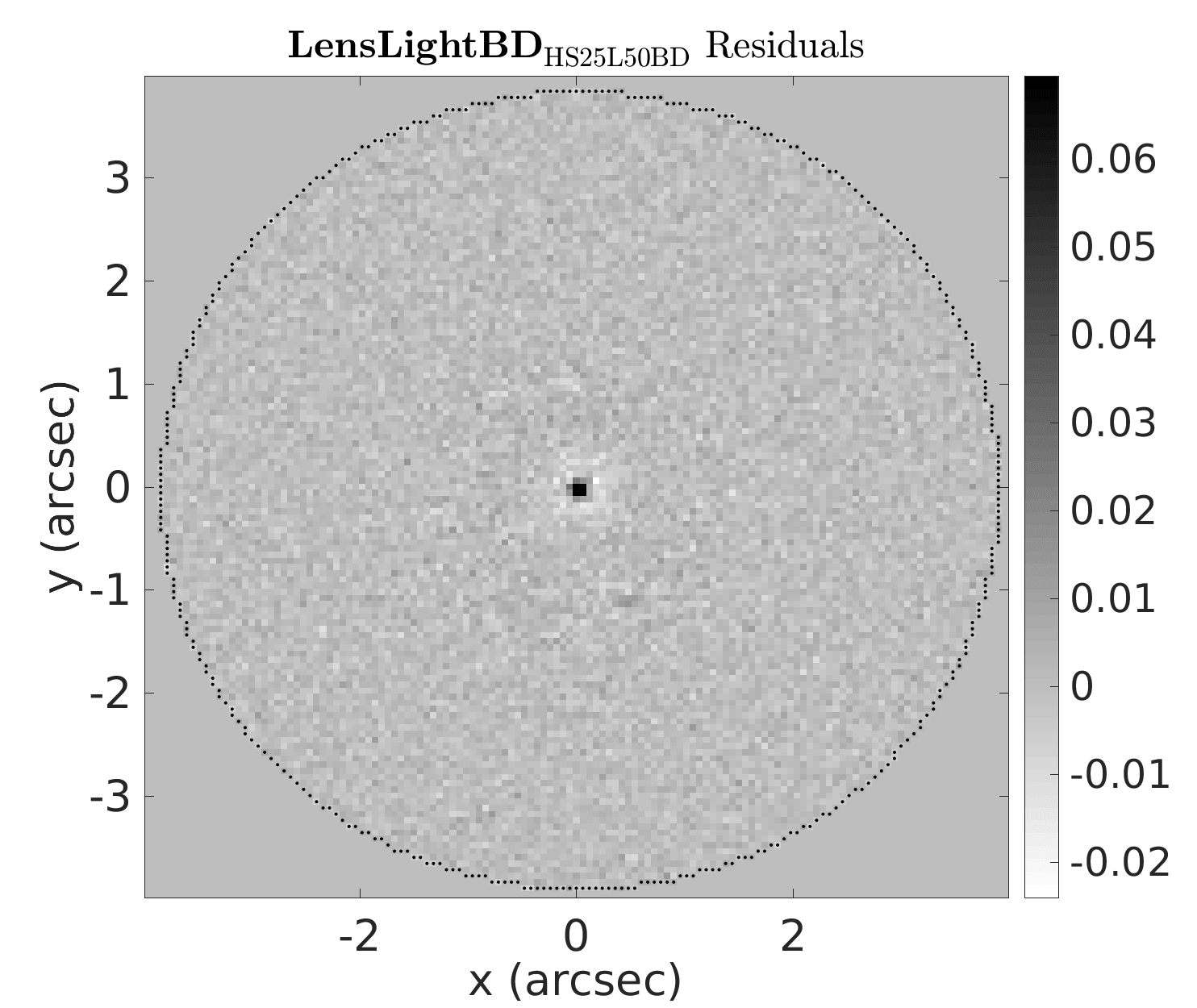}
\includegraphics[width=0.24\textwidth]{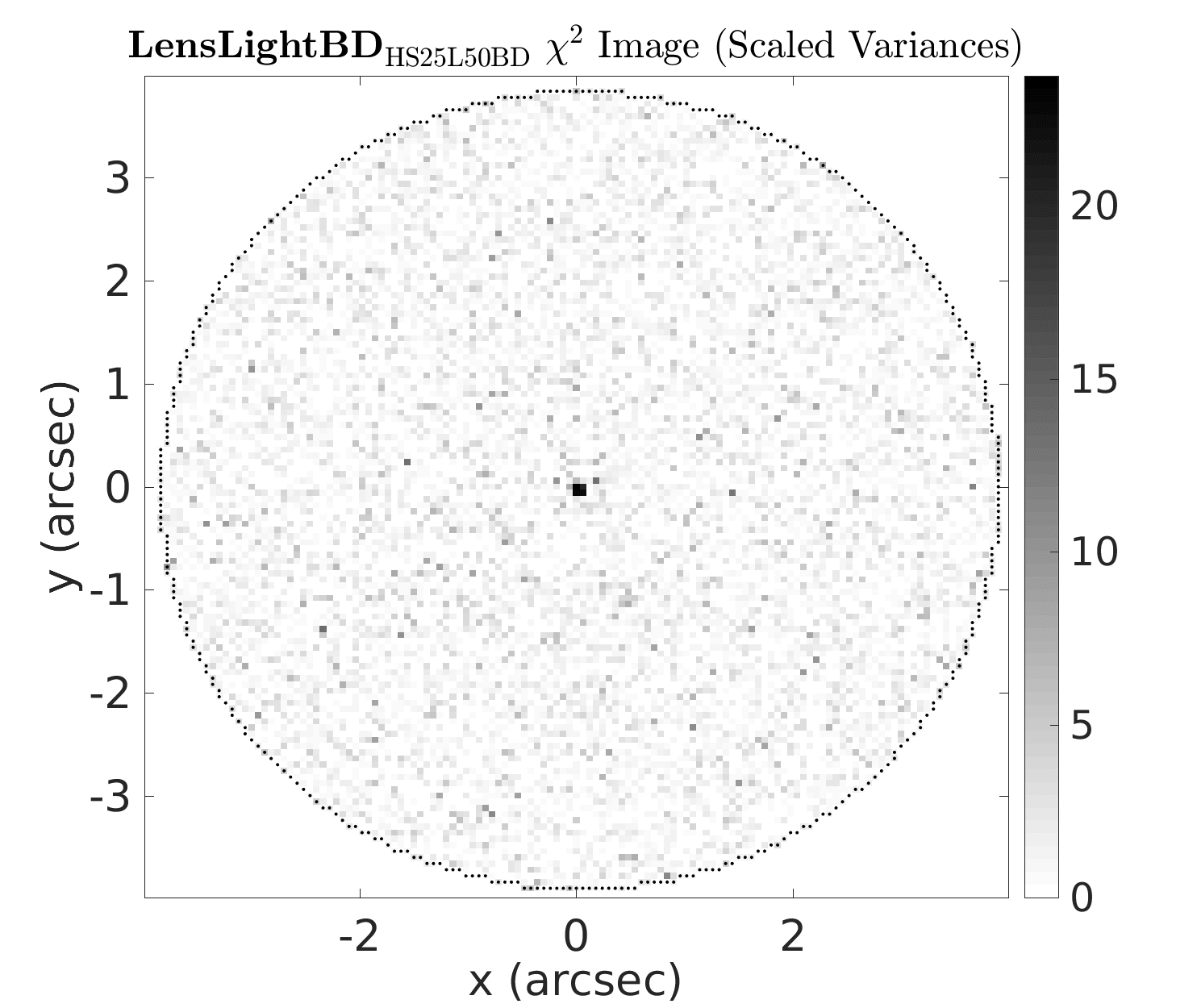}
\includegraphics[width=0.232\textwidth]{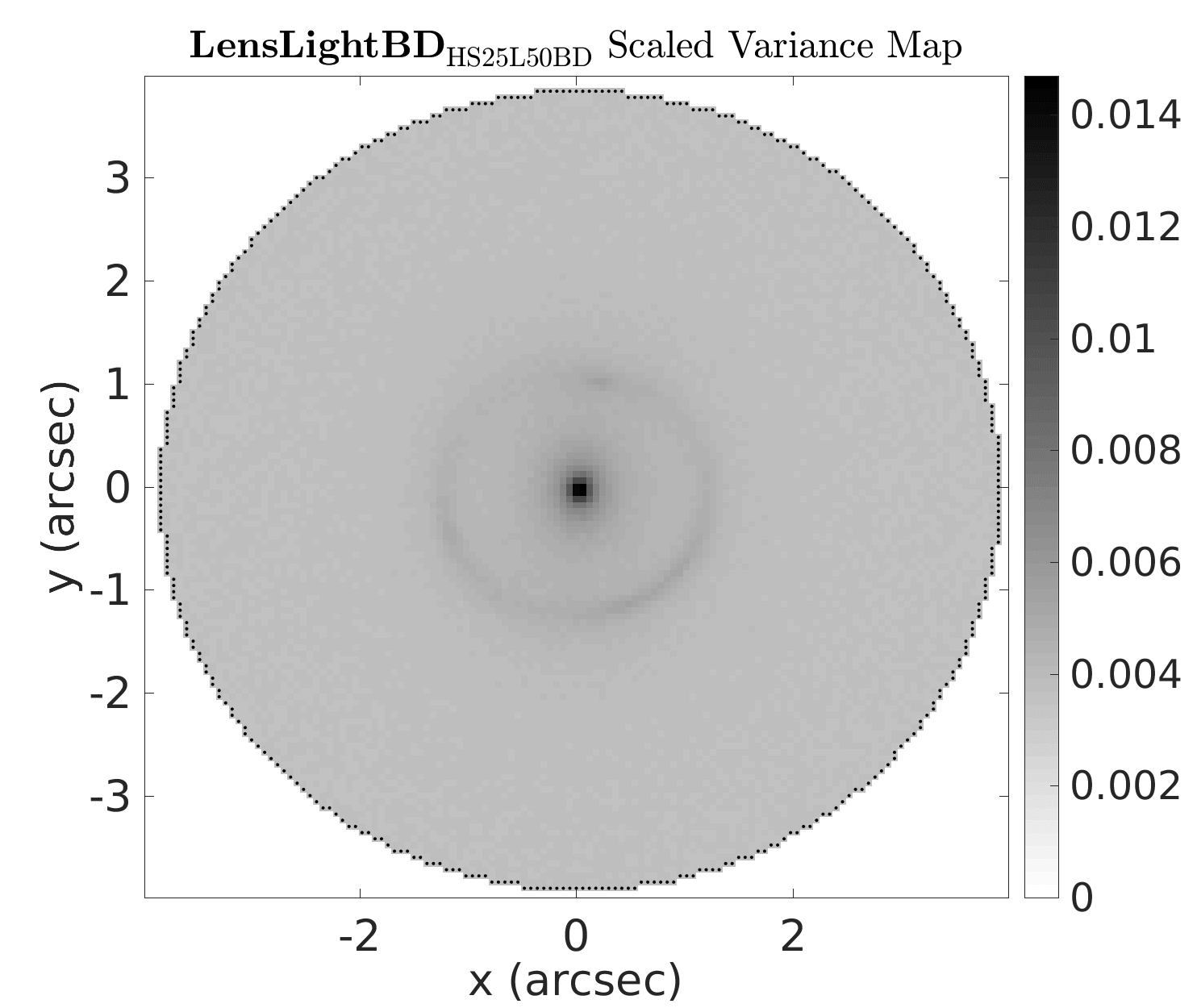}
\caption{The model images, residuals, $\chi^2$ images and variance maps for the analysis of the image $\textbf{LensLightBD}_{\rm  HS25L50BD}$, fitted using a $Sersic$ light model even though its true underlying profile is a $Sersic$ + $Exp$. Two fits are shown, using the basic implementation (top row, adaptive image and source analysis switched off) and adaptive implementation (middle row, adaptive image and source analysis turned on). The adaptive implementation can be seen to increase the variances where the lens and source are located. The resulting lens models are given by the sixth to ninth rows of table \ref{table:AdaptDemoModels}. The basic implementation can be seen to fit the image poorly, leaving residuals around both the lens and source, which dominate the model's overall $\chi^2$ value. The adaptive implementation gives poor residuals around the lens (as expected because the $Sersic$ model cannot provide an accurate fit), however the source residuals are mostly removed and its $\chi^2$ image shows a significantly less skewed distribution.} 
\label{figure:AdaptDemoLens}
\end{figure*}

Further issues arise when the lens and source are modeled simultaneously, in particular when the lens light subtraction leaves residuals. Figure \ref{figure:AdaptDemoLens} illustrates this particular circumstance, showing the results of fitting the image $\textbf{LensLightBD}_{\rm  HS25L50BD}$ with a $Sersic$ light profile, even though the simulated lens was generated using the more complex $Sersic$ + $Exp$ profile. As expected, the lens subtraction for both implementations leaves significant residuals, as the $Sersic$ profile is simply unable to provide a good fit to the lens's more complex morphology. However, comparison of the residuals and $\chi^2$ image of the two implementations show they differ in two ways: (i) the basic implementation leaves residuals after fitting the lensed source, whereas the adaptive implementation does not (the much smoother source morphology for this object means this is not due to the issue above); (ii) the $\chi^2$ image for the adaptive implementation almost fully realizes the image's Gaussian noise, except for a few pixels in the centre of the image, whereas the basic implementation again suffers the skewed $\chi^2$ distribution discussed above, however now also in the central pixels where the lens is located. 

The fundamental problem here is that when simultaneously modeling both the lens and source the  source reconstruction cannot distinguish between lensed source flux and residual lens light. Problematically, it treats the latter as if they are part of the source, corrupting the image reconstruction and ultimately biasing the inferred lens model, as shown in table \ref{table:AdaptDemoModels}. This issue impacts the image reconstruction in two ways, both of which the adaptive implementation was developed specifically to tackle.

The first is the impact of central residual light on the source reconstruction. The linear inversion will attempt to fit these pixels like any other, but fail to do so, given that they map to the exterior regions of the source-plane where all of the other traced image pixels map to the background sky (see figure \ref{figure:CenIms1}). This by itself is acceptable, as the method shouldn't reconstruct these pixels as if they are part of the source. The problem, however, is regularization, as (in an analogous manner to the cuspy source above) these pixels lead the basic implementation to set a compromised higher value of $\lambda$ which leads the source reconstruction to be over-smoothed, producing the source residuals seen in figure \ref{figure:AdaptDemoLens} and inflating the lens model parameter errors as shown in table \ref{table:AdaptDemoModels}.

The adaptive implementation does not suffer this issue because of luminosity-weighted regularization, which allows the source-reconstruction to simultaneously smooth over the exterior regions of the source-plane that map to the lens subtraction residuals (and background sky) whilst simultaneously fitting the detailed structure of the source with an appropriate and reduced level of regularization. Therefore, even in the presence of a poor lens subtraction the adaptive implementation can still fit the source accurately, as shown by the removal of source residuals in figure \ref{figure:AdaptDemoLens}. When the method is able to smooth over the source-plane's exterior regions with very high levels of regularization, the issues discussed in section \ref{CenIms} related to the source reconstruction fitting residual flux in central image pixels are circumvented.

The second problem is also due to central image pixels, but instead how the lens light model fits and subtracts them. These pixels are the highest S/N pixels in the data (typically by a large margin) and therefore have the potential to overwhelm the model's $\chi^2$ contribution if the lens subtraction is not perfect. When this occurs, the light model concentrates its flux into these central regions so as to accurately fit only these high S/N image pixels, over-concentrating the inferred light profile and failing to give a global representation of the lens's morphology. This is shown by table \ref{table:AdaptDemoModels}, where the basic implementation can be seen to go to a much higher value of $n_{\rm l}$ compared to the adaptive implementation. 

For real lenses, no light subtraction will ever be perfect and many lenses will posses detailed structures (e.g. bars, dust lanes) a smoothly parametrized profile cannot fit completely. Thus, it is paramount this issue is removed from the lens analysis, which the bottom row of figure \ref{figure:AdaptDemoLens} shows is exactly what variance scaling achieves, by increasing the variances around the lens galaxy such that the $\chi^2$ image reverts to being (almost) Gaussian, thereby again giving a global fit to the imaging data. 

\subsection{Variance Overscaling}\label{VarOverSca}

The $\chi^2$ images shown in figures \ref{figure:AdaptDemoSrcWrong} and \ref{figure:AdaptDemoLens} for the adaptive implementation (with variance scaling on) are not fully consistent with the image's Gaussian noise. A small subset of pixels retaining values of $\chi^2 \sim 10$ can be seen. This is because of the upper limits placed on the values of $\omega_{\rm Src}$ and $\omega_{\rm Lens}$, which restricted their maximum values such that $\chi^2_{\rm sca}$ could not be scaled below $10$. When these limits are not imposed the $\chi^2$ image becomes fully consistent with Gaussian noise. However, during testing of {\tt AutoLens}, it emerged that giving these hyper-parameters the freedom to go to arbitrarily high values lead to `variance overscaling'. 

Here, the scaled variances were increased to such large values that the effective S/N (and $\chi_{\rm sca}^2$ values) of their corresponding image pixels were $\sim 0$. This decreased ln$\epsilon$, as the overall S/N of the observed image was reduced (recall the description of the Bayesian evidence in section \ref{HyperParams}). However, a net increase in ln$\epsilon$ was still possible, because the source reconstruction and lens model could change so as to fit other regions of the image better. These solutions are not desirable, as the method is essentially ignoring the central regions of the lens and source galaxies in order to better fit more exterior regions of each. The limits on $\omega_{\rm Src}$ and $\omega_{\rm Lens}$ are therefore imposed to prevent this from happening, by ensuring the $\chi_{\rm sca}^2$ values of these image pixels cannot be reduced to $0$. The target values of $\chi_{\rm base}^2 = 10$ are chosen to give a balance between reducing the $\chi^2$s such that the source reconstruction and lens light profiles do not over-fit the central regions, whilst also ensuring that variance over-scaling does not occur. For real lenses, it will be important to investigate the impact of changing these target scalings on the inferred lens model.

\subsection{Decomposed Mass Modeling}

Equation \ref{eqn:Sersickap}, used for decomposed mass modeling, assumes that light perfectly traces mass, an assumption which will hold only approximately for real lenses. However, the lens's light profile is constrained by two aspects of a decomposed analysis: (i) the quality of the light subtraction; (ii) the decomposed mass model's source and image reconstructions. For decomposed mass modeling it is therefore desirable to give the lens's light matter distribution the freedom to deviate from the lens's light profile, if doing so improves the mass model. Variance scaling facilitates this, such that the $\chi^2$ contribution of central image pixels can be down weighted to allow the light profile to deviate from its true profile. Doing so decreases ln$\epsilon$, but could potentially produce a net ln$\epsilon$ increase by improving the mass model. Thus, {\tt AutoLens} does not strictly assume that light traces mass and is able to deviate from this assumption within a Bayesian context. Comparison to models assuming a total-mass profile, like the $SPLE$, can offer insight into whether this is occurring and to what degree. However, the simulated images used in this work assume that light traces mass, thus this is not tested explicitly here.

\section{Pipeline Automation}\label{SLPipeline}

\begin{figure*}
\centering
\includegraphics[width=0.8\textwidth]{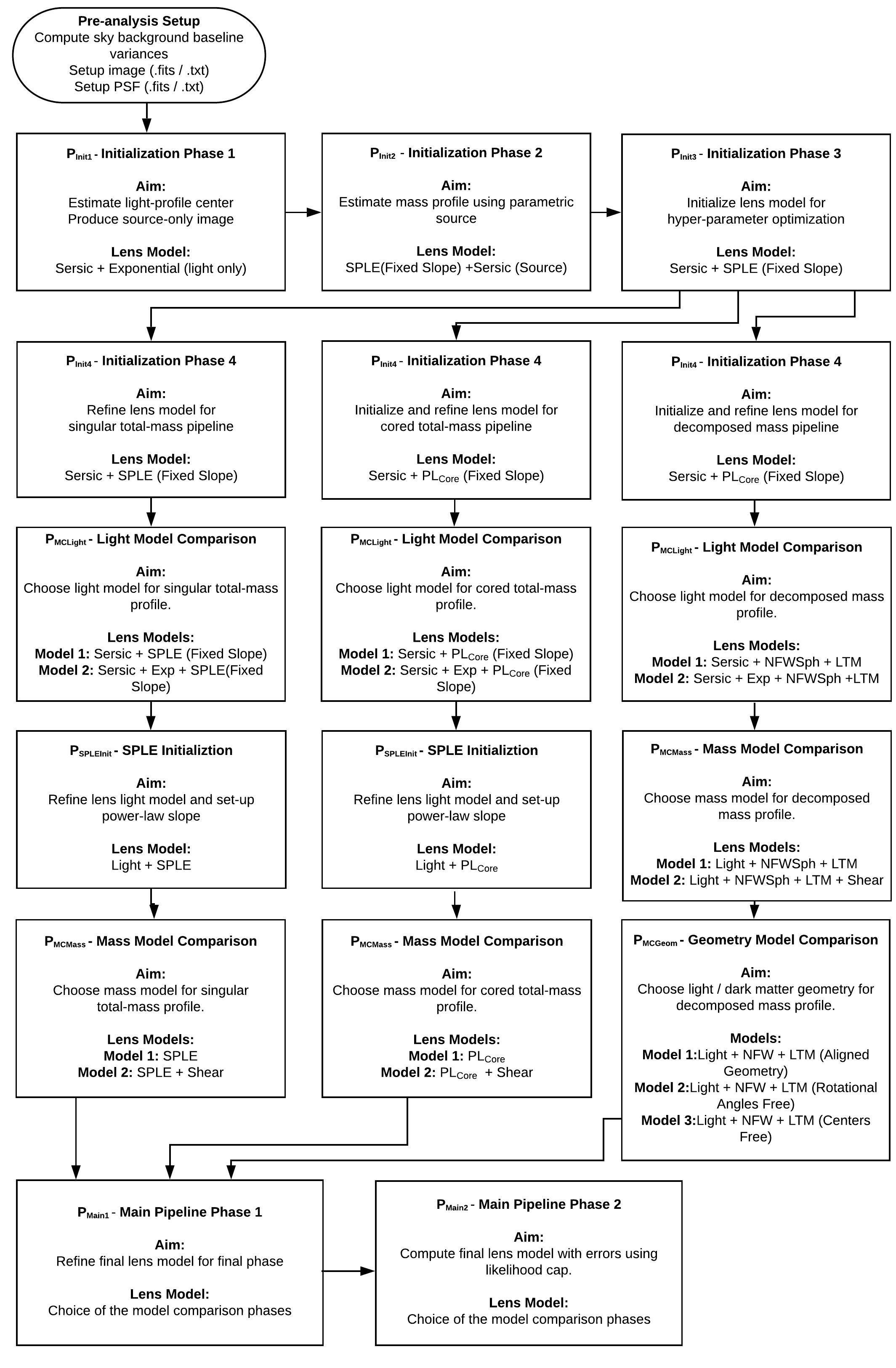}
\caption{A flow diagram of the {\tt AutoLens} automated analysis framework, which is described in detail throughout section \ref{SLPipeline}. The method begins with four initialization phases, which aim to accurately compute an accurate lens model alongside a robust initialization of the hyper-parameters. Before the fourth initialization phase, the pipeline splits into one of three routes, depending on whether a singular total-mass, cored total-mass or decomposed mass profile is desired. Each of these pipelines then runs a set of model comparison phases which set the complexity of the lens light profile, mass profile and light and dark matter geometry. The main analysis pipeline then begins which computes high precision estimates of every lens model parameter. Images without a lens light component use a reduced version of this pipeline described in section \ref{NoLensLight}.}
\label{figure:AutoLensFlow}
\end{figure*}

{\tt AutoLens} is a multi-phase automated analysis pipeline, designed with scalability to very large lens datasets in mind. Each phase involves a separate {\tt MultiNest} search, but generates initial points from priors derived from the highest likelihood regions of the previous phase's posterior distributions. Many tasks required to set up {\tt AutoLens} are performed automatically between phases, most notably optimizing the hyper-parameters of the adaptive image and source reconstruction features. Figure \ref{figure:AutoLensFlow} provides a flow diagram of {\tt AutoLens}, showing the different phases used throughout the automated analysis framework. \textbf{The figure shows that the pipeline incorporates three parallel routes, a particular route being chosen depending on whether a singular total-mass profile, cored total-mass profile or decomposed mass profile is being fitted. Also shown is the primary aim of each phase and lens model that is fitted.}

\subsection{Pipeline Phase Linking}

In the initial phases of the automated analysis pipeline broad uniform priors are assigned to all lens model parameters, since they have no expectation values computed for them. However, once estimated, this information is used to set that parameter's priors in the subsequent phase of the pipeline. The motivation behind this is that the more complex lens models used by {\tt AutoLens} have a large and highly degenerate non-linear parameter space within which accurate sampling and location of the global maximum is unattainable if broad priors are assumed on all parameters simultaneously. Therefore, the initial phases of {\tt AutoLens} accurately estimate a less complex lens model, with later phases using these results to gradually increase the lens model complexity whilst ensuring the non-linear parameter space is sampled accurately. To accompany this, the image and source reconstructions also gradually adapt to the properties of the lens and source being analyzed, facilitating further the fitting of more complex lens models.

The lens models used in different phases of the pipeline are linked via Gaussian priors centered on each parameter's high-likelihood regions, as estimated in the previous phase. Although it is possible to choose narrow priors to expedite the exploration of parameter space, the prior scaling values chosen for this work sample very broad regions, ensuring no results are simply a consequence of overly restrictive priors (but offering enough information to ensure parameter space is sampled robustly). Nevertheless, the freedom offered by the ability to scale the degree of sampling will be key to scaling the method up to large lens samples in the future. Appendix C gives a full description of how each phase is linked, along with the priors used to link every mass and light model from one phase to the next.
\subsection{Pipeline Initialization}\label{PLInit}
Initialization involves four automated tasks, the aim of which is to compute an accurate light ($Sersic$) and mass ($SPLE$ $\alpha = 2.2$) model alongside a robust initialization of the hyper-parameters.

\begin{itemize}

\item $\textbf{P}_{\rm  Init1}$ - Lens Light Subtraction - This phase fits a $Sersic$ + $Exponential$ light profile to the observed strong lens with the lensing analysis turned off. Shown in figure \ref{figure:PLInit} (top row), the resulting light model and subtraction are poor. However, they are sufficient for the centre of the lens's light profile to be estimated and to provide a lens-subtracted image for the next phase. No other information is used from this phase.

\begin{figure*}
\centering
\includegraphics[width=0.245\textwidth]{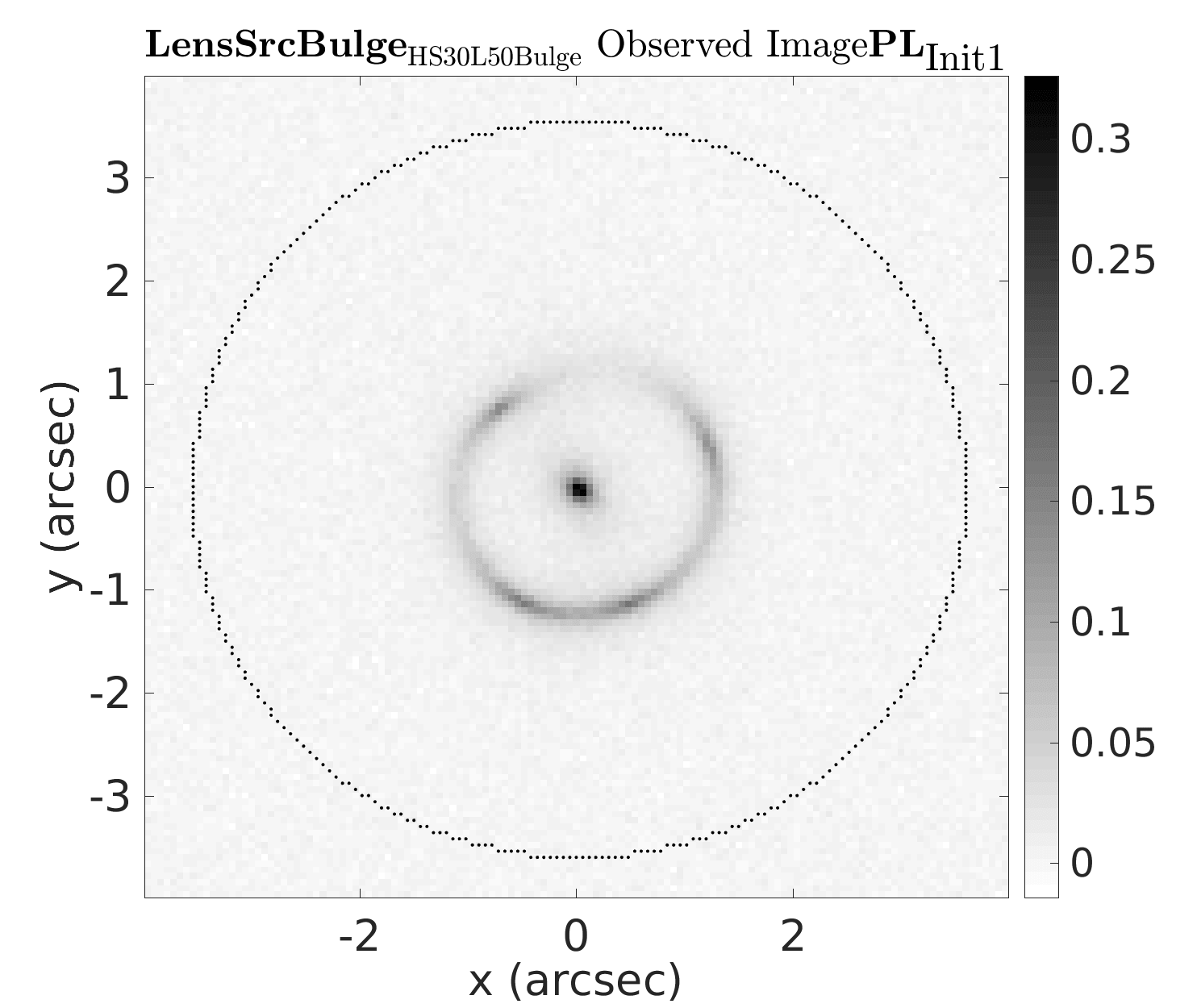}
\includegraphics[width=0.245\textwidth]{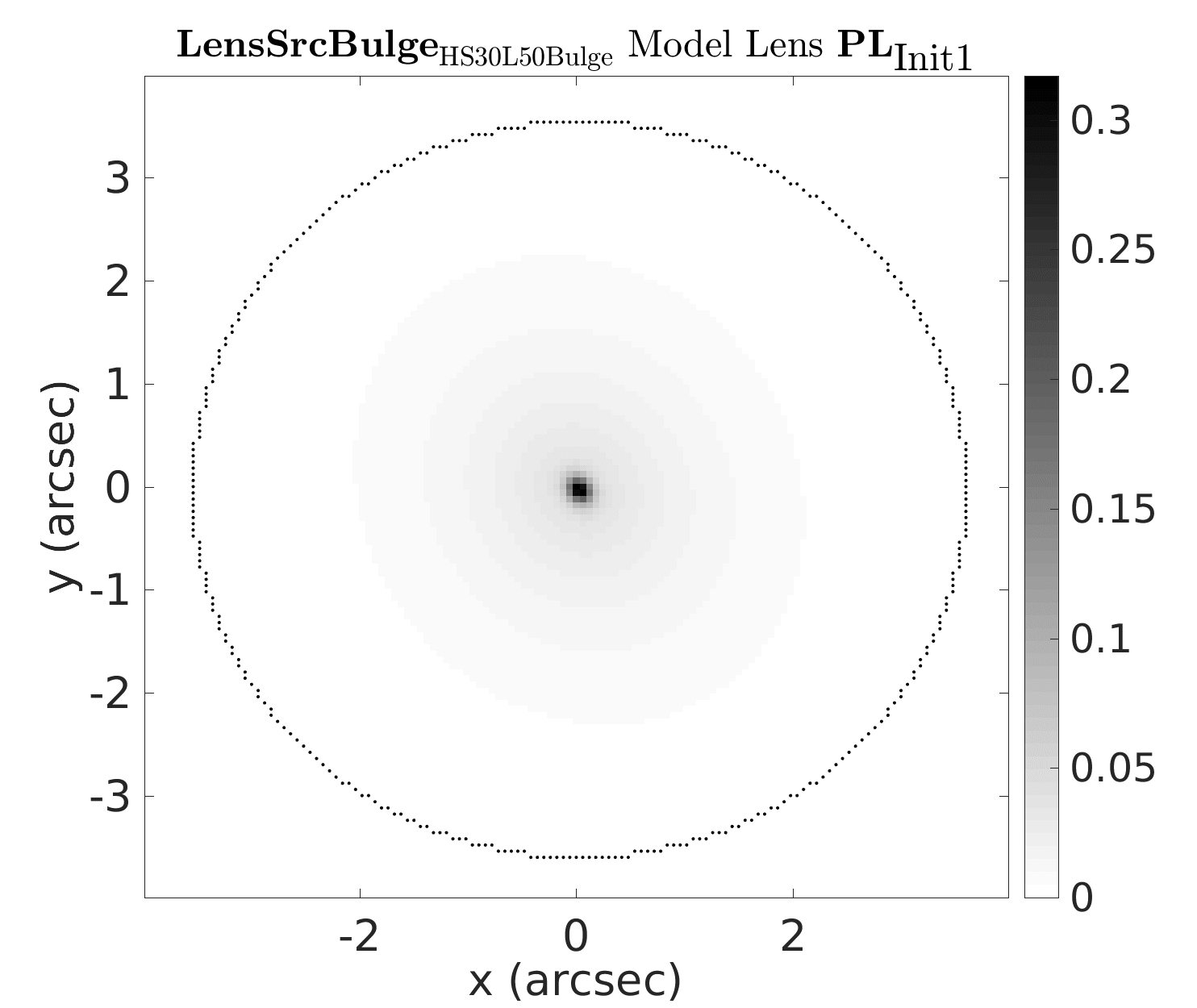}
\includegraphics[width=0.245\textwidth]{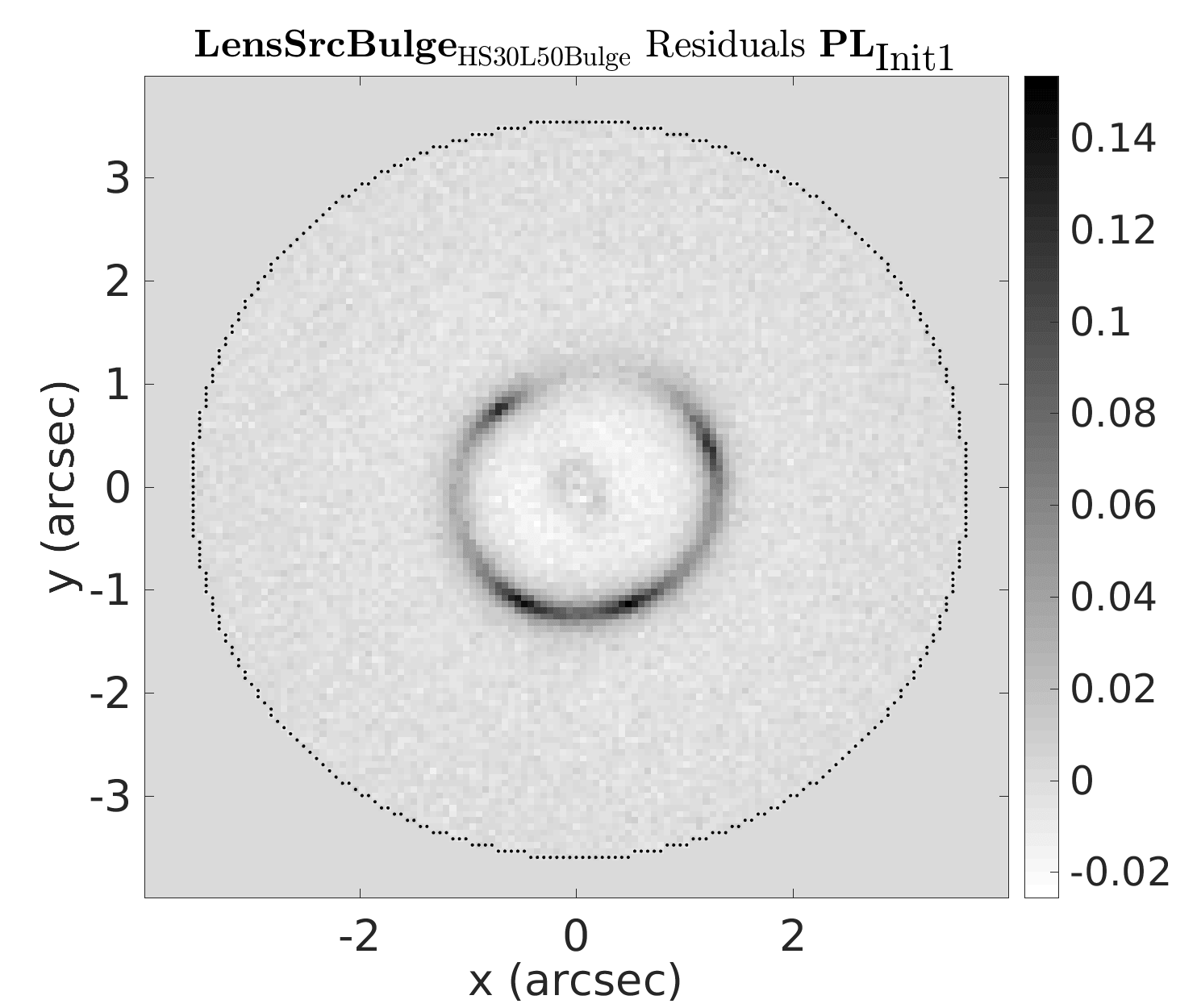}
\includegraphics[width=0.245\textwidth]{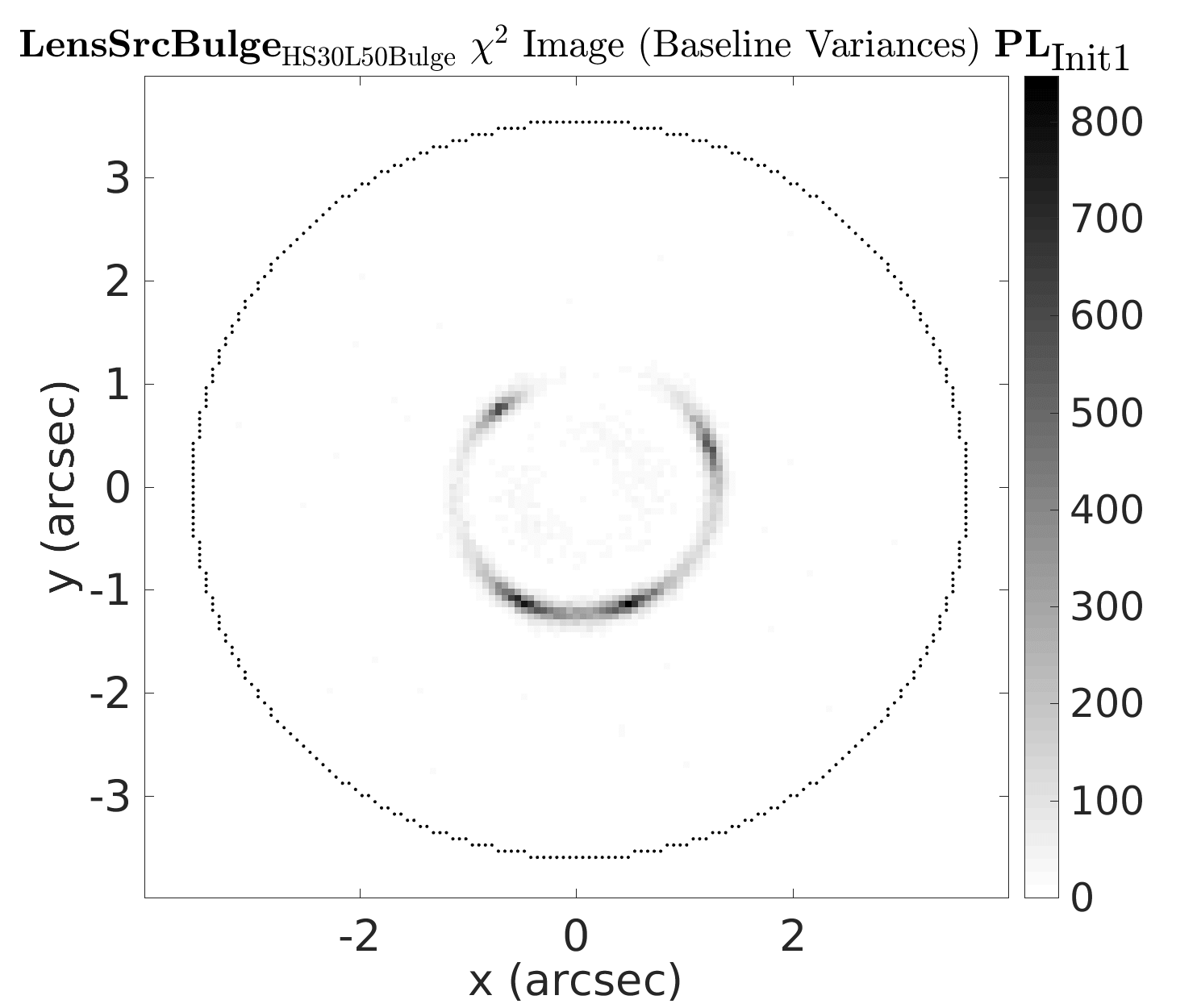}
\includegraphics[width=0.245\textwidth]{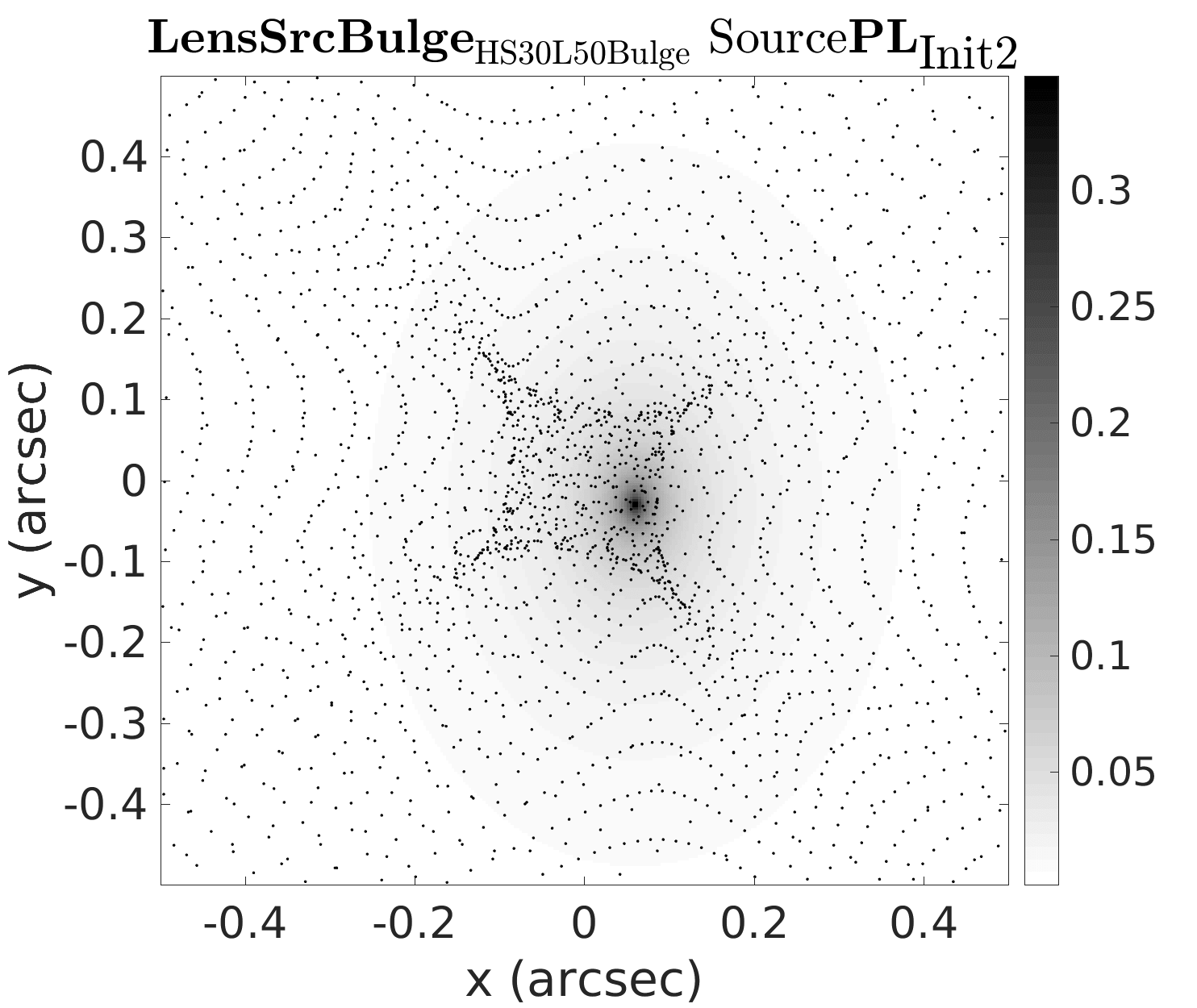}
\includegraphics[width=0.245\textwidth]{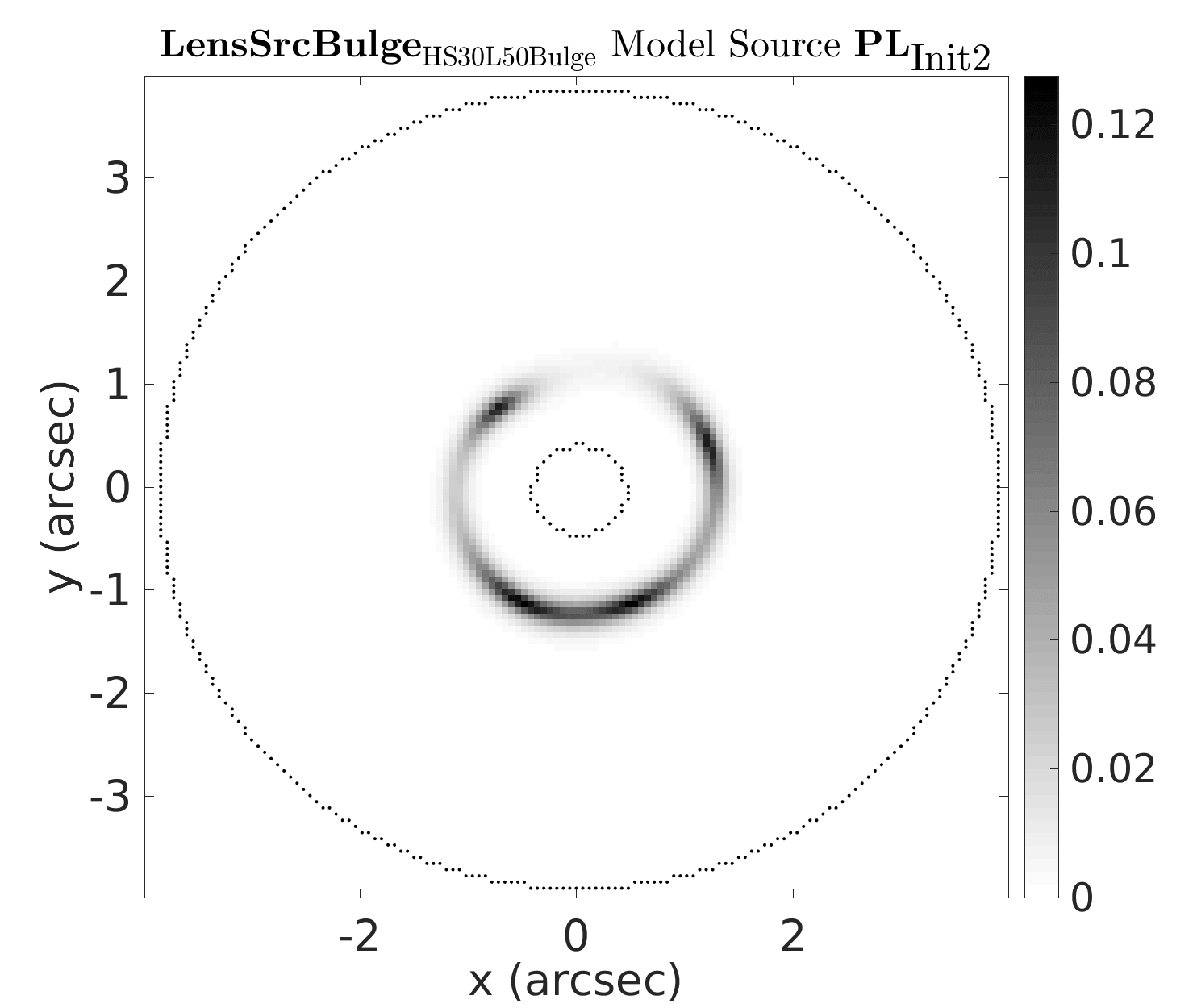}
\includegraphics[width=0.245\textwidth]{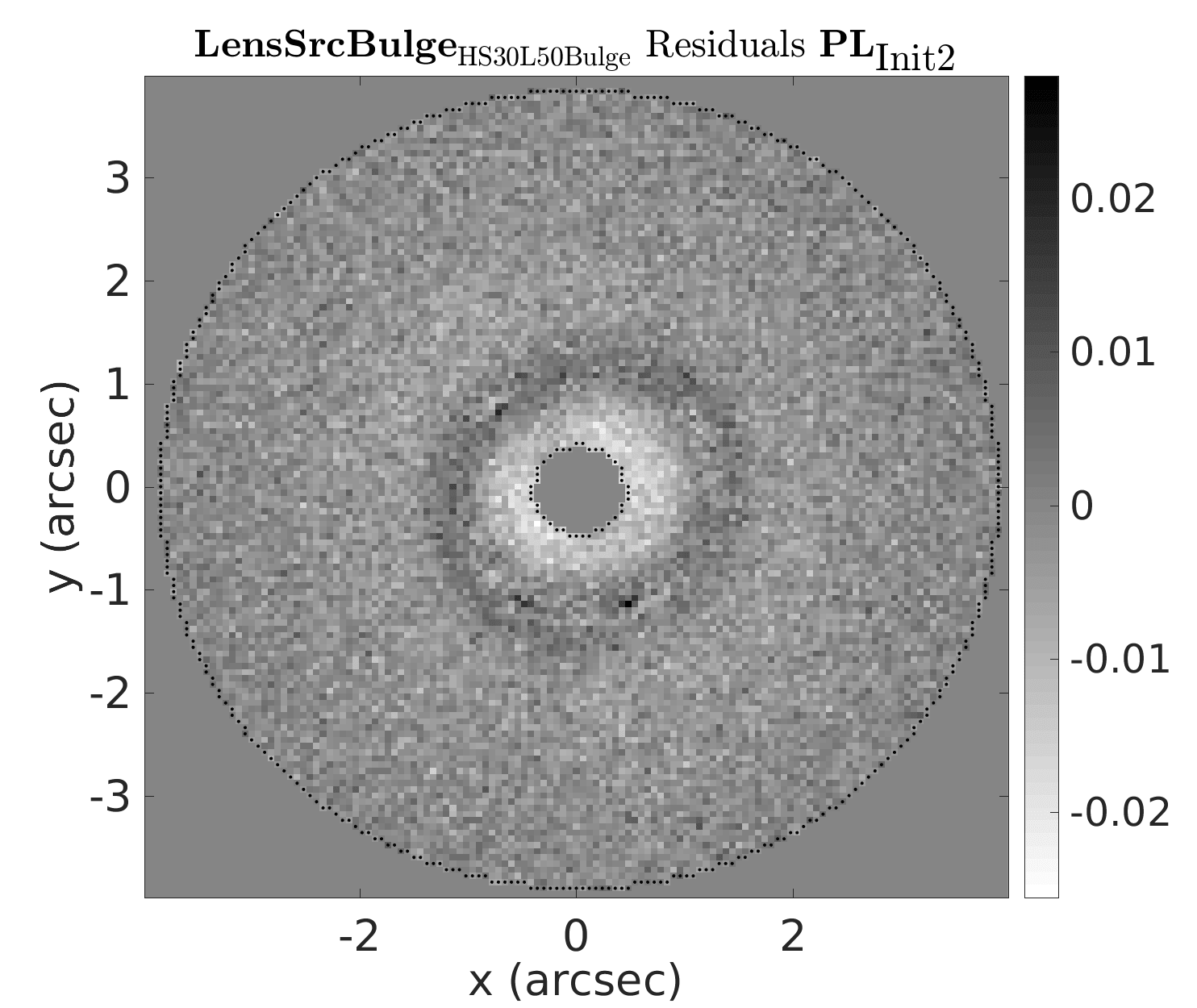}
\includegraphics[width=0.245\textwidth]{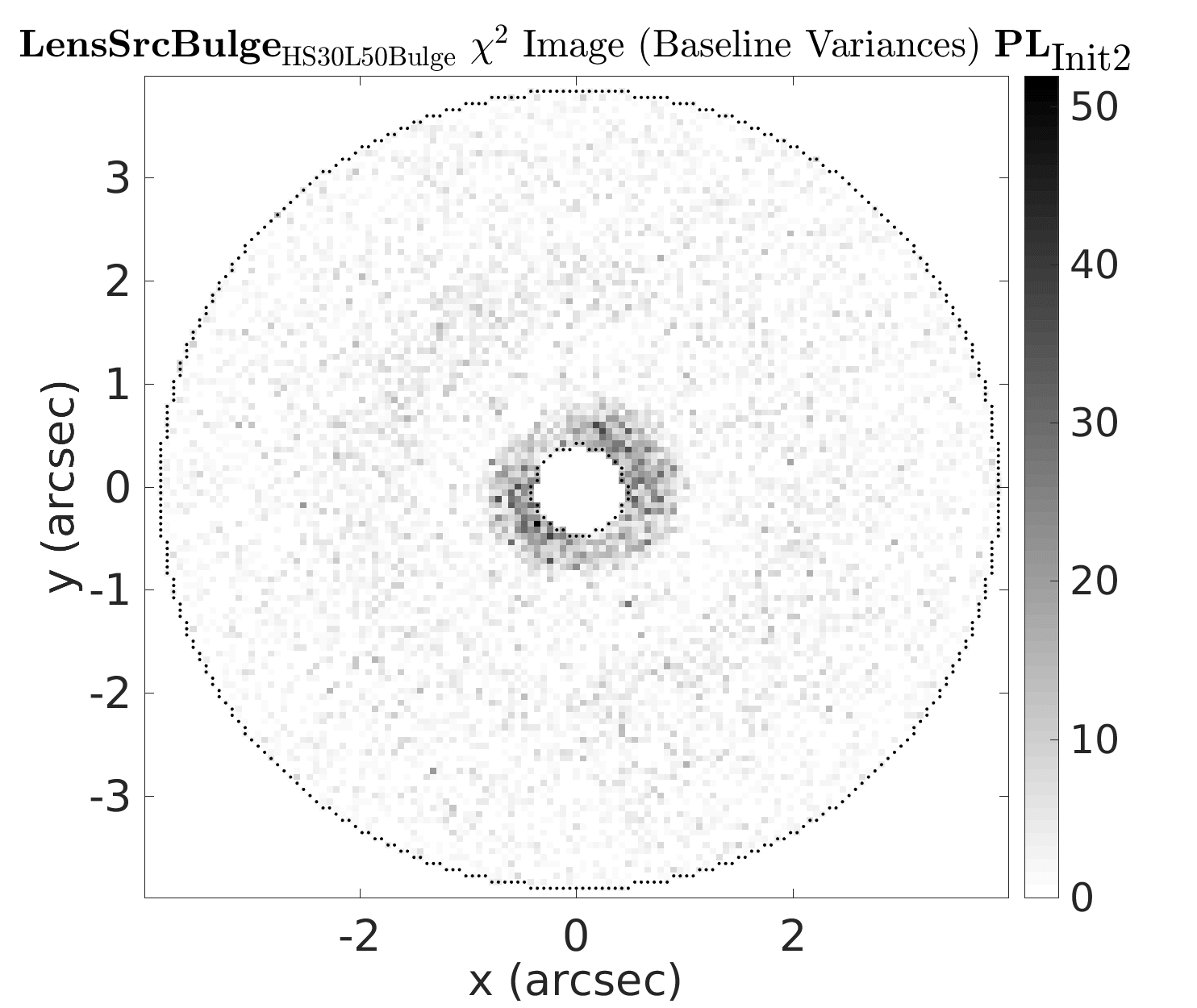}
\includegraphics[width=0.245\textwidth]{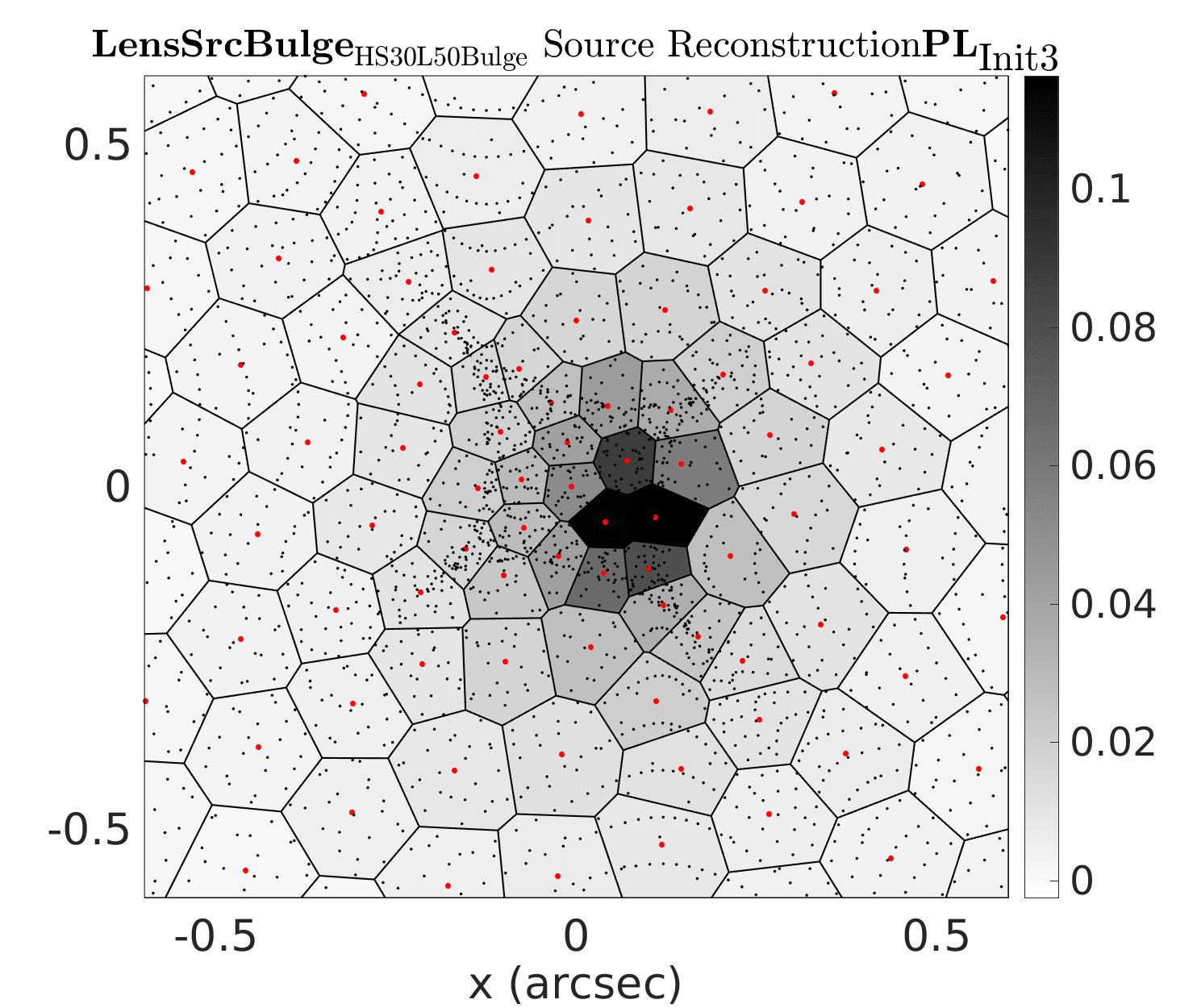}
\includegraphics[width=0.245\textwidth]{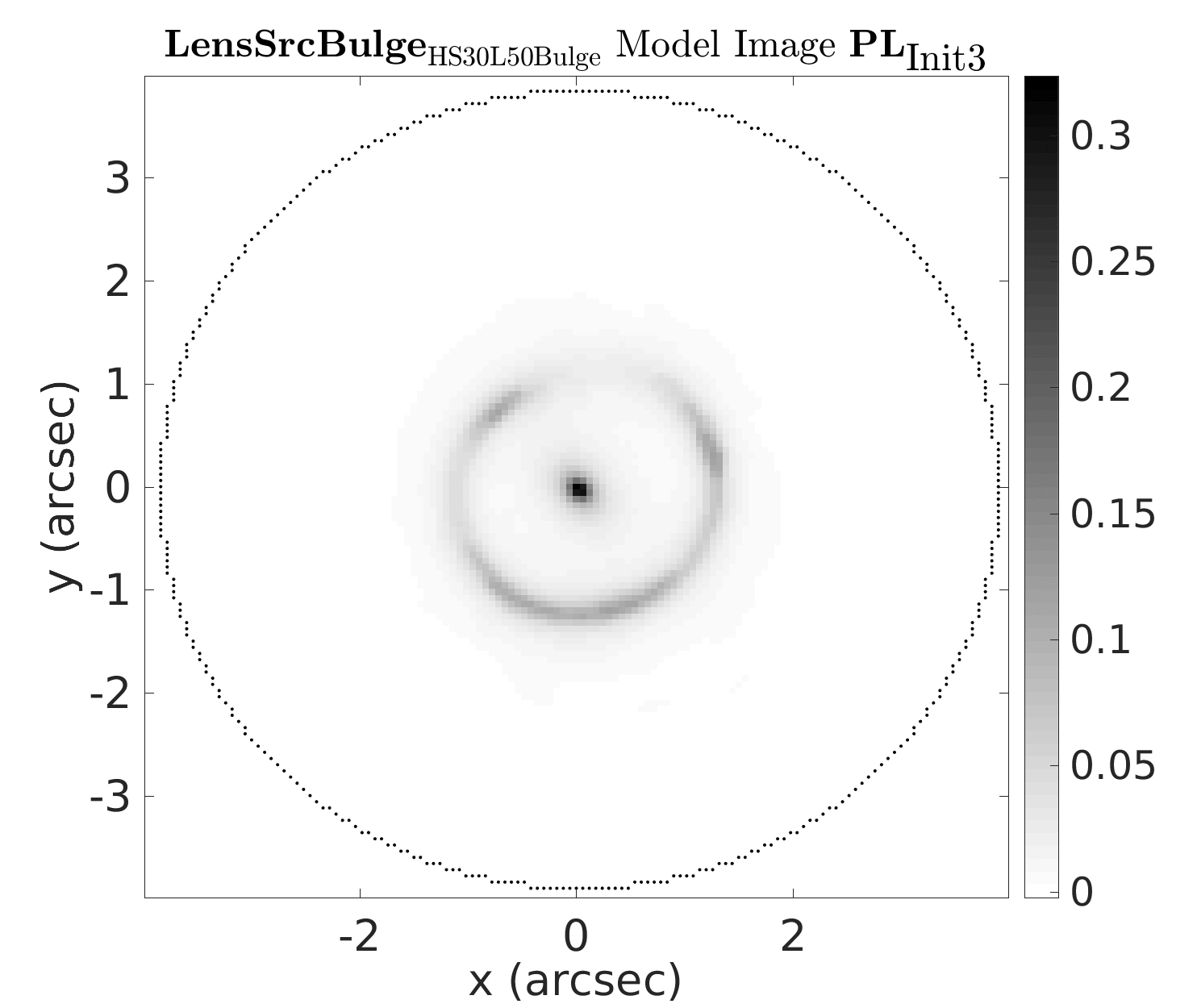}
\includegraphics[width=0.245\textwidth]{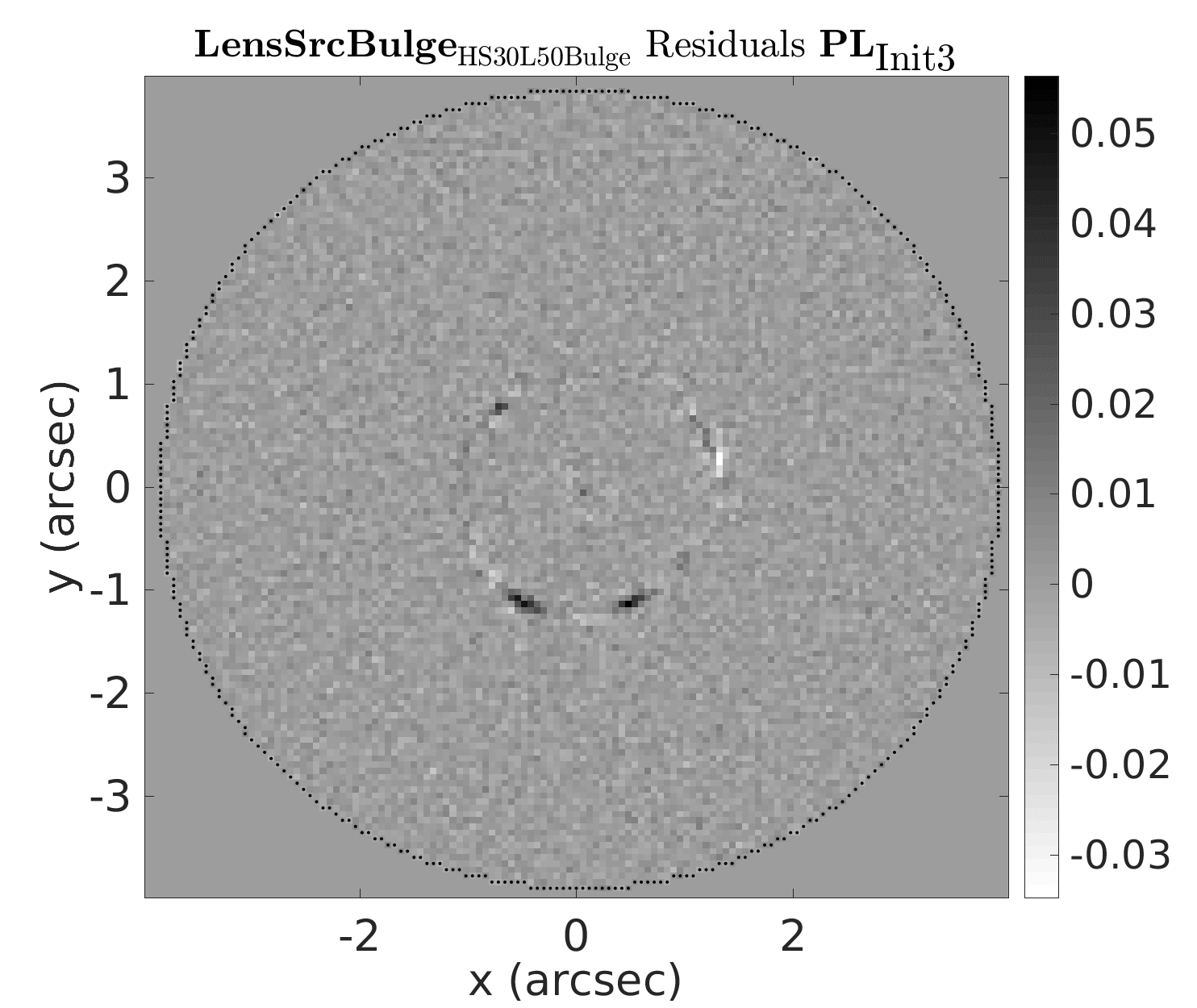}
\includegraphics[width=0.245\textwidth]{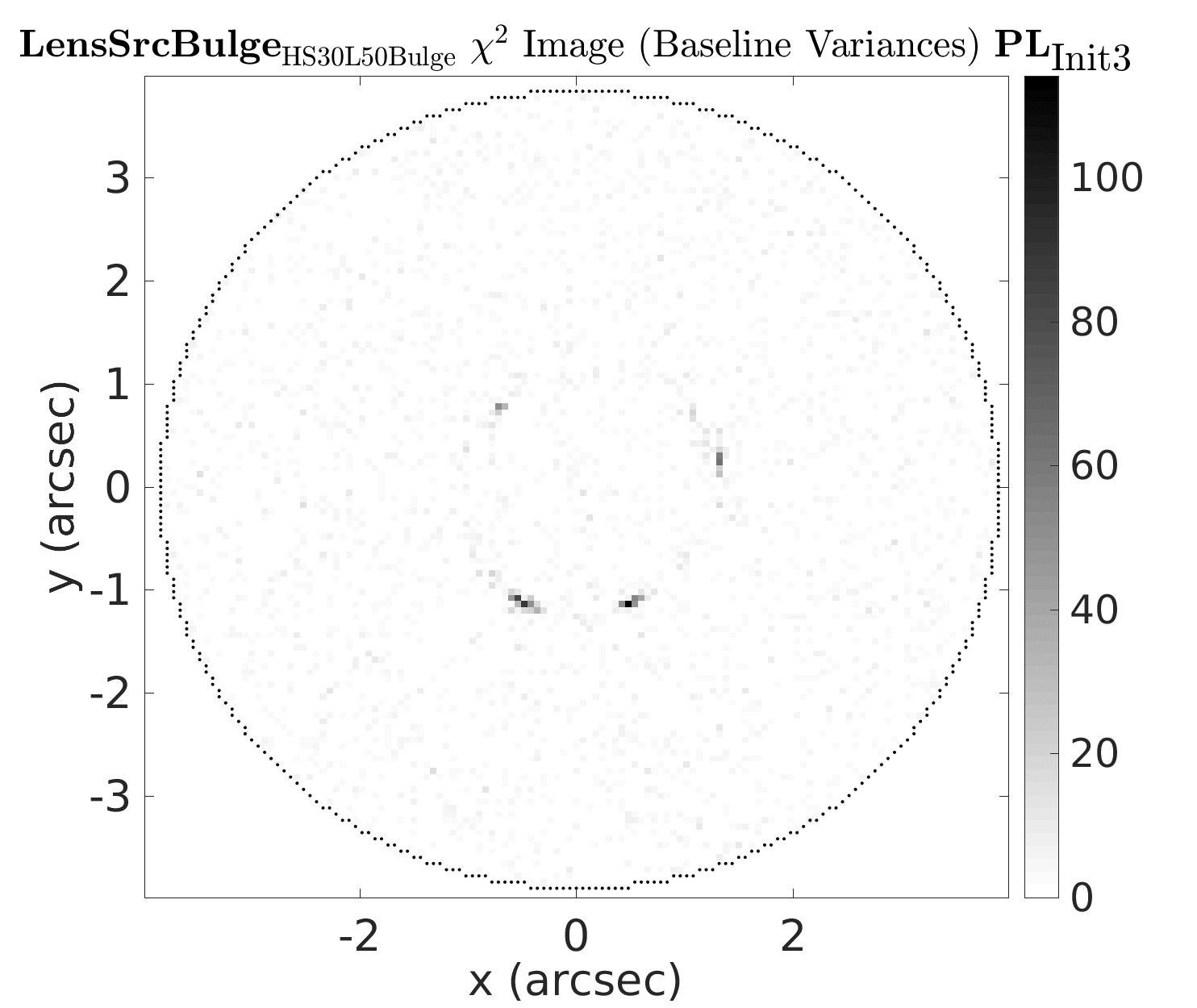}
\includegraphics[width=0.245\textwidth]{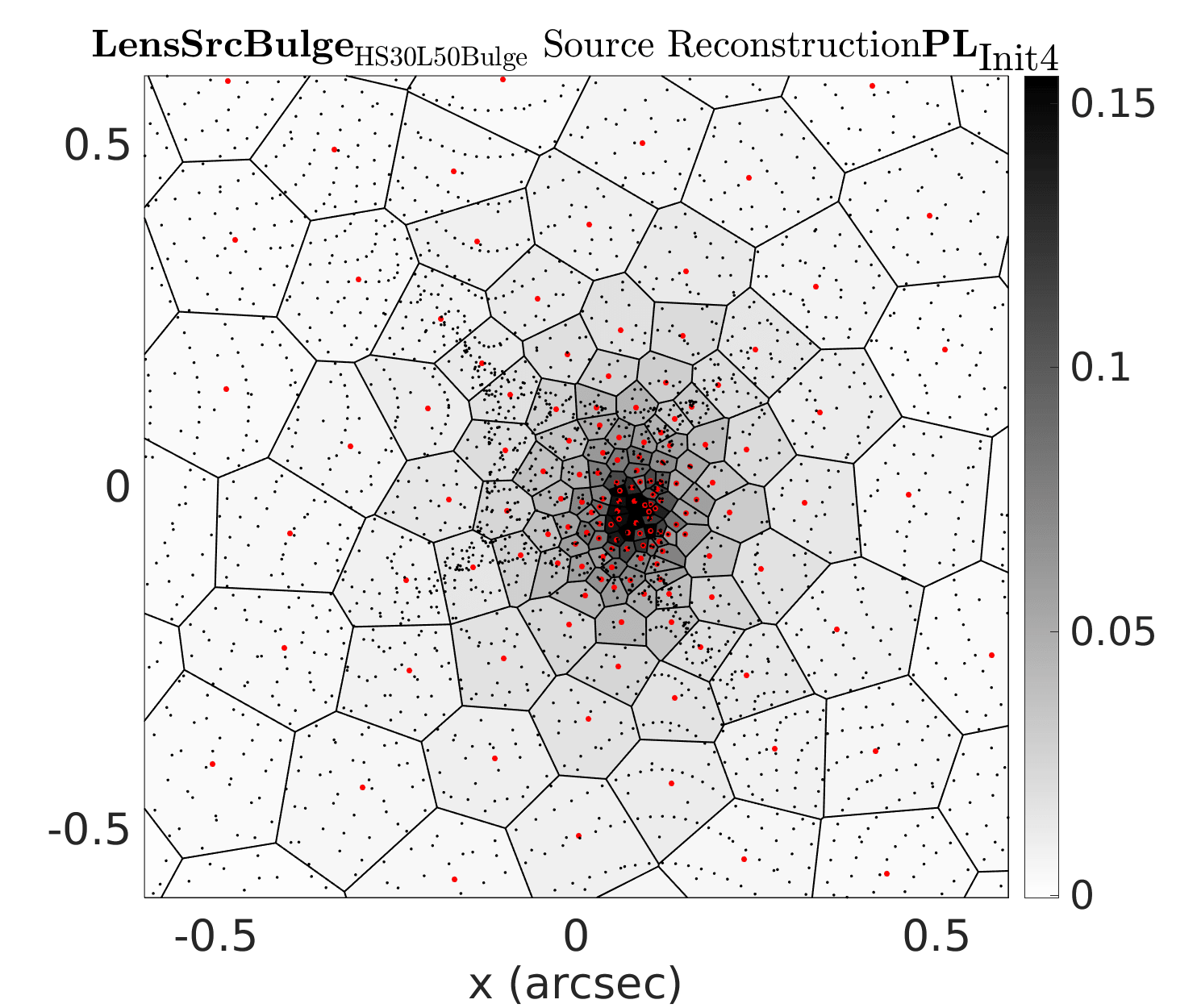}
\includegraphics[width=0.245\textwidth]{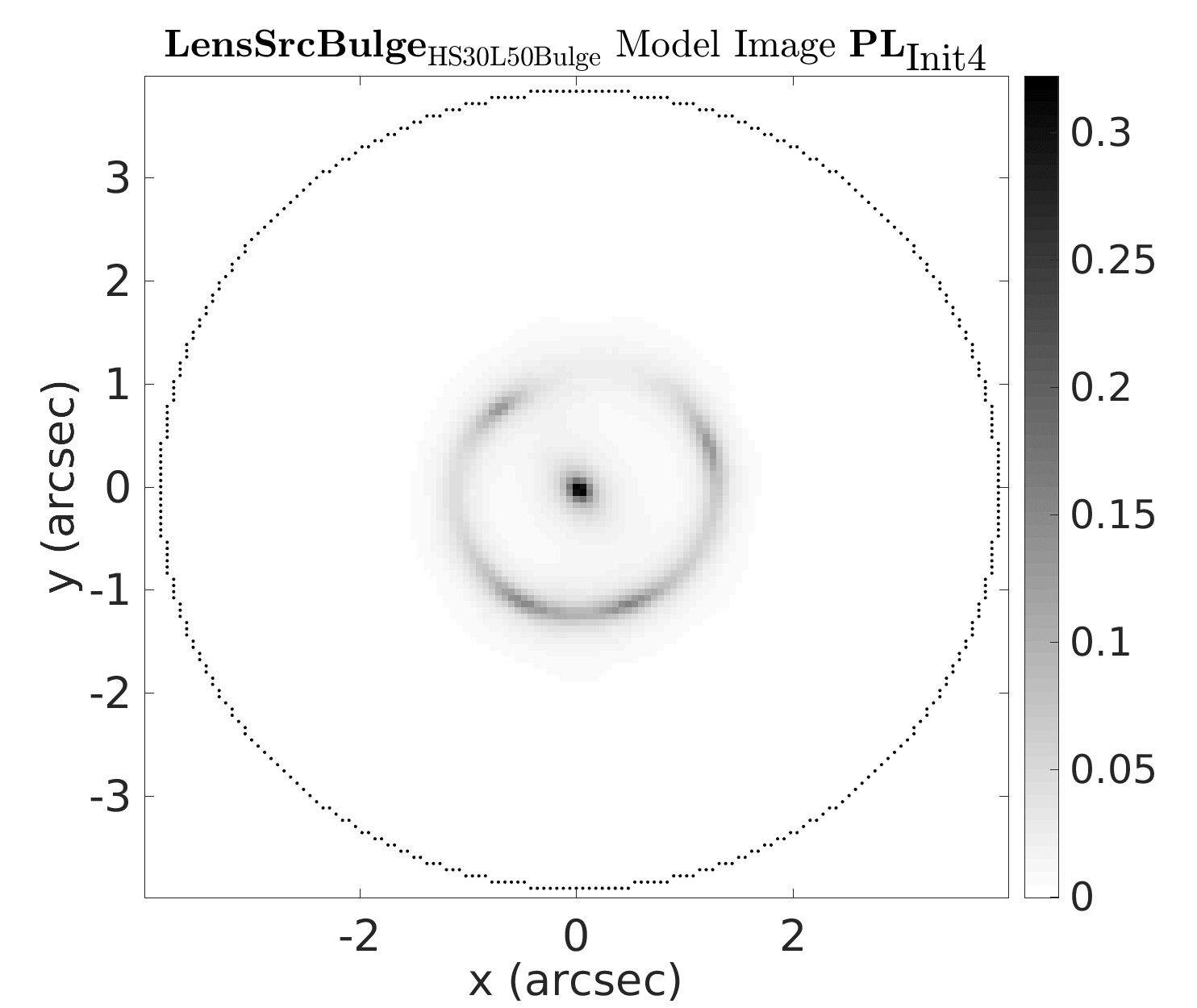}
\includegraphics[width=0.245\textwidth]{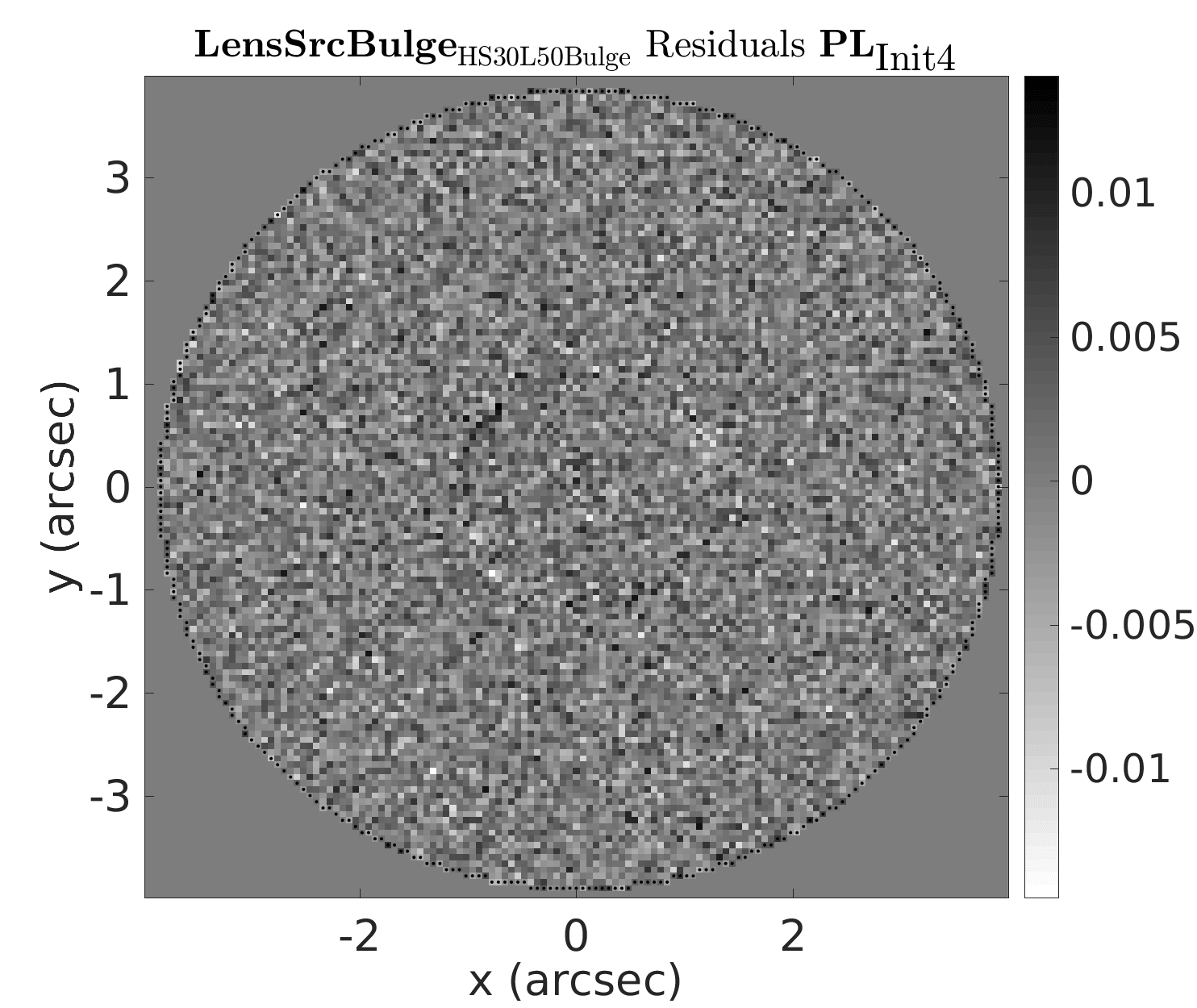}
\includegraphics[width=0.245\textwidth]{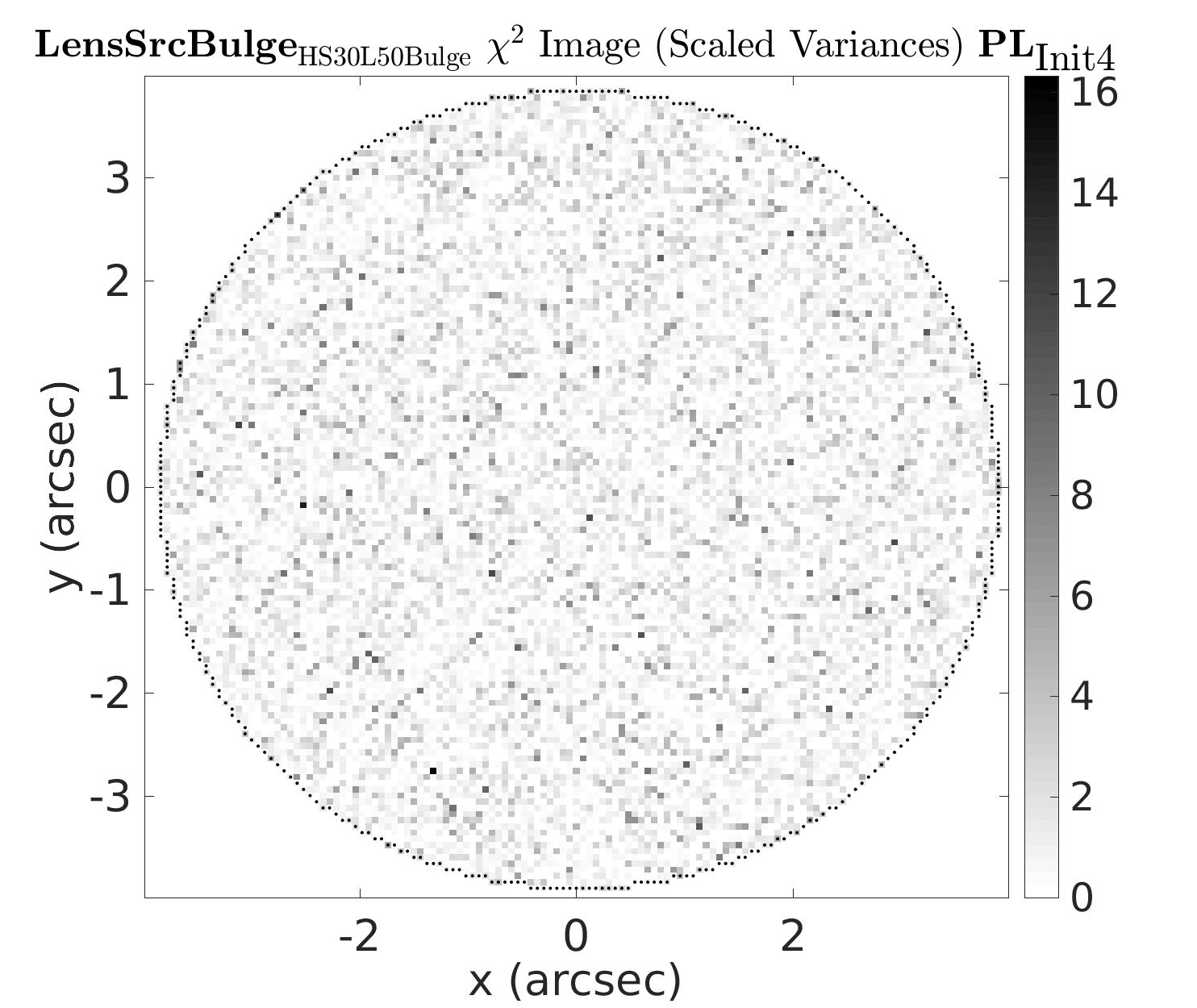}
\caption{A demonstration of the analysis performed in phases $\textbf{P}_{\rm  Init1}$ (top row), $\textbf{P}_{\rm  Init2}$ (top-middle row), $\textbf{P}_{\rm  Init3}$ (bottom-middle row)  and $\textbf{P}_{\rm  Init4}$ (bottom row) using the image $\textbf{LensSrcBulge}_{\rm  HS30L50Bulge}$. The left panels show the observed image and mask (top-row) and source reconstructions of each phase (remaining rows), the middle-left panels the reconstructed model images, the middle-right panels the residuals and right most panels the $\chi^2$ images (equation \ref{eqn:ChiSqSrc}). \textit{top row} ($\textbf{P}_{\rm \mathrm{Init1}}$) - A $Sersic$ + $Exp$ light profile is fitted to the observed image with lensing analysis turned off. The light model gives a poor overall fit to the observed image but gives an accurate estimate of $x_{\rm l}$ and $y_{\rm l}$ and reveals the lensed source for the next phase. \textit{top-middle row} ($\textbf{P}_{\mathrm{Init2}}$) - A $SPLE$ ($\alpha = 2.2$) + $Sersic$ model is fitted to the lens subtracted image generated in the previous phase, where the source is modeled using a smoothly parametrized profile which is sampled simulataneously with the mass model. This gives a robust initialization of the mass model's parameters for the next phase and initializes the positional image pixels. \textit{bottom-middle row} ($\textbf{P}_{\rm \mathrm{Init3}}$) - A $Sersic$ + $SPLE$ ($\alpha = 2.2$) model is fitted to the observed image with the adaptve image and source analaysis features switched off and $\lambda$ included as a free parameter. The light and source models give a reasonable fit to the image, but the residuals and $\chi^2$ image show the issues demonstrated in section \ref{AdaptDemo} are present. Nevertheless, the model is sufficient to optimize the hyper-parameters so that the next phase can use the adaptive image and souce features. \textit{bottom row} ($\textbf{P}_{\rm \mathrm{Init3}}$) - A $Sersic$ + $SPLE$ ($\alpha = 2.2$) model is again fitted to the observed image but now with the adaptive image and source features. The light and source models give an accurate fit and the residuals and $\chi^2$ are improved from the previous phase (and will be further improved after this phase's hyper-parameter optimization).} 
\label{figure:PLInit}
\end{figure*}

\item $\textbf{P}_{\rm  Init2}$ - Parametric Source Model - Initializing a lens model with a pixelized source reconstruction is a surprisingly non-trivial task. This is because of the over / under fit solutions discussed in N15, which reconstruct the source as a demagnifed version of the lensed image. In general, these solutions have a lower ln$\epsilon$ than the input model, but occupy much larger volumes of non-linear parameter, causing {\tt MultiNest} to get stuck in their local maxima. To circumvent this issue a parametric source profile is used first, for which these unwanted and unphysical solutions do not exist. This allows an initial estimate of the mass profile to be computed in a completely general way, which can then be used in the next phase to prevent {\tt MultiNest} from sampling these over / under fit solutions. Thus, this phase fits the $SPLE$ ($\alpha = 2.2$) + $Sersic$ (source) model to the image, omitting lens light modeling and masking the central regions where the poor lens subtraction leaves residuals. This is illustrated in figure \ref{figure:PLInit} (top-middle row), where the lensed source models can be seen to fit the observed image well enough to ensure the mass model has been estimated reliably. Finally, the positional image pixels and threshold value are updated (see appendix \ref{AppCalc}).

\item $\textbf{P}_{\rm  Init3}$ - Initial Lens Model and Hyper-Parameters - This phase now uses a pixelized source-plane, fitting the $Sersic$ +  $SPLE$ ($\alpha = 2.2$) model to the image ($\alpha = 2.2$ to remove the biases described in section \ref{CenIms}). Restrictive priors are placed on the mass model parameters, ensuring the method does not sample the unphysical solutions corresponding to a demagnified image reconstruction. The baseline variance-map is used, source-light adaptation is turned off and a constant regularization scheme is applied with $\lambda$ included as a free parameter in the non-linear search. This is illustrated in figure \ref{figure:PLInit} (middle row), where both the light and source models can be seen to fit the observed image reasonably well, but with the residuals and $\chi^2$ image showing the issues discussed previously (as is expected given the adaptive source / image features are not implemented yet). However, the model is of sufficient accuracy to initialize the hyper-parameters, which is performed before phase $\textbf{P}_{\rm  Init4}$. The positional image pixels and threshold are again updated.

\item $\textbf{P}_{\rm  Init4}$ - Model Refinement - This phase fits one of three models: (i) the $Sersic$ + $SPLE$ ($\alpha = 2.2$) model (with priors relaxed compared to the previous phase); (ii) the $Sersic$ + $PL_{\rm Core}$ ($\alpha = 2.0$) model (using priors from the previous phase and a broad prior on the core radius) or; (iii) the $Light$ + $NFWSph$ model (with broad uniform priors). The models correspond to the singular total mass pipeline, cored total mass pipeline or decomposed mass pipeline, respectively. This phase benefits from the adaptive image and source features following the hyper-parameter initialization of the previous phase. The aim of this phase is to refine the lens model and ensure an excellent optimization of the hyper-parameters for the model comparison phases. This is illustrated in figure \ref{figure:PLInit} (bottom row), where an accurate model for both the light and mass components is shown, alongside much improved residuals and $\chi^2$ values. Following this phase, hyper-parameters are re-optimized and positional image pixels are recomputed.

\end{itemize}
\subsection{Bayesian Model Comparison}\label{PLMC}
The next stage of the pipeline `builds' the lens model, by performing Bayesian model comparison. A number of publications have already detailed the hierarchical Bayesian formalism of pixel-based lens analysis methods like ${\tt AutoLens}$ (e.g. S06, \citealt{Vegetti2009,Tagore2014}). Hence, only a brief overview is given here. Bayes's theorem is given by
\begin{equation}
\label{eqn:Bayes1}
P(m|d,M) =P(m|M) \frac{P(d|m,M)}{P(d|M)} \, \, ,
\end{equation}
where $d$ is the data and $m$ is a particular realization of the overall model $M$ which comprises all linear source parameters, hyper-parameters, lens model parameters and $\tens{H}_{\rm  \Lambda}$. $P(d|m,M)$ gives the likelihood, $P(m|M)$ the priors on model parameters and $P(m|d,M)$ the posterior probability. The Bayesian evidence is given by $P(d|M)$ (which readers should note is different to the Bayesian evidence given by equation \ref{eqn:evidence2}, which ranks the source reconstruction) and can be obtained by integrating over all possible models $m$ in the set of models $M$ as
\begin{equation}
\label{eqn:Bayes2}
P(d|M) = \int P(m|M)P(d|m,M) dm \, \, .
\end{equation}
This expression has the principle of Occam's razor built into it, whereby overly complex models are penalized if they do not give a justifiably improved fit to the observed data. Thus, maximizing this quantity objectively chooses the model which best fits the data without being overly complex. The ratio of the evidence of two models (e.g. $P(d|M_1)/P(d|M_2$)) gives their Bayes factor and to accept a more complex model, a Bayes factor greater than twenty is required  (considered `strong' evidence in  Bayesian statistics). The integral given in equation \ref{eqn:Bayes2} is estimated by {\tt MultiNest} and therefore is a natural byproduct of {\tt AutoLens}'s analysis. It should be noted that during the model comparison phases, {\tt MultiNest}'s non-linear parameter space comprises a subset of $M$'s parameters (because certain hyper-parameters are left fixed) and the evidence is estimated for only these parameters. The parameters which are omitted have no impact on the value of evidence and are omitted for efficiency. In practice, model comparison with {\tt AutoLens} simply amounts to fitting different light or mass models at various stages of the pipeline and choosing a more complex model when the evidence increases over the previous (simpler) model by a threshold value, which is set to $20$.

When transitioning to a more complex lens model, the setup of the adaptive image and source features may be problematic. These features adapt to a specific source morphology and suppress the $\chi^2$ contribution of poorly fitted image pixels by increasing their variances. If a more complex model changes the reconstructed source’s morphology or accurately fits pixels which previously had their variances increased, there is a risk that using the hyper-parameters of the simpler model may prevent the more complex model from making a sufficiently high gain in ln$\epsilon$ to be correctly favoured by model comparison. Therefore, the relevant hyper-parameters are included as free parameters in each model comparison's non-linear search. This allows the model to change the source reconstruction and undo the suppression of image pixels $\chi^2$ values if and when the new lens model begins to accurately fit them, which in turn allows the correct ln$\epsilon$ values to be sampled.

The model comparison phases (and an intermediate linking phase) follow the initialization phases above and are (noting that following all the phases below is a hyper-parameter re-optimization):

\begin{itemize}

\item $\textbf{P}_{\rm  MCLight}$ - Light Model - This phase chooses the light model. For the singular total mass pipeline a $SPLE$ ($\alpha = 2.2$) mass model is used, for a cored total-mass pipeline a $PL_{\rm Core}$ ($\alpha = 2.0$) model and decomposed pipeline a $NFWSph$ model. The mass model's parameters are initialized using the results of phase $\textbf{P}_{\rm  Init4}$. The hyper-parameters $\lambda_{\rm  Src}$, $\omega_{\rm  Lens}$ and $\omega_{\rm Src}$ are not fixed. First, the $Sersic$ + $Mass$ model is refitted, to compute the Bayesian evidence now the hyper-parameters have been re-optimized. This is compared to the $Sersic$ + $Exp$ + $Mass$ model, thus determining whether a two-component light profile is required. For the total-mass pipelines the light model does not contribute to the mass model, whereas for the decomposed pipeline it does.

The simplified mass profiles used during this phase often leave residuals in the lensed source. The two-component light model was found to make gains in Bayesian evidence by subtracting these source residuals. This behaviour is undesirable, therefore the upper limit on the hyper-parameter $\omega_{Src}$ is increased to a target value $\chi_{\rm sca}^2 = 1.0$ (as opposed to the value $\chi_{\rm sca}^2 = 10.0$ used everywhere else in the pipeline). This ensures these residuals are not fitted by the light profile, as they are down-weighted by variance scaling. After this phase, the method reverts to a target value $\chi_{\rm sca}^2 = 10.0$.

\item $\textbf{P}_{\rm  SPLEInit}$ - SPLE Initialization - For the total-mass pipelines, if the lens's slope deviates from the fixed value previously assumed for $\alpha$ it is beneficial to refine the mass model and adaptive source features to reflect. This phase does exactly this, by fitting the $SPLE$ or $PL_{\rm Core}$ mass model with free $\alpha$, alongside the light profile just chosen. If variance scaling is on, $\omega_{\rm Lens}$ and $\omega_{\rm Src}$ are included as free parameters, to ensure the new light profile (if chosen) and mass profile are able to fit regions of the image that may have previously had their $\chi^2$ suppressed.

\item $\textbf{P}_{\rm  MCMass}$ - Mass Model - For the total-mass pipelines, the most probable light model computed in $\textbf{P}_{\rm  SPLEInit}$ is subtracted from the observed image to create a source-only image. This image is then fitted to choose the mass model with light modeling turned off for computational speed. For the decomposed mass pipeline, the original image is used with light modeling turned on. The hyper-parameters $\lambda_{\rm  Src}$ and $\omega_{\rm  Src}$ (and $\omega_{\rm  Lens}$ for the decomposed pipeline) are not fixed. First, the $SPLE$, $PL_{\rm Core}$ or $Light$ + $NFWSph$ model is fitted, which is subsequently compared to the same model with the inclusion of a $Shear$ term, thus determining if an external shear component is necessary.

\item $\textbf{P}_{\rm  MCGeom}$ - Light / Dark Matter Geometry - For the decomposed model pipeline this phase determines the light and dark matter geometries, where the $Light$ model is the light model chosen in the phase $\textbf{P}_{\rm  MCLight}$. The hyper-parameters $\lambda_{\rm  Src}$, $\omega_{\rm  Src}$ and $\omega_{\rm  Lens}$ are not fixed. First, the $Light$ + $NFW$ model is fitted assuming geometric alignment ($\theta_{\rm l} = \theta_{\rm d}$, $x_{\rm l} = x_{\rm d}$, $y_{\rm l} = y_{\rm d}$). This is compared to a $Light$ + $NFW$ model which allows rotational misalignment ($\theta_{\rm l}$ and $\theta_{\rm d}$ both free) but retains the assumption of a common center. A third comparison is then performed, which allows the centers to vary ($x_{\rm l}$, $y_{\rm l}$, $x_{\rm d}$ and $y_{\rm d}$ all free) and assuming the rotational alignment determined from the previous result. 
\end{itemize}

\subsection{Separate Pipelines}\label{PLCore}

It is worth noting the importance of using separate pipelines for different mass profiles. For example, attempts to fit images generated using a cored mass profile with a singular mass profile were found to give poor results, because lensed source features specific to a cored mass model (radial arcs and a central image) cannot be replicated accurately by a singular model. On the other hand, if a cored model is wrongly assumed, it will wrongly include some of the lens's light in the source reconstruction, biasing the lens model. Singular total-mass and decomposed mass models were generally found to produce lensed sources with the same overall structure. However, geometric offsets between a decomposed model's light and dark matter components produce unique features a singular model cannot replicate (see section \ref{Results}).

Therefore, it is important to use separate pipelines for models which produce different and unique lensed source features. To choose between these models, Bayesian model comparison is again used, now using the final results of each pipeline. Whilst this is computationally expensive, the splitting of each pipeline means they can run in parallel.

\subsection{Main Pipeline}\label{PLMain}

Following the initialization and model comparison, phase one of the main pipeline begins, using the lens model previously chosen. The lens's light and mass are modeled simultaneously, with all initial parameters sampled via Gaussian priors initialized using the previous phase's results. The main pipeline comprises two phases: (i) the lens model is estimated and the hyper-parameters re-optimized; (ii) the same model is estimated again, but using more computationally intensive settings providing more accurate parameter estimates and errors. A third phase is possible, which includes all hyper-parameters in the non-linear search. However, this final phase is computationally expensive and was found to have negligible influence on the inferred lens model.

\begin{table*}
\resizebox{\linewidth}{!}{
\begin{tabular}{ l | l | l | l | l | l | l | l | l} 
\multicolumn{1}{p{1.2cm}|}{Phase} 
& \multicolumn{1}{p{1.5cm}|}{Method} 
& \multicolumn{1}{p{1.6cm}|}{\centering $N_{\rm s}$ Likelihood \\ Tolerance} 
& \multicolumn{1}{p{1.4cm}|}{\centering $\vec{{\alpha}}_{\rm x,y}$ Subgrid \\ Degree} 
& \multicolumn{1}{p{1.3cm}|}{\centering PSF \\ Trim}  
& \multicolumn{1}{p{2.2cm}|}{\centering {\tt MultiNest} \\ Live Points} 
& \multicolumn{1}{p{2.2cm}|}{\centering {\tt MultiNest} \\ Tolerance} 
& \multicolumn{1}{p{2.2cm}|}{\centering {\tt MultiNest} \\ Reduction Factor} 
& \multicolumn{1}{p{2.2cm}}{\centering Stopping \\ Factor} 
\\ \hline
& & & & & & & &\\[-4pt]
$\textbf{PL}_{\rm  Init1}$         & Light Only   & N/A  & N/A            & $10\%$  & 50  & 1000. & 0.2 & N/A \\[2pt]
$\textbf{PL}_{\rm  Init2}$         & Mass + Source & N/A  & $2 \times 2$   & $10\%$  & 200 & 100.0 & 0.2 & 5.0 \\[2pt]
$\textbf{PL}_{\rm  Init3}$         & Light + Mass & 15.0 & $4 \times 4$   & $10\%$  & 125 & 50.  & 0.2 & 2.0 \\[2pt]
$\textbf{PL}_{\rm  Init4}$         & Light + Mass & 15.0 & $4 \times 4$   & $10\%$  & 125 & 50.  & 0.2 & 2.0 \\[2pt]
$\textbf{PL}_{\rm  MCLight}$       & Light + Mass & 8.0  & $8 \times 8$   & $0\%$   & 150 & 1.0   & 0.2 & 2.0 \\[2pt]
$\textbf{PL}_{\rm  SPLEInit}$       & Light + Mass & 8.0  & $8 \times 8$   & $0\%$   & 150 & 1.0   & 0.2 & 2.0 \\[2pt]
$\textbf{PL}_{\rm  MCMass}$        & Mass Only    & 5.0  & $8 \times 8$   & $0\%$   & 125 & 1.0   & 0.2 & 2.0 \\[2pt]
$\textbf{PL}_{\rm  MCGeom}$       & Light + Mass & 8.0  & $8 \times 8$   & $0\%$   & 175 & 1.0   & 0.2 & 2.0 \\[2pt]
Phase One                     & Light + Mass & 2.0  & $8 \times 8$   & $0\%$   & 150 & 0.8   & 0.15  & 2.0 \\[2pt]
Phase Two                     & Light + Mass & 0.5  & $8 \times 8$   & $0\%$   & 400 & 0.8   & 0.1 & 4.0 \\[2pt]
Hyper                         & Fixed        & Next & Next           & Next    & 150 & 100.  & 0.2  & 2.0 \\[2pt]
\end{tabular}
}
\caption[Automated pipeline's analysis settings]{Settings used in each phase of the {\tt AutoLens} pipeline described in section \ref{SLPipeline}. Column one shows the pipeline phase. Column two shows whether the light and mass are modeled independently or simultaneously. Column three shows the likelihood tolerance on the selection of $N_{\rm s}$, whereby the hyper-parameter with the lowest value of $N_{\rm s}$ is chosen within this likelihood range to give a faster overall run-time. Column four shows the degree of image subgridding used by the analysis. Column five shows the fraction of the PSF that is trimmed about its centre. Columns six to eight show the {\tt MultiNest} settings for the number of live points, tolerance and reduction factor. Column nine shows the stopping factor discussed in section \ref{StopCrit}, which determines at what acceptance rate {\tt MultiNest} terminates. Settings are changed to trade-off fast run time in the early phases to high precision accuracy in the later phases. The final row shows the settings used when optimizing the hyper-parameters between each phase.}
\label{table:PLSettings}
\end{table*}
\subsection{No Lens Light}\label{NoLensLight}

{\tt AutoLens} uses a simplified pipeline for modeling images where there is no lens light component. This applies the phases $\textbf{P}_{\rm  Init2}$, $\textbf{P}_{\rm  Init3}$ and $\textbf{P}_{\rm  MCMass}$ without lens light modeling. The main pipeline then runs with only the mass model, using only phase one. 

\subsection{Pipeline Settings}\label{PLBenefits}

The analysis has a number of settings which are changed throughout the pipeline. In the early phases these are chosen to give a fast run-time, since only an estimate of which models reasonably fit the data is necessary. More computationally intensive settings are used later on once high accuracy parameter, error and Bayesian evidence estimation is required. These are shown in table \ref{table:PLSettings}, where settings like the PSF trimming and image subgridding are altered to give fast computation early on and high-accuracy later. Early hyper-parameter initialization is also restricted to lower values of $N_{\rm  s}$, given how this drives the computational run-time (provided doing so does not decrease ln$\epsilon$ significantly). The setup of {\tt MultiNest} is also changed, such that earlier pipeline phases converge more quickly towards a solution, with more thorough sampling employed in later phases. These settings can be altered to allow {\tt AutoLens} to scale up to larger lens samples whilst keeping the overall run-time feasible. 

\subsection{Parameter Estimation}

Unless otherwise stated, estimates for each parameter are the median of their one-dimensional marginalized posterior probably distribution, which is calculated by weighting each accepted sample in {\tt MultiNest} by its sampling probability\footnote{The sampling probability is a quantity output by {\tt MultiNest} corresponding to the sample prior mass multiplied by likelihood and normalized by the evidence.}. This set of parameters then constitutes what is referred to as the `most probable' lens model and the set which corresponds to the maximum overall likelihood gives the `most likely' lens model. Errors correspond to the $3\sigma$ confidence bounds on each parameter's marginalized one dimensional posterior distribution function (PDF) unless otherwise stated and 2D PDF contours are calculated by marginalizing over all other parameters. The results presented in this work use only the {\tt MultiNest} samples generated from the second phase of the main pipeline unless stated otherwise (or the final phase of the no lens light pipeline).

\subsection{Stopping Criteria}\label{StopCrit}

A {\tt MultiNest} search stops when its estimate of the global posterior log-evidence exceeds a user-defined threshold accuracy, which corresponds to the point where all active points have roughly the same likelihood values. However, as shown in N15, the changing discretization of the source pixelization leads to a noisy and non-smooth likelihood function in non-linear parameter space. Therefore, whilst {\tt MultiNest} does an adequate job sampling this, its stopping criterion is ill-defined, as it tries to fully map out all of the noise in parameter space. In practise this means that once the lens model is estimated accurately there are one or two active {\tt MultiNest} points with anomalously high likelihood values (due to discretization noise) which prevent {\tt MultiNest} from stopping. This leads {\tt MultiNest}'s acceptance rate to plummet, as it can no longer maintain a high acceptance rate by further reducing the lens model's iso-density contours around the high-likelihood regions. At this point, any further increases in likelihood (or the Bayesian evidence) comes from randomly producing a `good' source-plane discretization, information which is of no practical use in terms of actually constraining the lens model. 

Therefore, to circumvent this issue and offer a meaningful stopping criterion, {\tt MultiNest} is automatically terminated once its acceptance rate falls below the target sampling rate divided by a user-specified value, which are both given in table \ref{table:PLSettings}. This division value is never below two, ensuring all phases end only when noise in the parameter space is all that is left being fitted ({\tt MultiNest} consistently maintains its target sampling efficiency otherwise). 

In the final analysis phase a different approach is used. Instead, a `likelihood cap' is imposed, such that any samples with a higher likelihood are reduced to this cap's value. This cap is calculated by taking the previous phase's most likely lens model and hyper-parameter set and computing the mean likelihood of $100$ different source reconstructions, corresponding to the value above which {\tt MultiNest} begins fitting noise (relying on the fact that the lens model is unchanged from the previous phase and already estimated accurately). {\tt MultiNest} then runs until all active points hit this value, thus preventing it from fitting the parameter space's noise. This is important for ensuring the errors of the most probable lens model are estimated accurately, as noise-fitting can bias this towards a few points which gain anomalously high likelihoods due to favourable discretizations. A low value for the likelihood cap will only lead the method to over-estimate parameter errors, given that it exclusively trims the highest likelihood regions of parameter space.

\section{Results}\label{Results}
\begin{table*}
\resizebox{\linewidth}{!}{
\begin{tabular}{ l | l | l l l | l} 
\multicolumn{1}{p{1.6cm}|}{Image} 
& \multicolumn{1}{p{1.8cm}|}{\centering \textbf{Component}} 
& \multicolumn{1}{p{2.2cm}}{\textbf{Parameters ($3\sigma$)}} 
& \multicolumn{1}{p{2.2cm}}{}  
& \multicolumn{1}{p{2.2cm}|}{} 
& \multicolumn{1}{p{2.2cm}}{\textbf{Parameters ($1\sigma$)}} 
\\ \hline
& & & & & \\[-4pt]
$\textbf{SrcBulge}_{\mathrm{HS50NLBulge}}$ & Mass (\textbf{SPLE}) & $\Delta \theta_{\mathrm{Ein}}=0.0043^{+0.0125}_{\rm -0.0136}$($\theta_{\mathrm{Ein}}=1.20$) & $\Delta q=-0.0094^{+0.0329}_{\rm -0.0300}$($q=0.80$) & $\Delta \alpha=0.0227^{+0.0692}_{\rm -0.0789}$($\alpha=2.00$) & $\Delta \alpha=0.0227^{+0.0290}_{\rm -0.0248}$($\alpha=2.00$)\\[2pt]
$\textbf{SrcBulge}_{\mathrm{HS30NLBulge}}$ & Mass (\textbf{SPLE}) & $\Delta \theta_{\mathrm{Ein}}=0.0087^{+0.0248}_{\rm -0.0223}$($\theta_{\mathrm{Ein}}=1.20$) & $\Delta q=-0.0194^{+0.0499}_{\rm -0.0551}$($q=0.80$) & $\Delta \alpha=0.0456^{+0.1196}_{\rm -0.1189}$($\alpha=2.00$) & $\mathbf{\Delta \alpha=0.0456^{+0.0422}_{\rm -0.0427}}$($\alpha=2.00$)\\[2pt]
$\textbf{SrcBulge}_{\mathrm{HS10NLBulge}}$ & Mass (\textbf{SPLE}) & $\Delta \theta_{\mathrm{Ein}}=0.0228^{+0.0337}_{\rm -0.0320}$($\theta_{\mathrm{Ein}}=1.20$) & $\Delta q=-0.0495^{+0.0713}_{\rm -0.0702}$($q=0.80$) & $\Delta \alpha=0.1092^{+0.1378}_{\rm -0.1524}$($\alpha=2.00$) & $\mathbf{\Delta \alpha=0.1092^{+0.0527}_{\rm -0.0493}}$($\alpha=2.00$)\\[2pt]
$\textbf{SrcBulge}_{\mathrm{ES50NLBulge}}$ & Mass (\textbf{SPLE}) & $\Delta \theta_{\mathrm{Ein}}=0.0096^{+0.0171}_{\rm -0.0165}$($\theta_{\mathrm{Ein}}=1.20$) & $\Delta q=-0.0217^{+0.0427}_{\rm -0.0423}$($q=0.80$) & $\Delta \alpha=0.0468^{+0.0852}_{\rm -0.0891}$($\alpha=2.00$) & $\mathbf{\Delta \alpha=0.0468^{+0.0317}_{\rm -0.0307}}$($\alpha=2.00$)\\[2pt]
$\textbf{SrcBulge}_{\mathrm{ES30NLBulge}}$ & Mass (\textbf{SPLE}) & $\Delta \theta_{\mathrm{Ein}}=0.0028^{+0.0253}_{\rm -0.0247}$($\theta_{\mathrm{Ein}}=1.20$) & $\Delta q=-0.0072^{+0.0655}_{\rm -0.0663}$($q=0.80$) & $\Delta \alpha=0.0040^{+0.1425}_{\rm -0.1559}$($\alpha=2.00$) & $\Delta \alpha=0.0040^{+0.0564}_{\rm -0.0507}$($\alpha=2.00$)\\[2pt]
$\textbf{SrcBulge}_{\mathrm{ES10NLBulge}}$ & Mass (\textbf{SPLE}) & $\Delta \theta_{\mathrm{Ein}}=-0.0020^{+0.0308}_{\rm -0.0275}$($\theta_{\mathrm{Ein}}=1.20$) & $\Delta q=0.0131^{+0.0705}_{\rm -0.0720}$($q=0.80$) & $\Delta \alpha=-0.0259^{+0.1810}_{\rm -0.2001}$($\alpha=2.00$) & $\Delta \alpha=-0.0259^{+0.0688}_{\rm -0.0576}$($\alpha=2.00$)\\[-4pt]
& & & & & \\[-4pt]
\hline
& & & & & \\[-4pt]
$\textbf{SrcBD}_{\mathrm{HS50NLBD}}$ & Mass (\textbf{SPLE}) & $\Delta \theta_{\mathrm{Ein}}=0.0047^{+0.0196}_{\rm -0.0198}$($\theta_{\mathrm{Ein}}=1.20$) & $\Delta q=-0.0038^{+0.0335}_{\rm -0.0332}$($q=0.70$) & $\Delta \alpha=0.0082^{+0.0505}_{\rm -0.0527}$($\alpha=1.70$) & $\Delta \alpha=0.0082^{+0.0183}_{\rm -0.0178}$($\alpha=1.70$)\\[2pt]
$\textbf{SrcBD}_{\mathrm{HS30NLBD}}$ & Mass (\textbf{SPLE}) & $\Delta \theta_{\mathrm{Ein}}=0.0015^{+0.0255}_{\rm -0.0242}$($\theta_{\mathrm{Ein}}=1.20$) & $\Delta q=0.0012^{+0.0425}_{\rm -0.0440}$($q=0.70$) & $\Delta \alpha=-0.0005^{+0.0675}_{\rm -0.0673}$($\alpha=1.70$) & $\Delta \alpha=-0.0005^{+0.0239}_{\rm -0.0238}$($\alpha=1.70$)\\[2pt]
$\textbf{SrcBD}_{\mathrm{HS10NLBD}}$ & Mass (\textbf{SPLE}) & $\Delta \theta_{\mathrm{Ein}}=0.0106^{+0.0586}_{\rm -0.0568}$($\theta_{\mathrm{Ein}}=1.20$) & $\Delta q=-0.0101^{+0.0962}_{\rm -0.0927}$($q=0.70$) & $\Delta \alpha=0.0203^{+0.1422}_{\rm -0.1537}$($\alpha=1.70$) & $\Delta \alpha=0.0203^{+0.0561}_{\rm -0.0514}$($\alpha=1.70$)\\[2pt]
$\textbf{SrcBD}_{\mathrm{ES50NLBD}}$ & Mass (\textbf{SPLE}) & $\Delta \theta_{\mathrm{Ein}}=0.0038^{+0.0210}_{\rm -0.0214}$($\theta_{\mathrm{Ein}}=1.20$) & $\Delta q=-0.0001^{+0.0407}_{\rm -0.0393}$($q=0.70$) & $\Delta \alpha=0.0041^{+0.0586}_{\rm -0.0615}$($\alpha=1.70$) & $\Delta \alpha=0.0041^{+0.0221}_{\rm -0.0211}$($\alpha=1.70$)\\[2pt]
$\textbf{SrcBD}_{\mathrm{ES30NLBD}}$ & Mass (\textbf{SPLE}) & $\Delta \theta_{\mathrm{Ein}}=0.0109^{+0.0336}_{\rm -0.0295}$($\theta_{\mathrm{Ein}}=1.20$) & $\Delta q=-0.0147^{+0.0479}_{\rm -0.0530}$($q=0.70$) & $\Delta \alpha=0.0234^{+0.0802}_{\rm -0.0751}$($\alpha=1.70$) & $\Delta \alpha=0.0234^{+0.0269}_{\rm -0.0288}$($\alpha=1.70$)\\[2pt]
$\textbf{SrcBD}_{\mathrm{ES10NLBD}}$ & Mass (\textbf{SPLE}) & $\Delta \theta_{\mathrm{Ein}}=0.0144^{+0.0720}_{\rm -0.0534}$($\theta_{\mathrm{Ein}}=1.20$) & $\Delta q=-0.0113^{+0.0920}_{\rm -0.1058}$($q=0.70$) & $\Delta \alpha=0.0159^{+0.1504}_{\rm -0.1397}$($\alpha=1.70$) & $\Delta \alpha=0.0159^{+0.0492}_{\rm -0.0506}$($\alpha=1.70$)\\[-4pt]
& & & & & \\[-4pt]
\hline
& & & & & \\[-4pt]
$\textbf{SrcDisk}_{\mathrm{HS50NLDisk}}$ & Mass (\textbf{SPLE}) & $\Delta \theta_{\mathrm{Ein}}=0.0020^{+0.0204}_{\rm -0.0191}$($\theta_{\mathrm{Ein}}=1.20$) & $\Delta q=-0.0071^{+0.0555}_{\rm -0.0552}$($q=0.75$) & $\Delta \alpha=0.0062^{+0.0541}_{\rm -0.0537}$($\alpha=2.30$) & $\Delta \alpha=0.0062^{+0.0187}_{\rm -0.0196}$($\alpha=2.30$)\\[2pt]
$\textbf{SrcDisk}_{\mathrm{HS30NLDisk}}$ & Mass (\textbf{SPLE}) & $\Delta \theta_{\mathrm{Ein}}=-0.0045^{+0.0286}_{\rm -0.0218}$($\theta_{\mathrm{Ein}}=1.20$) & $\Delta q=0.0158^{+0.0674}_{\rm -0.0809}$($q=0.75$) & $\Delta \alpha=-0.0186^{+0.0785}_{\rm -0.0762}$($\alpha=2.30$) & $\Delta \alpha=-0.0186^{+0.0262}_{\rm -0.0271}$($\alpha=2.30$)\\[2pt]
$\textbf{SrcDisk}_{\mathrm{HS10NLDisk}}$ & Mass (\textbf{SPLE}) & $\Delta \theta_{\mathrm{Ein}}=-0.0100^{+0.0546}_{\rm -0.0369}$($\theta_{\mathrm{Ein}}=1.20$) & $\Delta q=0.0205^{+0.1263}_{\rm -0.1440}$($q=0.75$) & $\Delta \alpha=-0.0249^{+0.1521}_{\rm -0.1536}$($\alpha=2.30$) & $\Delta \alpha=-0.0249^{+0.0561}_{\rm -0.0547}$($\alpha=2.30$)\\[2pt]
$\textbf{SrcDisk}_{\mathrm{ES50NLDisk}}$ & Mass (\textbf{SPLE}) & $\Delta \theta_{\mathrm{Ein}}=0.0029^{+0.0352}_{\rm -0.0305}$($\theta_{\mathrm{Ein}}=1.20$) & $\Delta q=-0.0046^{+0.0926}_{\rm -0.0951}$($q=0.75$) & $\Delta \alpha=0.0118^{+0.0732}_{\rm -0.0819}$($\alpha=2.30$) & $\Delta \alpha=0.0118^{+0.0285}_{\rm -0.0262}$($\alpha=2.30$)\\[2pt]
$\textbf{SrcDisk}_{\mathrm{ES30NLDisk}}$ & Mass (\textbf{SPLE}) & $\Delta \theta_{\mathrm{Ein}}=-0.0042^{+0.0335}_{\rm -0.0291}$($\theta_{\mathrm{Ein}}=1.20$) & $\Delta q=0.0228^{+0.0944}_{\rm -0.1029}$($q=0.75$) & $\Delta \alpha=-0.0144^{+0.1032}_{\rm -0.1045}$($\alpha=2.30$) & $\Delta \alpha=-0.0144^{+0.0375}_{\rm -0.0367}$($\alpha=2.30$)\\[2pt]
$\textbf{SrcDisk}_{\mathrm{ES10NLDisk}}$ & Mass (\textbf{SPLE}) & $\Delta \theta_{\mathrm{Ein}}=0.0128^{+0.1009}_{\rm -0.0784}$($\theta_{\mathrm{Ein}}=1.20$) & $\Delta q=-0.0229^{+0.2111}_{\rm -0.2175}$($q=0.75$) & $\Delta \alpha=0.0653^{+0.1885}_{\rm -0.2548}$($\alpha=2.30$) & $\mathbf{\Delta \alpha=0.0653^{+0.0840}_{\rm -0.0646}}$($\alpha=2.30$)\\[-4pt]
& & & & & \\[-4pt]
\hline
& & & & & \\[-4pt]
$\textbf{SrcMulti}_{\mathrm{HS50NLMulti}}$ & Mass (\textbf{SPLE}) & $\Delta \theta_{\mathrm{Ein}}=0.0007^{+0.0140}_{\rm -0.0137}$($\theta_{\mathrm{Ein}}=1.20$) & $\Delta q=-0.0010^{+0.0281}_{\rm -0.0282}$($q=0.75$) & $\Delta \alpha=0.0006^{+0.0640}_{\rm -0.0658}$($\alpha=2.10$) & $\Delta \alpha=0.0006^{+0.0236}_{\rm -0.0230}$($\alpha=2.10$)\\[2pt]
$\textbf{SrcMulti}_{\mathrm{HS30NLMulti}}$ & Mass (\textbf{SPLE}) & $\Delta \theta_{\mathrm{Ein}}=-0.0044^{+0.0192}_{\rm -0.0193}$($\theta_{\mathrm{Ein}}=1.20$) & $\Delta q=0.0092^{+0.0391}_{\rm -0.0397}$($q=0.75$) & $\Delta \alpha=-0.0251^{+0.0892}_{\rm -0.0949}$($\alpha=2.10$) & $\Delta \alpha=-0.0251^{+0.0345}_{\rm -0.0316}$($\alpha=2.10$)\\[2pt]
$\textbf{SrcMulti}_{\mathrm{HS10NLMulti}}$ & Mass (\textbf{SPLE}) & $\Delta \theta_{\mathrm{Ein}}=-0.0049^{+0.0399}_{\rm -0.0337}$($\theta_{\mathrm{Ein}}=1.20$) & $\Delta q=0.0065^{+0.0667}_{\rm -0.0736}$($q=0.75$) & $\Delta \alpha=-0.0284^{+0.1697}_{\rm -0.1700}$($\alpha=2.10$) & $\Delta \alpha=-0.0284^{+0.0612}_{\rm -0.0611}$($\alpha=2.10$)\\[2pt]
$\textbf{SrcMulti}_{\mathrm{ES50NLMulti}}$ & Mass (\textbf{SPLE}) & $\Delta \theta_{\mathrm{Ein}}=0.0028^{+0.0158}_{\rm -0.0149}$($\theta_{\mathrm{Ein}}=1.20$) & $\Delta q=-0.0073^{+0.0311}_{\rm -0.0309}$($q=0.75$) & $\Delta \alpha=0.0143^{+0.0679}_{\rm -0.0715}$($\alpha=2.10$) & $\Delta \alpha=0.0143^{+0.0230}_{\rm -0.0219}$($\alpha=2.10$)\\[2pt]
$\textbf{SrcMulti}_{\mathrm{ES30NLMulti}}$ & Mass (\textbf{SPLE}) & $\Delta \theta_{\mathrm{Ein}}=0.0162^{+0.0358}_{\rm -0.0346}$($\theta_{\mathrm{Ein}}=1.20$) & $\Delta q=-0.0315^{+0.0667}_{\rm -0.0671}$($q=0.75$) & $\Delta \alpha=0.0624^{+0.1246}_{\rm -0.1330}$($\alpha=2.10$) & $\mathbf{\Delta \alpha=0.0624^{+0.0472}_{\rm -0.0453}}$($\alpha=2.10$)\\[2pt]
$\textbf{SrcMulti}_{\mathrm{ES10NLMulti}}$ & Mass (\textbf{SPLE}) & $\Delta \theta_{\mathrm{Ein}}=0.0008^{+0.1147}_{\rm -0.0831}$($\theta_{\mathrm{Ein}}=1.20$) & $\Delta q=-0.0047^{+0.1650}_{\rm -0.1834}$($q=0.75$) & $\Delta \alpha=-0.0191^{+0.3940}_{\rm -0.4142}$($\alpha=2.10$) & $\Delta \alpha=-0.0191^{+0.1479}_{\rm -0.1413}$($\alpha=2.10$)\\[-4pt]
\end{tabular}
}
\caption{Results of fitting the $\textbf{SrcBulge}$, $\textbf{SrcBD}$, $\textbf{SrcDisk}$ and $\textbf{SrcMulti}$ images using the (no lens light) automated analysis pipeline, corresponding to results generated at end of phase one of the main pipeline. Each image's name is given in the first column and the mass model fitted in the second column. The third to sixth columns show parameter estimates, where each parameter is offset by $\Delta P = P_{\rm  true} - P_{\rm  model}$, such that zero corresponds to the input lens model. The input lens model values are given in brackets to the right of each parameter estimate. Columns three to five show $\Delta \theta_{\rm  Ein} $, $\Delta q$ and $\Delta \alpha$ within $3 \sigma$ confidence and column six $\Delta \alpha$ at $1 \sigma$. Parameter estimates in bold text are inconsistent with the input lens model at their stated error estimates. The other parameters not shown ($x$, $y$ and $\theta$) are all estimated accurately within $3\sigma$.}
\label{table:TableSrcOnly}
\end{table*}

This section presents {\tt AutoLens}'s automated analysis of the full suite of simulated images. Given the large library of results, this section focuses on only a subset of lens model parameters that best summarize the accuracy of each analysis. For the $Sersic$ light component of a lens model, the effective radius $R_{\rm l}$, sersic index $n_{\rm l}$ and axis ratio $q_{\rm l}$ are used, with multi-component light models using $n_{\rm  l1}$, $R_{\rm  l1}$ $q_{\rm  l1}$, $R_{\rm l2}$ and $q_{\rm  l2}$. For $SPLE$ mass components the Einstein radius $\theta_{\rm  Ein}$, axis ratio $q$ and density slope $\alpha$ are used, with the core radius $S$ included for the $PL\textsubscript{Core}$ model and shear orientation $\theta_{\rm  sh}$ and magnitude $\gamma_{\rm  sh}$ for a $Shear$ component. The $NFW$ model is summarized with its normalization $\kappa_{\rm d}$ and axis-ratio $q_{\rm d}$ whereas a light profile's mass component uses instead its mass-to-light ratio $\Psi_{\rm l}$. The geometry of light and dark matter are also investigated using their centroids and rotation angles $x_{\rm l}$, $y_{\rm l}$, $\theta_{\rm l}$, $x_{\rm d}$, $y_{\rm d}$ and $\theta_{\rm d}$. 

To ease the reader's comparison to the input values, all results are presented as the difference between the estimated value and simulated lens's true input value, $\Delta P = P_{\rm  true} - P_{\rm  model}$, where $P$ is a given parameter. The parameters that have been omitted are those that are generally `easy' to estimate and share no degeneracies with the other parameters (e.g. $x_{\rm l}$, $x$, $\theta$). Again, for brevity, only a sub-set of images for each analysis is presented, choosing results that show the general trends and exceptions. 

\subsection{Source-Only}\label{ResultsSrc}

\begin{figure*}
\centering
\includegraphics[width=0.195\textwidth]{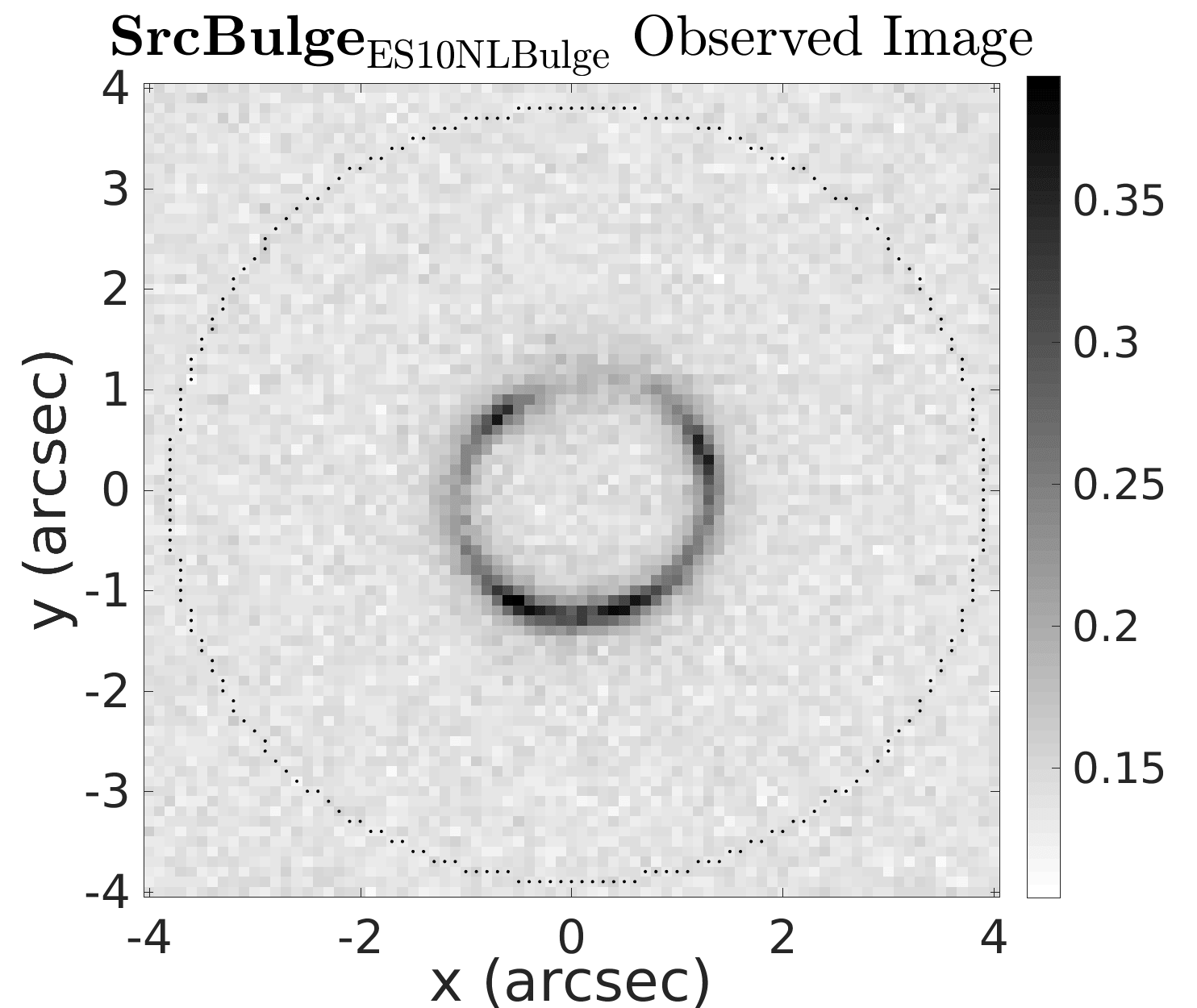}
\includegraphics[width=0.195\textwidth]{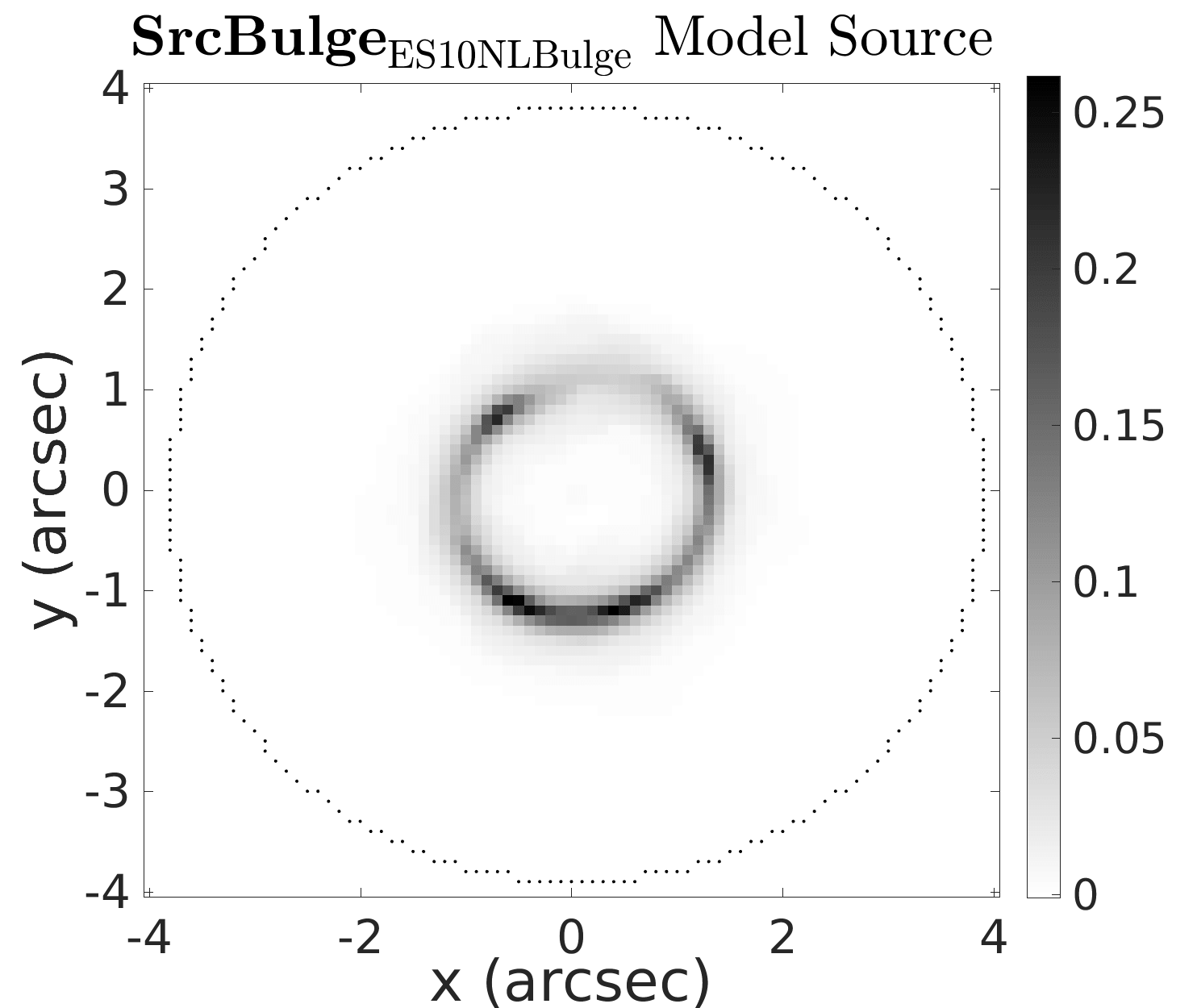}
\includegraphics[width=0.195\textwidth]{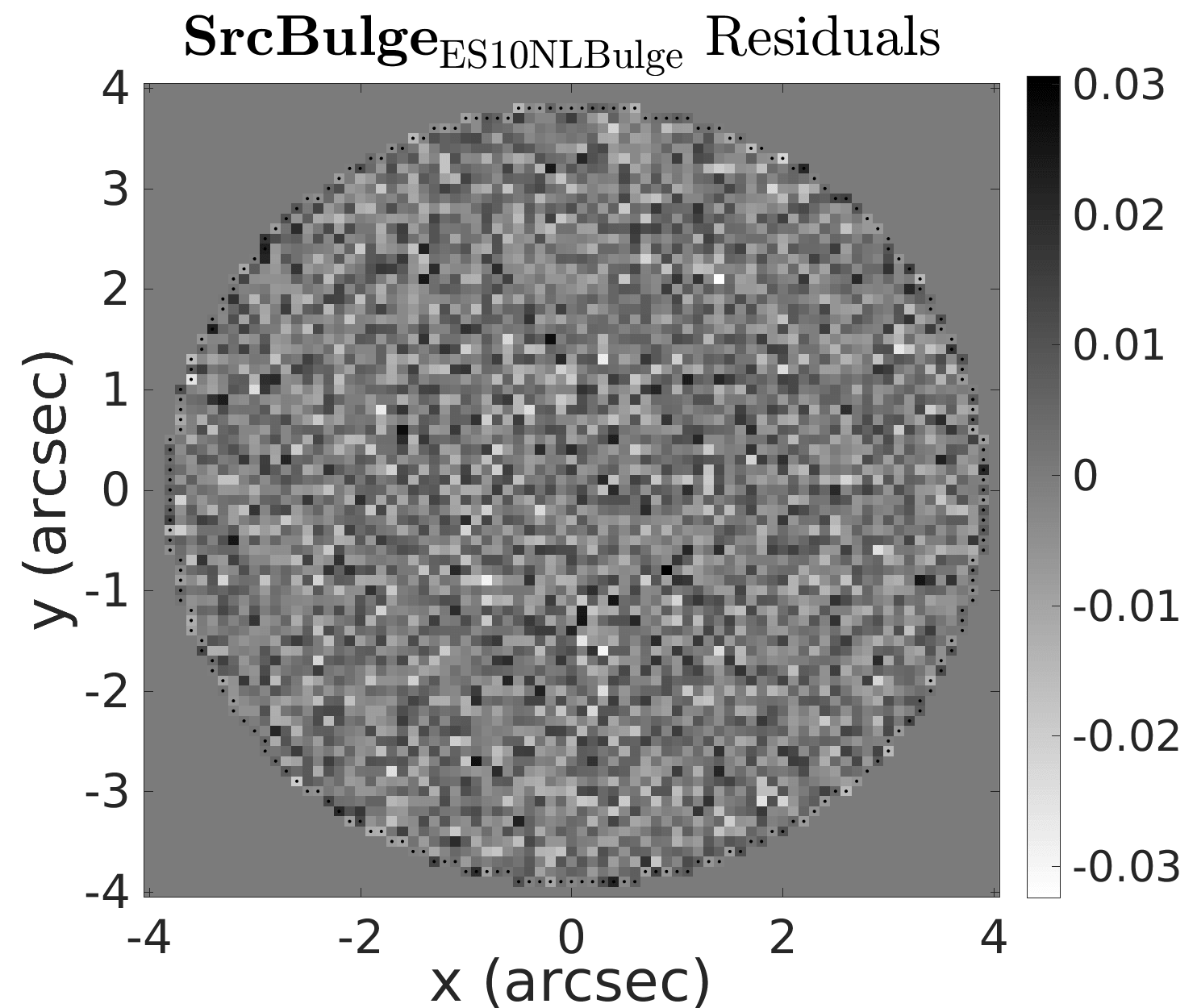}
\includegraphics[width=0.195\textwidth]{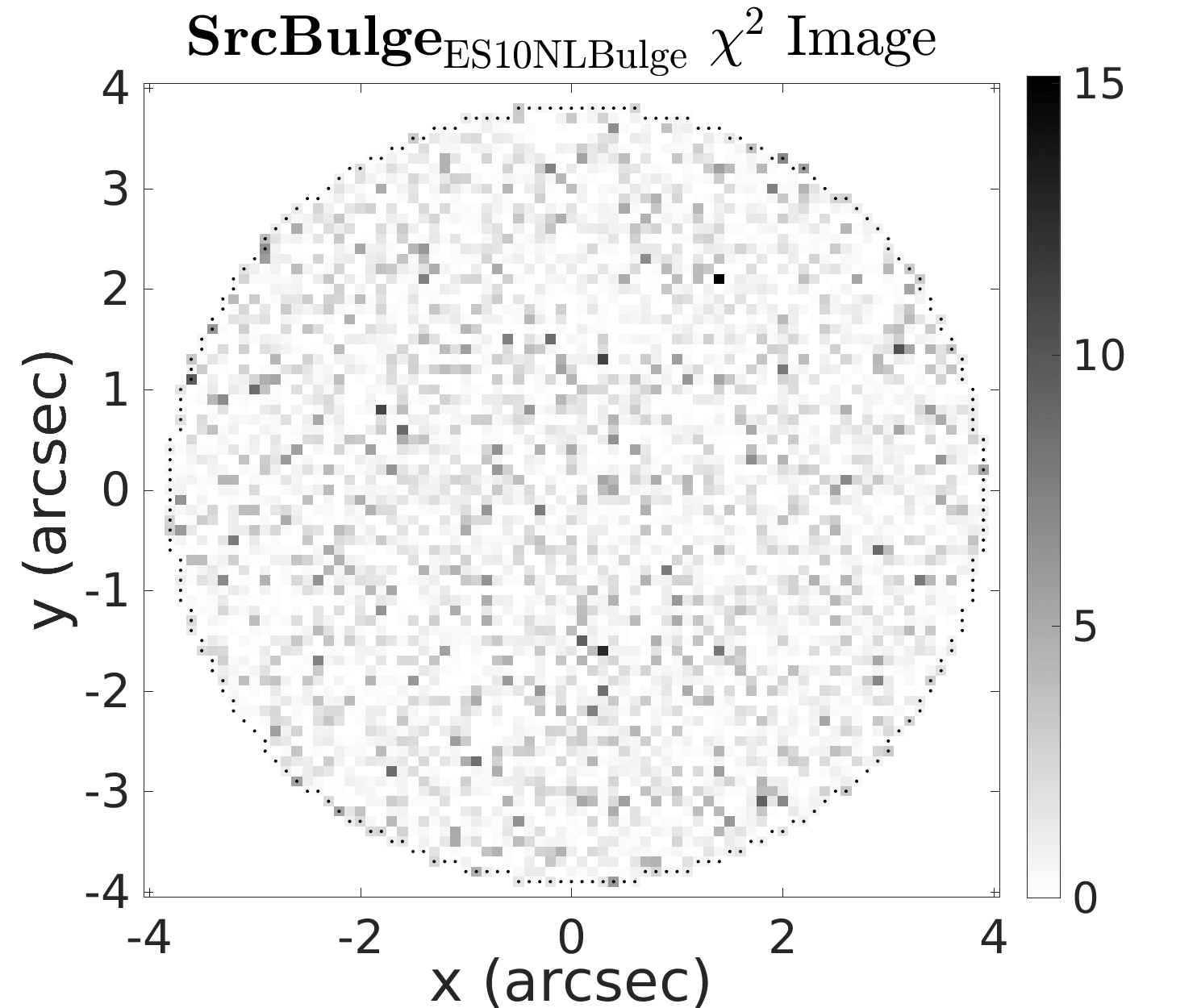}
\includegraphics[width=0.195\textwidth]{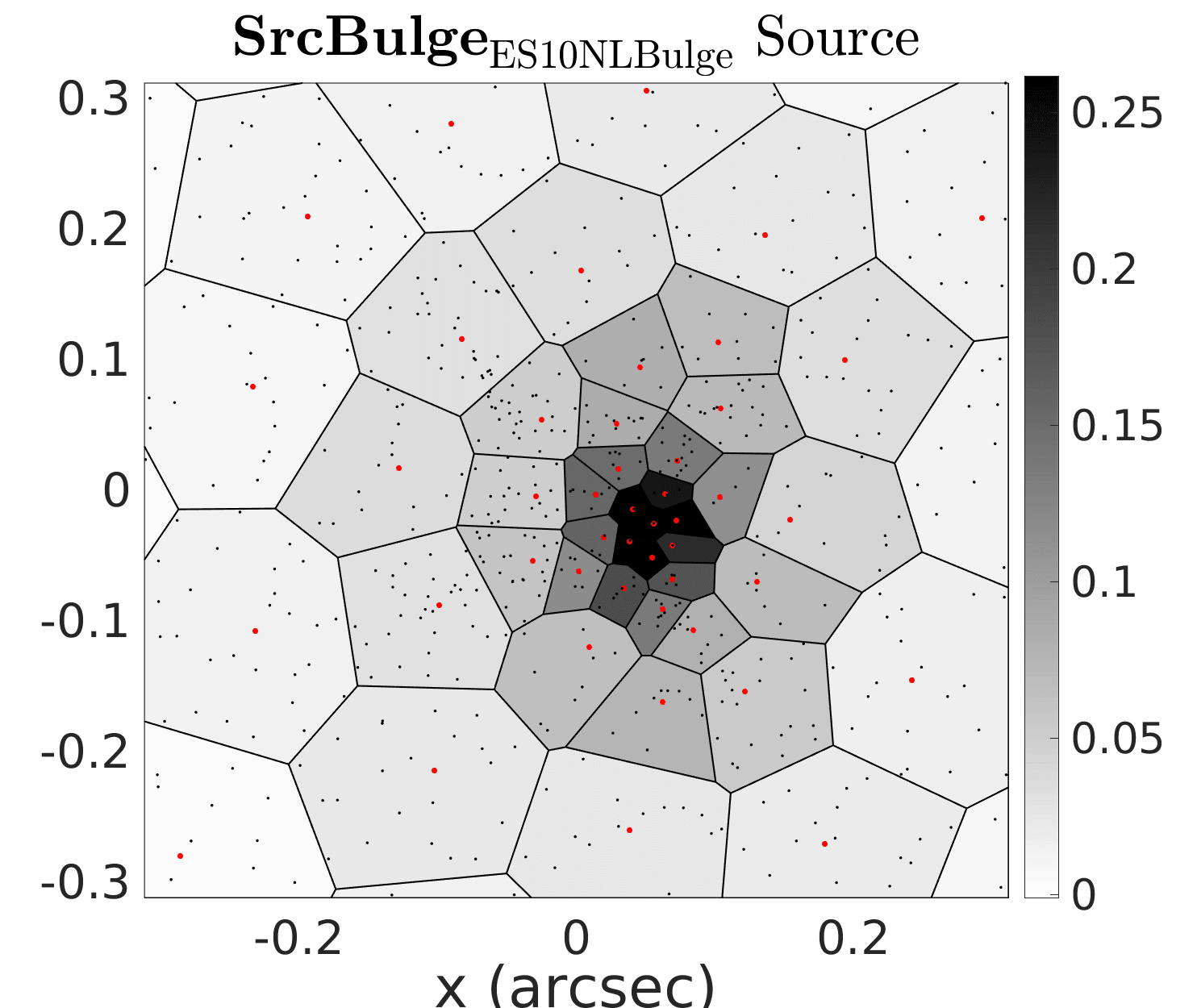}
\includegraphics[width=0.195\textwidth]{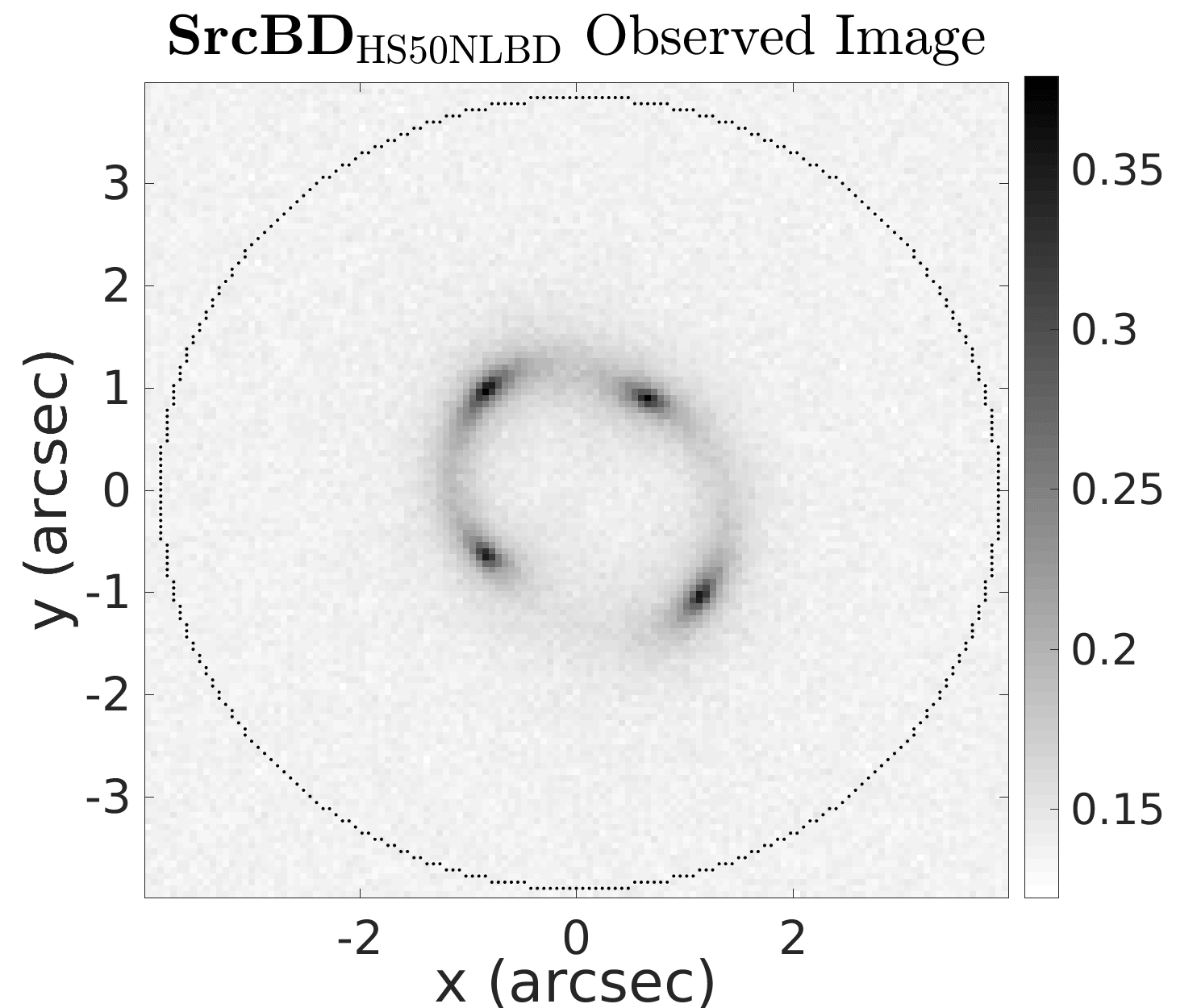}
\includegraphics[width=0.195\textwidth]{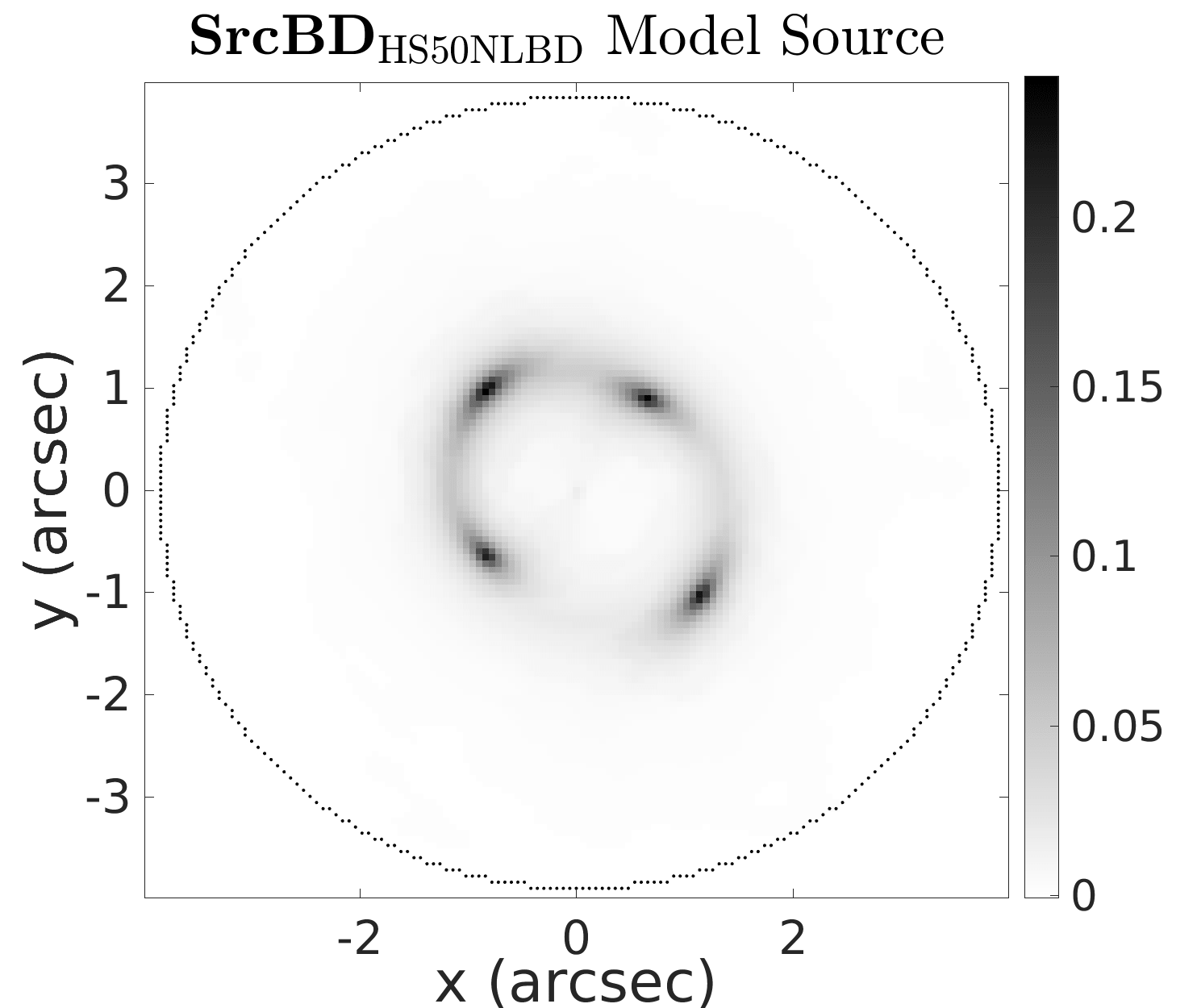}
\includegraphics[width=0.195\textwidth]{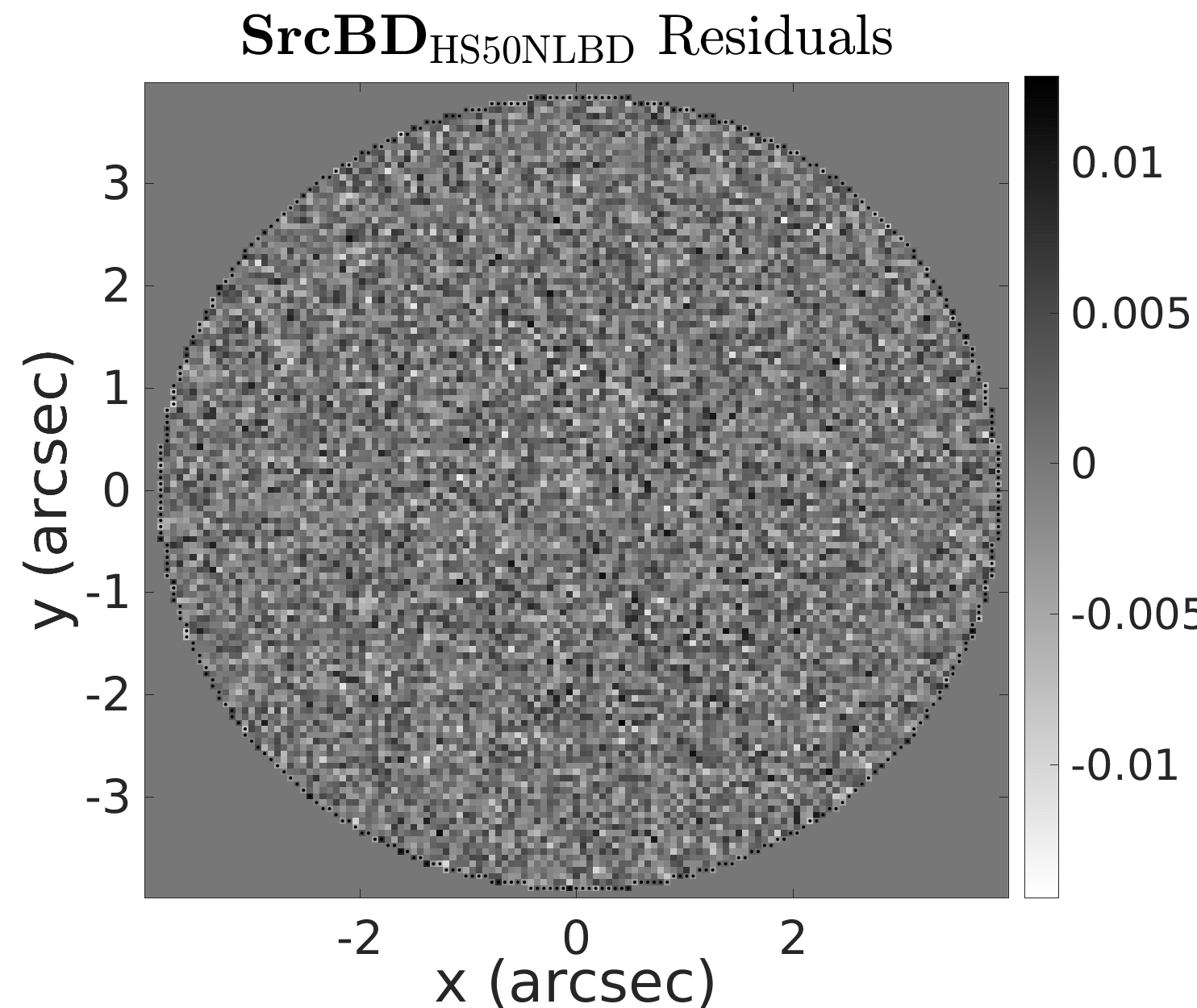}
\includegraphics[width=0.195\textwidth]{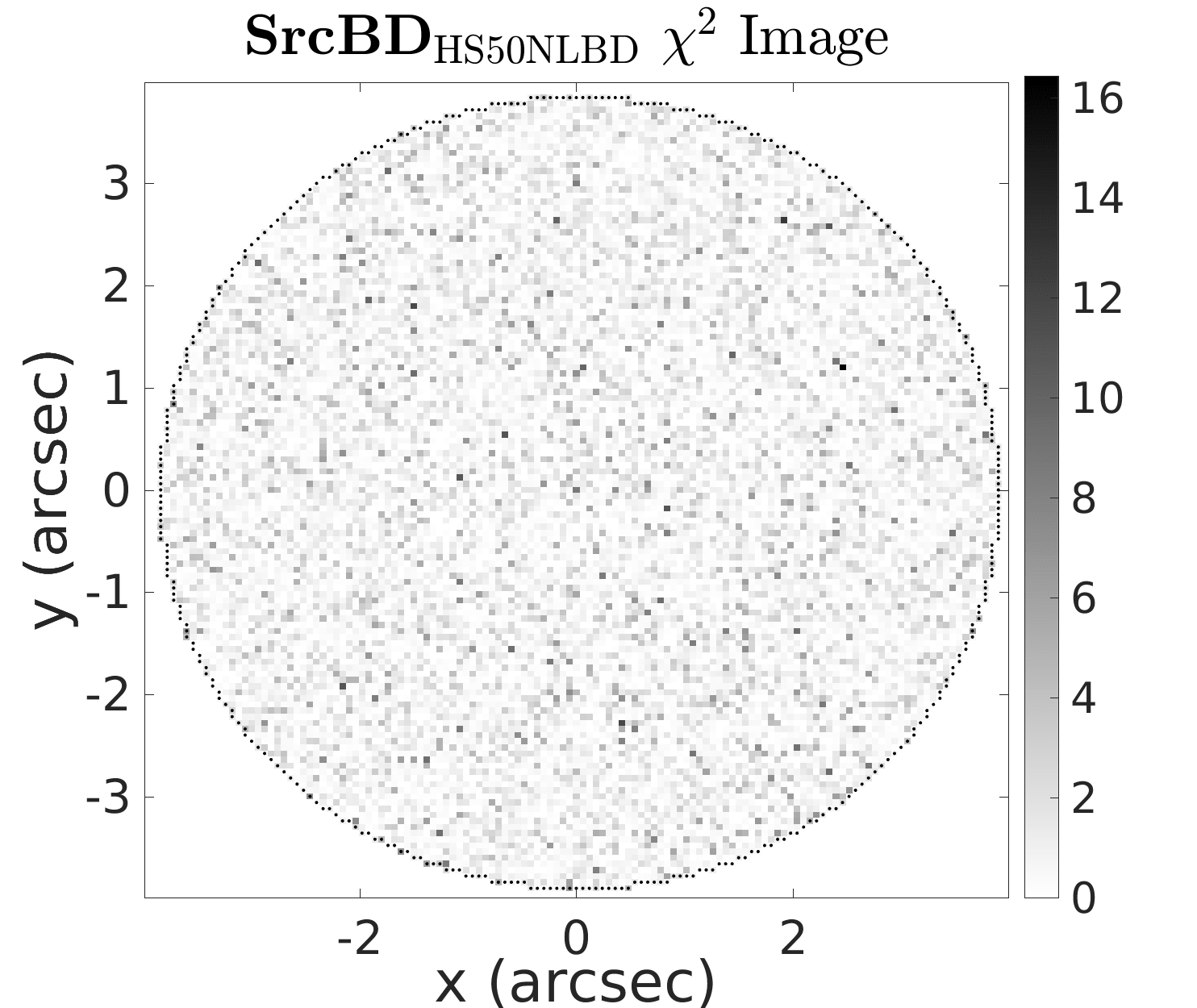}
\includegraphics[width=0.195\textwidth]{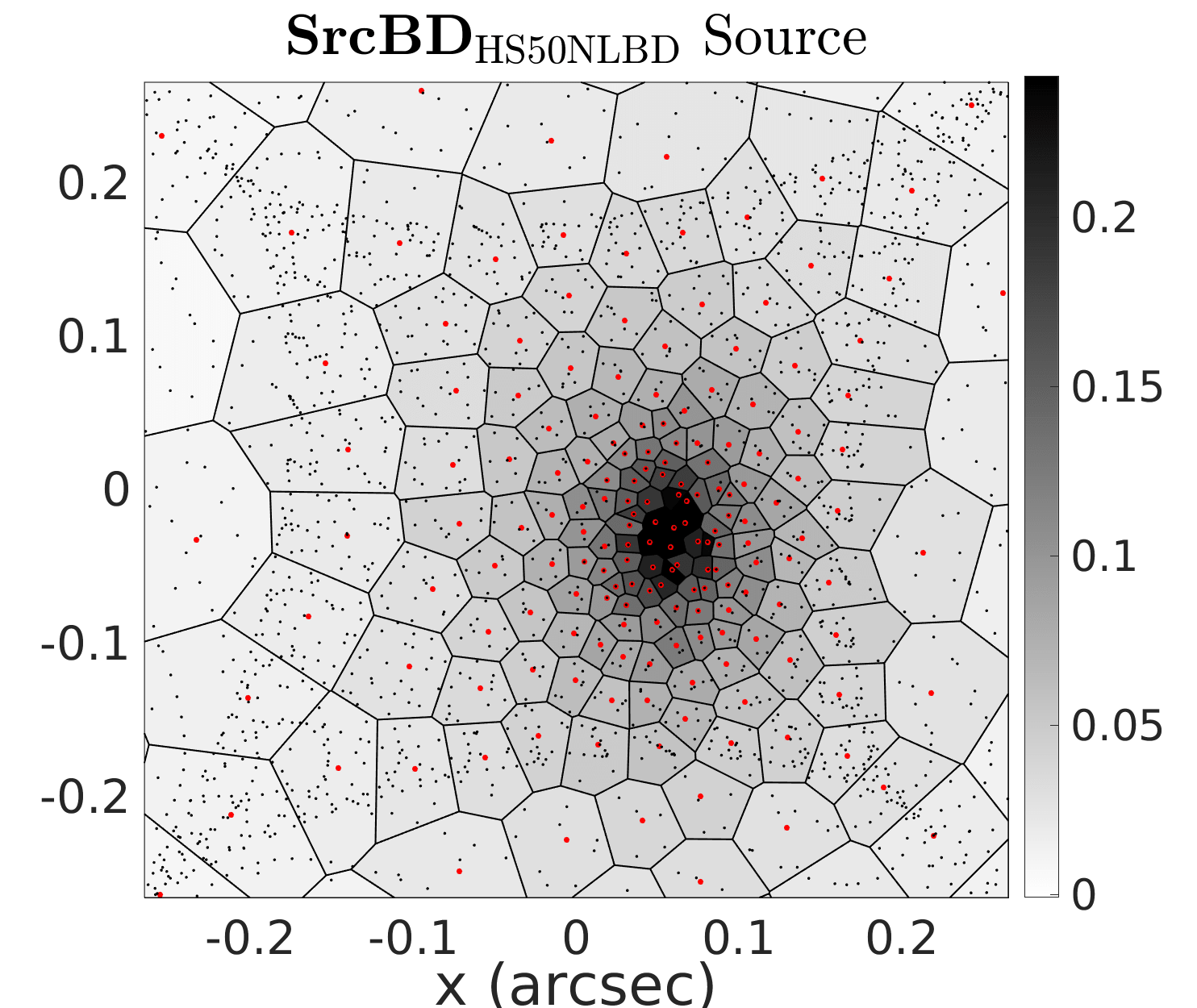}
\caption{The simulated images, model images, residuals, $\chi^2$ images and source reconstructions for the no lens light analysis of $\textbf{SrcBulge}_{\rm  ES10NLBulge}$ (top row) and $\textbf{SrcBD}_{\rm  HS50NLBD}$ (bottom row). Images correspond to the most likely model at the end of the main pipeline, corresponding to the models given by rows six, seven, thirteen and twenty-one of table \ref{table:TableSrcOnly} respectively.} 
\label{figure:ResultsSrcOnlyIms}
\end{figure*}

The source-only simulation suite consists of four unique lens and source models, each of which are used to generate six images, three at Hubble resolution with a source $S/N = 50, 30, 10$ and three at Euclid resolution at $S/N = 50, 30, 10$, giving a total of twenty-four images. The analysis of each image uses the reduced pipeline for objects without a lens light component, therefore also omitting the $\omega_{\rm  Lens}$ hyper-parameters. The shear model comparison phase is also omitted for brevity. The results are summarized using the mismatch parameters $\Delta \theta_{\rm  Ein}$, $\Delta q$ and $\Delta \alpha$ in table \ref{table:TableSrcOnly}, where all parameters are correctly estimated within $3 \sigma$ confidence. Figure \ref{figure:ResultsSrcOnlyIms} shows the observed image, model image, residuals, $\chi^2$ images and source reconstructions for the images $\textbf{SrcBulge}_{\rm  ES10NLBulge}$ and $\textbf{SrcBD}_{\rm  HS50NLFlat}$, which cover different lens models, source morphologies and image resolutions and S/N ratios. The residuals are nearly featureless and $\chi^2$ image realizes the image's noise, as section \ref{AdaptDemo} discussed is the desired solution. 

\begin{figure*}
\centering
\includegraphics[width=0.242\textwidth]{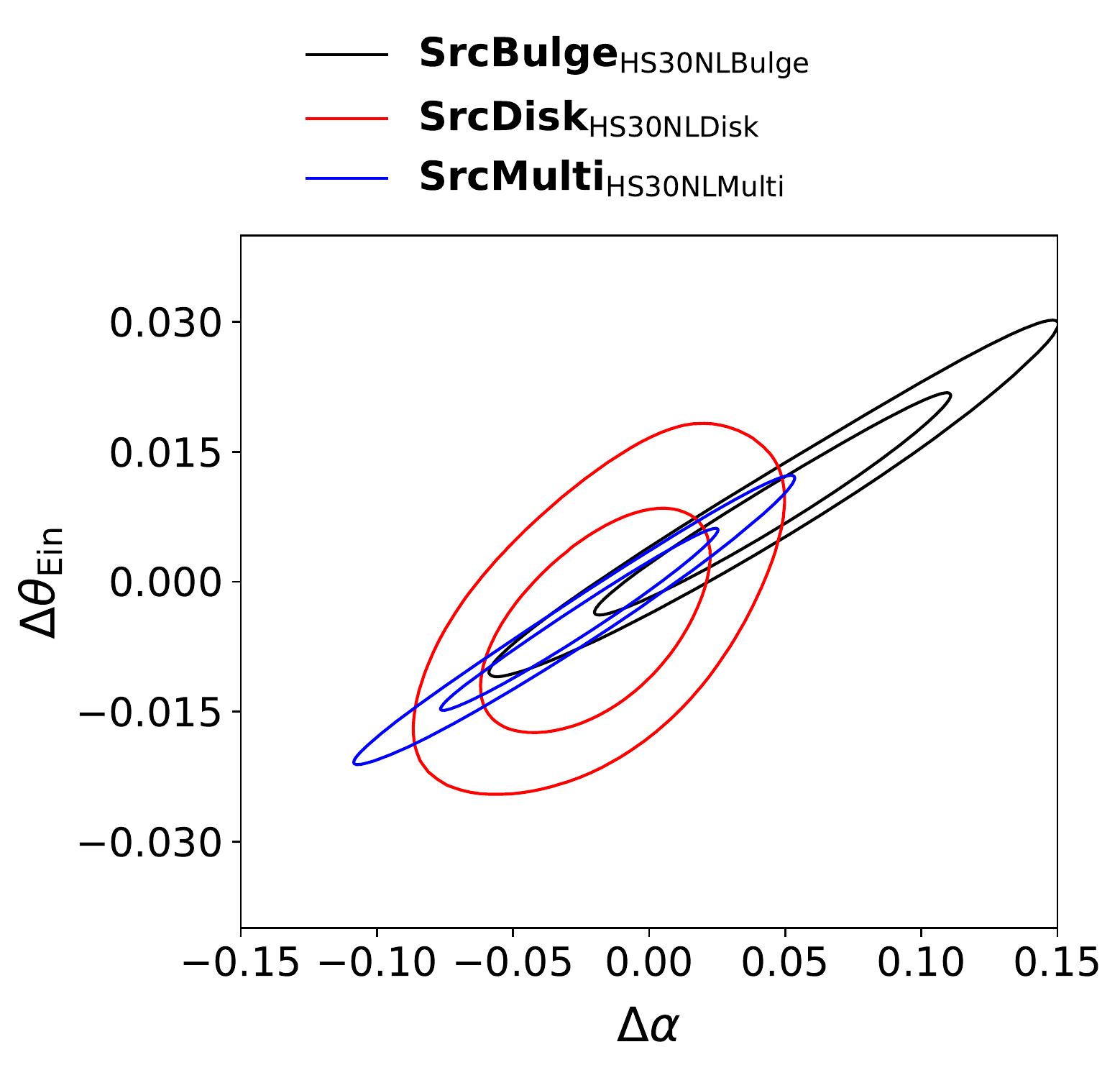}
\includegraphics[width=0.244\textwidth]{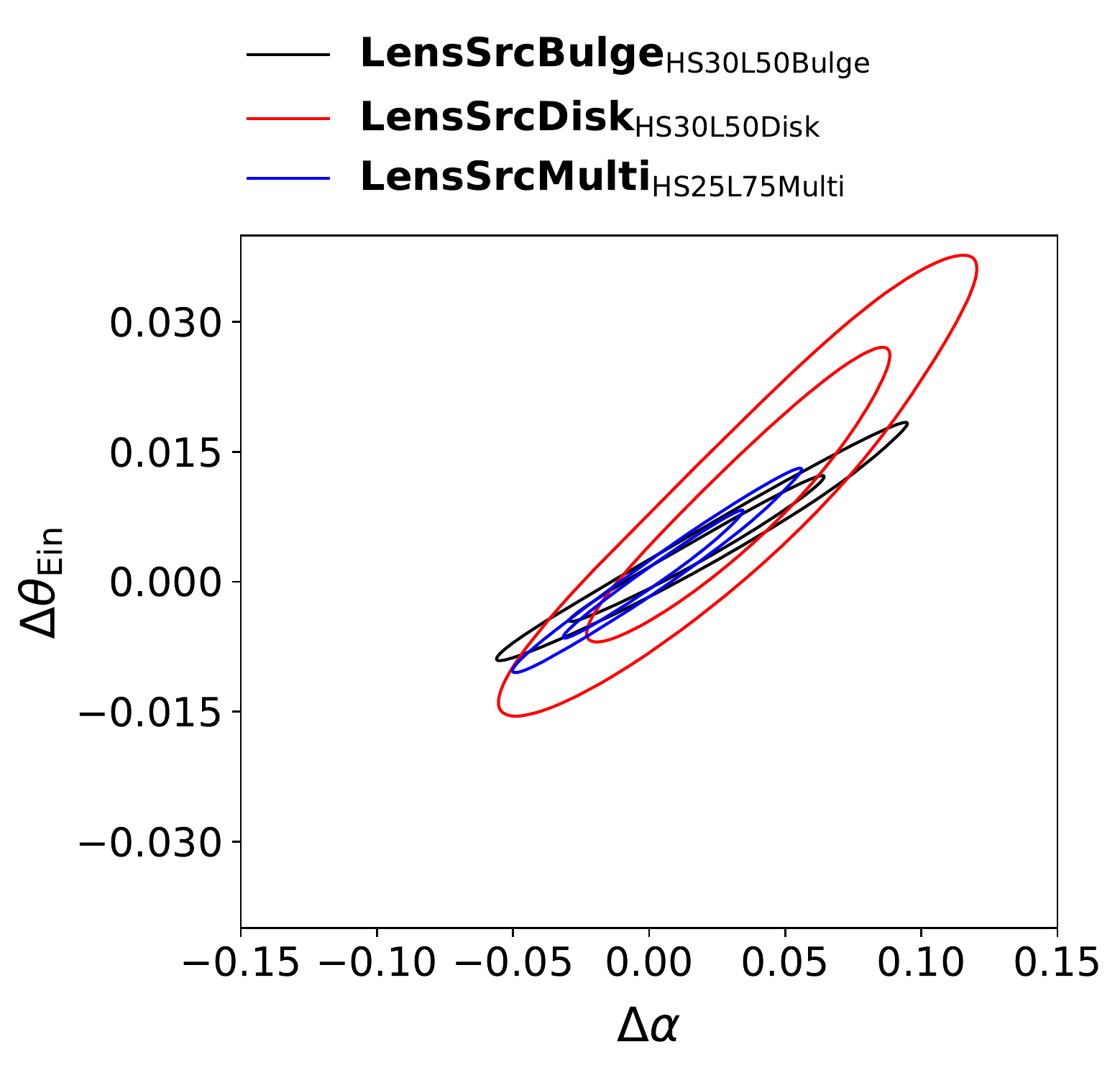}
\includegraphics[width=0.242\textwidth]{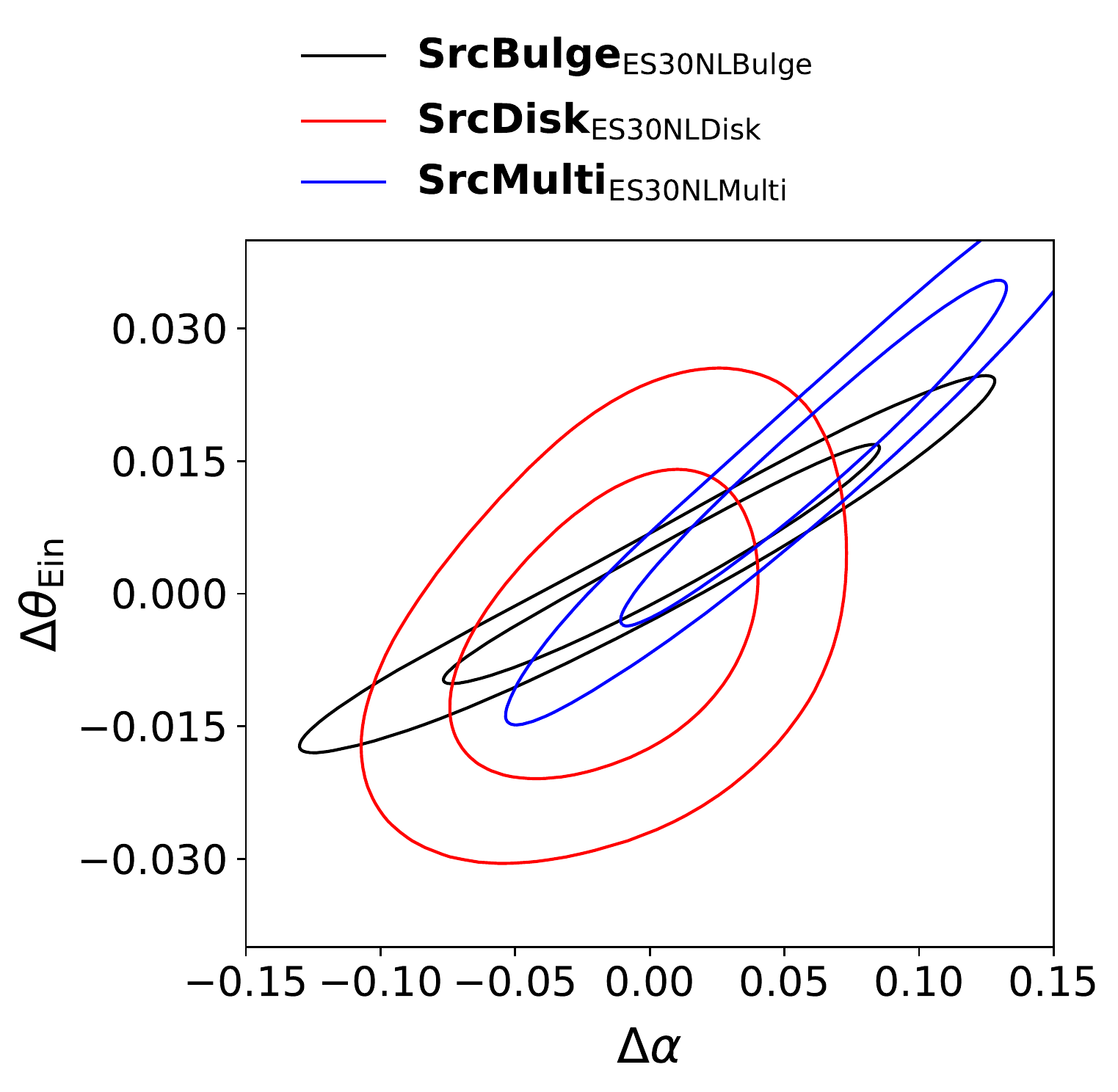}
\includegraphics[width=0.242\textwidth]{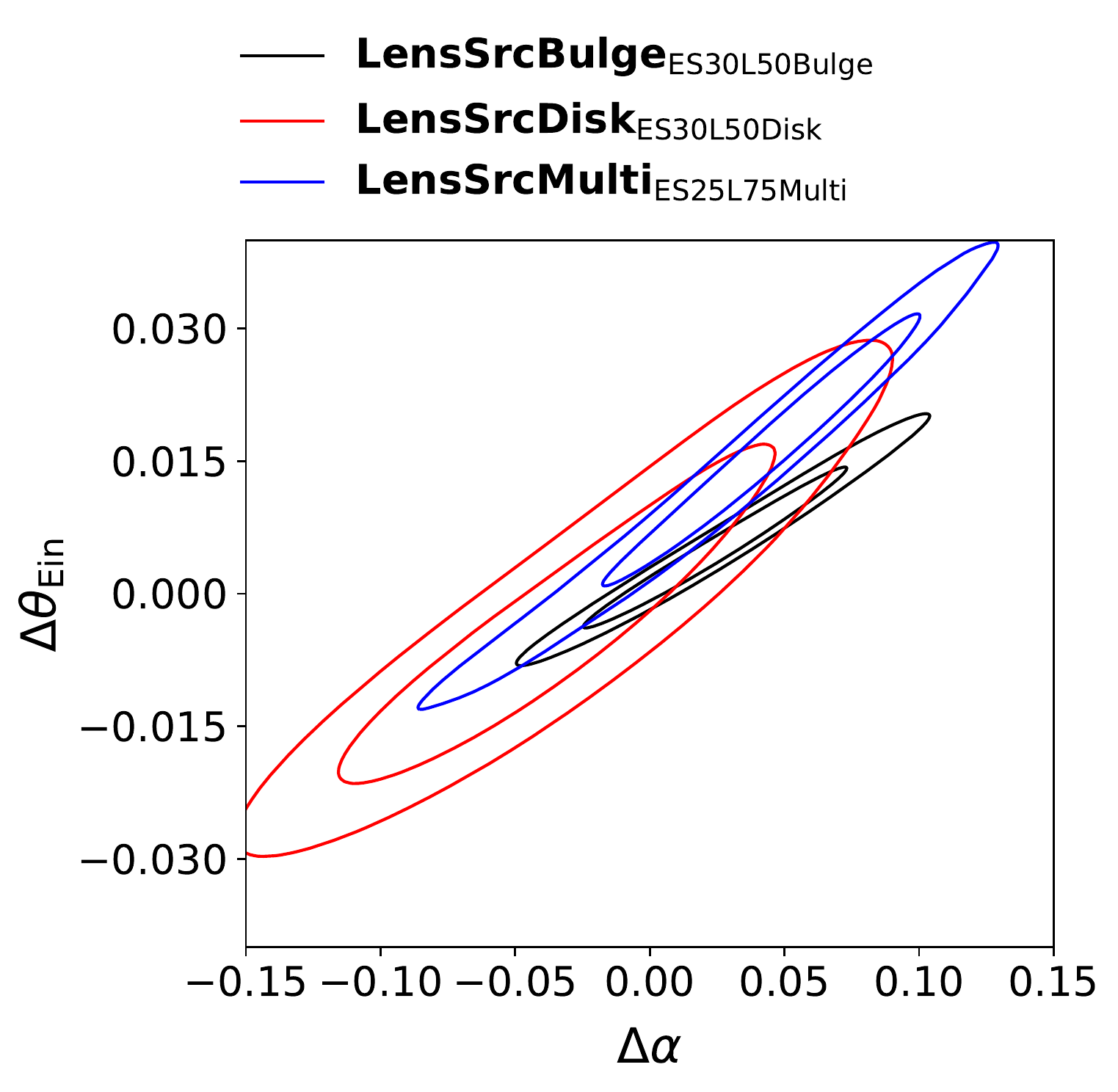}
\caption{Marginalized two-dimensional PDF's of the mismatch parameters $\Delta P = P_{\rm True} - P_{\rm Model}$ for the mass model Einstein radius $\Delta \theta_{\rm  Ein}$ and power-law density slope $\Delta \alpha$ for images of the $\textbf{SrcBulge}$, $\textbf{SrcDisk}$ and $\textbf{SrcMulti}$ models (first and third panels) and $\textbf{LensSrcBulge}$, $\textbf{LensSrcDisk}$ and $\textbf{LensSrcMulti}$ models (second and fourth panels). The legend at the top of each panel indicates the image each contour corresponds to, where the $\textbf{Src}$ images have been chosen to closely match the S/N of the $\textbf{Lens}$ images, such that the left two panels compare the PDFs of similiar lensed sources at Hubble resolution and right panels at Euclid resolution. Contours give the $1\sigma$ (interior) and $3 \sigma$ (exterior) confidence regions. For $\textbf{SrcBulge}$ and $\textbf{LensSrcBulge}$ the input values of each parameter are $\theta_{\rm  Ein} = 1.2"$, $q = 0.8$ and $\alpha = 2.0$, for $\textbf{SrcDisk}$ and $\textbf{LensSrcDisk}$ $\theta_{\rm  Ein} = 1.2$, $q = 0.75$ and $\alpha = 2.1$. Panels one and two, or three and four, therefore compare the analysis of nearly identical lensed sources but for a model which either does or does not include lens light modeling. With lens light modeliing included, the PDFs do not appear broader.} 
\label{figure:PDFsSrc2D}
\end{figure*}

Figure \ref{figure:PDFsSrc2D} shows the two-dimensional PDFs of $\Delta \alpha$ against $\Delta \theta_{\rm  Ein}$ for the Hubble resolution (first panel) and Euclid resolution (third panel) images of the $S/N = 30$ images of the $\textbf{SrcBulge}$, $\textbf{SrcDisk}$ and $\textbf{SrcMulti}$ models. The same degeneracy discussed in N15 is seen between the parameters governing the lens's mass distribution, where the degenerate models shown by the contours each integrate to give approximately the same $M_{\rm  Ein}$. Accompanying this (but not shown) is the source-plane scaling effect, demonstrated in figure 4 of N15, whereby steeper mass profiles lead to a more expanded source reconstruction. As expected, the posterior probability distribution function broadens for lower resolution imaging. 

\begin{figure*}
\centering
\includegraphics[width=0.4\textwidth]{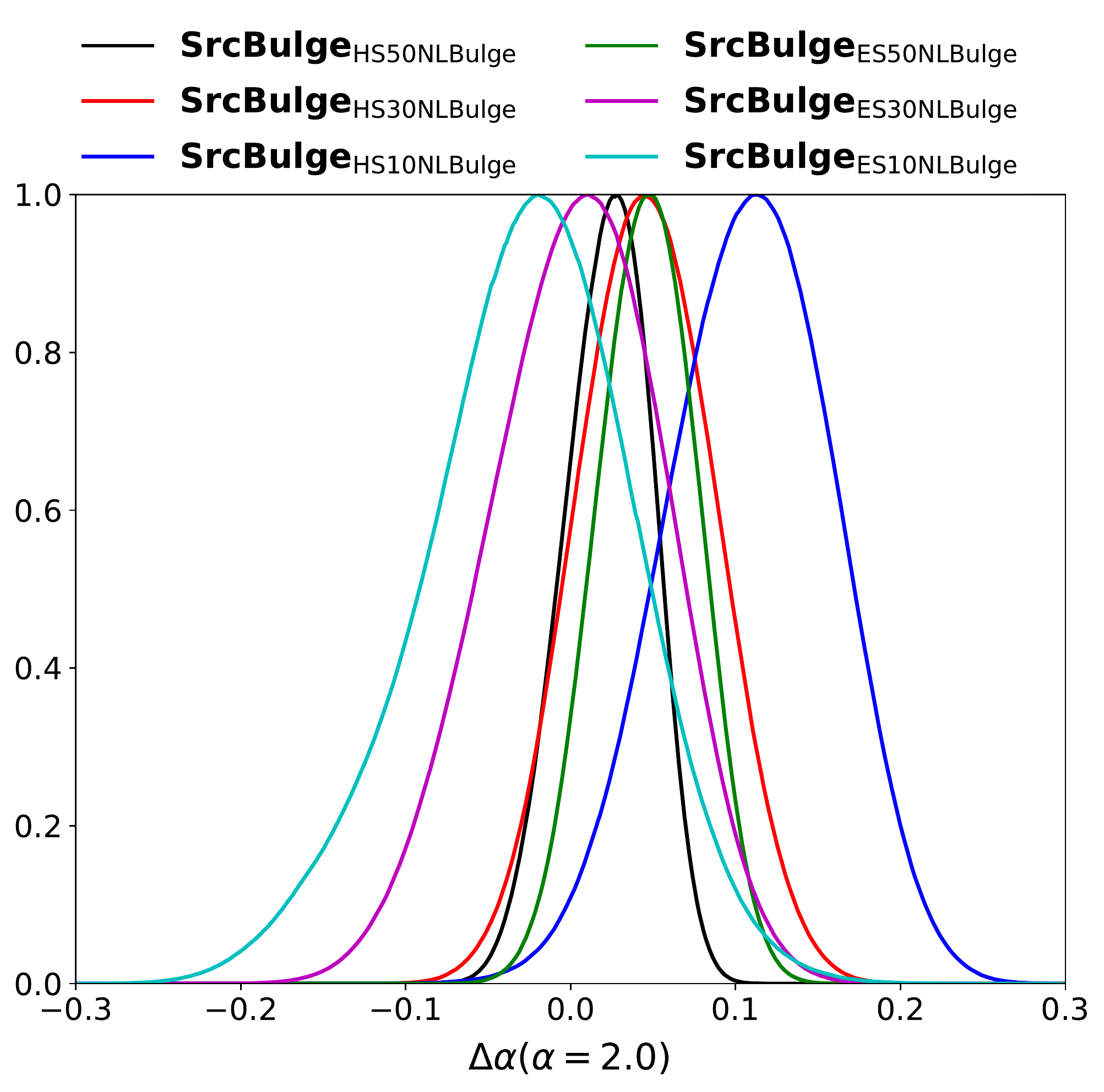}
\includegraphics[width=0.4\textwidth]{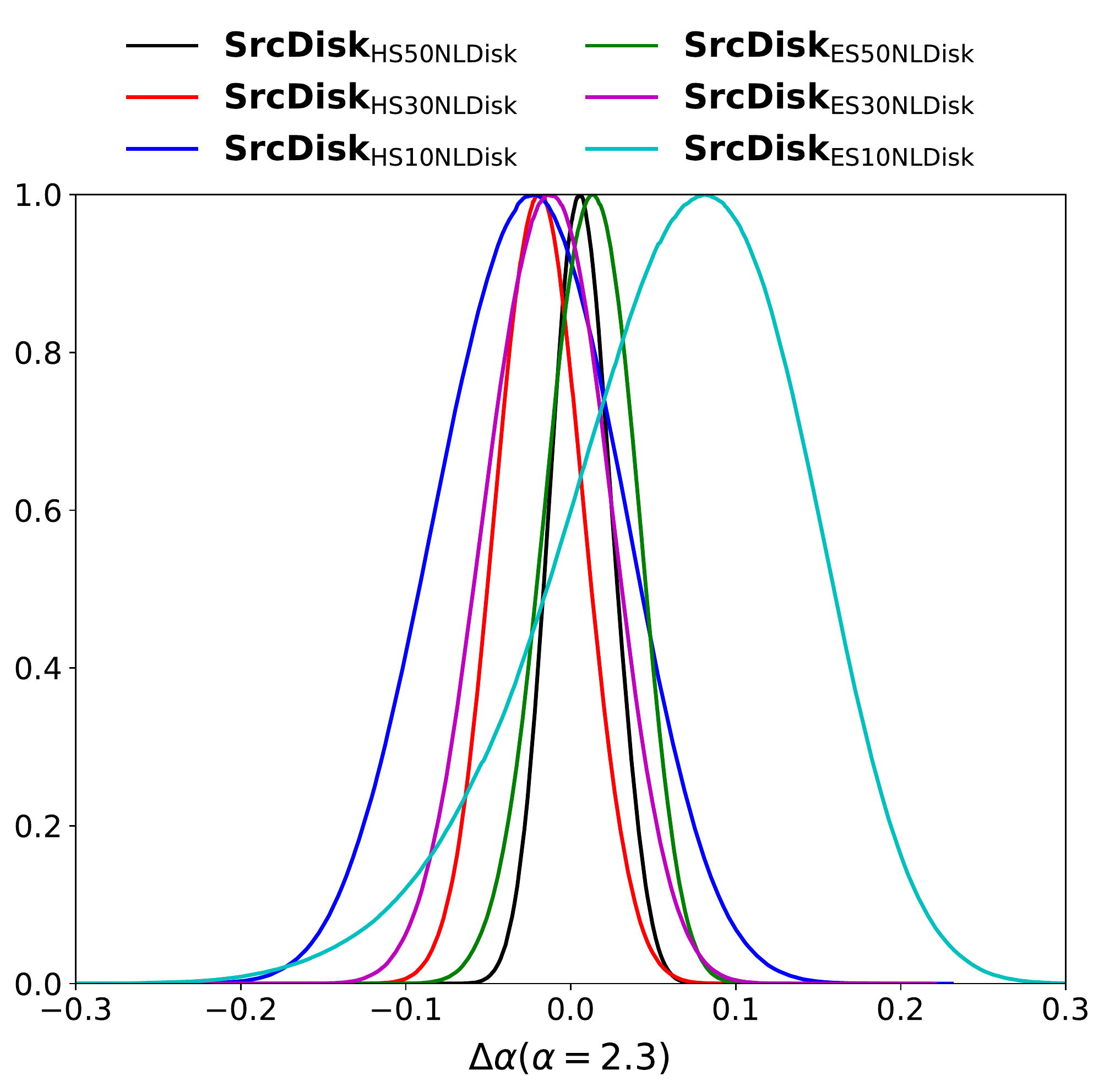}
\includegraphics[width=0.4\textwidth]{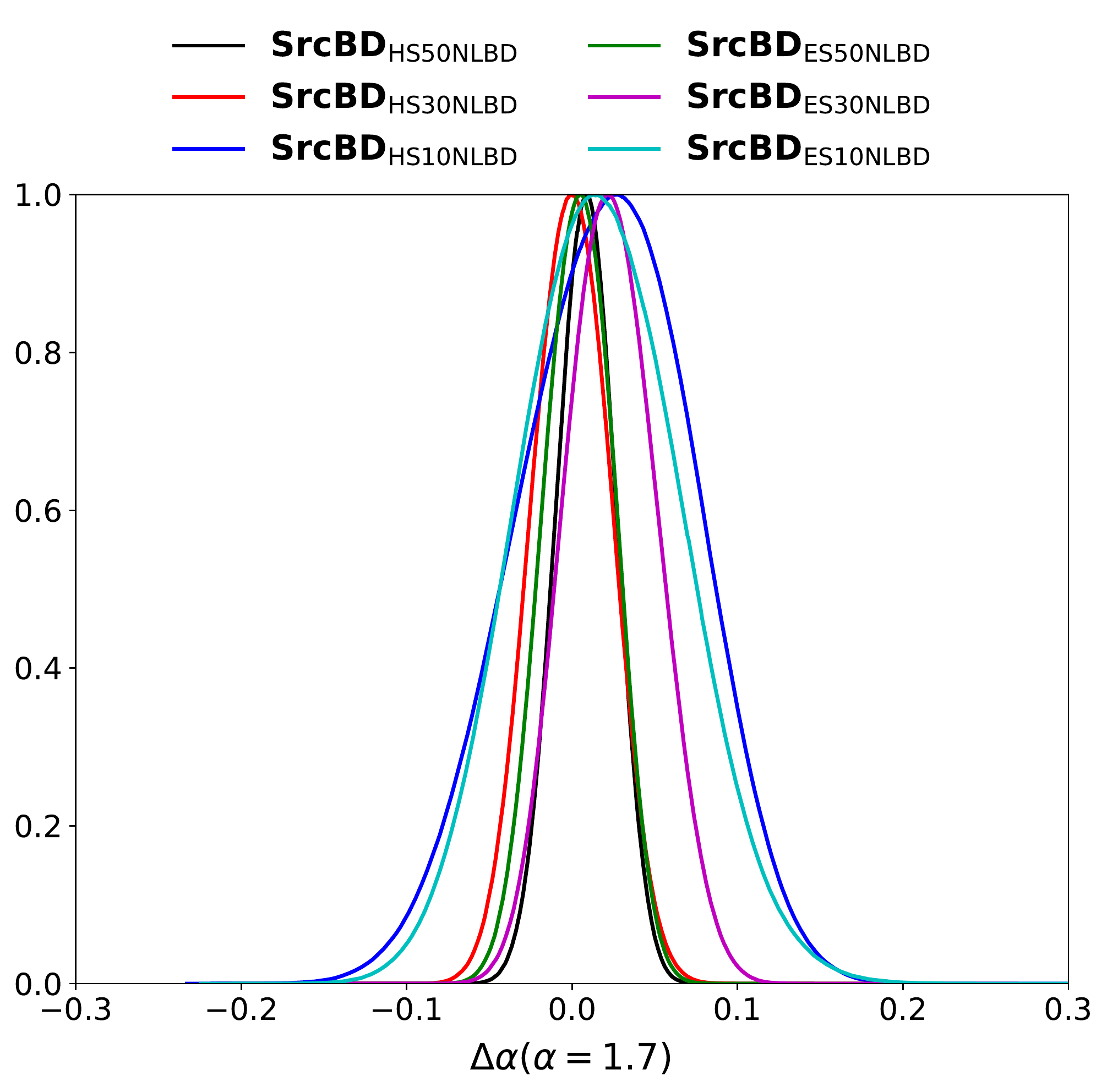}
\includegraphics[width=0.4\textwidth]{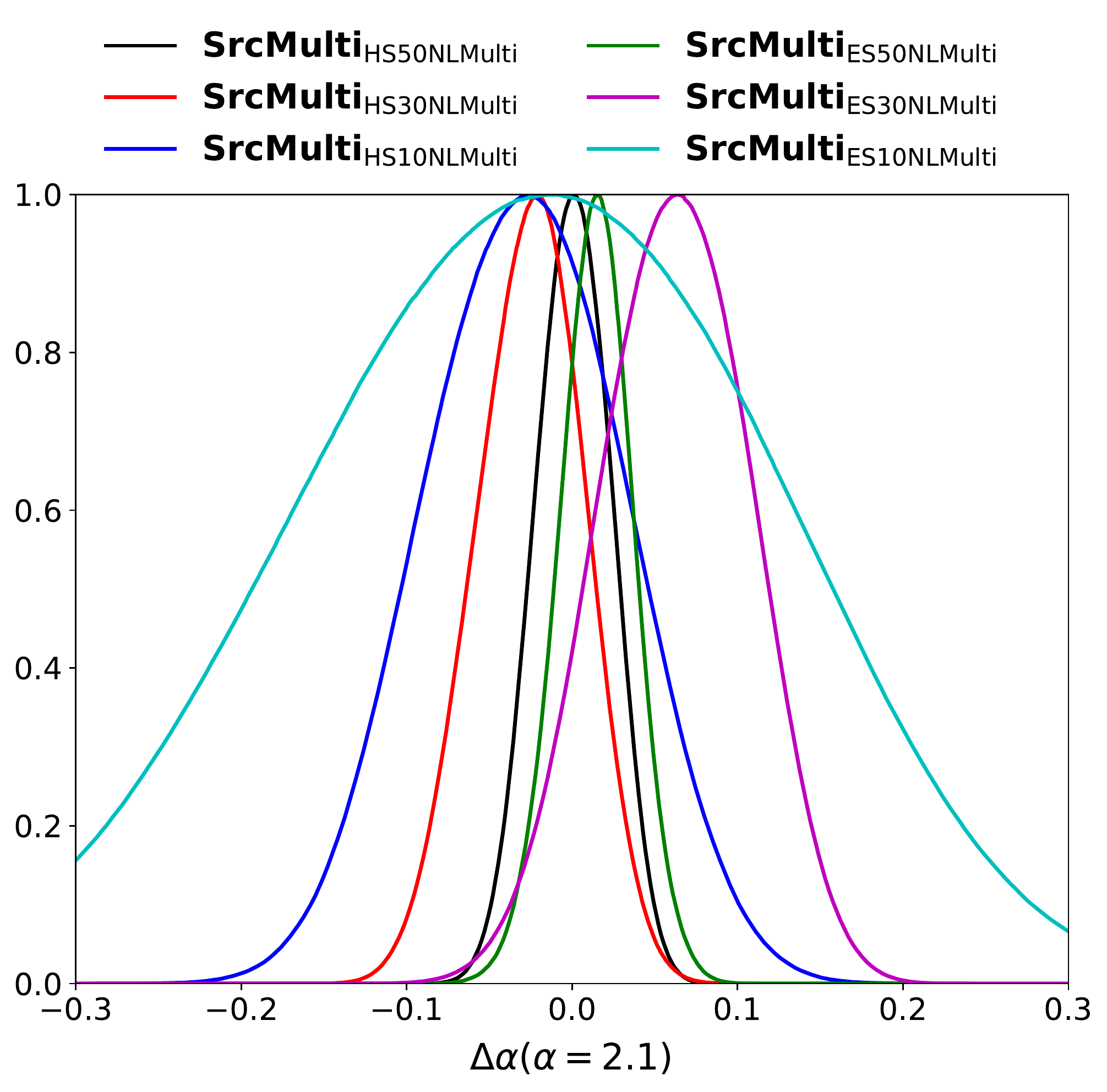}
\caption{Marginalized one-dimensional PDF's of the mass-profile density mismatch $\Delta \alpha = \alpha_{\rm True} - \alpha_{\rm Model}$ for the twenty-four mass models given in table \ref{table:TableSrcOnly}. The top-left panel corresponds to the six images of the $\textbf{SrcBulge}$ model, the top-right $\textbf{SrcDisk}$, bottom-left $\textbf{SrcBD}$ and bottom-right $\textbf{SrcMulti}$. Each graph's legend indicates the image that each line corresponds to, where black / red / blue lines give Hubble resolution images at S/N = $50$ / $30$ / $10$ and green / purple / cyan give Euclid resolution images at S/N = $50$ / $30$ / $10$, respectively. The input value of $\alpha$ for each model is given in brackets by the x-axis label. All PDF's are consistent with the input lens model ($\Delta \alpha = 0.0$).} 
\label{figure:PDFsSrcOnly}
\end{figure*}

Figure \ref{figure:PDFsSrcOnly} shows $\Delta \alpha$'s one-dimensional PDF for each image of every lens model. There is no systematic deviation of $\Delta \alpha$ with varying image resolution, S/N ratio, mass model or source morphology, confirming {\tt AutoLens}'s source-only analysis is free of systematic bias. As expected, images at higher spatial resolution or S/N give tighter lens model constraints. This figure also gives a sense of how precisely images of different image resolution or S/N constrain $\alpha$, suggesting that Euclid wide-field imaging will be able to estimate $\alpha$ to a precision $\pm 0.1$, or better, at $3 \sigma$ confidence.

By comparing each panel, one can also see how the lens model's precision depends on the source morphology. The bulge-disk morphology is marginally the most-well constrained, benefiting from how its source has both a smooth extended disk component and cuspy central light profile. In contrast, the bulge-only morphology offers the loosest constraints, suggesting that an extended envelope of source light is important to reducing errors. However, for images with the same resolution and S/N, the differences in error estimates are marginal, thus for smoothly parametrized source morphologies the profile shape and number of components appears to play no major role in determining how precisely the mass model is constrained. This contradicts discussions by the authors \citep{Vegetti2012, Lagattuta2012}, who argue that multiple sources with non-symmetric morphologies offer much tighter constraints, as they produce a less degenerate set of possible image reconstructions from which there is a smaller sub-set of mass models that are able to reconstruct them accurately. Such a trend is not seen for the analysis of the $\textbf{SrcMulti}$ images. However, this is most likely a reflection of the fact that these simulated images are modeled with their input mass profile and the tighter constraints offered by complex sources are more readily observed on real strong lens imaging, where the mass model offers only an approximate fit. 

\subsection{Non-cored Lens and Source}\label{ResultsLensSrc}

\begin{figure*}
\centering
\includegraphics[width=0.162\textwidth]{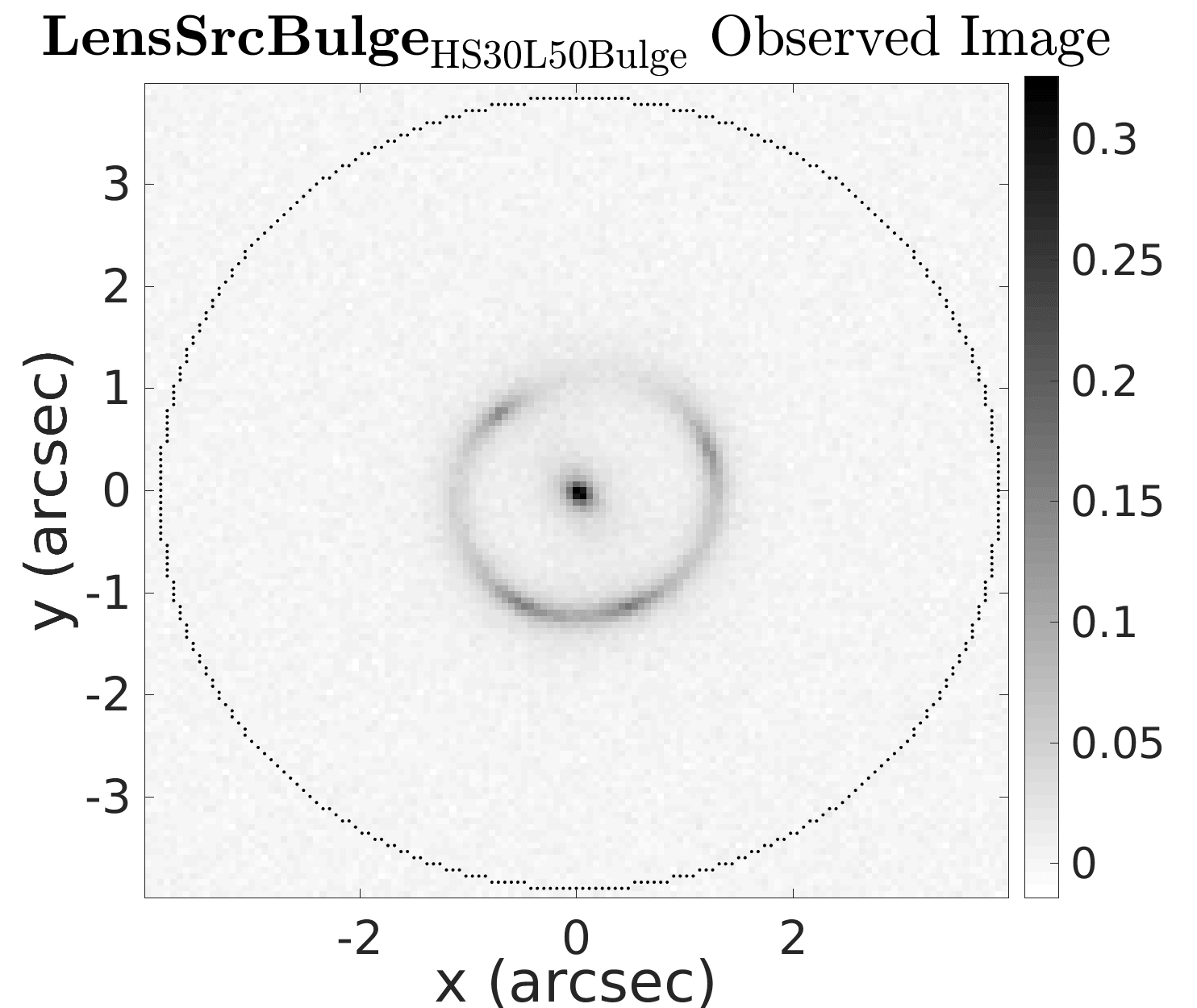}
\includegraphics[width=0.162\textwidth]{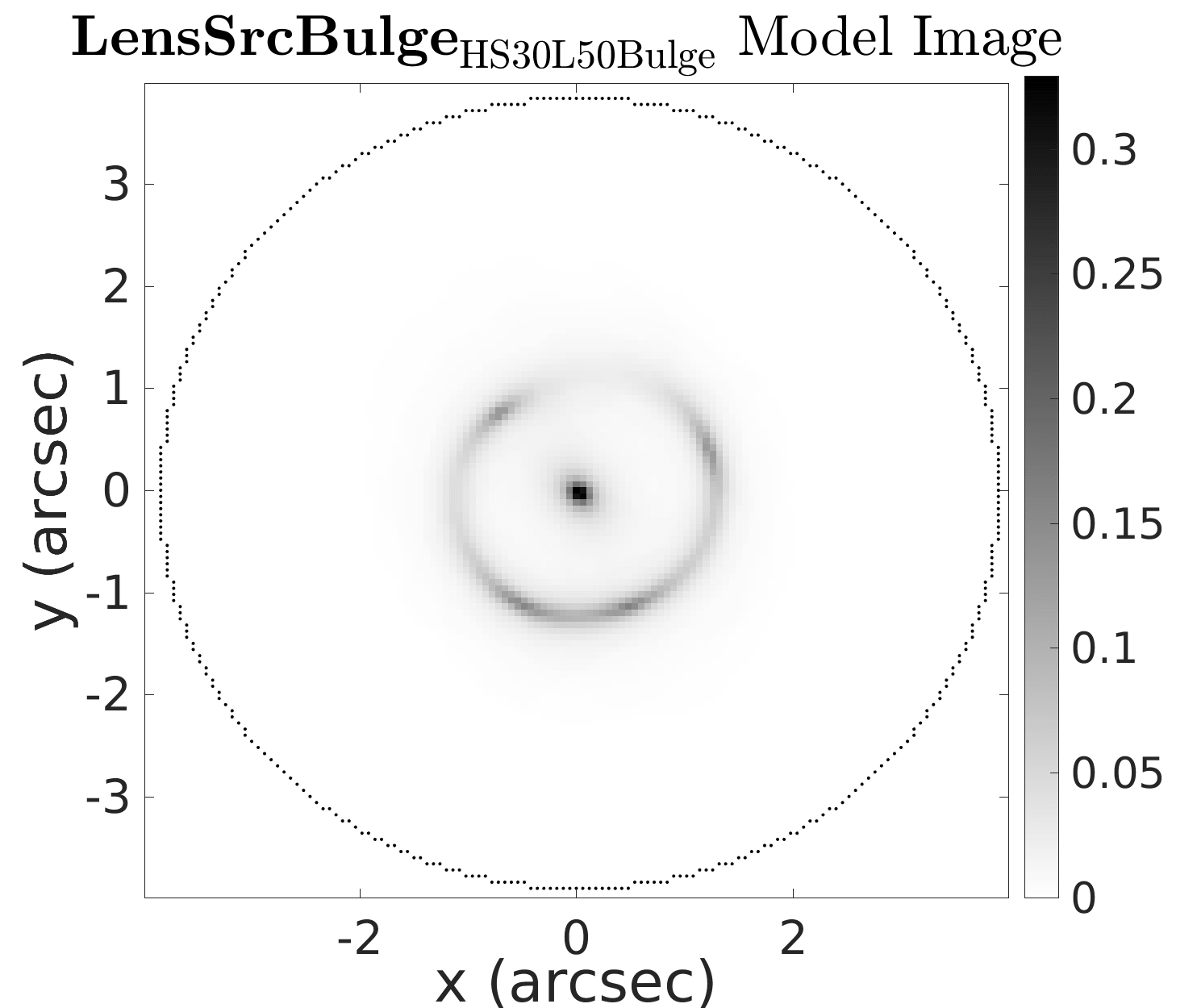}
\includegraphics[width=0.162\textwidth]{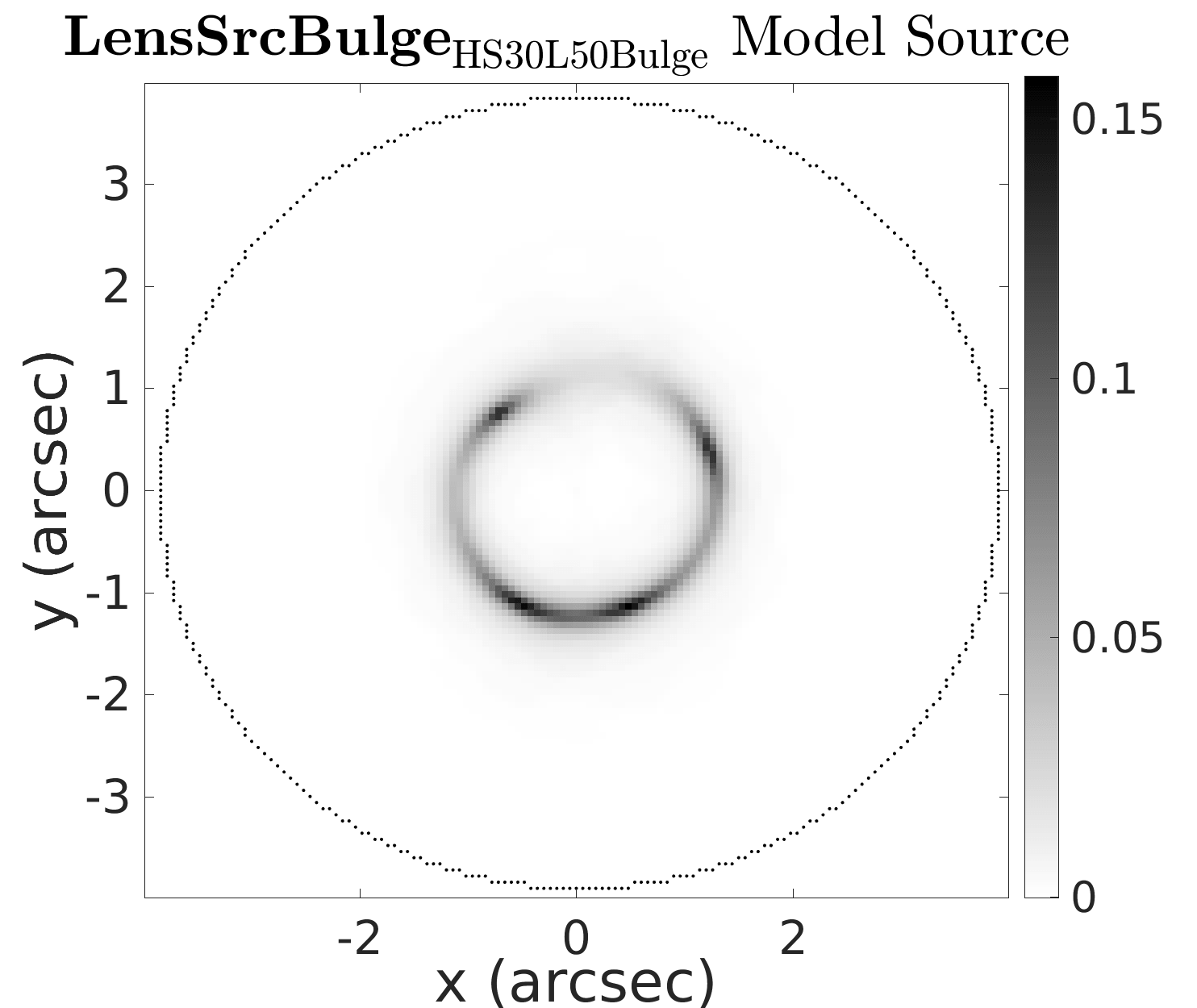}
\includegraphics[width=0.162\textwidth]{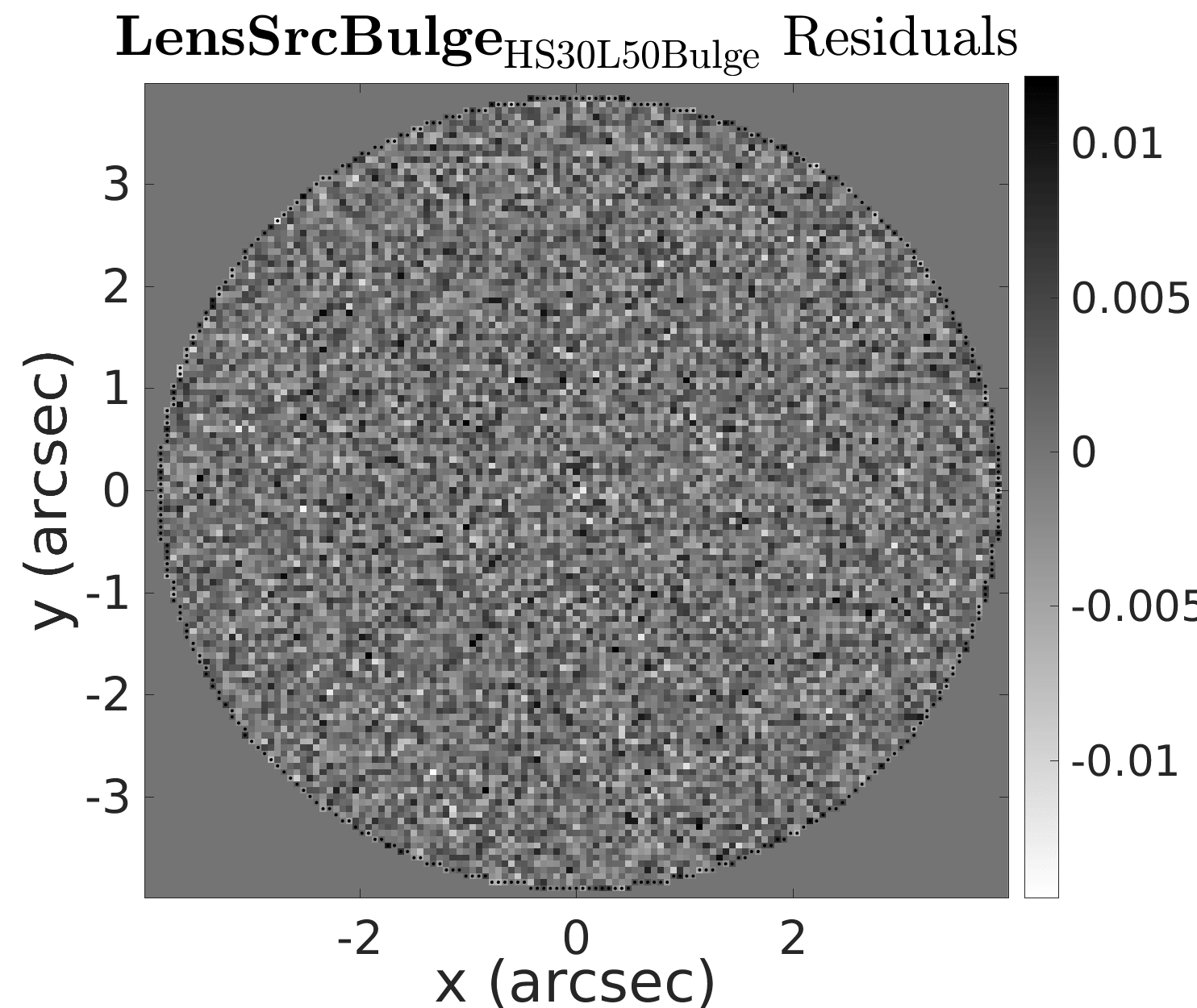}
\includegraphics[width=0.162\textwidth]{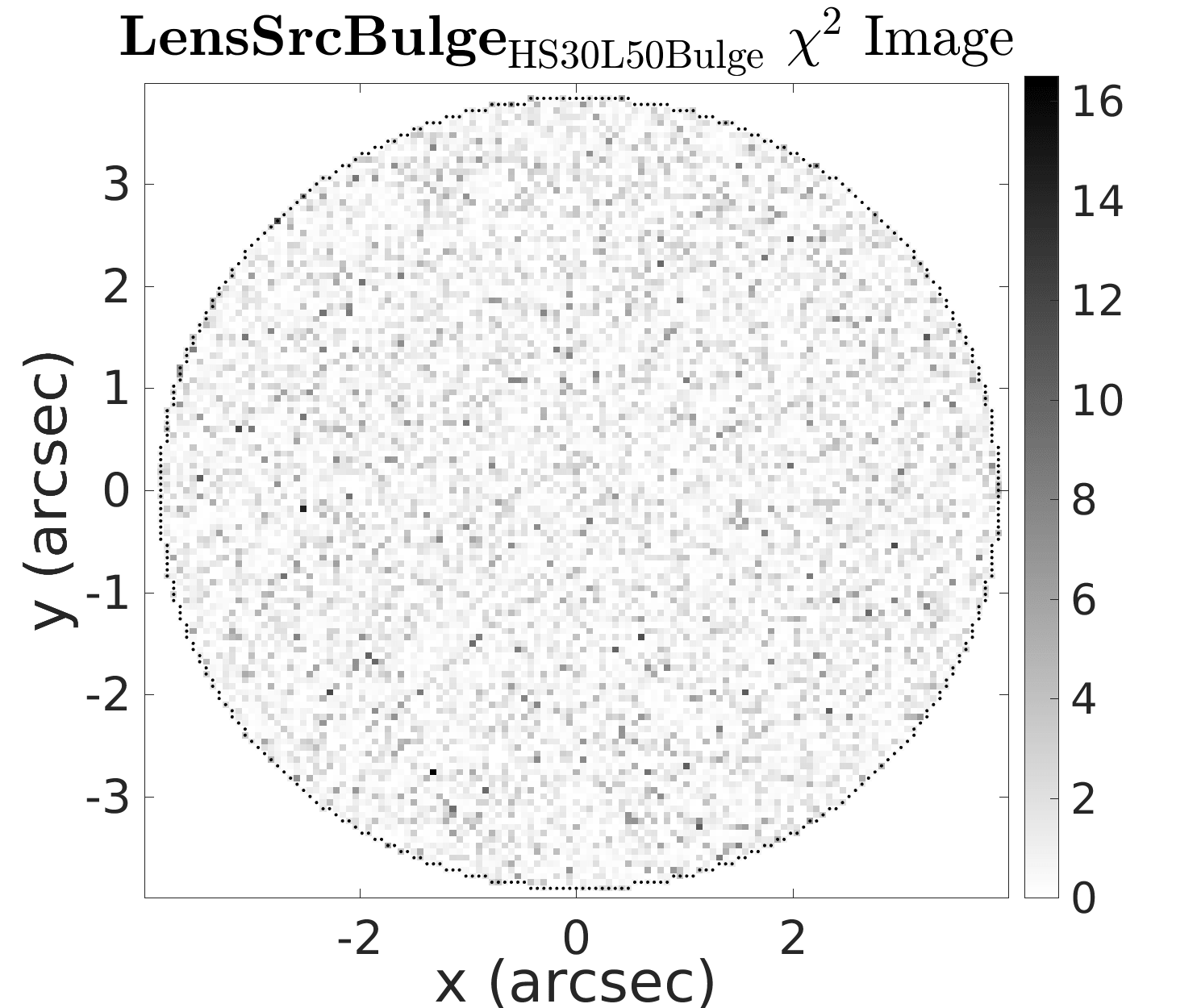}
\includegraphics[width=0.162\textwidth]{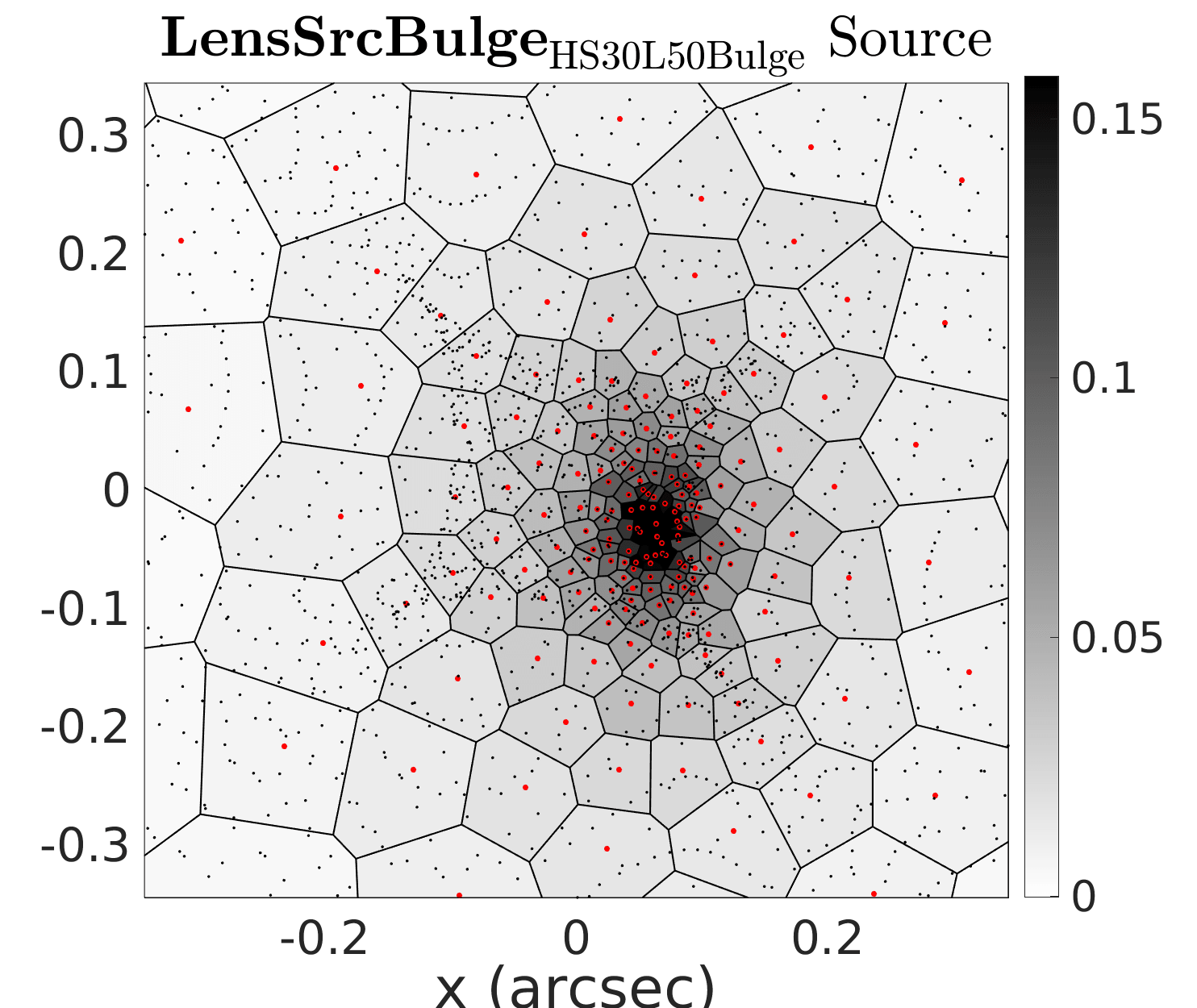}
\includegraphics[width=0.162\textwidth]{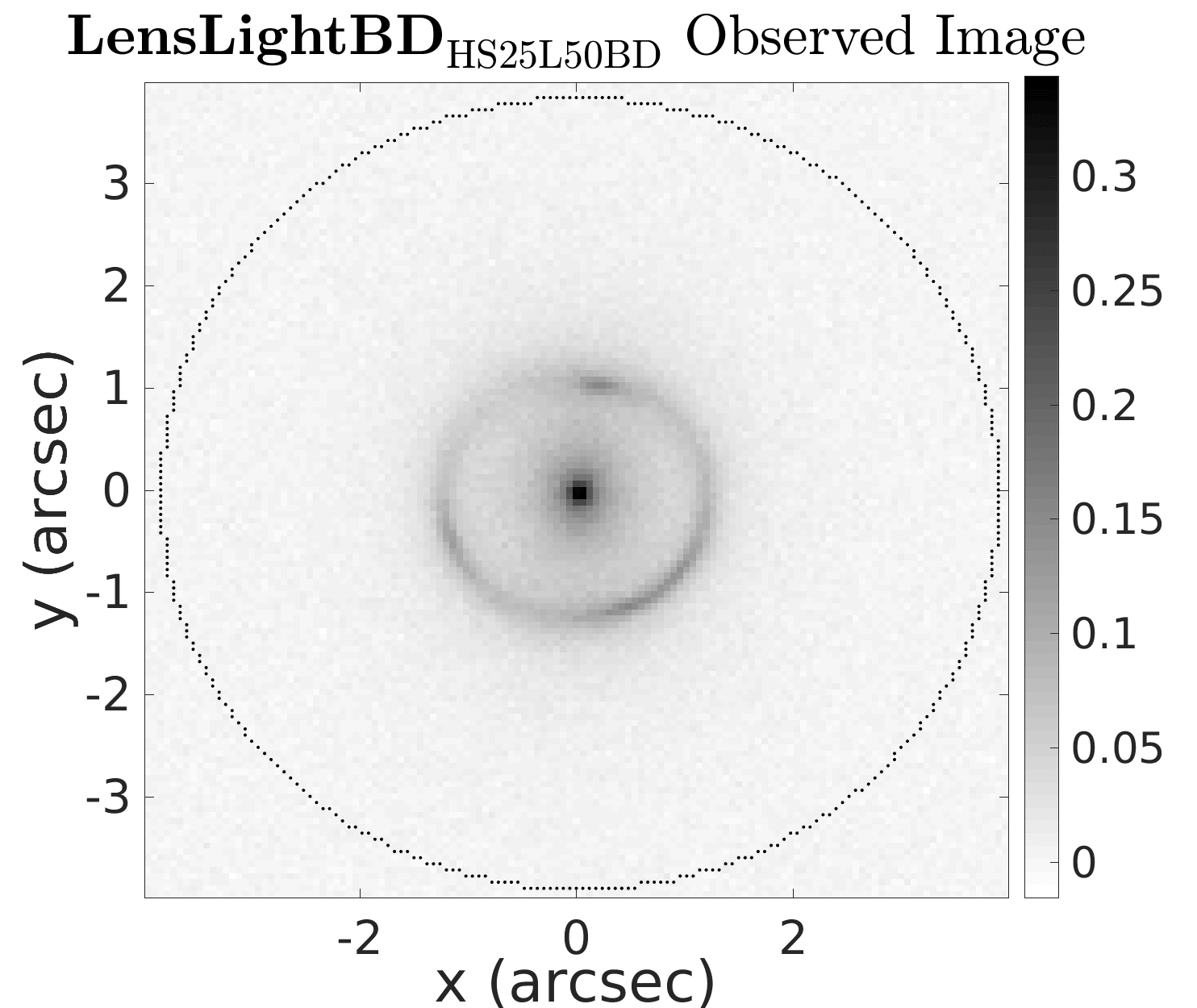}
\includegraphics[width=0.162\textwidth]{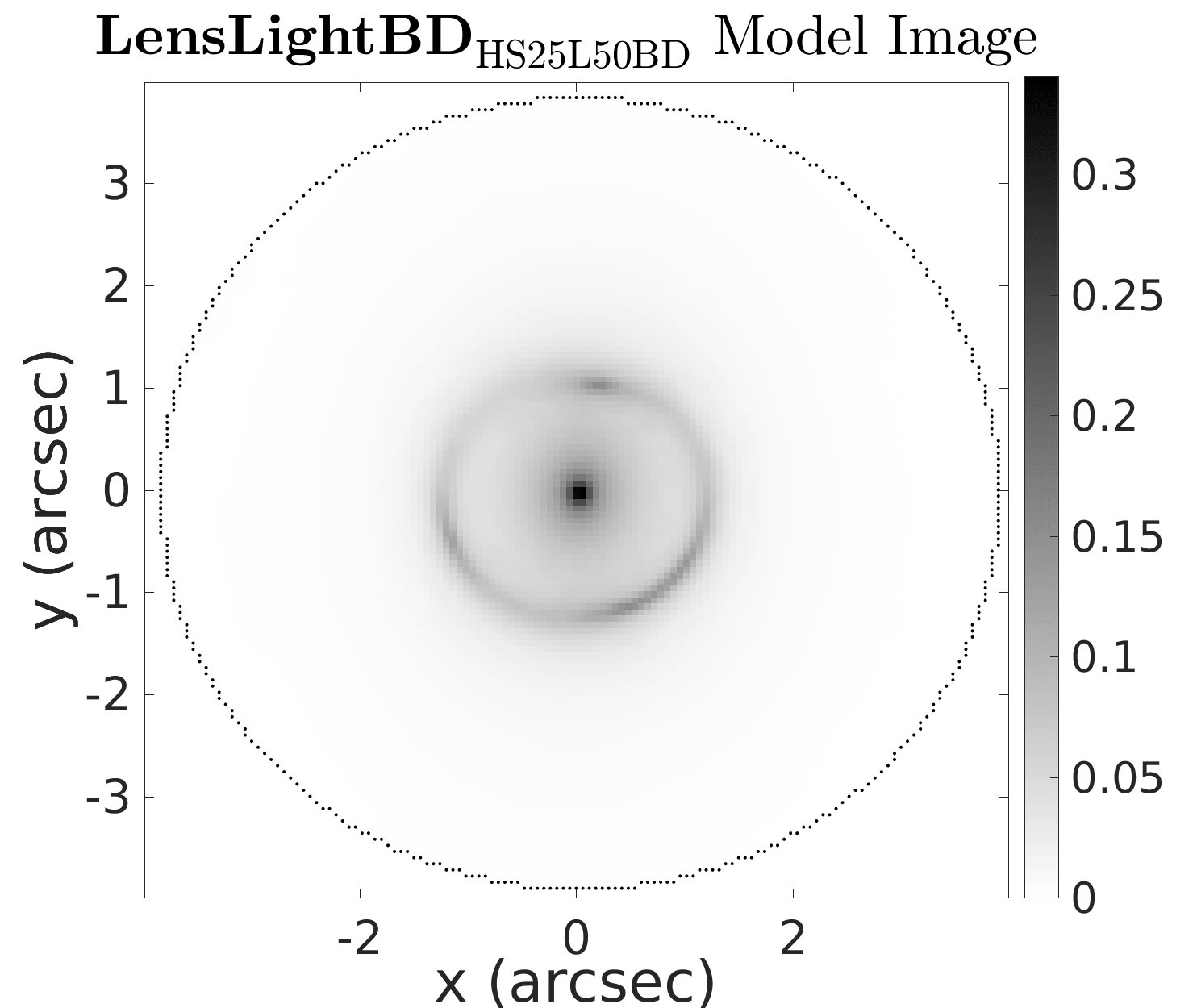}
\includegraphics[width=0.162\textwidth]{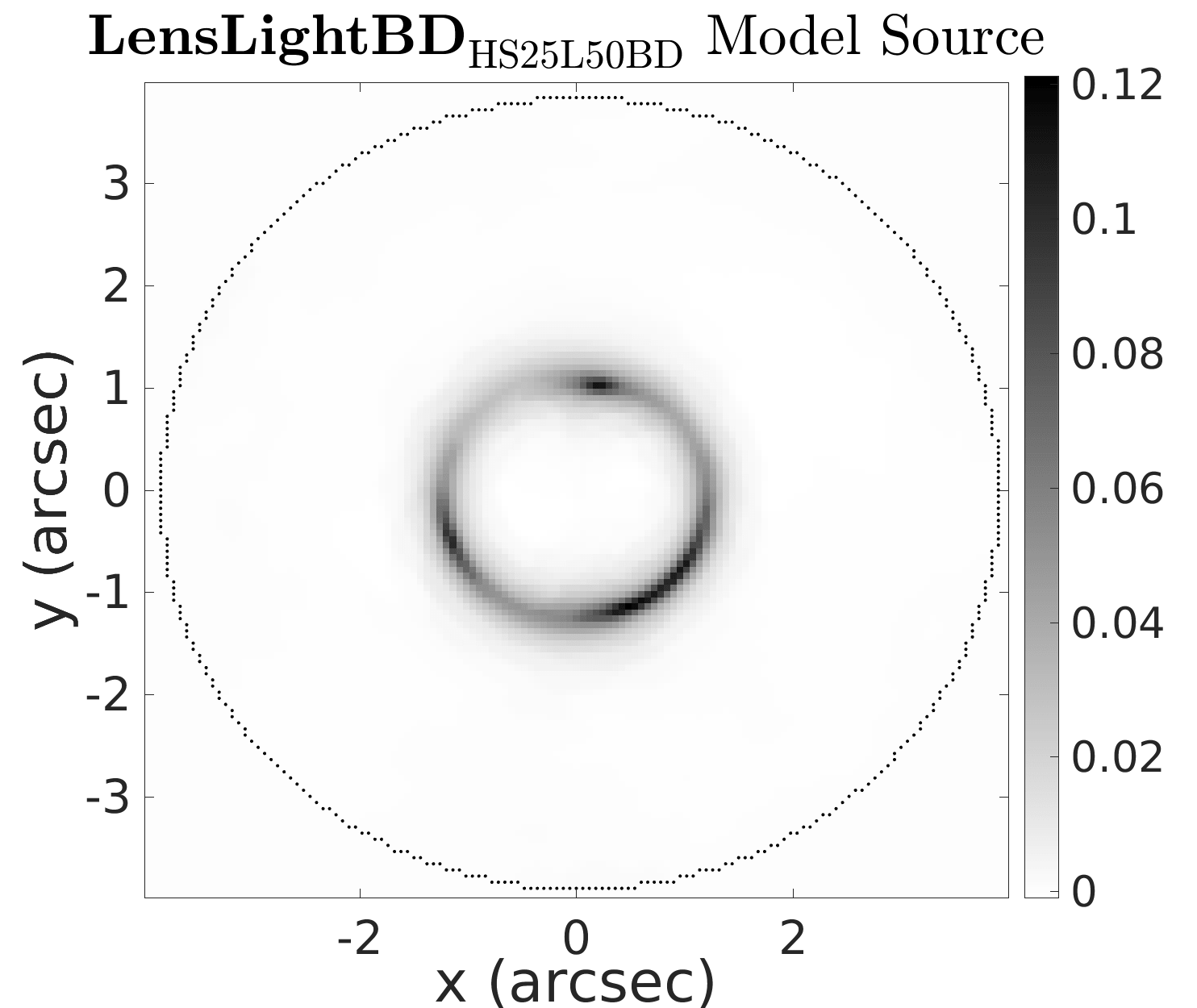}
\includegraphics[width=0.162\textwidth]{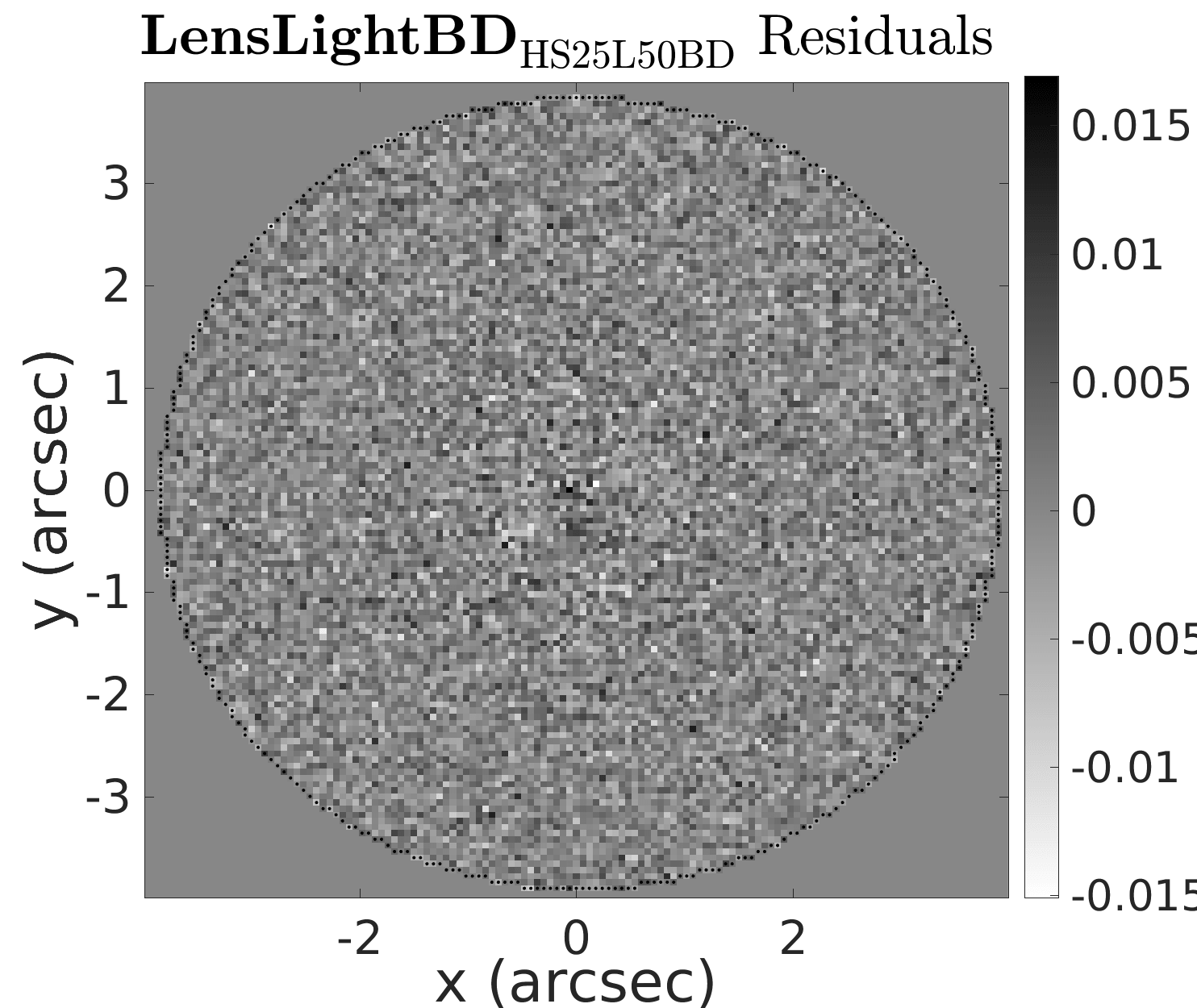}
\includegraphics[width=0.162\textwidth]{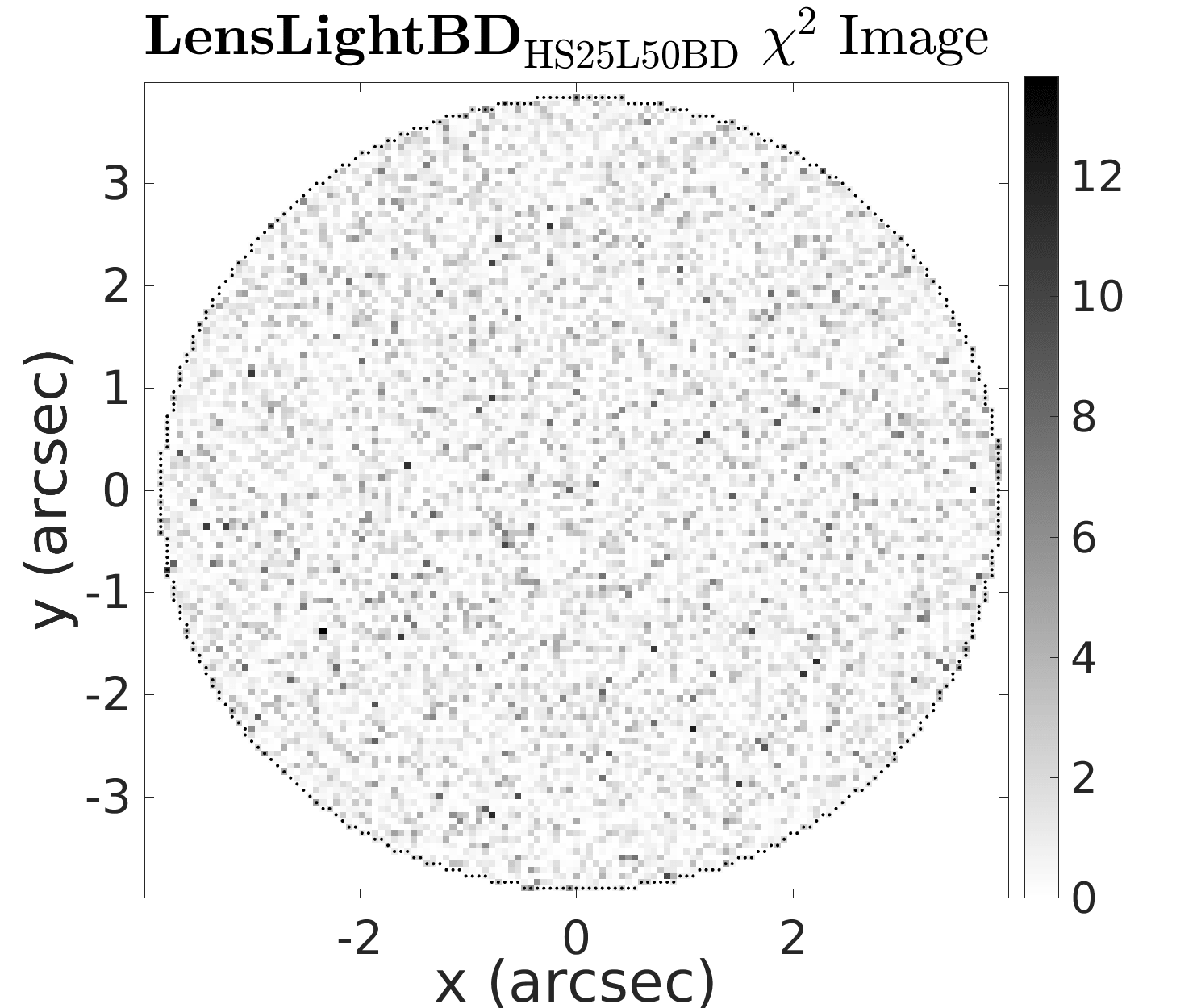}
\includegraphics[width=0.162\textwidth]{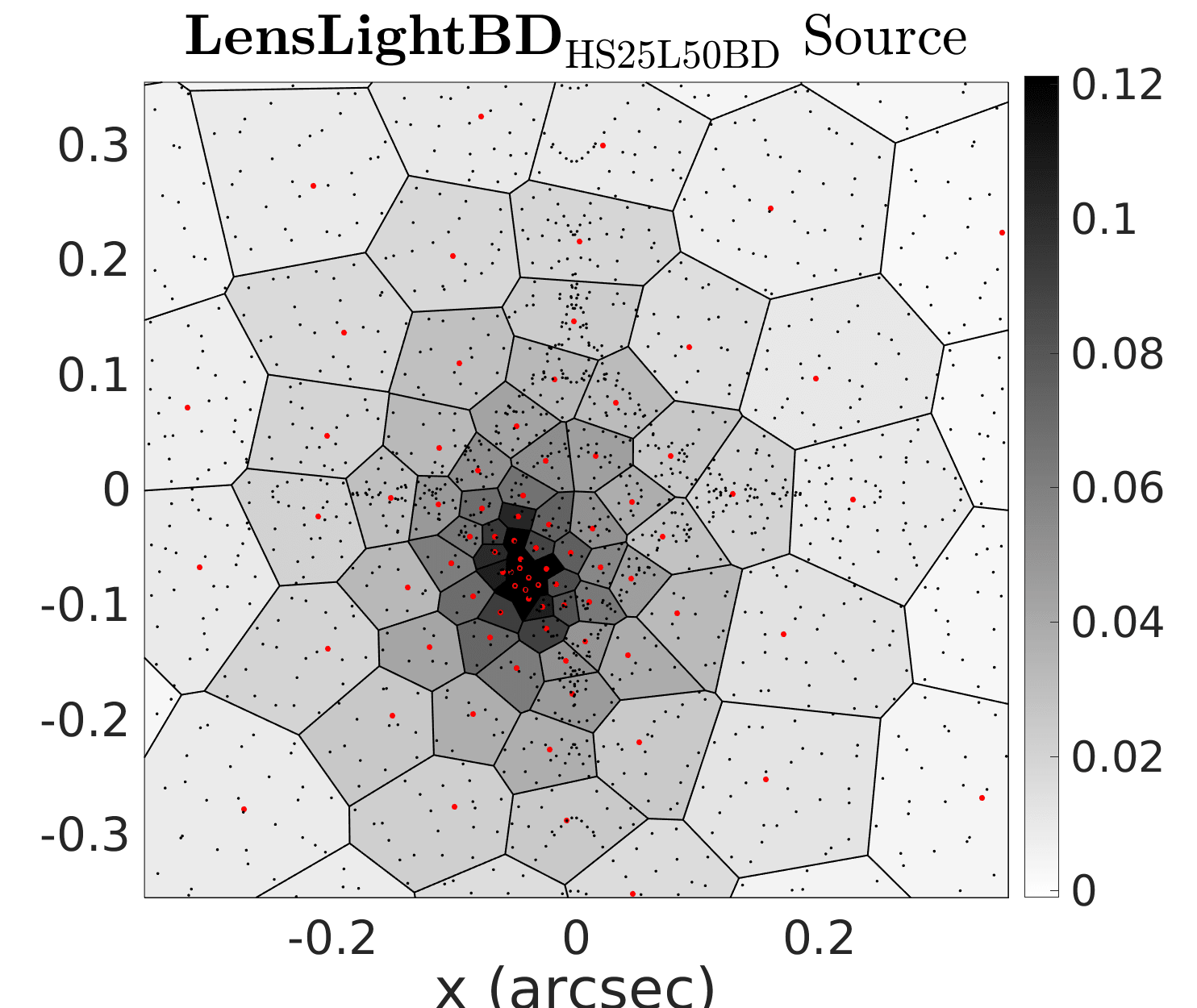}
\includegraphics[width=0.162\textwidth]{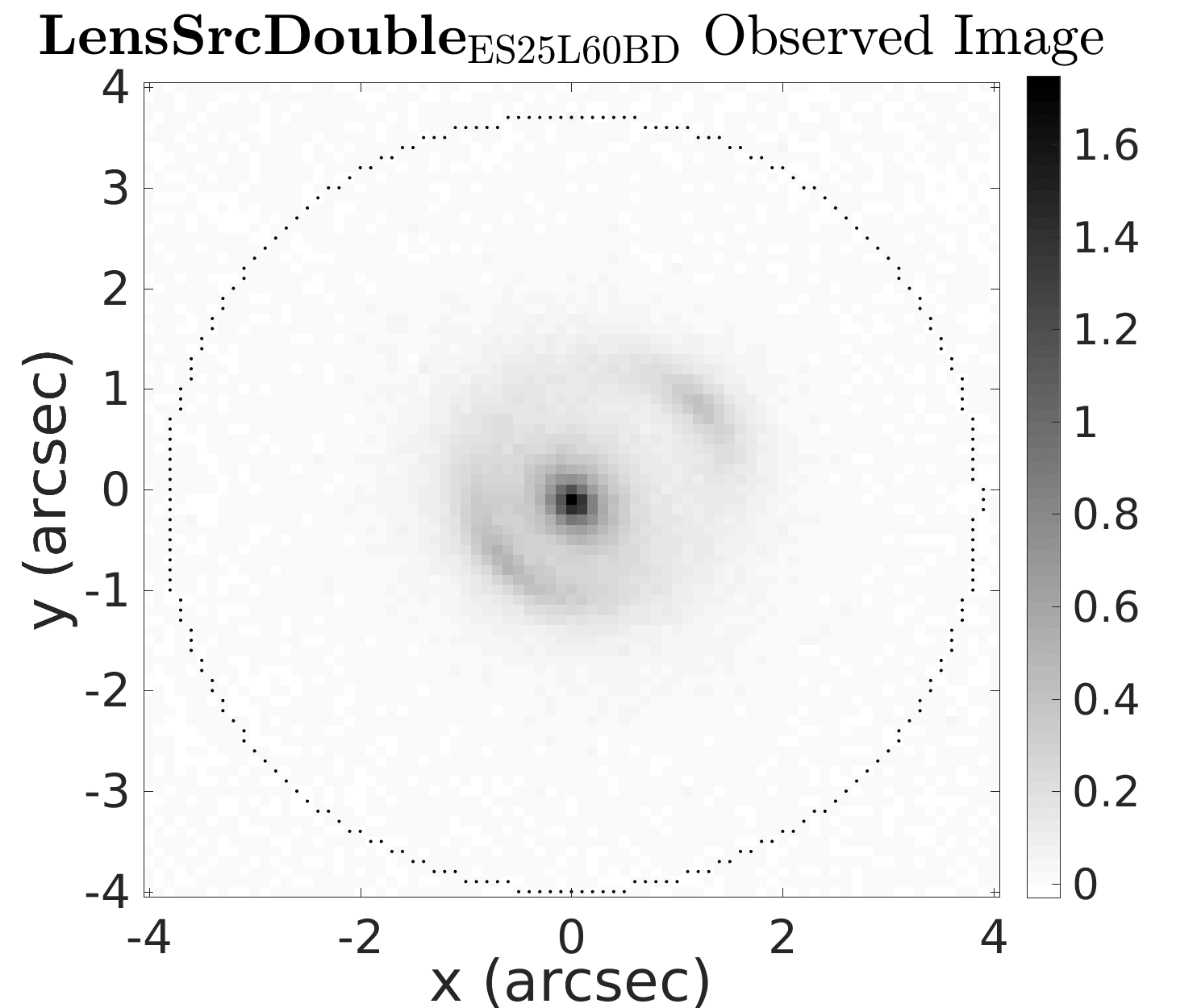}
\includegraphics[width=0.162\textwidth]{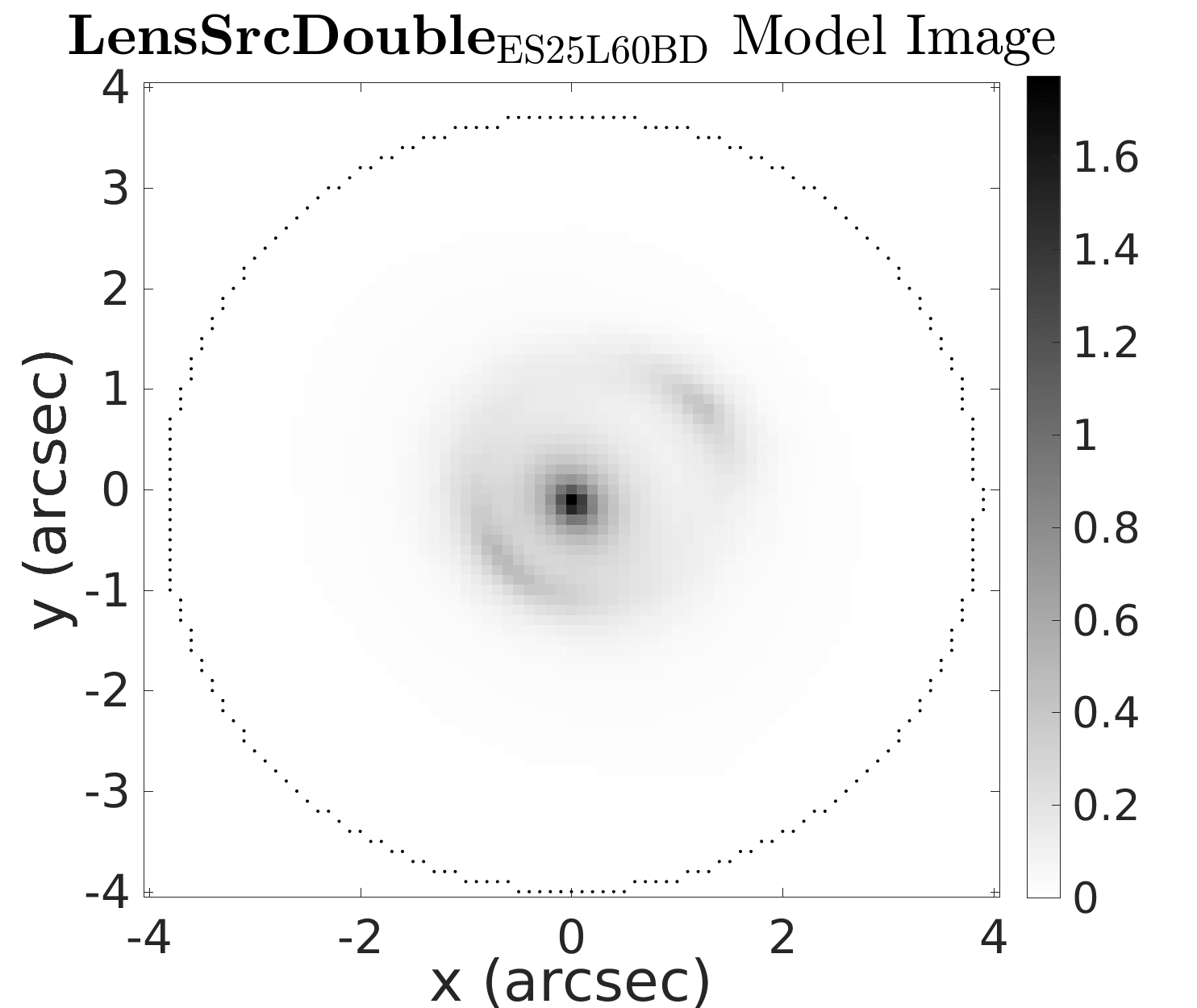}
\includegraphics[width=0.162\textwidth]{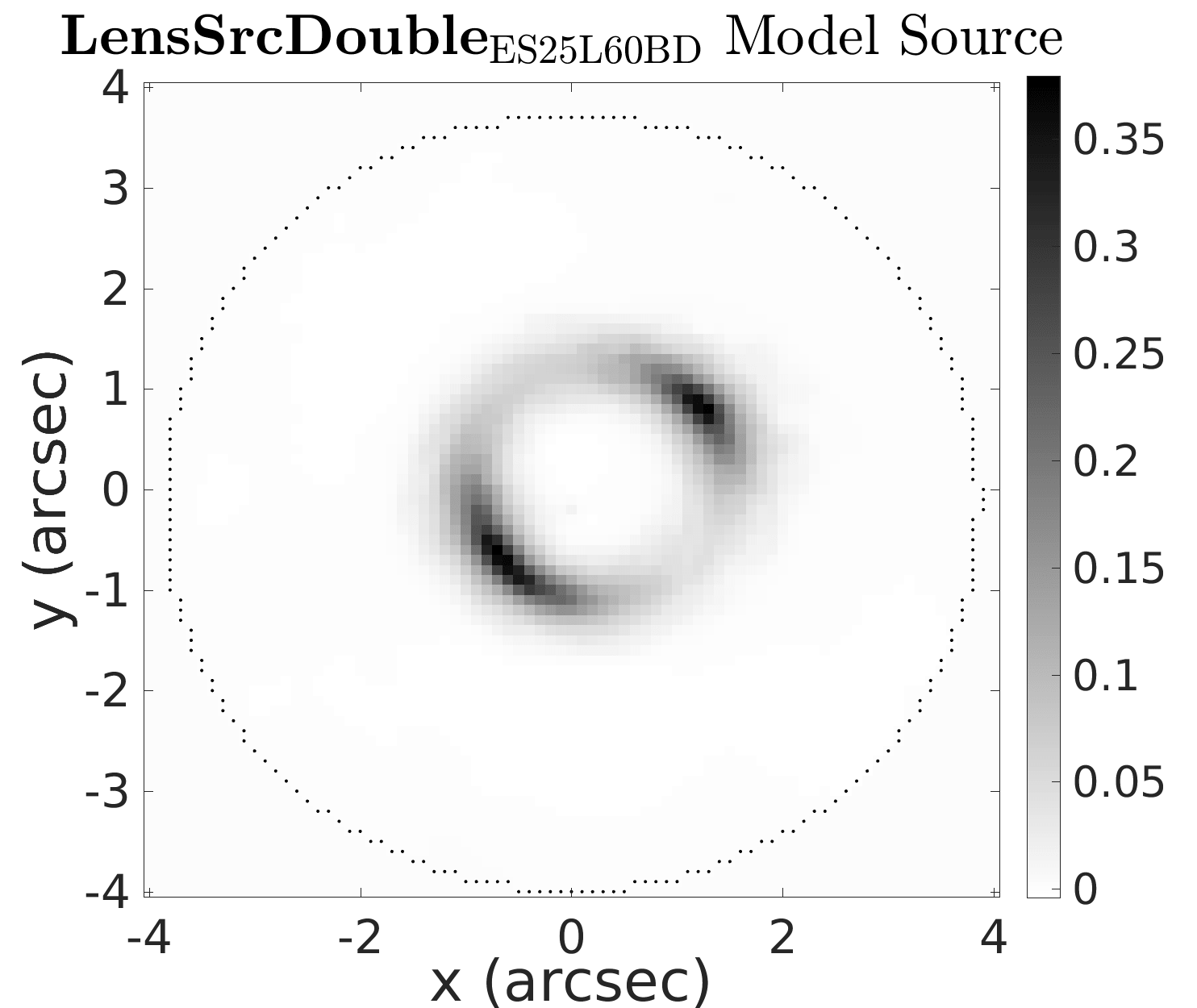}
\includegraphics[width=0.162\textwidth]{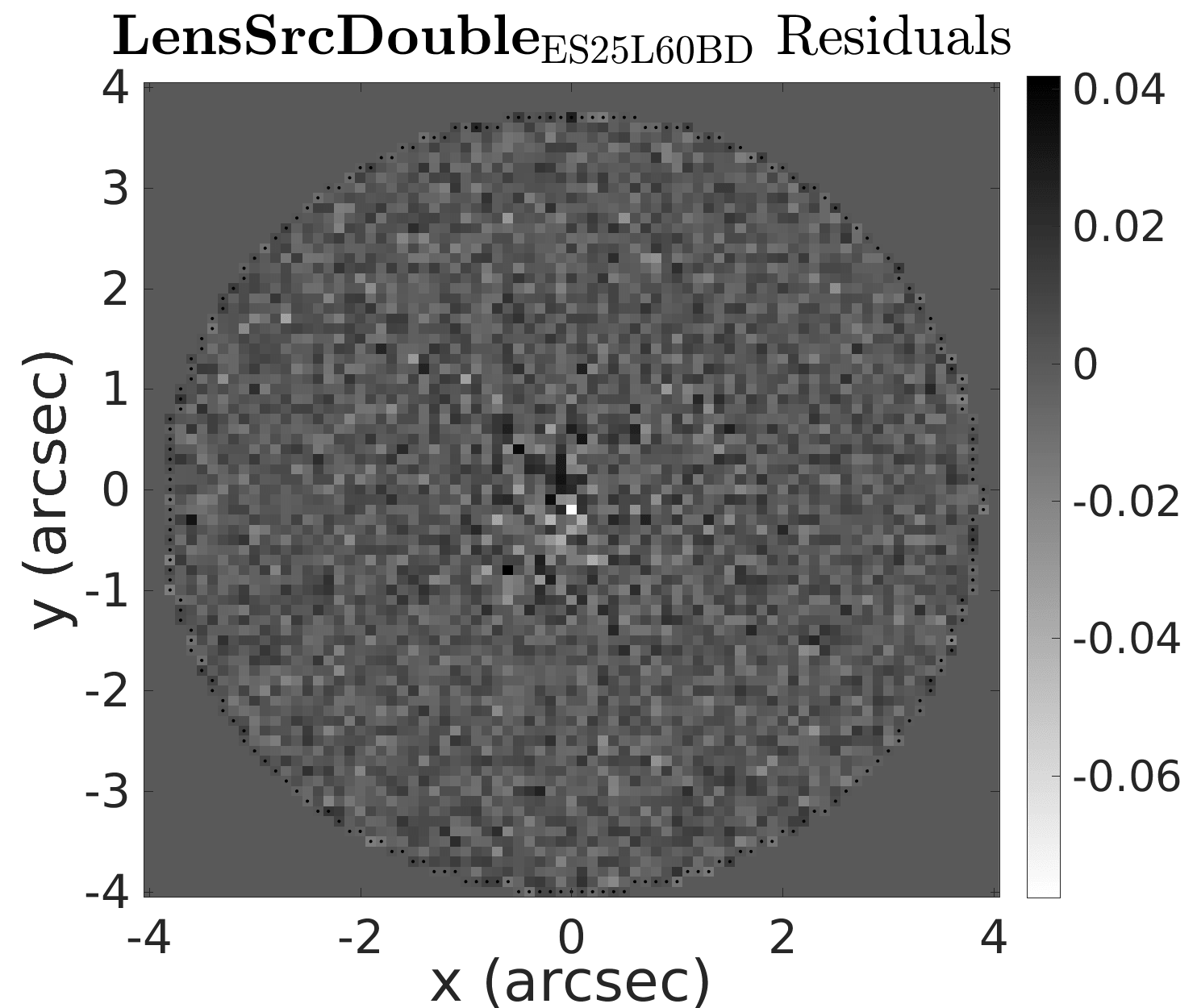}
\includegraphics[width=0.162\textwidth]{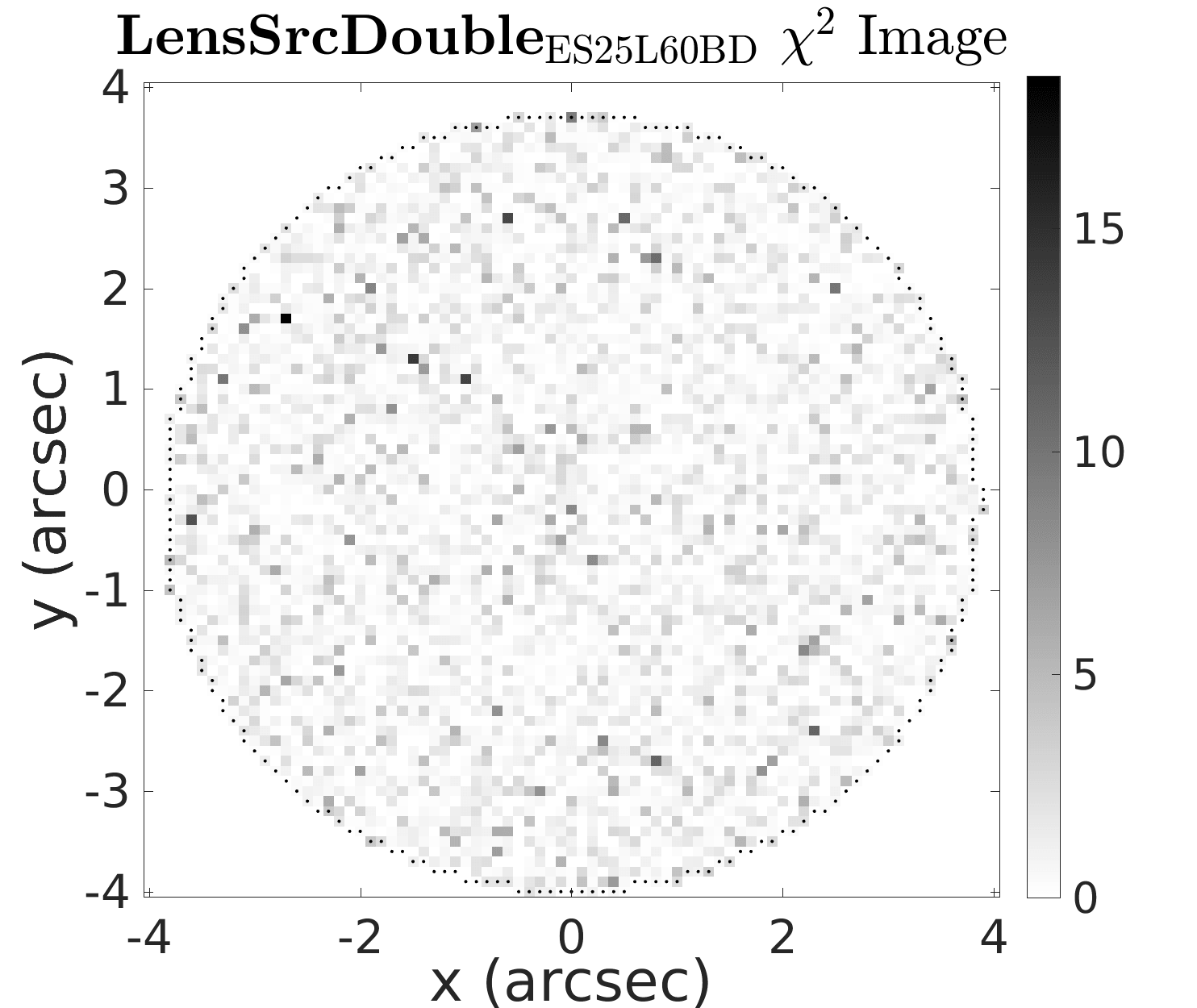}
\includegraphics[width=0.162\textwidth]{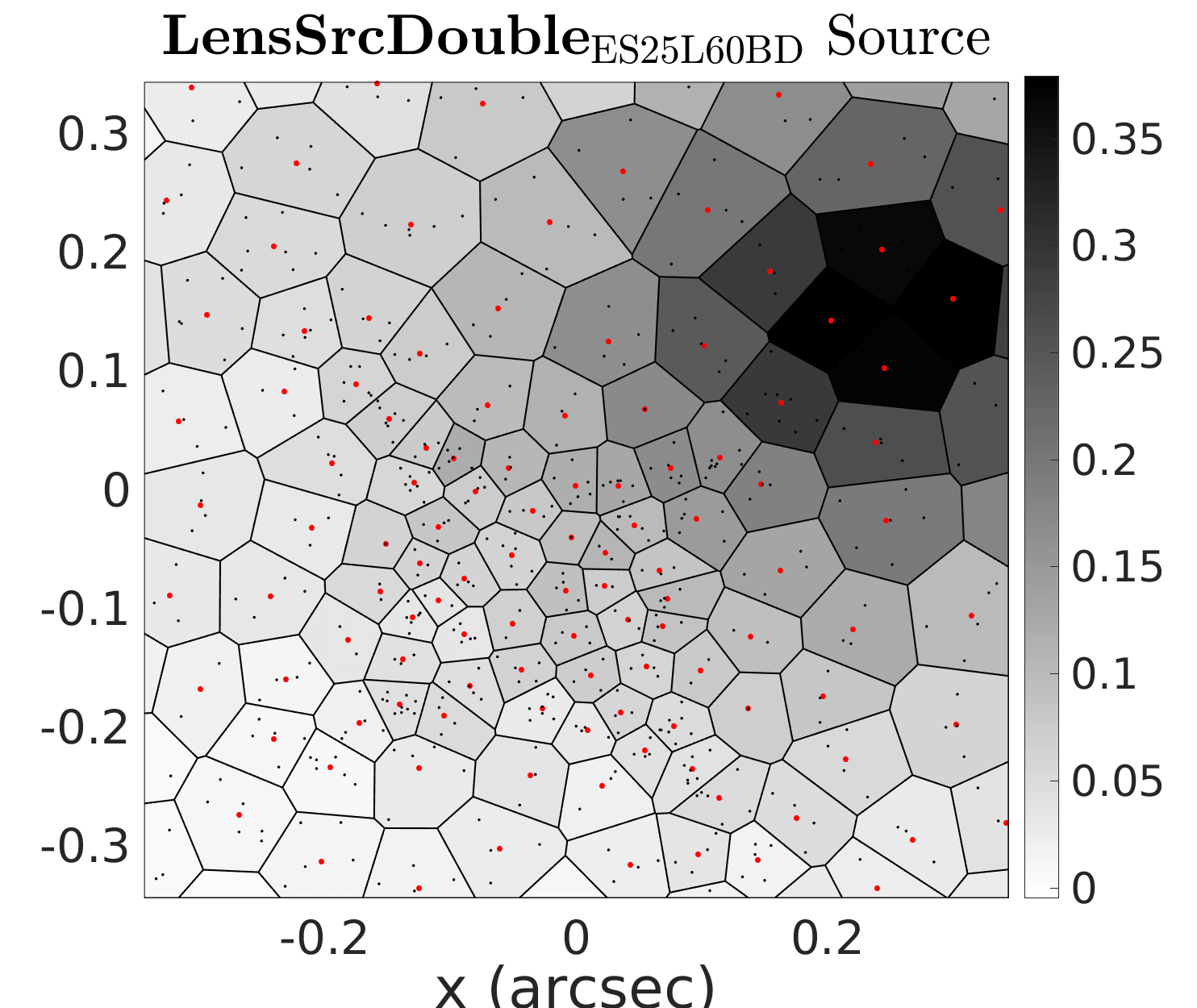}
\caption{The simulated images, model images, source models, residuals, $\chi^2$ images and source reconstructions for the images $\textbf{LensSrcBulge}_{\mathrm{HS30L50Bulge}}$ (top row), $\textbf{LensLightBD}_{\mathrm{HS25L50BD}}$ (middle row) and $\textbf{LensSrcDouble}_{\mathrm{ES25L60BD}}$ (bottom row). Images correspond to the most likely model at the end of phase 2 of the main pipeline, corresponding to the models given by rows $1-2$, $27-29$, $15-16$ and $11-12$ of table \ref{table:TableSPLEFg} respectively.} 
\label{figure:ResultsSPLEFgIms}
\end{figure*}

The non-cored lens simulation suite consists of seven unique lens and source models, which are each used to generate two images at Hubble and Euclid resolution with a range of source and lens S/N ratios, giving a total of fourteen images. The analysis of each image uses the singular total-mass pipeline, with parameter estimates and results corresponding to the end of phase two. The results of using Bayesian model comparison to choose the light and mass profiles are shown in table \ref{table:SPLEFgMC}, with the input model correctly chosen for all images. The $Sersic$ light model is summarized using the mismatch parameters $\Delta R_{\rm l}$, $\Delta n_{\rm l}$ an $\Delta q_{\rm l}$, whereas the $Exp$ model use $\Delta R_{\rm  l1}$, $\Delta q_{\rm l1}$, $\Delta R_{\rm  l2}$ and $\Delta q_{\rm l2}$. These are listed, alongside $\Delta \theta_{\rm  Ein}$, $\Delta q$ and $\Delta \alpha$, in table \ref{table:TableSPLEFg}, showing the majority of parameter estimates are accurate within $3 \sigma$ confidence. Figure \ref{figure:ResultsSPLEFgIms} shows the observed images, model images, model sources, residuals, $\chi^2$ images and source reconstructions for three images. The image residuals can be seen to realize each image's noise whereas the $\chi^2$ images are Gaussian, as desired. These images are indicative of the analysis of all non-cored images.

\subsubsection{Model Comparison}\label{ResultsLenSrcMC}

\begin{table*}
\scriptsize
\centering
\begin{tabular}{ l | l | l || l | l } 
\multicolumn{1}{p{1.0cm}|}{\centering \textbf{Image}} 
& \multicolumn{1}{p{2.1cm}|}{\centering \textbf{Sersic} \\ ($\textbf{P}_{\rm MCLight}$) } 
& \multicolumn{1}{p{2.1cm}||}{\centering \textbf{Sersic + Exp} \\ ($\textbf{P}_{\rm MCLight}$) } 
& \multicolumn{1}{p{2.1cm}|}{\centering \textbf{SPLE} \\ ($\textbf{P}_{\rm MCMass}$)  } 
& \multicolumn{1}{p{2.1cm}}{\centering \textbf{SPLE + Shear} \\ ($\textbf{P}_{\rm MCMass}$) } 
\\ 
& & & & \\[-5pt]
\hline
& & & & \\[-4pt]
$\textbf{LensSrcBulge}_{\mathrm{HS30L50Bulge}}$ & \textbf{55549.5448} & 55552.1930 & \textbf{55602.9633} & 55604.7013\\[2pt]
$\textbf{LensSrcBulge}_{\mathrm{ES30L50Bulge}}$ & \textbf{15895.6885} & 15894.4006 & \textbf{15915.2358} & 15917.5771\\[2pt]
\hline
& & & & \\[-4pt]
$\textbf{LensSrcDisk}_{\mathrm{HS30L50Disk}}$ & \textbf{55113.8123} & 55122.6322 & \textbf{55154.4289} & 55153.1476\\[2pt]
$\textbf{LensSrcDisk}_{\mathrm{ES30L50Disk}}$ & \textbf{15567.8761} & 15571.8895 & \textbf{15580.8484} & 15573.1699\\[2pt]
\hline
& & & & \\[-4pt]
$\textbf{LensSrcCusp}_{\mathrm{HS20L60Cusp}}$ & \textbf{54989.4958} & 54981.9373 & \textbf{55025.2533} & 55025.8970\\[2pt]
$\textbf{LensSrcCusp}_{\mathrm{ES20L60Cusp}}$ & \textbf{15111.1705} & 15101.0015 & \textbf{15137.4186} & 15137.7187\\[2pt]
\hline
& & & & \\[-4pt]
$\textbf{LensSrcDouble}_{\mathrm{HS25L60BD}}$ & \textbf{55450.3420} & 55426.1254 & \textbf{55478.9227} & 55480.2748\\[2pt]
$\textbf{LensSrcDouble}_{\mathrm{ES25L60BD}}$ & \textbf{15588.8998} & 15600.0188 & \textbf{15623.4948} & 15621.5500\\[2pt]
\hline
& & & & \\[-4pt]
$\textbf{LensSrcMulti}_{\mathrm{HS25L75Multi}}$ & \textbf{55487.6488} & 55491.7482 & \textbf{55522.6763} & 55522.9063\\[2pt]
$\textbf{LensSrcMulti}_{\mathrm{ES25L75Multi}}$ & \textbf{15609.1851} & 15576.4738 & \textbf{15629.4569} & 15633.4754\\[2pt]
\hline
& & & & \\[-4pt]
$\mathbf{LensMassShear}_{\mathrm{HS40L80Disk}}$ & \textbf{54300.2109} & 54314.5680 & 54528.1859 & \textbf{54722.6101}\\[2pt]
$\mathbf{LensMassShear}_{\mathrm{ES40L80Disk}}$ & \textbf{14787.7869} & 14797.0800 & 14841.6874 & \textbf{14951.8195}\\[2pt]
\hline
& & & & \\[-4pt]
$\textbf{LensLightBD}_{\mathrm{HS25L50BD}}$ & 55177.7344 & \textbf{55327.3654} & \textbf{55402.3462} & 55403.6771\\[2pt]
$\textbf{LensLightBD}_{\mathrm{ES25L50BD}}$ & 15319.9874 & \textbf{15407.3859} & \textbf{15487.9352} & 15488.5821\\[2pt]
\end{tabular}
\caption{The results of Bayesian model comparison in the phases $\textbf{P}_{\rm  LightMC}$ and $\textbf{P}_{\rm  MassMC}$ for the fourteen $\textbf{Lens}$ images, where the image's listed in the first column are generated using a variety of source morphologies, lens profile and mass models. The third and fourth columns show the Bayesian evidence values (equation \ref{eqn:Bayes2}) computed for the $Sersic$ and $Sersic$ + $Exp$ light models and the fifth and sixth columns the values for the $SPLE$ and $SPLE$+$Shear$ mass models. Values in bold correspond to those chosen by the pipeline, noting that a threshold of twenty must be exceeded to favour a more complex model.}
\label{table:SPLEFgMC}
\end{table*}

The results of the model comparison phases $\textbf{P}_{\rm  LightMC}$ and $\textbf{P}_{\rm  MassMC}$ for all images are given in table \ref{table:SPLEFgMC}. For all images, model comparison correctly chooses the input light and mass models. Therefore, at Euclid resolution or higher, multi-component light profiles and detection of an external shear are possible. Model comparison also never wrongly favours a more complex model, reaffirming that model comparison functions exactly as expected. For many of these comparisons the more complex model is an extension of the true model (e.g. all $SPLE$ + $Shear$ models can reproduce the input $SPLE$ model if $\gamma_{\rm  sh} = 0$), the scenario which acts as the most stringent test of model comparison.  For many model comparisons, the highest likelihood values found in the (rejected) more complex models were higher than the highest likelihoods of the simpler model, by values of approximately ln$\epsilon = 5-15$. This shows a maximum likelihood based approach is not well suited to determining the lens model complexity and demonstrates {\tt AutoLens}'s use of Occam's Razor.

\subsubsection{Modeling Results}\label{ResultsLensModel}

\begin{figure*}
\centering
\includegraphics[width=0.42\textwidth]{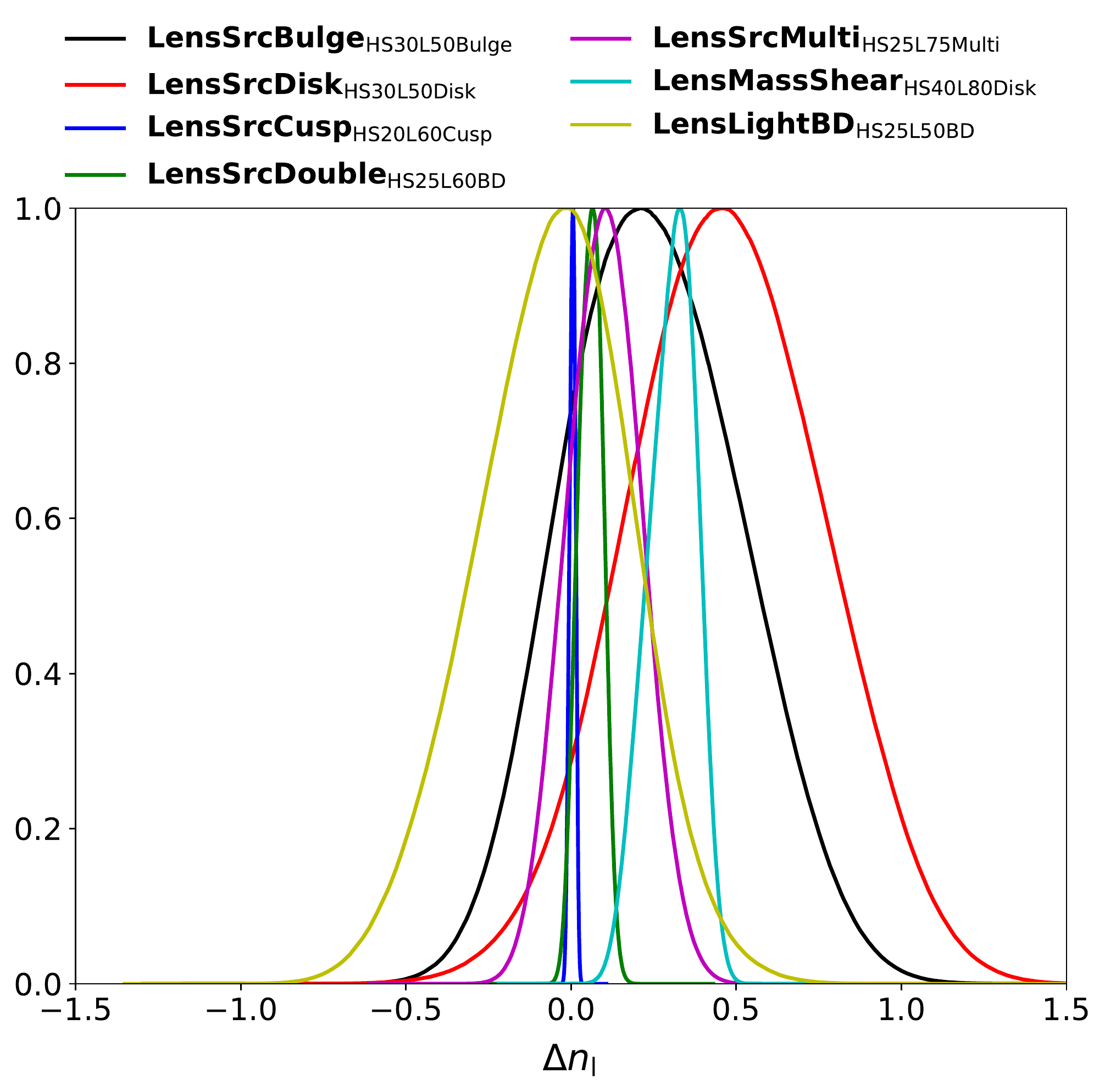}
\includegraphics[width=0.42\textwidth]{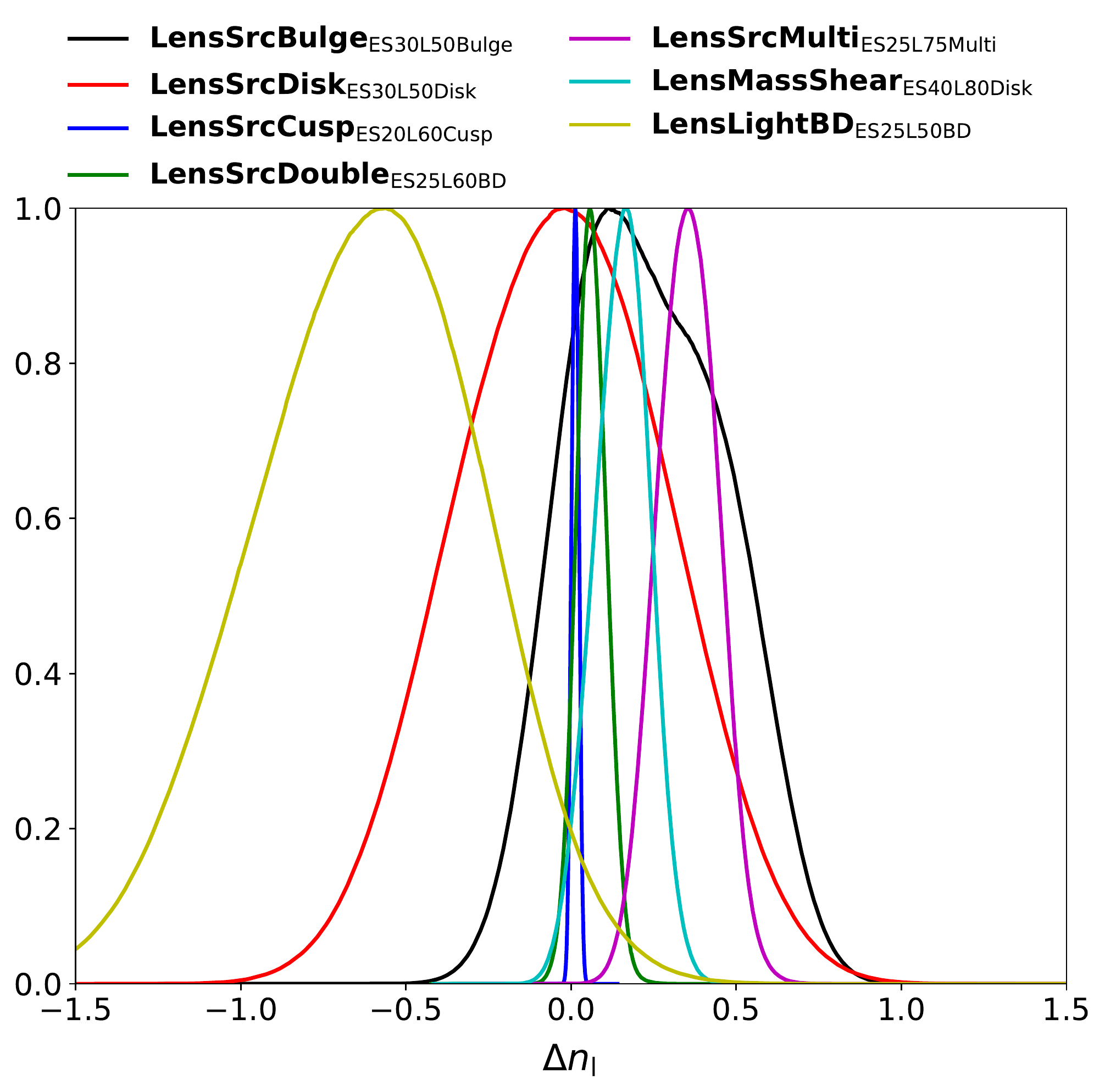}
\caption{Marginalized one-dimensional PDF's of $\Delta n_{\rm l}$ (or $\Delta n_{\rm l1}$ for images of the model $\textbf{LensLightBD}$) for the seven Hubble resolution (left) and seven Euclid resolution (right) images of the $\textbf{Lens}$ model images, corresponding to the results given in table \ref{table:TableSPLEFg}. Each graph's legend indicates the colour each line corresponds to. The input value of $n_{\rm l}$ for each model are give in table \ref{table:SimModels}. The majority of PDF's are consistent with the input lens model ($\Delta n_{\rm l} = 0.0$). The width of PDFs shows no trend with image S/N or resolution, instead it is the size of the lens galaxy which drives the precision of parameter estimates.} 
\label{figure:PDFsSPLEFg}
\end{figure*}

The modeling results for the non-cored simulation suite, given in table \ref{table:TableSPLEFg}, are positive, with all but one mass model parameter estimated incorrectly within $3 \sigma$ confidence and their estimates at $1 \sigma$ consistent with expectations. The incorrect mass model parameter is the density-slope $\alpha$ for the image $\mathbf{LensMassShear}_{\mathrm{ES40L80Disk}}$. Given the accurate parameter estimates for the equivalent high-resolution image, it appears that this image is simply too low resolution to accurately constrain the density-slope simultaneously with an external shear.

The majority of light model parameters are also estimated correctly within $3 \sigma$ confidence and estimates at $1 \sigma$ are again consistent with Gaussian expectations. However, there is a trend throughout the parameter estimates whereby the light model's $R_{\rm l}$ and $n_{\rm l}$ values are over-estimated compared to their true input value. In most cases, this is consistent within their errors (at $1 - 2 \sigma$ confidence), however there are five cases which are incorrect within $3 \sigma$. These offsets are most prevalent for simulated images which are: (i) generated with a low $R_{\rm l}$; (ii) generated with a high $n_{\rm l}$ and; (iii) generated at Euclid resolution. Thus, these offsets are most significant in cases where there is more blending between the source and lens light, especially when this occurs towards its central regions where the strongest constraints on $R_{\rm l}$ and $n_{\rm l}$ are placed. This is a limitation of any analysis which tries to deblend the lens from the source using single-waveband imaging without a sufficiently high resolution.

\begin{table*}
\resizebox{\linewidth}{!}{
\begin{tabular}{ l | l | l l l | l} 
\multicolumn{1}{p{1.6cm}|}{\textbf{Image}} 
& \multicolumn{1}{p{1.8cm}|}{\centering \textbf{Component}} 
& \multicolumn{1}{p{2.2cm}}{\textbf{Parameters ($3\sigma$)}} 
& \multicolumn{1}{p{2.2cm}}{}  
& \multicolumn{1}{p{2.2cm}|}{} 
& \multicolumn{1}{p{2.2cm}}{\textbf{Parameters ($1\sigma$)}} 
\\ \hline
$\textbf{LensSrcBulge}_{\mathrm{HS30L50Bulge}}$ & Light (\textbf{Sersic}) & $\Delta R_{\mathrm{l}}=0.1133^{+0.2837}_{\rm -0.2232}$($R_{\mathrm{l}}=0.60$) & $\Delta q_{\mathrm{l}}=-0.0039^{+0.0465}_{\rm -0.0442}$($q_{\mathrm{l}}=0.72$) & $\Delta n_{\mathrm{l}}=0.2463^{+0.7624}_{\rm -0.6917}$($n_{\mathrm{l}}=4.00$) & $\Delta n_{\mathrm{l}}=0.2463^{+0.2534}_{\rm -0.2852}$($n_{\mathrm{l}}=4.00$)\\[2pt]
$\textbf{LensSrcBulge}_{\mathrm{HS30L50Bulge}}$ & Mass (\textbf{SPLE}) & $\Delta \theta_{\mathrm{Ein}}=0.0039^{+0.0202}_{\rm -0.0151}$($\theta_{\mathrm{Ein}}=1.20$) & $\Delta q=-0.0098^{+0.0353}_{\rm -0.0433}$($q=0.80$) & $\Delta \alpha=0.0193^{+0.1005}_{\rm -0.0874}$($\alpha=2.00$) & $\Delta \alpha=0.0193^{+0.0300}_{\rm -0.0312}$($\alpha=2.00$)\\[2pt]
$\textbf{LensSrcBulge}_{\mathrm{ES30L50Bulge}}$ & Light (\textbf{Sersic}) & $\Delta R_{\mathrm{l}}=0.0786^{+0.1810}_{\rm -0.1523}$($R_{\mathrm{l}}=0.60$) & $\Delta q_{\mathrm{l}}=0.0060^{+0.0364}_{\rm -0.0387}$($q_{\mathrm{l}}=0.72$) & $\Delta n_{\mathrm{l}}=0.2296^{+0.6116}_{\rm -0.5765}$($n_{\mathrm{l}}=4.00$) & $\Delta n_{\mathrm{l}}=0.2296^{+0.2292}_{\rm -0.2637}$($n_{\mathrm{l}}=4.00$)\\[2pt]
$\textbf{LensSrcBulge}_{\mathrm{ES30L50Bulge}}$ & Mass (\textbf{SPLE}) & $\Delta \theta_{\mathrm{Ein}}=0.0052^{+0.0175}_{\rm -0.0151}$($\theta_{\mathrm{Ein}}=1.20$) & $\Delta q=-0.0078^{+0.0361}_{\rm -0.0417}$($q=0.80$) & $\Delta \alpha=0.0248^{+0.0917}_{\rm -0.0863}$($\alpha=2.00$) & $\Delta \alpha=0.0248^{+0.0302}_{\rm -0.0341}$($\alpha=2.00$)\\[-4pt]
& & & & & \\[-4pt]
\hline
& & & & & \\[-4pt]
$\textbf{LensSrcDisk}_{\mathrm{HS30L50Disk}}$ & Light (\textbf{Sersic}) & $\Delta R_{\mathrm{l}}=0.1939^{+0.3672}_{\rm -0.3250}$($R_{\mathrm{l}}=0.60$) & $\Delta q_{\mathrm{l}}=-0.0114^{+0.0563}_{\rm -0.0572}$($q_{\mathrm{l}}=0.72$) & $\Delta n_{\mathrm{l}}=0.4707^{+0.8360}_{\rm -0.8643}$($n_{\mathrm{l}}=4.00$) & $\mathbf{\Delta n_{\mathrm{l}}=0.4707^{+0.2931}_{\rm -0.3025}}$($n_{\mathrm{l}}=4.00$)\\[2pt]
$\textbf{LensSrcDisk}_{\mathrm{HS30L50Disk}}$ & Mass (\textbf{SPLE}) & $\Delta \theta_{\mathrm{Ein}}=0.0097^{+0.0323}_{\rm -0.0280}$($\theta_{\mathrm{Ein}}=1.20$) & $\Delta q=-0.0189^{+0.0645}_{\rm -0.0685}$($q=0.75$) & $\Delta \alpha=0.0347^{+0.0988}_{\rm -0.1009}$($\alpha=2.30$) & $\Delta \alpha=0.0347^{+0.0368}_{\rm -0.0356}$($\alpha=2.30$)\\[2pt]
$\textbf{LensSrcDisk}_{\mathrm{ES30L50Disk}}$ & Light (\textbf{Sersic}) & $\Delta R_{\mathrm{l}}=-0.0029^{+0.2580}_{\rm -0.2306}$($R_{\mathrm{l}}=0.60$) & $\Delta q_{\mathrm{l}}=-0.0064^{+0.0519}_{\rm -0.0541}$($q_{\mathrm{l}}=0.72$) & $\Delta n_{\mathrm{l}}=-0.0257^{+0.8801}_{\rm -0.8826}$($n_{\mathrm{l}}=4.00$) & $\Delta n_{\mathrm{l}}=-0.0257^{+0.3230}_{\rm -0.3230}$($n_{\mathrm{l}}=4.00$)\\[2pt]
$\textbf{LensSrcDisk}_{\mathrm{ES30L50Disk}}$ & Mass (\textbf{SPLE}) & $\Delta \theta_{\mathrm{Ein}}=-0.0030^{+0.0353}_{\rm -0.0284}$($\theta_{\mathrm{Ein}}=1.20$) & $\Delta q=0.0108^{+0.0745}_{\rm -0.0827}$($q=0.75$) & $\Delta \alpha=-0.0331^{+0.1406}_{\rm -0.1345}$($\alpha=2.30$) & $\Delta \alpha=-0.0331^{+0.0507}_{\rm -0.0559}$($\alpha=2.30$)\\[-4pt]
& & & & & \\[-4pt]
\hline
& & & & & \\[-4pt]
$\textbf{LensSrcCusp}_{\mathrm{HS20L60Cusp}}$ & Light (\textbf{Sersic}) & $\Delta R_{\mathrm{l}}=0.0151^{+0.0246}_{\rm -0.0244}$($R_{\mathrm{l}}=1.20$) & $\Delta q_{\mathrm{l}}=0.0009^{+0.0060}_{\rm -0.0061}$($q_{\mathrm{l}}=0.60$) & $\Delta n_{\mathrm{l}}=0.0043^{+0.0233}_{\rm -0.0253}$($n_{\mathrm{l}}=1.25$) & $\Delta n_{\mathrm{l}}=0.0043^{+0.0096}_{\rm -0.0086}$($n_{\mathrm{l}}=1.25$)\\[2pt]
$\textbf{LensSrcCusp}_{\mathrm{HS20L60Cusp}}$ & Mass (\textbf{SPLE}) & $\Delta \theta_{\mathrm{Ein}}=0.0279^{+0.0477}_{\rm -0.0403}$($\theta_{\mathrm{Ein}}=1.40$) & $\Delta q=-0.0409^{+0.0599}_{\rm -0.0664}$($q=0.70$) & $\Delta \alpha=0.0694^{+0.0833}_{\rm -0.0884}$($\alpha=2.35$) & $\mathbf{\Delta \alpha=0.0694^{+0.0324}_{\rm -0.0302}}$($\alpha=2.35$)\\[2pt]
$\textbf{LensSrcCusp}_{\mathrm{ES20L60Cusp}}$ & Light (\textbf{Sersic}) & $\Delta R_{\mathrm{l}}=0.0240^{+0.0281}_{\rm -0.0312}$($R_{\mathrm{l}}=1.20$) & $\Delta q_{\mathrm{l}}=0.0113^{+0.0084}_{\rm -0.0145}$($q_{\mathrm{l}}=0.60$) & $\Delta n_{\mathrm{l}}=0.0125^{+0.0297}_{\rm -0.0300}$($n_{\mathrm{l}}=1.25$) & $\mathbf{\Delta n_{\mathrm{l}}=0.0125^{+0.0110}_{\rm -0.0108}}$($n_{\mathrm{l}}=1.25$)\\[2pt]
$\textbf{LensSrcCusp}_{\mathrm{ES20L60Cusp}}$ & Mass (\textbf{SPLE}) & $\Delta \theta_{\mathrm{Ein}}=-0.0137^{+0.0617}_{\rm -0.0422}$($\theta_{\mathrm{Ein}}=1.40$) & $\Delta q=0.0298^{+0.0771}_{\rm -0.0956}$($q=0.70$) & $\Delta \alpha=-0.0217^{+0.1690}_{\rm -0.1687}$($\alpha=2.35$) & $\Delta \alpha=-0.0217^{+0.0555}_{\rm -0.0590}$($\alpha=2.35$)\\[-4pt]
& & & & & \\[-4pt]
\hline
& & & & & \\[-4pt]
$\textbf{LensSrcDouble}_{\mathrm{HS25L60BD}}$ & Light (\textbf{Sersic}) & $\mathbf{\Delta R_{\mathrm{l}}=0.0811^{+0.0776}_{\rm -0.0741}}$($R_{\mathrm{l}}=0.90$) & $\Delta q_{\mathrm{l}}=0.0081^{+0.0226}_{\rm -0.0237}$($q_{\mathrm{l}}=0.80$) & $\Delta n_{\mathrm{l}}=0.0589^{+0.0997}_{\rm -0.1002}$($n_{\mathrm{l}}=2.00$) & $\mathbf{\Delta n_{\mathrm{l}}=0.0589^{+0.0389}_{\rm -0.0364}}$($n_{\mathrm{l}}=2.00$)\\[2pt]
$\textbf{LensSrcDouble}_{\mathrm{HS25L60BD}}$ & Mass (\textbf{SPLE}) & $\Delta \theta_{\mathrm{Ein}}=0.0119^{+0.0220}_{\rm -0.0226}$($\theta_{\mathrm{Ein}}=1.20$) & $\Delta q=-0.0228^{+0.0540}_{\rm -0.0498}$($q=0.75$) & $\Delta \alpha=0.0576^{+0.0937}_{\rm -0.1001}$($\alpha=2.10$) & $\mathbf{\Delta \alpha=0.0576^{+0.0349}_{\rm -0.0328}}$($\alpha=2.10$)\\[2pt]
$\textbf{LensSrcDouble}_{\mathrm{ES25L60BD}}$ & Light (\textbf{Sersic}) & $\Delta R_{\mathrm{l}}=0.0828^{+0.1065}_{\rm -0.0947}$($R_{\mathrm{l}}=0.90$) & $\Delta q_{\mathrm{l}}=0.0122^{+0.0261}_{\rm -0.0250}$($q_{\mathrm{l}}=0.80$) & $\Delta n_{\mathrm{l}}=0.0624^{+0.1483}_{\rm -0.1398}$($n_{\mathrm{l}}=2.00$) & $\mathbf{\Delta n_{\mathrm{l}}=0.0624^{+0.0444}_{\rm -0.0481}}$($n_{\mathrm{l}}=2.00$)\\[2pt]
$\textbf{LensSrcDouble}_{\mathrm{ES25L60BD}}$ & Mass (\textbf{SPLE}) & $\Delta \theta_{\mathrm{Ein}}=0.0160^{+0.0319}_{\rm -0.0253}$($\theta_{\mathrm{Ein}}=1.20$) & $\Delta q=-0.0303^{+0.0629}_{\rm -0.0668}$($q=0.75$) & $\Delta \alpha=0.0676^{+0.1244}_{\rm -0.1132}$($\alpha=2.10$) & $\mathbf{\Delta \alpha=0.0676^{+0.0378}_{\rm -0.0421}}$($\alpha=2.10$)\\[-4pt]
& & & & & \\[-4pt]
\hline
& & & & & \\[-4pt]
$\textbf{LensSrcMulti}_{\mathrm{HS25L75Multi}}$ & Light (\textbf{Sersic}) & $\Delta R_{\mathrm{l}}=0.0633^{+0.1790}_{\rm -0.1578}$($R_{\mathrm{l}}=0.80$) & $\Delta q_{\mathrm{l}}=0.0086^{+0.0220}_{\rm -0.0227}$($q_{\mathrm{l}}=0.70$) & $\Delta n_{\mathrm{l}}=0.1005^{+0.3237}_{\rm -0.3123}$($n_{\mathrm{l}}=3.00$) & $\Delta n_{\mathrm{l}}=0.1005^{+0.1137}_{\rm -0.1136}$($n_{\mathrm{l}}=3.00$)\\[2pt]
$\textbf{LensSrcMulti}_{\mathrm{HS25L75Multi}}$ & Mass (\textbf{SPLE}) & $\Delta \theta_{\mathrm{Ein}}=0.0009^{+0.0141}_{\rm -0.0128}$($\theta_{\mathrm{Ein}}=1.20$) & $\Delta q=0.0011^{+0.0262}_{\rm -0.0278}$($q=0.75$) & $\Delta \alpha=0.0022^{+0.0626}_{\rm -0.0595}$($\alpha=2.10$) & $\Delta \alpha=0.0022^{+0.0210}_{\rm -0.0218}$($\alpha=2.10$)\\[2pt]
$\textbf{LensSrcMulti}_{\mathrm{ES25L75Multi}}$ & Light (\textbf{Sersic}) & $\mathbf{\Delta R_{\mathrm{l}}=0.1998^{+0.1452}_{\rm -0.1379}}$($R_{\mathrm{l}}=0.80$) & $\Delta q_{\mathrm{l}}=-0.0135^{+0.0183}_{\rm -0.0191}$($q_{\mathrm{l}}=0.70$) & $\mathbf{\Delta n_{\mathrm{l}}=0.3531^{+0.2701}_{\rm -0.2601}}$($n_{\mathrm{l}}=3.00$) & $\mathbf{\Delta n_{\mathrm{l}}=0.3531^{+0.0937}_{\rm -0.0915}}$($n_{\mathrm{l}}=3.00$)\\[2pt]
$\textbf{LensSrcMulti}_{\mathrm{ES25L75Multi}}$ & Mass (\textbf{SPLE}) & $\Delta \theta_{\mathrm{Ein}}=0.0153^{+0.0330}_{\rm -0.0375}$($\theta_{\mathrm{Ein}}=1.20$) & $\Delta q=-0.0278^{+0.0744}_{\rm -0.0588}$($q=0.75$) & $\Delta \alpha=0.0389^{+0.1184}_{\rm -0.1595}$($\alpha=2.10$) & $\mathbf{\Delta \alpha=0.0389^{+0.0476}_{\rm -0.0310}}$($\alpha=2.10$)\\[-4pt]
& & & & & \\[-4pt]
\hline
& & & & & \\[-4pt]
$\mathbf{LensMassShear}_{\mathrm{HS40L80Disk}}$ & Light (\textbf{Sersic}) & $\Delta R_{\mathrm{l}}=0.0253^{+0.0433}_{\rm -0.0674}$($R_{\mathrm{l}}=1.50$) & $\Delta q_{\mathrm{l}}=-0.0034^{+0.0083}_{\rm -0.0073}$($q_{\mathrm{l}}=0.60$) & $\Delta n_{\mathrm{l}}=0.0321^{+0.0773}_{\rm -0.0633}$($n_{\mathrm{l}}=2.50$) & $\Delta n_{\mathrm{l}}=0.0321^{+0.0437}_{\rm -0.0412}$($n_{\mathrm{l}}=2.50$)\\[2pt]
$\mathbf{LensMassShear}_{\mathrm{HS40L80Disk}}$ & Mass (\textbf{SPLE}) & $\Delta \theta_{\mathrm{Ein}}=0.0031^{+0.0050}_{\rm -0.0040}$($\theta_{\mathrm{Ein}}=1.15$) & $\Delta q=-0.0093^{+0.0156}_{\rm -0.0160}$($q=0.95$) & $\Delta \alpha=0.0467^{+0.0754}_{\rm -0.0732}$($\alpha=1.92$) & $\mathbf{\Delta \alpha=0.0467^{+0.0165}_{\rm -0.0197}}$($\alpha=1.92$)\\[2pt]
$\mathbf{LensMassShear}_{\mathrm{HS40L80Disk}}$ & Mass (\textbf{Shear}) & $\Delta \theta_{\mathrm{sh}}=-1.3846^{+3.1279}_{\rm -2.4625}$($\theta_{\mathrm{sh}}=40.00$) & $\Delta \gamma_{\mathrm{sh}}=0.0019^{+0.0044}_{\rm -0.0055}$($\gamma_{\mathrm{sh}}=0.03$) &  & $\mathbf{\Delta \gamma_{\mathrm{sh}}=0.0019^{+0.0025}_{\rm -0.0018}}$($\gamma_{\mathrm{sh}}=0.03$)\\[2pt]
$\mathbf{LensMassShear}_{\mathrm{ES40L80Disk}}$ & Light (\textbf{Sersic}) & $\mathbf{\Delta R_{\mathrm{l}}=0.1845^{+0.1433}_{\rm -0.1416}}$($R_{\mathrm{l}}=1.50$) & $\Delta q_{\mathrm{l}}=-0.0029^{+0.0175}_{\rm -0.0177}$($q_{\mathrm{l}}=0.60$) & $\Delta n_{\mathrm{l}}=0.0777^{+0.1237}_{\rm -0.1254}$($n_{\mathrm{l}}=2.50$) & $\mathbf{\Delta n_{\mathrm{l}}=0.0777^{+0.0437}_{\rm -0.0412}}$($n_{\mathrm{l}}=2.50$)\\[2pt]
$\mathbf{LensMassShear}_{\mathrm{ES40L80Disk}}$ & Mass (\textbf{SPLE}) & $\Delta \theta_{\mathrm{Ein}}=0.0040^{+0.0060}_{\rm -0.0045}$($\theta_{\mathrm{Ein}}=1.15$) & $\Delta q=-0.0101^{+0.0262}_{\rm -0.0270}$($q=0.95$) & $\mathbf{\Delta \alpha=0.2562^{+0.1829}_{\rm -0.1761}}$($\alpha=1.92$) & $\mathbf{\Delta \alpha=0.2562^{+0.0573}_{\rm -0.0589}}$($\alpha=1.92$)\\[2pt]
$\mathbf{LensMassShear}_{\mathrm{ES40L80Disk}}$ & Mass (\textbf{Shear}) & $\Delta \theta_{\mathrm{sh}}=-3.8668^{+5.0033}_{\rm -4.3041}$($\theta_{\mathrm{sh}}=40.00$) & $\Delta \gamma_{\mathrm{sh}}=0.0049^{+0.0064}_{\rm -0.0068}$($\gamma_{\mathrm{sh}}=0.03$) &  & $\mathbf{\Delta \gamma_{\mathrm{sh}}=0.0049^{+0.0025}_{\rm -0.0023}}$($\gamma_{\mathrm{sh}}=0.03$)\\[-4pt]
& & & & & \\[-4pt]
\hline
& & & & & \\[-4pt]
$\textbf{LensLightBD}_{\mathrm{HS25L50BD}}$ & Light (\textbf{Sersic}) & $\Delta R_{\mathrm{l1}}=0.0202^{+0.2174}_{\rm -0.2099}$($R_{\mathrm{l1}}=0.40$) & $\Delta q_{\mathrm{l1}}=0.0064^{+0.0902}_{\rm -0.0827}$($q_{\mathrm{l1}}=0.74$) & $\Delta n_{\mathrm{l1}}=-0.0535^{+0.6778}_{\rm -0.6973}$($n_{\mathrm{l1}}=3.00$) & $\Delta n_{\mathrm{l1}}=-0.0535^{+0.2493}_{\rm -0.2215}$($n_{\mathrm{l1}}=3.00$)\\[2pt]
$\textbf{LensLightBD}_{\mathrm{HS25L50BD}}$ & Light (\textbf{Exp}) & $\mathbf{\Delta R_{\mathrm{l2}}=0.0849^{+0.0783}_{\rm -0.0681}}$($R_{\mathrm{l2}}=1.15$) & $\Delta q_{\mathrm{l2}}=-0.0112^{+0.0360}_{\rm -0.0343}$($q_{\mathrm{l2}}=0.80$) &  & $\mathbf{\Delta R_{\mathrm{l2}}=0.0849^{+0.0218}_{\rm -0.0235}}$($R_{\mathrm{l2}}=1.15$)\\[2pt]
$\textbf{LensLightBD}_{\mathrm{HS25L50BD}}$ & Mass (\textbf{SPLE}) & $\Delta \theta_{\mathrm{Ein}}=0.0092^{+0.0190}_{\rm -0.0148}$($\theta_{\mathrm{Ein}}=1.20$) & $\Delta q=-0.0173^{+0.0397}_{\rm -0.0408}$($q=0.80$) & $\Delta \alpha=0.0448^{+0.0980}_{\rm -0.0850}$($\alpha=2.05$) & $\mathbf{\Delta \alpha=0.0448^{+0.0270}_{\rm -0.0347}}$($\alpha=2.05$)\\[2pt]
$\textbf{LensLightBD}_{\mathrm{ES25L50BD}}$ & Light (\textbf{Sersic}) & $\Delta R_{\mathrm{l1}}=-0.1312^{+0.1404}_{\rm -0.1260}$($R_{\mathrm{l1}}=0.40$) & $\Delta q_{\mathrm{l1}}=-0.0197^{+0.1134}_{\rm -0.1045}$($q_{\mathrm{l1}}=0.74$) & $\Delta n_{\mathrm{l1}}=-0.6174^{+0.9480}_{\rm -1.0253}$($n_{\mathrm{l1}}=3.00$) & $\mathbf{\Delta n_{\mathrm{l1}}=-0.6174^{+0.3687}_{\rm -0.3285}}$($n_{\mathrm{l1}}=3.00$)\\[2pt]
$\textbf{LensLightBD}_{\mathrm{ES25L50BD}}$ & Light (\textbf{Exp}) & $\Delta R_{\mathrm{l2}}=0.0618^{+0.0713}_{\rm -0.0636}$($R_{\mathrm{l2}}=1.15$) & $\Delta q_{\mathrm{l2}}=0.0002^{+0.0363}_{\rm -0.0307}$($q_{\mathrm{l2}}=0.80$) &  & $\mathbf{\Delta R_{\mathrm{l2}}=0.0618^{+0.0219}_{\rm -0.0225}}$($R_{\mathrm{l2}}=1.15$)\\[2pt]
$\textbf{LensLightBD}_{\mathrm{ES25L50BD}}$ & Mass (\textbf{SPLE}) & $\Delta \theta_{\mathrm{Ein}}=0.0092^{+0.0330}_{\rm -0.0280}$($\theta_{\mathrm{Ein}}=1.20$) & $\Delta q=-0.0190^{+0.0654}_{\rm -0.0744}$($q=0.80$) & $\Delta \alpha=0.0372^{+0.1628}_{\rm -0.1654}$($\alpha=2.05$) & $\Delta \alpha=0.0372^{+0.0595}_{\rm -0.0646}$($\alpha=2.05$)\\[-4pt]
\end{tabular}
}
\caption{Results of fitting the fourteen $\textbf{Lens}$ model images using the automated analysis pipeline, corresponding to results generated at end of phase two of the main pipeline. Each image's name is given in the first column and the light or mass model component in the second column. The third to sixth columns show parameter estimates, where each parameter is offset by $\Delta P = P_{\rm  true} - P_{\rm  model}$, such that zero corresponds to the input lens model. The input lens model values are given in brackets to the right of each parameter estimate. Parameters estimates are shown using $\Delta R_{\rm l}$, $\Delta q_{\rm l}$ and $\Delta n_{\rm l}$ for $Sersic$ light models, $\Delta R_{\rm l1}$, $\Delta q_{\rm l1}$, $\Delta n_{\rm n1}$, $\Delta R_{\rm l2}$ and $\Delta q_{\rm l2}$ for $Sersic$ + $Exp$ light models, $\Delta \theta_{\rm  Ein} $, $\Delta q$ and $\Delta \alpha$ for $SPLE$ mass models and $\Delta \theta_{\rm sh}$ and $\Delta \gamma_{sh}$ for a $Shear$ component. Columns three to five show parameter estimates within $3 \sigma$ confidence and column six at $1 \sigma$. Parameter estimates in bold text are inconsistent with the input lens model at their stated error estimates. The other parameters not shown (e.g. $x_{\rm l}$, $\theta_{\rm l}$, $\theta$) are all estimated accurately within $3\sigma$ or above.}
\label{table:TableSPLEFg}
\end{table*}

\begin{figure*}
\centering
\includegraphics[width=0.42\textwidth]{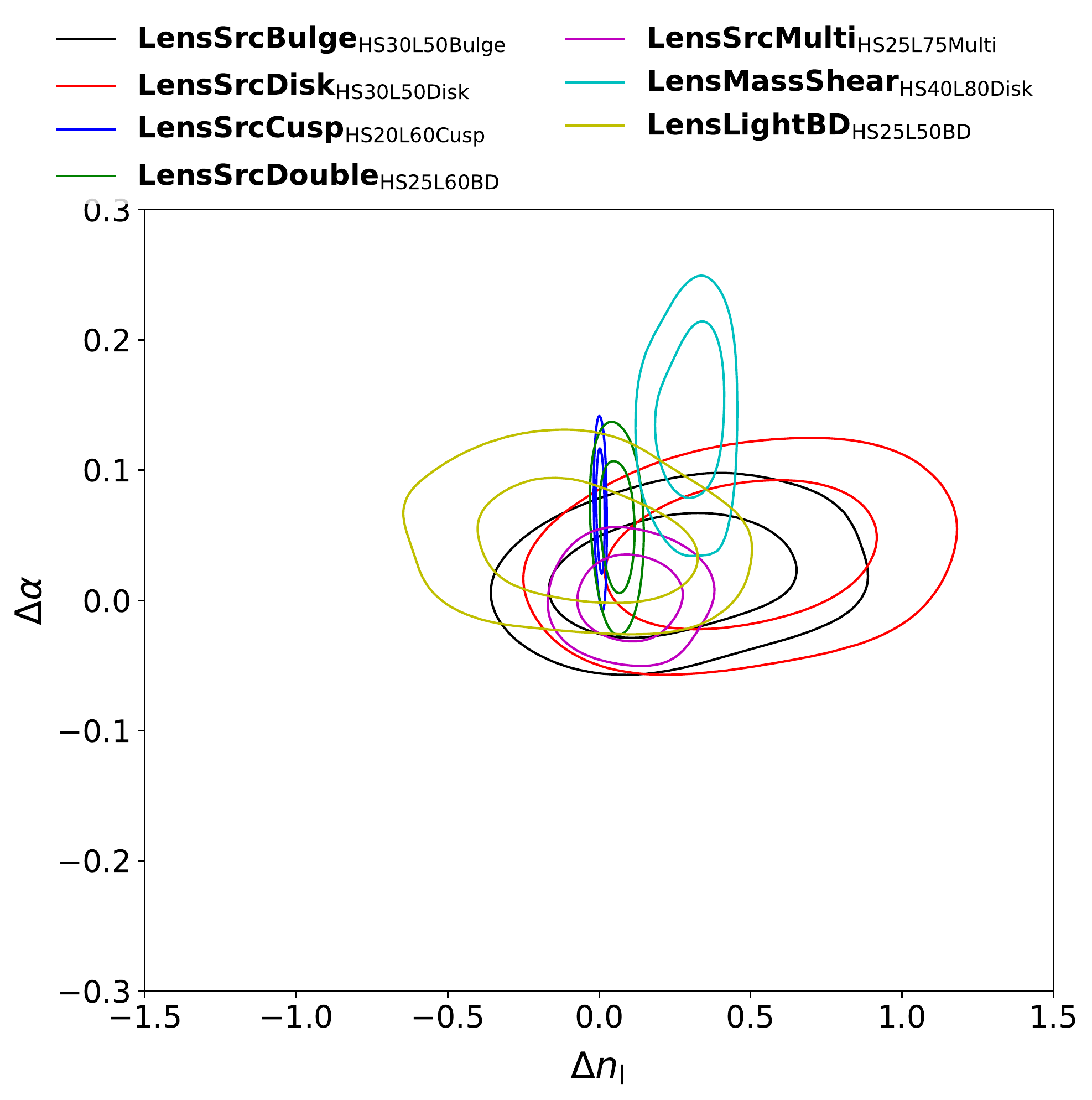}
\includegraphics[width=0.42\textwidth]{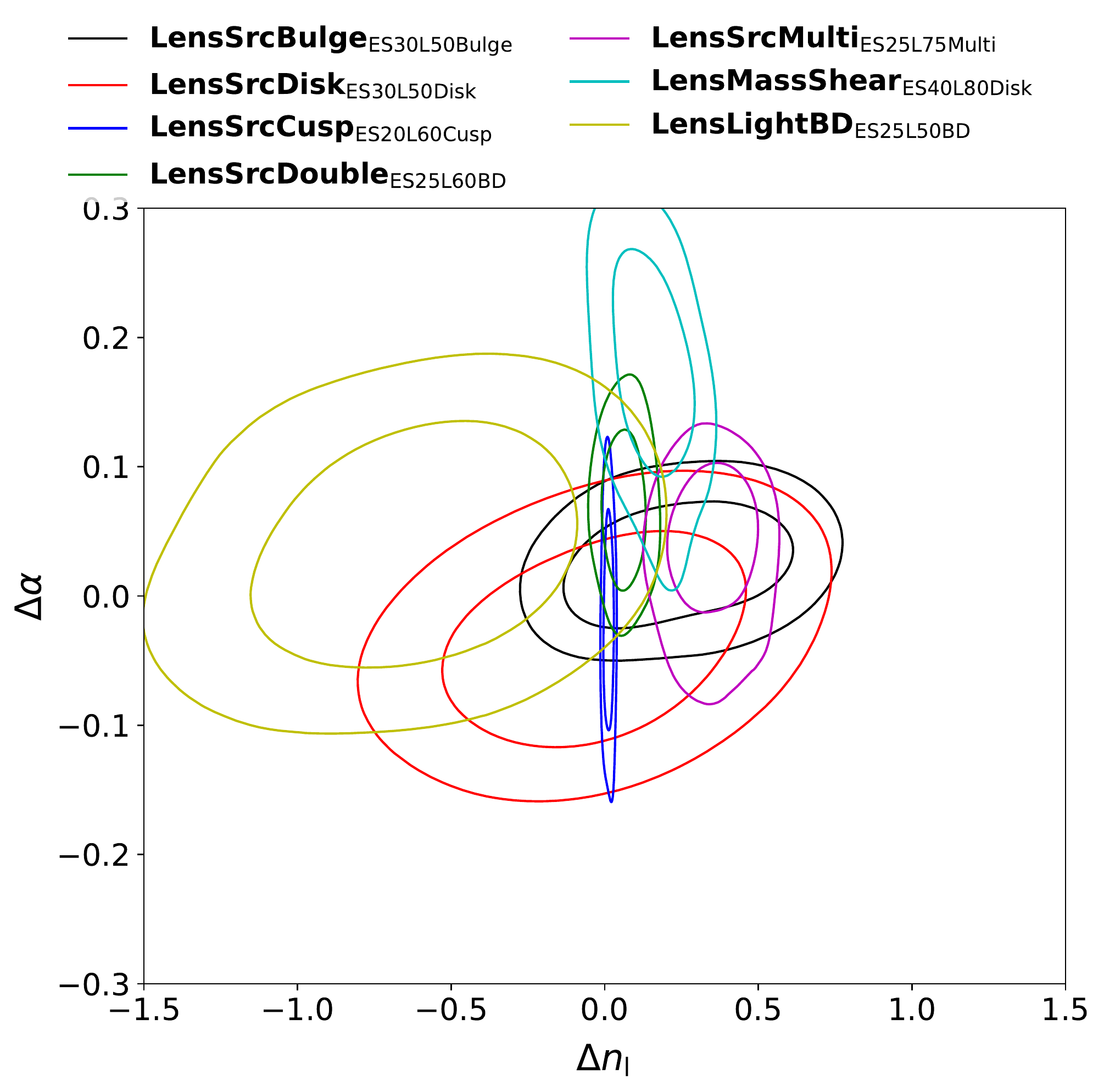}
\caption{Marginalized two-dimensional PDF's of $\Delta n_{\rm l}$ versus $\Delta \alpha$ for the Hubble resolution (left panels) and Euclid resolution (right panels) models of the $\textbf{Lens}$ images. Legends at the top of each panel indicate the image that each line corresponds too. Contours give the $1\sigma$ (interior) and $3 \sigma$ (exterior) confidence regions. The input value of $n_{\rm l}$ and $\alpha$ for each model are give in table \ref{table:SimModels}. Little to no degeneracy is observed in any of the PDFs, demonstrating that the mass and light models are essentially independent of one another.} 
\label{figure:PDFsLensSrc2D}
\end{figure*}

\begin{figure*}
\centering
\includegraphics[width=0.157\textwidth]{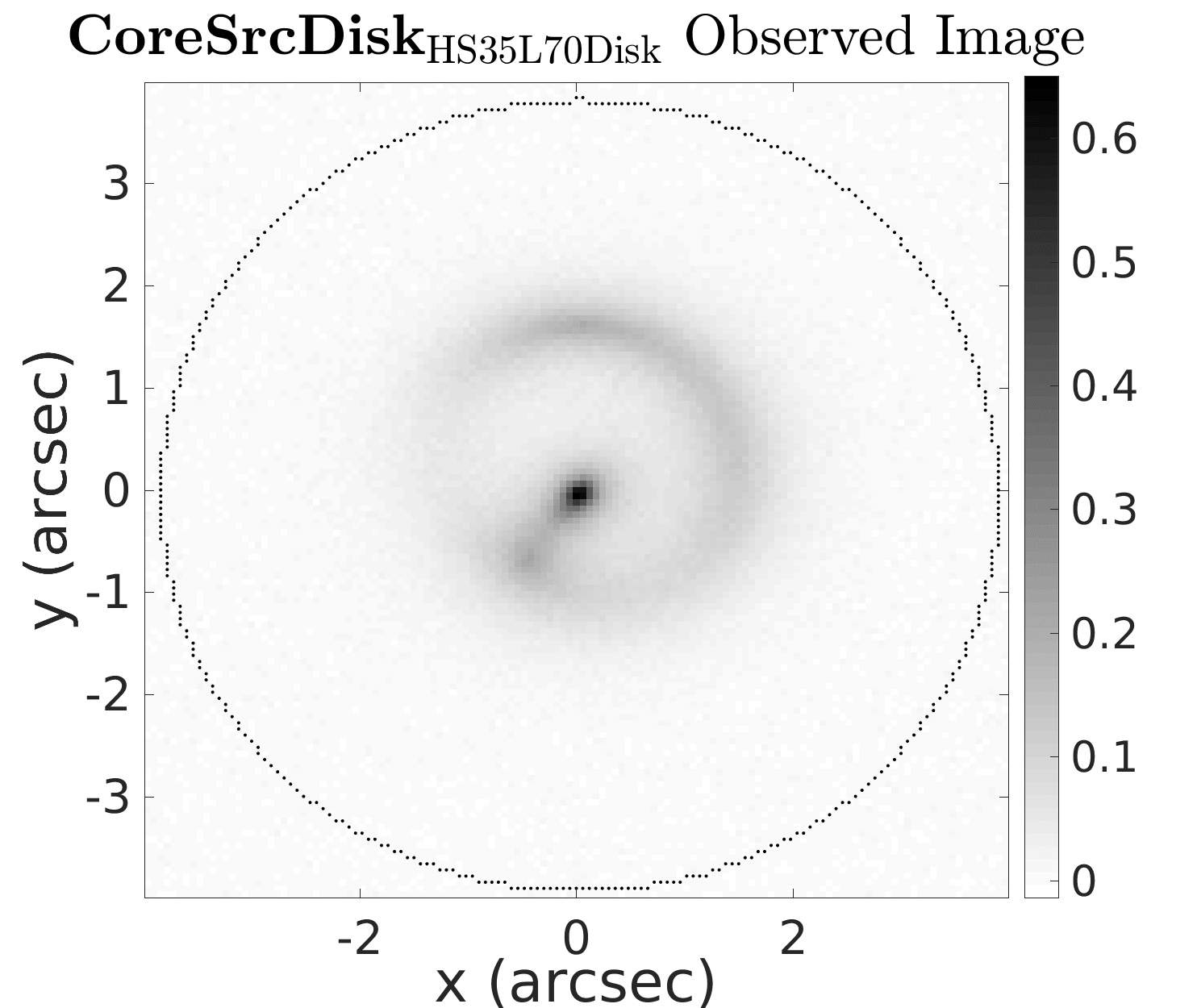}
\includegraphics[width=0.157\textwidth]{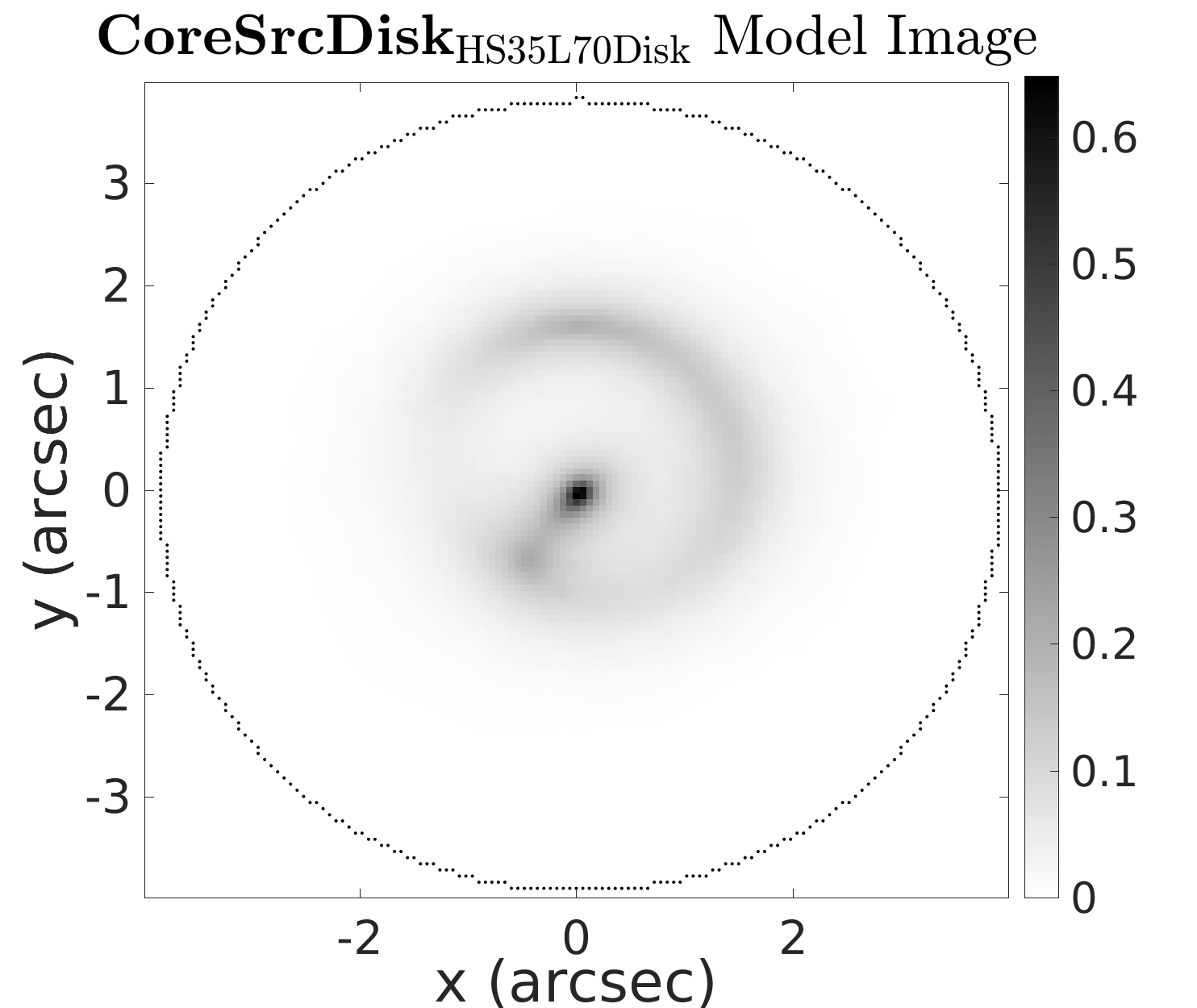}
\includegraphics[width=0.157\textwidth]{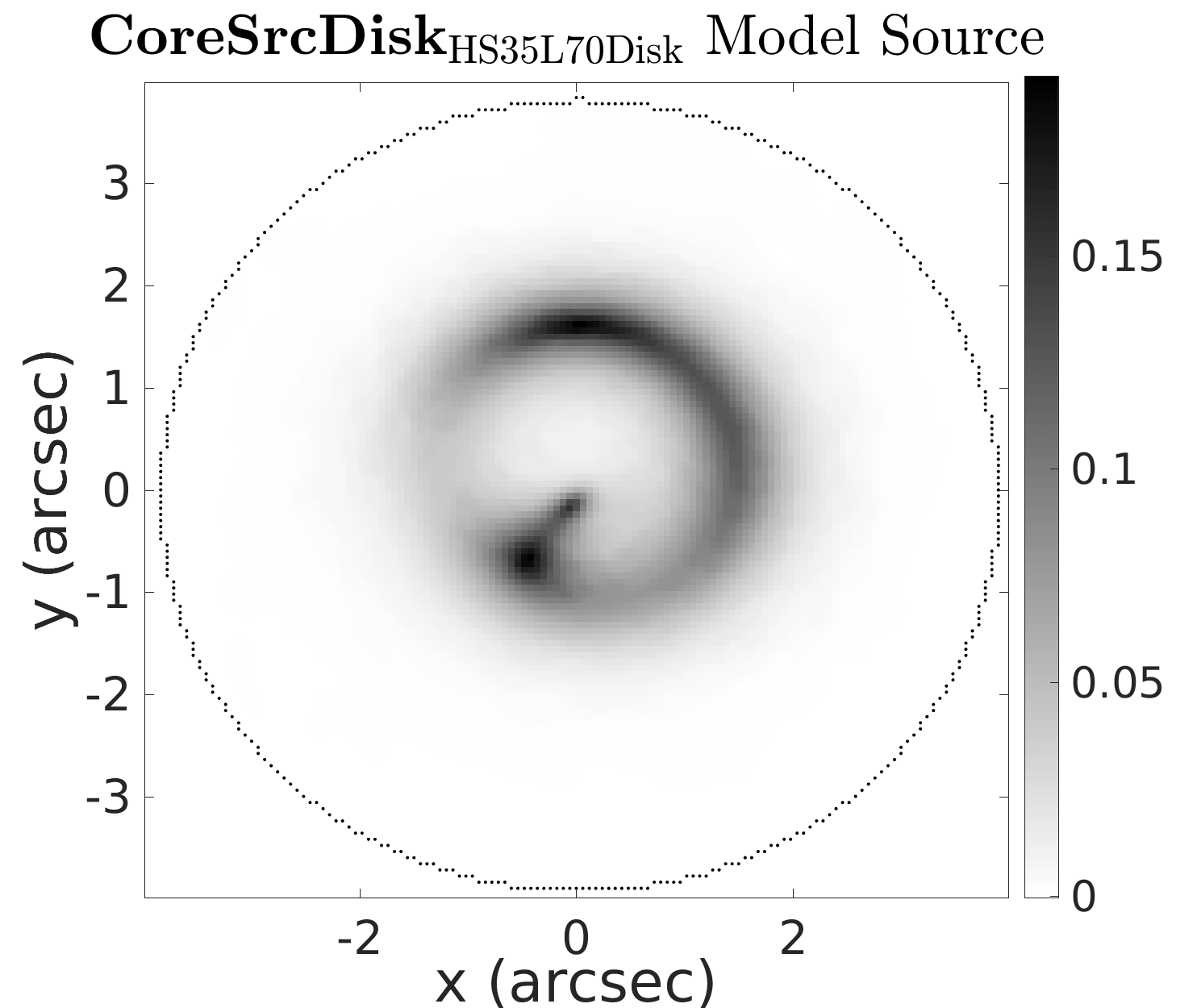}
\includegraphics[width=0.157\textwidth]{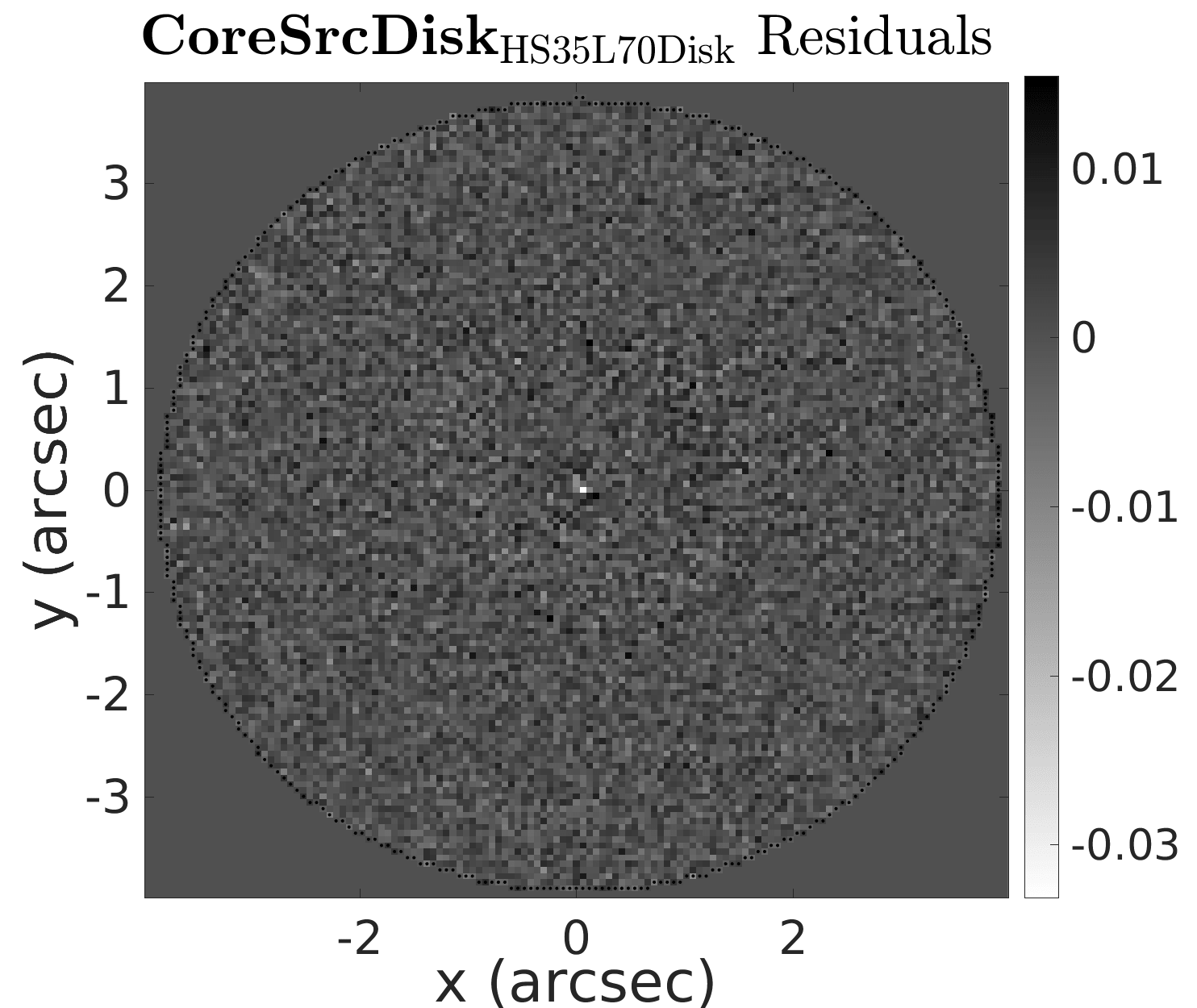}
\includegraphics[width=0.157\textwidth]{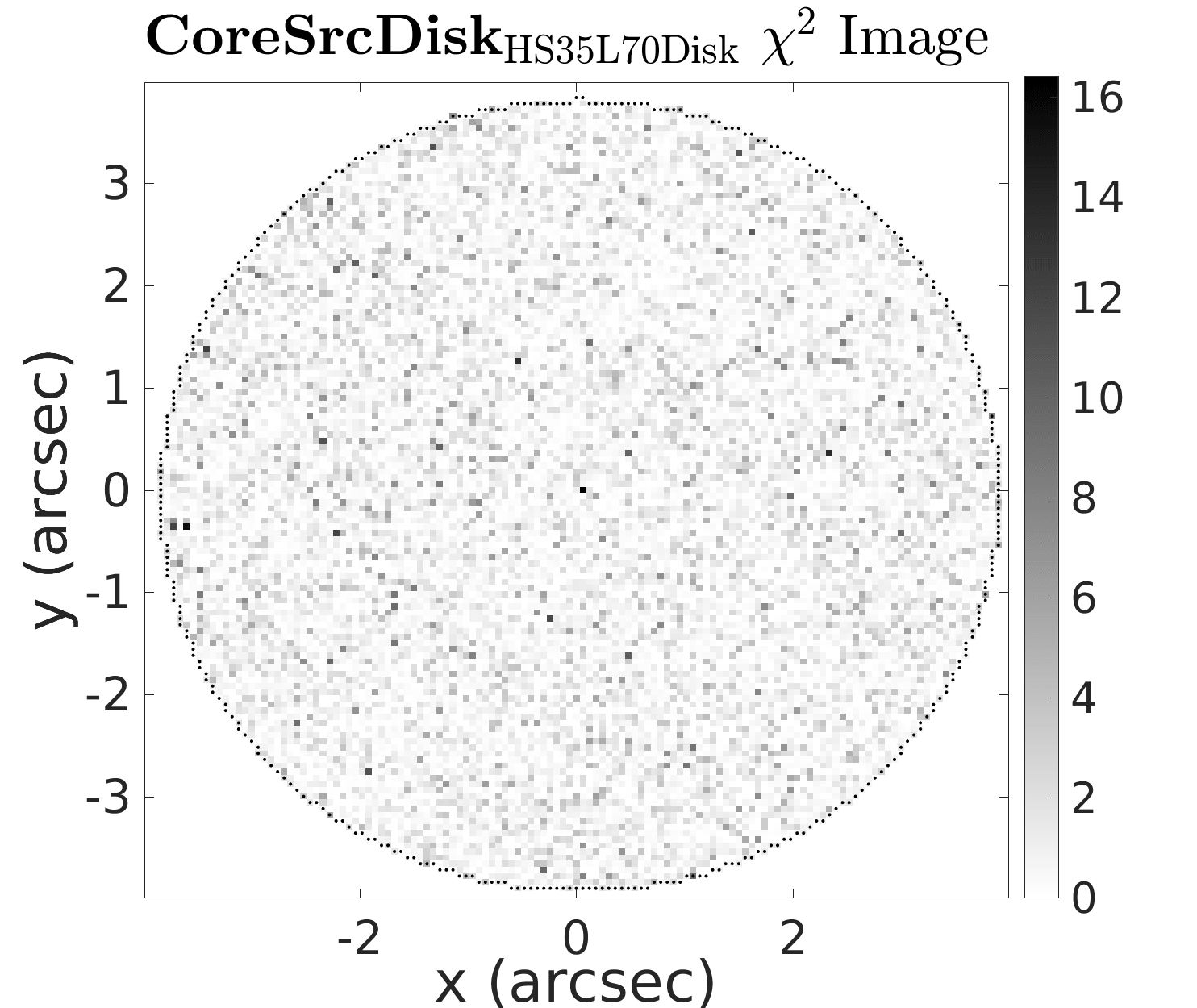}
\includegraphics[width=0.157\textwidth]{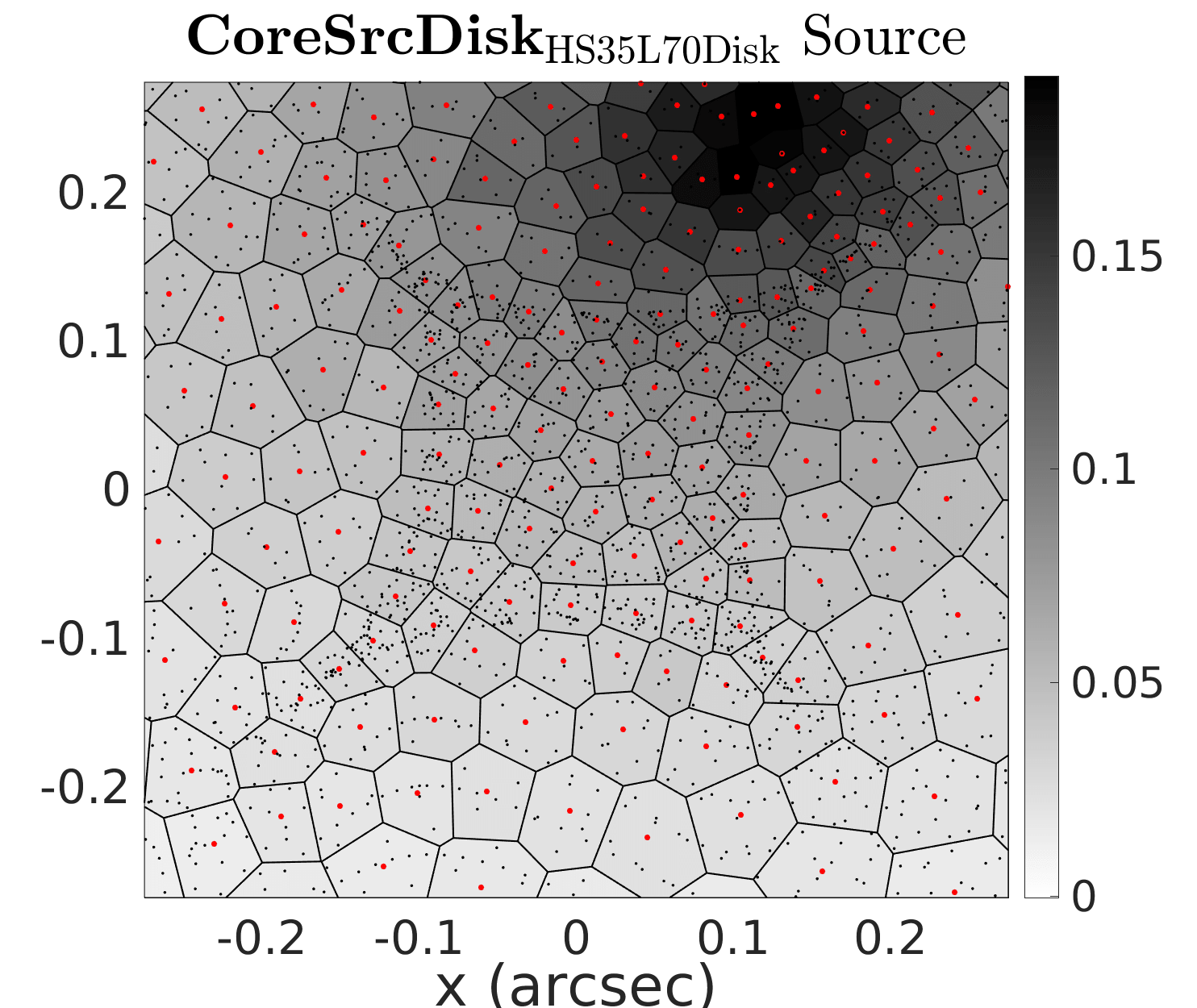}
\includegraphics[width=0.157\textwidth]{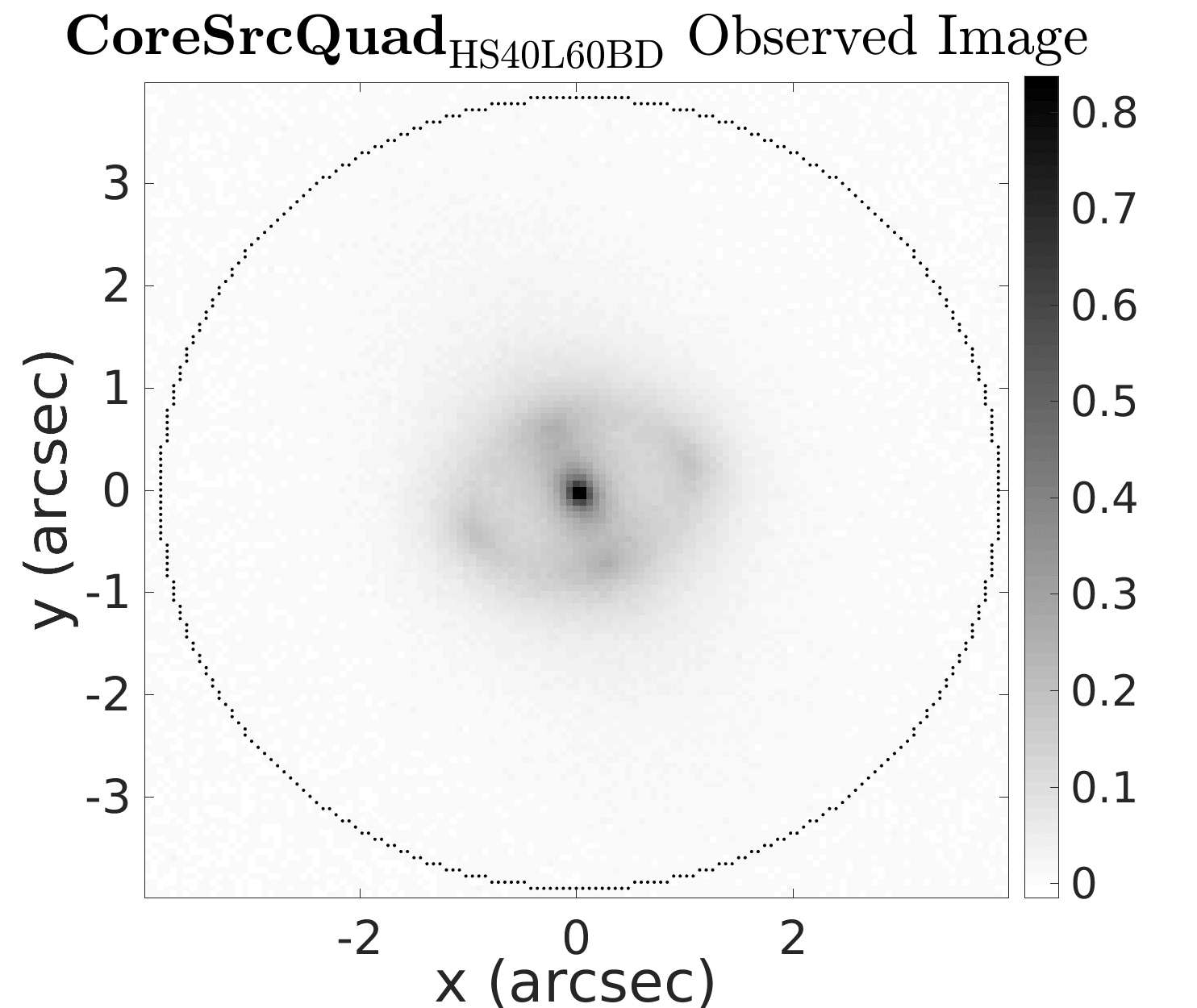}
\includegraphics[width=0.157\textwidth]{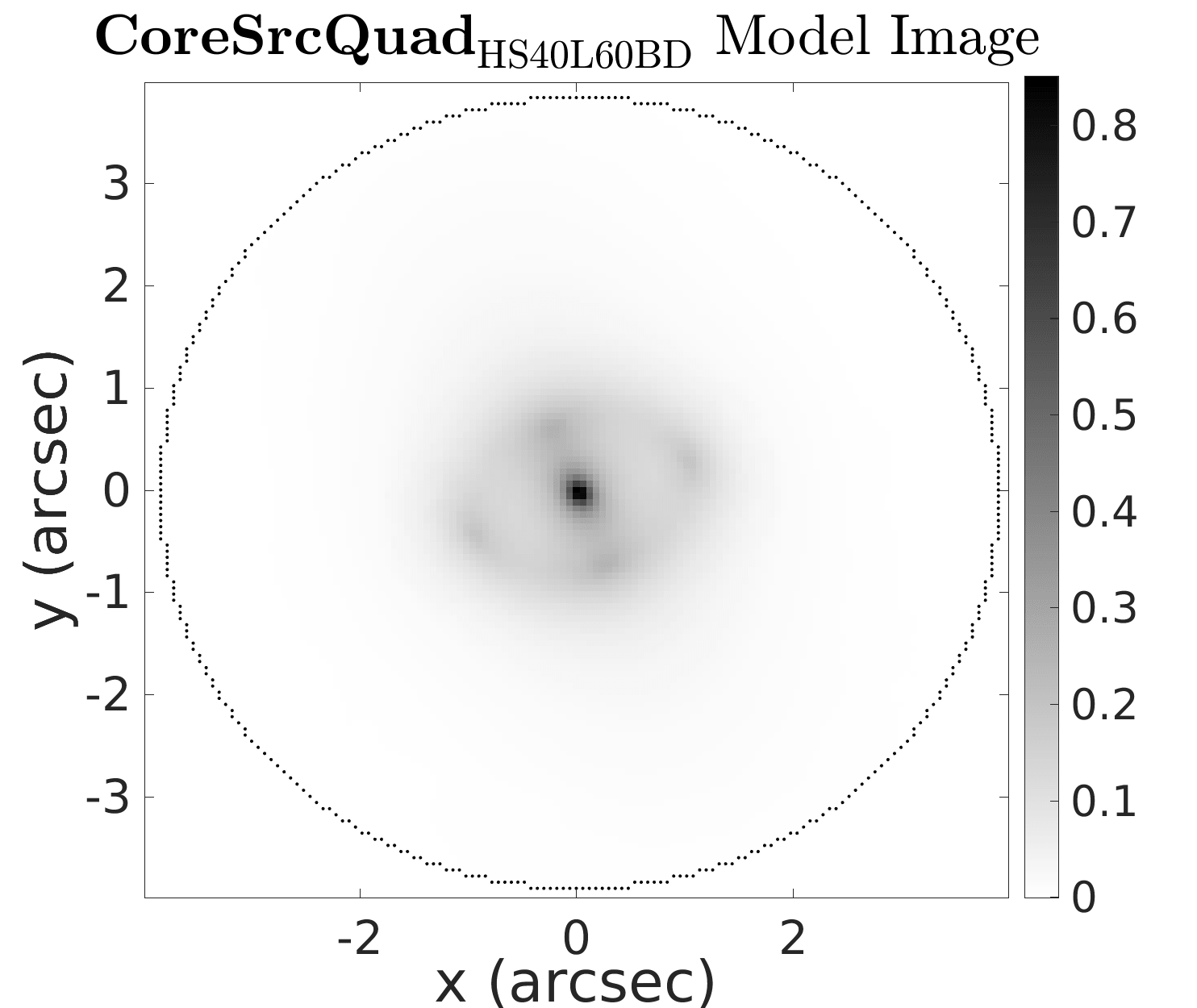}
\includegraphics[width=0.157\textwidth]{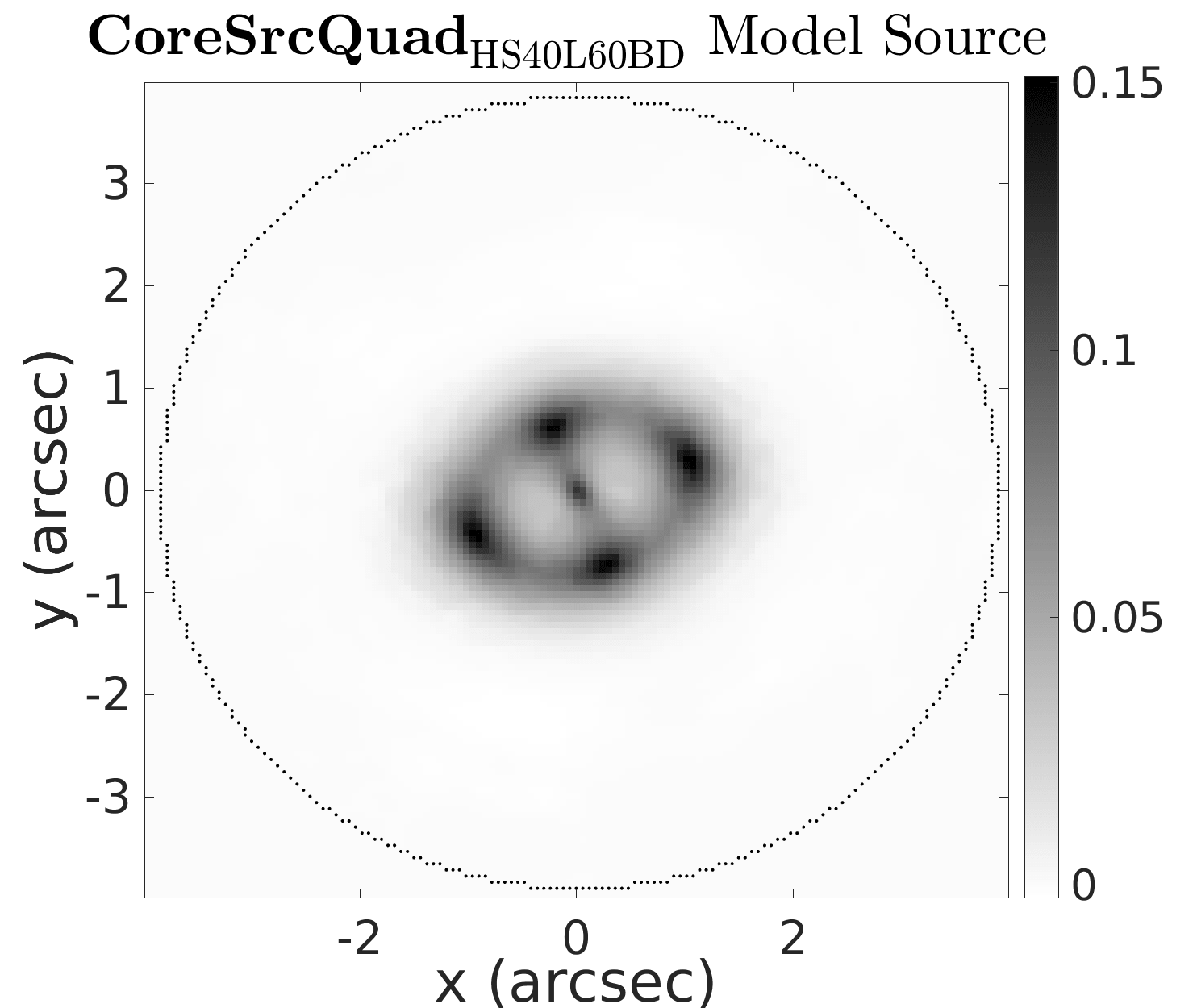}
\includegraphics[width=0.157\textwidth]{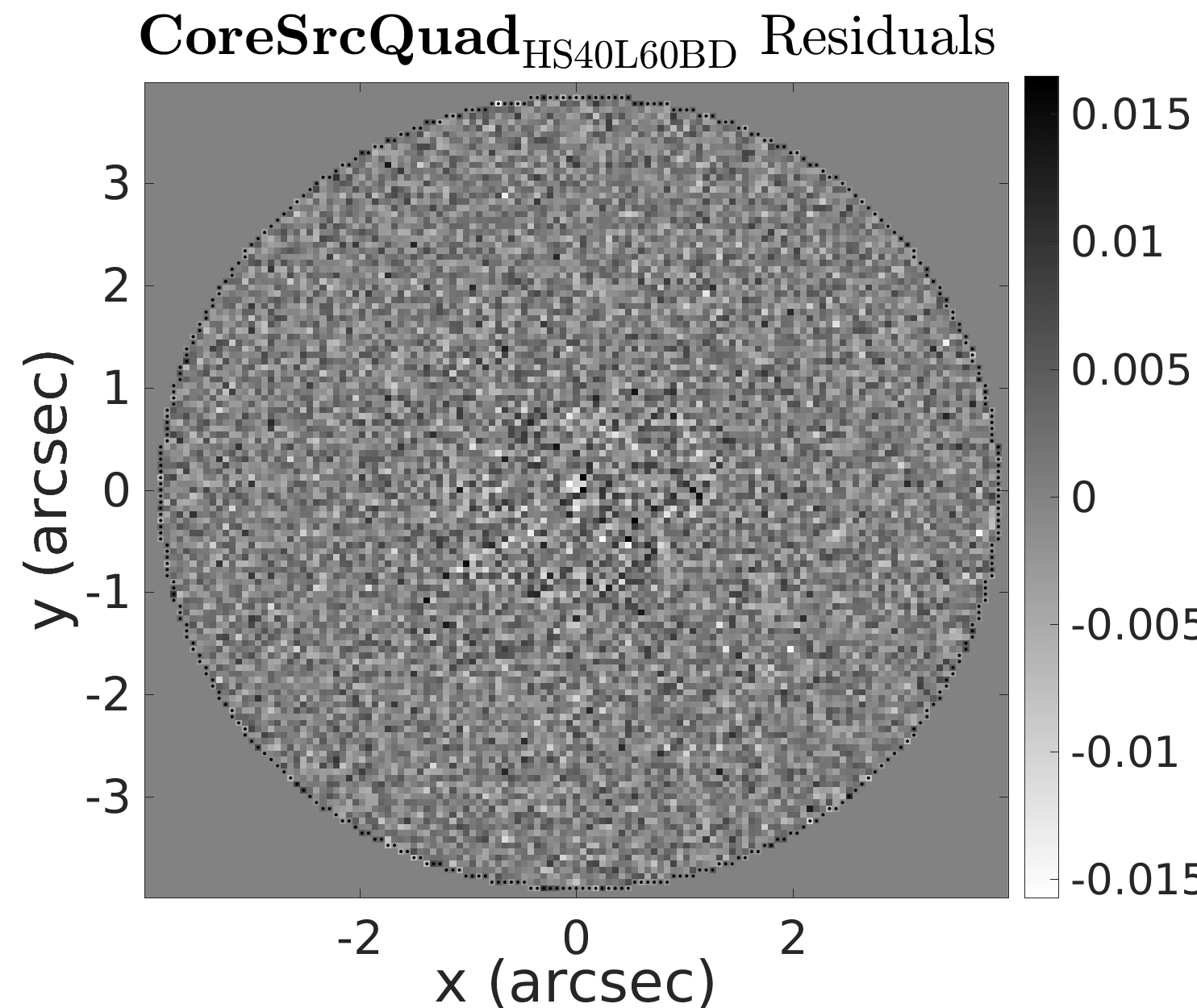}
\includegraphics[width=0.157\textwidth]{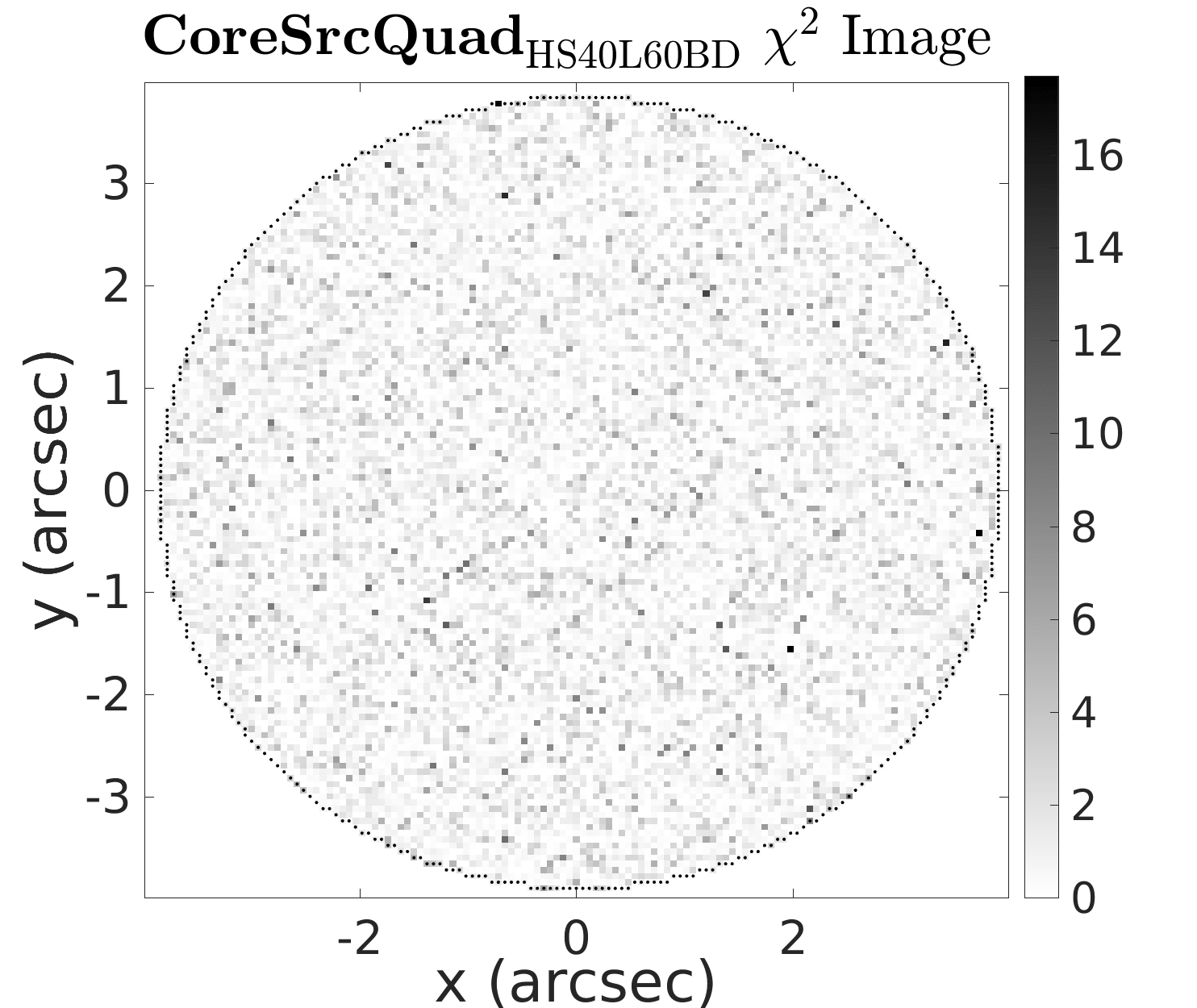}
\includegraphics[width=0.157\textwidth]{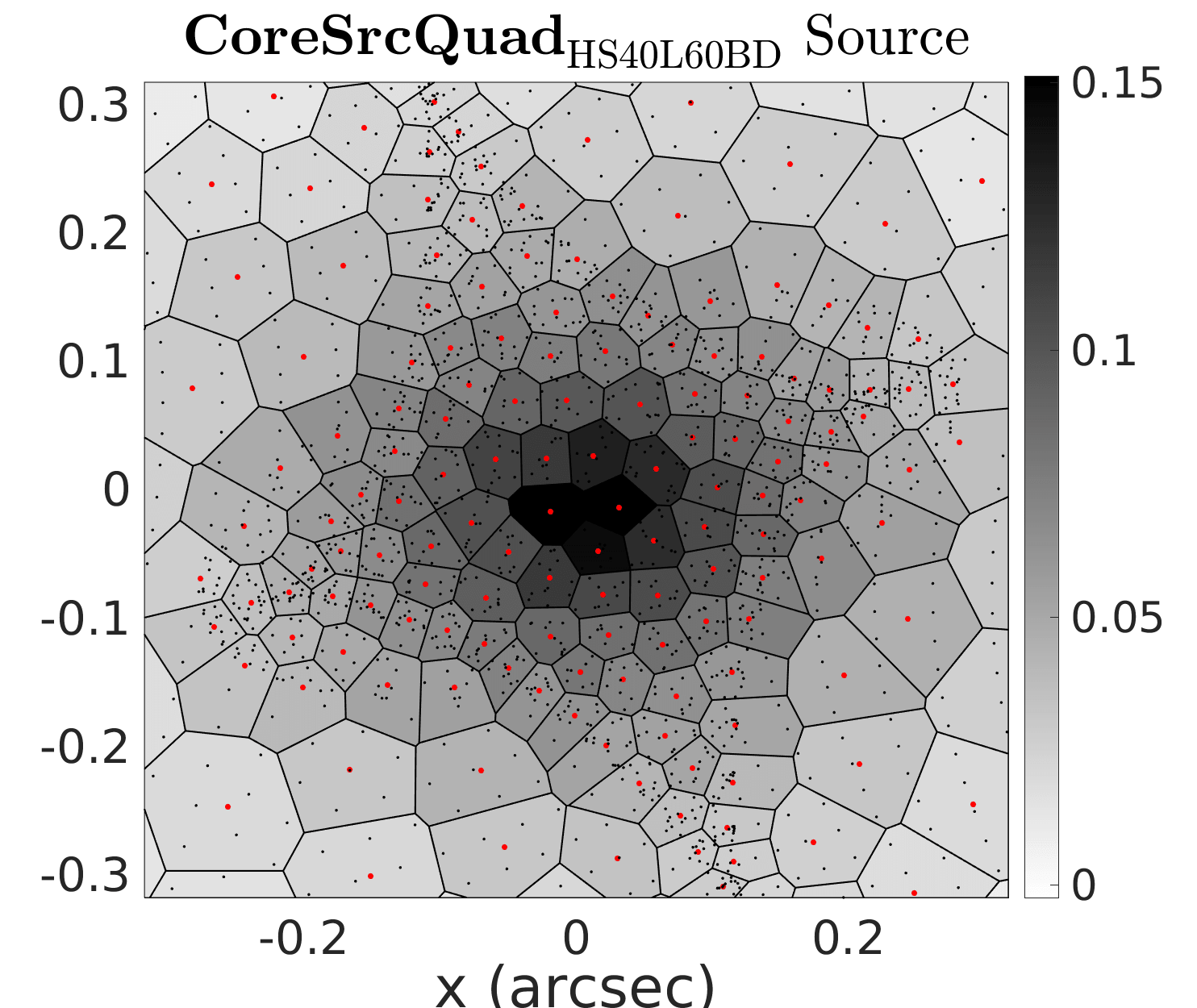}
\includegraphics[width=0.157\textwidth]{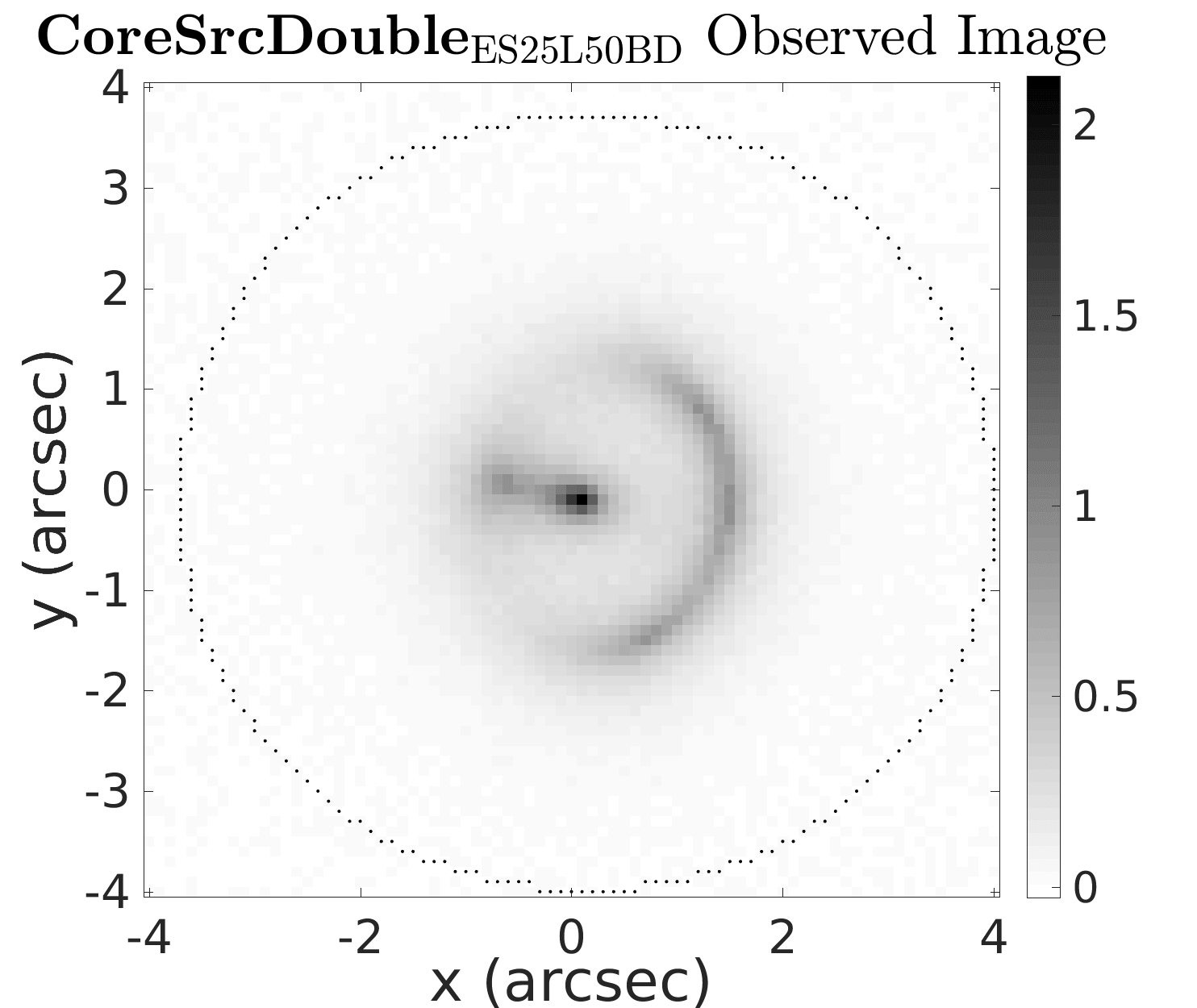}
\includegraphics[width=0.157\textwidth]{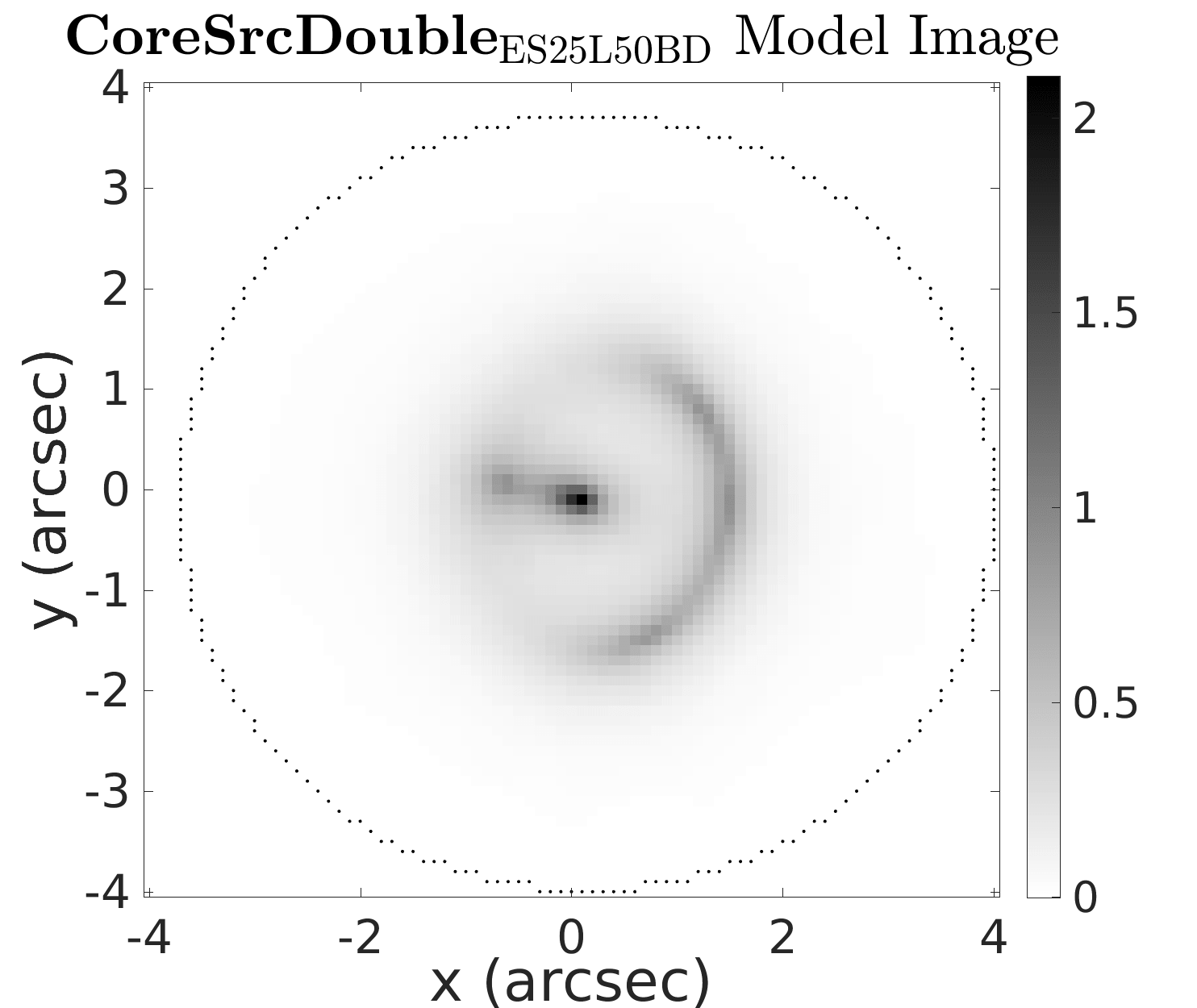}
\includegraphics[width=0.157\textwidth]{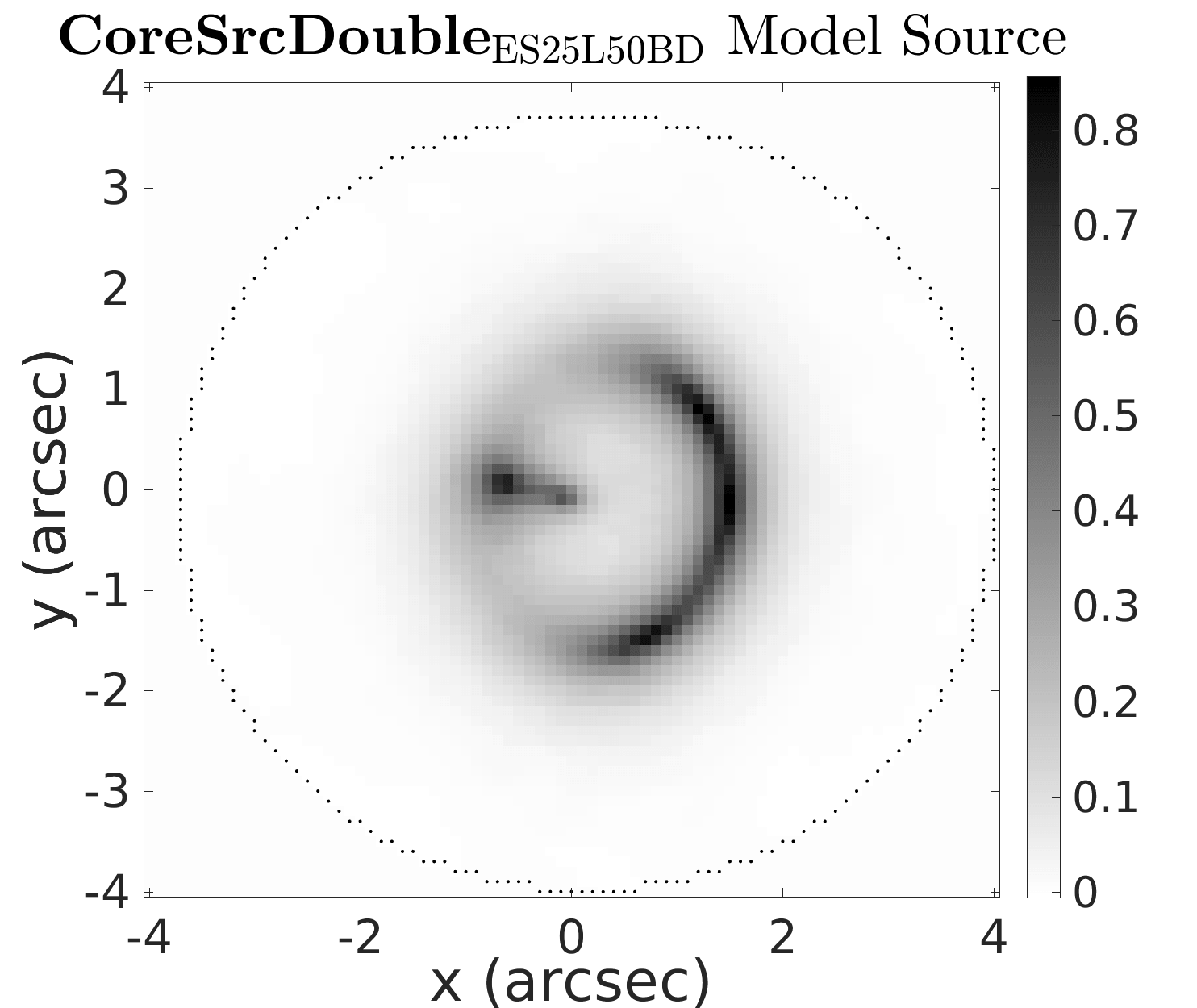}
\includegraphics[width=0.157\textwidth]{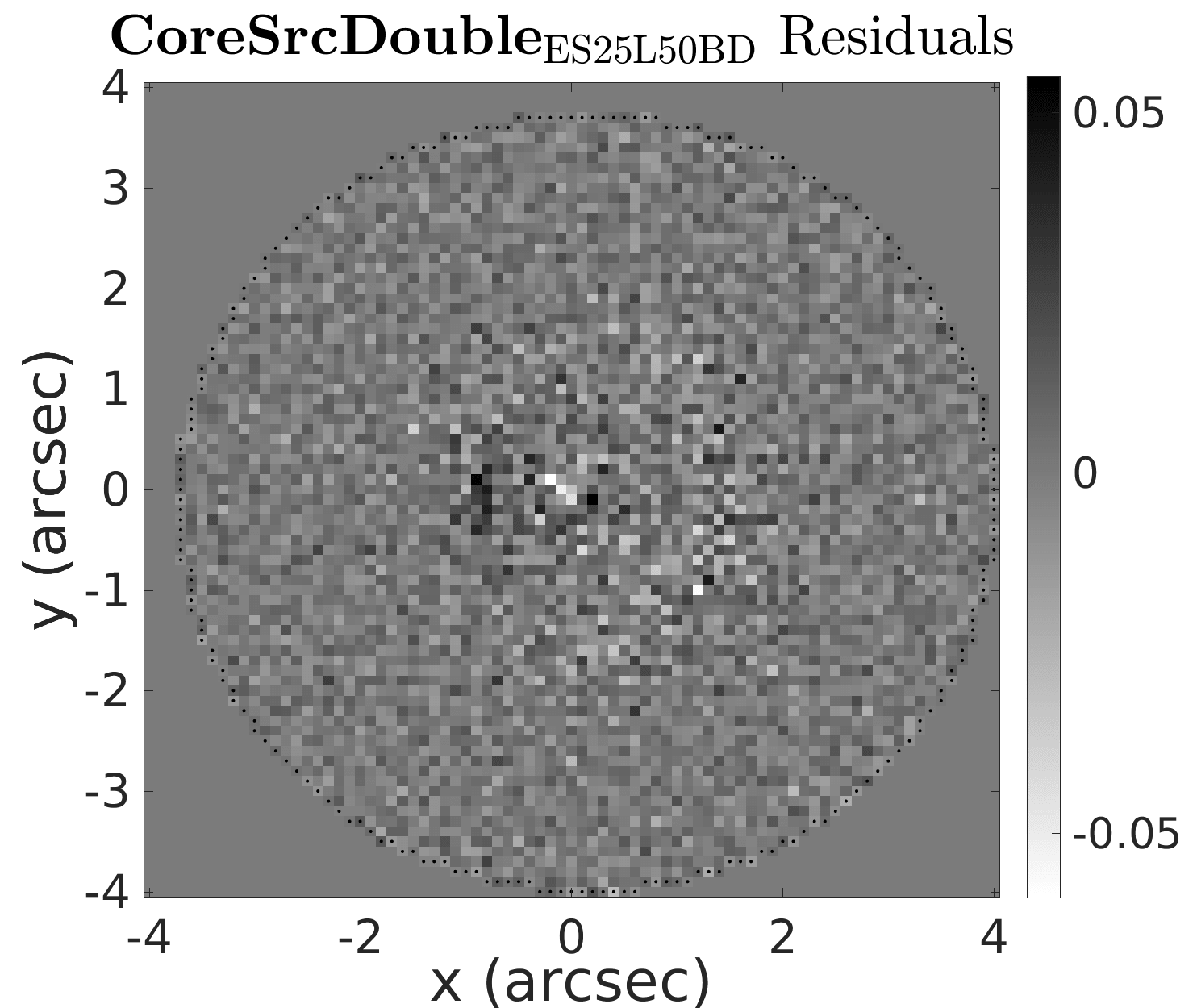}
\includegraphics[width=0.157\textwidth]{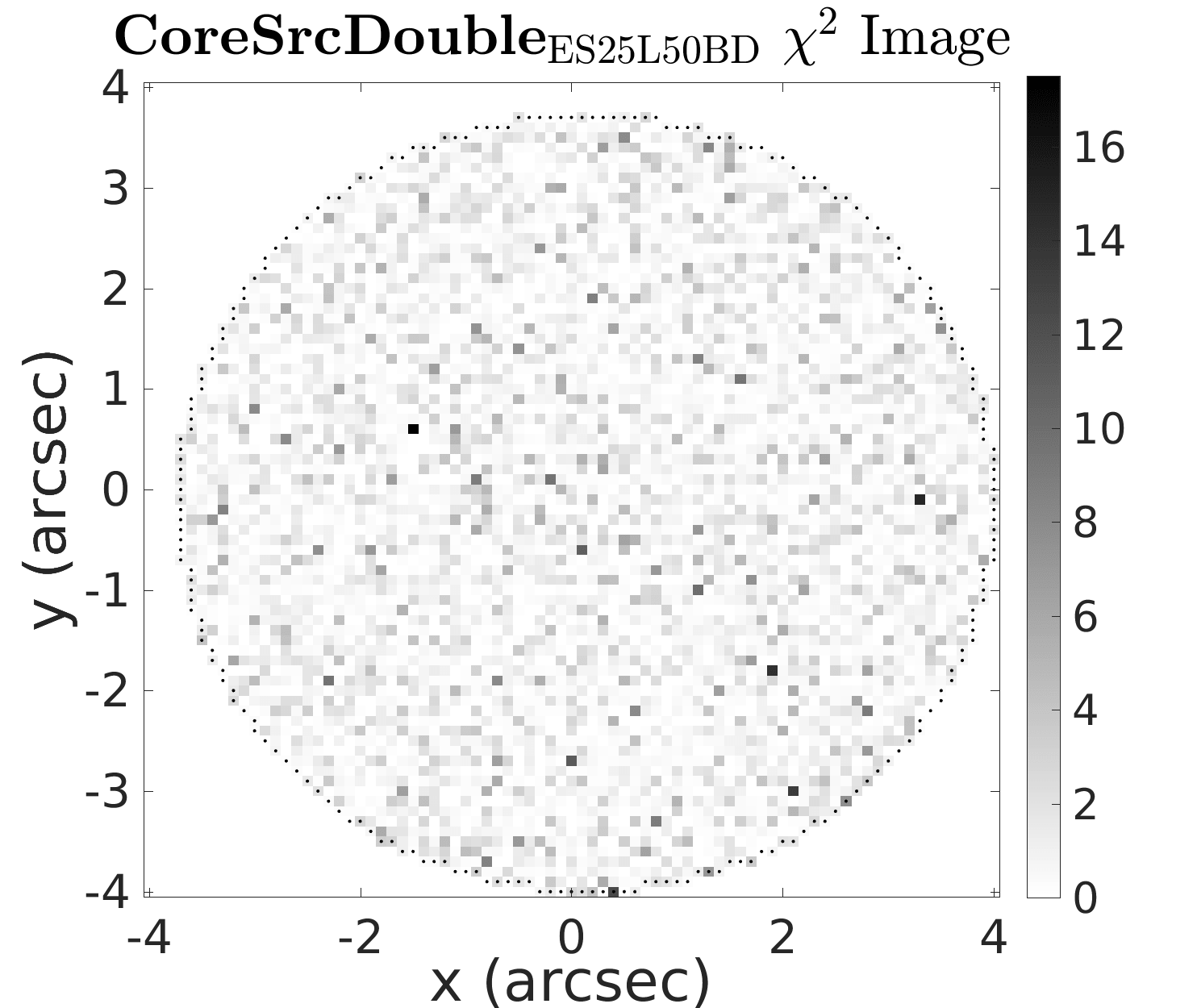}
\includegraphics[width=0.157\textwidth]{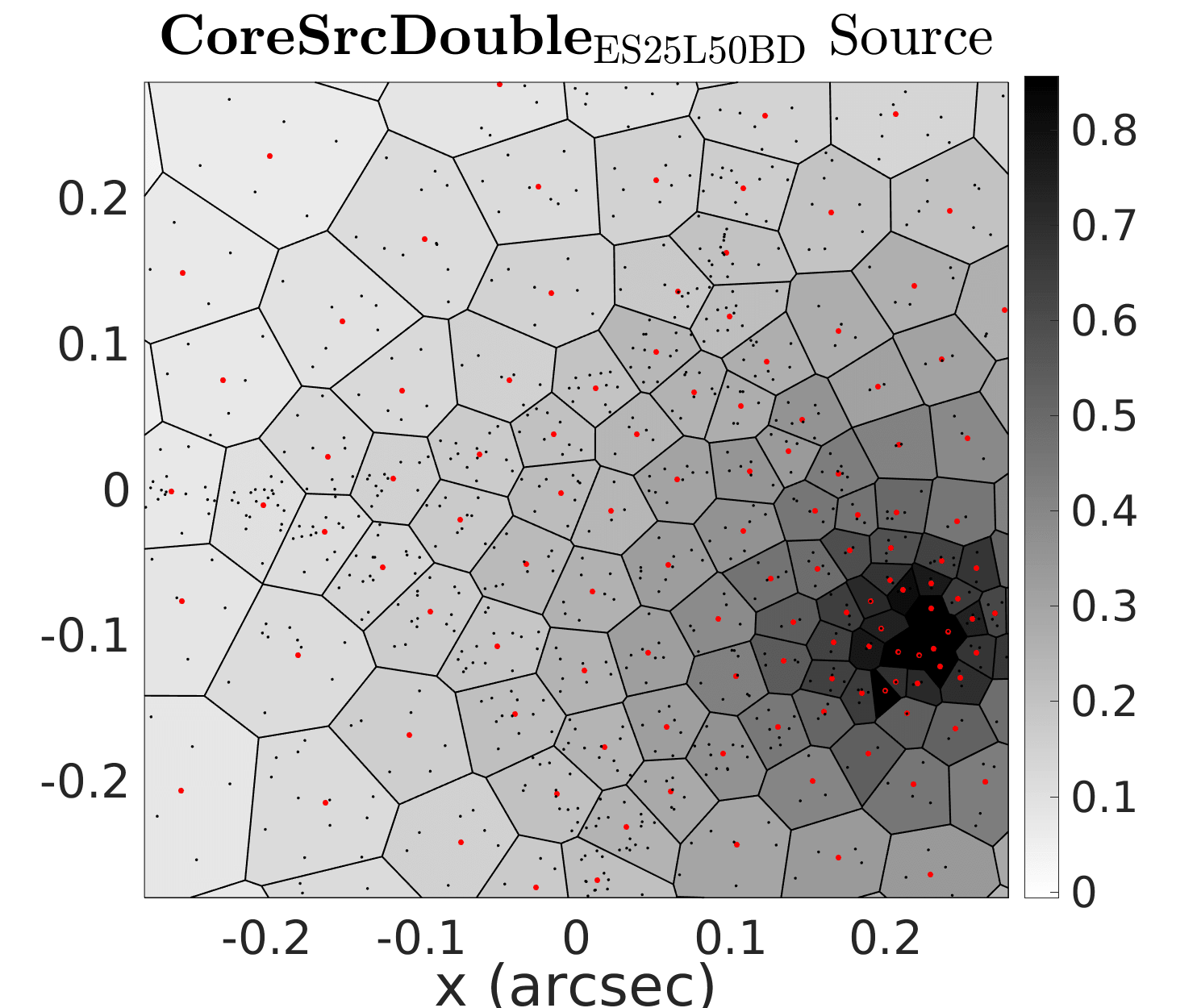}
\caption{The simulated image, model image, source model image, residuals, $\chi^2$ image and source reconstructions for the analysis of the images $\textbf{CoreSrcDisk}_{\mathrm{HS35L70Disk}}$ (top row), $\textbf{CoreSrcQuad}_{\mathrm{HS40L60BD}}$ (middle row) and $\textbf{CoreSrcDouble}_{\mathrm{ES50L50BD}}$ (bottom row). Images correspond to the most probable model at the end of the main pipeline, corresponding to the models given by rows $1-2$, $5-8$ and $13-14$ of table \ref{table:TableSPLEFgCore} respectively.} 
\label{figure:ResultsSPLEFgCoreIms}
\end{figure*}

Figure \ref{figure:PDFsSPLEFg} shows the one-dimensional PDF of $\Delta n_{\rm l}$ for the Hubble resolution (left panel) and Euclid resolution (right panel) images of each lens model, corresponding to the models given by table \ref{table:TableSPLEFg}. The majority of $n_l$ values are estimated correctly, with those that are not discussed above. The tendency to be shifted above the lens's input value of $n_{\rm l}$ is also visible in this figure. The size of each PDF and therefore the precision inferred on the value of $n_{\rm l}$ can be seen to vary greatly, but without an obvious dependence on the image resolution or S/N. Thus, these factors do not appear to be the most important for constraining the lens's light profile. Instead, the model's precision is most heavily dependent on the lens's input values of $R_{\rm l}$ and $n_{\rm l}$, where larger effective radii and lower Sersic indexes provide a more precise light model. In these cases, a greater amount of lens light (that is not obscured by the source's light) is visible and available to constrain the lens's light profile. Therefore, it is the degree of lens and source light blending that drives how well the light profile can be measured.

Now the lens's light and mass are modeled simultaneously, it is interesting to investigate what interplay, if any, there is between the two components. Figure \ref{figure:PDFsLensSrc2D} shows the marginalized two-dimensional PDF's between $\Delta \alpha$ and $\Delta n_{\rm l}$ for the Hubble resolution (left panel) and Euclid resolution (right panel) images. Contours are orthogonal, demonstrating there is no noticeable degeneracy between the mass and light models, where visual inspection of other parameter pairs (e.g. $I_{\rm  l}$, $R_L$, $\theta_{\rm  Ein}$, etc.) reveals this holds in general. 

The precision of the mass model also appears to show no dependence on whether the lens's light is included, as can be inferred by contrasting the errors on $\theta_{\rm Ein}$, $q$ and $\alpha$ for the $S/N = 30$ images of the $\textbf{LensSrcBulge}$ / $\textbf{SrcBulge}$, $\textbf{LensSrcDisk}$ / $\textbf{SrcDisk}$ and $\textbf{LensSrcMulti}$ / $\textbf{SrcMulti}$ models. Each pair of models share the same spatial resolution, S/N ratio, mass and source profiles, with the only difference between them being the inclusion of the lens's light. The magnitude of errors for the source-only case and lens and source case show no systematic increase across all images, demonstrating that the light subtraction (provided it is accurate) does not impact the mass model precision. Figure \ref{figure:PDFsSrc2D} reinforces this further, showing that the mass-profile degeneracies between $\theta_{\rm Ein}$ and $\alpha$ are similar for these images, regardless of whether the lens is included or not.

Whilst the lens's light has no impact on the mass model precision, the reverse is not true. That is, the presence of the source's light has a huge impact on the precision of the inferred light profile. This was confirmed by comparing the errors on the light profile parameters shown in figure \ref{table:TableSPLEFg} to the errors computed by fitting lens-only variants of each image. For example, for the image $\textbf{LensSrcBulge}_{\rm HS30L50Bulge}$, errors on $n_{\rm l}$ were approximately $\pm 0.05$ when no source was present and $\pm 0.7$ when it was (see table \ref{table:TableSPLEFg}). This occurs because of the smooth and symmetric nature of the lens's light profile, which has a sizable fraction of its light obscured by the ring-like source, such that much looser constraints are possible. Furthermore, for the simulated lenses used in this work, the source obscures the lens in and around its half-light radius, where the largest impact on the estimation of $R_{\rm l}$ (and $n_{\rm l}$ due to their degeneracy) can be expected. In contrast, the source morphology is asymmetric and irregular and therefore looks similar regardless of the smooth lens profile used for the light subtraction, such that similar constraints are offered across a range of lens light models. 

\subsection{Cored Lens and Source}\label{ResultsCore}

\begin{table*}
\scriptsize
\centering
\begin{tabular}{ l | l | l | l || l | l || l } 
\multicolumn{1}{p{1.0cm}|}{\centering \textbf{Image}} 
& \multicolumn{1}{p{1.0cm}|}{\centering \textbf{Pipeline \\ Run}} 
& \multicolumn{1}{p{1.6cm}|}{\centering \textbf{Sersic} \\ ($\textbf{P}_{\rm MCLight}$) } 
& \multicolumn{1}{p{1.6cm}||}{\centering \textbf{Sersic + Exp} \\ ($\textbf{P}_{\rm MCLight}$) } 
& \multicolumn{1}{p{1.6cm}|}{\centering \textbf{SPLE} \\ ($\textbf{P}_{\rm MCMass}$)  } 
& \multicolumn{1}{p{1.6cm}||}{\centering \textbf{SPLE + Shear} \\ ($\textbf{P}_{\rm MCMass}$) } 
& \multicolumn{1}{p{1.6cm}}{\centering \textbf{Core Model Chosen?} } 
\\ \hline
& & & & & & \\[-4pt]
$\textbf{CoreSrcDisk}_{\mathrm{HS35L70Disk}}$ & Non-cored & \textbf{53835.2452} & 53849.6432 & 54459.7417 & 54482.4440 & \\[2pt]
$\textbf{CoreSrcDisk}_{\mathrm{HS35L70Disk}}$ & Cored & \textbf{54505.0945} & 54500.4490 & \textbf{54549.9967} & 54554.9393 & \\[2pt]
$\textbf{CoreSrcDisk}_{\mathrm{ES35L70Disk}}$ & Non-cored & \textbf{14791.5954} & 14769.7912 & 15013.6825 & 15021.6143 & \textbf{Yes} \\[2pt]
$\textbf{CoreSrcDisk}_{\mathrm{ES35L70Disk}}$ & Cored & \textbf{15025.1912} & 15032.7236 & \textbf{15048.2671} & 15047.9044 & \textbf{Yes}  \\[2pt]
\hline
& & & & & & \\[-4pt]
$\textbf{CoreSrcQuad}_{\mathrm{HS40L60BD}}$ & Non-cored & 54019.3266 & \textbf{54125.3480} & 54551.2353 & 54554.2885 & \\[2pt]
$\textbf{CoreSrcQuad}_{\mathrm{HS40L60BD}}$ & Cored & 54376.6692 & \textbf{54632.3307} & \textbf{54706.6518} & 54709.8142 & \\[2pt]
$\textbf{CoreSrcQuad}_{\mathrm{ES40L60BD}}$ & Non-cored & 14697.6901 & \textbf{14834.0991} & 14937.6529 & 14938.3421 & \textbf{Yes} \\[2pt]
$\textbf{CoreSrcQuad}_{\mathrm{ES40L60BD}}$ & Cored & 14801.0601 & \textbf{14998.6605} & \textbf{15105.6929} & 15110.8025 & \textbf{Yes}\\[2pt]
\hline
& & & & & & \\[-4pt]
$\textbf{CoreSrcDouble}_{\mathrm{HS25L50BD}}$ & Non-cored & 53375.7084 & \textbf{53425.0099} & 54429.5815 & 54431.3582 & \\[2pt]
$\textbf{CoreSrcDouble}_{\mathrm{HS25L50BD}}$ & Cored & \textbf{54236.5839} & 54234.0987 & \textbf{54563.2273} & 54564.5934 & \\[2pt]
$\textbf{CoreSrcDouble}_{\mathrm{ES25L50BD}}$ & Non-cored & \textbf{14454.5533} & 14458.9280 & 14867.2487 & 14868.3826 & \textbf{Yes} \\[2pt]
$\textbf{CoreSrcDouble}_{\mathrm{ES25L50BD}}$ & Cored & \textbf{14795.4639} & 14809.6472 & 14903.2335 & \textbf{14931.0893} & \textbf{Yes} \\[2pt]
\end{tabular}
\caption{The results of Bayesian model comparison in the phases $\textbf{P}_{\rm  LightMC}$ and $\textbf{P}_{\rm  MassMC}$ for the six $\textbf{Core}$ images, using both the non-cored and cored mass model pipelines. Image names are listed in the first column, with the second column listing whether each row corresponds to a non-cored or cored pipeline run. The third and fourth columns show the Bayesian evidence values (equation \ref{eqn:Bayes2}) computed for each light model and the fifth and sixth columns evidences for comparison of the $SPLE$ and $SPLE$+$Shear$ mass model. These are then used to determine whether the cored or non-cored model is favoured, as shown by the final column. Values in bold correspond to those chosen by the pipeline, noting that a threshold of twenty must be exceeded to favour a more complex model.}
\label{table:SPLEFgCoreMC}
\end{table*}
\begin{table*}
\resizebox{\linewidth}{!}{
\begin{tabular}{ l | l | l l l | l} 
\multicolumn{1}{p{1.6cm}|}{Image} 
& \multicolumn{1}{p{1.8cm}|}{\centering \textbf{Component}} 
& \multicolumn{1}{p{2.2cm}}{\textbf{Parameters ($3\sigma$)}} 
& \multicolumn{1}{p{2.2cm}}{}  
& \multicolumn{1}{p{2.2cm}|}{} 
& \multicolumn{1}{p{2.2cm}}{\textbf{Parameters ($1\sigma$)}} 
\\ \hline
& & & & & \\[-4pt]

$\textbf{CoreSrcDisk}_{\mathrm{HS35L70Disk}}$ & Light (\textbf{Sersic}) & $\Delta R_{\mathrm{l}}=0.0647^{+0.2325}_{\rm -0.1977}$($R_{\mathrm{l}}=0.7$) & $\Delta q_{\mathrm{l}}=0.0085^{+0.0599}_{\rm -0.0542}$($q_{\mathrm{l}}=0.80$) & $\Delta n_{\mathrm{l}}=0.1021^{+0.4380}_{\rm -0.4239}$($n_{\mathrm{l}}=2.50$) & $\Delta n_{\mathrm{l}}=0.1021^{+0.1532}_{\rm -0.1617}$($n_{\mathrm{l}}=2.50$)\\[2pt]
$\textbf{CoreSrcDisk}_{\mathrm{HS35L70Disk}}$ & Mass (\textbf{SPLE}) & $\Delta S=-0.0191^{+0.0787}_{\rm -0.0809}$($S=0.20$) &  & $\Delta \alpha=-0.0379^{+0.1781}_{\rm -0.1497}$($\alpha=1.85$) & $\Delta S=-0.0191^{+0.0283}_{\rm -0.0278}$($S=0.20$)\\[2pt]
$\textbf{CoreSrcDisk}_{\mathrm{ES35L70Disk}}$ & Light (\textbf{Sersic}) & $\Delta R_{\mathrm{l}}=0.1761^{+0.3080}_{\rm -0.2814}$($R_{\mathrm{l}}=0.7$) & $\Delta q_{\mathrm{l}}=-0.0079^{+0.0652}_{\rm -0.0640}$($q_{\mathrm{l}}=0.80$) & $\Delta n_{\mathrm{l}}=0.3441^{+0.5634}_{\rm -0.5754}$($n_{\mathrm{l}}=2.50$) & $\mathbf{\Delta n_{\mathrm{l}}=0.3441^{+0.2015}_{\rm -0.2006}}$($n_{\mathrm{l}}=2.50$)\\[2pt]
$\textbf{CoreSrcDisk}_{\mathrm{ES35L70Disk}}$ & Mass (\textbf{SPLE}) & $\Delta S=0.0319^{+0.1282}_{\rm -0.1335}$($S=0.20$) &  & $\Delta \alpha=0.0556^{+0.2383}_{\rm -0.2510}$($\alpha=1.85$) & $\Delta S=0.0319^{+0.0481}_{\rm -0.0469}$($S=0.20$)\\[-4pt]
& & & & & \\[-4pt]
\hline
& & & & & \\[-4pt]
$\textbf{CoreSrcQuad}_{\mathrm{HS40L60BD}}$ & Light (\textbf{Sersic}) & $\mathbf{\Delta R_{\mathrm{l1}}=-0.0682^{+0.0520}_{\rm -0.0435}}$($R_{\mathrm{l1}}=0.25$) & $\Delta q_{\mathrm{l1}}=0.0116^{+0.0583}_{\rm -0.0522}$($q_{\mathrm{l1}}=0.77$) & $\mathbf{\Delta n_{\mathrm{l1}}=-0.7565^{+0.4355}_{\rm -0.3844}}$($n_{\mathrm{l1}}=2.50$) & $\mathbf{\Delta n_{\mathrm{l1}}=-0.7565^{+0.1333}_{\rm -0.1480}}$($n_{\mathrm{l1}}=2.50$)\\[2pt]
$\textbf{CoreSrcQuad}_{\mathrm{HS40L60BD}}$ & Light (\textbf{Exp}) & $\Delta R_{\mathrm{l2}}=-0.0307^{+0.0990}_{\rm -0.0830}$($R_{\mathrm{l2}}=1.35$) & $\mathbf{\Delta q_{\mathrm{l2}}=0.0802^{+0.0278}_{\rm -0.0335}}$($q_{\mathrm{l2}}=0.60$) &  & $\mathbf{\Delta R_{\mathrm{l2}}=-0.0307^{+0.0288}_{\rm -0.0356}}$($R_{\mathrm{l2}}=1.35$)\\[2pt]
$\textbf{CoreSrcQuad}_{\mathrm{HS40L60BD}}$ & Mass (\textbf{SPLE}) & $\Delta S=-0.0060^{+0.1476}_{\rm -0.1412}$($S=0.30$) &  & $\Delta \alpha=0.0487^{+0.1450}_{\rm -0.1007}$($\alpha=1.75$) & $\Delta S=-0.0060^{+0.0538}_{\rm -0.0568}$($S=0.30$)\\[2pt]
$\textbf{CoreSrcQuad}_{\mathrm{ES40L60BD}}$ & Light (\textbf{Sersic}) & $\mathbf{\Delta R_{\mathrm{l1}}=-0.0941^{+0.0372}_{\rm -0.0236}}$($R_{\mathrm{l1}}=0.25$) & $\Delta q_{\mathrm{l1}}=-0.0125^{+0.0663}_{\rm -0.0576}$($q_{\mathrm{l1}}=0.77$) & $\mathbf{\Delta n_{\mathrm{l1}}=-1.1441^{+0.3980}_{\rm -0.3459}}$($n_{\mathrm{l1}}=2.50$) & $\mathbf{\Delta n_{\mathrm{l1}}=-1.1441^{+0.1090}_{\rm -0.1336}}$($n_{\mathrm{l1}}=2.50$)\\[2pt]
$\textbf{CoreSrcQuad}_{\mathrm{ES40L60BD}}$ & Light (\textbf{Exp}) & $\mathbf{\Delta R_{\mathrm{l2}}=-0.0934^{+0.0482}_{\rm -0.0489}}$($R_{\mathrm{l2}}=1.35$) & $\mathbf{\Delta q_{\mathrm{l2}}=0.0673^{+0.0163}_{\rm -0.0176}}$($q_{\mathrm{l2}}=0.60$) &  & $\mathbf{\Delta R_{\mathrm{l2}}=-0.0934^{+0.0167}_{\rm -0.0170}}$($R_{\mathrm{l2}}=1.35$)\\[2pt]
$\textbf{CoreSrcQuad}_{\mathrm{ES40L60BD}}$ & Mass (\textbf{SPLE}) & $\Delta S=-0.1168^{+0.1323}_{\rm -0.1468}$($S=0.30$) &  & $\Delta \alpha=-0.0457^{+0.1072}_{\rm -0.1067}$($\alpha=1.75$) & $\mathbf{\Delta S=-0.1168^{+0.0439}_{\rm -0.0404}}$($S=0.30$)\\[-4pt]
& & & & & \\[-4pt]
\hline
& & & & & \\[-4pt]
$\textbf{CoreSrcDouble}_{\mathrm{HS25L50BD}}$ & Light (\textbf{Sersic}) & $\Delta R_{\mathrm{l}}=0.1938^{+0.7768}_{\rm -0.6068}$($R_{\mathrm{l}}=1.50$) & $\Delta q_{\mathrm{l}}=0.0006^{+0.0461}_{\rm -0.0482}$($q_{\mathrm{l}}=0.70$) & $\Delta n_{\mathrm{l}}=0.1238^{+0.5826}_{\rm -0.5375}$($n_{\mathrm{l}}=3.50$) & $\Delta n_{\mathrm{l}}=0.1238^{+0.1995}_{\rm -0.2133}$($n_{\mathrm{l}}=3.50$)\\[2pt]
$\textbf{CoreSrcDouble}_{\mathrm{HS25L50BD}}$ & Mass (\textbf{SPLE}) & $\Delta S=-0.0099^{+0.0349}_{\rm -0.0350}$($S=0.25$) &  & $\Delta \alpha=-0.0096^{+0.0631}_{\rm -0.0691}$($\alpha=1.65$) & $\Delta S=-0.0099^{+0.0124}_{\rm -0.0114}$($S=0.25$)\\[2pt]
$\textbf{CoreSrcDouble}_{\mathrm{ES25L50BD}}$ & Light (\textbf{Sersic}) & $\Delta R_{\mathrm{l}}=0.1850^{+0.8773}_{\rm -0.8788}$($R_{\mathrm{l}}=1.50$) & $\Delta q_{\mathrm{l}}=0.0057^{+0.0754}_{\rm -0.0716}$($q_{\mathrm{l}}=0.70$) & $\Delta n_{\mathrm{l}}=0.0663^{+0.7554}_{\rm -0.8303}$($n_{\mathrm{l}}=3.50$) & $\Delta n_{\mathrm{l}}=0.0663^{+0.3127}_{\rm -0.2861}$($n_{\mathrm{l}}=3.50$)\\[2pt]
$\textbf{CoreSrcDouble}_{\mathrm{ES25L50BD}}$ & Mass (\textbf{SPLE}) & $\Delta S=-0.0106^{+0.0979}_{\rm -0.1035}$($S=0.25$) &  & $\Delta \alpha=-0.0071^{+0.1498}_{\rm -0.1654}$($\alpha=1.65$) & $\Delta S=-0.0106^{+0.0358}_{\rm -0.0348}$($S=0.25$)\\[-4pt]
\end{tabular}
}
\caption{Results of fitting the six $\textbf{Core}$ model images using cored mass profile automated analysis pipeline, corresponding to results generated at end of phase two. Each image's name is given in the first column and the light or mass model component in the second column. The third to sixth columns show parameter estimates, where each parameter is offset by $\Delta P = P_{\rm  true} - P_{\rm  model}$, such that zero corresponds to the input lens model. The input lens model values are given in brackets to the right of each parameter estimate. Parameters estimates are shown using $\Delta R_{\rm l}$, $\Delta q_{\rm l}$ and $\Delta n_{\rm l}$ for $Sersic$ light models, $\Delta R_{\rm l1}$, $\Delta q_{\rm l1}$, $\Delta n_{\rm l1}$, $\Delta R_{\rm l2}$ and $\Delta q_{\rm l2}$ for $Sersic$ + $Exp$ light models and $\Delta S$ and $\Delta \alpha$ for $PL\textsubscript{Core}$ mass models. Columns three to five show parameter estimates within $3 \sigma$ confidence and column six at $1 \sigma$. Parameter estimates in bold text are inconsistent with the input lens model at their stated error estimates. The other parameters not shown (e.g. $x_{\rm l}$, $\theta_{\rm l}$, $\theta$) are all estimated accurately within $3\sigma$.}
\label{table:TableSPLEFgCore}
\end{table*}

\begin{figure*}
\centering
\includegraphics[width=0.95\textwidth]{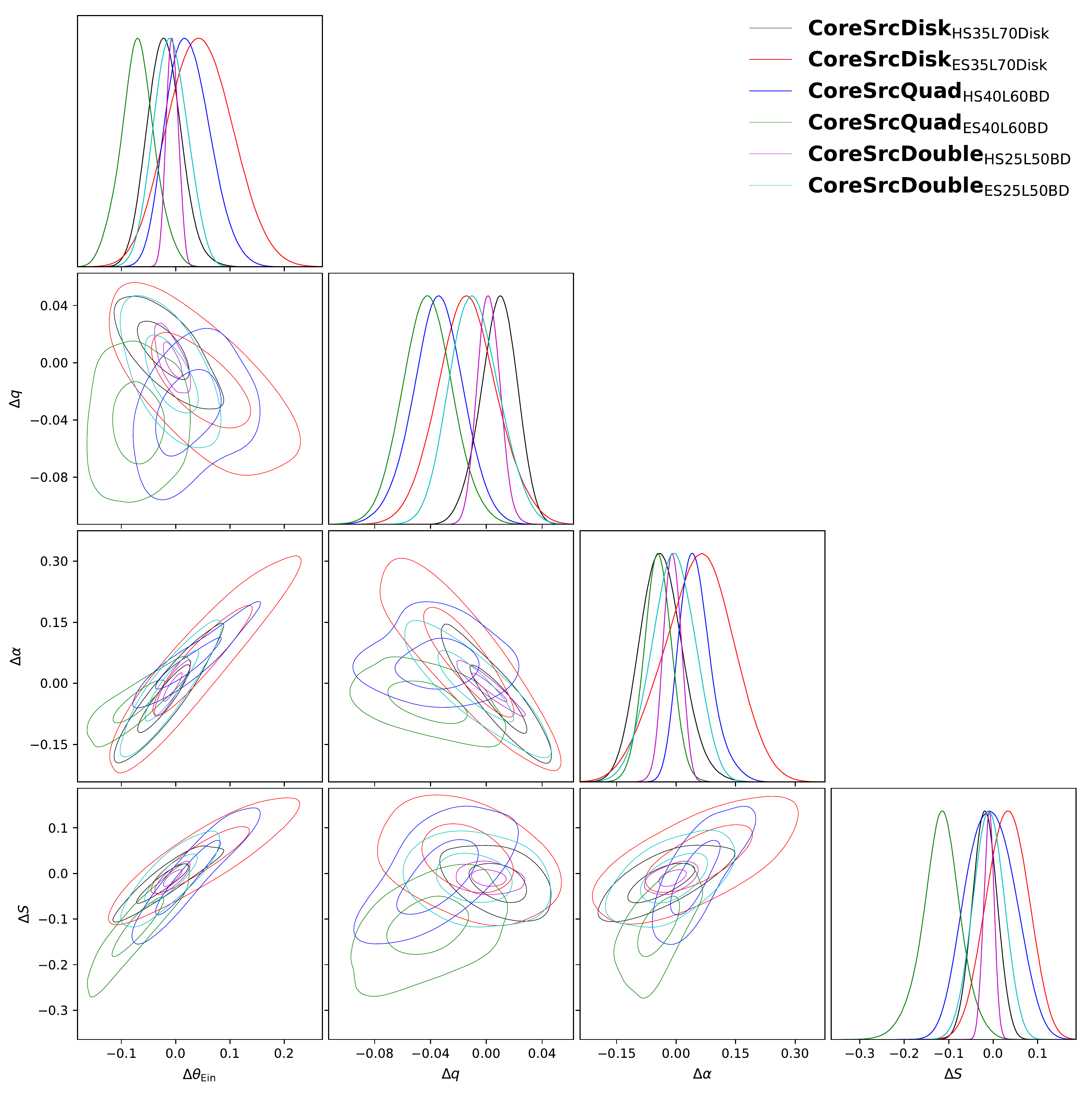}
\caption{Marginalized two-dimensional PDF's of $\Delta \theta_{\rm Ein}$, $\Delta q$, $\Delta \alpha$ and $\Delta S$ for the images of the $\textbf{Core}$ models. The top-right legend indicates the image that each coloured line corresponds top. Contours give the $1\sigma$ (interior) and $3 \sigma$ (exterior) confidence regions. For $\textbf{CoreSrcDisk}$ the input values of each parameter are $\theta_{\rm  Ein} = 1.4"$, $q = 0.8$, $\alpha = 1.85$ and $S = 0.2"$, for $\textbf{CoreSrcQuad}$ $\theta_{\rm  Ein} = 1.0"$, $q = 0.7$, $\alpha = 1.75$ and $S = 0.3"$ and for $\textbf{CoreSrcDouble}$ $\theta_{\rm  Ein} = 1.3"$, $q = 0.8$, $\alpha = 1.65$ and $S = 0.25"$. A degeneracy can be seen between all parameters, which is an extension of the degeneracy between mass, ellipticty and slope described in N15, but including also the core radius $S$.} 
\label{figure:PDFsCore2D}
\end{figure*}

The cored lens simulation suite consists of three unique lens and source models, each of which are again used to generate an image at Hubble resolution and Euclid resolution both with a source $S/N = 35-50$ and lens $S/N$ spanning $50-70$, giving a total of six images. Each image is analyzed using {\tt AutoLens}'s singular total-mass profile and cored total-mass profile pipelines. The results of model comparison are shown in table \ref{table:SPLEFgCoreMC}, showing that the light profiles and cored models are correctly chosen for all images. Parameter estimates for each image are summarized in table \ref{table:TableSPLEFgCore}, using $\Delta S$ and the same mismatch parameters as before, where many parameters are estimated correctly within $3 \sigma$ confidence but one image, discussed next, has clear problems. Figure \ref{figure:ResultsSPLEFgCoreIms} shows the observed images, model images, model sources, residuals $\chi^2$ images and source reconstruction for the high-resolution images, showing that the image and source reconstructions successfully reproduce the features of a cored mass profile, like radial arcs or a central image, and again gives residuals and $\chi^2$ images consistent with the noise.

\subsubsection{Model Comparison}\label{ResultsCoreMC}

The results of the model comparison between the $SPLE$ and $PL\textsubscript{Core}$ models,  using a full {\tt AutoLens} analysis for each, are given for all six images in table \ref{table:SPLEFgCoreMC}. For all images model comparison correctly chooses the cored model, demonstrating the method's success at modeling a cored profile. The correct lens light profile is also chosen for all images. There is one case, given by the bottom row of table \ref{table:SPLEFgCoreMC}, which is incorrect, where the image $\textbf{CoreSrcDouble}_{\mathrm{ES25L50BD}}$ includes a $Shear$ component. There is no obvious explanation for this occurrence and the shear magnitude $\gamma_{\rm sh}$ reverts to approximately zero in the main pipeline. Thus, its inclusion is disregarded as non-consequential.
\subsubsection{Modeling Results}\label{ResultsCoreModel}

\begin{figure*}
\centering
\includegraphics[width=0.95\textwidth]{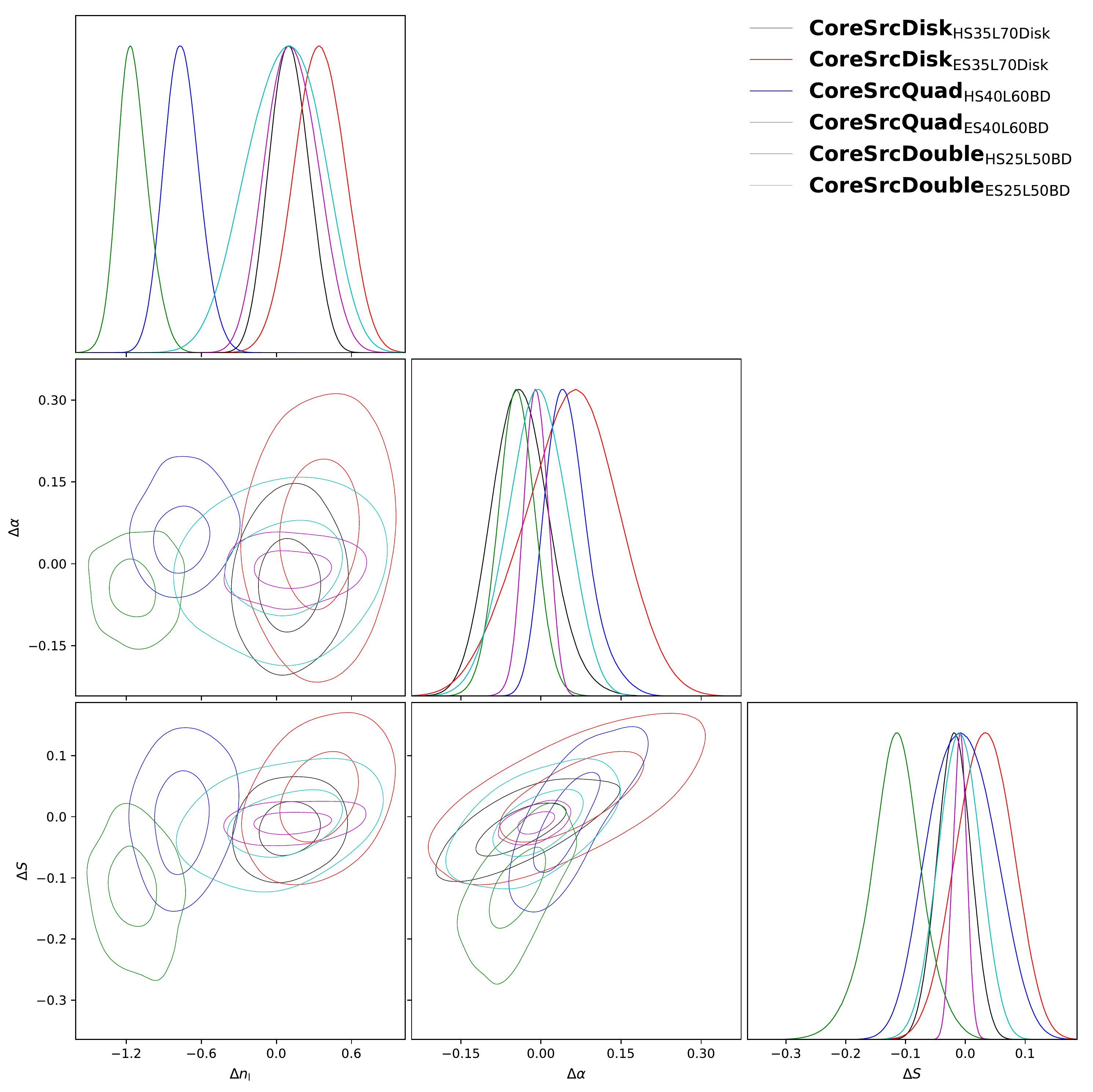}
\caption{Marginalized two-dimensional PDF's of $\Delta n_{\rm l}$, $\Delta \alpha$ and $\Delta S$ for the images of the $\textbf{Core}$ models. The top-right legend indicates the image that each coloured line corresponds top. Contours give the $1\sigma$ (interior) and $3 \sigma$ (exterior) confidence regions. For $\textbf{CoreSrcDisk}$ the input values of each parameter are $n_{\rm l} = 2.5$, $\alpha = 1.85$ and $S = 0.2"$, for $\textbf{CoreSrcQuad}$ $n_{\rm l1} = 2.5$ $\alpha = 1.75$ and $S = 0.3"$ and for $\textbf{CoreSrcDouble}$ $n_{\rm l} = 3.5$, $\alpha = 1.65$ and $S = 0.25"$. No degeneracy can be seen between $n_{\rm l}$ and the mass-profile parameters, suggesting there is no degeneracy between a light model and cored mass model.} 
\label{figure:PDFsLightCore2D}
\end{figure*}

The parameter estimates for the cored images are given in table \ref{table:TableSPLEFgCore}, where parameter estimates for four out of six images are all accurate within $3 \sigma$ confidence. However, for images of the $\textbf{CoreSrcQuad}$ model, the light profile parameter estimates are poor, with the values of $n_{\rm l}$ significantly offset from the input value. The middle row of figure \ref{figure:ResultsSPLEFgCoreIms} shows the image $\textbf{CoreSrcQuad}_{\mathrm{HS40L60BD}}$, where it can be seen this an example of an image where the source's central image perfectly overlaps the centre of the lens's light profile. This is the case of maximum blending and it is no surprise that the light model is inaccurate, as the lens's central light profile is completely obscured. This means that, if lenses of this configuration are found in nature, care must be taken in ensuring their lens subtraction is accurate and their inferred light profiles should be viewed with caution. Multi-wavelength imaging may be able to decouple the lens and source. The mass model for this configuration is still estimated accurately.

Figure \ref{figure:PDFsCore2D} shows the one and two-dimensional PDF's of $\Delta \theta_{\rm  Ein}$, $\Delta q$, $\Delta \alpha$ and $\Delta S$ for all of the $\textbf{Core}$ images. In general, lower-resolution imaging gives wider parameter estimates compared to their higher resolution counterparts and the $\textbf{CoreSrcQuad}$ images are less precisely constrained, due to the lens light blending discussed above. $S$ becomes another parameter in the mass-profile degeneracy discussed in section \ref{Method} and in N15, offering the mass model an additional means by which to change its mass distribution whilst still integrating to give an accurate $M_{\rm  Ein}$. The additional freedom introduced by $S$ is constrained by the very central regions of the lensed source, either its central image or radial arcs. 

Having established that the lens's light and mass profiles are separable for non-cored models, it is interesting to ask whether a degeneracy emerges between the light profile and cored mass model. Figure \ref{figure:PDFsLightCore2D} shows two-dimensional PDFs between $n_{\rm  l}$, $\Delta \alpha$ and $\Delta S$. Once again, no degeneracy is observed between the mass and light model parameters (with inspection of other parameter pairs confirming this trend is general). This is initially surprising, but builds on the discussion above that because the lensed source's appearance is non-symmetric and irregular it shares no degeneracy with the subtraction of a smooth light profile. 

\subsection{Decomposed Mass Models}\label{ResultsDecomp}

\begin{figure*}
\centering 
\includegraphics[width=0.157\textwidth]{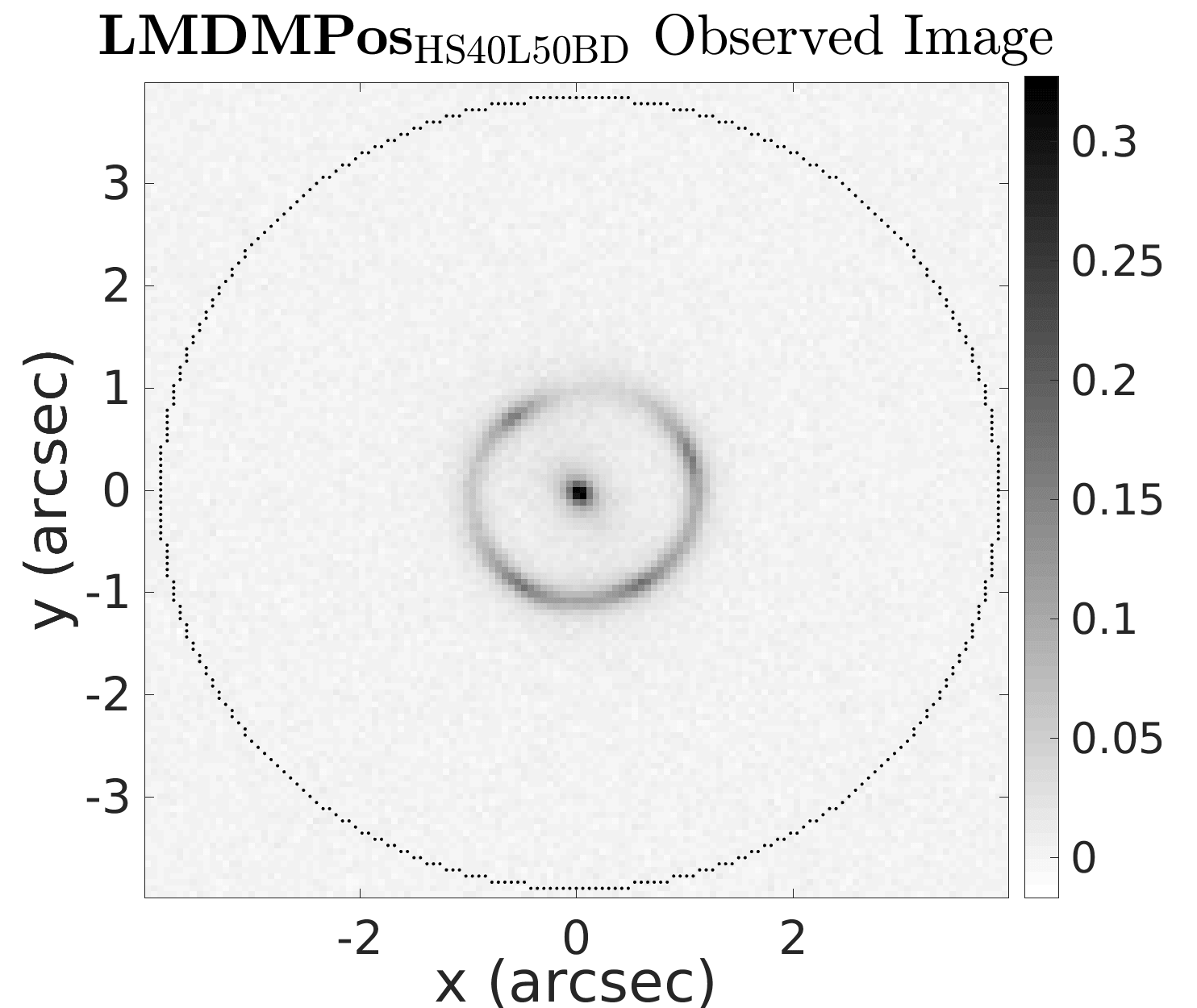}
\includegraphics[width=0.157\textwidth]{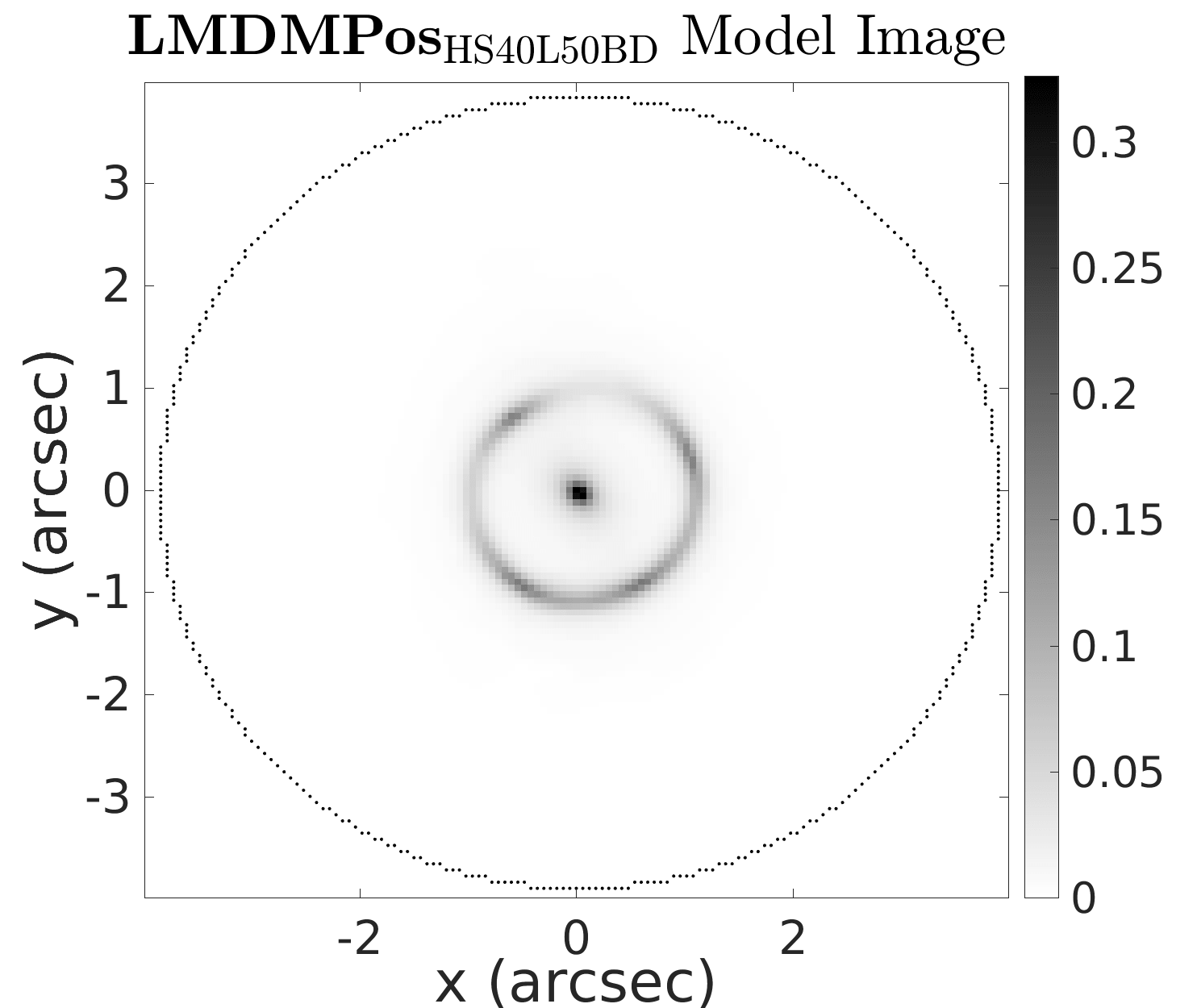}
\includegraphics[width=0.157\textwidth]{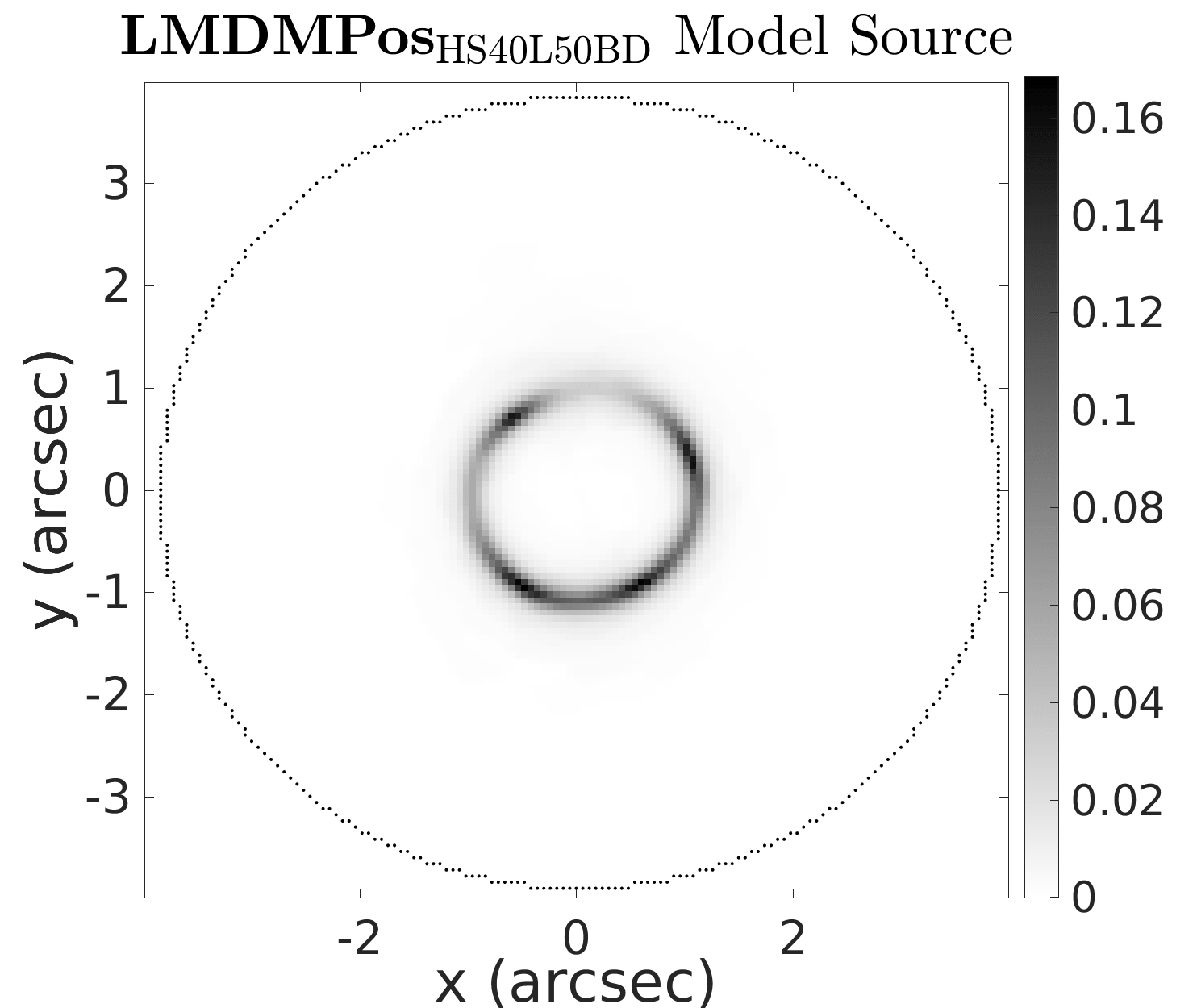}
\includegraphics[width=0.157\textwidth]{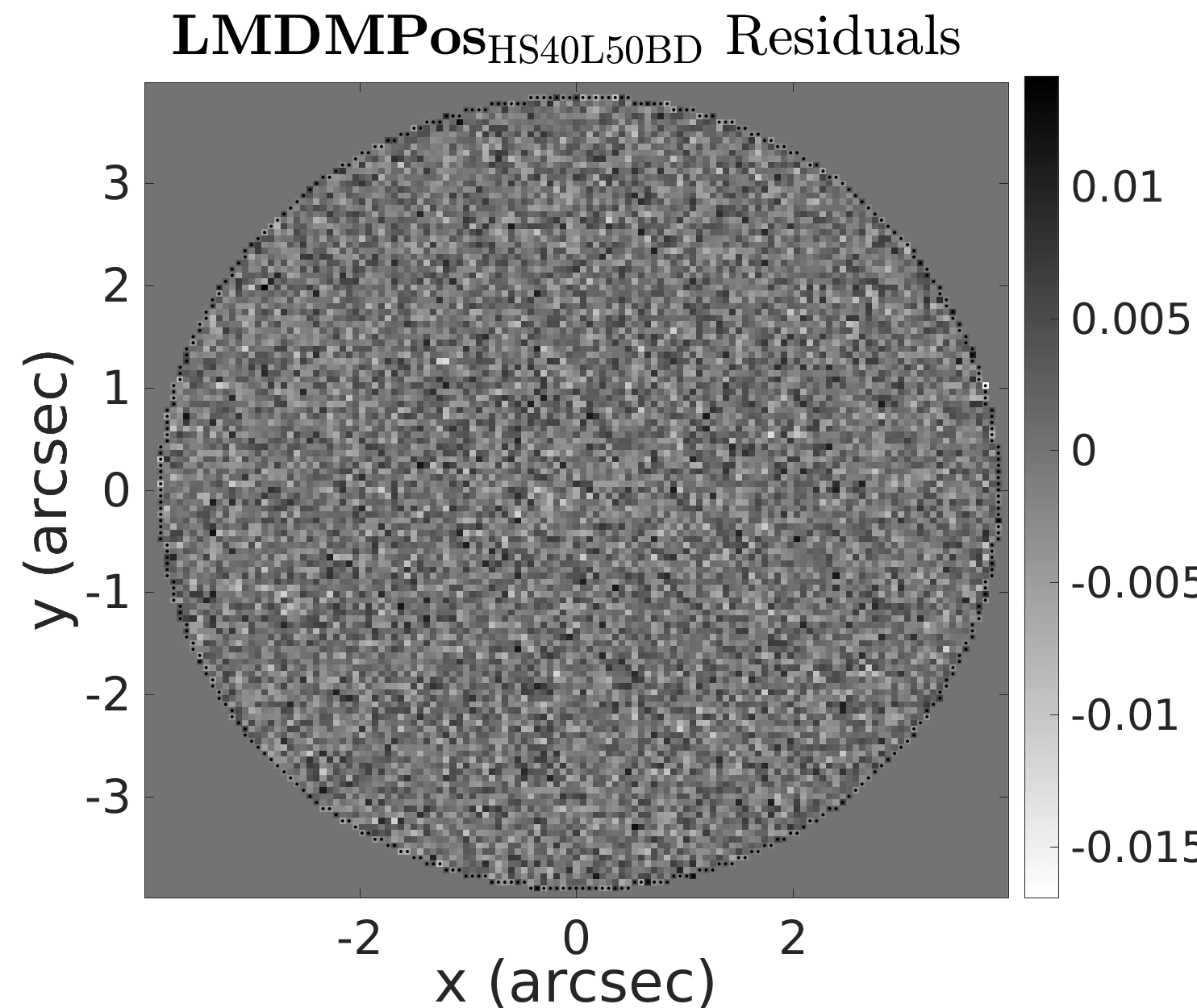}
\includegraphics[width=0.157\textwidth]{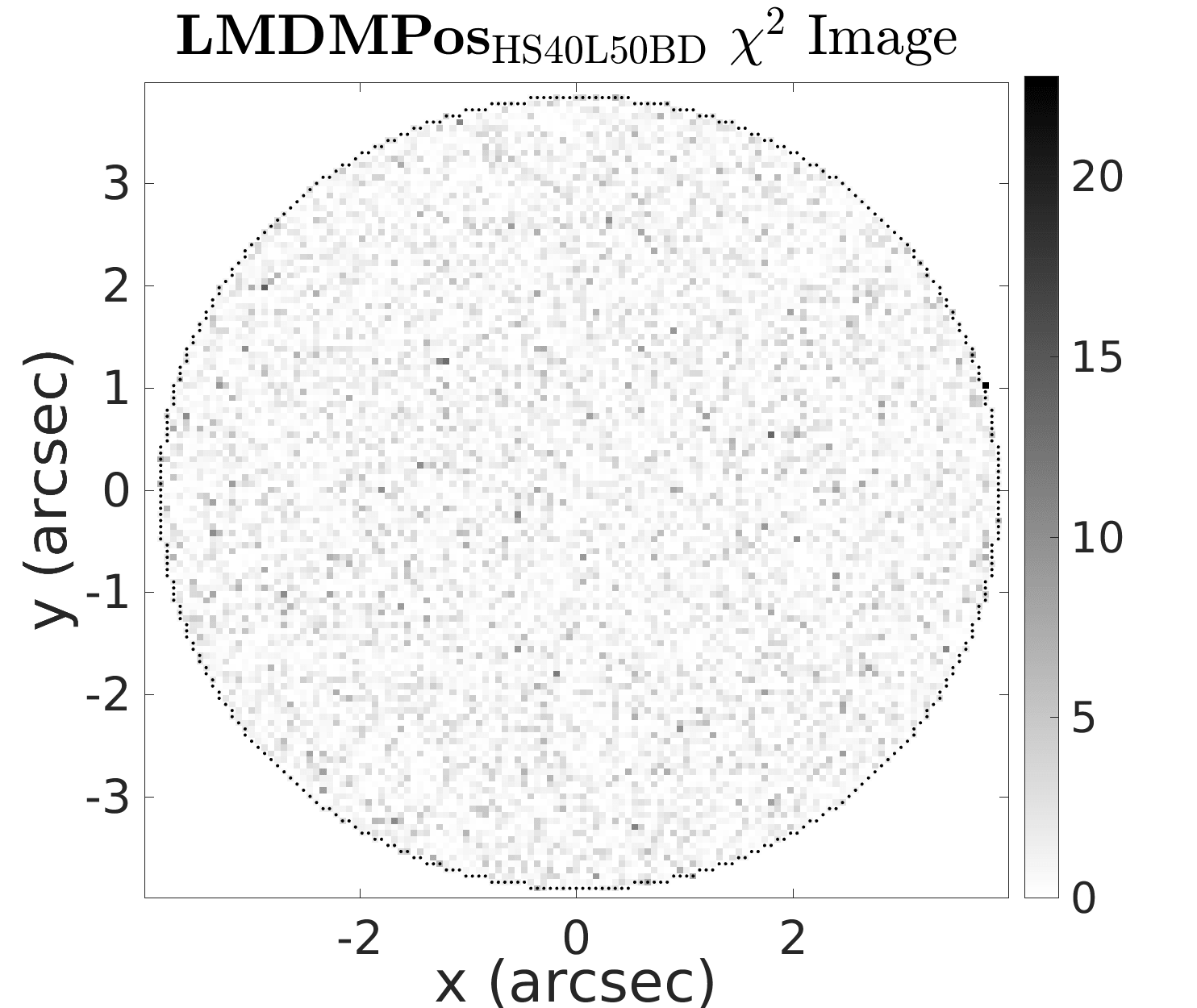}
\includegraphics[width=0.157\textwidth]{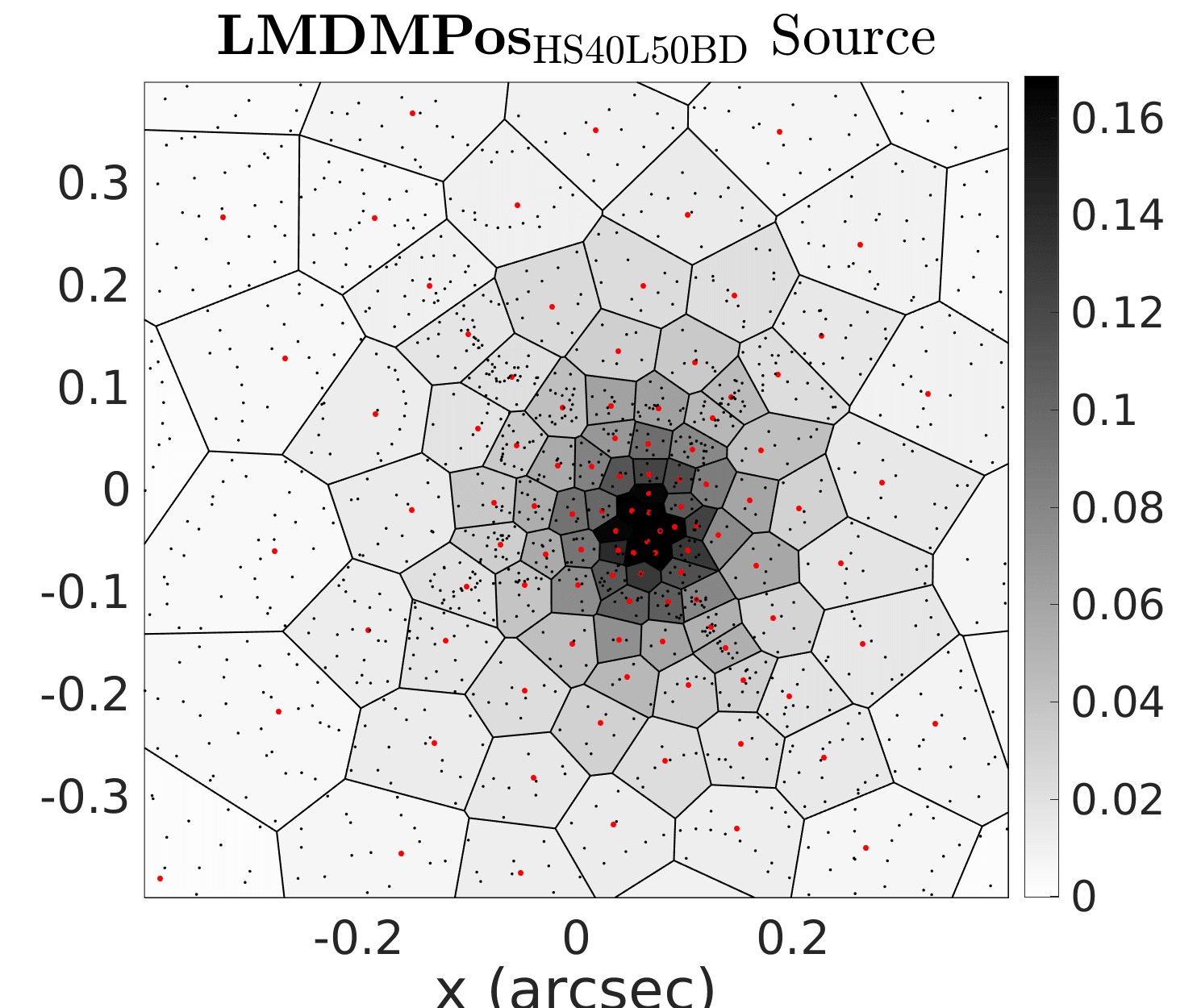}
\includegraphics[width=0.157\textwidth]{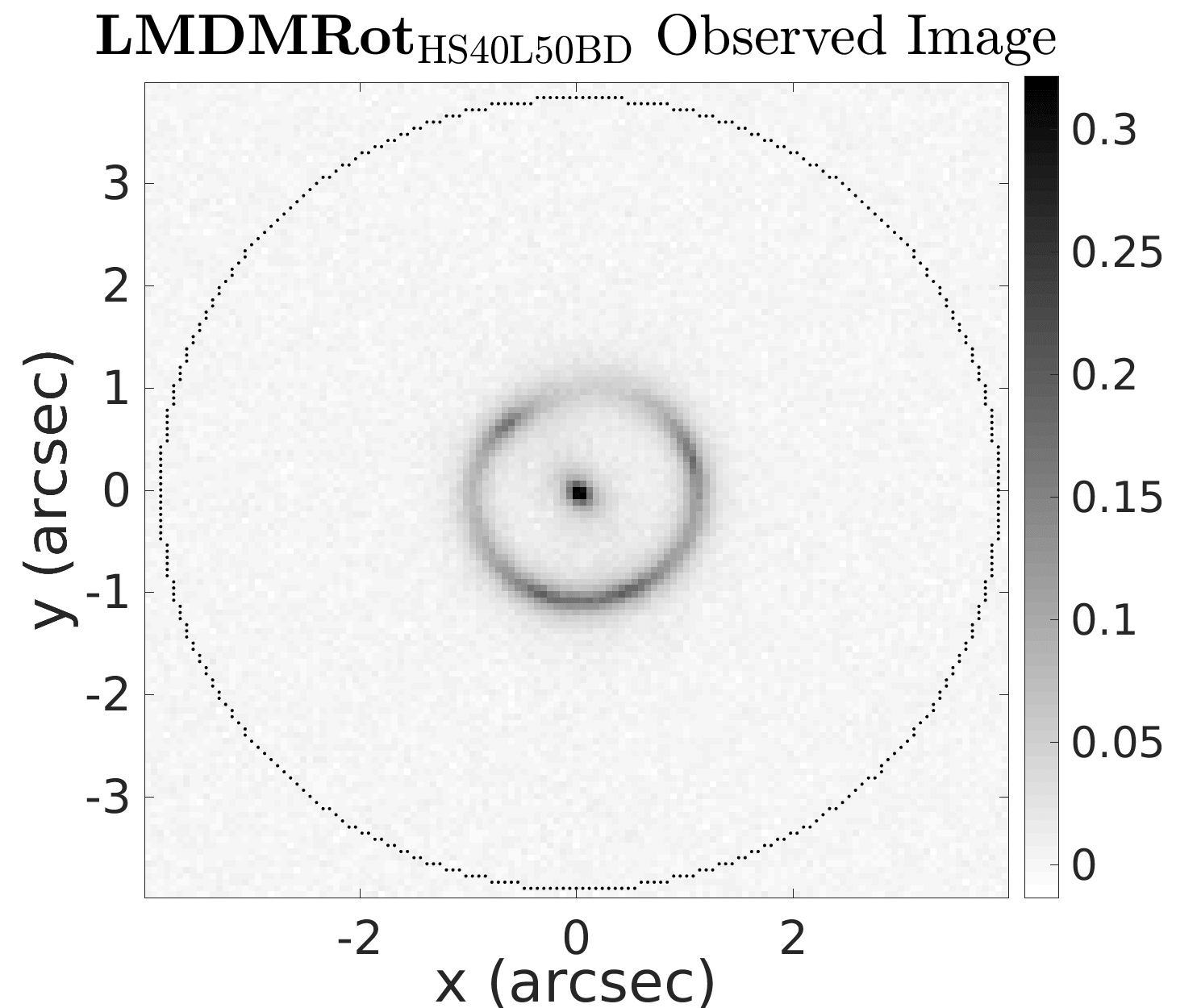}
\includegraphics[width=0.157\textwidth]{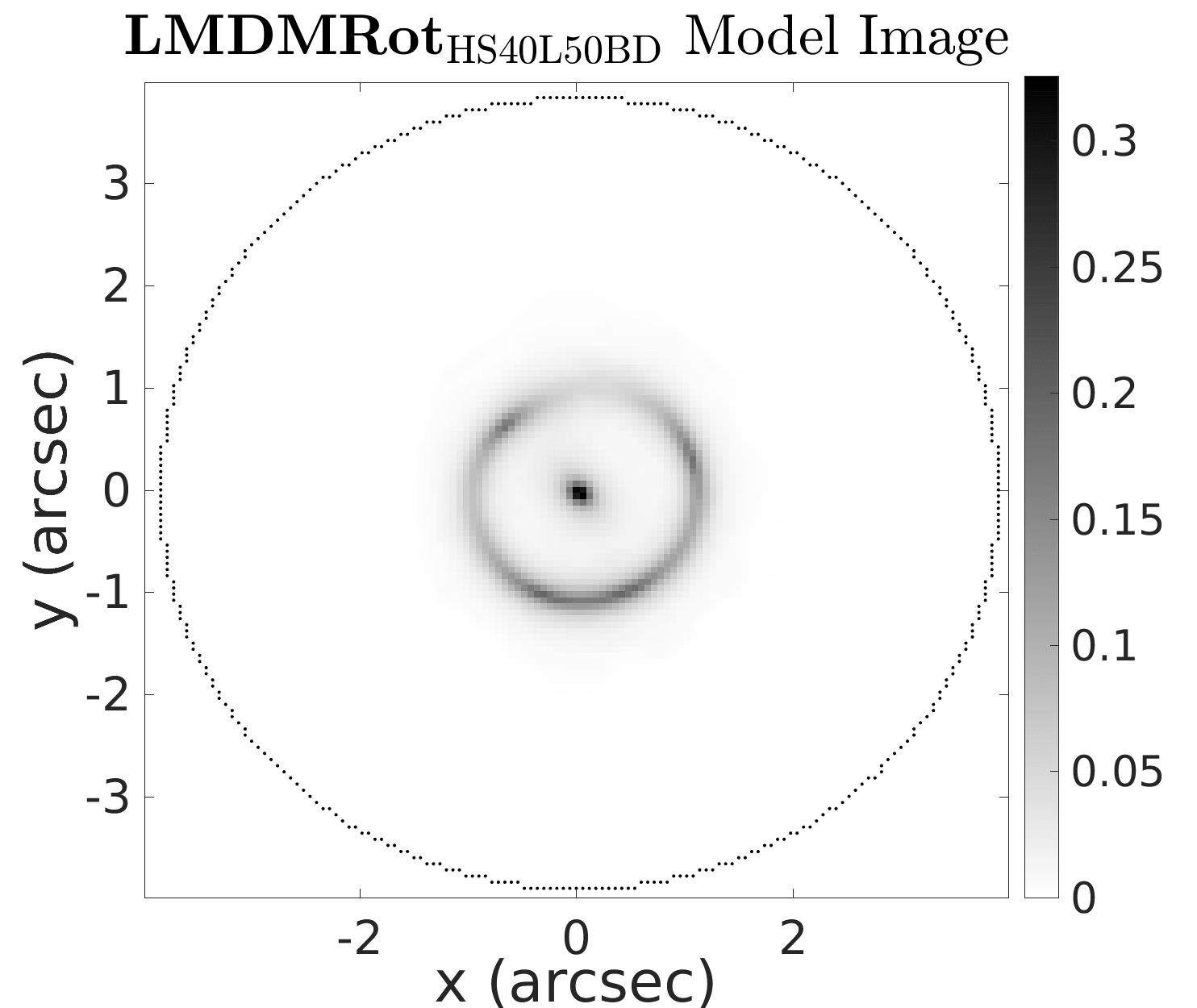}
\includegraphics[width=0.157\textwidth]{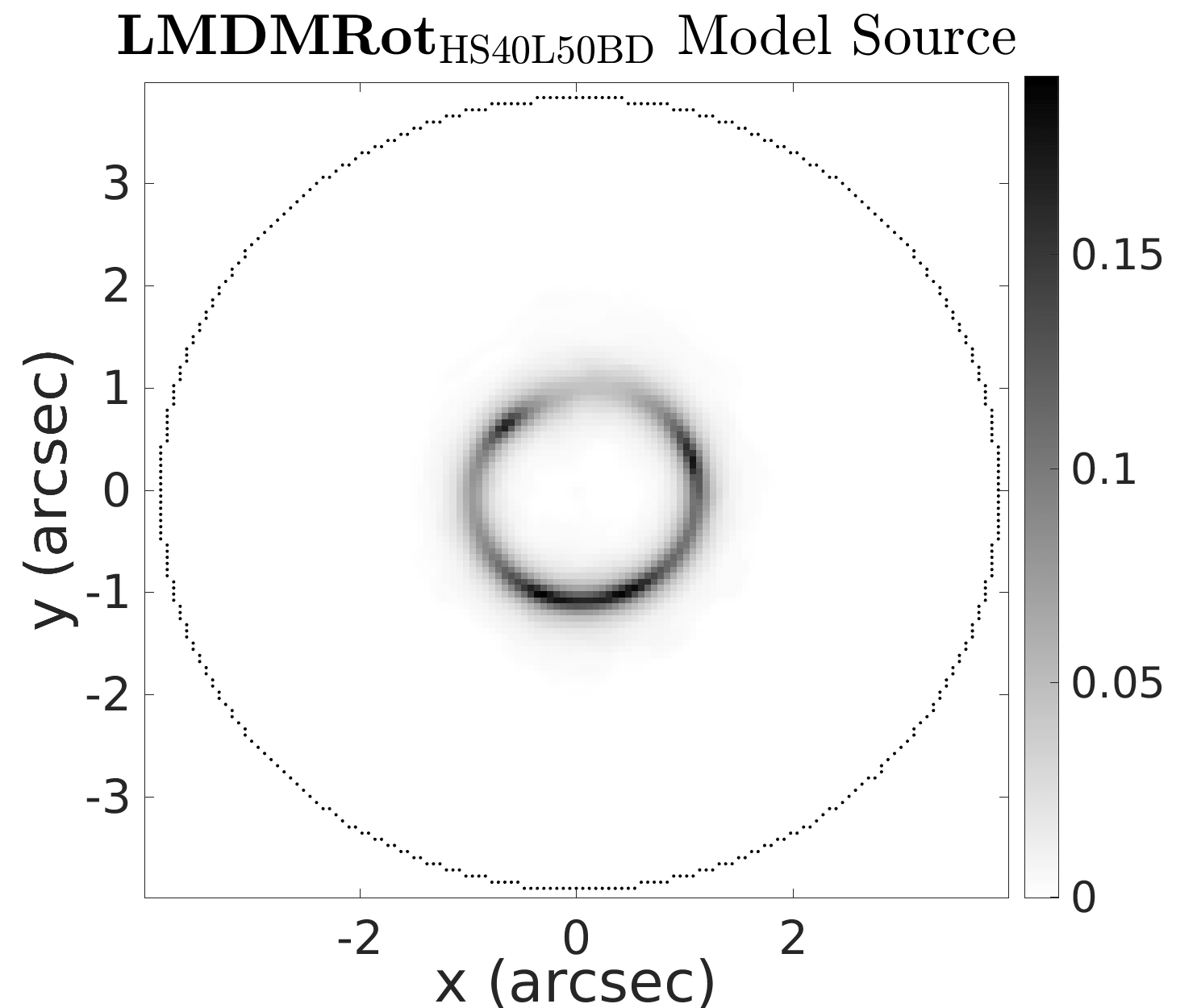}
\includegraphics[width=0.157\textwidth]{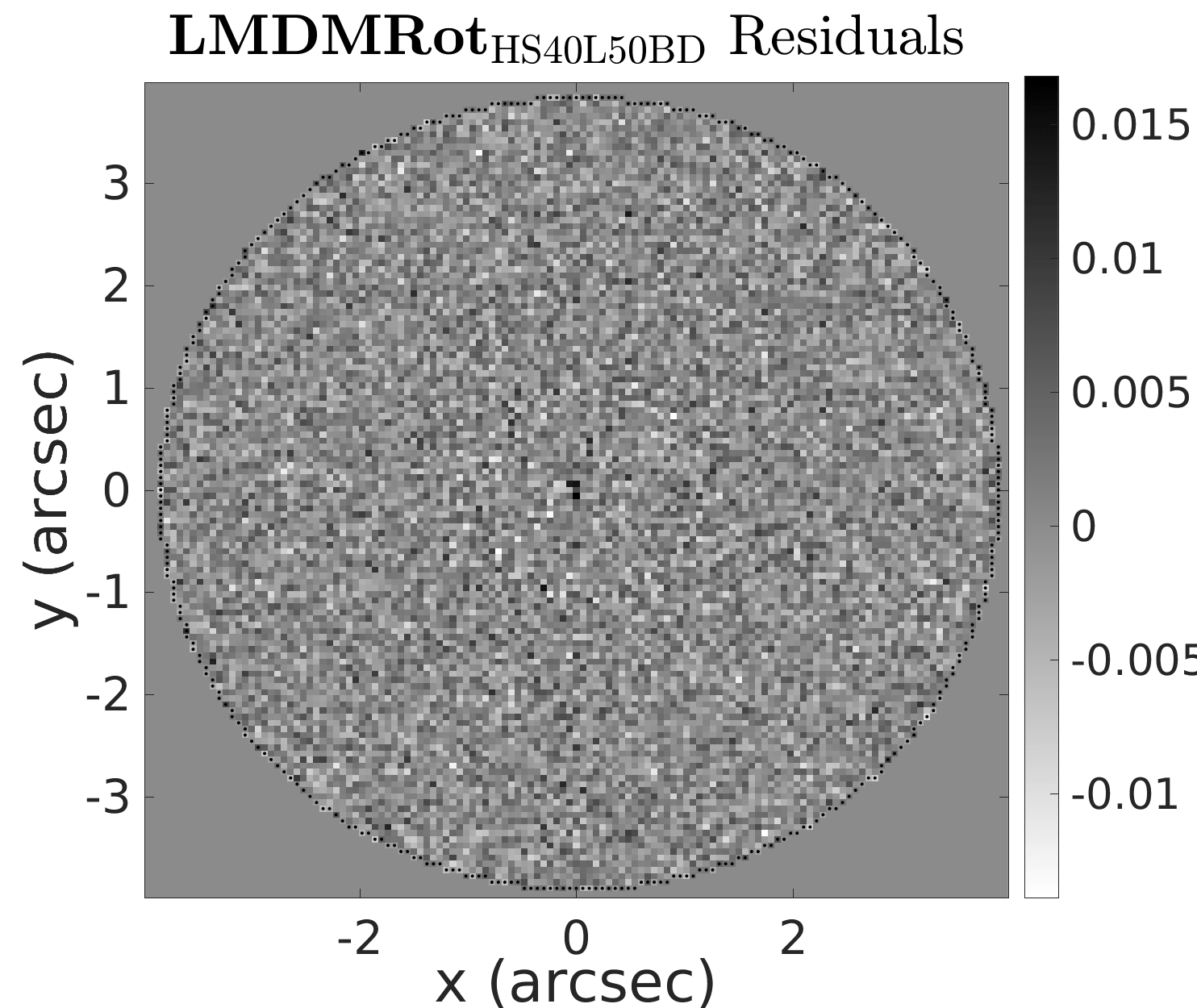}
\includegraphics[width=0.157\textwidth]{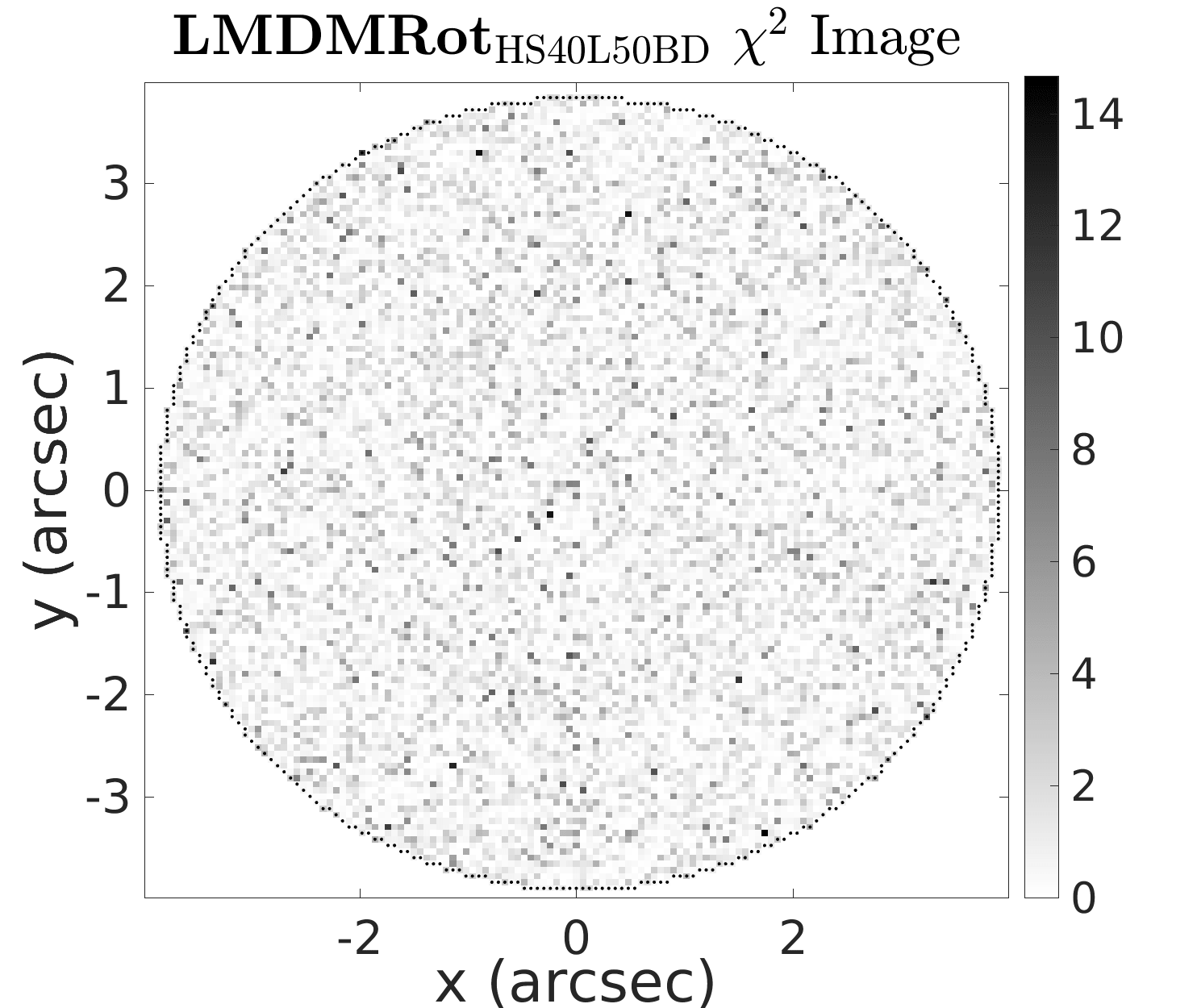}
\includegraphics[width=0.157\textwidth]{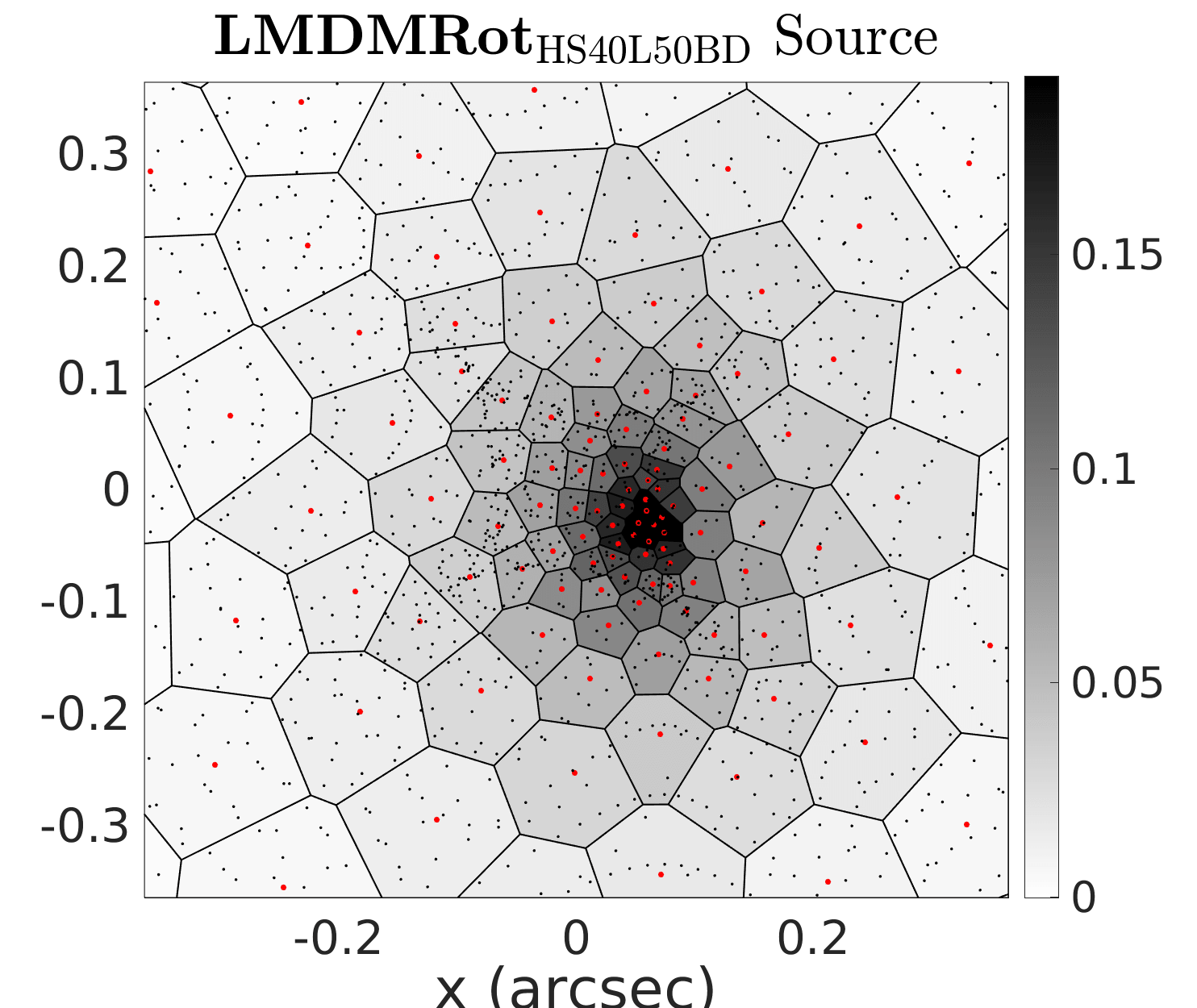}
\includegraphics[width=0.157\textwidth]{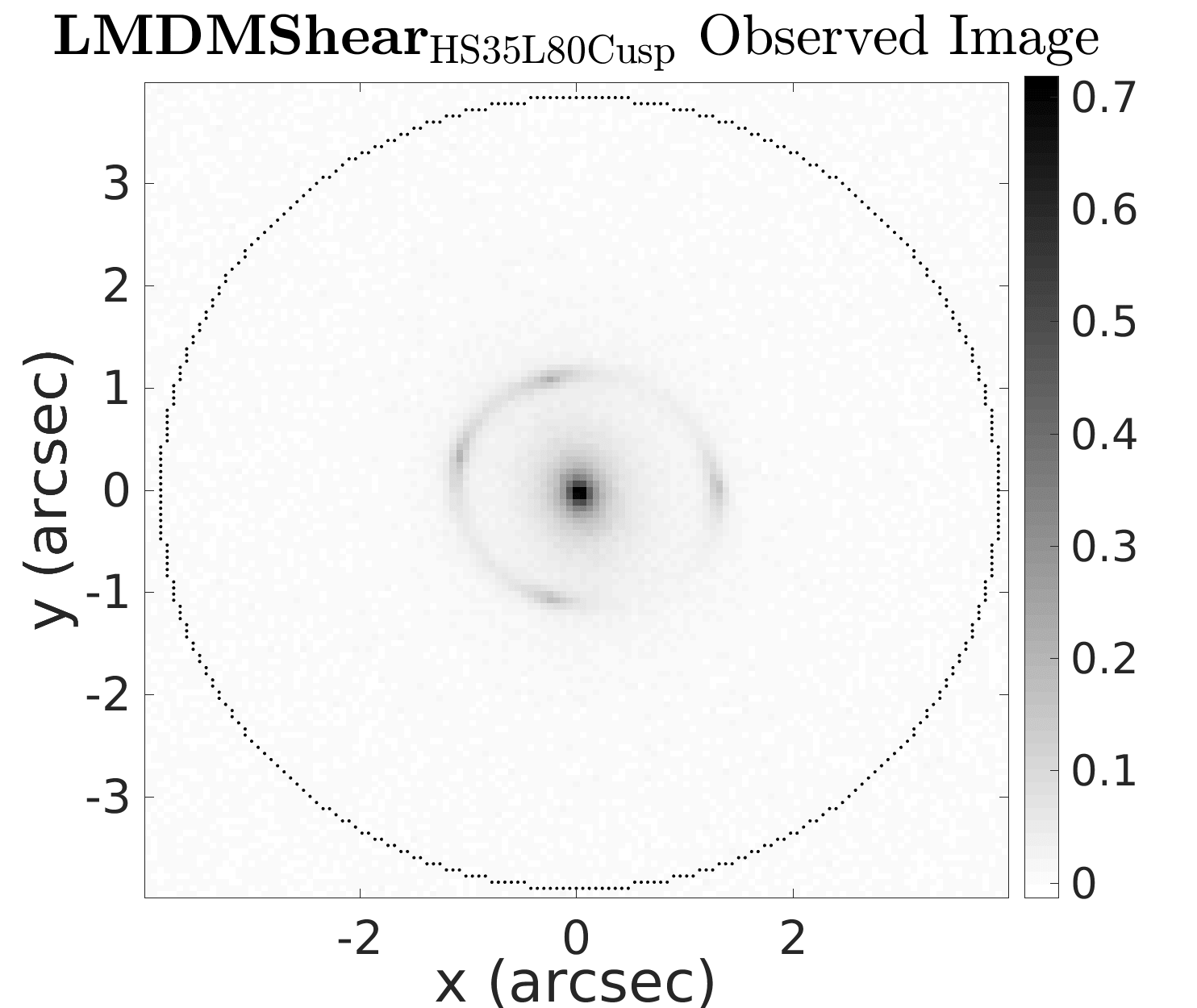}
\includegraphics[width=0.157\textwidth]{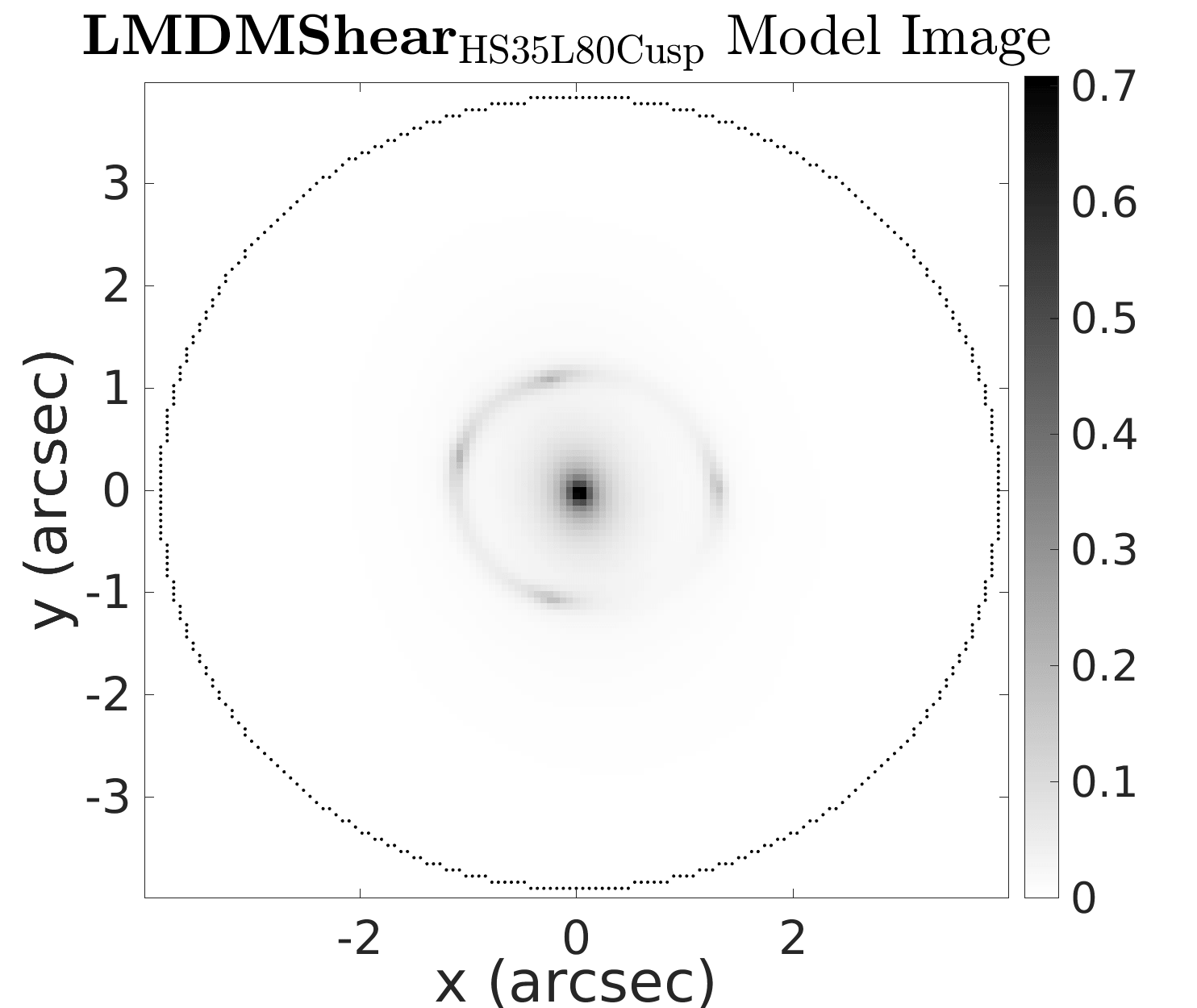}
\includegraphics[width=0.157\textwidth]{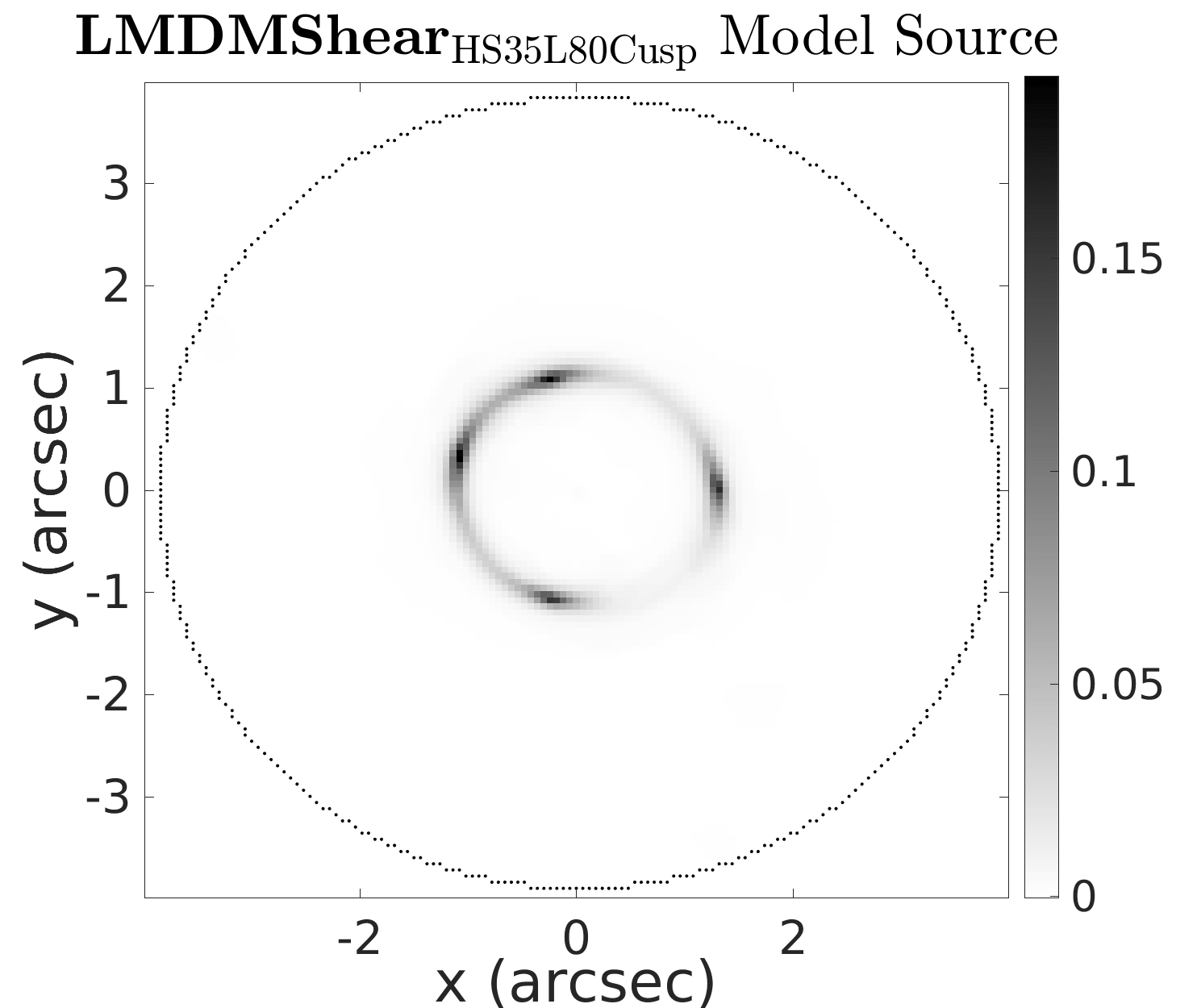}
\includegraphics[width=0.157\textwidth]{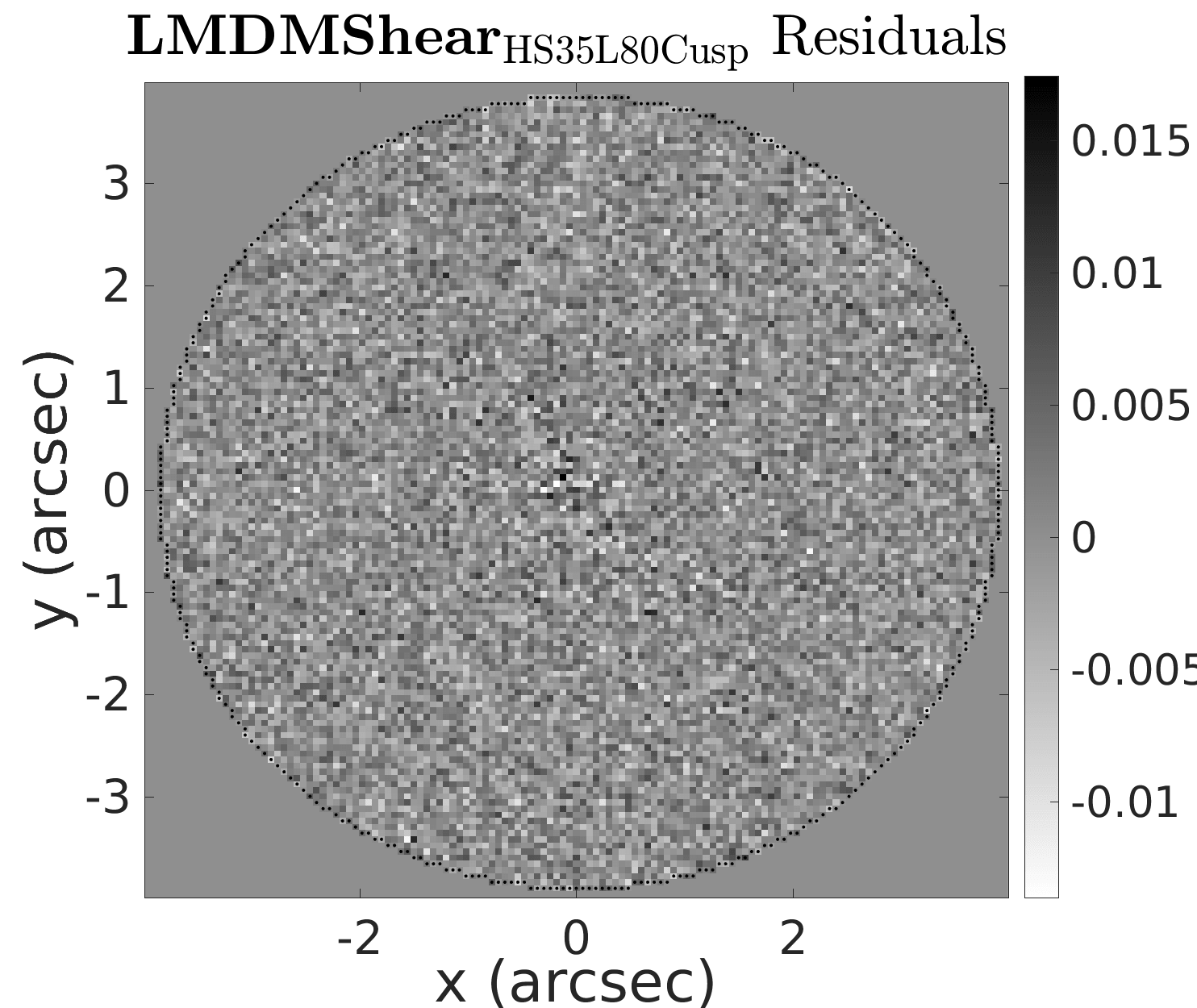}
\includegraphics[width=0.157\textwidth]{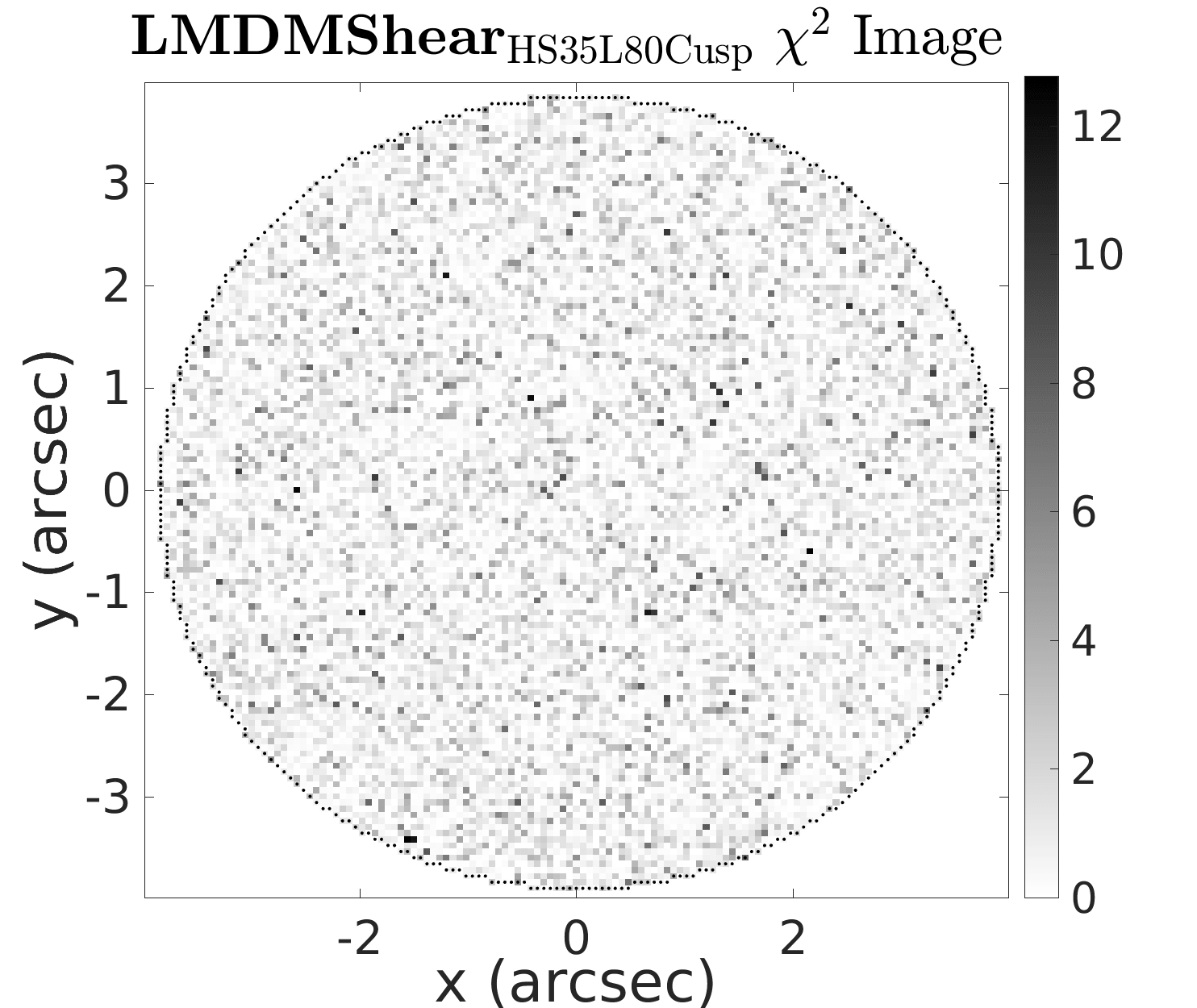}
\includegraphics[width=0.157\textwidth]{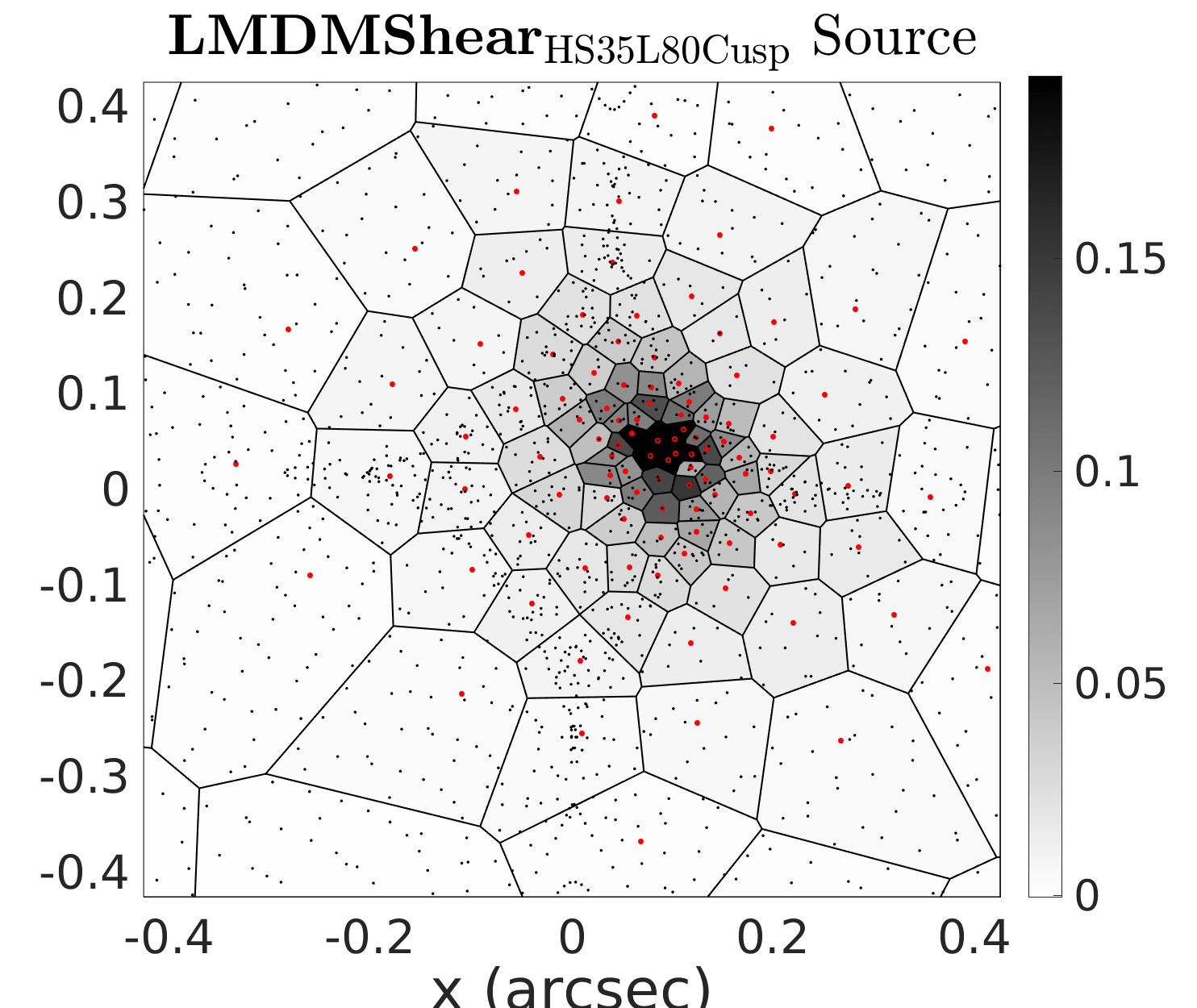}
\caption{The simulated image, model image, source model image, residuals, $\chi^2$ image and source reconstructions for analysis of the images $\mathbf{LMDMPos}_{\mathrm{HS40L50BD}}$, $\mathbf{LMDMRot}_{\mathrm{ES40L50Multi}}$ and $\mathbf{LMDMShear}_{\mathrm{HS35L80Cusp}}$. Images correspond to the most probable model at the end of the main pipeline, corresponding to the models given by table \ref{table:TableLMDM}.} 
\label{figure:ResultsLMDMIms}
\end{figure*}

\begin{table*}
\scriptsize
\centering
\begin{tabular}{ l | l | l || l | l || l | l | l } 
\multicolumn{1}{p{1.0cm}|}{\centering \textbf{Image}} 
& \multicolumn{1}{p{1.3cm}|}{\centering \textbf{Sersic} \\ ($\textbf{P}_{\rm MCLight}$) } 
& \multicolumn{1}{p{1.3cm}||}{\centering \textbf{Sersic} \\  \textbf{+ Exp} \\ ($\textbf{P}_{\rm MCLight}$) } 
& \multicolumn{1}{p{1.3cm}|}{\centering \textbf{SPLE} \\ ($\textbf{P}_{\rm MCMass}$)  } 
& \multicolumn{1}{p{1.3cm}||}{\centering \textbf{SPLE + Shear} \\ ($\textbf{P}_{\rm MCMass}$) } 
& \multicolumn{1}{p{1.3cm}|}{\centering \textbf{NFW (Align)} \\ ($\textbf{P}_{\rm MCGeom}$)  } 
& \multicolumn{1}{p{1.3cm}|}{\centering \textbf{NFW (Rot)} \\ ($\textbf{P}_{\rm MCGeom}$)  } 
& \multicolumn{1}{p{1.3cm}}{\centering \textbf{NFW (Pos)} \\ ($\textbf{P}_{\rm MCGeom}$)  } 
\\ 
& & & & & & & \\[-5pt]
\hline
& & & & & & & \\[-4pt]
$\mathbf{LMDMAlign}_{\mathrm{HS40L50BD}}$ & \textbf{55633.3189} & 55640.7631 & \textbf{55647.3234} & 55651.7835 & \textbf{55643.8825} & 55644.4506 & 55651.3071\\[2pt]
$\mathbf{LMDMAlign}_{\mathrm{ES40L50BD}}$ & \textbf{15820.3812} & 15824.0408 & \textbf{15843.2371} & 15841.4428 & \textbf{15831.0632} & 15833.0704 & 15829.2484\\[2pt]
\hline
& & & & & & & \\[-4pt]
$\mathbf{LMDMPos}_{\mathrm{HS40L50BD}}$ & \textbf{55628.4640} & 55632.5936 & \textbf{55655.0241} & 55657.4802 & 55612.5394 & 55612.7578 & \textbf{55659.5510}\\[2pt]
$\mathbf{LMDMPos}_{\mathrm{ES40L50BD}}$ & \textbf{15919.0649} & 15920.2272 & \textbf{15944.3584} & 15948.5535 & \textbf{15913.5722} & 15910.0842 & 15928.2960\\[2pt]
\hline
& & & & & & & \\[-4pt]
$\mathbf{LMDMRot}_{\mathrm{HS40L50BD}}$ & \textbf{55413.7926} & 55424.4787 & \textbf{55469.5119} & 55467.4304 & \textbf{55449.3334} & 55456.2422 & 55443.2313\\[2pt]
$\mathbf{LMDMRot}_{\mathrm{ES40L50BD}}$ & \textbf{15840.9690} & 15842.9032 & \textbf{15851.6601} & 15852.1060 & \textbf{15840.3070} & 15842.4563 & 15841.4788\\[2pt]
\hline
& & & & & & & \\[-4pt]
$\mathbf{LMDMRot90}_{\mathrm{HS30L75Multi}}$ & \textbf{54794.3025} & 54795.2551 & 55195.8635 & \textbf{55265.1315} & 55263.0546 & \textbf{55287.5303} & 55262.7623\\[2pt]
$\mathbf{LMDMRot90}_{\mathrm{ES30L75Multi}}$ & \textbf{15070.9989} & 15066.1369 & \textbf{15367.6957} & 15379.6356 & \textbf{15372.5886} & 15371.7799 & 15377.4355\\[2pt]
\hline
& & & & & & & \\[-4pt]
$\mathbf{LMDMShear}_{\mathrm{HS35L80Cusp}}$ & \textbf{55363.3473} & 55341.4098 & 55462.1340 & \textbf{55593.4590} & 55538.4963 & 55548.8675 & \textbf{55577.3864}\\[2pt]
$\mathbf{LMDMShear}_{\mathrm{ES35L80Cusp}}$ & \textbf{15621.6493} & 15620.9123 & 15654.5684 & \textbf{15695.2625} & \textbf{15686.0568} & 15690.4766 & 15695.1630\\[2pt]
\end{tabular}
\caption{The results of Bayesian model comparison in the phases $\textbf{P}_{\rm  LightMC}$, $\textbf{P}_{\rm  MassMC}$ and $\textbf{P}_{\rm  GeomMC}$ for the ten $\textbf{LMDM}$ images, which test decomposed mass modeling. Image names are listed in the first column. The second and third columns show the Bayesian evidence values (equation \ref{eqn:Bayes2}) computed for the $Sersic$ and $Sersic$ + $Exp$ light models, the fourth and fifth columns evidences for comparison of the $SPLE$ and $SPLE$+$Shear$ mass models and final three columns the evidences for the geometrically aligned, rotationally offset and / or positionally offset $Light$ + $NFW$ model. Values in bold correspond to those chosen by the pipeline, noting that a threshold of twenty must be exceed to favour a more complex model.}
\label{table:LMDMMC}
\end{table*}

The decomposed lens simulation suite consists of five unique lens and source models, each of which are used to generate an image at Hubble and Euclid resolution, with source $S/N = 30-40$ and lens $S/N = 50-80$, giving a total of ten images. Each image is analyzed using the decomposed profile pipeline, producing a $Light$ + $NFW$ model. The results of model comparison are shown in table \ref{table:LMDMMC}. The correct light model and inclusion of an external shear is correct for all models, however a number of light and dark matter geometries are not consistent with their input models. Parameter estimates for each image are summarized in table \ref{table:TableLMDM}, using $\Delta R_{\rm l}$, $\Delta n_{\rm l}$ $\Delta q_{\rm l}$, $\Delta \kappa_{\rm d}$, $\Delta q_{\rm d}$ and $\Delta \Psi_{\rm l}$. Figure \ref{figure:ResultsLMDMIms} shows the observed images, model images, model sources, residuals, $\chi^2$ images and source reconstructions for three images, showing residuals and $\chi^2$ images which realize the image's noise.

\subsubsection{Model Comparison}

\begin{table*}
\resizebox{\linewidth}{!}{
\begin{tabular}{ l | l | l l l | l} 
\multicolumn{1}{p{1.6cm}|}{Image} 
& \multicolumn{1}{p{1.8cm}|}{\centering \textbf{Component}} 
& \multicolumn{1}{p{2.2cm}}{\textbf{Parameters ($3\sigma$)}} 
& \multicolumn{1}{p{2.2cm}}{}  
& \multicolumn{1}{p{2.2cm}|}{} 
& \multicolumn{1}{p{2.2cm}}{\textbf{Parameters ($1\sigma$)}} 
\\ \hline
& & & & & \\[-4pt]
$\mathbf{LMDMAlign}_{\mathrm{HS40L50BD}}$ & Light (\textbf{Sersic}) & $\Delta R_{\mathrm{l}}=0.0675^{+0.1676}_{\rm -0.1667}$($R_{\mathrm{l}}=0.60$) & $\Delta q_{\mathrm{l}}=-0.0033^{+0.0396}_{\rm -0.0357}$($q_{\mathrm{l}}=0.72$) & $\Delta n_{\mathrm{l}}=0.0587^{+0.5377}_{\rm -0.5343}$($n_{\mathrm{l}}=4.00$) & $\Delta n_{\mathrm{l}}=0.0587^{+0.2059}_{\rm -0.2118}$($n_{\mathrm{l}}=4.00$)\\[2pt]
$\mathbf{LMDMAlign}_{\mathrm{HS40L50BD}}$ & Mass (\textbf{NFW + $\Psi_{\mathrm{l}}$}) & $\Delta \kappa_{\mathrm{d}}=-0.0271^{+0.0294}_{\rm -0.0297}$($\kappa_{\mathrm{d}}=0.13$) & $\Delta q_{\mathrm{d}}=-0.0523^{+0.0821}_{\rm -0.1042}$($q_{\mathrm{d}}=0.82$) & $\Delta \Psi_{\mathrm{l}}=5.3773^{+6.5113}_{\rm -6.6825}$($\Psi_{\mathrm{l}}=42.00$) & $\mathbf{\Delta \Psi_{\mathrm{l}}=5.3773^{+3.0125}_{\rm -2.3066}}$($\Psi_{\mathrm{l}}=42.00$)\\[2pt]
$\mathbf{LMDMAlign}_{\mathrm{ES40L50BD}}$ & Light (\textbf{Sersic}) & $\Delta R_l=0.1544^{+0.2302}_{\rm -0.2259}$($R_l=0.60$) & $\Delta q_l=-0.0167^{+0.0419}_{\rm -0.0370}$($q_l=0.72$) & $\Delta n_l=0.3100^{+0.7137}_{\rm -0.7622}$($n_l=4.00$) & $\mathbf{\Delta n_l=0.3100^{+0.2798}_{\rm -0.2705}}$($n_l=4.00$)\\[2pt]
$\mathbf{LMDMAlign}_{\mathrm{ES40L50BD}}$ & Mass (\textbf{NFW + $\Psi_{\mathrm{l}}$}) & $\Delta \kappa_{d}=-0.0368^{+0.0568}_{\rm -0.0512}$($\kappa_{d}=0.13$) & $\Delta q=-0.0843^{+0.1523}_{\rm -0.2575}$($q=0.82$) & $\Delta \Psi_{l}=1.3238^{+2.0777}_{\rm -2.2659}$($\Psi_{l}=6.74$) & $\mathbf{\Delta \Psi_{l}=1.3238^{+0.8504}_{\rm -0.7660}}$($\Psi_{l}=6.74$)\\[-4pt]
& & & & & \\[-4pt]
\hline
& & & & & \\[-4pt]
$\mathbf{LMDMRot}_{\mathrm{HS40L50BD}}$ & Light (\textbf{Sersic}) & $\Delta R_{\mathrm{l}}=0.0833^{+0.1777}_{\rm -0.1639}$($R_{\mathrm{l}}=0.60$) & $\Delta q_{\mathrm{l}}=-0.0090^{+0.0418}_{\rm -0.0396}$($q_{\mathrm{l}}=0.72$) & $\Delta n_{\mathrm{l}}=0.1747^{+0.5322}_{\rm -0.6237}$($n_{\mathrm{l}}=4.00$) & $\Delta n_{\mathrm{l}}=0.1747^{+0.2314}_{\rm -0.1820}$($n_{\mathrm{l}}=4.00$)\\[2pt]
$\mathbf{LMDMRot}_{\mathrm{HS40L50BD}}$ & Mass (\textbf{NFW + $\Psi_{\mathrm{l}}$}) & $\Delta \kappa_{\mathrm{d1}}=0.0014^{+0.0334}_{\rm -0.0408}$($\kappa_{\mathrm{d1}}=0.13$) & $\Delta q_{\mathrm{d1}}=0.0151^{+0.0531}_{\rm -0.0784}$($q_{\mathrm{d1}}=0.82$) & $\Delta \Psi_{\mathrm{l2}}=-1.6112^{+9.5929}_{\rm -8.1477}$($\Psi_{\mathrm{l2}}=42.00$) & $\Delta \Psi_{\mathrm{l2}}=-1.6112^{+2.6403}_{\rm -2.8220}$($\Psi_{\mathrm{l2}}=42.00$)\\[2pt]
$\mathbf{LMDMRot}_{\mathrm{ES40L50BD}}$ & Light (\textbf{Sersic}) & $\Delta R_l=0.0371^{+0.1695}_{\rm -0.1512}$($R_l=0.60$) & $\Delta q_l=-0.0135^{+0.0307}_{\rm -0.0309}$($q_l=0.72$) & $\Delta n_l=0.1219^{+0.5845}_{\rm -0.5974}$($n_l=4.00$) & $\Delta n_l=0.1219^{+0.2239}_{\rm -0.2083}$($n_l=4.00$)\\[2pt]
$\mathbf{LMDMRot}_{\mathrm{ES40L50BD}}$ & Mass (\textbf{NFW + $\Psi_{\mathrm{l}}$}) & $\Delta \kappa_{d}=-0.0257^{+0.0410}_{\rm -0.0433}$($\kappa_{d}=0.13$) & $\Delta q=-0.0572^{+0.0935}_{\rm -0.1600}$($q=0.82$) & $\Delta \Psi_{l}=1.0416^{+1.7427}_{\rm -1.6689}$($\Psi_{l}=6.74$) & $\mathbf{\Delta \Psi_{l}=1.0416^{+0.5385}_{\rm -0.5935}}$($\Psi_{l}=6.74$)\\[-4pt]
& & & & & \\[-4pt]
\hline
& & & & & \\[-4pt]
$\mathbf{LMDMPos}_{\mathrm{HS40L50BD}}$ & Light (\textbf{Sersic}) & $\Delta R_l=0.0142^{+0.1021}_{\rm -0.1114}$($R_l=0.60$) & $\Delta q_l=0.0013^{+0.0302}_{\rm -0.0332}$($q_l=0.72$) & $\Delta n_l=0.0451^{+0.4004}_{\rm -0.4506}$($n_l=4.00$) & $\Delta n_l=0.0451^{+0.2154}_{\rm -0.1641}$($n_l=4.00$)\\[2pt]
$\mathbf{LMDMPos}_{\mathrm{HS40L50BD}}$ & Mass (\textbf{NFW + $\Psi_{\mathrm{l}}$}) & $\mathbf{\Delta \kappa_{d}=-0.0318^{+0.0251}_{\rm -0.0273}}$($\kappa_{d}=0.13$) & $\mathbf{\Delta q=-0.0859^{+0.0827}_{\rm -0.1089}}$($q=0.82$) & $\mathbf{\Delta \Psi_{l}=8.6911^{+6.7850}_{\rm -6.8201}}$($\Psi_{l}=42.00$) & $\mathbf{\Delta \Psi_{l}=8.6911^{+2.5766}_{\rm -2.4224}}$($\Psi_{l}=42.00$)\\[2pt]
$\mathbf{LMDMPos}_{\mathrm{HS40L50BD}}$ & Geometry & $\Delta x_l=0.0005^{+0.0023}_{\rm -0.0022}$($x_l=0.00$) &  & $\Delta x=0.0105^{+0.0292}_{\rm -0.0247}$($x=0.05$) & $\mathbf{\Delta x=0.0105^{+0.0088}_{\rm -0.0104}}$($x=0.05$)\\[2pt]
$\mathbf{LMDMPos}_{\mathrm{HS40L50BD}}$ & Geometry & $\Delta y_l=0.0004^{+0.0023}_{\rm -0.0021}$($y_l=0.00$) &  & $\Delta y=-0.0086^{+0.0271}_{\rm -0.0322}$($y=0.00$) & $\Delta y=-0.0086^{+0.0099}_{\rm -0.0083}$($y=0.00$)\\[2pt]
$\mathbf{LMDMPos}_{\mathrm{ES40L50BD}}$ & Light (\textbf{Sersic}) & $\Delta R_l=0.1853^{+0.2576}_{\rm -0.2174}$($R_l=0.60$) & $\Delta q_l=0.0097^{+0.0410}_{\rm -0.0386}$($q_l=0.72$) & $\Delta n_l=0.6120^{+0.7859}_{\rm -0.7706}$($n_l=4.00$) & $\mathbf{\Delta n_l=0.6120^{+0.2903}_{\rm -0.3108}}$($n_l=4.00$)\\[2pt]
$\mathbf{LMDMPos}_{\mathrm{ES40L50BD}}$ & Mass (\textbf{NFW + $\Psi_{\mathrm{l}}$}) & $\Delta \kappa_{d}=-0.0041^{+0.0479}_{\rm -0.0555}$($\kappa_{d}=0.13$) & $\Delta q=0.0010^{+0.0888}_{\rm -0.1307}$($q=0.82$) & $\Delta \Psi_{l}=0.0345^{+2.1667}_{\rm -1.9210}$($\Psi_{l}=6.74$) & $\Delta \Psi_{l}=0.0345^{+0.7048}_{\rm -0.7664}$($\Psi_{l}=6.74$)\\[-4pt]
& & & & & \\[-4pt]
\hline
& & & & & \\[-4pt]
$\mathbf{LMDMRot90}_{\mathrm{HS30L75Multi}}$ & Light (\textbf{Sersic}) & $\Delta R_{\mathrm{l}}=0.0346^{+0.1457}_{\rm -0.1774}$($R_{\mathrm{l}}=0.60$) & $\Delta q_{\mathrm{l}}=0.0037^{+0.0242}_{\rm -0.0280}$($q_{\mathrm{l}}=0.72$) & $\Delta n_{\mathrm{l}}=0.1452^{+0.3286}_{\rm -0.3463}$($n_{\mathrm{l}}=4.00$) & $\mathbf{\Delta n_{\mathrm{l}}=0.1452^{+0.1807}_{\rm -0.1424}}$($n_{\mathrm{l}}=4.00$)\\[2pt]
$\mathbf{LMDMRot90}_{\mathrm{HS30L75Multi}}$ & Mass (\textbf{NFW + $\Psi_{\mathrm{l}}$}) & $\Delta \kappa_{\mathrm{d1}}=0.0456^{+0.0543}_{\rm -0.0589}$($\kappa_{\mathrm{d1}}=0.13$) & $\Delta q_{\mathrm{d1}}=0.0873^{+0.0678}_{\rm -0.0937}$($q_{\mathrm{d1}}=0.82$) & $\Delta \Psi_{\mathrm{l2}}=-6.743^{+8.2456}_{\rm -9.456}$($\Psi_{\mathrm{l2}}=42.0$) & $\mathbf{\Delta \Psi_{\mathrm{l2}}=-6.743^{+4.1238}_{\rm -4.4573}}$($\Psi_{\mathrm{l2}}=42.0$)\\[2pt]
$\mathbf{LMDMRot90}_{\mathrm{HS30L75Multi}}$ & Mass (\textbf{Shear}) & $\mathbf{\Delta \gamma_{\mathrm{sh}}=0.0022^{+0.0020}_{\rm -0.0021}}$ &  & $\mathbf{\Delta \theta_{\mathrm{sh}}=111.3445^{+27.5354}_{\rm -23.4699}}$ & $\mathbf{\Delta \theta_{\mathrm{sh}}=111.3445^{+7.3562}_{\rm -7.2346}}$ \\[2pt]
$\mathbf{LMDMRot90}_{\mathrm{HS30L75Multi}}$ & Geometry & $\Delta \theta_{\mathrm{l}}=0.2368^{+0.6835}_{\rm -0.3861}$($\theta_{\mathrm{l}}=127.0$) & & $\Delta \theta_{\mathrm{d}}=1.253^{+3.3573}_{\rm -3.3842}$($\theta_{\mathrm{d}}=17.0$) & $\Delta \theta_{\mathrm{d}}=1.253^{+1.4686}_{\rm -1.4921}$($\theta_{\mathrm{d}}=17.0$)\\[2pt]
$\mathbf{LMDMRot90}_{\mathrm{ES30L75Multi}}$ & Light (\textbf{Sersic}) & $\Delta R_{\mathrm{l}}=0.0665^{+0.0920}_{\rm -0.1174}$($R_{\mathrm{l}}=0.60$) & $\Delta q_{\mathrm{l}}=0.0043^{+0.0340}_{\rm -0.0380}$($q_{\mathrm{l}}=0.72$) & $\Delta n_{\mathrm{l}}=0.1952^{+0.4069}_{\rm -0.4900}$($n_{\mathrm{l}}=4.00$) & $\mathbf{\Delta n_{\mathrm{l}}=0.1952^{+0.1807}_{\rm -0.1424}}$($n_{\mathrm{l}}=4.00$)\\[2pt]
$\mathbf{LMDMRot90}_{\mathrm{ES30L75Multi}}$ & Mass (\textbf{NFW + $\Psi_{\mathrm{l}}$}) & $\mathbf{\Delta \kappa_{\mathrm{d1}}=0.0508^{+0.0289}_{\rm -0.0342}}$($\kappa_{\mathrm{d1}}=0.13$) & $\mathbf{\Delta q_{\mathrm{d1}}=0.1691^{+0.0152}_{\rm -0.0339}}$($q_{\mathrm{d1}}=0.82$) & $\mathbf{\Delta \Psi_{\mathrm{l2}}=-2.4666^{+1.3591}_{\rm -1.1391}}$($\Psi_{\mathrm{l2}}=6.74$) & $\mathbf{\Delta \Psi_{\mathrm{l2}}=-2.4666^{+0.3866}_{\rm -0.3988}}$($\Psi_{\mathrm{l2}}=6.74$)\\[2pt]
$\mathbf{LMDMRot90}_{\mathrm{ES30L75Multi}}$ & Mass (\textbf{Shear}) & $\mathbf{\Delta \gamma_{\mathrm{sh}}=0.0239^{+0.0070}_{\rm -0.0049}}$ &  & $\mathbf{\Delta \theta_{\mathrm{sh}}=108.0721^{+5.6003}_{\rm -4.6170}}$ & $\mathbf{\Delta \theta_{\mathrm{sh}}=108.0721^{+1.5561}_{\rm -1.9700}}$ \\[-4pt]
& & & & & \\[-4pt]
\hline
& & & & & \\[-4pt]
$\mathbf{LMDMShear}_{\mathrm{HS35L80Cusp}}$ & Light (\textbf{Sersic}) & $\mathbf{\Delta R_l=0.0572^{+0.0565}_{\rm -0.0518}}$($R_l=0.75$) & $\Delta q_l=-0.0020^{+0.0118}_{\rm -0.0105}$($q_l=0.75$) & $\mathbf{\Delta n_l=0.1083^{+0.1030}_{\rm -0.0996}}$($n_l=2.50$) & $\mathbf{\Delta n_l=0.1083^{+0.0366}_{\rm -0.0376}}$($n_l=2.50$)\\[2pt]
$\mathbf{LMDMShear}_{\mathrm{HS35L80Cusp}}$ & Mass (\textbf{NFW + $\Psi_{\mathrm{l}}$}) & $\mathbf{\Delta \kappa_{d}=-0.0989^{+0.0218}_{\rm -0.0157}}$($\kappa_{d}=0.13$) & $\Delta q=-0.0733^{+0.2089}_{\rm -0.2052}$($q=0.80$) & $\mathbf{\Delta \Psi_{l}=5.8475^{+0.9455}_{\rm -1.2747}}$($\Psi_{l}=11.52$) & $\mathbf{\Delta \Psi_{l}=5.8475^{+0.4979}_{\rm -0.3113}}$($\Psi_{l}=11.52$)\\[2pt]
$\mathbf{LMDMShear}_{\mathrm{HS35L80Cusp}}$ & Geometry & $\Delta x_l=0.0001^{+0.0014}_{\rm -0.0014}$($x_l=0.00$) &  & $\mathbf{\Delta x=0.1648^{+0.1530}_{\rm -0.1101}}$($x=0.03$) & $\mathbf{\Delta x=0.1648^{+0.0399}_{\rm -0.0522}}$($x=0.03$)\\[2pt]
$\mathbf{LMDMShear}_{\mathrm{HS35L80Cusp}}$ & Geometry & $\Delta y_l=0.0010^{+0.0019}_{\rm -0.0019}$($y_l=0.00$) &  & $\mathbf{\Delta y=0.1721^{+0.1918}_{\rm -0.1400}}$($y=0.03$) & $\mathbf{\Delta y=0.1721^{+0.0485}_{\rm -0.0630}}$($y=0.03$)\\[2pt]
$\mathbf{LMDMShear}_{\mathrm{ES35L80Cusp}}$ & Light (\textbf{Sersic}) & $\Delta R_l=0.0454^{+0.0673}_{\rm -0.0678}$($R_l=0.75$) & $\Delta q_l=-0.0064^{+0.0186}_{\rm -0.0184}$($q_l=0.75$) & $\Delta n_l=0.0839^{+0.1495}_{\rm -0.1597}$($n_l=2.50$) & $\mathbf{\Delta n_l=0.0839^{+0.0564}_{\rm -0.0533}}$($n_l=2.50$)\\[2pt]
$\mathbf{LMDMShear}_{\mathrm{ES35L80Cusp}}$ & Mass (\textbf{NFW + $\Psi_{\mathrm{l}}$}) & $\mathbf{\Delta \kappa_{d}=-0.0687^{+0.0604}_{\rm -0.0630}}$($\kappa_{d}=0.13$) & $\Delta q=-0.1507^{+0.2686}_{\rm -0.3081}$($q=0.80$) & $\mathbf{\Delta \Psi_{l}=0.7474^{+0.6398}_{\rm -0.6199}}$($\Psi_{l}=2.00$) & $\mathbf{\Delta \Psi_{l}=0.7474^{+0.2185}_{\rm -0.2259}}$($\Psi_{l}=2.00$)\\[2pt]
$\mathbf{LMDMShear}_{\mathrm{ES35L80Cusp}}$ & Mass (\textbf{Shear}) & $\mathbf{\Delta \gamma_{sh}=0.0185^{+0.0147}_{\rm -0.0121}}$($\gamma_{sh}=0.03$) &  & $\Delta \theta_{sh}=0.5427^{+13.9894}_{\rm -15.3828}$($\theta_{sh}=150.00$) & $\Delta \theta_{sh}=0.5427^{+5.5488}_{\rm -5.1392}$($\theta_{sh}=150.00$)\\[-4pt]
\end{tabular}
}
\caption{Results of fitting the ten $\textbf{LMDM}$ model images using the automated analysis pipeline, corresponding to results generated at the end of phase two. Each image's name is given in the first column and the light or mass model component in the second column. The third to sixth columns show parameter estimates, where each parameter is offset by $\Delta P = P_{\rm  true} - P_{\rm  model}$, such that zero corresponds to the input lens model. The input lens model values are given in brackets to the right of each parameter estimate. Parameters estimates are shown using $\Delta R_{\rm l}$, $\Delta q_{\rm l}$ and $\Delta n_{\rm l}$ for $Sersic$ light models, $\Delta R_{\rm l1}$, $\Delta q_{\rm l1}$, $\Delta R_{\rm l2}$ and $\Delta q_{\rm l2}$ for $Dev$ + $Exp$ light models, $\Delta \kappa_{\rm d}$ and $\Delta q_{\rm d}$ for $NFW$ mass models, $\Psi_{\rm l}$ for light mass models and $\gamma_{\rm sh}$ and $\theta_{\rm sh}$ for a $Shear$ component. Columns three to five show parameter estimates within $3 \sigma$ confidence and column six at $1 \sigma$. Parameter estimates in bold text are inconsistent with the input lens model at their stated error estimates. The other parameters not shown (e.g. $x_{\rm l}$, $\theta_{\rm l}$) are all estimated accurately within $3\sigma$.}
\label{table:TableLMDM}
\end{table*}

The results of the model comparison phase $\textbf{P}_{\rm  MCGeom}$ for all images is shown in table \ref{table:LMDMMC}. The input light profiles are chosen for every image, and the shear is chosen correctly for nine out of ten images. However, for five images, an offset in their light and dark matter geometry is not chosen, even though they are present in their lens's input profiles. In all cases, model comparison opts to choose the simpler (aligned) model over the input geometrically offset models. Therefore, the non-inclusion of these components can simply be attributed to the observed image's resolution and S/N being insufficient to offer a large enough increase in evidence to favour the more complex model. In many cases, the input model does increase the Bayesian evidence, but does not meet the threshold value of $20$. Nevertheless, detections are made for the high-resolution images of the $\mathbf{LMDMRot90}$ and $\mathbf{LMDMPos}$ images, demonstrating that geometric offsets can detected in strong lens imaging, but that such a detection requires higher quality data than the models discussed previously.

Whilst image quality is a driving factor in detecting geometric offsets, another important aspect is the presence of an external shear. In the absence of an external shear, no degeneracy is observed between the geometric parameters $\Delta x_{\rm d}$, $\Delta y_{\rm d}$ and $\Delta \theta_{\rm d}$ and those governing the lens's light or mass profiles. This is important, because it suggests that when inferring a geometric misalignment any assumptions related to the lens's mass distribution (e.g. the form of the dark matter profile or use of a constant mass to light ratio) may not be very important. However, when a shear is present, this is found to no longer be the case and a degeneracy emerges between the dark matter geometry parameters, the shear parameters $\gamma_{\rm sh}$ and $\theta_{\rm sh}$ and the mass profile's other parameters. Degeneracies between ellipticity, rotational misalignments and an external shear have been long established \citep{Keeton1997} and they are particularly problematic for decomposed mass modeling as it becomes the dark matter profile which makes up a smaller fraction of overall mass where one is trying to detect them.

The degeneracy between dark matter geometry and an external shear also explains why analysis of the high-resolution $\textbf{LMDMRot90}$ image incorrectly includes an external shear in the mass model. The $Shear$ component is chosen earlier in the pipeline using the axisymmetric $NFWSph$ profile, thus the $Shear$ is included as it mimics the effect of the rotational misalignment that the $NFWSph$ cannot capture. Later in the pipeline, when the $NFW$ + $Light$ model with a rotational offset is chosen, the shear magnitude $\gamma_{\rm sh}$ reduces to nearly zero, effectively removing the shear and giving an accurate lens model. In the future, it may prove beneficial to choose an external shear via model comparison independently in the $\textbf{P}_{\rm MCGeom}$ phase. This strategy will be considered in the future, where independent constraints on the shear from weak-lensing will also be considered.

In summary, {\tt AutoLens} can successfully detect geometrically misaligned light and dark matter profiles. This is because, in an analogous fashion to the cored profile earlier, geometrically misaligned mass components impart unique features into the lensed source's extended light profile which an axisymmetric model cannot fit, especially given the constraints placed on the light profile's geometry due to lens light fitting. However, based on these results, it is clear that the prospects for detecting geometric offsets (if present in nature) are heavily dependent on the quality of the imaging data, the size of the offset and the presence and magnitude of an external shear. Such an analysis may be beyond the reach of Euclid wide-field imaging, but within the realms of possibility for currently available Hubble imaging. Additional information from an independent measurement, such as weak-lensing, may be crucial, as it can offer independent constraints on the shear magnitude and direction.
\subsubsection{Modeling Results}\label{LMDMResults}

The results of the parameter estimates for the decomposed mass models are given in table \ref{table:TableLMDM}. For the first seven images, the results are positive, with the majority of lens model parameters estimated correctly within $3 \sigma$. However, the image $\mathbf{LMDMRot90}_{\mathrm{ES30L75Multi}}$, and both of the $\textbf{LMDMShear}$ images, are poorly estimated. In all cases, a rotational misalignment was not chosen when present in the input image, which likely contributes to this result. The large and inaccurate values of $x_{\rm d}$ and $y_{\rm d}$ for the high-resolution $\textbf{LMDMShear}$ image also suggest the analysis is unable to accurately resolve the degeneracy between geometric offset and external shear. This result reaffirms the caution that must be taken when attempting to model an external shear and geometric offset.

\begin{figure*}
\centering
\includegraphics[width=0.95\textwidth]{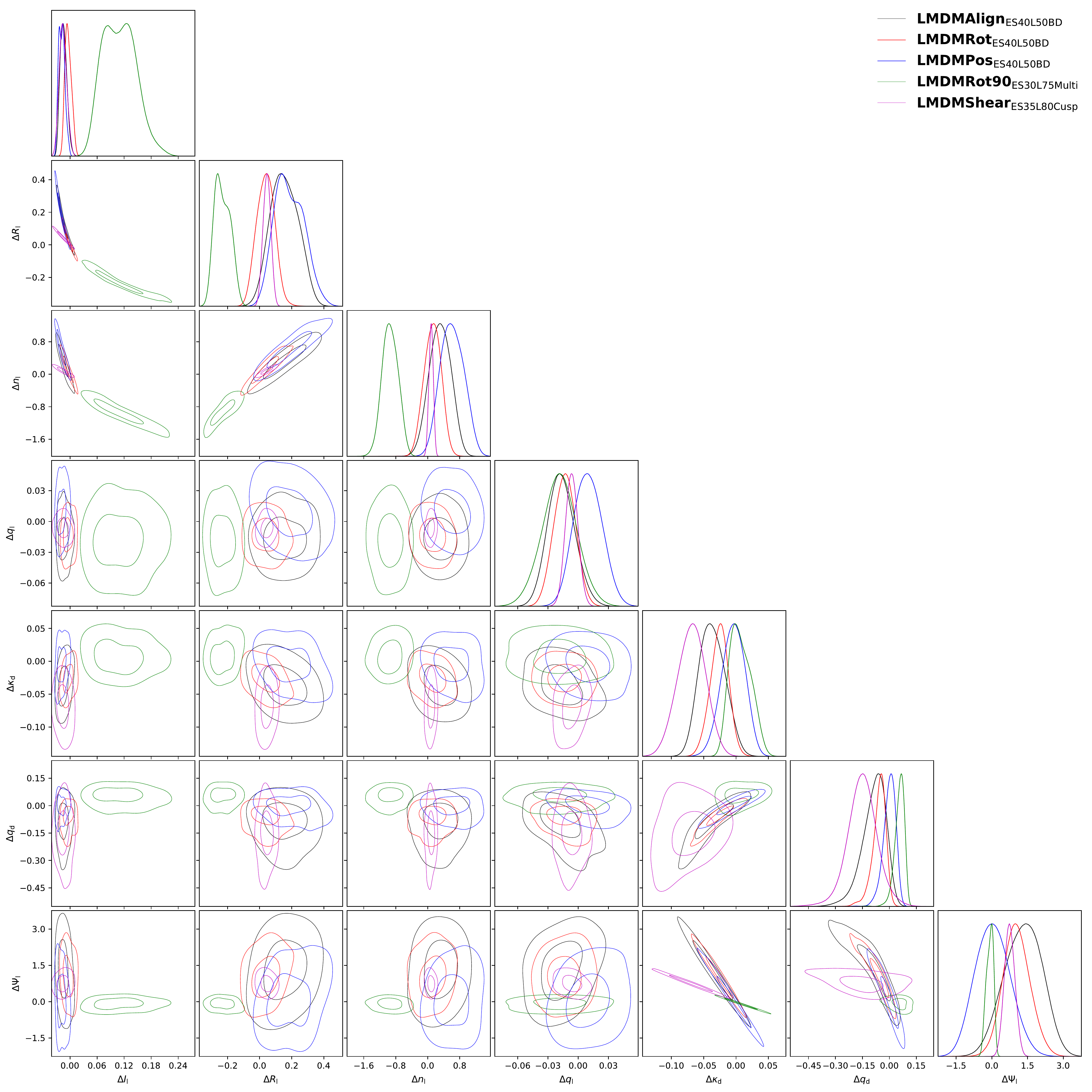}
\caption{Marginalized one and two-dimensional PDF's of the parameters governing the lens's mass distribution for decompossed mass modeling, $\Delta I_{\rm l}$, $\Delta R_{\rm l}$, $\Delta n_{\rm l}$, $\Delta q_{\rm l}$, $\Delta \kappa_{\rm d}$, $\Delta q_{\rm d}$ and $\Delta \Psi_{\rm l}$, for the low-resolution image's of the $\textbf{LMDM}$ models. The legends at the top indicates the image that each coloured line corresponds top. Contours give the $1\sigma$ (interior) and $3 \sigma$ (exterior) confidence regions. The input values for each model can be found in table \ref{table:SimModels}. A degeneracy can be seen between all parameters, an extension of the degeneracy between mass, ellipticity and slope described in N15, but now including components for both the light and dark matter profiles. However, fitting of the lens's light reduces this degeneracy thereby permitting accurate sampling.} 
\label{figure:PDFsLMDMHRes2D}
\end{figure*}

The most significant parameter degeneracies of the previous mass models were found between the parameters governing their mass distributions, which consisted of either three or four parameters. For a decomposed mass model, seven (or more for a multi-component light profile) parameters determine the lens's mass distribution; $I_{\rm l}$, $R_{\rm l}$, $n_{\rm l}$, $q_{\rm l}$, $\kappa_{\rm d}$, $q_{\rm d}$ and $\Psi_{\rm l}$. This would create an extremely complex and degenerate non-linear parameter space, from which lens models constraints are not possible, if it were not for the constraints placed on the light model parameters by the light profile's fit to the lens galaxy. Shown by figure \ref{figure:PDFsLMDMHRes2D} for the low-resolution images, this results in essentially no degeneracy emerging between the light profile parameters ($I_{\rm l}$, $R_{\rm l}$, $n_{\rm l}$) and dark matter parameters ($\kappa_{\rm d}$, $q_{\rm d}$), with their degeneracy instead folded into the mass-to-light ratio $\Psi_{\rm l}$. 

\section{Discussion \& Summary}\label{Diss3}

\subsection{Automated Analysis}

All results were generated in this work without user intervention, demonstrating that {\tt AutoLens} successfully automates the lens modeling process. High quality imaging of many hundreds of strong lenses exist in the HST archive, a data-set that has not been fully exploited due to the time overheads historically associated with lens modeling. Therefore, in the short-term, {\tt AutoLens} can significantly increase the total number of strong lenses with a complete lens model and source reconstruction. In the long term, consideration must be given to how this automated framework will be expanded to samples of lenses in the tens of thousands, which will be provided by surveys such as Euclid in less than five years \citep{Collett2014}. 

Of course, one can envisage scenarios where the automated analysis, as presented here, breaks down. These include the presence of luminous or dark high-mass substructures in the lens galaxy or lens morphologies whose light profile cannot be fitted accurately with the models used here (e.g. late-type lenses with complicated structure due to star formation and dust lanes). Therefore, as {\tt AutoLens} is expanded to larger samples it will inevitably require new functionality and the design philosophy, so far, has been to always develop automated solutions which offer a choice between precision and run-time. {\tt AutoLens}'s modular structure facilitates this, by breaking down the complex and iterative nature of lens modeling into a simple set of self-contained phases ensuring that new functionality can be introduced. Equally, the model comparison framework means that adding more complex light and mass profiles does not require a large reworking of each {\tt AutoLens} pipeline. However, it remains to be seen whether the most complex lens modeling tasks can be performed in a truly automated fashion.

The total run-time for a full {\tt AutoLens} analysis depends critically on the total number of image pixels in the analysis and is therefore driven by two factors: (i) the image resolution; (ii) the overall size of the lens and lensed source. The total run time for a unified lens and source analysis in this work ranged from 15-300 CPU hours, which when run in parallel across 8 cores translates to 2-40+ real-time hours. The large samples that will be provided by Euclid will fall in the faster range of run-times, whereas currently available high-resolution HST imaging could exceed these figures if drizzled to resolutions higher than those used in this work. Thus, processing large lens data-sets is already feasible on modern high performance computing facilities, even for samples in the tens of thousands. Increasing the analysis speed further would be beneficial and is possible, through either advances in computational processing like graphical processing units or by using a reduced and simpler analysis pipeline with more restrictive lens model priors. 

Another pressing issue is continued testing of {\tt AutoLens}. The simulated images used throughout this work were generated using the same mass models that were then ultimately fitted, circumventing issues like the MST and not testing assumptions like a constant mass-to-light ratio. Thus, they are somewhat uninformative in revealing what a strong lens analysis actually measures for a real lens galaxy. The assumptions underlying lens modeling with {\tt AutoLens} will be discussed in future publications, where they can be considered in more detail by comparing and contrasting a range of different lens models with different underlying assumptions, as well as looking for additional guidance from independent mass probes. However, the ideal means of testing this analysis will use simulated lenses generated via ray-tracing through realistic cosmological hydrodynamical simulations \citep{Schaye2014} and it is anticipated {\tt AutoLens} will take part in such work in the future. This work also bypassed a number of instrumental effects that could have the potential to bias the lens model, like a poor PSF sampling or image artefacts.

\subsection{Unified Modeling}

The results of unified modeling were highly successful and motivate existing high resolution lens samples for {\tt AutoLens}'s first analyses of known strong lenses. These samples has superior spatial resolution, S/N (e.g. \citep{Bolton2008}) and more source complexity \citep{Newton2011} than the images simulated in this work, as well as average lens and source redshifts below this work's fiducial values of $z_{\rm lens} = 0.5$ and $z_{\rm src} = 1.0$ . Altogether, this should improve the precision of each lens model compared to the values quoted in this work and may even allow for more complex lens models to be invoked, for example assuming a non-constant mass-to-light ratio. The lower redshifts also reduce the average external shear across the sample \citep{Jaroszynski2012}, thus reducing its degeneracy with the mass profile ellipticity and, if present, any geometric offsets.

For the $SPLE$ profile, the mass and light models were independent of one another, confirming that the lens light subtraction has little impact, if any, on the accuracy of the mass model inferred alongside only a minor impact on its precision. This validates approaches in the literature which infer a mass model from an image whose lens is subtracted before lens modeling (e.g. via a B-Splines interpolation). However, the reverse is not true and this approach cannot yield an accurate model for the lens's light, as there is no analogous way to accurately subtract the source before fitting the lens. Therefore, a selling point of unified modeling is its ability to accurately characterize the lens's visual morphology, which underpins the method’s ability to perform decomposed mass modeling.

Features such as radial arcs or a central image were successfully detected, demonstrating the method is able to diagnose a cored mass profile. Therefore, {\tt AutoLens} will also begin searching for cored mass profiles in existing lens data-sets. In the future, this effort will benefit from data-sets which resolve the source and lens at an increased spatial resolution as well as multi-wavelength imaging which contrast the relative brightnesses of the lens and source relative. It remains to be seen whether dust in the lens galaxy will prevent any such effort.

Although the mass model was not degenerate with the light subtraction, there are cases where the improved light subtraction offered by unified modeling could improve or even change the mass model altogether, when it reveals faint features in the lensed source that other approaches wrongly subtracted. This happened recently when a number of teams reassessed the light subtraction of HST imaging of the lens ID81 \citep{Dye2014, Dye2015, Rybak2015, Swinbank2015}, finding additional structure in the lensed source that changed the mass model and led to a more complete source reconstruction. Following the same argument, the non-detection of source light is an equally powerful means by which to constrain the mass model and it is not uncommon for the image reconstruction to place extraneous flux where it isn't observed, information which a poor lens subtraction and restrictive source-only masking will struggle to exploit. An extreme of this arises when a mass model goes to particularly shallow density profiles and begins to predict a central image in the region other approaches will have most likely masked before the analysis began.

Finally, and most importantly, unified modeling enables use of decomposed mass models which allow a number of unique measurements to be made for both a lens's stellar and dark components. These models offer a significant improvement to lens modelng in general, because \textbf{fitting of the lens's light profile is exploited to place constraints on its underlying stellar mass distribution}. For example, in the majority of SLACS lenses, the stellar component is dominant, making up over $90$ per cent of the total mass within $R_{Ein}$ \citep{Barnabe2011}. Therefore, lenses with a fainter, less extended or doubly imaged source, which offer loose constraints on an $SPLE$ mass model, will no doubt benefit from the additional information extracting by a decomposed mass model. Care must, of course, be taken to understand the impact of assuming a constant mass-to-light ratio.

\subsection{Comparison To Other Methods}

Other methods in the literature \citep{Dye2005, Vegetti2009, Suyu2012, Collett2014, Tagore2014, Birrer2015} use approaches similar to {\tt AutoLens}'s image and source analysis. The key differences are: (i) the amorphous nature of {\tt AutoLens}'s source pixelization, which ensures the method can truly achieve the simplest solution using the fewest correlated source pixels; (ii) the freedom given to {\tt AutoLens}'s variances and source regularization, which are key to correctly fitting the lens's light profile and (iii) the method's removal of the discreteness biases described in N15, which (although not shown explicitly shown in this work) continued to have a significant impact on both mass and lens light modeling if not handled using the approach advocated in N15. Furthermore, {\tt AutoLens} changes its image and source analysis in an automated and fully self-consistent manner, in contrast to other methods that follow a more ad-hoc approach. This makes the results of {\tt AutoLens} reproducible.
\subsection{Summary}
This paper presents {\tt AutoLens}, the first automated modelling suite for strong gravitational lenses. Our key results are:
\begin{itemize}
\item The image and source analysis adapt automatically to the properties of the strong lens being analysed. This includes an amorphous source-plane pixelization which adapts to the source's unlensed surface-brightness profile, a source regularization scheme which adapts to the source's intrinsic morphology and scaling of the observed image's variance-map to ensure the method provides a global fit to the strong lens imaging. These are all performed self-consistently within the Bayesian framework of \citet{MacKay1992, Suyu2006}. 

\item Fitting of the lens's light profile is fully integrated into {\tt AutoLens} and performed simultaneously with the mass and source. Light profiles comprise both single and multi component models, allowing the method to perform a bulge-disk decomposition and model lens galaxies of arbitrary morphology. The adaptive image and source features above are key to ensuring that the lens's light profile is inferred accurately, something other methods are not well suited to.
\item The improved lens subtraction allows {\tt AutoLens} to model and detect faint features in a lensed source that previous methods may omit. This includes, but is not limited to, features indicative of a cored density profile, such as radial arcs or the source's third or fifth central image.
\item Lens light fitting allows the method to invoke decomposed mass models which model separately the lens's light and dark matter. These offer a significant improvement to the inferred mass model because they exploit how the lens galaxy's light traces its underlying stellar mass distribution. Thus, by incorporating its light profile into the mass model new information is exploited about the lens's mass distribution; information which previous approaches to lens modeling omit completely.
\item The complexity of the light and mass models, including the detection of radial arcs, a central image or a geometrically offset light / dark matter profile, is decided objectively via Bayesian model comparison. This is performed by estimating the Bayesian evidence of each unique lens model, by running a non-linear search using the nested sampling algorithm {\tt MultiNest} \citep{Skilling2006, Feroz2009} for each. 
\item A single lens therefore provides a diverse range of observations. Mass models which assume a total (light and dark summed) profile infer its inner density slope, can detect a central core and fully characterize the lens galaxy's light profile. Decomposed mass models offer a stellar mass-to-light ratio, dark matter halo ellipticity and dark matter fraction as a function of radius, as well as determining if the light and dark matter are geometrically aligned with or offset from one another. The highly-magnified source galaxy is also fully reconstructed. 
\item {\tt AutoLens} is fully automated, such that all results presented in this work are generated without any user-intervention. This successfully automates the lens modeling process. 
\item {\tt AutoLens} is demonstrated on a suite of fifty four simulated images which span a variety of lens and source morphologies, mass profiles, lensing geometries and imaging data resolutions and signal-to-noise ratios. The method performs well for all of the observables listed above, choosing the correct model complexity in the majority of cases and inferring most parameters accurately within $3\sigma$ confidence.  
\end{itemize}
\subsection{Concluding Remarks}
Over the past two decades an outstanding and diverse multi-wavelength data-set of high quality strong lens images has been amassed throughout the literature. However, the complex and iterative nature of lens modeling has historically restricted their analysis to small samples, simplified mass models and little to no consideration of how the lens galaxy's light profile can benefit their analysis. {\tt AutoLens} addresses these issues, enabling the application of decomposed mass profiles that fully exploit the information contained within the lens's light on large lens samples. Future work will see {\tt AutoLens} applied to these data-sets, with the lens galaxy morphologies, density profiles and dark matter geometries key topics of interest. This work will lay the foundations for study of the anticipated large lens datasets comprising tens of thousands of strong lenses to ultimately give an unprecedented understanding of the nature of galaxy formation, dark matter and the Universe in general.

\section*{Acknowledgements}

JN acknowledges support from STFC grant ST/N00149/1 and the University of Nottingham. SD acknowledges support from the Midlands Physics Alliance and STFC Ernest Rutherford Fellowship Scheme. RJM acknowledges support from the Royal Society University Research Fellowship. We are grateful for access to the University of Nottingham High Performance Computing Facility. This paper used the DiRAC Data Centric system at Durham University, operated by the Institute for Computational Cosmology on behalf of the STFC DiRAC HPC Facility (www.dirac.ac.uk). This equipment was funded by BIS National E-infrastructure capital grant ST/K00042X/1, STFC capital grant ST/H008519/1, and STFC DiRAC Operations grant ST/K003267/1 and Durham University. DiRAC is part of the National e-Infrastructure. We acknowledge use of the software program (getdist.readthedocs.io/en/latest/).

\appendix

\section{Efficiency Tricks and Algorithms}\label{AppCalc}

\subsection{Light Profile}

High levels of oversampling are required to accurately compute a two-dimensional Sersic light profile. This is especially true when evaluated at low radii, where it diverges. Therefore, adaptive oversampling is applied, to ensure the light profile is computed fast and accurately. This routine first acts on all pixels within $R_{\rm  l}$. The intensity of the pixel, $I(\xi_{\rm l})$, is computed at its center. This pixel is then divided into a $2 \times 2$ sub-grid and the mean of intensities at the sub-pixel centres is computed. If the fractional change in $I(\xi_{\rm l})$ is $< 0.0001$ then sufficient accuracy has been achieved and the value just computed is used. However if the fractional change is $> 0.0001$, this process is iteratively repeated for higher levels of oversampling, up to a subgrid of degree $2000 \times 2000$. Finally, if all pixels within $R_{\rm  l}$ required oversampling, then pixels outside this radius are re-evaluated until the accuracy threshold is met at least once without oversampling. If the initial $I(\xi_{\rm l})$ value is below the numerical precision of {\tt AutoLens} ($10^{-16}$) it is not oversampled, as rounding errors prevent convergence (and the flux is negligible anyway). These occurrences are rare and happen when $\xi_{\rm l}$ is large and $R_{\rm l}$ is small.

\subsection{Deflection Angles}

Numerical integration is used to compute $\vec{{\alpha}}_{\rm x,y}$ from the above $\kappa (\vec{x})$ profiles. {\tt AutoLens} uses an adaptive numerical integration technique, which iteratively refines the subintervals over which the integral is evaluated until a threshold accuracy is achieved. The expressions for $\phi_r$ can be found in K01, with equation (55) giving the NFW profile, equation (45) the de Vaucouleurs profile and equation (74) the exponential profile. For every profile the $\kappa (\vec{x})$ and $\vec{{\alpha}}_{\rm x,y}$ maps generated by {\tt AutoLens} have been compared with those given by the lensing software {\tt gravlens} \citep{Keeton2003}, ensuring all are implemented in {\tt AutoLens} correctly.

N15 showed that image oversampling (termed subgridding in N15) is required to remove aliasing effects which lead to inaccurate lens modeling. Oversampling splits each image pixel into a set of square sub-pixels, the centers of which are all traced to the source-plane and used by the inversion. However, high levels of oversampling requires $\vec{{\alpha}}_{\rm x,y}$ to be computed for each additional sub-pixel, which can prove computationally expensive. 

A bi-linear interpolation scheme is therefore now applied to greatly increase efficiency, whereby deflection angles are computed at the center of image pixels and interpolated to give the sub-pixel deflection angles. In the central regions of the mass profile (where the density profile is rapidly increasing) this interpolation scheme becomes inaccurate, thus in these central regions each sub-pixel deflection angle is computed explicitly. Further out (where the density profile is much flatter) deflection angles are interpolated from a grid of computed deflection angles, thus reducing the number of overall deflection angle calculations. This grid becomes more coarse as one reaches the flatter regions of the mass profile's density. This interpolation scheme calculates $\vec{{\alpha}}_{\rm x,y}$ at sub-pixels to a fractional accuracy of $10^{-4}$, which is more than sufficient given the systematics associated with source-plane discretization. Whilst a $4$ $\times$ $4$ sub-pixel grid was found to be sufficient in N15, higher levels of over-sampling ($8$ $\times$ $8$) are used in this work, given it is now computationally feasible.

\subsection{Positional Information}

The initial calculation of lens models that accurately fit the image data involves searching large portions of non-linear parameter space, which is prohibitively expensive. Positional information is therefore used to increase speed, by requiring that any lens model must first meet the requirement that four image pixels in the lensed source must trace to within a threshold value of one another in the source-plane. If this criterion fails, a new lens model is sampled. This approach was introduced by \citet{Brewer2008} for a strongly lensed quasar, the point source nature of which allowed them to impose that image pixels trace to within $10^{-5}$ arcsec of one another. Here, a much larger threshold is used throughout, because unlike \citet{Brewer2008} this is not imposed to constrain the lens model but simply to improve the speed of the initial non-linear sampling. 

However, due to the complex source morphologies of real strong lens imaging, one can never be sure which image-plane pixels actually neighbour one another in the source-plane. Therefore, positional information is only exploited once an accurate model for the source has been computed, which in the automated analysis pipeline is after the phase $\textbf{P}_{\rm Init2}$ where a parametric source is fitted. The positional image pixels are calculated as the four image pixels which trace closest to the parametric source's center ($x_{s}$, $y_{s}$) and also: (i) are separated by over $20\%$ of the lens's $R_{Ein}$ value in the image-plane and (ii) have one pixel rotationally offset from the others in the image-plane by at least $120\,^{\circ}-240\,^{\circ}$ degrees around the lens center (to ensure multiple images are sampled as opposed to just one image's extended arc). To compute positional image pixels for a pixelied source-plane these requirements are followed using image pixels which trace to the brightest pixel in the source reconstruction. If they are not met using only this source pixel its closest traced image pixels are iteratively used until four image pixels are chosen. 

The threshold value is then reduced to a value of $3\times$ the maximum source-plane separation of these newly allocated image pixels or $0.3”$, whichever is smaller, thereby giving significant efficiency gains whilst ensuring no feasible lens models are wrongly discarded. Positional information is also key to removing the unwanted over / under fit solutions described in N15 and section \ref{SLPipeline}.

\section{Constant Regularization}\label{AppReg}

The linear regularization matrix $\tens{H}$ used in \citealt{Warren2003} and N15 is derived following the formalism given in \citet{Press2001}. This computes $\tens{H}$ as $\tens{H} = \tens{B}^T \tens{B}$, where the matrix $\tens{B}$ stores the regularization pattern of source pixels with one another. For example, to regularize each source pixel with its neighbor, assuming the numbering scheme is such that pixel one is a neighbour of pixel two, and two of three, etc., the matrix $\tens{B_x}$ is given as
\begin{equation}
\begin{bmatrix}
-1  &  1 & 0  & 0 & ...\\ 
 0  & -1 & 1  & 0 & ...\\ 
 0  &  0 & -1 & 1 & ... \\ 
... & ... & ... & ... & ...
\end{bmatrix}
\, \, .
\end{equation}
For gradient regularization on an $N \times N$ square grid, this matrix gives the regularization of source pixels across the x-direction, where every N elements will be a row of zeros. This matrix then gives a regularization matrix $\tens{H_x} = \tens{B_x}^T \tens{B_x}$. For regularization in the y-direction, a second $\tens{B_y}$ matrix is generated, where the negative ones are again across the diagonal and the positive ones every N elements across from this, with the final N rows all zeros. $\tens{B_y}$ is then used to compute a second regularization matrix $\tens{H_y} = \tens{B_y}^T \tens{B_y}$, which is added to the first to give the overall regularization matrix $\tens{H} = \tens{H_x} + \tens{H_y}$. For {\tt AutoLens}'s Voronoi regularization scheme the same pattern is followed, using around 5-10 $\tens{H}$ matrices corresponding to regularization across all of the Voronoi vertex indices.
\subsection{Non-Constant Regularization}
Formally, $\lambda$ can be included in the $\tens{B}$ matrices above. However, because it is a fixed single value, it is convention to take it outside $\tens{B}$. For example, in N15, this saw $\lambda$ included in three terms in the expression for the Bayesian evidence (e.g. $\lambda \vec{s}^T \tens{H} \vec{s}$). For the non-constant regularization scheme used to weight regularization by the source's luminosity, a 1D vector of regularization coefficients $\vec{\rm \Lambda}$ must be employed and incorporated into the computation of $\tens{H}$. 

Therefore, the $\vec{B}$ matrices above are redefined to include each pixel's effective regularization coefficient, $\lambda_{\rm eff}$, as $\tens{B}_{\rm  \Lambda} = \vec{\rm \Lambda}\tens{B}$, where $\vec{\Lambda}$ is computed as described in section \ref{RegAdapt}. The corresponding regularization matrix is then $\tens{H}_{\rm  \Lambda} = \tens{B}_{\rm  \Lambda}^T \tens{B}_{\rm  \Lambda}$.
\section{Pipeline Priors and Linking}\label{AppPrior}
This appendix presents a more detailed overview of pipeline phase linking, describing the priors given to the different light and mass profiles used to initialize that profile at different points in the pipeline.

For each phase, initial parameter sampling is performed using one of two priors:
\begin{itemize}
\item \textbf{Uniform Prior (UP)} – Draws points randomly from a uniform distribution defined by a maximum and minimum value.
\item \textbf{Gaussian Prior (GP)} – Draws points randomly from a normal distribution defined by a mean $\mu_g$ and width $\sigma_g$. The value for $\mu_g$ is estimated from the previous phase, using the median of the same parameter or a related parameter's marginalized 1D probability distribution function (PDF). This PDF is used to estimate that parameters $3\sigma$ confidence bounds.
\end{itemize}

The value of $\sigma_g$ is defined such that $68.2\%$ of points sampled (on average) are between that parameter's previously estimated $3\sigma$ confidence bounds. These confidence bounds are generally not symmetric and the value furthest from the mean is used. This ensures that when linking phases the wider area of the previous phase's posterior is sampled. For example, if a parameter is estimated as $n = 4^{+0.2}_{\rm  -0.3}$ at $3\sigma$ confidence, $68.2\%$ of samples in the next phase will (on average) lie between $3.7$ and $4.3$. However, for high quality imaging data, parameters can be estimated to a very high accuracy and their errors could therefore be very small. Using these errors to set $\sigma_g$ therefore runs a risk of biasing an analysis by placing overly restrictive priors. To overcome this, each parameter has a minimum $\sigma_g$ value, which replaces the previous analysis's error estimate if it below this minimum. Extending the previous example, if this minimum value were $0.6$, then $68.2\%$ of samples in the next phase will be between $3.4$ and $4.6$ despite its errors only corresponding to a size of $0.3$. These minimum values are given for each phase in the tables described next.

For the first optimization of the hyper-parameters uniform priors are assumed on all parameters, except $\lambda$ which uses a broad Gaussian prior centered on phase $\textbf{P}_{\rm  Init3}$'s initial estimate of $\lambda$. These uniform priors are broad, but may not be sufficient to capture the optimum value of all of the hyper-parameters. However, re-optimization of the hyper-parameters uses Gaussian priors centered on their previous phase's most probable values with $\sigma_g = \mu_g / 2$, thus the optimum values will be reached after two or three hyper-parameter optimizations. 

The following tables give the priors used in each pipeline phase for the different light and mass models. It should be noted some quantities which depend on the image properties, like $I_{\rm  l}$ and $\Psi_{\rm l}$, have priors which depend on initial estimates of their values from the image data. All tables follow the same notation, where \textbf{UP} $^{a}_{\rm  b}$ corresponds to a uniform prior between the values $a$ and $b$ and \textbf{GP} $a$ corresponds to a Gaussian prior with minimum value $\sigma_g$ given by $a$. Initialization of parameters in the main pipeline uses exclusively Gaussian priors, retaining the minimum $\sigma_g$ values given for the model comparison phases.

\begin{table*}
\resizebox{\linewidth}{!}{
\centering
\begin{tabular}{ l | l | l | l l l l l l l } 
 \multicolumn{1}{p{1.5cm}|}{\textbf{Model}} 
& \multicolumn{1}{p{1.1cm}|}{\centering \textbf{Comp-} \\ \textbf{onent}} 
& \multicolumn{1}{p{1.0cm}|}{\textbf{Prior Load}} 
& \multicolumn{1}{p{1.1cm}}{\textbf{Parameters}} 
& \multicolumn{1}{p{1.1cm}}{} 
& \multicolumn{1}{p{1.1cm}}{} 
& \multicolumn{1}{p{1.1cm}}{}  
& \multicolumn{1}{p{1.1cm}}{} 
& \multicolumn{1}{p{1.1cm}}{} 
& \multicolumn{1}{p{1.1cm}}{} 
\\ \hline
& & & & & & & & & \\[-4pt]
$Sersic$       & Light & None & $x_{\rm  l1}$ UP $^{0.5}_{\rm  -0.5}$  &  $y_{\rm  l1}$ UP $^{0.5} _{\rm  -0.5}$ & $I_{\rm  l1}$ UP $^{2}_{\rm  0}$ & $R_{\rm  l1}$ UP $^{4} _{\rm  0}$ & $n_{\rm  l1}$ UP $^{9.5}_{\rm  0.7}$ & $q_{\rm  l1}$ UP $^{1.0}_{\rm  0.3}$ & $\theta_{\rm  l1}$ UP $^{180}_{\rm  0}$ \\[2pt]
\hline
& & & & & & & \\[-4pt]
$Exponential$  & Light & None & $x_{\rm  l2} = x_{\rm  l1}$                     &  $y_{\rm  l2} = y_{\rm  l1}$                     & $I_{\rm  l2}$ UP $^{2}_{\rm  0}$ & $R_{\rm  l2}$ UP $^{4} _{\rm  0}$ & $n_{\rm  l2} = 1$                        & $q_{\rm  l2}$ UP $^{1.0}_{\rm  0.3}$ & $\theta_{\rm  l2} = \theta_{\rm  l1}$ \\[2pt]
\end{tabular}
}
\caption[Priors used to initalize phase $\textbf{P}_{\rm  Init1}$]{Priors used to initialize the Sersic + Exponential fit used in $\textbf{P}_{\rm  Init1}$}
\label{table:PLInita}
\end{table*}

\begin{table*}
\resizebox{\linewidth}{!}{
\centering
\begin{tabular}{ l | l | l | l l l l l l l } 
 \multicolumn{1}{p{1.5cm}|}{\textbf{Model}} 
& \multicolumn{1}{p{1.1cm}|}{\centering \textbf{Comp-} \\ \textbf{onent}} 
& \multicolumn{1}{p{1.0cm}|}{\textbf{Prior Load}} 
& \multicolumn{1}{p{1.1cm}}{\textbf{Parameters}} 
& \multicolumn{1}{p{1.1cm}}{} 
& \multicolumn{1}{p{1.1cm}}{} 
& \multicolumn{1}{p{1.1cm}}{}  
& \multicolumn{1}{p{1.1cm}}{} 
& \multicolumn{1}{p{1.1cm}}{} 
& \multicolumn{1}{p{1.1cm}}{} 
\\ \hline
& & & & & & & & \\[-4pt]
$SPLE$    & Mass  & $\textbf{P}_{\rm  Init1}$ & $x$ GP $0.1"$ &  $y$ GP $0.1"$  & $\theta_{\rm  E}$ UP $^{4} _{\rm  0}$ & $q$ UP $^{1.0}_{\rm  0.3}$ & $\theta$ UP $^{180}_{\rm  0}$ $\alpha = 2.2$ & \\[2pt]
\hline
& & & & & & & & \\[-4pt]
$Sersic$ & Source & None &  $x_{\rm  s}$ UP $^{0.5}_{\rm  -0.5}$  &  $y_{\rm  s}$ UP $^{0.5} _{\rm  -0.5}$ & $I_{\rm  s}$ UP $^{2}_{\rm  0}$ & $R_{\rm  s}$ UP $^{4} _{\rm  0}$ & $n_{\rm  s}$ UP $^{9.5}_{\rm  0.7}$ & $q_{\rm  s}$ UP $^{1.0}_{\rm  0.3}$ & $\theta_{\rm  s}$ UP $^{180}_{\rm  0}$ \\[2pt]
\end{tabular}
}
\caption{Priors used to initialize the $SPLE$ + $Sersic$ (Source) or $PL\textsubscript{Core}$ + $Sersic$ (Source) profile fit used in $\textbf{P}_{\rm  Init2}$. The Sersic and $SPLE$ and $PL\textsubscript{Core}$ parameters are initialized from the results of $\textbf{P}_{\rm  Init1}$.}.
\label{table:PLInitc}
\end{table*}

\begin{table*}
\resizebox{\linewidth}{!}{
\centering
\begin{tabular}{ l | l | l | l l l l l l l } 
 \multicolumn{1}{p{1.5cm}|}{\textbf{Model}} 
& \multicolumn{1}{p{1.1cm}|}{\centering \textbf{Comp-} \\ \textbf{onent}} 
& \multicolumn{1}{p{1.0cm}|}{\textbf{Prior Load}} 
& \multicolumn{1}{p{1.1cm}}{\textbf{Parameters}} 
& \multicolumn{1}{p{1.1cm}}{} 
& \multicolumn{1}{p{1.1cm}}{} 
& \multicolumn{1}{p{1.1cm}}{}  
& \multicolumn{1}{p{1.1cm}}{} 
& \multicolumn{1}{p{1.1cm}}{} 
& \multicolumn{1}{p{1.1cm}}{} 
\\ \hline
& & & & & & & & &  \\[-4pt]
$Sersic$ & Light &  None  & $x_{\rm  l}$ UP $^{0.5}_{\rm  -0.5}$  &  $y_{\rm  l}$ UP $^{0.5} _{\rm  -0.5}$ & $I_{\rm  l1}$ UP $^{2}_{\rm  0}$ & $R_{\rm  l1}$ UP $^{4} _{\rm  0}$ & $n_{\rm  l1}$ UP $^{9.5}_{\rm  0.7}$ & $q_{\rm  l1}$ UP $^{1.0}_{\rm  0.3}$ & $\theta_{\rm  l1}$ UP $^{180}_{\rm  0}$ \\[2pt]
\hline
& & & & & & & & & \\[-4pt]
$SPLE$     & Mass   & $\textbf{P}_{Init2}$   & $x$ ($x_{\rm l}$) GP $0.02$     &  $y$ ($y_{\rm l}$) GP $0.02$  & $\theta_{\rm  E}$ GP $\theta_{\rm Ein}/4$ &  $q$ GP $0.05$ & $\theta$ GP $20.0$  & $\alpha = 2.2$ & \\[2pt]
\hline
& & & & & & & & & \\[-4pt]
Hyper Params  & Hyper & None   & $\lambda$ UP $^{1000}_{\rm  0}$ & & & & & & \\[2pt]
\end{tabular}
}
\caption{Priors used to initialize the $Sersic$ light model and $SPLE$ mass model used in $\textbf{P}_{\rm  Init3}$. The $SPLE$ and $PL\textsubscript{Core}$ parameters $x$, $y$ and $\theta$ are initialized from the results of $\textbf{P}_{\rm  Init2}$.}
\label{table:PLInitb}
\end{table*}

\begin{table*}
\resizebox{\linewidth}{!}{
\centering
\begin{tabular}{ l | l | l | l l l l l l l } 
 \multicolumn{1}{p{1.5cm}|}{\textbf{Model}} 
& \multicolumn{1}{p{1.1cm}|}{\centering \textbf{Comp-} \\ \textbf{onent}} 
& \multicolumn{1}{p{1.0cm}|}{\textbf{Prior Load}} 
& \multicolumn{1}{p{1.1cm}}{\textbf{Parameters}} 
& \multicolumn{1}{p{1.1cm}}{} 
& \multicolumn{1}{p{1.1cm}}{} 
& \multicolumn{1}{p{1.1cm}}{}  
& \multicolumn{1}{p{1.1cm}}{} 
& \multicolumn{1}{p{1.1cm}}{} 
& \multicolumn{1}{p{1.1cm}}{} 
\\ \hline
& & & & & & & & \\[-4pt]
$Sersic$ & Light & $\textbf{P}_{\rm  Init3}$ & $x_{\rm  l}$ GP $0.02"$ & $y_{\rm  l}$ GP $0.02"$ & $I_{\rm  l}$ GP $I_{\rm l}/2$ & $R_{\rm  l}$ GP $R_{\rm l}/2$ & $n_{\rm  l}$ GP $0.8$ & $q_{\rm  l}$ GP $0.1$ & $\theta_{\rm  l}$ GP $20.0$ \\[2pt]
\hline
& & & & & & & & \\[-4pt]
$SPLE$    & Mass  & $\textbf{P}_{\rm  Init3}$ & $x$ GP $0.1"$ &  $y$ GP $0.1"$  & $\theta_{\rm  E}$ GP $\theta_{\rm E}/2$ & $q$ GP $0.1$ & $\theta$ GP $20.0$ & $\alpha = 2.2$ & \\[2pt]
\hline
& & & & & & & & \\[-4pt]
$PL\textsubscript{Core}$   & Mass  & $\textbf{P}_{\rm  Init3}$  & $x$ GP $0.1"$ &  $y$ GP $0.1"$  & $\theta_{\rm  E}$ GP $\theta_{\rm E}/2$  & $q$ GP $0.1$ & $\theta$ GP $20.0$  & $\alpha = 2.0$ & $s$ GP $0.1"$ \\[2pt]
\hline
& & & & & & & & \\[-4pt]
$NFWSph$                   & Mass  & $\textbf{P}_{\rm  MCMass}$ & $x_d$ ($x$) GP $0.2"$     &  $y$ ($y$) GP $0.2"$     & $\kappa_{\rm s}$ UP $^{3} _{\rm  0}$   & $q = 1.0 $  & & & \\[2pt]
\hline
& & & & & & & \\[-4pt]
$\Psi_{\rm l}$             & Mass  & & $\Psi_l$ UP $^{5.0}_{\rm  0.0}$ & & & & & & \\[2pt]
\end{tabular}
}
\caption{Priors used to initialize the $Sersic$ light profile and mass profile used in $\textbf{P}_{\rm  Init4}$. The $Sersic$, $SPLE$, $PL\textsubscript{Core}$ and $NFWSph$ parameters are initialized from the results of $\textbf{P}_{\rm  Init3}$.}.
\label{table:PLInitd}
\end{table*}

\begin{table*}
\resizebox{\linewidth}{!}{
\centering
\begin{tabular}{ l | l | l | l l l l l l l } 
 \multicolumn{1}{p{1.5cm}|}{\textbf{Model}} 
& \multicolumn{1}{p{1.1cm}|}{\centering \textbf{Comp-} \\ \textbf{onent}} 
& \multicolumn{1}{p{1.0cm}|}{\textbf{Prior Load}} 
& \multicolumn{1}{p{1.1cm}}{\textbf{Parameters}} 
& \multicolumn{1}{p{1.1cm}}{} 
& \multicolumn{1}{p{1.1cm}}{} 
& \multicolumn{1}{p{1.1cm}}{}  
& \multicolumn{1}{p{1.1cm}}{} 
& \multicolumn{1}{p{1.1cm}}{} 
& \multicolumn{1}{p{1.1cm}}{} 
\\ \hline
& & & & & & & & \\[-4pt]
$Sersic$          & Light & $\textbf{P}_{\rm  Init4}$ & $x_{\rm  l}$ GP $0.02"$ &  $y_{\rm  l}$ GP $0.02"$ & $I_{\rm  l}$ UP $^{2}_{\rm  0}$ & $R_{\rm  l}$ UP $^{4} _{\rm  0}$ & $n_{\rm  l}$ UP $^{9.5}_{\rm  0.7}$ & $q_{\rm  l}$ UP $^{1.0}_{\rm  0.3}$ & $\theta_{\rm  l}$ GP $20.0$ \\[2pt]
\hline
& & & & & & & \\[-4pt]
$Sersic + Exp$          & Light & $\textbf{P}_{\rm  Init4}$ & $x_{\rm  l}$ GP $0.02"$ &  $y_{\rm  l}$ GP $0.02"$   & $I_{\rm  l1}$ UP $^{2}_{\rm  0}$ & $R_{\rm  l1}$ UP $^{4} _{\rm  0}$ & $n_{\rm  l1}$ UP $^{9.5}_{\rm  0.7}$ & $q_{\rm  l1}$ UP $^{1.0}_{\rm  0.3}$ & $\theta_{\rm  l}$ GP $20.0$ \\[2pt]
                        &                       &       & $x_{\rm  l2} = x_{\rm  l1}$ &  $y_{\rm  l2} = y_{\rm  l1}$ & $I_{\rm  l2}$ UP $^{2}_{\rm  0}$ & $R_{\rm  l2}$ UP $^{4} _{\rm  0}$ & $n_{\rm  l2} = 1.0$             & $q_{\rm  l2}$ UP $^{1.0}_{\rm  0.3}$ & $\theta_{\rm  l2} = \theta_{\rm  l1}$ \\[2pt]
\hline
& & & & & & & \\[-4pt]
$SPLE$             & Mass  & $\textbf{P}_{\rm  Init4}$ & $x$ GP $0.01"$     &  $y$ GP $0.01"$     & $\theta_{\rm  E}$ GP $0.05"$   & $q$ GP $0.01$  & $\theta$ GP $2.0$  & $\alpha = 2.2$ & \\[2pt]
\hline
& & & & & & & \\[-4pt]
$PL\textsubscript{Core}$        & Mass  & $\textbf{P}_{\rm  Init4}$ & $x$ GP $0.05"$     &  $y$ GP $0.05"$     & $\theta_{\rm  E}$ GP $\theta_{\rm E}/2$   & $q$ GP $0.1$ & $\theta$ GP $10.0$ & $\alpha = 2.0$ & $S$ UP $^{0.2"}_{\rm 0.0"}$ \\[2pt]
\hline
& & & & & & & \\[-4pt]
$NFWSph$                      & Mass  & $\textbf{P}_{\rm  MCMass}$ & $x_d$ ($x$) GP $0.2"$     &  $y$ ($y$) GP $0.2"$     & $\kappa_{\rm s}$ GP $\kappa_{\rm s}/2$   & $q = 1.0$ & & & \\[2pt]
\hline
& & & & & & & \\[-4pt]
$\Psi_{\rm l}$             & Mass  & & $\Psi_l$ GP $\Psi_l/2$ & & & & & & \\[2pt]
\hline
& & & & & & & \\[-4pt]
Hyper Params  & Hyper & Prev   & $\lambda_{Src}$ GP $\lambda_{Src}/2$ & $\omega_{Lens}$ UP $\omega_{Lens}/2$ & $\omega_{Src}$ UP $\omega_{Src}/2$ & & \\[2pt]
\end{tabular}
}
\caption{Priors used to initialize the Mass + Light profile fits used for model comparison in phase $\textbf{P}_{\rm  MCLight}$. Parameter initializations are derived from the results of $\textbf{P}_{\rm  Init4}$. The hyper parameters $\omega_{Lens}$ and $\omega_{Src}$ are initialized using the $\chi^2_{\rm Base}$ values of the observed image. }.
\label{table:PLMCLight}
\end{table*}

\begin{table*}
\resizebox{\linewidth}{!}{
\centering
\begin{tabular}{ l | l | l | l l l l l l l } 
 \multicolumn{1}{p{1.5cm}|}{\textbf{Model}} 
& \multicolumn{1}{p{1.1cm}|}{\centering \textbf{Comp-} \\ \textbf{onent}} 
& \multicolumn{1}{p{1.0cm}|}{\textbf{Prior Load}} 
& \multicolumn{1}{p{1.1cm}}{\textbf{Parameters}} 
& \multicolumn{1}{p{1.1cm}}{} 
& \multicolumn{1}{p{1.1cm}}{} 
& \multicolumn{1}{p{1.1cm}}{}  
& \multicolumn{1}{p{1.1cm}}{} 
& \multicolumn{1}{p{1.1cm}}{} 
& \multicolumn{1}{p{1.1cm}}{} 
\\ \hline
& & & & & & & \\[-4pt]
$Light$                     & Light & $\textbf{P}_{\rm  MCLight}$ & $x_l$ GP $0.02"$     &  $y_l$ GP $0.02"$   & $I_{\rm  l}$ GP $I_{\rm l}/2$ & $R_{\rm  l}$ GP $R_{\rm l}/2$ & $n_{\rm  l}$ GP $0.8$ & $q_{\rm  l}$ GP $0.1$ & $\theta_{\rm  l}$ GP $20.0$ \\[2pt]
\hline
& & & & & & & \\[-4pt]
$SPLE$                      & Mass  & $\textbf{P}_{\rm  MCLight}$ & $x$ GP $0.05"$     &  $y$ GP $0.05"$     & $\theta_{\rm  E}$ UP $^{4} _{\rm  0}$   & $q$ UP $^{1.0} _{\rm  0.3}$  & $\theta$ GP $60.0$  & $\alpha$ UP $^{2.5}_{\rm  1.5}$ & \\[2pt]
\hline
& & & & & & & \\[-4pt]
$PL\textsubscript{Core}$        & Mass  & $\textbf{P}_{\rm  Init4}$ & $x$ GP $0.05"$     &  $y$ GP $0.05"$     & $\theta_{\rm  E}$ GP $\theta_{\rm E}/2$   & $q$ GP $0.1$ & $\theta$ GP $10.0$ & $\alpha = 2.0$ & $S$ UP $^{0.2"}_{\rm 0.0"}$ \\[2pt]
\hline
& & & & & & & \\[-4pt]
Hyper Params  & Hyper & Prev   & $\omega_{Lens}$ GP $\omega_{Lens}/2$ & $\lambda_{Src}$ GP $\lambda_{Src}/2$ & & & \\[2pt]
\end{tabular}
}
\caption{Priors used to initialize the light and mass profile fits used for model comparison phase $\textbf{PL}_{\rm  SPLEInit}$. All components of the light model use the same priors given in the row labled the $Light$ model, regardless of whether it is a single $Sersic$ or multiple $Exp$ and / or $Sersic$ profile. These are derived from the results of the $\textbf{P}_{\rm  MCLight}$ phase, as are the $SPLE$, $PL\textsubscript{Core}$ and $Shear$ parameter initializations.}.
\label{table:PLSPLEInit}
\end{table*}

\begin{table*}
\resizebox{\linewidth}{!}{
\centering
\begin{tabular}{ l | l | l | l l l l l l l } 
 \multicolumn{1}{p{1.5cm}|}{\textbf{Model}} 
& \multicolumn{1}{p{1.1cm}|}{\centering \textbf{Comp-} \\ \textbf{onent}} 
& \multicolumn{1}{p{1.0cm}|}{\textbf{Prior Load}} 
& \multicolumn{1}{p{1.1cm}}{\textbf{Parameters}} 
& \multicolumn{1}{p{1.1cm}}{} 
& \multicolumn{1}{p{1.1cm}}{} 
& \multicolumn{1}{p{1.1cm}}{}  
& \multicolumn{1}{p{1.1cm}}{} 
& \multicolumn{1}{p{1.1cm}}{} 
& \multicolumn{1}{p{1.1cm}}{} 
\\ \hline
& & & & & & & \\[-4pt]
$SPLE$                      & Mass  & $\textbf{P}_{\rm  MCLight}$ & $x$ GP $0.05"$     &  $y$ GP $0.05"$     & $\theta_{\rm  E}$ UP $^{4} _{\rm  0}$   & $q$ UP $^{1.0} _{\rm  0.3}$  & $\theta$ GP $60.0$  & $\alpha$ UP $^{2.5}_{\rm  1.5}$ & \\[2pt]
\hline
& & & & & & & \\[-4pt]
$PL\textsubscript{Core}$  & Mass  & $\textbf{P}_{\rm  MCLight}$ & $x$ GP $0.05"$     &  $y$ GP $0.05"$     & $\theta_{\rm  E}$ UP $^{4} _{\rm  0}$   & $q$ UP $^{1.0} _{\rm  0.3}$ & $\theta$ GP $60.0$  & $\alpha$ UP $^{2.5}_{\rm  1.5}$ & $s$ UP $^{0.0}_{\rm  2.0}$ \\[2pt]
\hline
& & & & & & & \\[-4pt]
$PL\textsubscript{Core}$        & Mass  & $\textbf{P}_{\rm  Init4}$ & $x$ GP $0.05"$     &  $y$ GP $0.05"$     & $\theta_{\rm  E}$ GP $\theta_{\rm E}/2$   & $q$ GP $0.1$ & $\theta$ GP $10.0$ & $\alpha = 2.0$ & $S$ UP $^{0.2"}_{\rm 0.0"}$ \\[2pt]
\hline
& & & & & & & \\[-4pt]
$NFWSph$                      & Mass  & $\textbf{P}_{\rm  MCMass}$ & $x_d$ ($x$) GP $0.2"$     &  $y$ ($y$) GP $0.2"$     & $\kappa_{\rm s}$ GP $\kappa_{\rm s}/2$   & $q = 1.0$ & & & \\[2pt]
\hline
& & & & & & & \\[-4pt]
$\Psi_{\rm l}$             & Mass  & & $\Psi_l$ GP $\Psi_l/2$ & & & & & & \\[2pt]
\hline
& & & & & & & \\[-4pt]
$Shear$                     & Mass  & $\textbf{P}_{\rm  MCLight}$ & $x_{\rm  sh} = x$     &  $y_{\rm  sh} = y$    & $\gamma_{\rm  sh}$ UP $^{0.4}_{\rm  0.0}$ & $\theta_{\rm  sh}$ UP $^{180}_{\rm  0}$  & & & \\[2pt]
\hline
& & & & & & & \\[-4pt]
Hyper Params  & Hyper & Prev   & $\lambda_{Src}$ GP $\lambda_{Src}/2$ & $\omega_{Src}$ UP $\omega_{Src}/2$ & $\omega_{Lens}$ UP $\omega_{Lens}/2$ & & & \\[2pt]
\end{tabular}
}
\caption{Priors used to initialize the mass profile fits used for model comparison phase $\textbf{PL}_{\rm  MCMass}$. Parameter initializations are derived from the results of $\textbf{P}_{\rm  MCLight}$.}.
\label{table:PLMCMass}
\end{table*}

\begin{table*}
\resizebox{\linewidth}{!}{
\centering
\begin{tabular}{ l | l | l | l l l l l l l } 
 \multicolumn{1}{p{1.5cm}|}{\textbf{Model}} 
& \multicolumn{1}{p{1.1cm}|}{\centering \textbf{Comp-} \\ \textbf{onent}} 
& \multicolumn{1}{p{1.0cm}|}{\textbf{Prior Load}} 
& \multicolumn{1}{p{1.1cm}}{\textbf{Parameters}} 
& \multicolumn{1}{p{1.1cm}}{} 
& \multicolumn{1}{p{1.1cm}}{} 
& \multicolumn{1}{p{1.1cm}}{}  
& \multicolumn{1}{p{1.1cm}}{} 
& \multicolumn{1}{p{1.1cm}}{} 
& \multicolumn{1}{p{1.1cm}}{} 
\\ \hline
& & & & & & & \\[-4pt]
$Light$                     & Light & $\textbf{P}_{\rm  MCLight}$ & $x_l$ GP $0.02"$     &  $y_l$ GP $0.02"$   & $I_{\rm  l}$ GP $I_{\rm l}/2$ & $R_{\rm  l}$ GP $R_{\rm l}/2$ & $n_{\rm  l}$ GP $0.8$ & $q_{\rm  l}$ GP $0.1$ & $\theta_{\rm  l}$ GP $20.0$ \\[2pt]
\hline
& & & & & & & \\[-4pt]
$NFW$                      & Mass  & $\textbf{P}_{\rm  MCMass}$ & $x_d$ ($x$) GP $0.2"$     &  $y$ ($y$) GP $0.2"$     & $\kappa_{\rm s}$ UP $^{3} _{\rm  0}$   & $q$ UP $^{1.0} _{\rm  0.2}$  & $\theta$ UP $^{180.0}_{\rm  0.0}$  & & \\[2pt]
\hline
& & & & & & & \\[-4pt]
$\Psi_{\rm l}$             & Mass  & & $\Psi_l$ UP $^{5.0}_{\rm  0.0}$ & & & & & & \\[2pt]
\hline
& & & & & & & \\[-4pt]
$Shear$                     & Mass  & $\textbf{P}_{\rm  MCLight}$ & $x_{\rm  sh} = x_{\rm d}$     &  $y_{\rm  sh} = y_{\rm d}$ & $\gamma_{\rm  sh}$ GP $0.1$ & $\theta_{\rm  sh}$ GP $90.0$  & & & \\[2pt]
\hline
& & & & & & & \\[-4pt]
Hyper Params  & Hyper & Prev   & $\lambda_{Src}$ GP $\lambda_{Src}/2$ & $\omega_{Lens}$ UP $\omega_{Lens}/2$ & $\omega_{Src}$ UP $\omega_{Src}/2$ & & & \\[2pt]
\end{tabular}
}
\caption{Priors used to initialize the light and mass profile fits used for model comparison phase $\textbf{PL}_{\rm  MCGeom}$. All components of the light model use the same priors given in the row labeled the $Light$ model, regardless of whether it is a single $Sersic$ or multiple $Dev$, $Exp$ and / or $Sersic$ profile. These are derived from the results of the $\textbf{P}_{\rm  MCLight}$ phase, whereas the $Shear$ parameter initializations are derived from the results of $\textbf{P}_{\rm  MCMass}$.}.
\label{table:PLMCGeom}
\end{table*}

\begin{table*}
\resizebox{\linewidth}{!}{
\centering
\begin{tabular}{ l | l | l  l l l l} 
 \multicolumn{1}{p{1.5cm}|}{\textbf{Model}} 
& \multicolumn{1}{p{1.5cm}|}{\textbf{Feature}} 
& \multicolumn{1}{p{1.1cm}}{\textbf{Parameters}} 
& \multicolumn{1}{p{1.1cm}}{} 
& \multicolumn{1}{p{1.1cm}}{} 
& \multicolumn{1}{p{1.1cm}}{} 
& \multicolumn{1}{p{1.1cm}}{} 
\\ \hline
& & & & & & \\[-4pt]
 $Hyper$       & Source Adaption   & $L_{\rm  Clust1}$ UP $^{0.0}_{\rm  10.0}$ & $L_{\rm  Clust2}$ UP $^{0.0}_{\rm  10.0}$ & $N_{\rm s}$ UP $^{80.0}_{\rm  800.0}$ & & \\[2pt]
\hline
& & & & & & \\[-4pt]
                & Contribution Maps & $\omega_{\rm Frac}$ UP $^{0.0}_{\rm  10.0}$ & & & & \\[2pt]
\hline
& & & & & & \\[-4pt]
                & Variance Scaling  & $\omega_{\rm  BG}$ UP $^{0.0}_{\rm  2.0}$ & $\omega_{\rm  Lens}$ UP $^{0.0}_{\rm  6.0}$ & $\omega_{\rm  Lens2}$ UP $^{0.0}_{\rm  6.0}$ & $\omega_{\rm  Src}$ UP $^{0.0}_{\rm  6.0}$ & $\omega_{\rm  Src2}$ UP $^{0.0}_{\rm  6.0}$ \\[2pt]
\hline
& & & & & & \\[-4pt]
                & Regularization & $\lambda_{\rm  Src}$ UP $^{0.0}_{\rm  1000.0}$ & $\lambda_{\rm  Bg}$ UP $^{0.0}_{\rm  1000.0}$ & $L_{\rm  Lum}$ UP $^{0.0}_{\rm  5.0}$ & & \\[2pt]
\hline
& & & & & & \\[-4pt]
                & Sky Subtraction & $\omega_{\rm  Sky}$ UP $^{-1.0}_{\rm  1.0}$ & & & & \\[2pt]
\end{tabular}
}
\caption{Priors used to initialize the hyper-parameter optimization. After the first hyper-parameter initialization, hyper-parameters are initialized using Gaussian priors with a width half their median value.}.
\label{table:PLHyper}
\end{table*}

\section{Parameter Table}

\begin{table*}
\begin{tabular}{ l | l } 
 \multicolumn{1}{p{2.0cm}|}{} 
& \multicolumn{1}{p{8.0cm}}{} 
 \\
 & \\[-20pt] 
\textbf{Lensing Quantities} & \\[2pt]
\hline
 & \\[-4pt] 
 $M_{\rm Ein}$ & Einstein mass \\[2pt]
 $R_{\rm Ein}$ & Einstein radius \\[2pt]
 $\vec{x}$ & Image-plane coordinate (Image-plane reference frame) \\[2pt]
 $\vec{x'}$ & Image-plane coordinate (Lens reference frame) \\[2pt]
 $\kappa (\vec{x})$ & Lens convergence profile \\[2pt]
 $\phi$ & Lens deflection potential \\[2pt]
 $\vec{{\alpha}}_{\rm x,y}$ & Deflection angle map (x and y dimensions) \\[2pt]
 $\Sigma_{\rm cr}$ & Critical surface mass density \\[2pt]
  & \\[-2pt] 
 \textbf{Light Profiles} & \\[2pt]
 \hline
 & \\[-4pt] 
 $x_{\rm l}, y_{\rm l}$  & Centers (arc seconds) [$Exp$, $Sersic$] \\[2pt]
 $\theta_{\rm l}$        & Rotation angle (clockwise from north) [$Exp$, $Sersic$] \\[2pt]
 $q_{\rm l}$             & Axis ratio [$Exp$, $Sersic$] \\[2pt]
 $\xi_{\rm l}$           & Elliptical coordinate ($\xi_{\rm l} = \sqrt{{x_{\rm l}}^2 + y_{\rm l}^2/q_{\rm l}^2}$) [$Exp$, $Sersic$] \\[2pt]
 $I_{\rm l}$             & Intensity (electrons per second) [$Exp$, $Sersic$] \\[2pt]
 $R_{\rm l}$             & Effective radius (circular) [$Exp$, $Sersic$] \\[2pt]
 $n_{\rm l}$             & Sersic index [$Exp$ ($n_{\rm l} = 1.0$), $Sersic$] \\[2pt]
 $k_{\rm l}$             & Function of Sersic index [$Exp$, $Sersic$] \\[2pt]
 & \\[-2pt]
 \textbf{Mass Profiles} & \\[2pt]
 \hline
  & \\[-4pt]
 $x, y$                  & Centers (arc seconds) [$SIE$, $SPLE$, $PL\textsubscript{Core}$] \\[2pt]
 $\theta$                & Rotation angle (clockwise from north)   [$SIE$, $SPLE$, $PL\textsubscript{Core}$] \\[2pt]
 $q$                     & Axis ratio [$SIE$, $SPLE$, $PL\textsubscript{Core}$] \\[2pt]
 $\xi$                   & Elliptical coordinate ($\xi = \sqrt{{x}^2 + y^2/q^2}$)   [$SIE$, $SPLE$, $PL\textsubscript{Core}$] \\[2pt]
 $\theta_{\rm Ein}$      & Einstein radius (arc seconds) [$SIE$, $SPLE$, $PL\textsubscript{Core}$] \\[2pt]
 $\alpha$                & Power-law density slope ($ \rho(r) = \rho_o (r / r_o)^{-\alpha}$. $\vec{{\alpha}}_{\rm x,y}$) [$SIE$, $SPLE$, $PL\textsubscript{Core}$] \\[2pt]
 $S$                     & Core radius (arc seconds) [$PL\textsubscript{Core}$] \\[2pt]
 $x_{\rm d}, y_{\rm d}$  & Centers (arc seconds) [$NFW$] \\[2pt]
 $\theta_{\rm d}$        & Rotation angle (clockwise from north) [$NFW$] \\[2pt]
 $q_{\rm d}$             & Axis ratio [$NFW$] \\[2pt]
 $\xi_{\rm d}$           & Elliptical coordinate ($\xi_{\rm d} = \sqrt{{x_{\rm d}}^2 + y_{\rm d}^2/q_{\rm d}^2}$) [$NFW$] \\[2pt]
 $\rho_{\rm s}$          & Halo scale normalization [$NFW$] \\[2pt]
 $r_{\rm s}$             & Halo scale radius ($r_{\rm s} = 30$ kpc) [$NFW$] \\[2pt] 
 $\kappa_{\rm d}$        & Halo normalization $\kappa_{\rm  d} = \rho_{\rm s} r_{\rm s} / \Sigma_{\rm  cr}$) [$NFW$] \\[2pt]
 $\eta_{\rm d}$          & Scaled elliptical coordinate ($\eta_{\rm d} = \xi_{\rm d} / r_{\rm s}$) [$NFW$] \\[2pt]
 $\Psi_{\rm l}$          & Mass-to-light ratio (electrons per second) [$Exp$, $Sersic$] \\[2pt]
 $x_{\rm sh}, y_{\rm sh}$& Centers (arc seconds) [$Shear$] \\[2pt]
 $\gamma_{\rm sh}$       & Magnitude [$Shear$] \\[2pt]
 $\theta_{\rm sh}$       & Rotation angle (clockwise from north) [$Shear$] \\[2pt] 
  & \\[-2pt]
 \textbf{Source Profiles} & \\[2pt]
 \hline
  & \\[-4pt]
 $x_{\rm s}, y_{\rm s}$  & Centers (Arc seconds) [$Exp$, $Sersic$] \\[2pt]
 $\theta_{\rm s}$        & Rotational angle (clockwise from north) [$Exp$, $Sersic$] \\[2pt]
 $q_{\rm s}$             & Axis ratio [$Exp$, $Sersic$] \\[2pt]
 $\xi_{\rm s}$           & Elliptical coordinate ($\xi_{\rm s} = \sqrt{{x_{\rm s}}^2 + y_{\rm s}^2/q_{\rm s}^2}$) [$Exp$, $Sersic$] \\[2pt]
 $I_{\rm s}$             & Intensity (electrons per second) [$Exp$, $Sersic$] \\[2pt]
 $R_{\rm s}$             & Effective radius (circular) [$Exp$, $Sersic$] \\[2pt]
 $n_{\rm s}$             & Sersic index [$Exp$ ($n_{\rm s} = 1.0$), $Sersic$] \\[2pt]
 $k_{\rm s}$             & Function of Sersic index [$Exp$, $Sersic$] \\[2pt]
 
\end{tabular}
\caption{Parameter symbols and descriptions for all parameters used in this work.}
\label{table:Parameters}
\end{table*}

\begin{table*}
\begin{tabular}{ l | l } 
 \multicolumn{1}{p{2.0cm}|}{} 
& \multicolumn{1}{p{8.0cm}}{} 
 \\
 & \\[-20pt] 
 \textbf{Semi-linear Inversion} & \\[2pt]
 \hline
  & \\[-4pt]
  $I$ & Total source pixels \\[2pt]
  $i$ & Source pixel number \\[2pt]
  $J$ & Total image pixels \\[2pt]
  $j$ & Image pixel number \\[2pt]
  $f_{\rm i,j}$ & Matrix mapping image pixels to source pixels \\[2pt]
  $\vec{d}$ & Observed image values in (electrons per second) \\[2pt]
  $\vec{\sigma}$ & Observed image statistical uncertainties (electrons per second) \\[2pt]
  $\vec{b}$ & Model lens light profile values \\[2pt]
  $\vec{s}$ & Model reconstructed source surface-brightness values \\[2pt]
  $\vec{D}$ & Observed image mapping vector used for linear inversion (See \citet{Warren2003}) \\[2pt]
  $\vec{F}$ & Image-source plane mapping matrix used for linear inversion (See \citet{Warren2003}) \\[2pt]
  $\vec{H_{\rm \Lambda}}$ & Regularization matrix \\[2pt]
  $\chi^2$ & Residuals over uncertainties squared \\[2pt]
  $\epsilon$ & Linear inversion bayesian evidence \\[2pt]
 & \\[-2pt] 
\textbf{Hyper Parameters} & \\[2pt]
\hline
 & \\[-4pt]
 $N_s$  & Source-plane resolution (number of source pixels) \\[2pt] 
 $L_{\rm Clust1}$, $L_{\rm Clust2}$ & Control the source pixelization \\[2pt]
 $\omega_{LensFrac}$, $\omega_{SrcFrac}$ & Control the lens and source contribution maps \\[2pt]
 $\omega_{BG}$ & Scale the background sky variances \\[2pt] 
 $\omega_{Lens1}$, $\omega_{Lens2}$ & Scale the lens light variances \\[2pt] 
 $\omega_{Src1}$, $\omega_{Src2}$ & Scale the lensed source variances \\[2pt]
 $\omega_{Sky}$ & Scale the background sky subtraction \\[2pt] 
 $\lambda$ & Regularization coefficient (constant regularization scheme) \\[2pt]
 $\lambda_{Src}$ & Source regularization coefficient (non-constant scheme) \\[2pt]
 $\lambda_{Bg}$ & Background regularization coefficient (non-constant scheme) \\[2pt] 
 $L_{\rm Lum}$ & Controls transition of non-constant regularization \\[2pt]
 & \\[-2pt] 
\textbf{Adaptive Image / Source Vectors} & \\[2pt]
\hline
 & \\[-4pt]
 $\vec{\Xi}$ & Preloaded model of lensed source (from previous pipeline phase) \\[2pt] 
 $\vec{\L}$ & Preloaded model of lens galaxy (from previous pipeline phase) \\[2pt]  
 $K$ & Number of image pixels allocated to a given source pixel \\[2pt] 
 $\vec{E}$ & Cluster energies used for k-means clustering \\[2pt]
 $\vec{r}$ & Distances of $K$ traced image pixels to allocated source pixel \\[2pt] 
 $\vec{W}$ & Weights of each source pixel, used for surface-brightness adaption \\[2pt] 
 $\vec{T}$ & Preloaded source and lens image $\vec{T} = \vec{\Xi} + \vec{\L}$ \\[2pt]  
 $\vec{\Omega_{Src}}$ & Lensed source flux contribution map \\[2pt]
 $\vec{\Omega_{Lens}}$ & Lens light flux contribution map \\[2pt] 
 $\vec{v}$ & Source flux contribution of each source-pixel (computed from $\vec{\Omega_{Src}}$) \\[2pt] 
 $\vec{V}$ & Weights of each source pixel, used for luminosity-weighted regularization \\[2pt]  
 $\vec{\Lambda}$, ($\lambda_{eff}$) & Effective regularization coefficients \\[2pt] 
 $f_{\rm Bg}$ & Background sky flux used for sky subtraction \\[2pt]
 $\vec{\sigma_{\rm base}}$ & Observed image variances without scaling (counts) \\[2pt]
 $\chi^2_{\rm base}$ & $\chi^2$ values generated using unscaled baseline variances \\[2pt]
 $\vec{\sigma_{\rm scale}}$ & Observed image variances including scaling (counts) \\[2pt]
 $\chi^2_{\rm scale}$ & $\chi^2$ Values generated using scaled variances \\[2pt]
\end{tabular}
\caption{Parameter symbols and descriptions for all parameters used in this work.}
\label{table:Parameters}
\end{table*}

\bibliography{library}            

\label{lastpage}

\end{document}